\pdfoutput=1

\newif\ifcameraready
\camerareadytrue

\newif\ifhidechange
\hidechangefalse

\newcommand{\thesisversionnum}[0]{10}

\newcommand{\thesistitlenormal}{Storage-Centric System Designs for Enabling Fast, Efficient, and Low-Cost Genomic and Metagenomic Analyses}
\newcommand{\thesisTitleFrontmatter}{STORAGE-CENTRIC SYSTEM DESIGNS\\FOR ENABLING FAST, EFFICIENT, AND LOW-COST\\GENOMIC AND METAGENOMIC ANALYSES}

\newcommand{\thesisTitlePlain}{\thesistitlenormal}
\newcommand{\thesisDissNumber}{31882}
\newcommand{\thesisAuthor}{Nika Mansourighiasi}
\newcommand{\thesisUni}{\protect{ETH Zurich}}

\newcommand{\thesisYear}{2026}
\newcommand{\thesisVersion}[0]{\theversion.0}

\documentclass[12pt,oneside,a4paper]{ethzthesis}

\usepackage{dblfloatfix}
\usepackage{multirow}
\usepackage{booktabs}
\usepackage{array}
\usepackage{xcolor}
\usepackage{algorithm}
\usepackage[noend]{algpseudocode}
\usepackage[rightComments=true,commentColor=gray]{algpseudocodex}
\usepackage{eucal}
\usepackage{datetime}
\usepackage{enumitem} %
\usepackage{tikz}
\usepackage{pifont}
\usepackage{graphicx}
\usepackage{fancyhdr}
\usepackage{datetime2}
\usepackage{fontawesome5} %
\usepackage{hyperref}
\usepackage{cleveref}
\usepackage{etoolbox}

\usepackage[colorinlistoftodos,prependcaption,textsize=scriptsize,backgroundcolor=orange!10, linecolor=orange, bordercolor=orange,textcolor=orange]{todonotes}

\usepackage[bottom]{footmisc}

\usepackage{pbox}

\newcolumntype{L}[1]{>{\raggedright\let\newline\\\arraybackslash\hspace{0pt}}m{#1}}
\newcolumntype{C}[1]{>{\centering\let\newline\\\arraybackslash\hspace{0pt}}m{#1}}
\newcolumntype{R}[1]{>{\raggedleft\let\newline\\\arraybackslash\hspace{0pt}}m{#1}}

\makeatletter
\def\thickhline{%
  \noalign{\ifnum0=`}\fi\hrule \@height \thickarrayrulewidth \futurelet
   \reserved@a\@xthickhline}
\def\@xthickhline{\ifx\reserved@a\thickhline
               \vskip\doublerulesep
               \vskip-\thickarrayrulewidth
             \fi
      \ifnum0=`{\fi}}
\makeatother
\newlength{\thickarrayrulewidth}
\newcommand{\circled}[1]{\tikz[baseline=(char.base)]{
           \node[shape=circle,draw,inner sep=0pt,fill=black, text=white] (char) {#1};}}

\definecolor{freakishgreen}{HTML}{0A982B}
\definecolor{urlblue}{HTML}{319dd6}
\hypersetup{
    colorlinks=true,
    linkcolor=black,
    filecolor=magenta,
    citecolor=violet,
    urlcolor=urlblue,
    pdfpagemode=FullScreen,
}

\crefname{section}{§\hspace{-2pt}}{§§}
\Crefname{section}{§}{§§}

\newcommand\kon[1]{}

\newcommand{\thesisversiontext}[0]{\textcolor{blue}{Thesis v\thesisversionnum~---~\today, \xxivtime \ UTC}}

\ifcameraready

\else
    \ifhidechange

    \fi

    \fancypagestyle{plain}
    {
    	\fancyhead[C]{\thesisversiontext}
    	\fancyfoot[C]{\thepage}
    }
    \fancypagestyle{headings}
    {
    	\fancyhead[C]{\thesisversiontext}
    	\fancyfoot[C]{\thepage}
    }
\fi

\newcommand\mganalysis{metagenomic analysis\xspace}

\definecolor{BLACK}{rgb}{0.0, 0.0, 0.0}

\definecolor{darkviolet}{rgb}{0.58, 0.0, 0.83}
\definecolor{darkpastelred}{rgb}{0.76, 0.23, 0.13}
\definecolor{flame}{rgb}{0.89, 0.35, 0.13}
\newcommand\tomi[1]{{\color{orange}{#1}}}
\newcommand\tomii[1]{{\color{magenta}{#1}}}

\newcommand\asp[1]{{\color{black}{#1}}}

\renewcommand\tomi[1]{{\color{orange}{#1}}}
\renewcommand\tomii[1]{{\color{magenta}{#1}}}
\definecolor{aoenglish}{rgb}{0.0, 0.5, 0.0}
\newcommand\tomiii[1]{{\color{aoenglish}{#1}}}

\newcommand\tomiv[1]{{\color{red}{#1}}}
\definecolor{cornflowerblue}{rgb}{0.39, 0.58, 0.93}

\definecolor{tiffanyblue}{rgb}{0.04, 0.73, 0.71}

\newcommand\aooo[1]{{\color{cornflowerblue}{#1}}}

\newcommand\omiii[1]{{\color{black}{#1}}}

\newcommand\irevminor[1]{{\color{black}{#1}}}

\renewcommand\todo[1]{{}}
\renewcommand\aooo[1]{{\color{black}{#1}}}
\renewcommand\tomi[1]{{\color{black}{#1}}}
\renewcommand\tomii[1]{{\color{black}{#1}}}
\renewcommand\tomiii[1]{{\color{black}{#1}}}
\renewcommand\tomiv[1]{{\color{black}{#1}}}

\newcommand{\citesbs}{bentley_accurate_2008,margulies_genome_2005,shendure_accurate_2005,harris_single-molecule_2008,turcatti_new_2008,wu_termination_2007,fuller_rapid_2007,mckernan_reagents_2008,fuller_method_2011}
\newcommand{\citesmrt}{eid_real-time_2009}
\newcommand{\citenanopore}{menestrina_ionic_1986,cherf_automated_2012,manrao_reading_2012,laszlo_decoding_2014,deamer_three_2016,kasianowicz_characterization_1996,meller_rapid_2000,stoddart_single-nucleotide_2009,laszlo_detection_2013,schreiber_error_2013,butler_single-molecule_2008,derrington_nanopore_2010,song_structure_1996,walker_pore-forming_1994,wescoe_nanopores_2014,lieberman_processive_2010,bezrukov_dynamics_1996,akeson_microsecond_1999,stoddart_nucleobase_2010,ashkenasy_recognizing_2005,stoddart_multiple_2010,bezrukov_current_1993,zhang_single-molecule_2024,cali2017nanopore}

\newcommand{\citesequencing}{\citesbs,\citesmrt,\citenanopore,stoler2021sequencing,goodwin2016coming,davis2021sequencerr,sereika2022oxford,jain2018nanopore,payne2018bulkvis,amarasinghe2020opportunities,hon2020highly,ni2023benchmarking,wenger2019accurate}

\newcommand{\citepersonalized}{alkan2009personalized,lightbody_review_2019,morganti_next_2019,branco_bioinformatics_2021,quazi_artificial_2022,aronson_building_2015,f_lochel_comparative_2020,papadopoulou_application_2023,tafazoli_applying_2021,gambardella_personalized_2020,leary_development_2010,hamburg_margaret_a_path_2010,van_der_lee_technologies_2020,moon_precision_2022,mohan_profiling_2020,chung_rapid_2020,bielinski_preemptive_2014,ho_enabling_2020,hussen_emerging_2022,russell_pharmacogenomics_2021,verma_nanopore_2024,clark2019diagnosis,farnaes2018rapid,sweeney2021rapid,flores2013p4,ginsburg2009genomic,chin2011cancer,Ashley2016,hansen2026complete}

\newcommand{\citeoutbreakrapid}{dunn2021squigglefilter,bertelli_rapid_2013,arias_rapid_2016,comin_investigation_2020,Quick2016}

\newcommand{\citeoutbreak}{\citeoutbreakrapid,robinson_genomics_2013,fournier_clinical_2014,koser_routine_2012,eloit_diagnosis_2014,gardy_jennifer_l_whole-genome_2011,taylor_angela_j_characterization_2015,quainoo_scott_whole-genome_2017,goldberg_brittany_making_2015,besser_interpretation_2019,li_application_2021,deng_integrated_2021,kwong_whole_2015,deurenberg_application_2017,tang_infection_2017,croucher_application_2015,
bloom2021massively,yelagandula2021multiplexed,le2013selected,nikolayevskyy2016whole,qiu2015whole,gilchrist2015whole}

\newcommand{\citeagriculture}{the_arabidopsis_genome_initiative_analysis_2000,zhu_applications_2020,choi_nanopore_2020,stevens_sequence_2016,campos_high_2021,gao_genome_2021,van_dijk_machine_2021,sun_twenty_2022,kim_application_2020,thudi_genomic_2021,michael_building_2020,shen_omics-based_2022,shahroodi2022demeter,
prasad2021soil,Mascher2024,Schreiber2024}

\newcommand{\citeevolution}{kanehisa_toward_2019,qing_whole_2022,wittkopp_cis-regulatory_2012,romero_comparative_2012,wang_population_2020,hill_molecular_2021,vaishnav_evolution_2022,zhang_haplotype-resolved_2021,fay_evaluating_2008,kanzi_next_2020,wray_evolution_2003,wu_one_2021,signor_evolution_2018,whitehead_variation_2006,coolon_tempo_2014,
ellegren2014genome,Prado-Martinez2013,Prohaska2019human}

\newcommand{\citecancer}{lawrence_mutational_2013,vogelstein_cancer_2013,ramskold_full-length_2012,baslan_unravelling_2017,shapiro_single-cell_2013,sakamoto_new_2020,jia_high-throughput_2022,lawson_tumour_2018,liu_mrna-based_2023,van_de_sande_applications_2023,chakravarty_clinical_2021,cortes-ciriano_computational_2022,deveson_evaluating_2021,xiao_toward_2021,bolton_cancer_2020,szustakowski_advancing_2021,navin_future_2011,hong_rna_2020,lei_applications_2021,han_single-cell_2022,federici_variants_2020,zhang_singlecell_2021,ren_understanding_2018,tian_cicero_2020,malone_molecular_2020,tang_single-cell_2019,ellsworth_single-cell_2017,zhong_application_2021,stadler_therapeutic_2021,tan_targeted_2022,degasperi_substitution_2022,xu_single-cell_2022,horak_comprehensive_2021,zhang_single-cell_2016,bruno_next_2020,de_luca_fgfr_2020,waarts_targeting_2022,lim_advancing_2020,colomer_when_2020,saadatpour_single-cell_2015,dizman_sequencing_2020,buzdin_rna_2020,xiao_tumor_2021,nandwani_lncrnas_2021,marchetti_error-corrected_2023,chen_next-generation_2021,navin_first_2015}

\newcommand{\citemetagenomicsegs}{hhrlich2011metahit,sunagawa2015structure,fierer2017embracing,danko2021global,paoli2022biosynthetic,edgar2022petabase,prasad2021soil,biodigs2026,Ryon2022,Zhu2025,human2012structure}

\newcommand{\citemgprecision}{kintz2017introducing,dixon2020metagenomics,mousa2024gut,tegegne2025g,virgin2011metagenomics,zhao2024application}

\newcommand{\citemgurgentclinical}{taxt2020rapid,GRUMAZ2020405,sweeney2021rapid,clark2019diagnosis,farnaes2018rapid,gu2021rapid,charalampous2024routine,Heitz2023,Alcolea-Medina2025,LIANG2023101898,Chien2022,Neyton2023,GENG202181,Ren2021}

\newcommand{\citemgbiodiversity}{afshinnekoo2015geospatial,hsu2016urban,sunagawa2015structure,danko2021global,biodigs2026,fierer2017embracing,Zhu2025}

\newcommand{\citemgcommunicable}{john2021next,nagy2021targeted,nieuwenhuijse2017metagenomic,downie2023surveillance}

\newcommand{\citemgtracing}{hadfield2018nextstrain,parkins2024wastewater,cdcpulsenet,gilchrist2015whole}

\newcommand{\citealgoptimization}{zhang2000greedy,slater2005automated,li2018minimap2,myers1999fast,marco2021fast,marcosola2023optimal,grootkoerkamp2024apa2,xin2013accelerating,xin2015shifted,tseng2025ultrafast,walia2025ultrafast,kim_fastremap_2022,Ashyralyyev2026gencore,ashyralyyev2026lcpan,alicioglu2024pairwise,kim2019airlift,alser2025taming}
\newcommand{\citemgalgoptimization}{lapierre2020metalign,koslicki2016metapalette,Marcelino2020,piro2016dudes,piro2020ganon,pockrandt2022metagenomic,wood2014kraken,kim2016centrifuge,wood2019improved,muller2017metacache,song2024centrifuger,Dilthey2019,Fan2021,karasikov2020metagraph,fan2023fulgor}

\newcommand{\citegraphalgoptimization}{rautiainen2020graphaligner,kim2019graph,gao2020abpoa,jain2019pasgal,siren2021pangenomics,Rautiainen2019,Chandra2023,Ivanov2022,Ma2023,Darby2020vargas,Hwang2025MEMO,Romain2023svjedi,Li2020minigraph}

\newcommand{\citehwoptimization}{doblas2025smx,mutlu2023accelerating,alser2022molecules,lou2020helix,lou2018brawl,shahroodi2023swordfish,markus2020benchmarking,subramaniyan2021accelerated,huangfu2018radar,khatamifard2021genvom,gupta2019rapid,li2021pim,angizi2019aligns,zokaee2018aligner,turakhia2018darwin,fujiki2018genax,madhavan2014race,cheng2018bitmapper2,houtgast2018hardware,houtgast2017efficient,zeni2020logan,ahmed2019gasal2,nishimura2017accelerating,de2016cudalign,liu2015gswabe,liu2013cudasw++,liu2009cudasw++,liu2010cudasw++,wilton2015arioc,goyal2017ultra,chen2016spark,chen2014accelerating,chen2021high,fujiki2020seedex,banerjee2018asap,fei2018fpgasw,waidyasooriya2015hardware,chen2015novel,rucci2018swifold,haghi2021fpga,li2021pipebsw,ham2020genesis,ham2021accelerating,wu2019fpga,cali2020genasm,Zhang_2023_alignerD,soysal2025mars,kim2018grim,kaplan2020bioseal,mao2022genpip,dphls2026,wang20202,Walia2024talco,sadasivan2024genomic,Turakhia2025toward,Turakhia2019darwinwga,simon2026processing,simon2025processing,diab_framework_2022,firtina2025enabling,senol2021accelerating,koliogeorgi2023hardware,Lindegger2023scrooge,eudine2026genpairx,Pavon2024quetzal,alonso2024bimsa,firtina2024aphmm,Doblas2023gmx,kang2026lembas}
\newcommand{\citemghwoptimization}{jia2011metabing,kobus2021metacache,wang2023gpmeta,kobus2017accelerating,Su2012,su2013gpumetastorms,Yano2014,saavedra2020mining,zhang2023genomix,cervi2022metagenomic,wu2021sieve,shahroodi2022krakenonmem,shahroodi2022demeter,dashcam23micro,hanhan2022edam,zou2022biohd,dunn2021squigglefilter,shih2023efficient}

\newcommand{\citegraphhwoptimization}{cali2022segram,Zhang2024Harp,Zeng2024asgdp,Shen2024128parallel,Li2024,Mandal2020,Varma2013,Awan2021,Feng2021,Zhang2025,kim2025nmp,Huang2023meg2,Angizi2020Panda,Qiu2017,Zhou2021,Sarkar2021,Varma2017,Varma2016,Goswami2018,Galanos2021,Angizi2020,Sinha2022,Meng2014,Hu2016,Chen2023,Natarajan2018,Ren2018,li2024rapid}

\newcommand{\citegraphacc}{milanese2019microbial,salzberg2016next,gihawi2023major,pockrandt2022metagenomic,berger2023navigating,Ackelsberg2015,Nasko2018,meyer2021critical,alser2020technology,alser2022molecules}

\newcommand{\citefilteroptimizationadditional}{singh2021fpga,kim20111,hameed2021alpha,guo2019hardware}

\newcommand{\citemapping}{alser2020accelerating, huangfu2018radar, cali2020genasm, turakhia2018darwin, fujiki2018genax, fujiki2020seedex, banerjee2018asap, khatamifard2021genvom, gupta2019rapid, li2021pim, angizi2019aligns, zokaee2018aligner, madhavan2014race, cheng2018bitmapper2, houtgast2018hardware,houtgast2017efficient, goyal2017ultra, chen2016spark, chen2014accelerating, chen2021high, zeni2020logan, ahmed2019gasal2, nishimura2017accelerating, de2016cudalign, liu2015gswabe, liu2013cudasw++, wilton2015arioc, fei2018fpgasw, waidyasooriya2015hardware, chen2015novel, rucci2018swifold, haghi2021fpga, li2021pipebsw, ham2020genesis, ham2021accelerating, wu2019fpga,doblas2025smx,kim2018grim,alser2020sneakysnake,bingol2021gatekeeper,xin2016optimal,mao2022genpip,mansouri2022genstore,xin2015shifted,alser2017gatekeeper,alser2019shouji,alser2017magnet,Zhang_2023_alignerD,cali2022segram,mutlu2023accelerating,alser2022molecules,subramaniyan2021accelerated,xin2013accelerating,kaplan2020bioseal,angizi2020pim}

\newcommand{\citeasm}{cali2020genasm,vsovsic2017edlib,alser2017gatekeeper,alser2017magnet,alser2019shouji,alser2020sneakysnake,kim2018grim,kim2019airlift,kim2024airlifttcbb,needleman1970general,smith1981identification,gotoh1982improved}

\newcommand{\citegc}{tavakkol2018flin,kim2020evanesco,cai2017error,park-dac-2016, park-dac-2019,kim2023decoupled,cai-insidessd-2018}
\newcommand{\citewearleveling}{chang2007efficient,cai2017error,cai-insidessd-2018}
\newcommand{\citeflashrefresh}{cai2013error,luo2018improving, cai2017error,cai2012flash,
cai2017vulnerabilities,luo2018heatwatch,cai2015read, ha2015integrated, cai-insidessd-2018,luo2015warm}

\newcommand{\citebasecalling}{cock2009sanger,alser2022molecules,lou2020helix,shahroodi2023swordfish,Samarakoon2023accelerated}

\newcommand{\citebasecallnanodnn}{cavlak2022targetcall,cavlak_targetcall_2024,xu2021fast,peresini2021nanopore,boza_deepnano_2017,boza_deepnano-blitz_2020,oxford_nanopore_technologies_dorado_2024,oxford_nanopore_technologies_guppy_2017,lv_end--end_2020,singh2024rubicon,zhang_nanopore_2020,xu_lokatt_2023,zeng_causalcall_2020,teng_chiron_2018,konishi_halcyon_2021,yeh_msrcall_2022,noordijk_baseless_2023,huang_sacall_2022,miculinic_mincall_2019}
\newcommand{\citebasecallnanohmm}{loman_complete_2015,david_nanocall_2017,timp_dna_2012,schreiber_analysis_2015}

\newcommand{\citehwmapping}{alser2020accelerating, huangfu2018radar, cali2020genasm, turakhia2018darwin, fujiki2018genax, fujiki2020seedex, banerjee2018asap, khatamifard2021genvom, gupta2019rapid, li2021pim, angizi2019aligns, zokaee2018aligner, madhavan2014race, cheng2018bitmapper2, houtgast2018hardware,houtgast2017efficient, goyal2017ultra, chen2016spark, chen2014accelerating, chen2021high, zeni2020logan, ahmed2019gasal2, nishimura2017accelerating, de2016cudalign, liu2015gswabe, liu2013cudasw++, wilton2015arioc, fei2018fpgasw, waidyasooriya2015hardware, chen2015novel, rucci2018swifold, haghi2021fpga, li2021pipebsw, ham2020genesis, ham2021accelerating, wu2019fpga,doblas2025smx,kim2018grim,alser2020sneakysnake,bingol2021gatekeeper,xin2016optimal,mao2022genpip,mansouri2022genstore,xin2015shifted,alser2017gatekeeper,alser2019shouji,alser2017magnet,Zhang_2023_alignerD,cali2022segram,mutlu2023accelerating,alser2022molecules,subramaniyan2021accelerated,xin2013accelerating,kaplan2020bioseal,angizi2020pim}

\newcommand{\citevariantcallers}{alkan_genome_2011,Sedlazeck2018,poplin_scaling_2018,weckx_novosnp_2005,kwok_comparative_1994,nickerson_polyphred_1997,marth_general_1999,Poplin2018,conrad_origins_2010,mills_mapping_2011,cooper_systematic_2008,eichler_widening_2006,cameron_comprehensive_2019,han_functional_2020,mandiracioglu_ecole_2024,dou_accurate_2020,smolka_detection_2024,lin_svision_2022,popic_cue_2023,narzisi_genome-wide_2018,zheng_symphonizing_2022,zhang_improved_2012,layer_lumpy_2014,karaoglanoglu_valor2_2020,jiang_long-read-based_2020,ahsan_nanocaller_2021,minoche_clinsv_2021,baird_rapid_2008,zarate_parliament2_2020,odonnell_mumco_2020,xu_smcounter2_2019,heller_svim_2019,pedersen_cyvcf2_2017,li_fermikit_2015,eisfeldt_tiddit_2017,zheng_svsearcher_2023,medvedev_detecting_2010,garrison_haplotype-based_2012,alser2022molecules,Olson2023,li2008mapping,Yu2015,Park2025,dunn2023n,wu2020high}

\newcommand{\citeassembly}{kececioglu_combinatorial_1995,Cheng2021,ekim_minimizer-space_2021,nurk_hicanu_2020,fleischmann_whole-genome_1995,myers_whole-genome_2000,Iqbal2012,chin_phased_2016,xiao_mecat_2017,chen_efficient_2021,li_genome_2024,jarvis2022semi,kolmogorov2019assembly,shafin_nanopore_2020,di_genova_efficient_2021,bankevich_multiplex_2022,Cheng2022,rautiainen2023telomere,ruan_fast_2020,cheng_scalable_2024,eche_bos_2023,vaser_time-_2021,chen_accurate_2021,pevzner_eulerian_2001,lin_assembly_2016,bonfield_new_1995,peltola_seqaid_1984,kamath_hinge_2017,butler_allpaths_2008,koren_canu_2017,myers_fragment_2005,
alser2022molecules,turakhia2018darwin,cali2017nanopore,li2016minimap,Haghshenas2020,Zimin2017hybrid,gupta2025accurate,firtina2020apollo}

\newcommand{\citetaxclassification}{wu2021sieve,wood2019improved,wood2014kraken,truong2015metaphlan2,kim2016centrifuge,song2024centrifuger,ounit2015clark,piro2016dudes,Fan2021,piro2020ganon,Marcelino2020,milanese2019microbial,lapierre2020metalign,pockrandt2022metagenomic,Dilthey2019,lemane2023kmindex,shen2022kmcp,sun2021challenges,lu2017bracken,koslicki2016metapalette,dimopoulos2022haystac,meyer2021critical}

\newcommand{\citemappairalign}{marco2021fast,Lindegger2023scrooge,marcosola2023optimal,baeza-yates_new_1992,myers1999fast,needleman1970general,smith1981identification,cali2020genasm,papamichail_improved_2009,suzuki_introducing_2018,waterman_biological_1976,wu_onp_1990,gotoh1982improved,wu_fast_1992,wagner_string--string_1974,sankoff_matching_1972,groot_koerkamp_exact_2024,sellers_theory_1974,ukkonen_algorithms_1985}

\newcommand{\citedp}{alser2020accelerating, alser2017gatekeeper, alser2017magnet, alser2019shouji, alser2020technology,cali2020genasm,kim2018grim,needleman1970general, smith1981identification, alser2020sneakysnake}

\newcommand{\citetightattachedSCC}{lee2020smartssd,ajdari2019cidr,Jeong2025upp,mahapatra2024instoragedomainspecificaccelerationserverless,Kang2024sting,An2023baraddur,Khadirsharbiyani2024smartgraph,Lee2024presto,lee2022smartsage,jang2024smart}

\newcommand{\citegpisp}{gu2016biscuit,kang2013enabling,acharya1998active,keeton1998case,riedel1998active,riedel2001active,tiwari2013active,tiwari2012reducing,boboila2012active,bae2013intelligent,torabzadehkashi2018compstor,kang2021iceclave,zou2022assasin,koo2017summarizer}

\newcommand{\citeispaiml}{li2023ecssd,liang2019ins,wang2024beacongnn,chen2025reis,Niu2024flashgnn,Pan2025instattention,mailthody2019deepstore,Mahapatra2025isp_rag}

\newcommand{\citeispgraphs}{Wang2024ndsearch,Zhang2025taijigraph,jun2018grafboost,matam2019graphssd}

\newcommand{\citesccgraphs}{Wang2024ndsearch,lee2022smartsage,Niu2024flashgnn,Khadirsharbiyani2024smartgraph,Zhang2025taijigraph,An2023baraddur,Kang2024sting,jun2018grafboost,matam2019graphssd}

\newcommand{\citesscotherapps}{liang2019cognitive,kim2020reducing,lim2021lsm,li2021glist,wang2016ssd,lee2020neuromorphic,wang2018three,han2019novel,choi2020flash,pei2019registor,do2013query,seshadri2014willow,kim2016storage,jeong2019react,Wong2025anvil,mahapatra2024instoragedomainspecificaccelerationserverless}

\newcommand{\citeispotherapps}{liang2019cognitive,kim2020reducing,lim2021lsm,li2021glist,wang2016ssd,pei2019registor,do2013query,seshadri2014willow,kim2016storage,jeong2019react}

\newcommand{\citeispgpu}{cho2013xsd}

\newcommand{\citeispfpga}{jun2015bluedbm, jun2016bluedbm, torabzadehkashi2019catalina}

\newcommand{\citeifp}{chun2022pif,chen2024search,lee2025aif,Sun2025lincoln,Yu2024cambriconllm,Kim2023optimstore,choi2020flash}

\newcommand{\citeufp}{gao2021parabit,park2022flash,chun2024rif,Chen2024aresflash,Wong2025anvil,han2019novel,wang2018three,kang2021s,lee2020neuromorphic,kabra2025ciphermatch,kim2025crossbit}

\newcommand{\citenongenomegraphisp}{Wang2024ndsearch,lee2022smartsage,Niu2024flashgnn,Khadirsharbiyani2024smartgraph,Zhang2025taijigraph,An2023baraddur,Kang2024sting,jun2018grafboost,matam2019graphssd}

\newcommand{\citenongraphgenomeisp}{mansouri2022genstore,abakus23taco,megis,jun2016storage,kim2025nmp,soysal2025mars,zheng2025storage}

\newcommand{\citeflash}{micheloni2010inside,cai-insidessd-2018,meza_revisiting_2015,meza_case_2013,meza2015large,cai_data_2015,cai_error_2012,cai2017error,cai2013error,cai_program_2013,cai2018errorsarxiv,cai_threshold_2013,cai2017vulnerabilities,cai2012flash,luo2015warm,luo2018improving,cho2024aero,nadig2026conduit,nadig2023venice,kim2020evanesco,park-dac-2016,park-nvmsa-2018,park2021reducing,kim-dac-2017,tavakkol2018flin,cai2015read,ha2015integrated,wang2014enhanced,zhao2013ldpc,dong2010use,kim2021performance,kim2018ssdcheck,kim2017ssd,cai2014neighbor,luo2016enabling,park2026experimental,shim2019exploiting,chun2026straw,kim2024norns}

\newcommand{\citebasecallsbs}{kao_naivebayescall_2010,cacho_base-calling_2018,erlich_alta-cyclic_2008,wang_adaptive_2017,rougemont_probabilistic_2008,shen_particlecall_2012,ji_bm-bc_2012,das_base_2013,kircher_improved_2009,massingham_all_2012,ye_blindcall_2014,renaud_freeibis_2013,das_onlinecall_2012,menges_totalrecaller_2011,bravo_model-based_2010,kao_bayescall_2009}

\newcommand{\citesignalanalysis}{Bao2021Squigglenet,loose_real-time_2016,zhang2021real,kovaka2020targeted,senanayake_deepselectnet_2023,sam_kovaka_uncalled4_2024,lindegger_rawalign_2024,firtina2023rawhash,firtina2023rawhash2,firtina_rawsamble_2024,shih2023efficient,sadasivan2023rapid,dunn2021squigglefilter,shivakumar_sigmoni_2024,sadasivan_accelerated_2024,gamaarachchi_gpu_2020,samarasinghe_energy_2021,rawbench}

\newcommand{\citesinglecell}{tang2009mrnaseq,angerer2017single,lahnemann2020eleven,wen2022singlecell,ni2023benchmarking,trapnell2014dynamics,wolf2018scanpy,stuart2019comprehensive,hafemeister2019normalization,baslan_unravelling_2017,shapiro_single-cell_2013,ramskold_full-length_2012,jia_high-throughput_2022,lawson_tumour_2018,van_de_sande_applications_2023,navin_future_2011,lei_applications_2021,han_single-cell_2022,ren_understanding_2018,tang_single-cell_2019,ellsworth_single-cell_2017,xu_single-cell_2022,zhang_single-cell_2016,saadatpour_single-cell_2015,navin_first_2015}

\newcommand{\citetranscriptomics}{wang2009rna,lowe2017transcriptomics,angerer2017single,Stark2019,chen2023hitchhikers,weirather2017comprehensive,Sibbesen2023,haas_forensic_2021,liu_desalt_2019,lachmann2018massive,clough2023ncbigeo,bray2016near}

\newcommand{\citeauditandprivacy}{Pattengale2020Decentralized,ma2020efficient,bonomi2020privacy,akgun2015privacy}

\newcommand{\citeerrorcorrection}{firtina_hercules_2018,xiao_mecat_2017,stanojevic_telomere--telomere_2024,salmela_lordec_2014,kang_hybrid-hybrid_2023,holley_ratatosk_2021,morisse_hybrid_2018,wang_fmlrc_2018,zhu_lcat_2023,salmela_accurate_2017,bao_halc_2017,haghshenas_colormap_2016,goodwin_oxford_2015,hackl_proovread_2014,hu_lscplus_2016,salmela_correcting_2011,schroder_shrec_2009,salmela_correction_2010,koren_hybrid_2012,au_improving_2012}

\newcommand{\citepum}{chang2016low,seshadri2017ambit,hajinazarsimdram,seshadri2013rowclone,seshadri2019dram,seshadri2016processing,seshadri.bookchapter17,seshadri2016buddy,seshadri2015fast,angizi2019graphide,ferreira2021pluto,mimdramextended,missingnot,yuksel2024simultaneous,olgun2022pidram,angizi2018pima,angizi2018cmp,angizi2019aligns,levy.microelec14,kvatinsky.tcasii14,Shafiee2016,kvatinsky.iccd11,kvatinsky.tvlsi14,gaillardon2016plim,bhattacharjee2017revamp,hamdioui2015memristor,xie2015fast,hamdioui2017myth,yu2018memristive,yavits2021giraf, xi2020memory, zheng2016tcam, truong2021racer,li2017drisa, truong2022adapting,ma20232,slesazeck20192tnc,wang20211t2c,aga2017compute,eckert2018neural,dualitycache,kang2014energy,de2025proteus,tokuda2026clutch,yuksel2025pudhammer,tokuda2026pudghost,
deng2018dracc,Chi2016,xin2020elp2im,Song2018graphr,song2017pipelayer,gao2019computedram,Besta2021SISA,seshadri2018rowclone,li2016pinatubo,imani2019floatpim,he2020sparse,olgun2021quactrng,kim2019d,
kim2018dram,bostanci2022dr,ali2019memory,li2018scope,subramaniyan2017parallel,zha2020hyper,fujiki2018memory,orosa2021codic,sharad2013ultra,rezaei2020nom,simon2020blade,nag2019gencache,wang2019bit,al2020towards,kim2021colonnade,jiang2020c3sram,jeloka201628,wang2023infinity,kang2015energy,imani2020dual,deng2019lacc,sutradhar2021look,sutradhar2020ppim,peng2023chopper,shahroodi2023swordfish,sudarshan2022fefet,sudarshan2022weighted,sudarshan2022optimization,sudarshan2021novel}

\newcommand{\revmark}[1]{}

\newcommand{\squishlist}{
 \begin{list}{$\circ$}
  { \setlength{\itemsep}{0pt}
     \setlength{\parsep}{0pt}
     \setlength{\topsep}{3pt}
     \setlength{\partopsep}{0pt}
     \setlength{\leftmargin}{1em}
     \setlength{\labelwidth}{1em}
     \setlength{\labelsep}{0.5em} } }

\newcommand{\squishend}{
  \end{list}  }

\newcommand{\atb}[1]{\textcolor{black}{#1}}
\newcommand{\damlaa}[1]{{\color{black}#1}}

\algrenewcommand\algorithmicrequire{\textbf{Input:}}
\algrenewcommand\algorithmicensure{\textbf{Output:}}

\definecolor{darkspringgreen}{rgb}{0.09, 0.45, 0.27}
\definecolor{denim}{rgb}{0.08, 0.38, 0.74}
\definecolor{darkolivegreen}{rgb}{0.33, 0.42, 0.18}
\definecolor{tangerine}{rgb}{0.95, 0.52, 0.0}
\definecolor{mahogany}{rgb}{0.75, 0.25, 0.0}
\definecolor{coolblack}{rgb}{0.0, 0.18, 0.39}
\definecolor{darkpink}{rgb}{0.91, 0.35, 0.6}
\definecolor{darkblue}{rgb}{0.0, 0.0, 0.67}

\definecolor{seagreen}{rgb}{0.18, 0.55, 0.34}

\definecolor{pred}{rgb}{0.7843, 0.0039, 0.3137} 
\newcommand{\js}[1]{{\color{pred}{#1}}} %

\definecolor{darkpink}{rgb}{0.88, 0.28, 0.54}
\definecolor{forestgreen}{rgb}{0.0, 0.27, 0.13}
\definecolor{amber}{rgb}{1.0, 0.49, 0.0}

\newcommand{\joel}[1]{{\color{cyan}#1}}

\newcommand{\inum}[1]{(\textit{#1})\xspace}

\newcommand{\sects}[1]{{Sections~#1}\xspace} %
\newcommand{\sect}[1]{{Section~#1}\xspace} %
\newcommand{\head}[1]{{\vspace{0.5em}\noindent\textbf{#1.}\xspace}} %
\newcommand{\figs}[1]{{Figures~#1}\xspace} %
\newcommand{\fig}[1]{{Figure~#1}\xspace} %

\newcommand\base{\textsf{Base}\xspace}
\newcommand\swf{\textsf{SW-filter}\xspace}
\newcommand\acc{\textsf{ACC}\xspace}
\newcommand\isf{\textsf{Ideal-ISF}\xspace}
\newcommand\iof{\textsf{Ideal-OSF}\xspace}
\newcommand\isfacc{\textsf{Ideal-ISF+ACC}\xspace}
\newcommand\ssdl{\texttt{SSD-L}\xspace}
\newcommand\ssdm{\texttt{SSD-M}\xspace}
\newcommand\ssdh{\texttt{SSD-H}\xspace}
\newcommand\dram{\texttt{\omciv{DRAM}}\xspace}

\newcommand\gs{\textsf{GS}\xspace}
\newcommand\gsos{\textsf{GS-Ext}\xspace}
\newcommand\simd{\textsf{SIMD}\xspace}

\newcolumntype{Y}{>{\centering\arraybackslash}X}
\usepackage{tikz}

\DeclareRobustCommand\wcirc[1]{\tikz[baseline=(char.base)]{           \node[shape=circle,draw,inner sep=0pt,fill=white, text=black] (char) {#1};}}

\usepackage{blindtext,graphicx}
\usepackage[absolute]{textpos}
\usepackage{datetime}

\definecolor{seagreen}{rgb}{0.18, 0.55, 0.34}
\definecolor{ballblue}{rgb}{0.13, 0.67, 0.8}
\newcommand\omc[1]{{{#1}}}
\newcommand\omcc[1]{{{#1}}}
\newcommand\omccc[1]{{{#1}}}
\newcommand\omciv[1]{{{#1}}}
\newcommand\omcv[1]{{{#1}}}
\newcommand\omcvi[1]{{{#1}}}
\newcommand\omcvii[1]{{{#1}}}
\newcommand\omcviii[1]{{{#1}}}
\newcommand\omcix[1]{{{#1}}}
\newcommand\omcx[1]{{{#1}}}
\newcommand\jsr[1]{{{#1}}}
 \newcommand\rev[1]{{{#1}}}
 \newcommand\revp[1]{{{#1}}}
 \newcommand\jsiv[1]{{{#1}}}
 \newcommand\jk[1]{{{#1}}}

\definecolor{darkgreen}{rgb}{0.0, 0.44, 0.34}
 
 \newcommand\ssdc{\texttt{SSD-C}\xspace}
 \newcommand\ssdp{\texttt{SSD-P}\xspace}
 
\sloppypar

\newcommand\hm[1]{{\color{violet}{#1}}}

\newcommand\randomio{\textsf{Random-Qry}\xspace}
\newcommand\streamio{\textsf{Stream-Qry}\xspace}

\definecolor{dollarbill}{rgb}{0.52, 0.73, 0.4}

\newcommand\cmashopt{\textsf{KSS}\xspace}

\renewcommand\omc[1]{{\color{black}{#1}}}
\renewcommand{\js}[1]{{\color{pred}{#1}}} %
\renewcommand{\joel}[1]{{\color{black}#1}}
\renewcommand\hm[1]{{\color{denim}{#1}}}

\usetikzlibrary{patterns}
\usepackage[precision=2, unit=mm]{lengthconvert}

\definecolor{cyan(process)}{rgb}{0.0, 0.62, 0.82}

\renewcommand{\js}[1]{{\color{pred}{#1}}} %
\renewcommand\hm[1]{{\color{olive}{#1}}}

\newcommand\new[1]{{\color{blue}{#1}}}
\newcommand\proposal{SAGe\xspace}
\newcommand\proposals{SAGe's}

\renewcommand\new[1]{{\color{black}{#1}}}

\renewcommand{\js}[1]{{\color{black}{#1}}} %
\renewcommand\hm[1]{{\color{black}{#1}}}
\definecolor{cadmiumgreen}{rgb}{0.0, 0.50, 0.29}
\newcommand\irev[1]{{\color{cadmiumgreen}{#1}}}

\newcommand\revref[1]{\hyperref[rev:#1]{#1}}

\definecolor{dollarbill}{rgb}{0.52, 0.73, 0.4}
\definecolor{olive}{rgb}{0.5, 0.5, 0.0}
\definecolor{green(munsell)}{rgb}{0.0, 0.66, 0.47}
\definecolor{green(ryb)}{rgb}{0.4, 0.69, 0.2}
\definecolor{kellygreen}{rgb}{0.3, 0.73, 0.09}

\definecolor{acolor}{rgb}{0.0, 0.5, 1.0}
\definecolor{bcolor}{rgb}{0.54, 0.17, 0.89}
\definecolor{ccolor}{rgb}{0.4, 0.69, 0.2}
\definecolor{dcolor}{rgb}{0.92, 0.41, 0.12}
\definecolor{ecolor}{rgb}{0.6, 0.0, 0.156}

\definecolor{raspberry}{rgb}{0.89, 0.04, 0.36}

\definecolor{awesome}{rgb}{1.0, 0.13, 0.32}
\definecolor{cardinal}{rgb}{0.77, 0.12, 0.23}
\definecolor{cadet}{rgb}{0.33, 0.41, 0.47}
\definecolor{celadon}{rgb}{0.67, 0.88, 0.69}
\definecolor{persianblue}{rgb}{0.11, 0.22, 0.73}
\definecolor{ultramarine}{rgb}{0.07, 0.04, 0.56}
\definecolor{warmblack}{rgb}{0.0, 0.3, 0.3}

\newcommand\micro[1]{{\color{persianblue}{#1}}}
\newcommand\omcm[1]{{\color{orange}{#1}}}

\renewcommand\micro[1]{{\color{black}{#1}}}
\renewcommand\omcm[1]{{\color{black}{#1}}}

\newcommand\nh[1]{{\color{black}{#1}}}
\definecolor{tealish}{RGB}{0,187,161}
\definecolor{deepred}{RGB}{192,0,0}

\newcommand{\circg}[1]{\tikz[baseline=(char.base)]{
           \node[shape=circle,draw=none,inner sep=0.01pt,fill=tealish!100, text=white] (char) {#1};}}

\newcommand{\circr}[1]{\tikz[baseline=(char.base)]{
           \node[shape=circle,draw=none,inner sep=0.01pt,fill=deepred!100, text=white] (char) {#1};}}

\definecolor{lavenderrr}{rgb}{0.71, 0.49, 0.86}

\newcommand{\sgprop}[1]{
\colorbox{lavenderrr!25}{\textit{\textbf{(Property {#1})}}}}

\definecolor{burgundy}{rgb}{0.5, 0.0, 0.13}

\definecolor{dogwoodrose}{rgb}{0.84, 0.09, 0.41}

\newcommand\omcr[1]{{\color{black}{#1}}}

\definecolor{turquoise}{rgb}{0.19, 0.84, 0.78}
\definecolor{brightturquoise}{rgb}{0.03, 0.91, 0.87}
	\definecolor{cinnamon}{rgb}{0.82, 0.41, 0.12}
    
\newcommand\oii[1]{{\color{black}{#1}}}
\definecolor{azure}{rgb}{0.0, 0.5, 1.0}
\newcommand\oiii[1]{{\color{black}{#1}}}
\definecolor{emerald}{rgb}{0.20, 0.65, 0.38}

\newcommand\oiv[1]{{\color{black}{#1}}}
\definecolor{amethyst}{rgb}{0.85, 0.57, 0.0}
\definecolor{carminered}{rgb}{1.0, 0.0, 0.22}
\definecolor{harvestgold}{rgb}{0.85, 0.57, 0.0}
\definecolor{electricviolet}{rgb}{0.56, 0.0, 1.0}
\definecolor{ochre}{rgb}{0.8, 0.47, 0.13}

\newcommand\ov[1]{{\color{black}{#1}}}
\definecolor{cornflowerblue}{rgb}{0.39, 0.58, 0.93}
\definecolor{darkturquoise}{rgb}{0.0, 0.7, 0.7}
\newcommand\ovi[1]{{\color{black}{#1}}}
\makeatletter
\newcommand\requiredelimiter[2][########]{%
  \ifdefined#2%
    \def\@temp{\def#2#1}%
    \expandafter\@temp\expandafter{#2}%
  \else
    \@latex@error{\noexpand#2undefined}\@ehc
  \fi
}
\@onlypreamble\requiredelimiter
\makeatother

\begin{document}
\frenchspacing
\raggedbottom
\selectlanguage{english}
\pagenumbering{roman}
\pagestyle{plain}

\setbiblabelwidth{1000} %

\begin{titlepage}
    \large
    \begin{center}
        
        \ifcameraready
        \else
        \fi
        
        \begingroup
        \MakeUppercase{Diss. ETH No.}
        \thesisDissNumber{}
        \endgroup
    
        \hfill

        \vfill

        \begingroup
            \textbf{\thesisTitleFrontmatter}
        \endgroup

        \vfill

        \begingroup
            A thesis submitted to attain the degree of\\
            \vspace{0.5em}
            \MakeUppercase{Doctor of Sciences}\\
            \vspace{0.5em}
            (Dr. sc. \thesisUni) \\
            
        \endgroup

        \vfill

        \begingroup
            presented by\\
            \vspace{1.5em}
            Nika Mansourighiasi\\
            \vspace{0.5em}
            born on 03.06.1993
        \endgroup

        \vfill

        \begingroup
            \vspace{2em}
            accepted on the recommendation of\\
            \vspace{1em}
            Prof.\ Dr.\ Onur Mutlu, examiner\\
            \vspace{0.5em}
            Prof. Dr. Can Alkan, co-examiner \\
            \vspace{0.5em}
            Prof. Dr. Reetuparna Das, co-examiner \\
            \vspace{0.5em}
            Prof. Dr. Wen-mei Hwu, co-examiner \\
            \vspace{0.5em}
            Prof. Dr. Jangwoo Kim, co-examiner \\
            \vspace{0.5em}
            Prof. Dr. Yatish Turakhia, co-examiner \\
            
        \endgroup

        \vfill

        \thesisYear%

        \vfill
    \end{center}
\end{titlepage}

\thispagestyle{empty}

\hfill

\vfill

\noindent\thesisAuthor: \textit{\thesisTitlePlain,}
\textcopyright\ \thesisYear

\setstretch{1.3}
\thispagestyle{empty}

\vspace*{\fill}

\begin{center}
\emph{Expanding the boundaries of science is challenging in itself.\\
This dissertation is dedicated to\\
the resilient students and researchers
who push science forward\\
while navigating additional challenges beyond their scientific endeavors, such as\\
illness, poverty, bureaucratic hurdles, and the consequences of conflicts not of their making.}

\end{center}

\vspace*{\fill}

\setstretch{1}

\clearpage

\chapter*{Acknowledgments}
\addcontentsline{toc}{chapter}{Acknowledgments}

This PhD has been a challenging yet extremely rewarding journey for me, in which I not only worked to push the boundaries of science but also expanded the boundaries of my own capabilities.
This thesis is a snapshot of part of my research journey so far.
I will take this opportunity to thank the people who shared their knowledge, support, and kindness with me along the way. 

I first learned about computer architecture while following my interests in both engineering and art. While pursuing computer architecture further, I came across Professor~Onur Mutlu's online lectures as an embodiment of this intersection between engineering and art. Starting from the first lecture, Onur introduced computer architecture through analogies between ``how computers and computer architects evolve'' and ``how art and artists evolve.'' I immediately knew that this was someone from whom I wanted to learn much more. I am incredibly happy and grateful that I eventually had the opportunity to pursue my PhD with Onur. I arrived in Onur's group with a love for computer architecture and art, and I was pleasantly surprised by everything else I additionally encountered: freedom, borderless curiosity, courage, genuine excitement, strong rigor and care for the craft,  inclusion, and diversity. Now, looking back, I realize it was only natural, as these are the very same qualities that both art and science nurture. I extend my deepest gratitude to Onur for creating such a beautiful environment for learning, exploration, and growth. It has been an immense experience and honor to have him as a teacher, and I hope to continue learning from him throughout the rest of my journey.     

I am deeply grateful to my close mentors and collaborators. I thank Professor~Jisung Park for his strong technical mentorship, friendship, and for believing in me at times I could not do so myself. I am grateful to Dr.~Mohammad Sadrosadati for his profound insights and interesting discussions in computer architecture and many other realms. 
I extend my deep gratitude to Professor~Subhasish Mitra for advising me during my visiting research period in his group at Stanford University and for his continual guidance in pursuing bold and ambitious research topics. 
I am grateful to Professor~Nandita Vijaykumar for sharing deep insights about research and being a researcher, which played key roles in my academic journey. I am thankful to Dr.~Juan Gomez Luna for his mentorship during my master's studies and his great attention to technical details. I am deeply grateful to Professor~Zain Navabi, who was among the first to introduce me to computer architecture as an intersection of engineering and art during my undergraduate studies at the University of Tehran. 

I also extend my sincere gratitude to my doctoral exam committee members: Professor~Can Alkan, Professor~Reetuparna Das, Professor~Wen-mei Hwu, Professor~Jangwoo Kim, and Professor~Yatish Turakhia. Discussions with them and their constructive feedback greatly strengthened this thesis and also sparked interesting ideas for future research. I thank Professor~Morteza Aramesh for chairing my doctoral exam.

I thank all current and past members of the SAFARI Research Group for creating such a stimulating, inclusive, and family-like environment for doing research and experiencing life. I am especially thankful to my PhD buddies, Konstantinos Kanellopoulos and Rahul Bera. We started our PhDs at roughly the same time and had the chance to grow together as researchers and friends. Their friendship, support, genuine love for their research craft, and creativity made this long journey not only possible, but also pleasant. I thank Geraldo Francisco de Oliveira Junior for his pure heart and the light and joy he emanates to those around him. I learned a lot from him about processing-in-memory, gym exercises, and the joy of creating and sharing beautiful things (e.g., birthday cakes). I thank Can Firtina for many interesting collaborations, details about k-mers and methylation, and his sense of humor. I thank\linebreak Abdullah Giray Yaglikci, Nisa Bostanci, Ataberk Olgun, Sahand Divsalar, Haiyu Mao, and Ismail Yuksel for their pure kindness and friendship. 
I thank Lois Orosa and Yaohua Wang for their kindness, friendship, and wisdom in the early parts of my journey in SAFARI.
I thank Rakesh Nadig, Damla Senol Cali, Jeremie Kim, Konstantina Koliogeorgi, Rachata Ausavarungnirun, and Banu Cavlak for great collaborations and discussions. I also sincerely thank my great mentees for what they achieved and for teaching me how to teach. I am especially thankful to Talu Güloglu, whose hard work and strong motivation (along with frequently humming joyful Christmas songs at any time of the year) played a key role in realizing some of the mechanisms introduced in this thesis,  and to Eirini Tzermpou, Marc Rautmann, Timur Eke, and Wonyoung Cheon for the great time we have had working together so far. I also extend my deep gratitude to other past and present SAFARI members, 
especially 
Minesh Patel, Hasan Hassan, Christina Giannoula, Harshita Gupta, Mayank Kabra, Andreas Kosmas Kakolyris, Konstantinos Sgouras, Maria Makeenkova, 
Haocong Luo, Gagandeep Singh, and Zhiheng Yue. 
I thank Tracy Ewen for her kindness, wisdom, and help with many challenging administrative matters. I also thank Tulasi Blake for her administrative support. 

Beyond the lab, I am incredibly fortunate to have amazing friendships that not only supported me immensely throughout my PhD but also deeply enriched my experience of life. I thank Mehrasa for being such a pure, bright, and beautiful expression of friendship in this world. I feel extraordinarily lucky to experience a friendship like ours. Her unwavering presence, even from another continent, has been a key source of my persistence through the ups and downs of research and life.
I sincerely thank Yasmin for her kindness, patience, and for teaching me how to study deeply. I am grateful for our journey of studying, growth, and friendship. I thank Hashem for introducing me to computer architecture, many poems, and how computer architecture relates to some of those poems. I find his courage to wonder and his love to learn inspiring, in both research and life. I thank Julia for her joyful kindness, lively creativity, and for sharing her wisdom and support during some of the most difficult times of my life. I am also grateful to the avocado that initially sparked our friendship. 
I thank\linebreak Pascal. Without his presence, I might never have started my PhD journey; had I started, I might not have finished it; and had I finished it, I certainly would not have explored life with as much depth along the way.
I am grateful to Poulami for her positivity, inspiring strength, and generous advice, ranging from research journey in computer architecture to the housing market in Zurich. 
I thank Vesna, Michael, Aditi, and Niloofar, who taught me that friendship and profound connections persist through time and distance, even when life does not contain situations for us to meet so frequently. 

Most importantly, I thank my family. I am grateful to my mother, Mehrangiz, for everything. Her name in Persian means ``the one who inspires kindness and love,'' and I think it is such a fitting name. 
Like rays of sunshine, her love, kindness, strength, and wisdom bring light and warmth into my life. Yet unlike sunshine, which comes and goes, her light and warmth have always been there: constant, unwavering, and present. I thank my father, Amir, for supporting me throughout my education and for the experiences that led to me gaining a deeper understanding of myself. I thank my lovely brother, Ali, for his kindness and for tolerating me, as I was busy studying since the moment he opened his eyes to this world. While I hope to keep studying and learning in the rest of my journey, I dearly hope and will do my best to be a lot more present in our connection. I am grateful to my late grandmother, Narges, and Khaleh~Golshan for their pure and strong love. They showed me that true love does not die, and such a fundamental understanding this is. I thank my uncle~Masoud, aunt~Mina, and cousin~Mana for making me feel at home during my undergraduate studies in Tehran (particularly during the snowy seasons). I thank my cousins Arash and Azadeh for their kindness and support through the ups and downs of my life these past years. I thank Arash for introducing me to mindfulness, whose impact probably deserves its own chapter, if not a book. I am grateful to my husband's family, Tuncel, Niyazi, and Hazal, who supported me strongly and kindly throughout my PhD journey. 

Finally, I thank my husband, Harun, who is my best friend and collaborator in life and research. I am thankful for his pure kindness, patience, knowledge, imagination, morality, and for teaching me about what I consider the most important element in life:\linebreak \emph{unconditional love}.

\clearpage
\chapter*{\vspace{-2em}Abstract\vspace{-1em}}
\addcontentsline{toc}{chapter}{Abstract}

\emph{Genomic analysis} examines the genomic information of living organisms and other biological entities.
To analyze this information computationally, 
a sample of a biological entity's genetic material, typically DNA or RNA (that has been reverse-transcribed into DNA), undergoes a process called \emph{sequencing}.
Sequencing converts the information in DNA molecules to sequences of digital data. Traditional genomics focuses on the genome sequences of an individual (or a small group of individuals) from a single known species, whereas \emph{metagenomic analysis} examines genome sequences from multiple species in a shared environment.
Genomic and metagenomic analyses play critical roles in many fields, such as precision medicine, 
urgent clinical settings, and discovering early warnings of communicable diseases, ensuring food safety through pathogen monitoring, agriculture, and scientific discovery.
The adoption of genomic and metagenomic analyses has been rapidly increasing in recent years, driven by their critical importance and the rapid advancement of sequencing technologies. These factors have led to exponential growth in sequence data generation. Due to the challenges of analyzing and storing massive volumes of this data, significant efforts have been made in accelerating (meta)genomic analyses and storing sequence data in compressed forms. Despite the benefits of these techniques in improving the analysis and storage of large-scale sequence data, we identify two major outstanding problems in accessing stored sequence data and feeding it to the analysis units: \inum{i} the overhead of moving large amounts of low-reuse data from the storage system and the unnecessary burden on the rest of the system (e.g., main memory and computation units), and \inum{ii} the data preparation bottleneck, where compressed sequence data needs to be first decompressed and formatted before it can be analyzed.

In this dissertation, we aim to \inum{i}~alleviate the data movement overheads of genomics and metagenomics analyses from the storage system and reduce their overall computational burden, and \inum{ii}~mitigate the data preparation bottleneck while achieving high performance, energy efficiency, and compression ratios. We aim for our proposed designs to be lightweight so that they can seamlessly integrate with a broad range of (meta)genomic analysis systems.

To alleviate the overheads of moving large amounts of low-reuse (meta)genomic data from the storage system and reduce the overall computational burden, we propose \emph{storage-centric computing} (SCC) systems for (meta)genomics. SCC refers to processing data inside the storage device, either on the SSD controller or on the flash dies. SCC can be a fundamental solution by processing data where it originally resides. 
To mitigate the data preparation bottleneck, we propose an algorithm-architecture co-design for highly-compressed storage and high-performance access of sequence data.

First, we propose GenStore, the first in-storage processing system designed for genomic analysis. We design low-cost in-storage accelerators to accurately filter out genomic data that does not require expensive computation during genomic analysis, directly inside the storage device. Second, we introduce MegIS, the first in-storage processing system designed to significantly reduce the data movement overhead of the end-to-end metagenomic analysis pipeline. 
The key idea of MegIS is to enable  cooperative SCC for metagenomics, where we do not solely focus on processing inside the storage system, but instead, we capitalize on the strengths of processing \emph{both inside and outside} the storage system. Third, we propose GRAINS, the first system for analysis on large-scale (meta)genomic sequence graphs in storage. Through our detailed examination of typical analysis pipelines on large-scale sequence graphs, we perform storage-aware algorithm-architecture co-design to \inum{i}~make the pipelines more storage-friendly and \inum{ii}~improve performance, energy-efficiency, and cost-effectiveness via in-storage and in-flash processing. Fourth, to mitigate the data preparation bottleneck, we introduce SAGe, an algorithm-architecture co-design for highly-compressed storage and high-performance access of large-scale sequence data. SAGe's key insight is that the information encoded in genomic data follows specific trends, shaped by factors such as sequencing technology and common genetic phenomena. By carefully exploiting these characteristics to synergistically co-design algorithms and hardware, SAGe achieves high compression ratios, comparable to state-of-the-art genomic compressors, while enabling low decompression latencies, using only lightweight hardware and efficient streaming accesses. Finally, we demonstrate how our proposed designs can be put together in a unified system at low cost.

We demonstrate that by designing storage-centric systems that efficiently \inum{i}~analyze (meta)genomic data inside the storage system, and \inum{ii}~enable highly-compressed storage and high-performance access of large-scale sequence data, we can alleviate data movement overheads from the storage system, reduce the overall computational burden, and mitigate the data preparation bottleneck. We demonstrate that the proposed systems significantly improve system performance, energy efficiency, and system cost-efficiency of (meta)genomic analysis. We hope that the storage-centric systems proposed in this dissertation facilitate the broader adoption of (meta)genomic analyses and inspire future research to fundamentally improve the performance, energy efficiency, and cost-effectiveness of other data-intensive application domains related to health and life sciences.

\clearpage
\vspace{-3em}\chapter*{Zusammenfassung}
\addcontentsline{toc}{chapter}{Zusammenfassung}

Die Genomanalyse befasst sich mit der genomischen Information lebender Organismen sowie anderer biologischer Entitäten. Für die computergestützte Analyse wird zunächst eine Probe des Erbmaterials, typischerweise DNA oder RNA (die zuvor durch reverse Transkription in DNA überführt wurde), sequenziert. Die Sequenzierung überführt die in den DNA-Molekülen enthaltene Information in digitale Sequenzdaten. Die klassische Genomik befasst sich mit den Genomsequenzen eines Individuums (oder einer kleinen Gruppe von Individuen) einer einzelnen, bekannten Art. Die Metagenomik hingegen untersucht Genomsequenzen mehrerer Arten aus einer gemeinsamen Umgebung. Genom- und Metagenomanalysen spielen in vielen Bereichen eine entscheidende Rolle: in der Präzisionsmedizin, in der klinischen Akutversorgung, bei der frühzeitigen Erkennung von Warnsignalen für übertragbare Krankheiten, in der Lebensmittelsicherheit durch die Überwachung von Krankheitserregern, in der Landwirtschaft und beim wissenschaftlichen Erkenntnisgewinn.

Der Einsatz genomischer und metagenomischer Analysen hat in den vergangenen Jahren stark zugenommen, was sowohl auf ihre grosse Bedeutung in zahlreichen Anwendungsbereichen als auch auf die rasanten Fortschritte bei den Sequenzierungstechnologien zurückzuführen ist. Dadurch ist das Volumen der erzeugten Sequenzdaten exponentiell gewachsen. Angesichts der Herausforderungen bei der Analyse und Speicherung dieser enormen Datenmengen wurden zahlreiche Ansätze entwickelt, um (Meta-)Genomanalysen zu beschleunigen und Sequenzdaten komprimiert zu speichern. Trotz dieser Fortschritte in der Analyse und Speicherung (meta-)genomischer Daten bestehen beim Zugriff auf gespeicherte Sequenzdaten und bei deren Bereitstellung für die Analyseeinheiten weiterhin zwei wesentliche Probleme: \inum{i} der Aufwand für den Datentransfer grosser Mengen von Daten mit geringem Wiederverwendungsgrad aus dem Speichersystem, der andere Systemkomponenten, etwa Hauptspeicher und Recheneinheiten, unnötig belastet, und \inum{ii} der Engpass bei der Datenvorbereitung: Komprimierte Sequenzdaten müssen vor der Analyse zunächst dekomprimiert und formatiert\linebreak werden.

Diese Dissertation verfolgt daher zwei Ziele: \inum{i} den durch den Transfer genomischer und metagenomischer Daten aus dem Speichersystem entstehenden Aufwand sowie die gesamte Rechenlast zu reduzieren und \inum{ii} den Engpass bei der Datenvorbereitung zu entschärfen, bei gleichzeitig hoher Leistung, Energieeffizienz und hohen Kompressionsverhältnissen. Die vorgeschlagenen Ansätze sollen dabei leichtgewichtig sein, sodass sie sich nahtlos in eine Vielzahl von Systemen zur (Meta-)Genomanalyse integrieren lassen.

Um den Aufwand für den Transfer grosser Mengen (meta-)genomischer Daten mit geringem Wiederverwendungsgrad aus dem Speichersystem sowie die gesamte Rechenlast zu reduzieren, entwickeln wir speicherzentrierte Rechensysteme (Storage-Centric Computing, SCC) für die (Meta-)Genomanalyse. Bei SCC werden die Daten direkt im Speichergerät verarbeitet, entweder auf dem SSD-Controller oder direkt auf den Flash-Dies. SCC bietet hierfür einen grundlegenden Lösungsansatz, da die Daten dort verarbeitet werden, wo sie gespeichert sind. Um zudem den Engpass bei der Datenvorbereitung zu entschärfen, entwickeln wir einen Algorithmus-Architektur-Co-Design-Ansatz für hochkomprimierte Speicherung und leistungsfähigen Zugriff auf Sequenzdaten.

Erstens stellen wir GenStore vor, das erste In-Storage-Processing-System für die Genomanalyse. Wir entwerfen kostengünstige In-Storage-Beschleuniger, die direkt im Speichergerät genomische Daten zuverlässig herausfiltern, deren weitere Analyse keine aufwendigen Berechnungen erfordert. Zweitens präsentieren wir MegIS, das erste In-Storage-Processing-System, das den Datentransfer über die gesamte Metagenomik-Analysepipeline hinweg deutlich reduziert. Die Kernidee von MegIS ist ein kooperativer SCC-Ansatz für die Metagenomik: Anstatt die Verarbeitung ausschliesslich in das Speichersystem zu verlagern, nutzen wir gezielt die jeweiligen Stärken der Verarbeitung sowohl innerhalb als auch ausserhalb des\linebreak Speichersystems.

Drittens schlagen wir GRAINS vor, das erste System zur Analyse sehr grosser\linebreak (meta-)genomischer Sequenzgraphen direkt im Speichergerät. Auf Grundlage einer detaillierten Untersuchung typischer Analysepipelines auf grossen Sequenzgraphen verfolgen wir einen auf das Speichersystem abgestimmten Algorithmus-Architektur-Co-Design-Ansatz, um \inum{i} die Pipelines für die Verarbeitung im Speichersystem zu optimieren und \inum{ii} Leistung, Energieeffizienz und Kosteneffizienz durch In-Storage- und In-Flash-Processing zu verbessern.

Viertens stellen wir SAGe vor, einen Algorithmus-Architektur-Co-Design-Ansatz für die hochkomprimierte Speicherung und den leistungsfähigen Zugriff auf grosse Mengen von Sequenzdaten, mit dem wir den Engpass bei der Datenvorbereitung entschärfen. SAGe beruht auf der zentralen Erkenntnis, dass die in genomischen Daten kodierte Information charakteristische Muster aufweist, die etwa durch die Sequenzierungstechnologie und typische genetische Phänomene geprägt werden. Durch die gezielte Nutzung dieser Eigenschaften und das Co-Design von Algorithmen und Hardware erzielt SAGe Kompressionsverhältnisse, die mit denen modernster genomischer Kompressionsverfahren vergleichbar sind, und ermöglicht zugleich geringe Dekompressionslatenzen. Dabei kommt SAGe mit leichtgewichtiger Hardware und effizienten Streaming-Zugriffen aus. Schliesslich zeigen wir, wie sich unsere Ansätze kostengünstig in einem Gesamtsystem integrieren lassen.

Wir zeigen, dass speicherzentrierte Systeme, die \inum{i} (meta-)genomische Daten effizient direkt im Speichersystem analysieren und \inum{ii} eine hochkomprimierte Speicherung und einen leistungsfähigen Zugriff auf grosse Mengen an Sequenzdaten ermöglichen, den durch Datentransfers aus dem Speichersystem entstehenden Aufwand verringern, die gesamte Rechenlast reduzieren und den Engpass bei der Datenvorbereitung entschärfen. Die vorgeschlagenen Systeme verbessern Systemleistung, Energieeffizienz und Kosteneffizienz der\linebreak (Meta-)Genomanalyse erheblich. Wir hoffen, dass die in dieser Dissertation vorgestellten speicherzentrierten Systeme den breiteren Einsatz genomischer und metagenomischer Analysen erleichtern und künftige Forschung dazu anregen, auch die Leistung, Energieeffizienz und Kosteneffizienz anderer datenintensiver Anwendungsbereiche der Gesundheits- und Lebenswissenschaften grundlegend zu verbessern.

\setstretch{1.3}

\pagestyle{headings}
\cleardoublepage
\tableofcontents
\newpage
\listoffigures
\newpage
\listoftables

\cleardoublepage
\pagenumbering{arabic}%

\chapter{Introduction}

\emph{Genome sequence analysis}, which examines the \aooo{genomic information} of \aooo{living organisms and other biological entities}, plays an important role in many fields, such as personalized medicine\tomiv{~\cite{\citepersonalized}}, tracing outbreaks of communicable diseases~\cite{\citeoutbreak}, cancer research~\cite{\citecancer}, ensuring food safety through pathogen monitoring~\cite{e002244,TONG2021130}, agriculture~\cite{\citeagriculture}, scientific discovery~\cite{urbanek2018degradation,edgar2022petabase,paoli2022biosynthetic}, biodiversity conservation~\cite{Hogg2024,lewin2018earth}, evolutionary biology~\cite{\citeevolution}, and antimicrobial resistance surveillance~\cite{danko2021global,didelot2012transforming,marini2022towards}.
\aooo{To analyze genomic information computationally, a sample of an organism's or \aooo{a biological entity's} genetic material, typically DNA or RNA that has been reverse-transcribed into DNA~\cite{Houldcroft2017,Jansz2024viral}, undergoes a process called \emph{sequencing}~\cite{\citesequencing}.}
Sequencing converts the information from DNA molecules to digital data. Current sequencing technologies \emph{cannot} sequence long DNA molecules end-to-end. Instead, state-of-the-art sequencers generate randomly- and redundantly-sampled smaller and inexact DNA fragments, called \emph{reads}. Sets of genomic reads (called \emph{read sets}) are then used in genomic analysis. \tomi{Traditional genomics analyzes genome sequences of a genomic sample from an individual (or a small group of individuals) of the \emph{same known species}.}

\aooo{Since sometimes a sample contains organisms or biological entities with \emph{different species} present in a \emph{common environment} (e.g., human gut, soil, or oceans), genomic analysis is complemented by \emph{metagenomic analysis}~\cite{\citemetagenomicsegs}.} Metagenomic analysis refers to the study of the genome sequences of various organisms \aooo{or biological entities} with different species present in a common environment. Since metagenomics deals with genome sequences whose species are \emph{not known} in advance in many cases, it requires comparisons of the target sequences against large databases of many reference genomes. Metagenomics has led to groundbreaking advances in many fields, such as precision medicine~\cite{\citemgprecision}, urgent clinical settings~\cite{\citemgurgentclinical}, understanding microbial diversity of an environment~\cite{\citemgbiodiversity}, discovering early warnings of communicable diseases~\cite{\citemgcommunicable}, and outbreak tracing~\cite{\citemgtracing}.

\tomi{As genomic databases grow in complexity, \emph{graph-based (meta)genomic analysis} has emerged as a powerful approach for querying of massive and complex (meta)genomic databases~\cite{Iqbal2012,marchet2021data,karasikov2020metagraph,karasikov2022lossless,karasikov2019sparse,danciu2021topology,fan2023fulgor,bradley2019ultrafast}. 
Sequences in a graph are represented by graph walks~\cite{marchet2021data}. Genome graphs provide two fundamental benefits, making them indispensable, particularly in modern, population-scale genomics~\cite{danko2021global,karasikov2020metagraph,siren2021pangenomics,Sherman2020,taylor2024beyond}.
First, the graph topology, along with its associated metadata\footnote{Metadata can include the original species or samples of a graph node's sequence~\cite{Iqbal2012,marchet2021data}, associations with patient outcomes~\cite{karasikov2020metagraph}, and more.} offers vastly \emph{strong expressive power}~\cite{eizenga2020pangenome,Sherman2020,taylor2024beyond,Liao2023}. A graph naturally encodes the evolutionary history and diversity of the organisms in a database, revealing their shared and distinct sequences~\cite{eizenga2020pangenome,Sherman2020,Armstrong2020cactus,bradley2019ultrafast}. This leads to reduced bias and improved analysis accuracy. 
For example, genome graphs improve disease diagnosis by capturing population-specific variants often missed by traditional, linear sequences\tomiv{~\cite{Groza2024,Sherman2020,Holley2026}}.
Second, genome graphs leverage the inherent redundancy of genomic data to \emph{avoid redundant computation} on shared sequences. This is because shared genomic regions across database entries are stored only once and do not need to be queried separately. }

\begin{figure}[b]
         \centering
         \includegraphics[width=\columnwidth]{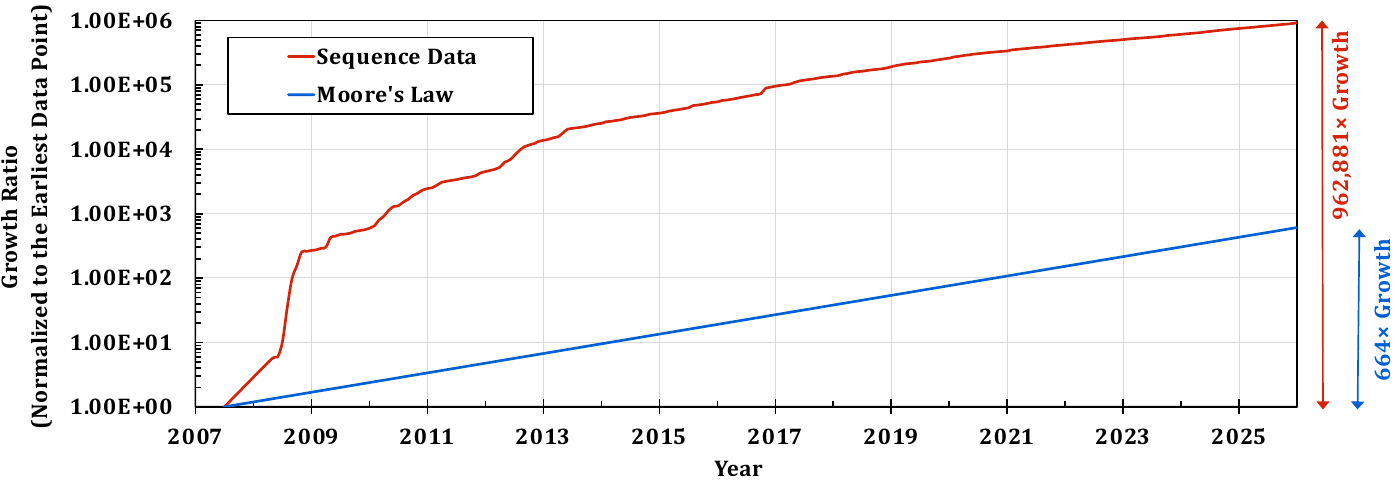}
         \caption{\aooo{Growth rate (log scale) of the Sequence Read Archive (SRA)~\cite{srastats} data volumes compared to Moore's Law projections~\cite{stephens2015big,moore1998cramming}.}}
         \label{fig:genomic-moore}
\end{figure}

The adoption of genomic and metagenomic analyses has been rapidly increasing in recent years~\cite{clark2019diagnosis,farnaes2018rapid,sweeney2021rapid,ginsburg2009genomic,chin2011cancer,Ashley2016,bloom2021massively,gilchrist2015whole, dixon2020metagenomics,chiang2019from,chiu2019clinical,kaplan2025pangenomicsbench,das2024systems}, driven by their critical importance and the rapid advancement of sequencing technologies (i.e., reduced costs and increased throughput~\cite{berger2023navigating}). These factors have led to exponential growth in genomic and metagenomic sequence data generation. \aooo{\fig{\ref{fig:genomic-moore}} demonstrates an example of the growth of the Sequence Read Archive (SRA)~\cite{srastats} data volumes after the breakthroughs in low-cost, high-throughput sequencing, highlighting the significantly larger growth rate of sequence data (by more than three orders of magnitude) compared to Moore's Law projections~\cite{stephens2015big,moore1998cramming}.\footnote{\aooo{Please note that this figure only shows data size in a major public repository (SRA). With advances in sequencing, a growing, large volume of data now resides in private repositories whose sizes are not publicly available.}} Analyzing large-scale sequence data on conventional, general-purpose systems poses significant challenges due to three reasons, as we describe next}.

\aooo{First, the sequencing and basecalling steps for a sample are usually one-time tasks~\cite{wang2021nanopore,hu2021next,alser2022molecules}. In many cases, the reads from a single sequenced sample can be analyzed by \emph{multiple studies} or \emph{at different times} in the same study. This is because
\inum{i}~there are many heuristics \tomi{and parameters} involved in (meta)genomics, \tomi{which need to be tuned to} achieve a desired sensitivity-specificity tradeoff~\cite{bokulich2020measuring}, \inum{ii}~there are different databases created with different parameters or genomes \tomi{that need to be used for achieving different accuracy goals}~\cite{schuele2021future}, and \inum{iii} a sample can be analyzed several times with reference genomes or databases that are regularly updated, personalized, or task-specific~\cite{schuele2021future,chen2024improved,Siren2024,vaddadi2023minimizing,aganezov2022complete,berger2023navigating,rhie2023complete,nurk2022complete,kim2024airlifttcbb}.} 

\aooo{Second, even when performing the (meta)genomic analysis step only once for a sample, the \emph{throughput} of this step is significantly lower than the sequencing throughput of modern sequencers (e.g.,~\cite{illuminax}). While sequencing one sample can take a long time,  a single sequencing machine can sequence \emph{many samples} from different sources in parallel~\cite{hu2021next,shokralla2015massively}, achieving very high throughput.  As an example, our analysis with a state-of-the-art metagenomic tool~\cite{lapierre2020metalign} shows that analyzing the data, sequenced and basecalled by a sequencer in 48 hours, takes 38 days on a high-end server node (detailed configurations in Chapter~\ref{chap:megis}). Such long analysis poses serious challenges, specifically for time-critical use cases \tomi{that require fast decision making} (e.g., clinical settings~\cite{taxt2020rapid} and timely surveillance of infectious diseases~\cite{hadfield2018nextstrain}). Since the growth rate of sequencing throughput is higher than Moore's Law\tomi{~\cite{berger2023navigating,moore1998cramming}}, this already large gap between sequencing and analysis throughput is widening~\cite{hu2021next,katz2021sra,enastats,leinonen2010sequence}, and simply scaling up traditional systems for analysis is not efficient.}

\aooo{Third, the development of sequencing technologies that enable analysis \emph{during} sequencing~\cite{zhang2021real,firtina2023rawhash,kovaka2020targeted, mutlu2023accelerating,Payne2021,Bao2021Squigglenet,ulrich2022readbouncer} increasingly necessitates the need for fast analysis that can keep up with sequencing throughput. \tomi{The possibility to perform analysis during sequencing enables new features, where sequencing can be dynamically stopped once a target pathogen is identified, or sufficient coverage is reached~\cite{\citesignalanalysis}. 
Therefore, to achieve this feature, the analysis throughput must match or exceed the sequencing throughput. Otherwise, the analysis step bottlenecks the entire process and makes this feature impractical in time-critical settings.}}

The analysis step is also the primary energy bottleneck in the workflow, and optimizing its efficiency is vital as sequencing technologies rapidly evolve. For example, a high-end sequencer~\cite{illuminax} uses 405 KJ to sequence and basecall 100 million reads, with 92.5 Mbp/s throughput and 2,500 W power consumption~\cite{illuminax}. In contrast, processing this dataset on a commodity server (detailed configurations in Chapter~\ref{chap:megis}) requires 675 KJ, accounting for 63\% of the total energy \tomi{of the workflow}. The need to enhance the analysis' energy efficiency is further increasing for two reasons.
First, sequencing efficiency has been continually improving. For example, a new version of Illumina sequencer from 2023~\cite{illuminax} provides 44$\times$ higher throughput at only 1.5$\times$ higher power consumption compared to an older version~\cite{illumina} from 2020, resulting in much better sequencing energy efficiency. Therefore, simply relying on scaling up commodity systems to improve the analysis throughput worsens the energy bottleneck. Second, the increased adoption of compact \emph{portable sequencers}~\cite{jain2016oxford} for on-site (meta)genomics (e.g., in remote locations~\cite{pomerantz2018real} or for personalized bedside care~\cite{chiang2019from}) offers high-throughput sequencing with low energy costs. \tomi{The drive towards portability} further amplifies the need for energy- and cost-effective analysis that can match the portability and convenience of these sequencers.

Due to the challenges of analyzing and storing massive volumes of genomic and metagenomic data, significant efforts have been made in two directions: \inum{i}~Accelerating genomic and metagenomic analys\tomi{e}s, and \inum{ii}~Storing genomic and metagenomic data in compressed forms.

\head{Accelerating Genomic and Metagenomic Analysis} There have been extensive efforts to \tomi{improve the performance and energy efficiency of} genomic and metagenomic analysis. 
For genomics, many works propose efficient heuristics and algorithmic optimizations (e.g.,~\cite{\citealgoptimization,\citegraphalgoptimization}), hardware accelerators (e.g.,~\cite{\citehwoptimization,\citegraphhwoptimization}), various filters that try to efficiently and accurately prune reads that do not require expensive computation (e.g.,~\cite{alser2020technology,kim2018grim,alser2020accelerating,cali2020genasm,singh2021fpga,nag2019gencache, kim20111, alser2017gatekeeper, alser2017magnet, alser2019shouji, alser2020sneakysnake, bingol2021gatekeeper, hameed2021alpha, guo2019hardware,xin2015shifted,xin2013accelerating}).
For metagenomics, many works propose algorithmic optimizations (e.g.,\tomiii{~\cite{\citemgalgoptimization}}), sampling to reduce database size\tomi{s}, \tomi{usually} at the cost of accuracy loss (e.g.,~\cite{kim2016centrifuge,wood2019improved,muller2017metacache,song2024centrifuger,Dilthey2019,Fan2021}), and hardware acceleration (e.g.,~\cite{\citemghwoptimization}).

\head{Storing Genomic and Metagenomic Data in Compressed Forms} It is common practice to store genomic and metagenomic sequence data in compressed forms\tomi{~\cite{berger2023navigating,zhu2013high,Deorowicz2013,giancarlo2013compressive,Betschart2025,Walia2026}} because storing uncompressed sequence data is impractical \omcr{due to its massive size}. In fact, due to the importance of storing sequence data in a space-efficient manner, there exist many compression techniques (e.g.,\tomiii{~\cite{chandak2018spring,Deorowicz2020,lan2021genozip,alyami2019lfastqc,kowalski2019pgrc,roguski2018fastore,chandak2017compression,cogo2021genodedup,Meng2023,kokot2022colord,dufort2020enano,dufort2021renano,karasikov2022lossless,vandamme2024tinted,dragenora,yang2025gpufastqlz,chen2023efficient,hach2012scalce,roguski2014dsrc2,grabowski2022mbgc,kowalski2026mbgc2,grabowski2026ffc,deorowicz2023agc,Kryukov2022,sousa2024jarvis3,Sun2023,Nazari2025}}) specialized for sequence data to achieve significantly higher compression ratios than state-of-the-art general-purpose compression methods (e.g.,\linebreak \cite{collet2018zstandard,pavlov20167,Brotli,Katz1991US5051745A,goyal2021dzip,goyal2018deepzip,chen2024ha,bartik2015lz4,liu2018data,fowers2015scalable,chen2021fpga,angerd2022gbdi,gao2024beezip,karandikar2023cdpu,9499902,abali2020data}).

\section{Problem Discussion}
\label{sec:intro.key_problem}

Although there have been \tomi{significant} efforts to improve the analysis and storage of large-scale genomic and metagenomic data, we identify \tomi{major outstanding} problems in accessing stored \tomi{sequence} data and feeding it to the analysis units.

\head{Overhead of Moving Large Amounts of Low-Reuse Data from the Storage System to Main Memory and Computation Units} Genomic and metagenomic analyses incur unnecessary data movement from the storage system for large amounts of \emph{low-reuse} data. \linebreak
In genomics, while existing filters prune many reads to avoid expensive computation, they still need to first read the entire read set from the storage system \tomi{all the way to the main memory, processor-side caches, and register files}, even though a large fraction of the reads would be filtered out and not be reused in the analysis. \tomi{Chapter~\ref{chap:genstore} analyzes the state-of-the-art genomic analysis tools and accelerators and finds that the  movement of large amounts of low-reuse genomic data from the storage system significantly hinders the end-to-end performance and energy-efficiency of genomic analysis due to storage I/O (input/output) overheads and unnecessary computational burden on the rest of the system.}
As we demonstrate in Chapter~\ref{chap:genstore},  these overheads can bottleneck the performance \tomi{and energy} of genomic analysis in both conventional (software-based) and emerging (hardware-accelerated) genomics systems, while having a larger impact on systems that reduce other \tomi{(e.g., computation and main memory)} bottlenecks\tomi{~\cite{\citehwoptimization,\citegraphhwoptimization}}.

In metagenomics, analysis also suffers from significant data movement overhead due to the need to access large amounts of low-reuse data. Since we do not know the species present in a metagenomic sample, \mganalysis requires searching large databases (e.g., several TBs~\cite{ncbi2023,karasikov2020metagraph,shiryev2023indexing,pebblescout,lemane2023kmindex,marchet2023scalable} or more than a hundred TBs in emerging databases~\cite{shiryev2023indexing,pebblescout}) that contain information on different organisms' genomes. Database sizes are expected to increase further in the future, and at a fast pace.\footnote{For example, based on recently published trends, the ENA assembled/annotated sequence database size currently \emph{doubles} every 19.9 months~\cite{enastats}, and the BLAST nt database size doubled from 2021 to 2022~\cite{ntdouble}.} Two notable reasons for this growth are 1) the rapid evolution of viruses and bacteria~\cite{Lynch2010}, which necessitates frequent updates with new reference genomes~\cite{Nasko2018,o2016reference}, and 2) the fact that databases may include sequences from both highly curated reference genomes and from less curated metagenomic sample sets~\cite{karasikov2020metagraph,shiryev2023indexing}. Particularly, as over 99\% of Earth's microbes remain unidentified and excluded from curated reference genome databases~\cite{jiao2020microbial,Li2024}, the expanded databases improve sensitivity~\cite{Li2024}. Recent advances in the automated and scalable construction of genomic data from more organisms have further contributed to database growth by enabling the rapid addition of new sequences to databases~\cite{rautiainen2023telomere,jarvis2022semi}. 
As we demonstrate in Chapter~\ref{chap:megis}, data movement overhead from the storage system to the rest of the system significantly impacts the end-to-end performance and \tomi{energy efficiency} of \tomi{the state-of-the-art \mganalysis tools}. Due to its low reuse, the data needs to move all the way from the storage system to the main memory, \tomi{processor-side caches, register files,} and processing units for its first use, and it will likely not be used again or reused very little during analysis. This unnecessary data movement, combined with the low computation intensity of \mganalysis and the limited I/O bandwidth, leads to large storage data movement overheads for \mganalysis. The impact of this overhead becomes even larger on systems that reduce other \tomi{(e.g., computation and main memory)} bottlenecks~\cite{\citemghwoptimization}.

\tomi{As we demonstrate in Chapter~\ref{chap:grains}, the overhead of moving large amounts of low-reuse data from the storage system and its burden on the rest of the system is also significant in graph-based (meta)genomic analysis. Our evaluation using the state-of-the-art tools for analysis with large-scale genome graphs shows that this overhead significantly hinders their end-to-end performance and energy efficiency.}

\begin{figure}[h]
         \centering
         \includegraphics[width=0.85\columnwidth]{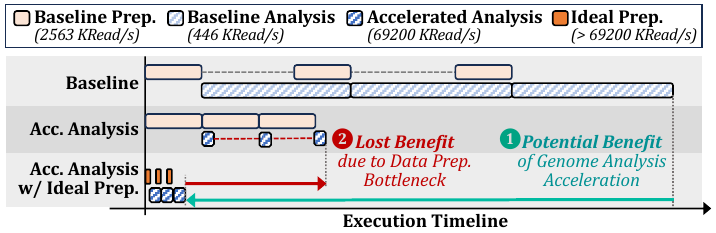}
         \caption{\tomi{Effect of data preparation (i.e., decompressing and formatting genomic sequence data before analysis) on genome analysis performance.}}
         \label{fig:intro-motivation}
\end{figure}

\head{Data Preparation Bottleneck} 
\tomi{In this dissertation, we introduce and extensively analyze
the \emph{data preparation bottleneck}, where compressed genomic sequence data needs to be first decompressed and formatted before it can be analyzed. As we demonstrate in Chapter~\ref{chap:sage}, the benefits of prior works on accelerating genome sequence analysis are \tomi{significantly} diminished due to this bottleneck.
For example, \fig{\ref{fig:intro-motivation}} shows the execution timeline of data preparation and genome analysis for a real-world genomic dataset (Chapter~\ref{chap:sage}) in three different configurations. The evaluated analysis task is read mapping, a fundamental process in genomics (\sect{\ref{sec:background-workflow}}). The configurations are 
\inum{i}~\textbf{Baseline:}~a state-of-the-art software analysis tool~\cite{li2018minimap2} with a state-of-the-art software genomic decompressor for data preparation~\cite{chandak2018spring};
\inum{ii}~\textbf{Acc. Analysis:} a state-of-the-art hardware-accelerated analysis tool~\cite{chen2023gem} with the same data preparation as Baseline;
\inum{iii}~\textbf{Acc. Analysis w/ Ideal Prep.:} the same accelerated analysis with \emph{ideal} preparation, where preparation is overlapped with analysis. 
For all configurations, preparation and analysis operate in a pipelined manner and in batches (i.e., when decompressing  batch $\#i$, the mapper
analyzes batch $\#i - 1$). Decompressed data batches are directly fed to the analysis stage.
We observe that \circg{1}~hardware acceleration of genome analysis can potentially offer substantial performance benefits; \circr{2}~however, as analysis gets faster, data preparation emerges as a critical bottleneck that hinders the full realization of these benefits. In Chapter~\ref{chap:sage}, we further demonstrate and extensively analyze this bottleneck across various real-world scenarios.}

\section{Goal}

The goal of this dissertation is to \inum{i}~alleviate the data movement overheads of genomics and metagenomics analyses from the storage system and reduce the computational burden from the rest of the system, and \inum{ii}~mitigate the data preparation bottleneck while achieving high performance, energy efficiency, and compression ratios. \tomi{We aim for our proposed designs to be lightweight such that they can} seamlessly integrate with a broad range of genomic and metagenomic analysis systems.

\section{Thesis Statement}

The following thesis statement encompasses our approach:

\begin{quote}
\emph{By designing customized storage-centric systems that efficiently}

\emph{1) analyze genomic and metagenomic data inside the storage system}

\emph{2) enable highly-compressed storage and high-performance access of large-scale sequence data,}

\emph{we can alleviate data movement overheads from the storage system, reduce the computational burden from the rest of the system, and mitigate the data preparation bottleneck,}

\emph{thereby significantly \tomi{(e.g., by one to two orders of magnitude)} improving system performance, energy efficiency, and cost of genomic and metagenomic analysis.}
\end{quote}

\section{Our Approach}

To alleviate the overheads of moving large amount\tomi{s} of low-reuse sequence data from the storage system and reduce the overall \tomi{computational} burden \tomi{from} the rest of the system, we propose \emph{storage-centric computing} (SCC) systems for 1) genomic analysis, 2) metagenomic analysis, and 3) graph-based (meta)genomic analysis, which plays critical roles in population-scale scenarios. \tomi{SCC refers to processing data inside the storage device, either on the SSD controller (in-storage processing, i.e., ISP) or on the flash dies (in-flash processing, i.e., IFP). SCC can be a fundamental solution for alleviating this data movement overhead by processing data where it originally resides.} 
To mitigate the data preparation bottleneck, we propose an algorithm-architecture co-design for highly-compressed storage and high-performance access of sequence data. In the rest of this section, we briefly overview our new mechanisms developed as part of this dissertation.

\subsection{In-Storage Filters for Genomic Analysis}

Read mapping~\cite{\citemapping} is a fundamental step in many genomics applications. It is used to identify potential matches and differences between reads of a sequenced genome and an already known genome (called a \emph{reference genome}).
Read mapping is  costly because it needs to perform \emph{approximate string matching (ASM)}~\cite{\citeasm} on large amounts
of data.
To address the computational challenges in genomic analysis, many prior works \tomi{(e.g.,~\cite{\citehwoptimization,\citealgoptimization,\citefilteroptimizationadditional})} propose various approaches such as accurate \emph{filters} that select the reads within a dataset of genomic reads (called a \emph{read set}) that \emph{must} undergo expensive computation, efficient heuristics, and hardware acceleration. 
While effective at reducing the amount of expensive computation, all such approaches still require the costly movement of a large amount of data from storage to the rest of the system, which significantly lower\tomi{s} the end-to-end performance of read mapping in conventional and emerging genomics systems. 

\tomi{This dissertation introduces} \textbf{GenStore}, the first in-storage processing system designed for genomic analysis that significantly reduces both \tomi{storage} data movement and computational overheads of genomic analysis by exploiting low-cost and accurate in-storage filters. 
GenStore leverages hardware/software co-design to address the challenges of in-storage processing,
supporting reads with  1)~different properties such as read lengths and error rates, which highly depend on the sequencing technology, and 2)~different \omcvii{degrees of} genetic variation compared to the reference genome, which highly depends on the genomes that are being compared.
Through rigorous analysis of read mapping processes of reads with different properties and degrees of genetic variation, we meticulously design low-cost hardware accelerators and data/computation flows inside a NAND flash-based solid-state drive (SSD).

Our evaluation using a wide range of real genomic datasets shows that GenStore, when implemented in three \omcvii{modern} NAND flash-based SSDs, significantly improves the read mapping performance of \tomi{three} state-of-the-art software (hardware) baselines~\tomi{\cite{li2018minimap2,nag2019gencache,turakhia2018darwin}} by 2.07-6.05$\times$ (\rev{1.52-3.32}$\times$) for \omccc{read sets with high similarity to the reference genome} and \rev{1.45-33.63}$\times$ (2.70-19.2$\times$) for \omccc{read sets with low similarity to the reference genome}. \tomi{Our energy analysis shows that GenStore reduces energy consumption by on average (up to) 3.92$\times$ (3.97$\times$) for read sets with high similarity to the reference genome, and on average (up to) 27.17$\times$ (29.25$\times$) for read sets with low similarity to the reference genome.}

\tomi{GenStore is published at the International Conference on Architectural Support for Programming Languages and Operating Systems (ASPLOS) in 2022~\cite{mansouri2022genstore}. It is fully open-sourced at~\cite{gssource}. An extended version of the ASPLOS 2022 paper is available on arXiv~\cite{arxivGS}. A significant amount of work (e.g.,~\cite{soysal2025mars,abakus23taco,zheng2025storage,kabra2025ciphermatch,chen2025reis,megis,grains,mansouri2026sage}) has already been influenced by GenStore, as we discuss in Chapter~\ref{chap:genstore}.}

\subsection{Cooperative In-Storage Processing for Metagenomic Analysis}

Since the species present in a metagenomic sample are not known in advance, metagenomic analysis commonly involves the key tasks of determining the species present in a sample and their relative abundances. These tasks require searching large metagenomic databases containing information on different species’ genomes. Metagenomic analysis suffers from significant data movement overhead due to moving large amounts of low-reuse data from the storage system to the rest of the system. In-storage processing can be a fundamental solution for reducing this overhead. However, designing an in-storage processing system for metagenomics is challenging because \tomi{none of the existing tools for} \mganalysis can be directly implemented in storage effectively due to the hardware limitations of modern SSDs.

\tomi{This dissertation proposes} \textbf{MegIS}, the \emph{first} in-storage processing system designed to significantly reduce the data movement overhead of the end-to-end metagenomic analysis pipeline. 
\tomi{The key idea of MegIS is to enable  \emph{cooperative} ISP for metagenomics, where we do not solely focus on processing inside the storage system but, instead, we capitalize on the strengths of processing \emph{both inside and outside} the storage system. We enable cooperative ISP via a synergistic hardware/software co-design between the storage system and the host system.
We design MegIS as an efficient pipeline between the SSD and the host system to \inum{i}~\emph{leverage} and \inum{ii}~\emph{orchestrate} the capabilities of both. Based on our rigorous analysis of the end-to-end \mganalysis pipeline, we propose a new hardware/software co-designed accelerator framework that consists of five aspects.
First, we partition and map different parts of the \mganalysis pipeline to the host and the ISP system such that each part is executed on the most suitable architecture. 
Second, we coordinate the data/computation flow between the host and the SSD such that MegIS \inum{i}~completely overlaps the data transfer time between them with computation time to reduce the communication overhead between different parts, \inum{ii}~leverages SSD bandwidth efficiently, and \inum{iii}~does not require large DRAM inside the SSD or a large number of writes to the flash chips. 
Third, we devise storage technology-aware metagenomics algorithm optimizations to enable efficient access patterns to the SSD. 
Fourth, we design lightweight in-storage accelerators to perform MegIS's ISP functionalities while minimizing the required SRAM/DRAM buffer spaces inside the SSD. 
Fifth, we design an efficient data mapping scheme and Flash Translation Layer (FTL) specialized to
the characteristics of \mganalysis to leverage the SSD's full internal bandwidth.} 
MegIS's design is flexible, capable of supporting different types of metagenomic input datasets, and can be integrated into various metagenomic analysis pipelines. 

Our evaluation shows that MegIS outperforms \tomi{two} state-of-the-art performance- and accuracy-optimized software metagenomic tools\tomi{~\cite{wood2019improved,lapierre2020metalign}} by 2.7$\times$--37.2$\times$ and 6.9$\times$--100.2$\times$, respectively,  while matching the accuracy of the accuracy-optimized tool. MegIS achieves 1.5$\times$--5.1$\times$ speedup compared to the state-of-the-art metagenomic hardware-accelerated  (using processing-in-memory~\cite{wu2021sieve}) tool, while achieving significantly higher accuracy. \tomi{MegIS provides large average energy reductions
of 5.4$\times$ and 1.9$\times$ compared to software and hardware performance-optimized baselines, respectively, and 15.2$\times$ compared to the accuracy-optimized baseline.}

\tomi{MegIS is published at the International Symposium on Computer Architecture (ISCA) in 2024~\cite{megis}. It is fully open-sourced at~\cite{megissource}. An extended version of the ISCA 2024 paper is available on arXiv~\cite{megisarxiv}. MegIS has already influenced several subsequent works (e.g.,~\cite{grains,mansouri2026sage,mansouri2026sagearxiv,chen2025reis,kabra2025ciphermatch}) as we discuss in Chapter~\ref{chap:megis}.}

\subsection{Storage-Aware Algorithm-Architecture Co-Design for Graph-Based Genomic and Metagenomic Analys\tomi{e}s}

Graph-based representations of genomic and metagenomic sequences have emerged as a powerful approach for representing massive sequence databases in \tomi{an} expressive and efficient manner~\cite{Iqbal2012,marchet2021data,karasikov2020metagraph,karasikov2022lossless,karasikov2019sparse,danciu2021topology,fan2023fulgor,bradley2019ultrafast}. Compared to traditional, linear sequences, sequence graphs enable more accurate and efficient genomic and metagenomic analyses, particularly in complex, population-scale settings, such as public health, precision medicine, and agriculture.   
Despite its benefits, analysis on large-scale sequence graphs incurs significant data movement overhead from the storage system due to accessing large amounts of low-reuse data. Processing data directly inside the storage device, where data originally resides, can be a fundamental solution for mitigating this overhead. However, none of the existing tools for graph-based genomic and metagenomic analys\tomi{e}s \tomi{(e.g.,~\cite{\citegraphhwoptimization,\citegraphalgoptimization})} can be efficiently \tomi{implemented} inside the storage system due to the limited internal hardware resources in modern SSDs. At the same time, prior storage-centric systems developed for \inum{i}~traditional, linear non-graph-based genomic and metagenomic analys\tomi{e}s \tomi{(e.g.,~\cite{\citenongraphgenomeisp})} or \inum{ii}~conventional, non-genomic graph analysis \tomi{(e.g.,~\cite{\citenongenomegraphisp})} are unsuitable for the unique data structures and access patterns of graph-based genomic and metagenomic analysis.

\tomi{This dissertation introduces} \textbf{GRAINS}, the \emph{first} system for analysis \tomi{on} large-scale \underline{\textbf{g}}enomic and metagenomic sequence g\underline{\textbf{ra}}phs \underline{\textbf{in}} \underline{\textbf{s}}torage. 
Through our detailed examination of typical analysis pipelines on large-scale sequence graphs, we perform storage-aware algorithm\tomi{-}architecture co-design to \inum{i}~make the pipelines more storage-friendly and \inum{ii}~improve performance, energy-efficiency, and cost-effectiveness via in-storage and in-flash processing. 
GRAINS's co-design is based on three key aspects.
First, we propose a new batching technique and execution flow, based on unique features of sequence graphs, that reduces the number of random \tomi{storage} accesses. Second, via in-flash and in-storage processing, we avoid transferring low-reuse or unused flash pages, preventing SSD channel and external I/O bandwidth waste. Third, to enable leveraging the full parallelism of all flash dies during in-flash processing, we design an effective, yet lightweight, scheduling technique, enabled by re-purposing the existing SSD structures in a new way.
GRAINS's design is versatile and flexible as it supports \tomi{major} operations on sequence graphs and can be integrated in various analysis pipelines. 

Our evaluation shows that GRAINS provides 2.7$\times$–47.8$\times$ speedup over the state-of-the-art software baselines\tomi{~\cite{karasikov2020metagraph,karasikov2022lossless,danciu2021topology,karasikov2019sparse}}, and 1.5$\times$–17.0$\times$ speedup over a hardware-accelerated (with processing-in-memory) baseline. \tomi{GRAINS provides significantly higher energy efficiency of on average 16.4$\times$ and 9.8$\times$ over software and hardware baselines, respectively.}

\tomi{GRAINS is published at the International Symposium on Computer Architecture (ISCA) in 2026~\cite{grains}. We plan to open-source GRAINS to facilitate future research. An extended version of the ISCA 2026 paper is available on arXiv~\cite{grainsextended}.}

\subsection{Algorithm-Architecture Co-Design for Highly-Compressed Storage and High-Performance Access of Sequence Data}

\tomi{Given the importance of genomics and the exponentially growing volumes of sequence data, there are extensive efforts to accelerate genomic analysis (e.g.,~\cite{\citehwoptimization}). In this work, we demonstrate a major bottleneck that significantly limits and diminishes the benefits of state-of-the-art genomic analysis accelerators: the data preparation bottleneck, where genomic sequence data is stored in compressed form and needs to be first decompressed and formatted before an accelerator can operate on it.} 

To mitigate the data preparation bottleneck, \tomi{this dissertation proposes} \textbf{SAGe}, an algorithm-architecture co-design for highly-compressed \textbf{\underline{s}}torage and high-performance \textbf{\underline{a}}ccess of large-scale \textbf{\underline{ge}}nomic sequence data. The key challenge is to improve data preparation performance while maintaining high compression ratios (comparable to genomic-specific compression algorithms) at low hardware cost. 
\tomi{SAGe addresses this challenge based on the \textbf{key insight} that the information encoded in genomic data follows specific trends, shaped by factors such as sequencing technology (e.g., error rates and read lengths) and common genetic phenomena (e.g., typical spatial distributions of genetic variations within genomes). By carefully exploiting these characteristics to synergistically co-design algorithms and hardware, SAGe achieves high compression ratios, comparable to state-of-the-art genomic compressors, while enabling low decompression latencies, using only lightweight hardware and efficient streaming accesses. SAGe's co-design consists of new}
\inum{i}~lossless (de)compression algorithm, \inum{ii}~hardware \nh{that decompresses data with lightweight operations and efficient streaming accesses}, \inum{iii}~storage data layout, and \inum{iv}~interface commands to access data. SAGe is highly versatile, as it supports datasets from different sequencing technologies and species.
Due to its lightweight design, SAGe can be seamlessly integrated with a broad range of hardware accelerators for genome sequence analysis to mitigate their data preparation bottlenecks.  

Our results demonstrate that SAGe improves the average end-to-end performance and energy efficiency of two state-of-the-art genome sequence analysis accelerators\tomi{~\cite{chen2023gem,mansouri2022genstore}} by 3.0$\times$--32.1$\times$ and \nh{13.0$\times$--34.0$\times$}, respectively, compared to when the accelerators rely on state-of-the-art \omcr{software and hardware} decompression tools.

\tomi{SAGe is published at the IEEE International Symposium on High-Performance Computer Architecture (HPCA) in 2026~\cite{mansouri2026sage}. We plan to open-source SAGe to facilitate future research. An extended version of the HPCA 2026 paper is available on arXiv~\cite{mansouri2026sagearxiv}.}

\tomi{\subsection{Putting It All Together} This dissertation proposes various storage-centric designs for addressing the problems of data movement overheads from the storage system and data preparation bottleneck in genomic and metagenomic analyses. 
In Chapter~\ref{chap:all-together}, we \tomii{describe how to combine} all proposed designs \tomii{(i.e., GenStore, MegIS, GRAINS, and SAGe)} into a unified system. First, we motivate the need for such a system by explaining how it \inum{i}~mitigates both data movement overheads and data preparation bottleneck, and enables SCC without relying on large-scale (meta)genomic data to be stored in uncompressed formats, and \inum{ii}~can facilitate wider adoption of the proposed designs since the unified system can support a wide range of critical genomic and metagenomic analysis tasks at low cost. Given that the unified system can merge common hardware components across designs (e.g., specific buffers used in each design), it has a smaller area compared to the sum of the areas of the individual designs. Second, we discuss the design details of the unified system. Finally, we show that the unified system provides significant benefits while still maintaining low costs.}

\begin{figure}[h]
         \centering
         \includegraphics[width=0.8\columnwidth]{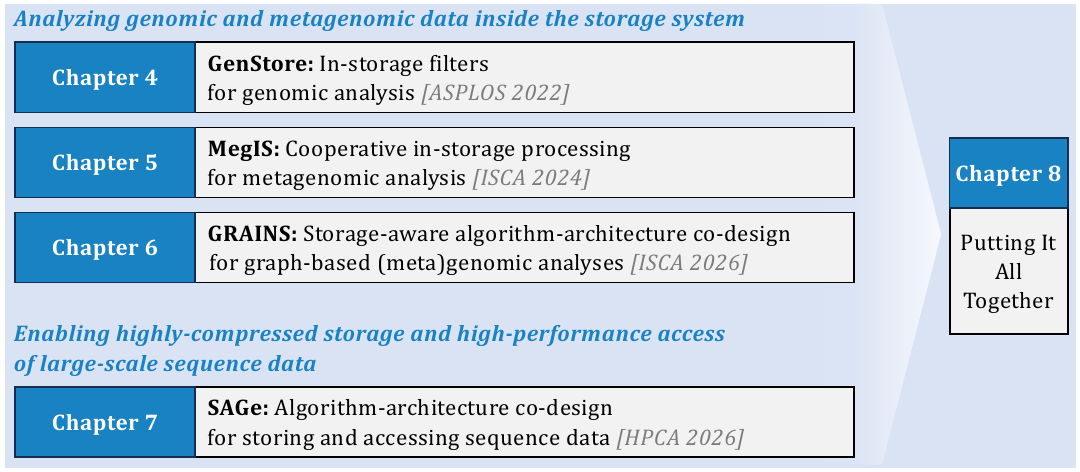}
         \caption{\aooo{Overview of the proposed designs and the chapters in which they are discussed.}}
         \label{fig:approach-thesis}
\end{figure}

\aooo{\fig{\ref{fig:approach-thesis}} provides an overview of the proposed designs and specifies the chapters in which they are discussed.}

\section{Contributions}

This dissertation makes the following contributions:

\begin{itemize}
    \item We identify that although there have been \tomi{significant} efforts to improve the analysis and storage of large-scale genomic and metagenomic data, there remain \tomi{two major} problems in accessing the stored data and feeding it to the analysis units: 1) the overhead of moving large amounts of low-reuse data from the storage system to main memory and computation units, and 2) the data preparation bottleneck, where compressed genomic sequence data needs to be first decompressed and formatted before it can be analyzed. 
    We develop a comprehensive understanding of how these \tomi{two major} problems hinder the end-to-end performance of genomic and metagenomic analys\tomi{e}s.  
    \begin{itemize}
        \item Chapters \tomiv{\ref{chap:genstore}, \ref{chap:megis}, and \ref{chap:grains}} demonstrate our \tomi{detailed and comprehensive} empirical analysis of the effects of storage data movement overheads.
        \item Chapter \ref{chap:sage} demonstrates \tomi{our detailed analysis of} the data preparation bottleneck.
    \end{itemize}
    \item We make the case for and quantitatively substantiate \tomi{the benefits of} storage-centric system designs to fundamentally alleviate data movement overheads from the storage system \tomi{(by analyzing data directly inside the storage system, where it originally resides)} and to mitigate the data preparation bottleneck \tomi{in genomics and metagenomics. Despite the benefits of storage-centric systems, designing such systems is challenging due to the hardware limitations of modern storage systems. By addressing these challenges, we demonstrate novel examples of storage-centric system designs: GenStore, MegIS, GRAINS, and SAGe.}
    \item We propose GenStore (Chapter~\ref{chap:genstore}), the first in-storage processing system designed for genomic analysis. We design low-cost in-storage accelerators to accurately filter out reads that do not require the expensive ASM computation during read mapping, directly inside the storage device.
    \begin{itemize}
        \item  \tomi{We address the challenges of storage-centric computing with hardware/software co-design in three key directions. First, based on our detailed analysis of read mapping, we design two different accelerators that can accelerate a wide range of read mapping applications. Each accelerator filters a large fraction of genomic read datasets using simple operations. Second, we develop storage technology-aware algorithmic optimizations to replace expensive random accesses with more efficient sequential accesses to storage devices. Third, we carefully design an efficient technique for data placement inside the storage device that takes full advantage of the internal SSD bandwidth.}
        \item \tomi{Using a wide range of real genomic datasets, we show that GenStore significantly improves the read mapping performance of \tomi{three} state-of-the-art software (hardware) baselines~\tomi{\cite{li2018minimap2,nag2019gencache,turakhia2018darwin}} by 2.07-6.05$\times$ (\rev{1.52-3.32}$\times$) for read sets with high similarity to the reference genome and 1.45-33.63$\times$ (2.70-19.2$\times$) for read sets with low similarity to the reference genome.} \tomi{Our energy analysis shows that GenStore reduces energy consumption by on average 3.92$\times$/29.25$\times$ for read sets with high/low similarity to the reference genome.}
    \end{itemize}
    \item We propose MegIS (Chapter~\ref{chap:megis}), the first in-storage processing system designed to reduce the data movement overheads inside the end-to-end metagenomic analysis pipeline. 
    \begin{itemize}
        \item To address the challenges of in-storage processing for metagenomics, we enable cooperative in-storage processing, where we do not solely focus on processing inside the storage system but, instead, we capitalize on the strengths of processing both inside and outside the storage system for metagenomic analysis. We realize cooperative in-storage processing via a synergistic hardware/software co-design between the storage system and the host system. 
        \item \tomi{We show that MegIS outperforms \tomi{two} state-of-the-art performance- and accuracy-optimized software metagenomic tools\tomi{~\cite{wood2019improved,lapierre2020metalign}} by 2.7$\times$--37.2$\times$ and 6.9$\times$--100.2$\times$, respectively,  while matching the accuracy of the accuracy-optimized tool. MegIS achieves 1.5$\times$--5.1$\times$ speedup compared to the state-of-the-art metagenomic hardware-accelerated  (using processing-in-memory~\cite{wu2021sieve}) tool, while achieving significantly higher accuracy. \tomi{MegIS provides large average energy reductions of 5.4$\times$ and 1.9$\times$ compared to software and hardware performance-optimized baselines, respectively, and 15.2$\times$ compared to the accuracy-optimized baseline.}}
    \end{itemize}
    \item We propose GRAINS (Chapter~\ref{chap:grains}), the first system for analysis with large-scale genomic and metagenomic sequence graphs in storage. 
    \begin{itemize}
        \item \tomi{We address the challenges of storage-centric computing} through our detailed examination of typical analysis pipelines on large-scale sequence graphs and performing storage-aware algorithm-architecture co-design with two key aspects. First, we make the graph-based (meta)genomic analysis pipelines more storage-friendly. Second, we improve performance, energy-efficiency, and cost-effectiveness via in-storage and in-flash processing. 
        \item \tomi{We show that GRAINS provides 2.7$\times$–47.8$\times$ speedup over the state-of-the-art software baselines\tomi{~\cite{karasikov2020metagraph,karasikov2022lossless,danciu2021topology,karasikov2019sparse}}, and 1.5$\times$–17.0$\times$ speedup over a hardware-accelerated (with processing-in-memory) baseline, while providing significantly higher energy efficiency at low cost. GRAINS provides significantly higher energy efficiency of on average 16.4$\times$ and 9.8$\times$ over software and hardware baselines, respectively.}
    \end{itemize}
    \item We propose SAGe (Chapter~\ref{chap:sage}) to mitigate the data preparation bottleneck, while maintaining high compression ratios (comparable to genomic-specific compression algorithms) and ensuring a lightweight design. To this end, we propose an algorithm-architecture co-design for highly-compressed storage and high-performance access of large-scale genomic sequence data. 
    \begin{itemize}
        \item \tomi{To address the challenge of improving data preparation performance while maintaining high compression ratios at low hardware cost, we leverage the key insight that the information encoded in genomic data follows specific trends, shaped by factors such as sequencing technology and common genetic phenomena. By carefully exploiting these characteristics to synergistically co-design algorithms and hardware, SAGe achieves high compression ratios, comparable to state-of-the-art genomic compressors, while enabling low decompression latencies, using only lightweight hardware and efficient streaming accesses.}
        \item\tomi{We demonstrate that SAGe improves the average end-to-end performance and energy efficiency of two state-of-the-art genome sequence analysis accelerators\tomi{~\cite{chen2023gem,mansouri2022genstore}} by 3.0$\times$--32.1$\times$ and 13.0$\times$--34.0$\times$, respectively, compared to when the accelerators rely on state-of-the-art software and hardware decompression tools.}
    \end{itemize}

    \item \tomi{We demonstrate in Chapters\tomiv{~\ref{chap:genstore}, \ref{chap:megis}, \ref{chap:grains}, and \ref{chap:sage}} that our storage-centric systems provide these performance and energy benefits without relying on costly hardware resources throughout the system (e.g., external I/O bandwidth, large DRAM capacity, extensive computational units), making genomics and metagenomics more accessible for wider adoption.}
    \item We describe how all storage-centric designs presented in this dissertation can be put all together \tomii{at low cost}. Chapter~\ref{chap:all-together} describes the combined design and its evaluations.

    \item We describe future opportunities and new directions enabled by the contributions presented in this dissertation. Chapter~\ref{chap:conclusion} discusses these opportunities and directions, including storage-centric computing for other biological data types, storage-centric designs for constructing and updating large-scale genomic databases and graphs, highly-compressed storage and high-performance access of other biological data types, and genomic privacy in the era of storage-centric computing.
\end{itemize}

\section{Dissertation Outline}

This dissertation is organized into \tomiv{nine} chapters. 
Chapter~\ref{chap:background} provides relevant background regarding genomic and metagenomic analyses, graph-based analysis, and storing sequence data. 
Chapter~\ref{chap:related-work} discusses related works on techniques for alleviating storage I/O overheads (including storage-centric computing), \tomiii{memory-centric computing}, accelerating genomic and metagenomic analyses, and genomic-specific compression \tomi{techniques}. 
Chapter~\ref{chap:genstore} presents GenStore and its evaluation. 
Chapter~\ref{chap:megis} presents MegIS and its evaluation. 
Chapter~\ref{chap:grains} presents GRAINS and its evaluation.
Chapter~\ref{chap:sage} introduces the data preparation bottleneck, along with presenting SAGe and its evaluations. Chapter~\ref{chap:all-together} discusses how we can put all the storage-centric designs proposed in this dissertation together. Finally, 
Chapter~\ref{chap:conclusion} provides future research directions and concluding remarks.

\chapter{Background}
\label{chap:background}

This chapter provides an overview of the background necessary for understanding the contributions, evaluations, and discussions in this dissertation. Section~\ref{sec:background-SSD} provides details on storage systems based on solid state drives (SSDs). Section~\ref{sec:background-workflow} explains the overall genomic \tomi{and metagenomic} workflow, which involves three key steps of sequencing, basecalling, and analysis \tomi{(which in itself consists of multiple steps)}. \tomi{Section~\ref{sec:genomic-analysis} explains genomic analysis in more detail,} and Section~\ref{sec:background-mg} explains metagenomic analysis in more detail.
Section~\ref{sec:background-graph} describes graph-based genomic and metagenomic analyses. Section~\ref{sec:background-compression} \tomi{explains genomics-specific compression}. 

\section{Solid State Drive Organization}
\label{sec:background-SSD}

In this thesis, we focus on storage systems based on solid state drives (SSDs) using NAND flash memory, the prevalent memory technology in modern storage systems\tomiv{~\cite{\citeflash,wu2025ssdtrain,qureshi2022bam,qureshi2023gpu,son2026exploring,\citeifp,\citeufp}}. We expect that our proposed designs would also provide performance and energy benefits with storage devices that are built using emerging  or older non-volatile memory technologies \tomiii{(e.g.,\linebreak \cite{kultursay2013evaluating,meena2014overview,lee2009architecting,akinaga2010resistive,tehrani1999progress,qureshi2009scalable,yoon2014efficient,meza2012enabling})}.

\begin{figure}[t]
         \centering
         \includegraphics[width=0.8\columnwidth]{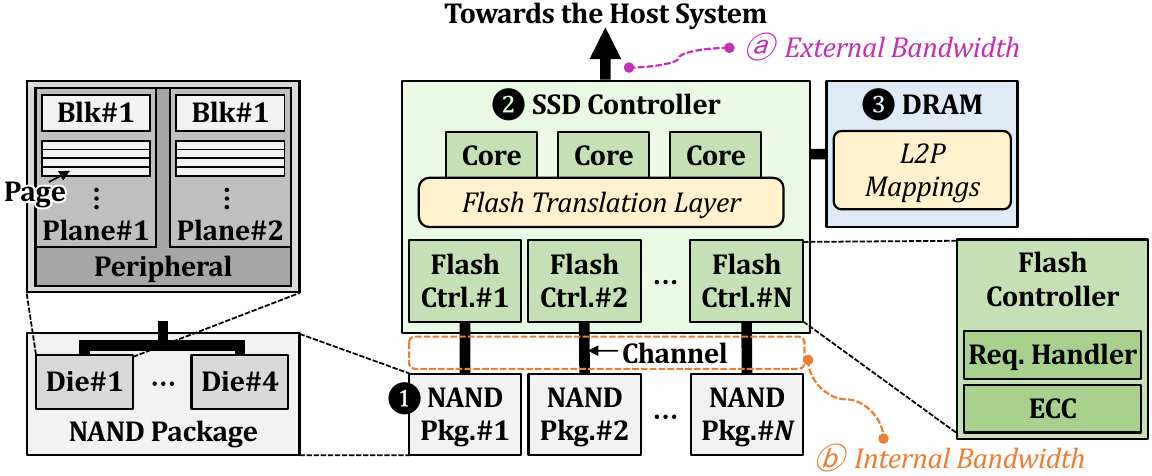}
         \caption{Organizational overview of a modern SSD.}
         \label{fig:ssd}
\end{figure}

\fig{\ref{fig:ssd}} depicts the internal organization of a modern NAND flash-based solid-state drive (SSD) that consists of three main components: 
\circled{1}~NAND flash packages, \circled{2} an SSD controller, and \circled{3} DRAM.

\head{NAND Flash Memory}
\tomi{An SSD consists of multiple \emph{NAND packages}, each} comprising multiple \emph{dies} (also called \emph{chips}) that share the package's I/O pins.  One or more packages share command/data busses (called \emph{channels}) to connect to the SSD controller. Dies sharing the same channel can operate independently of each other, but only one die can communicate with the SSD controller (e.g., for data transfer) at a time via the shared channel. A die has multiple (e.g., 2--8) \emph{planes}. Each plane contains thousands of blocks. A block includes hundreds to thousands of pages, each of which is 4--16 KiB in size. NAND flash memory performs read/write operations at page granularity but erase operations at block granularity. Planes in the same die share the peripheral circuitry used to access pages; as such, they can concurrently operate only when accessing pages (or blocks) at the same offset, which are called \emph{multi-plane} operations\tomi{~\cite{micheloni2010inside,hu2012exploring,gao2019parallel,gao2020boosting,park2022flash}}.

\head{SSD Controller}
An SSD controller has two main components: 
\inum{i}~multiple cores to run SSD firmware, commonly called the \emph{flash translation layer (FTL)}, 
and \inum{ii} per-channel hardware flash controllers for request handling and error-correcting codes (ECC) for underlying NAND flash chips. 
The FTL is responsible for communication with the host system, internal I/O scheduling, and various SSD management tasks required for hiding the unique characteristics of NAND flash memory from the host system\tomi{~\cite{\citeflash}}.
For example, a page of NAND flash memory needs to first be erased before it \tomi{can be} programmed,\footnote{This is called the \emph{erase-before-write} property~\tomi{\cite{micheloni2010inside,cho2024aero}}.} so the FTL always performs \emph{out-of-place} updates by writing the new data of a \emph{logical} page to a new \emph{physical} page that was erased previously. 
To this end, the FTL maintains logical-to-physical (L2P) address mappings\tomi{~\cite{micheloni2010inside,bjorling2017lightnvm}} for reads and performs garbage collection\tomi{~\cite{\citegc}} to reclaim new physical pages for writes.

\head{\tomi{SSD-}Internal DRAM} A modern SSD employs large low-power DRAM (e.g., 4GB LPDDR4 DRAM for a 4TB SSD~\cite{samsung860pro}) to store metadata for SSD management tasks. Most of the DRAM capacity is used to store the L2P mappings for address translation. It is common practice to maintain the L2P mappings at 4KiB granularity to provide high random access I/O performance~\cite{park-nvmsa-2018, kim-dac-2017}, so in a 32-bit architecture, the memory overhead for the L2P mappings is approximately 0.1\% of the SSD capacity (4 bytes per 4KiB data). 

\tomi{While the majority of this DRAM capacity is used to store the L2P mappings, SSD-internal DRAM is also used for several other critical functions. These include buffering and coalescing host write data, maintaining wear-leveling and garbage-collection metadata, storing bad block information, tracking open blocks and free space, and supporting firmware execution~\cite{micheloni2010inside}.}

\head{SSD I/O Bandwidth}
To mitigate the \omc{large} performance gap between main memory and the storage system, SSD manufacturers increase the external bandwidth of SSDs by employing advanced I/O interfaces between the host system and SSDs. For example, older SATA3 SSDs \tomi{from 2018} provide around 500MB/s sequential-read bandwidth~\cite{inteldcs4500, samsung860pro}, \tomi{but newer and} state-of-the-art PCIe-Gen5 SSDs \tomi{from 2025} provide significantly higher sequential-read bandwidth, up to 16 GB/s \tomi{for four-lane PCIe-Gen5 SSDs} (e.g., 14.8 GB/s in Samsung 9100 Pro SSD~\cite{samsung9100PRO}).

A modern SSD's \emph{internal} bandwidth, i.e., I/O bandwidth between NAND flash chips and SSD controller \tomi{(\wcirc{a} in \fig{\ref{fig:ssd}})} is usually higher than its external bandwidth, i.e., I/O bandwidth between the host and the SSD \tomi{(\wcirc{b})}. For example, an enterprise SSD controller~\cite{anandcontroller} \tomi{from 2020} supports 6,550MB/s external bandwidth and 19.2GB/s  internal bandwidth (16 channels, each with a bandwidth of 1.2 GB/s). In another example \tomi{from 2025}, a state-of-the-art SSD controller\cite{flashtecnvme5016} delivers 14~GB/s of external bandwidth and up to 57.6~GB/s of internal bandwidth (achieved via 16~channels \tomi{and} a peak \tomi{bandwidth} of 3.6~GB/s \tomi{per channel}). Over-provisioning the internal bandwidth is reasonable since 1) a modern SSD needs to perform various internal management tasks (e.g.,
garbage collection~\cite{\citegc}, wear-leveling~\cite{\citewearleveling}, and data refresh~\cite{\citeflashrefresh})
and 2) a higher number of channels reduces contention between requests by interleaving data between the channels.

\section{Genomic \tomi{and Metagenomic} Workflow\tomii{s}}
\label{sec:background-workflow}

\tomi{Traditional genomics analyzes genome sequences from a genomic sample, which contains an individual (or a small group of individuals) of the \emph{same known species}.
Metagenomics requires the analysis of the genome sequences of various organisms in a metagenomic sample, which contains \emph{different species} present in a common environment (e.g., human gut, soil, or oceans)~\cite{\citemetagenomicsegs}.} 

\begin{figure}[h]
  \centering
  \includegraphics[width=\columnwidth]{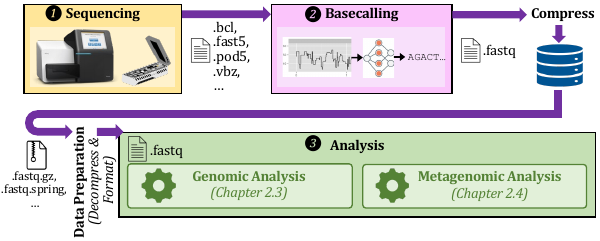}
  \caption{Overview of a typical genomic workflow.}
  \label{fig:workflow-overview} 
\end{figure}

Given a genomic \tomi{or a metagenomic} sample, a typical workflow consists of three key steps~\cite{alser2020technology,alser2020accelerating,berger2023navigating,alser2022molecules}: \inum{i}~sequencing, \inum{ii}~basecalling, and \inum{iii}~analysis. \fig{\ref{fig:workflow-overview}} shows an overview of these steps, as we will explain in the rest of this section. 

\subsection{Sequencing}

The first step, \emph{sequencing} (\circled{1} \tomi{in \fig{\ref{fig:workflow-overview}}})~\cite{\citesequencing}, converts DNA into \tomi{raw} digital signals. 
Sequencers cannot read many organisms' entire genomes as complete, error-free sequences. Instead, they produce many shorter overlapping sequences called \emph{reads} that are randomly sampled from different locations in the genome.  \tomi{To mitigate the impact of sequencing errors, multiple reads cover each genomic position.} The average number of reads per location is called the \emph{sequencing depth}. Deeper sequencing provides more information per location, which allows later analysis steps to better distinguish between true biological signals and sequencing errors.

Sequencers are distinguished by the characteristics of their reads.  Examples include sequencers that produce short reads (75 to 300 characters) with \tomi{very} high accuracy ($\sim$99.9\%)\oiii{~\cite{stoler2021sequencing,goodwin2016coming,davis2021sequencerr,sereika2022oxford}} and long reads (\oiii{typically 500} to 25k characters\oiii{, and up to $\sim$2.2M characters~\cite{jain2018nanopore,payne2018bulkvis,amarasinghe2020opportunities}}) with intermediate accuracy ($\sim$99\%)\oiii{~\cite{hon2020highly,ni2023benchmarking,wenger2019accurate, amarasinghe2020opportunities, sereika2022oxford}}. We describe the three major sequencing technologies in more detail.

\aooo{\noindent\textbf{Sequencing-by-synthesis (SBS)} sequencing~\cite{\citesbs} is a widely-used, accurate, and cost-effective sequencing technology. The SBS sequencing process has four key steps. First, the DNA molecule is fragmented into short pieces (e.g., approximately 200 bases), called \emph{reads}. To enable attaching these reads to a specialized glass slide called a \emph{flow cell}, a few nucleotides, called \emph{adapters}, are attached to each read. Second, these fragments are amplified into clusters of identical fragments to increase signal redundancy and improve sequencing reliability. Third, a complementary DNA strand from a read is synthesized base by base. To this end, during each sequencing cycle, an enzyme called \emph{DNA polymerase} incorporates a single fluorescently-labeled nucleotide into the growing complementary strand. Each base emits a distinct light color upon incorporation, thereby enabling the identification of each individual nucleotide. 
Fourth, the flow cell is imaged after every cycle, and the resulting series of images constitutes the raw sequencing data. This cycle of nucleotide incorporation and imaging is repeated for each position in the DNA fragment, progressively reconstructing its full sequence from the fluorescent signals recorded across successive cycles. Basecalling algorithms then decode these fluorescent signals into nucleotide sequences while correcting for sequencing errors~\cite{\citebasecallsbs}. SBS provides high accuracy. However, due to its cycling procedure, SBS is inherently limited to producing short reads. This limitation can pose challenges for downstream analyses such as large genome assembly and the resolution of complex genomic regions~\cite{alkan_limitations_2011, firtina_genomic_2016}.}

\aooo{\noindent\textbf{Single Molecule Real-Time (SMRT)} sequencing~\cite{\citesmrt} is a high-throughput sequencing technology that avoids the cycle-by-cycle approach of SBS, enabling the generation of longer reads (e.g., up to a few thousand bases). The SMRT sequencing process typically has five key steps. 
First, similar to SBS, DNA is fragmented into smaller pieces (i.e., reads). Second, these fragments are circularized using \emph{hairpin adapters} that form closed circular molecules, allowing the same fragment to be sequenced multiple times. Third, a single DNA polymerase enzyme is attached to each circularized template to initiate the synthesis of a complementary strand. Fourth, circularized fragments are loaded into \emph{Zero-Mode Waveguide (ZMW)} wells~\cite{levene_zero-mode_2003}, which are nano-scale structures that confine the sequencing reaction to a small volume, thereby reducing background noise and achieving a high signal-to-noise ratio. Fifth, unlike the discrete cycles in SBS, sequencing proceeds continuously in real time. As the polymerase incorporates fluorescently labeled nucleotides into the growing strand, each incorporation event emits a distinct light signal that is captured as a continuous stream of image data in movie format. SMRT sequencing supports two primary modes: \inum{i}~Circular Consensus Sequencing (CCS)~\cite{eid_real-time_2009, travers_flexible_2010}, in which the polymerase repeatedly sequences the same circular template to produce a highly accurate consensus sequence, and \inum{ii}~Continuous Long Read (CLR) mode~\cite{sharon_single-molecule_2013}, which prioritizes read length at the cost of higher error rates. Although SMRT sequencing images can be rapidly converted into nucleotide sequences, the inherent noise in the technology necessitates additional error-correction steps~\cite{\citeerrorcorrection}, including sequence alignment, consensus assembly construction, and assembly polishing.}

\aooo{\noindent\textbf{Nanopore} sequencing~\cite{\citenanopore} is a high-throughput sequencing technology that enables the sequencing of nucleic acid molecules (e.g., double-stranded DNA, dsDNA) as they pass through nano-scale pores, called \emph{nanopores}. The nanopores are embedded in a membrane that isolates the two sides. The nanopore sequencing process has three key steps. First, a voltage is applied across the membrane such that the upper part of the membrane (i.e., cis side) is negatively charged and the lower part (i.e., trans side) is positively charged, driving negatively charged DNA or RNA molecules through the nanopore from the cis to the trans side. A motor protein attached to one end of the molecule guides it toward a nanopore, separates the two strands (for dsDNA), and controls the translocation speed, which is a factor critical for sequencing accuracy. Second, as the molecule passes through the nanopore, specific sequences of k nucleotides (i.e., k-mers, with k typically between 6 and 9~\cite{samarakoon_leveraging_2024, deamer_three_2016}) partially disrupt the ionic current flowing through the pore's sensing regions, producing characteristic current changes that enable nucleotide identification. However, several sources of noise can affect measurement accuracy, including variations in translocation speed~\cite{bhattacharya_molecular_2012}, stochastic current fluctuations~\cite{deamer_three_2016}, and environmental factors such as temperature~\cite{kawano_controlling_2009}. Third, raw electrical signals are generated from each nanopore at a certain frequency (e.g., approximately 5{,}000 signals per second), typically with many reads sequenced simultaneously across multiple pores in the sequencer's flow cell. These electrical signals encode the characteristics and content of the nucleic acid molecules passing through the pores. Due to the nature of its sequencing method, nanopore sequencing enables the generation of ultra-long reads (e.g., up to a few million bases~\cite{jain2018nanopore,payne2018bulkvis,amarasinghe2020opportunities}).}

\subsection{Basecalling}
\label{sec:bg-basecalling}

The second step, \emph{basecalling} (\circled{2})~\cite{\citebasecalling,\citebasecallnanodnn,\citebasecallnanohmm}, converts the sequencer's raw signals to strings. \tomi{The basecalling process depends on the sequencing technology. For SBS sequencing signals, basecalling analyzes image intensities to identify the nucleotide sequences~\cite{\citebasecallsbs}. For SMRT sequencing signals, basecalling interprets fluorescence pulses to determine the nucleotide sequences~\cite{\citesmrt,wenger2019accurate}. In nanopore sequencing, basecalling converts the changes in the electrical signals into nucleotide sequences~\cite{\citebasecallnanodnn,\citebasecallnanohmm}.} 

Due to the noisy electrical signals, basecalling nanopore sequencing signals is more challenging compared to the more direct signal to nucleotide translation in SBS and SMRT. To improve the accuracy of nanopore basecalling, many modern basecallers propose techniques based on deep neural networks (DNNs)~\cite{\citebasecallnanodnn}.
\aooo{These models typically include convolutional neural networks (CNNs), recurrent neural networks (RNNs), or a combination of both. CNNs enable the extraction of structured features from raw signals that deep learning models can use to predict the nucleotide sequence. RNNs, often bidirectional Long short-term memory
(LSTM), capture temporal dependencies to improve prediction accuracy under noisy conditions of nanopore sequencing. To handle signal segments of varying length caused by varying translocation speed, basecalling models employ flexible decoder mechanisms such as Connectionist Temporal Classification (CTC) layers, Conditional Random Field (CRF) decoders, or a combination of both, thereby enabling accurate assignment of bases to variable-length signal windows.}

\tomi{At the end of the basecalling process,} each DNA read is converted to: \inum{i}~the DNA encoded using an alphabet representing the nucleotides (also called base pairs) A, C, G, T (along with N to represent unknowns), \inum{ii}~a quality score~\cite{ewing1998base} for each nucleotide, encoding its probability of being incorrect using an ASCII character \tomi{(the incorrect bases can arise due to the fact that sequencing signals are inherently noisy and must be computationally interpreted to infer the underlying nucleotide sequence)}, and \inum{iii}~a header. 
The strings from all reads in a sample \oii{(i.e., a read set)} are then written to a file. This \tomi{file} format is FASTQ~\cite{cock2009sanger}, the most common read set format~\cite{illuminafastq}. The file is then typically stored compressed  (\tomi{see}~\sect{\ref{sec:background-compression}}) for future analysis.

\subsection{Analysis}

The third step is \emph{analysis} (\circled{3}) on a collection of read sets.  \tomi{Since read sets are typically stored compressed\omcr{~\cite{berger2023navigating,zhu2013high,Deorowicz2013,giancarlo2013compressive,Betschart2025}}, they need to be prepared (i.e., decompressed and formatted) before analysis. A typical genomic analysis workflow quantifies mismatches between the reads and a reference genome. However, since metagenomics deals with genome
sequences whose species are not known in advance in many cases, it requires comparisons of
the target sequences against large databases of many reference genomes. \sect{\ref{sec:genomic-analysis}} and \sect{\ref{sec:background-mg}} explain genomic and metagenomic analysis in more detail, \tomiii{and \sect{\ref{sec:background-graph}} further details graph-based (meta)genomic analysis}.}

\section{\tomi{Genomic Analysis}}
\label{sec:genomic-analysis}

\tomi{In this section, we explain the key parts of genomic analysis and the importance of high-performance and energy-efficient genomic analysis.}

\subsection{Read Mapping Process}
\label{sec:background-readmapping}
A typical analysis workflow quantifies mismatches between the reads and a reference genome. Since the sequencing process does \emph{not} provide location information for reads, \tomi{genomic analysis} typically starts with the computationally-expensive \emph{read mapping} process~\cite{alser2020accelerating, alser2020technology,alser2022molecules,li2018minimap2,langmead2012fast,kim2019graph,li2013aligningsequencereadsclone}, which finds the potential matching locations of reads in the reference genome. 
For each \emph{matching location}, i.e., the location of each matching subsequence in the reference genome, a read mapper computes an \emph{alignment score}, \tomi{a numerical value that quantifies} the degree of similarity between the read and the region of the reference to which the read aligns.

Since each read is much shorter than the reference genome (e.g.,~the human reference genome contains $\sim$3.2~billion base pairs, \tomi{yet a read is usually 75--300/500-25k base pairs in short/long read sequencing technologies}), a~read mapper typically uses an index of the reference genome to reduce the search space for each read. The index is a dictionary, i.e.,~a~key-value store, where the keys are unique $k$-length subsequences (called \emph{$k$-mers}) extracted from the reference genome, and the values are \jsiv{the} exactly-matching locations of these $k$-mers in the reference genome~\cite{xin2013accelerating,xin2016optimal}. The value of $k$ is fixed during \tomi{the offline index generation step} and used for all subsequent steps.\footnote{$k$ is typically between 11 and 31~\cite{altschul1990basic,li2018minimap2,wood2019improved}, depending on the application.} To greatly reduce the storage overhead of the index and speed up queries against it, \omc{without significantly changing the final outcome of read mapping}, some read mappers index only a subset of reference genome $k$-mers called \emph{minimizers}~\cite{schleimer2003winnowing,roberts2004reducing,marccais2017improving}.  A minimizer is a \omcv{representative} $k$-mer of a set of $k$-mers according to a scoring mechanism. For example, some read mappers~\cite{li2016minimap, li2018minimap2} calculate hash values for all $k$-mers in a window of $w$ consecutive $k$-mers from an input sequence, and mark the $k$-mer with the smallest hash value as the minimizer $k$-mer.

\begin{figure}[h]
  \centering
  \includegraphics[width=0.95\columnwidth]{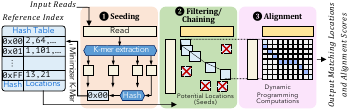}
  \caption{\tomi{Overview of a typical read mapping process.}}
  \label{fig:mapping-overview} 
\end{figure}

Read mapping is a three-step process. \tomi{\fig{\ref{fig:mapping-overview}} shows an overview of these steps, as we will explain in the rest of this section}. 

In the first step, \emph{seeding} (\circled{1} in \fig{\ref{fig:mapping-overview}}), the read mapper queries the \tomi{reference index} to determine potential locations in the reference genome where the read could map. To do so, the read mapper looks up \tomi{in the reference index} every minimizer k-mer fetched from a read. If the minimizer k-mer hits in the reference index, the read mapper marks the locations of such a k-mer in the reference genome as the read's \emph{potential matching locations}, also called \emph{seeds}.

In the second step, \emph{seed filtering} and/or \emph{chaining} (\circled{2}), the read mapper prunes \tomi{the} potential matching locations in the reference to which the read \tomi{cannot possibly} align \tomi{(due to large differences between the read and the reference at those locations)}. If all of the potential matching locations of the read get filtered \tomi{(i.e., excluded from further consideration)}, the read mapper discards the read from further analysis. The read mapper uses a dynamic programming (DP) algorithm\tomi{~\cite{\citedp}} to 1) merge overlapping seeds into longer regions, called \emph{chains}~\cite{li2018minimap2}, and 2) calculate their corresponding \emph{chaining scores,} which \tomi{are values that quantify} the approximation of the entire read's alignment score in these regions. 
If the read mapper finds one or more chains with a sufficiently high chaining score (indicating a high degree of similarity to the reference genome), then the read mapper performs the third step. If the read has no chain with a sufficiently high score, the read mapper \tomi{excludes} the read \tomi{from further consideration} and skips the third step. 

In the third step, \emph{sequence alignment}~\cite{\citemappairalign} (\circled{3}), the read mapper determines the exact differences between a read and the reference genome at the potential matching locations. Sequence alignment is done with a computationally-expensive DP algorithm~\cite{\citedp} to perform approximate string matching (ASM). Finally, the read mapper returns the locations in the reference genome with the best alignment scores for each read.

\subsection{Pre-Alignment Read Filtering}
\label{sec:background-prealignment}
To mitigate the high performance overhead of \tomi{sequence} alignment, read filtering approaches are widely used. Read filters can be incorporated at any stage of the process before alignment. There are two main filter types. 

The first filter type\tomi{~\cite{xin2013accelerating,xin2015shifted,alser2017gatekeeper, alser2017magnet,kim2018grim,alser2019shouji, bingol2021gatekeeper,alser2020sneakysnake}} aims to efficiently filter \tomi{out from consideration} potential matching locations in the reference genome that lead to a large number of edits (larger than a user-defined threshold) between the read and the reference genome at those locations. Doing so avoids a costly alignment step for potential locations at which the read would not match the subsequences of the reference genome. 

The second filter type~\cite{nag2019gencache} aims to detect if a read matches a subsequence of the reference genome with \emph{no} edits (i.e., exact-match) or \emph{very few} (e.g., 1-5) edits. Reads that satisfy this requirement are guaranteed to align to the reference genome without requiring the costly read alignment process, \tomi{and thus can be filtered (i.e., be excluded from the costly alignment process)}. This filter type is particularly effective for read sets with a large number of exactly-matching reads (e.g., ~80\% in human short read sets~\cite{10002015global,uk10k2015uk10k,nag2019gencache}).

While both filter types reduce computation overhead, they still require a large number of random memory accesses for each read.  In a typical read set of several gigabytes, read filters incur several random accesses per read 1)~to the reference index for seeding, and potentially, 2)~to the reference genome to compare the read with the subsequence of reference genome at each potential matching location.

\subsection{Further Steps in Analysis} 
\label{sec:bg-other-analysis-steps}

The result of read mapping \tomi{(matching locations and the alignment scores at these locations)} can be fed to various downstream analysis tasks (e.g., variant calling~\cite{\citevariantcallers}, genome assembly~\cite{\citeassembly}, and taxonomic classification~\cite{\citetaxclassification}). 
Analysis workflows typically access all base pairs \tomi{of each read}, but only a small fraction of \tomi{the corresponding} quality scores. This is because read mapping, while using all base pairs, typically ignores quality scores~\cite{li2018minimap2,vasimuddin2019efficient,Meng2023,kolmogorov2019assembly,Cheng2021}. The subsequent steps (e.g., variant calling) only need quality scores from the positions surrounding mismatches~\cite{Poplin2018,li2008mapping,Yu2015,Park2025} to distinguish true biological variation from sequencing errors.

\subsection{\tomi{Importance} of \tomi{High-Performance and Energy-Efficient} Genomic Analysis}
\label{sec:importance-genomics}

\tomi{The performance and energy efficiency of the analysis step is critical for three key reasons. First,}
it often begins with \emph{existing, already sequenced and basecalled read sets} because, in many cases, a single read set needs to be analyzed \emph{many times} and at \emph{different times}~\cite{aganezov2022complete,berger2023navigating,Siren2024,vaddadi2023minimizing,khayat2021hidden}. For example, some applications (e.g.,~measuring population genetic diversity~\cite{kostlbacher2021pangenomics,vandorp2020emergence,Logsdon2025,Zheng2017alignment,aganezov2022complete}) require analyzing a read set with many reference genomes at different times. Other applications require repeating the analysis many times (e.g.,~with updated or personalized references~\cite{chen2024improved,Siren2024,vaddadi2023minimizing,aganezov2022complete,berger2023navigating,rhie2023complete,nurk2022complete,kim2024airlifttcbb}) to improve accuracy. \tomi{Second, long analysis times pose critical challenges in time-critical use cases such as clinical settings~\cite{\citemgurgentclinical} and timely surveillance of diseases across a population~\cite{\citeoutbreakrapid}. Given the large throughput of modern sequencers and the growth rate of sequencing throughput being higher than Moore's Law~\cite{berger2023navigating},
the analysis throughput has emerged as a critical bottleneck~\cite{hu2021next,katz2021sra,enastats,leinonen2010sequence,wadden2022ultra}. Therefore, simply scaling up traditional systems for analysis is not efficient (both in terms of cost and energy). Third, the development of sequencing technologies that enable analysis \emph{during} sequencing~\cite{zhang2021real,firtina2023rawhash,kovaka2020targeted, mutlu2023accelerating,Payne2021,Bao2021Squigglenet,ulrich2022readbouncer,sadasivan2023rapid} increasingly necessitates the need for fast analysis that can keep up with sequencing throughput.}

Due to the criticality of the analysis step and the challenges of analyzing large amounts of sequencing reads on conventional systems, a large body of works (e.g.,~\cite{\citehwoptimization,\citegraphhwoptimization}) \tomi{propose techniques to improve performance and energy efficiency of genomic analysis}, particularly read mapping (e.g.,~\cite{\citehwmapping}), which is one of the most critical bottlenecks in many analysis applications~\cite{berger2023navigating}. \tomi{\sect{\ref{sec:related-acc-genomic}} provides a more comprehensive overview of works on accelerating genomic analysis.}

\section{Metagenomic Analysis}
\label{sec:background-mg}

\begin{figure}[h]
         \centering
         \includegraphics[width=0.8\columnwidth]{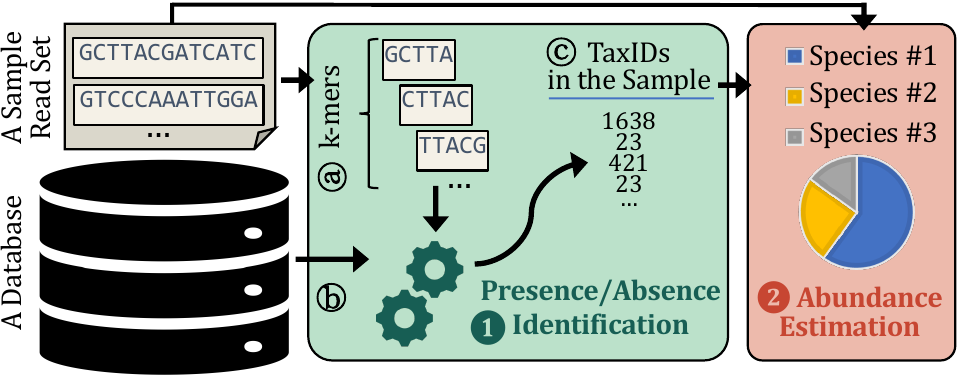}
         \caption{Overview of metagenomic analysis.}
         \label{fig:mg-overview}
\end{figure}

\tomi{In this section, we explain the key parts of genomic analysis and the importance of high-performance genomic analysis.}
While traditional genomics studies genome sequences from an individual (or a small group of individuals) of the \emph{same known species}, metagenomics requires the analysis of the genome sequences of various organisms of \emph{different species} present in a common environment (e.g., human gut, \tomi{public transport vehicles,} soil, or oceans)~\cite{\citemetagenomicsegs}. \fig{\ref{fig:mg-overview}} shows an overview of metagenomic analysis, which involves determining the species \emph{present/absent} in the sample \circled{1} and their \emph{relative abundances} (i.e., the relative frequencies of the occurrence of different species in the sample) \circled{2}.

\subsection{Presence/\tomi{A}bsence Identification} 

To find \tomi{the} species present in the sample \tomi{(\circled{1} in \fig{\ref{fig:mg-overview}})}, many tools (e.g.,\tomi{~\cite{wu2021sieve,wood2019improved,wood2014kraken,truong2015metaphlan2,kim2016centrifuge,song2024centrifuger,ounit2015clark,piro2016dudes,Fan2021,piro2020ganon,Marcelino2020,milanese2019microbial,lapierre2020metalign,pockrandt2022metagenomic,Dilthey2019,lemane2023kmindex,shen2022kmcp}}) extract \emph{k-mers} (i.e., subsequences of length $k$) from the input queries in a sample read set (\wcirc{a} in \fig{\ref{fig:mg-overview}}) and search for the k-mers in an input \emph{reference database} (\wcirc{b}). Each database contains k-mers extracted from reference genomes of a wide range of species. The database associates each indexed k-mer with a \emph{taxonomic identifier (taxID})\footnote{A taxID is an integer attributed to a cluster of related species.} of the reference genome(s) the k-mer comes from. At the end of the presence/absence identification process, the metagenomic tool outputs the taxIDs of the species present in the sample (\wcirc{c}). 

The GB- or TB-scale databases typically support random (e.g.,~\cite{wu2021sieve,wood2019improved,wood2014kraken,truong2015metaphlan2,kim2016centrifuge,song2024centrifuger,ounit2015clark,piro2016dudes,Fan2021,piro2020ganon,Marcelino2020,milanese2019microbial}) or streaming (e.g.,~\cite{lapierre2020metalign,pockrandt2022metagenomic,Dilthey2019,lemane2023kmindex,shen2022kmcp}) access patterns. \tomi{To gain a deeper understanding of data movement overheads in metagenomics, we discuss these access patterns in more detail.}

\head{Tools with Random Access Queries (\randomio)} Some \tomi{presence/absence identification} tools (e.g.,~\cite{wu2021sieve,wood2019improved,wood2014kraken,truong2015metaphlan2,kim2016centrifuge,song2024centrifuger,ounit2015clark,piro2016dudes,Fan2021,piro2020ganon,Marcelino2020,milanese2019microbial}) commonly perform random accesses to search their database. A state-of-the-art tool in this category is Kraken2~\cite{wood2019improved}, which maintains a hash table that maps each indexed k-mer to a taxID. To identify which species are present in a set of queries, Kraken2 extracts k-mers from the read queries and searches the hash table to retrieve the k-mers' associated taxIDs. For each read, Kraken2 collects the taxIDs of that read's k-mers and, based on the occurrence frequencies of these taxIDs, uses a classification algorithm to assign a single taxID to each read. Finally, Kraken2 identifies the species present in the sample based on the taxIDs of the reads in the sample.

\head{Tools with Streaming Access Queries (\streamio)}
Some \tomi{presence/absence identification} tools (e.g.,~\cite{lapierre2020metalign,pockrandt2022metagenomic,Dilthey2019,lemane2023kmindex,shen2022kmcp}) predominantly feature streaming accesses to their databases. A state-of-the-art tool in this category is Metalign~\cite{lapierre2020metalign}. Presence/absence identification in Metalign is done via 1) preparing the input read set queries, and 2) finding species present in them. To process the queries, the tool extracts k-mers from the reads and sorts them. Finding the species in the sample involves two steps.  First, the tool finds the \emph{intersecting k-mers}, which are k-mers that are common between the query k-mers and a pre-sorted reference database. In this step, the tool uses large k-mers (e.g., $k=60$) for both the queries and the database to maintain a low false positive rate. This is because large k-mers are more unique, and matching a long k-mer ensures that the queries have at least one long and specific match to the database. Second, the tool finds the taxIDs of the intersecting k-mers by searching for the intersecting k-mers or their prefixes in a smaller \emph{sketch database} of variable-sized k-mers. Each \emph{sketch} is a small representative subset of k-mers associated with a given taxID. Searching for both the intersecting k-mers and their prefixes in this step increases the true positive rate (i.e., species correctly identified as present in the sample out of all species actually present in the sample) by expanding the number of matches. 

\subsection{Abundance Estimation}

After finding the taxIDs of the species present in the sample, some applications require a more sensitive step to find the species' relative abundances~\cite{sun2021challenges,lu2017bracken,koslicki2016metapalette,piro2016dudes,truong2015metaphlan2,shen2022kmcp,Fan2021,lapierre2020metalign,kim2016centrifuge,song2024centrifuger,dimopoulos2022haystac} in the sample \tomi{(\circled{2} in \fig{\ref{fig:mg-overview}})}. Different tools implement their own approaches for estimating abundances, from lightweight statistical models~\cite{lu2017bracken,dimopoulos2022haystac,koslicki2016metapalette} to more accurate but computationally-intensive read mapping~\cite{lapierre2020metalign,kim2016centrifuge,milanese2019microbial,piro2016dudes,Fan2021,truong2015metaphlan2}. 
\tomi{Lightweight statistical models use the taxIDs assigned to each read, along with the relationships between taxIDs encoded in the taxonomic tree, to assign species to each read, determining the number of reads belonging to each species.}
Read mapping is the process of finding potential matching locations of reads against one or more reference genomes. Metagenomic tools can map the reads against reference genomes of species in the sample, accurately determining the number of reads belonging to each species.

\subsection{\tomi{Importance} of \tomi{High-Performance and Energy-Efficient} Metagenomic Analysis}
\label{sec:importance-metagenomics}

By enabling the analysis of the genomes of organisms from different species in a common environment, metagenomics overcomes a limitation of traditional genomics, which requires culturing individual known species in isolation. This limitation has been a major roadblock in many clinical and environmental use cases~\cite{national2007new}. The impact of metagenomics has been rapidly increasing in many areas that each has broad implications for society, such as
\tomi{precision medicine~\cite{\citemgprecision}, urgent clinical settings~\cite{\citemgurgentclinical}, understanding microbial diversity of an environment~\cite{\citemgbiodiversity}, discovering early warnings of communicable diseases~\cite{\citemgcommunicable}, outbreak tracing~\cite{\citemgtracing},} agriculture~\cite{cdcpulsenet}, and many other critical areas.
Due to its importance, metagenomics has attracted wide global attention, with medical and government health institutions heavily investing in metagenomic analysis~\cite{downie2023surveillance,cdcpulsenet,cdcamd}. The global amount of genomic data that is incorporated in metagenomic workflows is growing exponentially~\cite{stephens2015big,Nasko2018}, doubling every several months~\cite{CheckHayden2015,enastats,ntdouble,Nasko2018}, and is projected to surpass the data growth rate of various major internet media platforms~\cite{wu2021sieve,stephens2015big,CheckHayden2015}.

In metagenomics, the analysis step bottlenecks the end-to-end performance of the workflow, and therefore, poses a pressing need for acceleration~\cite{wu2021sieve,hanhan2022edam,chiang2019from,edgar2022petabase,taxt2020rapid,sereika2022oxford} for three reasons. First, the sequencing and basecalling steps for a sample are usually one-time tasks~\cite{wang2021nanopore,hu2021next,alser2022molecules}. In many cases, the reads from a single sequenced sample can be analyzed by \emph{multiple studies} or \emph{at different times} in the same study. This is because
\inum{i}~there are many heuristics \tomi{and parameters} involved in metagenomics, \tomi{which need to be tuned to} achieve a desired sensitivity-specificity tradeoff~\cite{bokulich2020measuring}, \inum{ii}~there are different databases created with different parameters or genomes \tomi{that need to be used for achieving different accuracy goals}~\cite{schuele2021future}, and \inum{iii} a sample can be analyzed several times with databases that are regularly updated with new genomes, or with syndrome-specific targeted databases~\cite{schuele2021future}. 

Second, even when performing the metagenomic analysis step only once for a sample, the \emph{throughput} of this step is significantly lower than the sequencing throughput of modern sequencers (e.g.,~\cite{illuminax}). While sequencing one sample can take a long time,  a single sequencing machine can sequence \emph{many samples} from different sources in parallel~\cite{hu2021next,shokralla2015massively}, achieving very high throughput.  Our analysis with a state-of-the-art metagenomic tool~\cite{lapierre2020metalign} shows that analyzing the data, sequenced and basecalled by a sequencer in 48 hours, takes 38 days on a high-end server node (detailed configurations in Chapter~\ref{chap:megis}). Such long analysis poses serious challenges, specifically for time-critical use cases \tomi{that require fast decision making} (e.g., clinical settings~\cite{taxt2020rapid} and timely surveillance of infectious diseases~\cite{hadfield2018nextstrain}). Since the growth rate of sequencing throughput is higher than Moore's Law\tomi{~\cite{berger2023navigating,moore1998cramming}}, this already large gap between sequencing and analysis throughput is widening~\cite{hu2021next,katz2021sra,enastats,leinonen2010sequence}, and simply scaling up traditional systems for analysis is not efficient.

Third, the development of sequencing technologies that enable analysis \emph{during} sequencing~\cite{zhang2021real,firtina2023rawhash,kovaka2020targeted, mutlu2023accelerating,Payne2021,Bao2021Squigglenet,ulrich2022readbouncer} increasingly necessitates the need for fast analysis that can keep up with sequencing throughput. \tomi{The possibility to perform analysis during sequencing enables new features, where sequencing can be dynamically stopped once a target pathogen is identified, or sufficient coverage is reached~\cite{\citesignalanalysis}. 
Therefore, to achieve this feature, the analysis throughput must match or exceed the sequencing throughput. Otherwise, the analysis step bottlenecks the entire process and makes this feature impractical in time-critical settings.}

The analysis step is also the primary energy bottleneck in the metagenomic workflow, and optimizing its efficiency is vital as sequencing technologies rapidly evolve. For example, a high-end sequencer~\cite{illuminax} uses 405 KJ to sequence and basecall 100 million reads, with 92.5 Mbp/s throughput and 2,500 W power consumption~\cite{illuminax}. In contrast, processing this dataset on a commodity server (detailed configurations in Chapter~\ref{chap:megis}) requires 675 KJ, accounting for 63\% of the total energy \tomi{of the metagenomic workflow}. The need to enhance the analysis' energy efficiency is further increasing for two reasons.
First, sequencing efficiency has been continually improving. For example, a new version of Illumina sequencer from 2023~\cite{illuminax} provides 44$\times$ higher throughput at only 1.5$\times$ higher power consumption compared to an older version~\cite{illumina} from 2020, resulting in much better sequencing energy efficiency. Therefore, simply relying on scaling up commodity systems to improve the analysis throughput worsens the energy bottleneck. Second, the increased adoption of compact \emph{portable sequencers}~\cite{jain2016oxford} for on-site metagenomics (e.g., in remote locations~\cite{pomerantz2018real} or for personalized bedside care~\cite{chiang2019from}) offers high-throughput sequencing with low energy costs. \tomi{The drive towards portability} further amplifies the need for energy- and cost-effective analysis that can match the portability and convenience of these sequencers.

\section{Graph-Based Genomic and Metagenomic Analysis}
\label{sec:background-graph}

\noindent\textbf{Sequence Graphs}
have emerged as a powerful approach for representing and querying massive and complex genomic databases \tomiii{(e.g., containing reference genomes from many species or previously sequenced samples)}\tomiv{~\cite{cali2022segram,marchet2021data,karasikov2020metagraph,fan2023fulgor,karasikov2022lossless,karasikov2019sparse,danciu2021topology,bradley2019ultrafast,Muggli2017SuccinctGraphs,Iqbal2012,turner2018integrating,hunt2024allthebacteria,bingol2026debruijn}}. Unlike traditional models \tomiii{of representing genomic sequences}, which represent genomic databases as a collection of disjoint, linear sequences, sequences in a graph are represented by graph walks\omiii{~\cite{marchet2021data,Iqbal2012}}. Genome graphs collapse subsequences shared across different database entries into single nodes and indicate adjacent subsequences with edges, compacting sequence sets (by up to thousand-fold~\cite{karasikov2020metagraph}).
A node's metadata then indicates which entries contain the represented sequence.

Sequence graphs provide two fundamental benefits, which make them indispensable, particularly in modern, population-scale genomics~\cite{danko2021global,karasikov2020metagraph,siren2021pangenomics,Sherman2020,taylor2024beyond}.
First, the graph topology, along with its metadata, offer \emph{\textbf{\omiii{vast} expressive power}}~\cite{eizenga2020pangenome,taylor2024beyond,Liao2023,Sherman2020,Mustafa2024MLA}. A graph naturally encodes the evolutionary history and diversity of a cohort of organisms, revealing their shared and distinct sequences~\cite{eizenga2020pangenome,Armstrong2020cactus,bradley2019ultrafast,Sherman2020}. 
This leads to reduced bias and improved accuracy in analyses. For example, genome graphs improve disease diagnosis by capturing population-specific variants often missed by analyses on traditional, linear sequences~\cite{Groza2024,Sherman2020,jain2019pasgal,Negi2025}. Second, genome graphs leverage the inherent redundancy of genomic data to \emph{\textbf{avoid redundant computation}}.  This is because shared genomic regions across database entries are stored only once and do not need to be queried separately. This is essential in settings such as population-scale pathogen surveillance\tomiii{~\cite{colquhoun2021pandora,alipanahi2021succinct,weisberg2021genomic}}, where public health systems must rapidly match new pathogen genomes against large national databases~\cite{danko2021global,karasikov2020metagraph,alipanahi2020metagenome,hunt2024allthebacteria,Vieira2024,gangwar2025wepp}, enabling timely detection of local outbreaks and limiting broader spread\tomiii{~\cite{Quick2016, besser_interpretation_2019, li_application_2021,parkins2024wastewater, polo2020making, downie2023surveillance,bloemen_development_2023, nieuwenhuijse2017metagenomic}}.

\begin{figure}[b]
         \centering
\includegraphics[width=0.93\columnwidth]{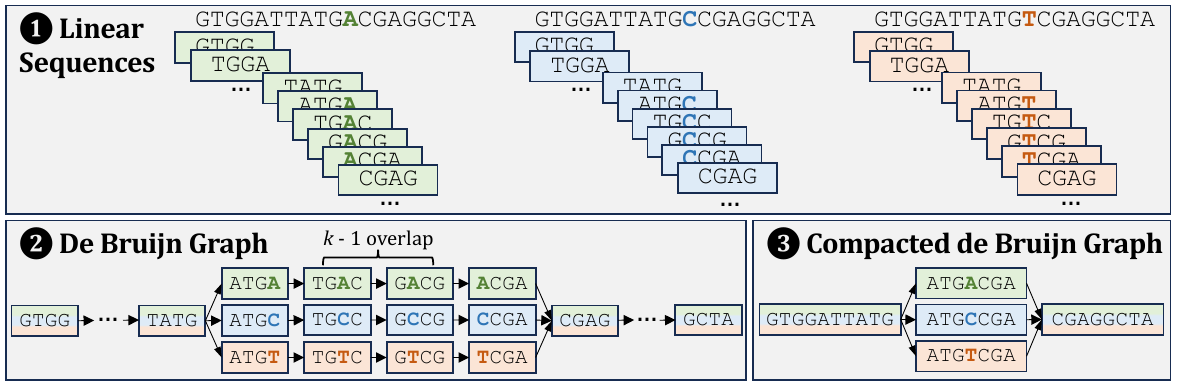}        
  \caption{\tomiii{An example de Bruijn graph \tomiii{(DBG)} and its compacted form.}}
\label{fig:dbg}
\end{figure}

\head{De Bruijn Graphs} To efficiently represent large-scale genome graphs, \emph{de Bruijn graphs} (DBGs), which are a type of overlap graph\omiii{~\cite{cali2022segram,marchet2021data,karasikov2020metagraph,fan2023fulgor,karasikov2022lossless,karasikov2019sparse,danciu2021topology,bradley2019ultrafast,Muggli2017SuccinctGraphs,Iqbal2012,turner2018integrating,pevzner_eulerian_2001,mustafa2022algorithms}}, have been adopted by many recent works~\cite{Muggli2017SuccinctGraphs,fan2023fulgor,alanko2023themisto,cracco2023extremely,karasikov2022lossless,danciu2021topology,karasikov2019sparse,karasikov2020metagraph,bradley2019ultrafast,Iqbal2012,turner2018integrating}.
While other sequence graph classes exist~\cite{Baaijens2022,paten2017genome,Armstrong2020cactus},
DBGs scale much better
 \omiii{with both the entries' size and genetic diversity of genomic} databases
\cite{karasikov2020metagraph,bvrinda2023efficient,Chikhi2013,karasikov2019sparse}, which is why they are preferred in these contexts. \fig{\ref{fig:dbg}} shows an example of \tomiii{\circled{1}~linear sequences, \circled{2}~their corresponding DBG, and \circled{3}~the corresponding compacted DBG.}
In a DBG, each node is a unique k-mer that is observed in the input, and there is a directed edge from node $u$ to node $v$ if the suffix $(k-1)$-mer of $u$ is equal to the prefix $(k-1)$-mer of $v$. 
Each node is then assigned a \emph{color}, representing its metadata. 
Example metadata can be the IDs of the database entries from which the k-mer originates, the occurrence frequencies of the k-mer~\cite{karasikov2022lossless,fan2023fulgor,marchet2020reindeer}, and \tomiii{the k-mer's} exact locations~\cite{karasikov2022lossless,vandamme2024tinted,almodaresi2018space,alipanahi2020metagenome}. 
A common strategy to reduce a DBG's size is to instead store its corresponding compacted DBG, where all $k$-mers from a maximal non-branching path are merged into a single node. 
There are two common ways to represent (compact\tomiii{ed}) DBGs~\cite{alanko2023small}.
A \emph{node-centric} DBG of order $k$ (e.g.,~\cite{fan2023fulgor,pibiri2022sparse,alanko2023themisto}) stores all $k$-mers observed in the entries and defines its edges implicitly (\fig{\ref{fig:dbg}}). An \emph{edge-centric} DBG of order $k$ (e.g.,~\cite{Bowe2012SuccinctGraphs,Muggli2017SuccinctGraphs,li2015megahit,karasikov2020metagraph,karasikov2022lossless,danciu2021topology,karasikov2019sparse}) stores observed $(k-1)$-mers as nodes and only stores the edges that represent observed $k$-mers. In the rest of this dissertation, we refer to compact\tomiii{ed} DBG as DBG for short.

\noindent\textbf{{\asp{Graph-based sequence analysis}}} follows \tomiii{the same high-level goal as the conventional analyses described in \sects{\ref{sec:genomic-analysis} and \ref{sec:background-mg}}: it involves 
querying reads \tomiii{from a sample} against a set of sequences (either genomes of the same species, or a database of genomes of different species or already sequenced samples)  to determine what species or genomic features are present in the sample~\cite{karasikov2020metagraph,alanko2023themisto,Muggli2017SuccinctGraphs,fan2023fulgor,piro2020ganon,mangul2016reference} and/or to find their differences from genomes in the database~\cite{aganezov2022complete,Groza2024}. The key difference lies in how the reference sequences and databases are represented and queried.  In conventional, linear sequence-based analysis, reference sequences and databases are stored as one or more independent linear sequences. Consequently, analysis identifies candidate positions of input query reads in these linear sequences and, when needed, aligns each read against the corresponding sequence regions. In graph-based analysis, reference sequences and databases are encoded as graph walks. Therefore, a read is matched not simply to a position in a linear sequence, but to one or more nodes or walks through the graph. The metadata associated with the matched nodes can additionally indicate which genomes, species, or database entries contain those sequences.}

\begin{figure}[b]
         \centering
         \includegraphics[width=0.9\columnwidth]{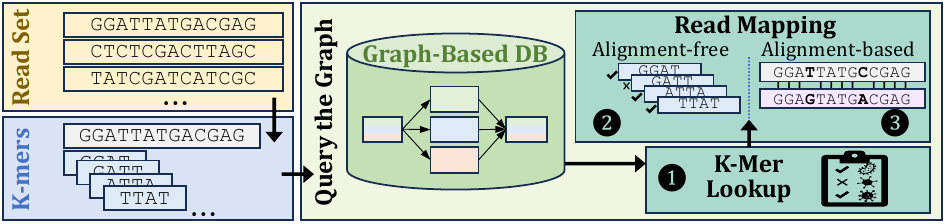}
         \caption{\asp{High-level overview of graph-based genome analysis.}}
         \label{fig:analysis-bg}
\end{figure}

As shown in \fig{\ref{fig:analysis-bg}}, queries fall into two categories: \circled{1} \emph{k-mer set lookups} (where k-mers in a read set are matched against the graph to determine the species present in a sample), and \emph{read mapping} (where each read is analyzed individually). Read mapping can be done by either \circled{2} with the alignment step, i.e., alignment-based \tomiii{(e.g.,~\cite{Mustafa2024MLA,karasikov2022lossless,karasikov2020metagraph,siren2021pangenomics,garrison2018vg,chang2025rapid,rautiainen2020graphaligner,Rautiainen2019})} or \circled{3} without the alignment step, i.e., alignment-free \tomiii{(e.g.,~\cite{karasikov2020metagraph,fan2023fulgor,campanelli2024where,Muggli2017SuccinctGraphs,Holley2020bifrost,marchet2021data,campanelli2025fast,alanko2023themisto})}. Alignment-free approaches (more common for large-scale studies~\cite{karasikov2020metagraph,karasikov2022lossless}) match all k-mers of a read to the graph, retrieve the metadata associated with the graph nodes to which the k-mers match, and classify the read using this metadata. Alignment-based approaches introduce an additional, computationally expensive approximate string matching step, where candidate regions identified by k-mer matches are refined via dynamic programming to determine precise differences between the read and the graph.

\subsection{\tomiii{Importance of High-Performance and Energy-Efficient Graph-Based Genomic Analysis}}
\label{sec:importance-graph-genomics}

\tomiii{Due to similar reasons explained in \sects{\ref{sec:importance-genomics} and \ref{sec:importance-metagenomics}}, the performance and energy efficiency of the analysis step in graph-based genome analysis is critical. Due to this criticality and the challenges of analyzing large amounts of
sequencing reads on conventional systems, a large body of work (e.g.,~\cite{\citegraphhwoptimization}) proposes techniques to improve the performance and energy efficiency of graph-based genomic analysis.}

\subsection{Unique Characteristics of Genome Graphs}
\label{sec:graph-uniqueness}
The type of data encoded in genome graphs and the underlying goals of querying genome graphs exhibit unique features not encountered in conventional, non-genome graphs (e.g., social networks, web graphs, road networks). General-purpose graph data structures and methods do not account for these features, leading to significant inefficiencies in storage and computation requirements and missed opportunities. This has led to significant efforts in both software (e.g.,~\cite{rautiainen2020graphaligner,kim2019graph,gao2020abpoa,jain2019pasgal,siren2021pangenomics,Rautiainen2019,Chandra2023,Ivanov2022,Ma2023,Darby2020vargas,Hwang2025MEMO,Romain2023svjedi,Li2020minigraph}) and hardware (e.g.,~\cite{cali2022segram,Zhang2024Harp,Zeng2024asgdp,Shen2024128parallel,Li2024,Mandal2020,Varma2013,Awan2021,Feng2021,Zhang2025,kim2025nmp,Huang2023meg2,Angizi2020Panda,Qiu2017}) to develop specialized techniques for genome graphs.

The unique semantic characteristics in genome graphs manifest in three key aspects regarding the graph structure,  storage (e.g., sparsity and degree distribution), and access patterns. First, in a traditional graph with $N$ nodes, there are $\mathcal O(N)$ to $\mathcal O(N^2)$ edges that must all be stored explicitly (e.g., adjacency lists\omiii{~\cite{cormen2009introduction,hopcroft1973algorithm}}, \omiii{compressed sparse row formats~\cite{ligra,merrill2012scalable}}). 
Genome graphs, most commonly represented as DBGs, are structurally constrained in that they either store nodes or edges and the other is obtained implicitly. This is a feature necessitated by the massive scale of genome graphs and enabled by the inherent overlap structure of the sequences they encode.

Second, traditional graphs typically exhibit power-law distributions, \omiii{in which a small number of nodes have very high degree and many graph-processing optimizations (e.g.,~\cite{chi2022accelerating,Dadu2021polygraph,yao2022scalagraph}) exploit this skew}. However, DBG node outdegrees and indegrees \omiii{are} at most four (determined by the fixed alphabet size, \texttt{A}, \texttt{C}, \texttt{G}, \texttt{T}, in the underlying data) regardless of the overall graph size. Therefore, optimizations that exploit skewed degree distributions, a common strategy in conventional graph processing systems, are \omiii{not well-suited} to genome graphs, necessitating different optimization approaches.

Third, due to the differences in underlying graph structures, access pattern optimization opportunities vary between genome graphs and non-genome graphs. The compressed data structures used in large-scale genome graphs enforce specific node orderings and, thus, \omiii{restrict} node reordering, a technique commonly used in non-genome graphs to improve access patterns~\cite{Bowe2012SuccinctGraphs,pibiri2022sparse}. However, the fact that genome graphs encode biological sequences, and that queries are themselves biological sequences, opens new avenues for improving access pattern locality that are unavailable in conventional graph workloads. For example, k-mers from different query reads that share common substrings map to nearby regions in the graph's index structures. In contrast, most conventional non-genome graph queries do not exhibit this kind of structural coupling between the queries and the graph, limiting opportunities for analogous locality-aware optimizations.

\section{Storing \tomiii{and Compressing} Sequence Data}
\label{sec:background-compression}

\nh{\head{\tomiii{Importance and Growth} of Sequence Data}} Driven by continuing exponential drops in sequencing costs~\cite{nhgri,pedro2021integration} and the pivotal importance of genomic and metagenomic sequence analysis, 
 \nh{sequence} data volumes 
in public~\cite{katz2021sra,srastats,enastats} and private~\cite{Bick2024,Li2023whole} repositories 
are growing by an \emph{order of magnitude} every few years, approaching and expected to exceed the data volumes generated by various major \oii{internet} media platforms~\cite{stephens2015big,srastats,katz2021sra,enastats}.
While \oii{genomic analysis also works on} other data types (e.g., assembled genomes\oii{~\cite{Garrison2024,Armstrong2020cactus,Minkin2020}}, epigenetic markers\oii{~\cite{Chen2025,Liu2025,Sigurpalsdottir2024,Li2011}}, and 3D chromosomal contact maps\oii{~\cite{Forcato2017,Pal2019,lieberman2009comprehensive}}), 
analyzing \tomiii{genomic} sequence data is one of the most critical processes in genomics, with sequence data constituting the largest fraction of data stored and analyzed~\cite{ncbi2025}. For example, sequence data is the \emph{largest} class of data maintained at institutions like the National Center for Biotechnology Information and the European Molecular Biology Laboratory~\cite{thakur2023embl,ncbi2025}. 
Note that many other genomic data types (e.g., assembled genomes, epigenetic markers, and 3D chromosomal contact maps) are also derived from sequence data, so it is critical to store this sequence data to ensure both reproducibility and to enable reanalysis with future improved workflows. 
For example, while some tasks (e.g., variant calling \tomiii{as explained in \sect{\ref{sec:bg-other-analysis-steps}}}) can start with already mapped reads, best-practice guidelines dictate that reads should be remapped during the course of a study to ensure the use of appropriate (updated\ov{~\cite{chen2024improved,aganezov2022complete,berger2023navigating,Siren2024,rhie2023complete,nurk2022complete}} or personalized\ov{~\cite{Siren2024,vaddadi2023minimizing}}) reference genomes and mapping parameters.
This is reflected in the fact that, as of October 2025, 75.9\% (18 peta DNA bases) of publicly-deposited whole-genome sequencing read sets~\cite{katz2021sra,ncbi2025} are in unmapped, i.e., FASTQ format~\cite{cock2009sanger}, and their overall size has been growing at a rate of 42.5\% on average per year over the last decade.

\head{Genomic-Specific Compression} Due to the importance of storing genomic data space-efficiently, there exist many compression techniques (e.g.,\tomiii{~\cite{chandak2018spring,chandak2017compression,kokot2022colord,Meng2023,roguski2018fastore,dufort2021renano,kowalski2019pgrc,dufort2020enano,dragenora,yang2025gpufastqlz,chen2023efficient,hach2012scalce,roguski2014dsrc2,Deorowicz2020,lan2021genozip,alyami2019lfastqc,cogo2021genodedup,Sun2023,Nazari2025}}) specialized for genomic \tomiii{read sets}. These techniques attain higher compression ratios (e.g.,~typically from $\sim$2 to $\sim$40~\cite{chandak2018spring,roguski2018fastore,dufort2020enano,dufort2021renano}) than general-purpose compressors (e.g.,~typically from $\sim$2 to $\sim$6~\cite{chandak2018spring}), because general-purpose approaches fail to leverage longer-range similarities common in DNA data~\cite{hernaez2019genomic,zhu2013high}. 
This large gap in compression ratios exists even with state-of-the-art general-purpose compressors in software (e.g.,~\cite{collet2018zstandard,pavlov20167,Brotli,Katz1991US5051745A,goyal2021dzip,goyal2018deepzip,chen2024ha}) or hardware (e.g.,~\cite{bartik2015lz4,liu2018data,fowers2015scalable,chen2021fpga,angerd2022gbdi,gao2024beezip,karandikar2023cdpu,9499902,abali2020data}), which is the reason for the large emphasis in genomics to use genomic-specific compressors~\cite{hernaez2019genomic,dragenora}.
For example, hardware-based \tomiii{general-purpose} Intel QAT~\cite{intelqat} and IBM zEDC~\cite{ibmzedc} compression accelerators lead to $2.6 \times$ and $2.7 \times$ lower average compression ratios, respectively, than genomic compressors~\cite{chandak2018spring,Meng2023} on our evaluated datasets (Chapter~\ref{chap:sage}). Similarly, state-of-the-art \tomiii{software-based} general\tomiii{-purpose} compression algorithms such as xz~\cite{xz} (a state-of-the-art LZMA-based compressor) and zstd~\cite{collet2018zstandard}, both at their highest levels, lead to 2.13$\times$ average (up to 6.7$\times$) worse compression ratios than genomic compressors~\cite{chandak2018spring,Meng2023} on our datasets. 

\tomiii{A genomic read set consists of two key parts: DNA bases and quality scores (as detailed in \sect{\ref{sec:bg-basecalling}}).} 
Since quality scores do not have the same redundancy patterns as DNA bases, state-of-the-art genomic compressors process them in separate streams and compress them using various techniques (e.g., lossless context models~\cite{dufort2020enano,kokot2022colord,Deorowicz2020,alyami2019lfastqc,cogo2021genodedup} or block sorting~\cite{chandak2018spring}, and lossy quantization~\cite{chandak2017compression}).
Some tools (e.g.,~\cite{Meng2023,vandamme2024tinted,karasikov2020metagraph}) discard quality scores altogether since many recent workflows (e.g.,~\cite{li2018minimap2,wood2019improved,song2024centrifuger,kolmogorov2019assembly,colquhoun2021pandora}), particularly those using accurate reads, do not use quality scores.

\begin{figure}[t]
  \centering
  \includegraphics[width=0.9\columnwidth]{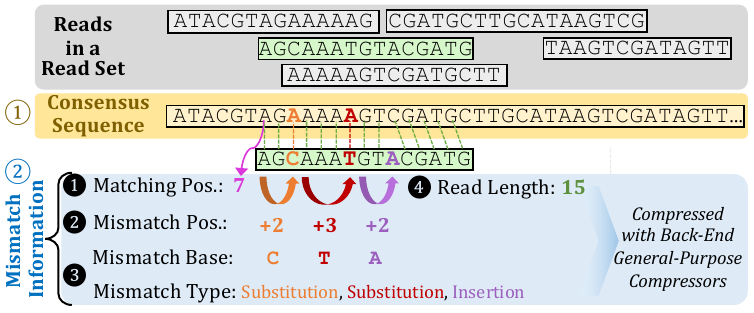}
  \caption{\new{Overview of genomics-specific compression.}}
  \label{fig:genomic-compression} 
\end{figure}

\fig{\ref{fig:genomic-compression}} shows an example of a typical genomic compression technique \tomiii{for DNA bases}. 
Genomic compressors typically represent \nh{the DNA bases in} each read set with a \wcirc{1}~\emph{consensus sequence} and \wcirc{2}~the \emph{mismatches} of each read in the read set compared to that consensus \tomiii{sequence}~\cite{chandak2017compression,dufort2021renano,kokot2022colord,kowalski2019pgrc}. A consensus sequence is an approximation of the organism's genome, and can either be a user-provided reference~\cite{dufort2021renano} or a de-duplicated string derived from the reads, representing the most likely character at each location~\cite{chandak2017compression,kowalski2019pgrc}. Note that it is not sufficient to store only the consensus \tomiii{sequence} to represent the read set because the consensus \tomiii{sequence} alone does not capture the \oii{variations} in the original reads. Individual reads often contain unique differences, including sequencing errors~\cite{goodwin2016coming,wang2021nanopore,stoler2021sequencing,Ma2019} or biological variation~\cite{Cheng2022}.

Given a consensus sequence, a lossless encoding of a read's DNA bases consists of \circled{1}~the read's matching position in the consensus sequence (many compressors, e.g.,~\cite{chandak2017compression,Meng2023,roguski2018fastore,kowalski2019pgrc}, store matching positions based on their order of appearance in the consensus, which enables space-efficient delta-encoding),
\circled{2}~mismatch positions (delta-encoded), 
\circled{3}~mismatch bases and types (i.e., substitution, insertion, deletion), and
\circled{4}~read length (especially for long reads since they have variable lengths). This encoding of the DNA bases is then more compressible using general-purpose compressors \tomiii{(e.g.,~\cite{chandak2017compression})}, which are then used by the state-of-the-art genomic compressors to further compress the mismatch information.

\head{Compression Does Not Replace Read Mapping} Note that identifying matching positions \tomiii{of a read} against the consensus \tomiii{sequence} during compression in genomic-specific compressors is independent from and does \emph{not} eliminate the need for read mapping operations during subsequent analysis stages~\cite{Siren2024,Sedlazeck2018,Abel2020}. This is due to three complementary reasons.  First, during compression, the exact reference genomes required for all potential future mapping analyses may not be known~\cite{Siren2024}.  Second, as explained in \sect{\ref{sec:background-workflow}}, a read set may undergo mapping multiple times or at different times, \nh{with different reference genomes or mapping parameters}. Therefore, simply having the matches against a single consensus \tomiii{sequence} is insufficient. Third, during compression, only one matching position for each read is stored (to minimize the size of the read set), whereas various mapping use cases require many matching positions of each read~\cite{Sedlazeck2018,Abel2020,li2018minimap2}.

\tomiii{\section{Summary \tomiv{and Outlook}}

Genomic analysis examines the genomic information of living organisms and other biological entities.
Traditional genomics focuses on the genome sequences of an individual (or a small group of individuals) from a single known species, whereas metagenomic analysis examines genome sequences from multiple species in a shared environment.
Genomic and metagenomic analyses play critical roles in many fields, such as precision medicine, 
urgent clinical settings, and discovering early warnings of communicable diseases, ensuring food safety through pathogen monitoring, agriculture, and scientific discovery.
The adoption of genomic and metagenomic analyses has been rapidly increasing in recent years, driven by their critical importance and the rapid advancement of sequencing technologies. These factors have led to exponential growth in sequence data generation. Due to the challenges of analyzing and storing massive volumes of this data, significant efforts have been made in accelerating (meta)genomic analyses and storing sequence data in compressed forms. Despite the benefits of these techniques in improving the analysis and storage of large-scale sequence data, we identify two major outstanding problems in accessing stored sequence data and feeding it to the analysis units: \inum{i} the overhead of moving large amounts of low-reuse data from the storage system and the unnecessary burden on the rest of the system (e.g., main memory and computation units), and \inum{ii} the data preparation bottleneck, where compressed sequence data needs to be first decompressed and formatted before it can be analyzed.

The goal of this dissertation is to \inum{i}~alleviate the data movement overheads of genomics and metagenomics analyses from the storage system and reduce the computational burden from the rest of the system, and \inum{ii}~mitigate the data preparation bottleneck while achieving high performance, energy efficiency, and compression ratios. We aim for our proposed designs to be lightweight such that they can seamlessly integrate with a broad range of genomic and metagenomic analysis systems.

To alleviate data movement overheads of moving large amounts of low-reuse sequence data from the storage system and reduce the overall computational burden from the rest of the system, we propose \emph{storage-centric computing} (SCC) systems for genomic analysis (Chapter~\ref{chap:genstore}), metagenomic analysis (Chapter~\ref{chap:megis}), and graph-based (meta)genomic analysis (Chapter~\ref{chap:grains}), which plays critical roles in population-scale scenarios. SCC refers to processing data inside the storage device, either on the SSD controller (in-storage processing, i.e., ISP) or on the flash dies (in-flash processing, i.e., IFP). SCC can be a fundamental solution for alleviating this data-movement overhead by processing data where it originally resides.
To mitigate the data preparation bottleneck, we propose an algorithm-architecture co-design for highly-compressed storage and high-performance access of sequence data (Chapter~\ref{chap:sage}).
Finally, we describe how all storage-centric designs presented in this dissertation can be put all together at low cost (Chapter~\ref{chap:all-together}) and describe future opportunities and new directions enabled by the contributions presented in this dissertation (Chapter~\ref{chap:conclusion}).
}

\chapter{Related Work}
\label{chap:related-work}

This chapter provides an overview of related work and describes where the prior work falls short. \sect{\ref{sec:related-io}} discusses related work on techniques for reducing storage I/O overheads, including storage-centric computing. 
Section~\ref{sec:related-mcc-overview} discusses an overview of related work on memory-centric computing.
\sect{\ref{sec:related-acc-genomic}} and \sect{\ref{sec:related-acc-metagenomics}} provide an overview of related work on accelerating genomic and metagenomic analyses, respectively. \sect{\ref{sec:related-acc-graphs}} discusses related work on accelerating graph-based (meta)genomics. \sect{\ref{sec:related-compression}} explains related work on genomic-specific compression.

\aooo{\section{Techniques for Reducing Storage I/O Overheads}
\label{sec:related-io}
The performance of many data-intensive applications is bottlenecked by storage I/O. A large body of work proposes techniques to reduce storage I/O overheads across different layers of the system stack. We summarize several key approaches.}

\aooo{\subsection{Reducing the I/O System Stack Overheads}

Several works aim to reduce the overhead of the software layers in the storage I/O stack (e.g., file system, block layer, device drivers, and various kernel subsystems). These layers introduce overheads related to context switches, data copies, lock contention, and interrupt handling. We discuss two broad classes of approaches that alleviate these overheads: \inum{i}~accelerating the I/O stack operations using specialized hardware or streamlined data paths, and \inum{ii}~bypassing the kernel via user-space I/O frameworks.}

\aooo{\subsubsection{Accelerating the I/O Stack Operations} 

Various approaches aim to accelerate the I/O stack by offloading or streamlining its operations. We explain several key examples in this section.} 

\aooo{\head{Zero-Copy Techniques} Some techniques (e.g.,~\cite{mmap,papagiannis2020optimizing,papagiannis2018efficient,splice,sendfile,io_uring}) reduce I/O overhead by eliminating unnecessary data copies between kernel and user space. For example, the mmap system call~\cite{mmap,papagiannis2020optimizing,papagiannis2018efficient} maps files directly into the application’s virtual address space, allowing applications to access stored data through memory load/store operations without relying on read/write system calls. While mmap avoids the cost of copying data into user-space buffers, it relies on the kernel’s page-fault handling mechanism, which can cause large overheads. Other zero-copy techniques (e.g.,~\cite{splice,sendfile}) reduce data copies for specific I/O patterns (e.g., transferring data between file descriptors) but still operate within the kernel I/O stack. io\_uring~\cite{io_uring} provides a modern asynchronous I/O interface that uses shared ring buffers between user and kernel space to reduce system call overheads, and supports zero-copy for both storage and network I/O.}

\aooo{\head{Data Processing Units} Some works (e.g.,~\cite{BlueField, NapatechF2070X,zhang2024dds,zhong2024dpc,zhu2025hidpu,gootzen2023dpfs,jeong2025mangoboost}) propose Data Processing Units (DPUs), which are specialized units designed to offload data-centric infrastructure tasks from the host system and reduce the additional burden on the host. For example, some DPUs can handle NVMe protocol processing, compression, deduplication, and encryption. DPU-based storage offloading is particularly beneficial in multi-tenant cloud environments and disaggregated storage, where infrastructure tasks compete with application tasks for hardware resources.}

\aooo{\head{Direct Storage Access for Accelerators} Some works (e.g.,~\cite{qureshi2023gpu,qureshi2022bam,Shainer2011,li2025managing,NVIDIA_GDS_Blog2022,Newburn2019GTC}) reduce the involvement of the host CPU from the storage access path by enabling hardware accelerators (e.g., GPUs) to access storage devices directly. Traditionally, GPU-based executions rely on the host CPU to manage all storage accesses, thereby introducing significant overheads. NVIDIA GPUDirect~\cite{NVIDIA_GDS_Blog2022,Shainer2011,Newburn2019GTC} and related technologies, such as Big accelerator Memory (BaM)~\cite{qureshi2023gpu,qureshi2022bam}, accelerate GPU storage access by enabling direct access between storage devices and GPUs, reducing CPU involvements and overheads. These approaches are particularly effective for emerging workloads that require fine-grained, data-dependent access to large data structures stored in the storage system.}

\aooo{\subsubsection{User-Space I/O Frameworks}
\label{sec:user-space-io}

Another approach for reducing I/O stack overheads is to bypass the kernel and enable applications to deal with storage devices directly from user space.}
\aooo{For example, kernel-bypass approaches (e.g.~\cite{yang2017spdk,kim2016nvmedirect,Sujay24BypassD,Kaesi2025LITESHIELD}) provide a set of user-space libraries and drivers that bypass the kernel storage layers. Key examples include Storage Performance Development Kit (SPDK)~\cite{yang2017spdk}, which unbinds NVMe devices from the kernel driver, and NVMeDirect~\cite{kim2016nvmedirect}, which provides a user-space I/O framework that leverages the kernel NVMe driver for control-plane operations while enabling user-space data-plane access.}

\aooo{\subsection{User-Controlled Data Placement}
\label{sec:user-controlled-placement}
Several storage interface standards (e.g.,~\cite{nvme2,min2023ezns,bjorling2017lightnvm,SamsungFDP2023,wang2014efficient}) expose the internal organization of the storage device to the host system. This enables users to optimize data placement and reduce internal device overheads. For example, Zoned Namespaces (ZNS) SSDs~\cite{nvme2,min2023ezns} partition the storage address space into zones that must be written sequentially and give the host control over data placement within each zone. By leveraging zones across applications, the host can reduce write amplification and garbage collection (GC) overhead within the SSD. Open-Channel SSDs~\cite{bjorling2017lightnvm,wang2014efficient} expose the physical configuration of the SSD (e.g., channels, dies, blocks) to the host, allowing host-side software to implement its own specialized flash translation layer (FTL). This enables users to optimize data mapping, wear leveling, and GC policies based on the specific needs of their workloads.}

\aooo{\subsection{Request Batching, Scheduling, and Prefetching}
Some techniques reduce I/O overheads by considering relationships across different I/O requests. Request batching techniques (e.g.,~\cite{do2021better,song2025cam,do2019improving}) amortize I/O request overheads (e.g., interrupts, command encoding) across multiple requests. I/O scheduling techniques (e.g.,~\cite{shen2013flashfq,mao2017improving,yang2019cars,wang2013novel}) reorder and merge requests to improve spatial locality and provide more storage-friendly access patterns. Prefetching techniques (e.g.,~\cite{nilakant2014prefedge,chakraborttii2020learning,li2022pattern}) predict future access patterns and proactively fetch data from the storage system into memory (or caches), overlapping I/O overheads with useful computation. However, the effectiveness of these techniques depends on the predictability of the applications' access patterns.}

\aooo{\subsection{Caching and Buffering}

Caching and buffering are fundamental techniques for reducing the number of storage accesses. By keeping frequently or recently accessed data in faster media (e.g., DRAM, SRAM), caches leverage temporal and spatial locality to serve some I/O requests without accessing the storage device itself. Operating systems maintain a page cache that transparently buffers recently accessed file data in main memory~\cite{da2012parallel,pham2024scalecache,brokhman2019gaia,li2024streamcache}. Database systems implement their own buffers with application-aware replacement policies that can outperform generic OS caching strategies~\cite{canim2010ssd,wang2014cache}. SSD controllers maintain internal DRAM caches for metadata (e.g., the flash translation layer mapping table) and for buffering write data before committing it to NAND flash. The effectiveness of caching techniques depends on the working set size relative to the cache size and the applications' access pattern locality.} 

\aooo{\subsection{Data Reduction}

Data reduction approaches reduce the volume of data transferred between storage and computational units, thereby lowering the demands for I/O bandwidth and storage capacity. We explain two key techniques for data reduction.}

\aooo{\head{Compression} Data compression (e.g.,~\cite{collet2018zstandard,pavlov20167,Brotli,Katz1991US5051745A,goyal2021dzip,goyal2018deepzip,chen2024ha,bartik2015lz4,liu2018data,fowers2015scalable,chen2021fpga,angerd2022gbdi,gao2024beezip,karandikar2023cdpu,9499902,abali2020data,pekhimenko2016case,pekhimenko2015energy,pekhimenko2013linearly,pekhimenko2012base,buyuktosunoglu2024enterprise,ekman2005robust,arelakis2014sc2,vijaykumar2015case}) reduces the size of stored data, thereby reducing the amount of data transferred during I/O operations. Compression can be applied at multiple levels: within the application, in the file system, block layer, or inside the storage device itself. Several works (e.g.,~\cite{bartik2015lz4,liu2018data,fowers2015scalable,chen2021fpga,angerd2022gbdi,gao2024beezip,karandikar2023cdpu,9499902,abali2020data,pekhimenko2016case,pekhimenko2015energy,pekhimenko2013linearly,pekhimenko2012base,buyuktosunoglu2024enterprise,ekman2005robust,arelakis2014sc2,vijaykumar2015case}) propose hardware compression to further improve compression performance. The trade-off is between compression ratio, performance, energy efficiency, and computational cost of compression/decompression. We further detail genomic-specific compression in \sect{\ref{sec:related-compression}}.} 

\aooo{\head{Data Deduplication} Data deduplication techniques (e.g.,~\cite{park2022deepsketch,ajdari2017scalable,ajdari2019cidr,ajdari2019fidr}) identify and eliminate redundant copies of data, storing only unique data blocks and replacing duplicates with references to the unique copy. Deduplication can operate at the file level (coarse-grained) or at the sub-file block level (fine-grained), and can be performed inline (during writes) or as a post-processing step. In storage systems with high data redundancy, deduplication can reduce both storage capacity requirements and I/O overheads. However, deduplication can introduce metadata management overhead (for tracking block references and hash indices) and can lead to fragmentation of the logical-to-physical mapping, potentially increasing read amplification.}

\aooo{\subsection{Hybrid Storage \tomiv{and Memory} Systems}

Hybrid storage \tomiv{and memory} systems combine multiple storage \tomiv{and memory} technologies with varying cost, capacity, and performance characteristics (e.g., HDDs, capacity-optimized SSDs, performance-optimized SSDs, persistent memory, and DRAM) to achieve different trade-offs. Various works (e.g.,\tomiv{~\cite{singh_sibyl_2022,nadig2025harmonia,yang2024term,niu2018hybrid,meza_case_2013,oliveira2023extending,meza2012enabling,li2017utility,sun2013hybrid,matsui2017design,ramos2011page}}) \tomiv{analyze these hybrid systems and/or}
propose placement and migration policies that determine where data resides across tiers based on access patterns and/or workload-specific heuristics. Tiered storage systems can operate transparently (managed by the OS or storage controller) or with the user's control.}

\aooo{\subsection{Memory and Storage Disaggregation and Expansion}
\label{sec:disaggregation}

Rather than reducing the cost of each I/O operation, another class of approaches changes where data resides and over which interconnect it is reached. We explain two key techniques: expanding the byte-addressable memory tier beyond a single node, and decoupling storage capacity from the compute node.}

\aooo{\head{CXL-Based Memory and Storage Expansion} Compute Express Link (CXL)~\cite{das2024introduction,van2019hoti} is a cache-coherent interconnect layered on top of PCIe that allows a host to attach memory devices accessible via load/store instructions rather than through the block I/O stack. CXL memory expanders and memory pools increase the capacity of the memory tier well beyond the DIMM slots of a single node and enable capacity to be shared across hosts~\cite{das2024introduction,lim-isca09}, thereby reducing how often an application's working set spills to the storage device. 
However, despite its benefits, CXL technology does not eliminate the need to store and access large, \emph{low-reuse} datasets in the storage system. We further evaluate the benefits of our proposed storage-centric designs compared to systems with large main memory in Chapters~\ref{chap:megis} and \ref{chap:grains}.}

\aooo{\head{Disaggregated Storage} Disaggregated storage decouples storage capacity from the compute node and accesses it over a network fabric. For example, NVMe over Fabrics (NVMe-oF)~\cite{nvmexpress_nvmeof} extends the NVMe command set across network fabrics, enabling a host to access remote SSDs with a latency approaching that of a locally attached device. This allows compute and storage to scale independently, improves device utilization, and simplifies capacity provisioning. However, every access additionally traverses the network stack and the fabric, which adds protocol-processing overhead, increases latency, and exposes requests to congestion and to interference from other tenants sharing the fabric and the remote device.}
\aooo{These techniques are orthogonal to those we propose in this dissertation. In fact, disaggregation increases the value of reducing the \emph{volume} of data that crosses the storage interface, since each byte that leaves the device also traverses the network fabric. Storage-centric computing (\sect{\ref{sec:related-scc}}) therefore becomes even more beneficial as the storage system moves further away from the computational units.}

\aooo{\subsection{I/O Amplification Reduction}

I/O amplification occurs when the amount of data actually read from or written to the storage device exceeds the amount logically requested by the application. Write amplification in NAND flash-based SSDs mainly stems from garbage collection. Read amplification can occur due to mismatches between the application’s access granularity and the storage device’s minimum read unit. Several works (e.g.~\cite{oh2024midas,lanyue2016wisckey,nvme2,bjorling2017lightnvm}) propose techniques to reduce I/O amplification via more efficient log-structured storage~\cite{oh2024midas}, key-value separation that stores large values separately from their keys to reduce compaction overhead~\cite{lanyue2016wisckey}, and alignment of application data structures with the device’s internal page and block sizes~\cite{nvme2,bjorling2017lightnvm}. As discussed in \sect{\ref{sec:user-controlled-placement}}, techniques such as ZNS SSDs~\cite{nvme2} and Open-Channel SSDs~\cite{bjorling2017lightnvm} enable host-managed data placement that can fundamentally reduce write amplification by ensuring that data with similar lifetimes is co-located in the same erase blocks.}

\aooo{\subsection{Improving the Performance of the Storage Device}

Orthogonal to system-level optimizations, improvements within the storage device itself can also reduce I/O latency and increase throughput. We describe several key examples.}

\aooo{\subsubsection{Reducing Contention}
Contention within the storage device arises when multiple requests compete for shared resources, such as channels, dies, planes, or internal buses. Some works aim to alleviate this contention by proposing efficient data placement (e.g.,~\cite{purandare2025valet,zheng2026solidattention}), lightweight network structures between the flash chips within the SSD (e.g.,~\cite{nadig2023venice,tavakkol2013nossd,kim2022networked}), or reducing the overhead of management tasks within the SSD, such as GC (e.g.,~\cite{yang2019reducing,park2017method,jung2012taking}).}

\aooo{\subsubsection{Improving NAND Flash Chip Latency}

Improvements in the NAND flash chip can directly reduce the fundamental read/write latencies of the storage device. Advances in NAND flash technology, such as transitions from SLC to MLC, TLC, and QLC, increase storage density but also increase read and write latencies~\cite{micheloni2010inside,cai-insidessd-2018,cai_threshold_2013,cai_program_2013,cai_data_2015,meza2015large}. Various techniques aim to improve NAND flash performance via \inum{i}~voltage optimizations that reduce read latency by tuning the read reference voltages to the specific wear state of each block (e.g.,~\cite{cai_error_2012,cai2013error,cai2017error}), \inum{ii}~optimizing read retry mechanisms (e.g.,~\cite{park2021reducing,ye2024achieving,chun2024rif}), and \inum{iii}~partial programming techniques that enable finer-grained writes, thereby reducing the effective write latency for small updates (e.g.,~\cite{cai2017vulnerabilities,kim2015subpage,kim2018utilizing}), \inum{iv}~erase operation optimizations (e.g.,~\cite{cho2024aero}). Continual advances in memory technologies (e.g., 3D NAND with improved cell architectures) offer denser storage with improved scalability compared to planar NAND~\cite{luo2018improving,luo2018heatwatch,compagnoni2017reviewing}.}

\aooo{\subsubsection{Other Storage Device Optimizations}

Device-level optimizations include a range of controller and firmware improvements that enhance I/O performance. Some works (e.g.,~\cite{tavakkol2018flin,hedayati2019multi,woo2021d2fq}) implement efficient command scheduling and queue management to improve performance, fairness across different applications/users, and quality of service.}

\subsection{Storage-Centric Computing}
\label{sec:related-scc}

\tomiii{Storage-centric computing (SCC)  is a computing
paradigm that aims to overcome data movement overheads between the storage systems and computational units by making storage systems compute-capable.} 
SCC can be a fundamental solution for alleviating \tomiii{storage I/O} data-movement overheads by processing data where it originally resides. The benefits of SCC stem from three key reasons. First, SCC eliminates unnecessary data movement of low-reuse data across the system. Second, SCC reduces the computational burden imposed by low-reuse data on the rest of the system (e.g., main memory and computational units), thereby freeing resources for other useful work. Third, SCC can leverage the high internal bandwidth of SSDs~\cite{li2023ecssd,mailthody2019deepstore,kang2021mithrilog,koo2017summarizer,mansouri2022genstore,wang2024beacongnn,jang2024smart,Kim2023optimstore,li2021glist}.
In modern SSDs, the internal bandwidth often surpasses the external bandwidth. 
Over-provisioning the internal bandwidth is reasonable since 1) a modern SSD needs to perform various internal management tasks (e.g., garbage collection~\cite{\citegc}, wearleveling~\cite{\citewearleveling}, and data refresh~\cite{\citeflashrefresh}) and 2) a higher channel count reduces contention between requests by interleaving data between the channels. The effective internal bandwidth increases even further when processing directly in the NAND flash dies, without the need to traverse the SSD channels~\cite{chun2022pif,park2022flash,lee2025aif}.

\tomiii{SCC approaches can be broadly divided into three categories based on where
computation is performed relative to the storage medium:
\inum{i} tightly attached processing, which places computational units outside
but in close proximity to the storage device;
\inum{ii} in-storage processing (ISP), which places computational units within
the storage device, on the storage controller; and
\inum{iii} in-flash processing (IFP), which performs computation within the flash dies.
We briefly describe these three approaches.}

\tomiii{\noindent\textbf{Tightly attached processing} places computational units outside the storage device but connects them closely to storage. Prior works, for example, tightly couple SSDs with computation units (e.g.,~\cite{\citetightattachedSCC}), enabling data accessed from storage to be processed without first moving to the conventional host processor and memory hierarchy. This approach benefits from the flexibility and computational capability of external accelerators. However, since computation remains outside the storage device, data must still cross the storage device's external interface before it can be processed, thereby limiting the amount of data movement that can be eliminated.}

\tomiii{\noindent\textbf{ISP} places computational units within the storage device, typically in or near the SSD controller, as general-purpose computing units (e.g.,~\cite{\citegpisp}), as FPGAs (e.g.,~\cite{\citeispfpga}) or GPUs (e.g.,~\cite{\citeispgpu}) tightly connected to the flash dies, or computing units specialized for different applications, such as artificial intelligence/machine learning (e.g.,~\cite{\citeispaiml}), pattern processing and k-mer counting (e.g.,~\cite{jun2016storage,abakus23taco}), graph analytics (e.g.,~\cite{\citeispgraphs}), and others (e.g.,~\cite{\citeispotherapps}). 
These works augment the computational capabilities already present in modern storage controllers with general-purpose or specialized processing units that operate directly on application data. Performing computation within the SSD avoids transferring all accessed data across the external storage interface and enables only the required results to be sent to the host. Compared with tightly attached processing, ISP eliminates data movement across the external SSD interface before computation. However, data must still be transferred from the flash dies through the SSD channels to the controller, where computation is performed.}

\tomiii{\noindent\textbf{IFP} moves computation even closer to the stored data by performing computation within the NAND flash dies. IFP approaches (e.g.,~\cite{\citeifp,\citeufp}) add computational logic to the flash dies or exploit the existing operational properties of NAND flash arrays to perform computation. IFP can avoid not only data movement between the storage device and the host but also a significant fraction of the data movement between the flash dies and the SSD controller. It can further exploit the high internal parallelism and bandwidth available within individual flash chips. However, compared with tightly attached processing and controller-level ISP, IFP generally requires more substantial modifications to the flash die's architecture. Several works (e.g.,~\cite{Kim2023optimstore, nadig2026conduit, liang2024hyqa}) propose SCC designs that combine ISP, IFP, and processing-in-memory (PIM).}

Various works propose SCC for different applications, such as artificial intelligence/machine learning (e.g.,~\cite{\citeispaiml,Sun2025lincoln,Yu2024cambriconllm,kang2021s}), pattern processing and k-mer counting (e.g.,~\cite{jun2016storage,abakus23taco,Hsu2024het3dnand}), graph analytics (e.g.,~\cite{\citesccgraphs}), and others (e.g.,~\cite{\citesscotherapps}). 
To our knowledge, our work, GenStore, is the first in-storage system designed for accelerating genome sequence analysis. GenStore works with both short and long reads, and different degrees of genetic variation between the compared genomes. A later work~\cite{zheng2025storage} proposes an additional filter for genomics in storage, which can be orthogonally combined with our design. Our work on SCC design for metagenomics, MegIS, is the first SCC system designed to significantly reduce the data movement overhead of the \emph{end-to-end} metagenomic analysis pipeline. By addressing the challenges of leveraging ISP for metagenomics, MegIS fundamentally alleviates its data movement overhead from the storage system via its efficient and cooperative pipeline between the host and the SSD. Our work, GRAINS, enables SCC for graph-based (meta)genomic analyses via storage-aware algorithm-architecture co-design to \inum{i}~make the pipelines more storage-friendly and \inum{ii}~further improve performance, energy-efficiency, and cost-effectiveness via ISP and IFP.

\aooo{Through storage-centric designs, our proposed techniques significantly alleviate the overhead of moving large amounts of low-reuse data from the storage system to the main memory and computational units, while reducing the computational burden on the rest of the system. Our techniques can be complementarily combined with various other approaches discussed in this section to further improve end-to-end performance. For example, our optimizations that make the genomic and metagenomic analysis pipelines more storage-friendly enable these pipelines to better leverage the available storage bandwidth provided by techniques such as direct storage access, user-space I/O frameworks, and storage device optimizations.}

\section{Memory-Centric Computing}
\label{sec:related-mcc-overview}

\tomii{Memory-centric computing (MCC) or Processing-in-Memory (PIM) is a computing paradigm that aims to overcome data movement overheads between main memory and computational units by making memory systems compute-capable\tomiv{~\cite{mutlu2025memory,mutlu2022modern,mutlu2025modern,mutlu2019processing,mutlu2019enabling, ghose2019processing,mutlu2013memory,mutlu2024memory,yuksel2026memory,oliveira2022accelerating}}. There are two general categories of MCC or PIM: \inum{i} Processing near memory (PNM), which places computational units on the same die as memory controllers, the logic layer of 3D-stacked memory, or the memory itself, and \inum{ii} Processing using memory (PUM), which leverages the analog properties of the memory technology to perform computation. We briefly describe these two approaches.}

\tomii{\noindent\textbf{PNM} can be divided into three general approaches. The first approach adds computational units to memory controllers (e.g.,~\cite{seshadri2015gather, lee2015decoupled, hashemi2016continuous, hashemi2016accelerating, singh2020nero}). These works enable the memory controller to perform more tasks than just scheduling the memory requests. For example, some works devise computational units on the memory controller to reduce the round-trip cost of data movement, avoiding the latency of moving data within the cache hierarchy. While these works provide the key benefit of seamless integration with conventional memory technologies such as DDRx, they do not eliminate the data movement overhead from the memory device itself. As the memory controller typically resides on the same chip as the CPU, these works still require moving data between computational units and main memory.}

\tomii{The second approach places computational units in close physical proximity to (but not within) the memory chip, enabling near-data processing without modifying the memory chip itself (e.g.,~\cite{asghari2016chameleon,sun2021abc,ke2021near,ke2019recnmp}). These designs typically rely on either \inum{i}~2.5D integration, in which logic is placed alongside high-bandwidth memory chips on a silicon interposer, providing low-latency, high-bandwidth communication between the logic and memory dies; or \inum{ii}~module-level integration, in which logic is embedded on the same printed circuit board (PCB) as the memory chips. While this approach benefits from avoiding off-chip data transfers between the host processor and main memory, it still suffers from data movement between memory chip(s) and computational chip(s). Such data movement consumes more energy than data movement only within each memory chip.}

\tomii{The third approach adds computational units to the memory chip itself, either \inum{i}~on the logic layer of a 3D-stacked memory (e.g.,\tomiv{~\cite{ ahn2015scalable,drumond2017mondrian,boroum2019conda,boroumand2017lazypim,NDC_ISPASS_2014,singh2019napel,azarkhish2016logic,azarkhish2018neurostream,top-pim,RVU,NIM,gao2017tetris,hsieh2016transparent,cali2020genasm,boroumand2021mitigating,boroumand2021google,boroumand2021polynesia,fernandez2020natsa,LiM_3D_FFT_MM,akin2014hamlet,gao2016hrl,farmahini2015nda,boroumand2018google,nai2017graphpim,kim2018grim,PEI,kwon202125,lee2021hardware,lee2016simultaneous,hsieh_accelerating_2016,pattnaik2016scheduling,syncron,ghiasi2022alp,bostanci2025revisiting,HBM,hbm2,hmc.spec.2.0,loh2008stacked,
gopireddy2019m3d,mitra2018vlse,hwang2018cmos,mitra2015nano,rich2020nano,sabry2015abundant,sabry2019n3xt,ghiasi2022revamp3d}}) or \inum{ii}~near memory banks (e.g.,\tomiv{~\cite{skhynixpim,kwon202125,lee2021hardware,devaux2019true,gomezluna2021benchmarking,gomez2021benchmarkingcut,gomez2022benchmarking,gu2025pim,chen2023simplepim,yang2026dcc,rhyner2024pimopt,giannoula2022sparsep,barkhordar2025alpha,giannoula2024pygim,gupta2023evaluating,zhao2026cosm}}) by leveraging manufacturing techniques that support mixing logic and memory within the chip. This approach benefits from the larger memory bandwidth inside the memory chip and higher energy efficiency compared to the two previous approaches.}

\tomii{\noindent\textbf{PUM} takes advantage of the existing interconnects and analog operational behavior of conventional memory architectures (e.g., SRAM, DDRx, LPDDRx, HBM, NVM, and NAND flash), without requiring dedicated computational units or a logic layer, and typically with low additional area and power overheads (e.g.,\tomiv{~\cite{\citepum}}). This approach takes advantage of the high internal bandwidth available within each memory cell array.}

\tomii{Our work on storage-centric design can be combined with memory-centric designs to provide large end-to-end benefits. For example, in Chapter~\ref{chap:genstore}, we demonstrate and evaluate how our proposed in-storage filter can be combined with a PIM-based read-mapping accelerator to remove I/O data movement overheads of the end-to-end read mapping process.}

\section{Accelerating Genomic Analysis}
\label{sec:related-acc-genomic}

There have been extensive efforts to accelerate genomic analysis. 
Many works propose efficient heuristics and algorithmic optimizations (e.g.,~\cite{\citealgoptimization,\citegraphalgoptimization}) and hardware accelerators (e.g.,~\cite{\citehwoptimization,\citegraphhwoptimization}). 

Several works (e.g.,~\cite{\citemapping}) accelerate read mapping, a commonly-used operation in genomics. These works tend to follow two general directions: 1) non-filtering and 2) filtering approaches.

\noindent\textbf{Non-filtering approaches} accelerate one or more non-filtering steps of read mapping (e.g., seeding and approximate string matching) using hardware accelerators. 
Examples of these accelerators include processing-in-memory architectures~\cite{huangfu2018radar,khatamifard2021genvom, cali2020genasm, gupta2019rapid,li2021pim,angizi2019aligns,zokaee2018aligner},
\omcv{ASICs}~\cite{turakhia2018darwin, fujiki2018genax, madhavan2014race},
GPUs~\cite{cheng2018bitmapper2,houtgast2018hardware,houtgast2017efficient, zeni2020logan,ahmed2019gasal2,nishimura2017accelerating,de2016cudalign,liu2015gswabe,liu2013cudasw++,liu2009cudasw++,liu2010cudasw++,wilton2015arioc},
and FPGAs~\cite{goyal2017ultra,chen2016spark,chen2014accelerating,chen2021high,fujiki2020seedex, banerjee2018asap,fei2018fpgasw,waidyasooriya2015hardware,chen2015novel,rucci2018swifold,haghi2021fpga,li2021pipebsw,ham2020genesis,ham2021accelerating,wu2019fpga}.

\noindent\textbf{Filtering approaches} accelerate the pre-alignment filtering step of read mapping. These works provide highly-parallel read filtering heuristics that quickly eliminate dissimilar sequences before invoking computationally-expensive alignment algorithms. 
Examples of these accelerators include
processing-near-memory architectures~\cite{nag2019gencache,kim20111,singh2021fpga,hameed2021alpha},
FPGAs~\cite{alser2017gatekeeper, alser2017magnet, alser2019shouji, alser2020sneakysnake,guo2019hardware},
GPUs~\cite{alser2020sneakysnake, bingol2021gatekeeper,guo2019hardware},
or traditional CPU-based acceleration~\cite{xin2015shifted,xin2013accelerating,alser2020sneakysnake}.

Our works on in-storage filters for genomics, GenStore, and on mitigating genomic data preparation bottleneck, SAGe, can be orthogonally combined with these works to improve the \emph{end-to-end} performance of genome analysis. We show examples of integrating GenStore and SAGe with state-of-the-art genome analysis accelerators in Chapters~\ref{chap:genstore} and \ref{chap:sage}, respectively.

\section{Accelerating Metagenomic Analysis}
\label{sec:related-acc-metagenomics}

There have been extensive efforts to accelerate metagenomic analysis.

\head{Software Optimizations} Several tools (e.g.,~\cite{lapierre2020metalign,song2024centrifuger,koslicki2016metapalette,Marcelino2020,piro2016dudes,piro2020ganon,pockrandt2022metagenomic,wood2014kraken}) use comprehensive databases for high accuracy, but usually incur significant computational and I/O costs.  
Some tools (e.g.,~\cite{kim2016centrifuge,wood2019improved,muller2017metacache,song2024centrifuger,Dilthey2019,Fan2021}) apply sampling to reduce database size, but at the cost of accuracy loss.

\head{Hardware Optimizations} Several works use GPUs (e.g.,~\cite{jia2011metabing,kobus2021metacache,wang2023gpmeta,kobus2017accelerating,Su2012,su2013gpumetastorms,Yano2014}),  FPGAs (e.g.,~\cite{saavedra2020mining,zhang2023genomix,cervi2022metagenomic}), and PIM (e.g.,~\cite{wu2021sieve,shahroodi2022krakenonmem,shahroodi2022demeter,dashcam23micro,hanhan2022edam,zou2022biohd}) to accelerate metagenomics by alleviating its computation or main memory overheads. 
These works do not reduce I/O overheads, whose impact on end-to-end 
performance becomes even larger when other bottlenecks are alleviated.
Some works~\cite{dunn2021squigglefilter,shih2023efficient} accelerate metagenomic analyses that use raw genomic signals in targeted sequencing~\cite{kovaka2020targeted,Payne2021,Bao2021Squigglenet,ahmed2021pan}. Targeted sequencing is not a focus of our work since this application looks for specific \emph{known} targets in a sample, while we focus on cases where the contents of the sample are \emph{not known} in advance and require looking up significantly larger databases.

Our work on SCC design for metagenomics, MegIS, is the first SCC system designed to significantly reduce the data movement overhead of the \emph{end-to-end} metagenomic analysis pipeline. By addressing the challenges of leveraging ISP for metagenomics, MegIS fundamentally alleviates its data movement overhead from the storage system via its efficient and cooperative pipeline between the host and the SSD. We compare MegIS against a state-of-the-art processing-in-memory baseline in Chapter~\ref{chap:megis}. 

\section{Accelerating Graph-Based Genomics and Metagenomics Analysis}
\label{sec:related-acc-graphs}

\head{Accelerating Operations on (Meta)Genome Sequence Graphs} Many works (e.g.,~\cite{rautiainen2020graphaligner,kim2019graph,gao2020abpoa,jain2019pasgal,siren2021pangenomics,Rautiainen2019,Chandra2023,Ivanov2022,Ma2023,Darby2020vargas,Hwang2025MEMO,Romain2023svjedi,Li2020minigraph}) 
propose software tools for (meta)genome sequence graphs. Several works propose hardware tools for querying sequence graphs by alleviating computation (e.g.,~\cite{cali2022segram,Zhang2024Harp,Zeng2024asgdp,Shen2024128parallel,Li2024,Mandal2020,Varma2013,Awan2021,Feng2021,Zhang2025}) and/or memory overheads (e.g.,~\cite{cali2022segram,kim2025nmp,Huang2023meg2,Angizi2020Panda,Qiu2017}). However, they do not alleviate I/O overheads, whose impact on end-to-end performance becomes even larger as other overheads are\linebreak alleviated.

As shown in our evaluations in Chapter~\ref{chap:grains}, our SCC system for sequence graphs, GRAINS, can flexibly integrate with these works to alleviate their I/O overhead. GRAINS enables SCC for graph-based (meta)genomic analyses via storage-aware algorithm-architecture co-design to \inum{i}~make the pipelines more storage-friendly and \inum{ii}~further improve performance, energy-efficiency, and cost-effectiveness via ISP and IFP. Some works (e.g.,~\cite{Zhou2021,Sarkar2021,Varma2017,Varma2016,Goswami2018,Galanos2021,Angizi2020,Sinha2022,Meng2014,Hu2016,Chen2023,Natarajan2018,Ren2018}) accelerate sequence graph construction, which is an important, but orthogonal task.

\head{Accelerating Non-Genomic Graph Analysis}
Many works accelerate graph analysis by alleviating computation (e.g.,~\cite{Dadu2021polygraph, Ham2016Graphicionado,Rahman2020graphpulse,Song2018graphr,Chen2022regraph,Yang2025IDGNN,Yan2025bingogcn,Peng2024maxkgnn,Yan2020hygcn,You2022gcod,Hwang2023grow,Sarkar2023flowgnn,Chen2022regnn,Li2021gcnax,Geng2020awbgcn,Chen2023metanmp,Zhao2025mehyper}), main memory (e.g.,~\cite{ahn2015scalable,Zhou2022GNNear,zhang2018graphp,Asiatici2021,Shin2025piccolo,Huang2022reflip,Besta2021SISA,Chen2023metanmp,Li2022hyperscale,Wang2024motionaccel,Dai2023cegma,Kim2025eod,Li2024celeritas}), or I/O (e.g.,~\cite{matam2019graphssd,Wang2024ndsearch,lee2022smartsage,Niu2024flashgnn,Lee2024presto,Khadirsharbiyani2024smartgraph,Zhang2025taijigraph,An2023baraddur,Kang2024sting}) overheads.
However, they are neither applicable to nor target the unique demands of genome graphs.

Our work, GRAINS, is the first SCC system designed to significantly reduce the data movement overhead of the end-to-end analysis pipelines on genome graphs.
Although some works propose SCC systems to mitigate I/O in conventional, non-genomic graph analytics (e.g.,\cite{matam2019graphssd,Wang2024ndsearch,lee2022smartsage,Niu2024flashgnn,Lee2024presto,Khadirsharbiyani2024smartgraph,Zhang2025taijigraph,An2023baraddur,Kang2024sting}), they are neither designed for nor directly applicable to genome graphs, which have fundamentally different structures, access patterns, and analysis requirements.
As detailed in \sect{\ref{sec:background-graph}}, large-scale genome graphs use fundamentally different data structures than non-genomic graphs (e.g., those for the web, social networks, or road networks). For example, DBGs often store only nodes (or edges) and encode edges (or nodes) implicitly via overlaps, since explicitly storing both is prohibitively wasteful at the genomic scale. Therefore, optimizations for conventional graphs are not only insufficient for genome graphs but also miss opportunities to exploit their distinctive structure.

Some techniques reduce I/O overheads with techniques other than SCC. For example, some works perform graph node reordering (e.g.,~\cite{esfahani2021locality,coleman2022graph}), but they cannot be applied to genome graphs since their state-of-the-art representations have stringent node ordering requirements~\cite{Bowe2012SuccinctGraphs,pibiri2022sparse}.
Some works (e.g.,~\cite{kyrola2012graphchi,bulucc2016recent}) propose data partitioning to partially alleviate I/O overheads in systems with relatively small main memory. Note that as shown in our motivational analysis in Chapter~\ref{chap:grains}, which is on a system with sufficient main memory capacity (i.e., where main memory is larger than the whole accessed data structures and working set),  large I/O overheads still exist due to the need for moving large amounts of low-reuse data from the SSD to main memory. In Chapter~\ref{chap:grains}, we further evaluate a system with limited main memory capacity while incorporating graph partitioning in the baselines, and show that GRAINS achieves significant speedups in this setting as well.

\section{Compress\tomi{ing} Sequence Data}
\label{sec:related-compression}

Efficient and compact compression techniques for genomics data are of critical importance due to the continuing exponential growth of genomics data repositories~\cite{srastats,katz2021sra,stephens2015big}.

\head{Genomics-Specific Compression}
Many works (e.g.,\tomiii{~\cite{chandak2018spring,Deorowicz2020,lan2021genozip,alyami2019lfastqc,kowalski2019pgrc,roguski2018fastore,chandak2017compression,cogo2021genodedup,Meng2023,kokot2022colord,dufort2020enano,dufort2021renano,karasikov2022lossless,vandamme2024tinted,dragenora,yang2025gpufastqlz,chen2023efficient,hach2012scalce,roguski2014dsrc2,grabowski2022mbgc,kowalski2026mbgc2,grabowski2026ffc,deorowicz2023agc,Kryukov2022,sousa2024jarvis3,Sun2023,Nazari2025}}) propose genomic compression.
As detailed in Chapter~\ref{chap:sage}, 
some works (e.g.,~\cite{guo2013gpu,qiao2019fpga,zhao2017streaming,jiang2021exma,arram2015fpga,wang2018accelerating,lim2025bancroft,chen2023efficient,leavline2013hardware}) accelerate certain computational kernels (e.g., BWT~\cite{zhao2017streaming,qiao2019fpga}, FM-Index search~\cite{jiang2021exma,arram2015fpga,wang2018accelerating}, and LZMA~\cite{chen2023efficient,leavline2013hardware})
widely used in genomic compressors. 
Despite their benefits, \ov{these approaches} 
\inum{i}~are unsuitable for resource-constrained environments and/or \inum{ii}~do \emph{not} fully mitigate the end-to-end preparation bottleneck (evaluated in Chapter~\ref{chap:sage}).
While some works~\cite{rajarajeswari2011dnabit,saada2016dna}
use algorithms that do not rely on expensive resources, they compress poorly.
As a strongly conservative evaluation, \texttt{0TimeDec} (evaluated in Chapter~\ref{chap:sage}) can serve as an idealized representation of decompressors that, despite their optimized performance, cannot be integrated in resource-constrained environments. As shown in our evaluations, by integration in a resource-constrained environment, our proposal, SAGe, leads to significant \ov{performance} benefits over \texttt{0TimeDec}.

\head{General-Purpose Compression} Many works propose general-purpose compression in software (e.g.,~\cite{collet2018zstandard,pavlov20167,Brotli,Katz1991US5051745A,goyal2021dzip,goyal2018deepzip,chen2024ha}) or hardware (e.g.,\ovi{~\cite{bartik2015lz4,liu2018data,fowers2015scalable,chen2021fpga,angerd2022gbdi,gao2024beezip,karandikar2023cdpu,9499902,abali2020data,pekhimenko2016case,pekhimenko2015energy,pekhimenko2013linearly,pekhimenko2012base,buyuktosunoglu2024enterprise,ekman2005robust,arelakis2014sc2,vijaykumar2015case}}). However, as analyzed in Chapter~\ref{chap:sage}, general-purpose compressors
achieve poor compression ratios for genomic data~\cite{hernaez2019genomic,zhu2013high}.

To our knowledge, our proposal for sequence data preparation, SAGe, is the \emph{first} system to mitigate the data preparation bottleneck in sequence analysis. SAGe achieves compression ratios comparable to state-of-the-art genomic compressors and maintains a lightweight design for integration with a wide range of genomic and metagenomic systems.

\chapter{GenStore: In-Storage Filtering for Genomic Analysis}
\label{chap:genstore}

\section{Motivational Studies}
\label{sec-gs:motivation}

We perform experimental studies to understand the potential of efficient in-storage accelerators for improving the performance of genome sequence analysis applications.

\subsection{Methodology} 

\head{Read Mappers} We evaluate five read mapping systems, each of which adopts different optimization techniques to accelerate read mapping:
\omc{1)~}\base uses Minimap2~\cite{li2018minimap2}, a state-of-the-art software tool for read mapping\omcvii{.} 
\omc{2)~}\swf extends Minimap2 to \emph{filter out exact\omc{ly}-matching reads} \omc{ (i.e., reads that  exactly \omcvii{match} \omciv{subsequences in} one or more locations in the reference genome) using simple single-instruction-multiple-data (SIMD) operations, without requiring costly ASM operations}\omcvii{.}
\omc{3)~}\isf uses an ideal In-Storage Filter \omc{(\emph{ISF})} that can \emph{concurrently} filter out exactly-matching reads inside the SSD while \omcvii{the} host CPU performs read mapping for non-filtered reads\omcvii{.}
\omc{4)~}\acc uses a state-of-the-art hardware accelerator for short read mapping, GenCache~\cite{nag2019gencache}\omcvii{.}
\omc{5)~}\isfacc use\omcvii{s} an ideal in-storage filter \omc{(ISF)} that can \emph{concurrently} filter out exact\omc{ly}-matching reads inside the SSD \omc{while a hardware accelerator (ACC)} performs read mapping for non-filtered reads. 

\omcc{\head{System Configuration}} To assess the impact of the storage subsystem on end-to-end application performance, we evaluate each \omciv{of the five} systems with four different configurations:
1)~a low-end SSD (\ssdl)~\cite{inteldcs4500} with a SATA3 interface\omciv{~\cite{SATA}}, 
2)~a mid-end SSD (\ssdm)~\cite{samsung980pro} using a PCIe Gen3 M.2 interface\omciv{~\cite{PCIE}}, 
3)~a high-end SSD (\ssdh)~\cite{samsungPM1735} with a PCIe Gen4 interface\omciv{~\cite{PCIE4}}, and
4)~a system where \emph{all} of the processed data is pre-loaded to DRAM \omc{with no performance cost for pre-loading}  (\dram), as the \emph{idealized} case where storage I/O overheads are completely eliminated (we do not evaluate \dram for \isf and \isfacc since using in-storage processing is contradictory to pre-loading \omc{all} the data to main memory).
\revp{We assume 8 channels for \ssdl and 16 channels for \ssdm and \ssdh, where the \omciv{maximum} bandwidth per channel is 1.2~GB/s. \omciv{The maximum internal bandwidth is calculated by 1.2~GB/s $\times$ channel count.} 
The external bandwidth of \ssdl, \ssdm, and \ssdh for sequential read\omc{s} is 500~MB/s, 3.5~GB/s, and 7~GB/s, respectively. 
\omcvii{Hence, t}he internal bandwidth of \ssdl, \ssdm, \omcvii{and} \ssdh~\omcvii{for sequential reads} is 19.2$\times$, 5.48$\times$, \omcvii{and} 2.74$\times$ \omciv{that of its external} bandwidth, respectively.}

We evaluate \base and \swf by running Minimap2 on a high-end server (AMD EPYC 7742 CPU~\cite{amdepyc} with \omc{1TB} DDR4 DRAM). \omciv{We simulate} the performance of the other \omciv{three} systems using our simulation environment that faithfully models system components including DRAM and storage devices (see \sect{\ref{sec-gs:methodology}}).
We map all reads of a short read dataset against the human reference genome, where 80\% of the reads have one or more \rev{exactly-}matching subsequences in the reference genome~\cite{10002015global,uk10k2015uk10k}.

\omcc{\head{Key Features of an Ideal In-Storage Filter}} \revp{We assume two key features for \isf and \isfacc. First, I/O overheads \omc{due to limited external SSD bandwidth} are \emph{completely} eliminated for filtered reads. Second, the system provides high in-storage filtering performance such that  
the filtering process can concurrently run in \omc{the} SSD, and the latency of this filtering process is \emph{fully hidden} by the read mapping of unfiltered reads in \omciv{the} host CPU (\isf) or hardware accelerator (\isfacc).
We assume that the accelerator or the CPU streams through the input reads in batches and analyzes a batch \emph{\omciv{concurrently}} with reading the next batch. \omciv{Thus}, the execution time of \textsf{Ideal-ISF} (+ACC) can be modeled as follows:
\begin{equation}
\label{eq-gs:ideal-isf}
T_\text{\textsf{Ideal-ISF}} = T_\text{I/O-Ref} + \max\left\lbrace T_\text{I/O-Unfiltered}, T_\text{RM-Unfiltered}\right\rbrace,
\end{equation}

\noindent
where $T_\text{I/O-Ref}$, $T_\text{I/O-Unfiltered}$, and $T_\text{RM-Unfiltered}$ are the latency \omciv{of} \omcc{reading} the reference genome \omcc{from the SSD}  \omciv{into main memory}, the latency of \omcc{reading} the unfiltered \omcc{genomic} reads from the SSD, and the latency of read mapping of the unfiltered reads, respectively. 
\omciv{For a given input size,} $T_\text{RM-Unfiltered}$ varies depending on the computation unit used for read mapping (i.e., the host CPU or \omciv{accelerator}), while the I/O-latency values \omc{only depend on the SSD configuration}.}

\subsection{Results \& Analysis}
\begin{figure}[h]
    \centering
    \includegraphics[width=.85\linewidth]{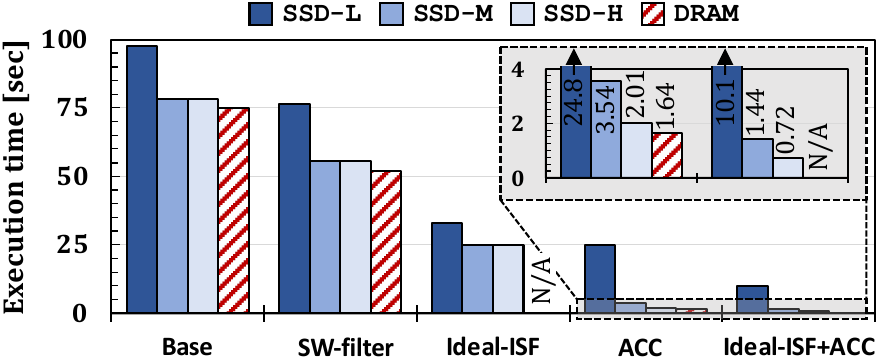}
    \caption{Execution time of read mapping \omciv{with four} different \omciv{storage} configurations.}
    \label{fig-gs:motivation1}
\end{figure}

\fig{\ref{fig-gs:motivation1}} shows the execution time of  read mapping in the five \omciv{evaluated} systems, \omciv{each} with \omciv{four} different storage \omciv{sub}system configurations. \omc{W}e make four key observations.

\head{Observation 1} The ideal in-storage filter  provides significant performance improvement\omcvii{s} over other systems.
\omcv{\isf  significantly outperforms \base and \swf (\omcvii{by} 3.12$\times$ and 2.21$\times$, respectively), and 
\isfacc provides a large speedup (2.78$\times$)
\omcv{over} \acc, when they all use \ssdh.}
\omcvi{These large improvements are} due to two key benefits \omcvi{provided by the ideal in-storage filter}: \omcc{1)} mitigati\omcvi{on of} data movement from the storage devices and \omcc{2)} remov\omcvi{al of} the burden of filtering out 80\% of the \omciv{input read set} from the rest of the system\omcc{, including processors and main memory}. 
\revp{To distinguish the effects of these two benefits, we analyze an ideal \emph{outside-storage} filter (\iof) \omcc{that provides only the second benefit;} \omciv{this filter} concurrently runs with the read mapper and fully overlaps the filtering process with the read mapping process of unfiltered \omcvii{reads}. The execution time of \iof(+ACC) can be formulated as follows: 
\begin{equation} 
\label{eq-gs:ideal-osf}
T_\text{\iof} = T_\text{I/O-Ref} + \max\left\lbrace T_\text{I/O-All-Reads},  T_\text{RM-Unfiltered}\right\rbrace,
\end{equation}

\noindent
where $T_\text{I/O-All-Reads}$ is the latency for reading all \omcc{genomic} reads from the SSD \omciv{into main memory}.
\omciv{Using} \ssdh, \iof leads to an execution time of 1.15 seconds, which is 60\% slower than the \isfacc. This is because  \omcc{$T_\text{I/O-All-Reads}$} is significantly larger than \omciv{both} $T_\text{RM-Unfiltered}$ and \omcc{$T_\text{I/O-Unfiltered}$ (in Equation~\eqref{eq-gs:ideal-isf})}.
}

\omcc{The remaining observations dive deeper into the effects of \omcvii{the} I/O bottleneck on  each read mapping \omciv{system}.}

\head{Observation 2} In \base and \swf, using high-end SSDs significantly improves read mapping performance over low-end SSDs, effectively \omciv{reducing} the storage \omciv{performance} bottleneck that exists in low-end SSDs. For example, using \ssdh instead of \ssdl reduces the execution time of \base and \swf by 24\% and 38\%, respectively, showing comparable performance to \dram, where all the data is \omcvii{pre-loaded to} main memory (i.e., \emph{no} I/O accesses). \omcc{This is because\omcvii{,} by using \ssdm and \ssdh,  the performance bottleneck of the application shift\omcvii{s} to parts of the system other than I/O \omciv{(e.g., CPU or main memory)}.}
This observation shows that I/O  has a significant impact on application performance but \omciv{this impact} can be \omcc{alleviated} at the cost of expensive storage devices and interfaces. 
Note that, while \ssdm and \ssdh provide an order-of-magnitude higher bandwidth for sequential reads
compared to \ssdl, 
\omcvii{i}t is challenging to scale a storage system's capacity using the high-end SSDs due to their significantly-higher prices and the \omcc{relatively smaller number} of the PCIe \omciv{slots} in a server.\footnote{\revp{The cost of the total storage system depends on both the price of each SSD and the available interconnection slots in the systems. High-bandwidth interconnects such as PCIe  take up very large  space in the system. \omciv{\omcv{As a result,} there are fewer PCIe slots than SATA slots in a system}. For example, building a 16-TB storage system with a single \omc{PCIe} SSD (Micron 9300 PRO~\cite{micros9300pro}) costs more than 3,000 USD, while \omc{the cost is}  less than 1,600 USD if we use \omc{four} 4-TB SATA SSDs (WD BLUE~\cite{wdblue}).}}

\head{Observation 3} \omciv{ Even though \swf outperforms \base, its filtering process is slow.
Potentially, \swf could provide significant performance benefits over \base due to two reasons;
1) as explained, 80\% of reads in the dataset exactly match  the reference genome, so only 20\% of the reads need to undergo the costly ASM \omcvii{computation}; 
2) exact-match filtering requires only simple computation, i.e., SIMD XOR operations used by \swf.
However, even with \dram, \swf's speedup over \base is \omciv{only} 41\%.
The limited speedup is mainly due to the large \omciv{number} of random memory accesses \revp{concurrently issued from all threads} to the reference index (\omcvii{explained in} Section~\ref{sec:background-readmapping}).
This observation highlights the potential of in-storage filtering. 
Even though both \swf and \isf filter out the same fraction of reads,  filtering outside the SSD must compete with the read mapping for the resources in the system (\omcv{e.g.,} the limited \omcv{main memory} bandwidth).
In contrast, \omcv{f}iltering of reads \omcvii{inside the SSD (where \omcvii{the reads} originally reside)} can remove the burden of filtering from the rest of \omcv{the} system.}

\head{Observation 4} \omcv{With a hardware accelerator (\acc)}, using the state-of-the-art SSD (\ssdh) does \emph{not} fully \omc{alleviate} the storage bottleneck, showing \jsr{23\%} longer execution time compared to when all the data is \omcvii{pre-loaded to} main memory (\dram). \omcc{While using \ssdm and \ssdh in \base and \swf \omciv{ shifts the bottleneck away from} I/O, \acc  turns I/O into a bottleneck again.}  
\omciv{This is because \acc greatly reduces the computational bottleneck, which increases the relative effect of the storage subsystem on \omcv{the} end-to-end execution time.
\omcvii{The} \acc and \isfacc} results clearly show that data movement between the storage devices and \omciv{the hardware} accelerator, which has not been properly considered in prior read mapping accelerators\omciv{~\cite{nag2019gencache, cali2020genasm, fujiki2018genax, turakhia2018darwin, madhavan2014race, cheng2018bitmapper2,houtgast2018hardware,houtgast2017efficient, goyal2017ultra,chen2016spark,chen2014accelerating,chen2021high,huangfu2018radar,khatamifard2021genvom}}, can significantly bottleneck the potential benefits of the accelerator.

\omcc{\head{Comparison to Other Near-Data \omcvii{Processing} Systems}} Even though read mapping could also benefit from other near-data processing (NDP) approaches such as processing-in-main memory (PIM)~\cite{kim2018grim, laguna2020seed, khatamifard2021genvom, kaplan2018rassa} or \omciv{processing}-in-caches~\cite{nag2019gencache}, in-storage processing  can \emph{fundamentally} address the data movement problem by filtering large, low-reuse data \emph{where the data initially resides}. 
As an extreme example, even if an \emph{ideal} accelerator achieved a \emph{zero} execution time for read mapping by addressing all of the computation and main memory overheads, there would still exist the need to bring the data from  storage to the accelerator.
In our motivational study, even \ssdh takes at least 1.55 seconds to read the entire dataset, which is 2.15$\times$ slower than the execution time that \isfacc provides (0.72 seconds).
Thus, even though solutions such as processing-in-memory can improve read mapping execution times, they still need to pay the cost of data movement from storage to the main memory\omciv{~\cite{singh2021fpga,turakhia2018darwin,cali2020genasm,kim2018grim,huangfu2018radar,khatamifard2021genvom,gupta2019rapid,li2021pim,angizi2019aligns,zokaee2018aligner}}. Therefore, \omciv{an} in-storage \omciv{filter} can be further integrated with any mapping accelerator, \omc{including PIM accelerators,} to alleviate their data movement overhead.

\vspace{-.5em}
\subsection{Our Goal}
Based on our observations, we \rev{conclude} that an efficient in-storage filter can be \omcc{a} key enabler for read mappers to achieve high performance in both \omcc{\omciv{conventional} software\omcv{-based}  (\omciv{e.g., \base and \swf)} and new hardware-accelerated \omciv{(e.g., \acc)}} genomics systems.
\omcvii{In particular, i}n-storage filtering \omcvii{enables} the system to take full advantage of the \omcvii{high} computation capability of \omcvii{hardware} accelerators by \omcvii{fundamentally} addressing the data movement bottleneck.
\linebreak
\omciv{\textbf{Our goal} is to design an in-storage filter for genome sequence analysis in a cost-effective manner.}

We have \textbf{three key objectives} in designing \omciv{our} new system.
\jsr{First, t}he system should provide high in-storage filtering performance to overlap the filtering  with the read mapping of unfiltered data (as \isf does in our motivation study).
\jsr{Second,} it should support \omcv{reads with \omcvi{1)} different properties (e.g., lengths and error rates) and \omcvi{2)} different degrees of genetic variation in the compared \omcvi{genomes}}.
\jsr{Third,} it should \emph{not} require significant additional hardware \omciv{overhead}, e.g., complicated logic circuits or large SRAM/DRAM memory.

\section{GenStore}
\label{sec-gs:mechanism}

\omcc{We propose \emph{GenStore}, the first in-storage processing system \jk{tailored} for genome sequence \jk{analysis. 
GenStore} greatly reduces both data movement and computational overhead\hm{s} \rev{of genome sequence analysis} by exploiting low-cost and accurate in-storage filters}.
GenStore supports \omciv{reads with different properties (lengths and error rates\omcv{)} and different degrees of genetic variation in the \omcvii{compared genomes}.}
\revmark{C1 and\\ F1}\revp{\omc{We primarily design GenStore as an in-storage accelerator, which is an \emph{extension} of the existing SSD controller and flash translation layer (FTL). GenStore is designed to be integrated into the system such that\omcvii{,} when the accelerator is not in use, the entire storage device is available to all other applications\jk{, just like} in a general-purpose system today.
}}

\omc{\subsection{Overview}}

\begin{figure}[h]
    \centering
    \includegraphics[width=0.9\linewidth]{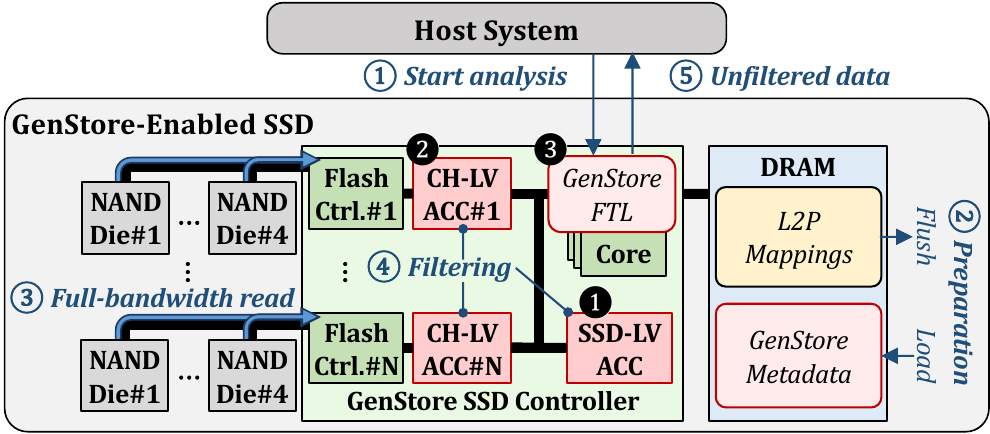}
    \caption{Overview of GenStore.}
    \label{fig-gs:genstore_overview}
\end{figure}

\omcc{The key idea of GenStore is to exploit low-cost \emph{in-storage} accelerators to accurately {\emph{filter out}} the reads that do not require the expensive alignment step in read mapping and \jk{thus} significantly reduce unnecessary data movement from the storage system to main memory and processors}. 
\fig{\ref{fig-gs:genstore_overview}} shows the overall architecture of GenStore and how it \hm{interacts} with the host system. GenStore employs two types of hardware accelerators: \circled{1}~a single \emph{SSD-level} accelerator and \circled{2}~\emph{channel-level} accelerators, each of which is dedicated to a channel. The GenStore-FTL (\circled{3}) communicates with the host system and manages the metadata and data flow over the \omcv{SSD} hardware components (i.e., NAND flash chips\jk{, \omcvii{internal} DRAM, and \omcvii{in-storage} accelerators}).

\jk{Once} the host system \omcv{indicates that the SSD should start analysis as required by} a read mapping application \omcvi{(\wcirc{1} in \fig{\ref{fig-gs:genstore_overview}})},  GenStore prepares \hm{for operation} as an accelerator \omcv{(\wcirc{2})}. It flushes the conventional FTL metadata necessary to operate as a regular SSD (e.g., L2P mappings~\cite{kim2020evanesco}), while loading the GenStore metadata necessary for each use case \rev{(Section~\ref{sec-gs:ftl} provides more details on GenStore FTL).}
After finishing the preparation, GenStore starts the filtering process.
\jsr{It}  keeps concurrently reading the data to process from \emph{all} \jk{NAND flash} chips \omcv{(\wcirc{3})} via multi-plane operations (i.e., it exploits the SSD's full internal bandwidth\omc{, which} is much higher than the I/O bandwidth between the SSD and host system~\cite{koo2017summarizer, mailthody2019deepstore}),  while  filtering out reads \omcv{(\wcirc{4})} that do not have to undergo \omciv{further analysis (\omcv{\omcvii{e.g.}, ASM computation})}.
Doing so is possible due to multiple channel-level accelerators that provide \jk{computational} throughput matching the SSD's internal bandwidth even for the most complicated computation required for filtering. 
The host system \omcv{performs} further computation as soon as GenStore sends unfiltered \omciv{reads} \omcv{(\wcirc{5})}, which removes GenStore's filtering process almost completely from the critical path of the application.

\rev{As explained in Section~\ref{sec:background-SSD}, most of the internal DRAM is occupied by the regular L2P mapping. Therefore, flushing the regular L2P mapping data into NAND flash memory enables GenStore} to exploit most of the GB-scale DRAM (e.g., 4GB DRAM in a 4TB SSD~\cite{samsung860pro}) \rev{for its operations}, which significantly reduces the overhead of additional internal DRAM that might \jk{otherwise} be required to store the GenStore metadata necessary for the filtering process. We carefully design \jk{the GenStore filtering algorithms} to only \emph{sequentially} access the underlying NAND flash \jk{chips, so GenStore} requires only a small amount of metadata to access the stored data. \omcc{Therefore, GenStore can use most of the internal DRAM space for \jk{such metadata.}} We envision that all GenStore metadata are built  \emph{offline} by the host or \omc{some} other system (e.g., \omc{by the sequencing machine} when the read set or reference genome are initially stored to the SSD). \omciv{Constructing GenStore metadata} is a \emph{one-time} \omc{preprocessing step} that can be performed independently of the read mapping process, while the result \omc{of the preprocessing step} can be used multiple times for different genomics applications.

There exist two main challenges in designing GenStore as an efficient in-storage filter for read mapping.
First, the behavior and data-access patterns in read mapping significantly vary depending on the read \omcv{properties} (\omccc{length and error rate) and genetic variation \omciv{between \omcvii{the} compared genomes}}.
Second, hardware resources (e.g., CPU and DRAM) are quite limited even in modern \omcvii{high-end} SSDs.
We address these challenges via thorough hardware/software co-design tailored for \omciv{filtering
1) exactly-matching reads, i.e., reads that exactly match subsequences of the reference genome  (Section~\ref{sec-gs:SRF}), and 2) \omcv{most of the} non-matching reads, i.e., reads that would not align to any subsequence of the reference genome (Section~\ref{sec-gs:LRF}).}

\vspace{-.3em}
\subsection{GenStore-\omccc{EM for Exactly-Matching Reads}}
\label{sec-gs:SRF}

\subsubsection{Approach Overview}
\label{sec-gs:sr-overview}

GenStore-EM accelerates  read mapping by using an efficient in-storage filter for reads that have at least one exact match in the reference genome. 
\omccc{Due to the low error rates of short reads, \jk{combined with low genetic variation between \omcvii{the} compared genomes}, a large fraction of short reads map \emph{exactly} to the reference genome~\cite{10002015global,uk10k2015uk10k,nag2019gencache}. For example, on average ~80\% of human short reads map \emph{exactly} to the human reference genome~\cite{10002015global,uk10k2015uk10k,nag2019gencache}.} 
Since exact-match detection is computationally cheaper than ASM, concurrently filtering exact-matching reads inside the SSD can significantly improve the runtime of read mapping \jk{(}as we demonstrate in  Section~\ref{sec-gs:motivation}\jk{)}.
\omciv{\omcvii{Note that,} GenStore-EM is not applicable to long reads due to their \omcviii{greater} length. For example, for a \omcv{10K base pair-long} human read\omcv{,} even with zero sequencing \omcv{error \omcviii{rate}},  the \omcvii{probability} of the read exactly matching a subsequence in the reference genome is very low (e.g., $<3.6\times10^{-6}$) due to natural genetic variation.\footnote{\revp{\omc{A typical human genome contains genetic variations at $\sim$4.1 to 5~million~\cite{10002015global} out of a total of $\sim$3.2~billion base pairs~\cite{schneider2017evaluation}. Thus, each 10K-bps read contains, on average, $\sim$12.5 to 15.3~\omcv{base pairs that are different from} the reference genome.}}}
}

\head{Key Challenges} 
The key challenge in designing GenStore-EM is \omcvii{the} large number of random accesses to large data structures inside the SSD. As explained in Section~\ref{sec:background-prealignment}, identifying exact matches for read mapping requires a number of random accesses \emph{\omcc{for each k-mer in a read}} to two large data structures: 1) a large k-mer index, to find \omciv{potential} \omcc{matching} locations  \omcc{of each k-mer in the reference genome}, and 2) the  reference genome, to find candidate matching sequences \omcc{at the candidate matching locations in the reference genome}.
\omcc{Handling random accesses to large data structures is challenging  for \omcvii{NAND flash-based} SSDs for two reasons.}
First, NAND flash memory  exhibits poor performance for  random \jk{access} reads. 
Second, \omcc{we cannot use in-SSD DRAM to store the \jk{large} data structures that are randomly accessed, since even}
in high-end SSDs, the size \omcvii{of internal DRAM} is \omcc{relatively small} (e.g., 4 GB~\cite{samsung860pro}) compared to the size of the data structures that GenStore-EM needs to handle \omcc{(e.g., 7 GB for \omcvii{the} human reference genome and its index~\cite{li2018minimap2})}.

\newcommand\readset{SRTable\xspace}
\newcommand\seedidx{\omcc{SKIndex}\xspace}

\head{Key Idea} 
The key idea in GenStore-EM is to \emph{sequentialize} most data accesses via a carefully designed metadata structure and data layout. \omcc{To do so, we \jk{design} a new \emph{sorted,} \emph{read-sized} k-mer index structure. This index enables \jk{\emph{sequential}} scanning \omccc{of} the read set and the new index, with only one index lookup per read while performing filtering.}

\begin{figure}[t]
        \centering
        \includegraphics[width=\linewidth]{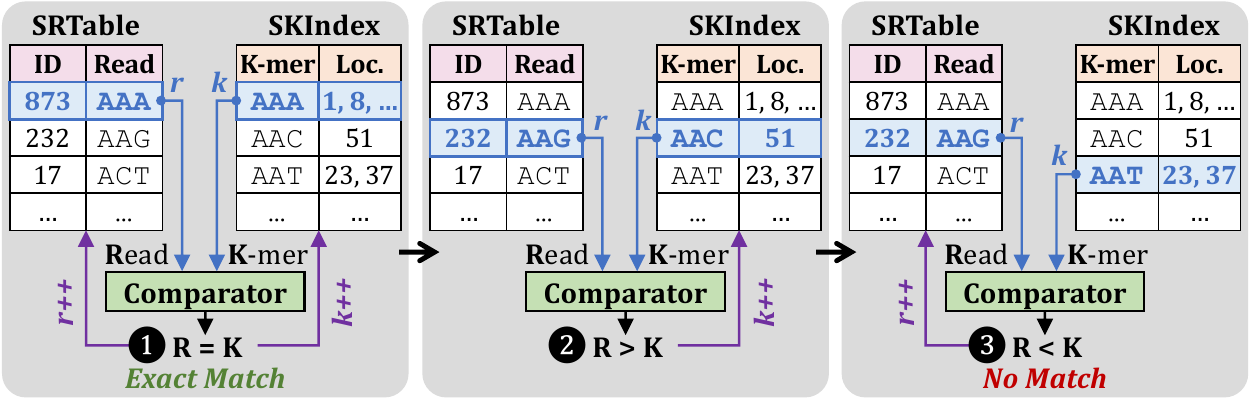}
        \caption{Overview of the key idea of GenStore-EM.}
        \label{fig-gs:sr_idea}
\end{figure}

Figure~\ref{fig-gs:sr_idea} shows \omcc{the key idea of GenStore-EM} with a simpl\omcc{ified} example in which each short read consists of \omcix{three} base pairs (bps).\footnote{\omcc{In a realistic scenario, the read \omcviii{length} is much larger (e.g., 150 bps).}}
Suppose \hm{that we have} two data structures: \rev{1)}~a \omcc{sorted read table (\emph{\readset})}, each entry of which stores a read and its unique ID, and \rev{2)}~a \omcc{sorted k-mer index (\emph{\seedidx})}, which contains \emph{all unique read-sized} k-mers of the reference genome, \omcc{along with each k-mer's corresponding locations in the reference genome.}
\omcx{Each} data structure \omcx{is} \emph{sorted} by read/k-mer in alphabetical order. 
With \jk{these two} data structures, it is possible to identify each read's exactly-matching locations \omcc{in the reference genome} by \emph{\omcix{streaming}} \jk{both} reads and k-mers \omcix{through a simple comparator}.
\omcix{We use two pointers, $r$ and $k$, which point to the current entries we are examining in \readset and \seedidx, respectively. 
We \emph{sequentially increment} the two pointers in three different ways based on the comparison result of the current read and k-mer.}
\omcix{First, when the current read and k-mer are identical (\circled{1} in Figure~\ref{fig-gs:sr_idea}),}
we record \omcix{the read} as an \omcix{exactly-matching read} and \omcix{increment $r$ and $k$}.
\omcix{Second, if the read is alphabetically larger than the k-mer (\circled{2}), we conclude that the k-mer does not match any \omcix{read} in \readset and increment $k$ \omcx{(so that we can examine if the next k-mer matches the read)}.}
\omcix{Third, if} the k-mer is alphabetically larger than the read (\circled{3}), \omcix{we conclude that the read does not match any k-mer in \seedidx.}
We record the read as \emph{not} an exact match
and \omcix{increment $r$ \omcx{(so that we can examine the next read)}}.

\jk{GenStore\omcvi{-EM}'s two} data structures and filtering algorithm enable an exact-match filter highly suitable for in-storage processing.
\omcc{First, }since \jk{these two} data structures are \jk{only} sequentially accessed, the filtering process can be done in a \emph{stream\omcc{ing}} manner, leveraging the high sequential read bandwidth of NAND flash memory. 
\omcc{Second, we can easily perform} exact-match detection of a read and a read-sized k-mer with simple comparator logic and \omcc{fully pipeline the} filtering process with sequential \jk{access to} the data structures.

A \emph{read-sized} k-mer index increases the total amount of accessed data for read mapping. \revmark{E4}\revp{The reason is that a read-length value of k (e.g., k=150) significantly increases the number of unique k-mers, \omcvii{compared to} \jk{\omcvii{the k values} commonly used in conventional read mappers} (e.g., k=15)}.
\omcv{For example, the size of an index structure for all unique k-mers in the human reference genome is 21 GB when k=15, while the size increases to 126 GB when k=150.} 
\jk{However, our \omcvii{proposal} (\omcvii{i.e.,}~a large yet sequentially-accessed SKIndex) is feasible and desirable} \omc{for in-storage processing}
due to the large capacity and high internal bandwidth of modern NAND flash-based SSDs.

\subsubsection{Design of GenStore-EM} 
\label{sec-gs:sr-design}

\begin{figure}[h]
        \centering
        \includegraphics[width=0.9\linewidth]{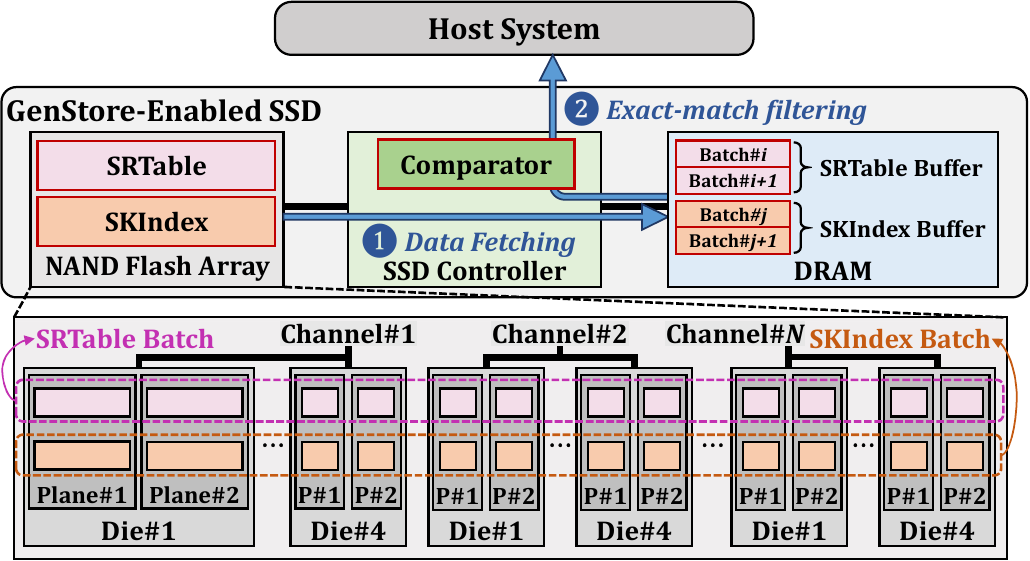}
        \caption{Overview of GenStore-EM.}
        \label{fig-gs:sr_overview}
\end{figure}

Figure~\ref{fig-gs:sr_overview} illustrates the overall \omcc{operational flow} of GenStore-EM, which consists of two steps: \textbf{Step~1.}~data fetching and \textbf{Step~2.}~exact-match filtering.  
GenStore-EM uses \jk{the} two data structures explained in Section~\ref{sec-gs:sr-overview}: 1) a sorted read table (\readset) for storing the read set and 2) a sorted k-mer index (\seedidx) for storing \jk{read-sized} k-mers from the reference genome.
Step~1 reads the two data structures from NAND flash chips to the SSD's internal DRAM in a \jk{batched manner} (\circled{1} in Figure~\ref{fig-gs:sr_overview}).
Step~2 performs exact-match filtering within each read batch, using simple comparator logic \omciv{in the SSD-level accelerator} (\circled{2}).
Steps 1 and 2 are performed in a pipelined manner. \omcc{\omciv{During filtering,} GenStore-EM sends the unfiltered reads to the host system \omciv{for} full read mapping. This enables the \omciv{concurrent} filtering of reads in \omcv{the} SSD and read mapping of the unfiltered reads in the host system.}

\head{Data Structures}
We carefully design \readset and \seedidx to minimize performance and storage overheads of GenStore-EM,
by extending the two data structures described in Figure~\ref{fig-gs:sr_idea} in two aspects. 
First, both \readset and \seedidx contain a \emph{strong} hash value (e.g., SHA-1~\cite{dang2015secure} or MD5~\cite{rivest1992rfc1321}) of each read and \omcc{read-sized} k-mer, respectively, which is used as both the sorting criterion of the data structures and a \emph{fingerprint} of each read and \omcc{k-mer} \jk{that is used by the comparator logic}.
Second, \omcc{\omcvii{unlike} the data structures described in Figure~\ref{fig-gs:sr_idea}}, \seedidx no longer contains the k-mers of the reference genome \jk{but only the fingerprints of the k-mers}.
Using strong hash values enables GenStore-EM to determine exact matches between \jk{a read and a k-mer by comparing only} their fingerprints, \omcc{which provides two benefits.}  \omcc{First, it} reduces the storage overhead of \seedidx by obviating the need to store the raw \omcc{read-sized} k-mers.\footnote{We design \readset to \omcc{store} the raw reads \omcc{so that we can} transfer \emph{unfiltered} reads to the host \omcc{for full read mapping} after \omcc{we detect them as non-exactly-matching reads}.} 
For \omcv{the human reference genome, the size of the optimized \seedidx \omcvi{ (}which stores fingerprints instead of read-sized k-mers) is \rev{32}~GB when the read size is 150 bps, which is 3.9$\times$ smaller than the size of the unoptimized \seedidx.} 
Second, \omcc{using fingerprints} significantly reduces the performance overhead of exact-match detection by avoiding comparisons of reads and k-mers \jk{that are hundreds of bytes in size}.

Note that such exact-match detection does \emph{not} affect the accuracy of read mapping due to the extremely low collision rate of strong hash functions. 
\revmark{C3}\revp{Even in an \jk{extremely  rare} case of a hash collision, \omc{the} impact \omc{of the collision} on GenStore’s accuracy will be negligible~\cite{arxivGS}, since the DNA information loss due to the falsely filtered read will highly likely be compensated by \omc{\jsr{other} reads generated from the neighboring locations in the DNA\jsr{, which almost \omciv{\emph{fully}} overlap with the falsely filtered read but have totally different hash values}.} This is because it is common practice to sequence each DNA fragment several times (i.e., with high coverage) to improve the accuracy of downstream genetic analyses~\cite{sims2014sequencing,quail2012tale,levy2016advancements,hu2021next}.
}

\omcc{We  envision that all GenStore data structures are built  \emph{offline} by the host or some other system (e.g., \jk{by the sequencing machine} when the read set or reference genome are initially \omcvii{written} to the SSD)}. \revmark{CQ2}\revp{\omcc{This} preprocessing overhead can be \jk{hidden} by two essential initial steps of the genome \jk{sequence} analysis pipeline: 1)~sequencing and 2) 
basecalling.
\omcv{For example, in the \omcv{current highest-throughput} Illumina \omcv{Nova}Seq \omcv{6000} sequencer~\cite{illumina}, sequencing and basecalling \omc{work in a pipelined manner and generate genomic read data at a} limited \omc{throughput of} \omcv{18.9} MB/s~\cite{illumina}.} 
\omcv{We analyze the throughput of GenStore's preprocessing step (i.e., generating hash values and sorting reads) and observe that even a personal laptop~\cite{lenovot740p} can provide 174 MB/s of preprocessing throughput.}
Therefore, GenStore’s preprocessing can be done in a pipeline\omc{d} manner with sequencing/basecalling, without decreasing \omc{the overall throughput of these steps}. 
This low preprocessing overhead can be \omc{further} amortized since the preprocessed data can be reused multiple times \omcc{in different read mapping experiments}.
}

\head{Step 1. Data Fetching}
GenStore-EM \jk{reads \readset and \seedidx in batches}, while exploiting the \emph{full} internal bandwidth of the SSD. \omcc{In Figure~\ref{fig-gs:sr_overview}, we refer to each batch of \readset as a\omcvi{n} \emph{\readset Batch} and each batch of \seedidx as a\omcvi{n} \emph{\seedidx batch}}. The batch size is equal to the size of data that can be read in parallel by a multi-plane read operation for each chip (i.e., \emph{Number of Planes in the SSD}~$\times$~\emph{Page Size}), 
which enables 100\% utilization of the NAND flash chips \jk{while} reading a batch. 
As shown in Figure~\ref{fig-gs:sr_overview} (bottom), GenStore-EM stores the \readset and \seedidx to NAND flash chips in an \emph{\jk{interleaved}} manner so that each of the data structures can be \emph{sequentially}, \emph{evenly} distributed across all the NAND flash chips.

GenStore-EM \omcc{exploits} the SSD's full internal bandwidth using double buffering.
As shown in Figure~\ref{fig-gs:sr_overview}, GenStore-EM employs two \omcc{sets of} batch buffers in the internal DRAM: \omcc{\emph{\readset Buffer}} (for \readset) and \omcc{\emph{\seedidx Buffer}} (for \seedidx), each of which can store two batches \omcc{of \jk{the respective} data structure}. 
\omcc{After Step 1 finishes fetching $Batch\#i$, it \omcvii{proceeds} to fetching $Batch\#i+1$, while Step 2 starts working on $Batch\#i$. If the two steps work with the same throughput, we only need to buffer two batches for each data structure. For example, to enable double-buffering} in an 8-channel SSD (with four 2-plane dies per channel and 16-KiB pages), the batch buffers require 8MB DRAM space in total.

\head{Step 2. Exact-Match Filtering}
Step 2 scans \omcc{through} each batch of \readset and \seedidx \omcc{ stored in \readset Buffer and \seedidx Buffer}, \omcvii{respectively,} comparing the fingerprints (strong hash values) with a simple hardware comparator.
When $FP(r_i)>FP(k_j)$ (where $FP(x)$ is the fingerprint of $x$, and $r_i$ and $k_j$ are the $i$-th read and $j$-th k-mer in the current batch, respectively), Step 2 scans \seedidx while increasing $j$ until it finds $k_j$ such that $FP(r_i)\leq{}FP(k_j)$.
If $FP(r_i)<FP(\omcviii{k}_j)$, it is guaranteed that no exact match exists \omcvii{for read~$r_i$}, so GenStore-EM sends \omcvii{$r_i$} to the host for the read mapping process.
When $FP(r_i)=FP(\omcviii{k}_j)$, i.e., the two fingerprints are identical, \omcc{GenStore-EM marks the read as an exactly matching read.}

\omcv{Due to the simple computation in Step 2, the execution time of GenStore-EM is bottlenecked by Step~1 (Data Fetching). As explained, Step~1 only streams the data structures in batches from the underlying NAND flash chips to the internal DRAM, leveraging the SSD's full internal bandwidth. 
Therefore, the performance of GenStore-EM can be easily scaled \omcvi{up} by increasing the SSD's internal parallelism (e.g., by deploying more channels or using low-latency NAND flash memory~\omcvi{\cite{cheong-isscc-2018, park-nvmsa-2018,park2021reducing}}).}

\subsection{\omciv{GenStore-NM for Non-Matching Reads}}
\label{sec-gs:LRF}

\subsubsection{Approach Overview}
\label{sec-gs:lr-overview}

    GenStore-NM filters \omciv{most of the} \emph{\underline{n}on-\underline{m}atching} reads, i.e., reads that \omciv{would not} align to any subsequence in the reference genome.
    \omcv{This is motivated by the fact that in read mapping, a \omcvi{large} fraction of reads might \emph{not} align to the reference genome due to} 1) the high sequencing error rate (in long reads) and/or 2) high genetic variation between \omcvii{the} compared genomes (in both short and long reads).
    \omcv{To \omcvi{illustrate} this, we analyze four \omcvi{read} mapping use cases, one with read sets with high sequencing error \omcvii{rates}, and three with high genetic variation in the compared \omcvi{genomes}. 
    Use cases with high genetic variation include 1) \omcvii{samples from rapidly-evolving species} (such as SARS-CoV-2~\omcvi{\cite{bhoyar2021high}}) that have high genetic variation compared to the reference genome,} 2)~samples with no known reference genomes, and
    3)~mapping \omcvi{a read set} against the human reference genome to filter out \omcvi{\emph{human contamination}, i.e.,~reads that have been sequenced from human-origin contaminant DNA in a non-human sample.\footnote{\omcvi{Contamination of non-human samples with human DNA is commonly observed~\cite{breitwieser2019human,human2012structure} and corrected for~\cite{danko2021global,knight2018best,human2012structure}}.}}
    \omcv{For each use case \omcvi{(except for the \omcvii{\emph{Contamination}} use case)}, we analyze \omcvi{two different combinations of \omcvi{the input} read set and \omcvi{the} reference}.} 
    Table~\ref{table-gs:lr-profile} \omcv{summarize\omcvi{s} the result of this analysis by showing \omcvi{input}} read sets, \omcv{the propert\omcvi{ies of each read set (e.g., \omcvii{read length and dataset size})}}, reference genomes, and the percentage of reads in each read set that align to subsequences in the reference genome.  
    We observe that a large fraction of reads \omciv{\omcvi{(31.7\%--99.6\%)} within a read set} does \emph{not align to \omcvii{any} subsequence} in the reference genome. 
    \omcv{\omcvi{Quickly f}iltering this large fraction of non-aligning reads in the SSD can reduce the \omcvi{large} data movement from the storage and \omcvi{expensive} ASM computation for reads that would not align.}

\begin{table}[h]
\centering
\caption{\omcv{Fraction of aligning reads in various read mapping use cases with short and long reads.}}
\resizebox{0.9\columnwidth}{!}{%
\begin{tabular}{c|c|c|c|c}
\toprule
\multirow{2}{*}{\textbf{\omcv{Use case}}} & \multirow{2}{*}{\textbf{Input read \omcvi{set} (S}hort/\textbf{L}ong\textbf{)}}  & \textbf{Size} & \multirow{2}{*}{\textbf{Reference}} & \omcv{\textbf{Align}} \\
& & \textbf{[GB]} & & \textbf{[\%]}\\
\midrule
\midrule
\multirow{2}{*}{Sequencing errors} & \omciv{ERR3988483 (L)~\cite{sayers2021database}} & 54 & \multirow{2}{*}{hg38~\cite{schneider2017evaluation}} & \omcv{{\bf 47.\omcvi{4}}}\\
& \omciv{HG002\_ONT\_20200204 (L)~\cite{zook2016extensive}} & 371 & & \omcv{{\bf 69.\omcvi{3}}}\\
\midrule
Rapidly evolving & SRR5413248 (L)~\cite{sayers2021database} & 1.69& NZ\_NJEX02~\cite{clark2016genbank} & \omcv{{\bf 60\omcvi{.0}}}\\
samples & \omciv{SRR12423642 (S)~\cite{sayers2021database}} & 0.466 & NC\_045512.2~\cite{wu2020new} & \omcv{{\bf 23.\omcvi{1}}}\\
\midrule
\multirow{2}{*}{No reference} & SRR6767727 (L)~\cite{sayers2021database} & 12.4& \multirow{2}{*}{NZ\_NJEX02~\cite{clark2016genbank}} & \omcv{{\bf 0.35 }}\\
& SRR9953689 (L)~\cite{sayers2021database} & 15.9&  & \omcv{{\bf 37\omcvi{.0} }} \\
\midrule
Contamination & SRR9953689 (L)~\cite{sayers2021database} & 15.9& hg38~\cite{schneider2017evaluation} & \omcv{{\bf  1\omcvi{.0} }} \\
\bottomrule   
\end{tabular}}
\label{table-gs:lr-profile}
\end{table}

\omciv{To avoid expensive ASM \omcv{computation} for reads that would not be aligned, state-of-the-art read mappers commonly employ a step called \emph{chaining} (described in Section~\ref{sec:background-readmapping}), which calculates each read's similarity score (called \emph{chaining score}) to the reference genome and filters out reads with a low score.
\omciv{GenStore-NM uses this basic idea of chaining to build an in-storage filter.}}

    \head{Key Challenges} Calculating a chaining score in the SSD is challenging because finding the best chaining score requires performing an expensive dynamic programming (DP) algorithm on every \omciv{potential matching location (i.e., seed)} within a read (explained in Section~\ref{sec:background-readmapping}).
    Normally this operation has $\mathcal{O}(N^2)$ time and space complexity, where $N$ is the number of seeds.
    \omcvii{Even though}
    \omciv{\omcvii{a} chaining score} can be approximated accurately \omcv{in} $\mathcal{O}(hN)$ time and space by considering only $h < 50$ seeds at a time~\cite{li2018minimap2}\omcvii{, w}e observe that some long reads can have up to thousands of seeds \omciv{due to their length}. Therefore, designing \omcv{a} chaining accelerator to \omciv{perform an expensive \omcvii{DP} algorithm on \omcvii{these} large \omcvii{reads} ($N > 1000$) can incur significant performance or area \omcv{overheads}.}

\begin{figure}[h]
        \centering
        \includegraphics[width=0.6\linewidth]{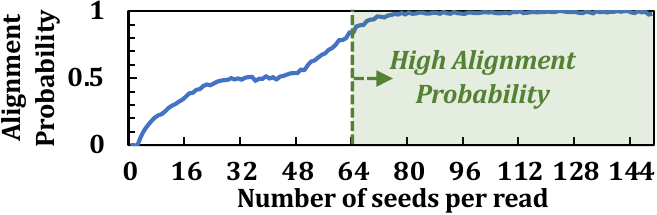}
        \caption{Alignment probability as a function of the number of seeds per read in a long read mapping use case.}
        \label{fig-gs:lr_align}
\end{figure}

    \head{Key Ideas} 
    \omcv{T}\omciv{o avoid such expensive chaining, \omcv{GenStore-NM} selectively perform\omcv{s} \omcv{a fast version of} chaining  \emph{only} on reads with a small number of seeds} and send\omcv{s} other reads to the host system for full read mapping (including \omcv{complete} chaining). This idea is based on \omciv{our} key observation that 1) a read with a large number of seeds most likely aligns to the reference \omciv{genome and does not require in-storage filtering, and 2) selective chaining \omcvi{can filter} many non-aligning long reads, without requiring costly hardware resources in the SSD}. \fig{\ref{fig-gs:lr_align}} shows the \omcvi{alignment} probability of a \omciv{read in a \omcv{long read dataset (SRR5413248 \cite{sichtig2019fda} in Table~\ref{table-gs:lr-profile})} \omcvi{to} subsequences} in the reference genome~\omcvii{(NZ\_NJEX02~\cite{clark2016genbank})}, \omcviii{as a function of the} \omcv{number of} seeds \omcv{per read \omcviii{($N$)}}.
    \omciv{The average read length in this dataset is 10K base pairs and the number of seeds goes up to several thousands for some reads.} 
    We observe that reads with \omcv{a sufficiently large number of seeds} are \omcv{very} likely to align \omciv{to subsequences in the reference genome} \omcv{(e.g., at least 85\% of reads with \omcvi{$N\geq64$} seeds align). Such reads }can be directly sent to the CPU for \omciv{full read mapping} (bypassing the \omciv{in-storage \omcviii{chaining-based}} filter). 
    
We extend this analysis to read datasets from various organisms commonly used in genomics studies: \textit{E. coli}~\cite{clark2016genbank}, yeast~\cite{engel2014reference}, thale cress~\cite{berardini2015arabidopsis}, fruit fly~\cite{larkin2020}, mouse~\cite{church2009lineage}, and human~\cite{schneider2017evaluation}. 
We observe that when a read has $N = 64$, $128$, or $256$ seeds, it aligns with average probabilities of $88.87\%$, $91.32\%$, and $93.84\%$, respectively. 
Based on these observations, we \omcvi{design GenStore-NM to \emph{selectively} perform chaining only on} reads with \omcvii{fewer} than $N$ seeds\omcvi{, while sending \omcvii{reads with \omcviii{at least} $N$ seeds} to the host system.}\footnote{\omcvi{GenStore-NM's design can be tuned based on different values of $N$.}} 
\omcv{This selective chaining} significantly reduces the chaining execution time and \omcv{additional hardware} area \omcv{cost,} while filtering most reads \omciv{that would not align to the reference genome}.

\newcommand\kindex{KmerIndex\xspace}
\subsubsection{Design of GenStore-NM}
\label{sec-gs:lr-design}

\begin{figure}[h]
        \centering
        \includegraphics[width=0.9\linewidth]{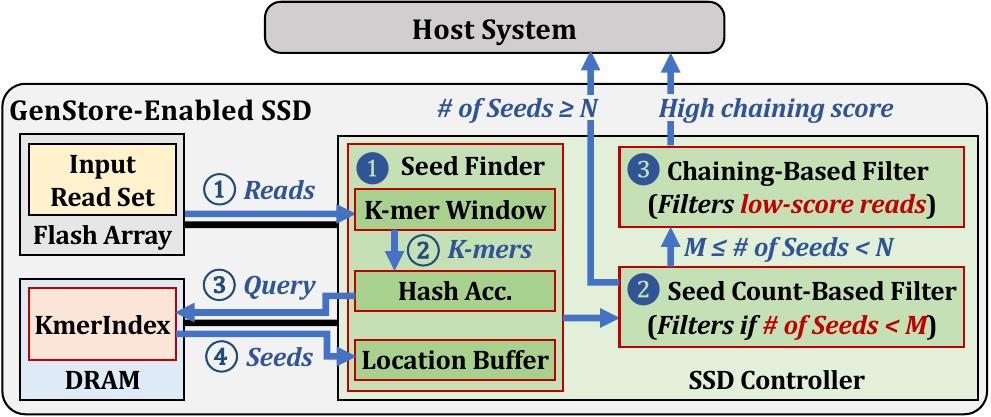}
        \caption{Overview of GenStore-NM.}
        \label{fig-gs:lr_overview}
\end{figure}

    Figure~\ref{fig-gs:lr_overview} shows the overview of GenStore-NM that filters out most of the non-matching reads in three steps. 
     In Step 1, \omciv{GenStore-NM} reads the input read set from the flash chips, generates minimizer \omciv{k-mers} for each read (as explained in Section~\ref{sec:background-readmapping}), and looks up each minimizer in a \emph{K-mer Index} (\kindex) to find \omciv{the potential matching locations, i.e., seeds  \omcv{(\circled{1} in \fig{\ref{fig-gs:lr_overview}})}}.  In Step 2, GenStore-NM \omciv{counts the number of seeds \omcv{in each} read \omcv{to decide} if the read needs to go through \omcv{chaining (\circled{2})}.} 
    \omcvi{To further improve \omcviii{overall} performance, Step 2 also filters out reads with \emph{too few} seeds \omcviii{(i.e., $< M$)}, which would not align to the reference and thus would be filtered anyway by the baseline read mapper~\cite{li2018minimap2}.}
    In Step 3, GenStore-NM filters reads based on their chaining scores using a \omcv{fast and} efficient chaining accelerator \omcv{(\circled{3})}.  If a read has at least one chain with a score above a specified threshold, it \omciv{is sent} to the CPU for mapping. Otherwise, the read \omciv{is} filtered. \omciv{All three} steps run in a pipelined manner.

    \head{Data Structures} We carefully design \omcv{the} \kindex to reduce the \omcv{SSD's internal} DRAM capacity required for storing it. \kindex, similar to the index used by the baseline read mapping tool (i.e., \omcvii{M}inimap2~\cite{li2018minimap2}), is a hierarchical hash table.
    To reduce the memory capacity required for storing the \kindex, we \omcv{make} three \omcv{modifications}: 1)~not stor\omc{e} the reference genome since it is not needed by our approach, 2)~not stor\omc{e} seeds with \omcv{\emph{many}} matching locations (\rev{e.g., \omcv{more than} 495 locations in the baseline read mapper~\cite{li2018minimap2}})\footnote{\revp{Each matching k-mer can map to one or more locations in the reference genome.}}
    \revmark{E5} \omcv{since} \omc{read mappers usually ignore such} seeds in the chaining process~\cite{li2018minimap2}, \omcv{and} 3)~increas\omcv{e} the number of \omciv{\omcv{hash table}} buckets so that each bucket holds one minimizer. \omciv{Increasing the number of buckets} increases the false positive rate in the \omciv{filter}, which leads to \omciv{finding extra seeds and performing} extra chaining operations (without loss of accuracy). We make this trade-off to reduce \omcv{\omcvi{the} \kindex's} size and the required memory space. \omciv{These optimization}s \omcvii{reduce} the size of the index for the human reference genome~\cite{li2018minimap2} from 5.8 GB to 2.9 GB, \omciv{which enables GenStore-NM to \omcv{easily} store} the \kindex in the \omcv{limited} DRAM \omcv{space} inside the SSD.\revmark{E7}
    
    \head{Step 1: Seed \omciv{Finding}} 
     \omcv{This} step find\omcv{s} the \omciv{potential} matching locations of a read \omciv{(i.e., seeds)} in the \kindex.
    \omciv{~GenStore-NM} \omcv{first} reads the input read set from all \omcvi{f}lash chips \omcv{(\wcirc{1} in Figure~\ref{fig-gs:lr_overview})}. 
    \omcvi{GenStore-NM} exploits the \emph{full} internal bandwidth of the SSD by reading the read set in parallel \omcv{via} a multi-plane read operation for each chip \omcv{(same as GenStore-EM, Section~\ref{sec-gs:sr-design})}. In a shifting buffer (\omcv{called the} \emph{K-mer Window}) \omciv{in the channel-level accelerator}, \omciv{GenStore-NM} stores the \omciv{$w$} most recently read k-mers of each read. 
     For each sliding window of $w$ k-mers\omcvi{,}\footnote{\omciv{The default value for $w$ in \omcvii{M}inimap2~\cite{li2018minimap2} is 10, but GenStore's design can be tuned for different values.}} GenStore-NM \omcvi{finds the minimizer of the window (described in \sect{\ref{sec:background-readmapping}})} using a 64-bit Integer Mix hash function (hash64)\omcv{~\cite{wang2007integer}} \omciv{in the SSD-level accelerator}
     \omcv{(\wcirc{2})}. 
     \omciv{GenStore\omcvi{-NM} queries} each minimizer in the \kindex until it reads $N$ seeds (\omciv{e.g.},  $N$ = 64) or it reaches the end of the read \omcv{(\wcirc{3})}. Each index query takes up to two memory accesses (to visit the two levels of the hash table). \omciv{GenStore-NM} stores the \omciv{seeds} in the \emph{Location Buffer} \omciv{in the channel-level accelerator} \omcv{(\wcirc{4})}, and moves to Step 2.

    \head{Step 2: Seed Count-Based Filtering} \omcv{This step} compare\omcv{s} the number of seeds \omciv{in the Location Buffer} with a lower bound $M$ and an upper bound $N$.
    If a read has \omcvii{fewer} than $M$ matching seeds, it is filtered since it will not meet the minimum chaining score requirement \omciv{of the baseline read mapper}\omcv{~\cite{li2018minimap2}}.\footnote{\omcv{We use $M$=3, as \omcvi{in}~\cite{li2018minimap2}, but GenStore-NM's design trivially supports different values of $M$.}} 
    If the read has more than $N$ seeds, it means that \omcvi{the read} will
    \omcv{very} likely \omcvi{align to the reference (e.g., reads with \omcviii{at least} 64 seeds in \fig{\ref{fig-gs:lr_align}})}
    and \omcvi{thus} require the full read mapping process. \omciv{GenStore-NM} sends such reads to the \omciv{host system for the full read mapping process}. This way, the read mapping process of the unfiltered reads can run concurrently with GenStore's \omciv{filtering} operations. For any other read, \omcv{GenStore-NM} performs chaining in the SSD in Step 3.
    
    \head{Step 3: Chaining-Based Filtering} \omcv{This} step performs chaining \omcv{(see \sect{\ref{sec:background-readmapping}), a step commonly used in state-of-the-art read mappers~\cite{li2018minimap2,dobin2012,li2016minimap}}} to filter out reads with low chaining scores and send reads with high chaining scores to the CPU to undergo the full read mapping process. 
    \omcv{The key difference in GenStore-NM's chaining process compared to existing read mappers is that GenStore-NM \emph{selectively} performs chaining only \omcvi{on} reads with lower seed counts than a threshold $N$. 
    Such selective chaining is based on our two key observations; 1) a long read with many \omcvi{seeds} most likely aligns to the reference genome and \omcvi{thus} does not require in-storage filtering, and 2)~selective chaining can filter many non-aligning long reads, without requiring costly hardware resources in the SSD (\omcvi{as shown in} Section~\ref{sec-gs:lr-overview}).
    
We design GenStore-NM's chaining unit based on the chaining algorithm used in Minimap2~\cite{li2018minimap2}, a state-of-the-art baseline read mapper.   }
A chain is computed from a sequence of seeds $S_1,\ldots, S_N$\footnote{\omcvii{Sorted based on their locations in the reference genome.}} \omciv{of lengths $w_1,\ldots, w_N$, respectively. For a given seed $S_i$, $x_i$ ($y_i$) denotes the ending position of the seed's matching location in the reference genome (the read).}
The chaining score acts as an approximation of an alignment score, \omciv{increasing linearly with the number of matching base pairs between the read and the reference, and decreasing with the size of gaps between the seeds. The chaining score is defined as follows (based on ~\cite{li2018minimap2}):}
\begin{equation}
\label{eq-gs:chain1}
f(i) = \max\left\lbrace\mathop{\max}_{i>j\geq 1}\left\lbrace f(j)+\alpha(j,i)-\beta(j,i)\right\rbrace, w_i\right\rbrace,
\end{equation}
\noindent where $f(i)$ is the best chaining score which can be computed with the seeds $S_1,\ldots, S_i$. Given a chain whose last seed is $S_j$, $\alpha(j,i)$ (also called the \emph{match score}) is the number of new base pairs added to the chain after adding $S_i$. $\beta(j,i)$ (also called the \emph{gap penalty}) subtracts from the chain score based on \omcv{the distance} between $S_i$ and $S_j$ \omcv{in the reference genome}.\footnote{\omcv{See Minimap2~\cite{li2018minimap2} for more details regarding $\alpha(j,i)$ and $\beta(j,i)$.}}

\begin{figure}[h]
        \centering
        \includegraphics[width=.9\linewidth]{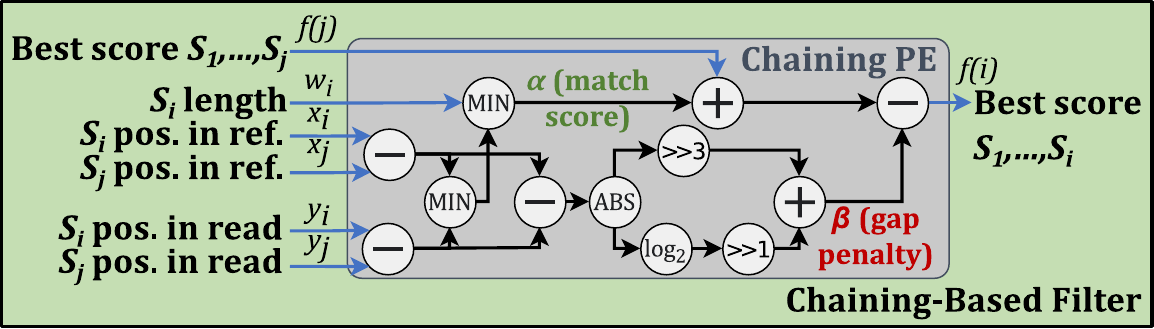}
        \caption{\omcv{Chaining \omcvi{processing element (PE)} in GenStore-NM}.}
        \label{fig-gs:chaining_unit}
\end{figure}

\omcv{GenStore-NM's Chaining-Based Filter (in Figure~\ref{fig-gs:lr_overview}) consists of 1)~a~\emph{Chaining Buffer} to store $x_i$, $y_i$, $w_i$, and $f(i)$ values and 2)~a~\emph{Chaining Processing Element (PE)} based on Equation~\eqref{eq-gs:chain1}. } Figure~\ref{fig-gs:chaining_unit} shows the \omcv{design of the Chaining PE} \omciv{in the channel-level accelerator}. We reduce the latency and size of this unit by approximating multiplication with shift operat\omciv{ions}. We ensure that \omcv{our hardware} optimizations always over-estimate the chaining score so that we do not \omciv{filter out} any potential read \omciv{mappings}. GenStore-NM does not affect the accuracy of the read mapper because the filter performs the same computations as the baseline chaining filter for reads with seeds $< N$ inside SSD. 
The chaining \omcv{PE} needs to be executed  $N \times M$ \omciv{times} per read where $N$ is the number of seeds per read and $M$ is the DP-iterations of each seed. Due to the \omcv{limited} number of seeds that go through the chaining step, we limit the number of chaining units that are needed to match the full internal bandwidth of SSD.

Similar to GenStore-EM, the steps of GenStore-NM \omciv{are} pipelined. The performance of GenStore-NM can be easily scaled \omcv{\omcvi{up} by increasing the SSD's internal parallelism (e.g., by deploying more channels or using low-latency NAND flash memory)~\omcvi{\cite{cheong-isscc-2018, park-nvmsa-2018,park2021reducing}}}.

\subsection{GenStore Flash Translation Layer (FTL)}\label{sec-gs:ftl}

GenStore requires simple changes to the existing FTL code\omciv{.}

\head{GenStore FTL Metadata} 
GenStore metadata includes the mapping information of the data structures necessary for read mapping acceleration, which enables access to the data structures without the L2P mapping table of the regular FTL. 
In accelerator mode, where GenStore operates as an accelerator, GenStore also keeps \omciv{in internal DRAM} other metadata structures of the regular FTL (e.g., the page status table and block read counts~\cite{park2021reducing,cai2017error}) which need to be updated \omciv{during} the filtering process.
We carefully design GenStore to only \emph{sequentially} access the underlying NAND flash chips while operating as an accelerator, so it requires only a small amount of metadata to access the stored data. 

\head{Data Placement} 
GenStore needs to properly \omciv{place its} data structures to enable the full utilization of \omciv{the} internal SSD bandwidth in accelerator mode. 
When \jsiv{each} data \jsiv{structure} is initially written \jsiv{to the SSD}, GenStore sequentially and evenly distribute\jsiv{s} it across NAND flash chips. 
We design GenStore to store \jsiv{every} data structure using multiple sets of NAND flash blocks \jsiv{such that} each block set consists of NAND flash blocks with the same block offset across the planes in a die\jsiv{.
For example,} a die’s block set \jsiv{$k$} includes all NAND flash blocks whose offset is \jsiv{$k$} in the die.
\jsiv{Such data placement} enables GenStore \jsiv{ not only} to always perform \omciv{\emph{multi-plane}} read operations \omciv{during} the filtering process, \jsiv{but also to significantly} reduce the size of GenStore metadata. 
For example, \jsiv{suppose that a GenStore-enabled SSD consists of} 128 2-plane NAND flash dies\jsiv{, and the block size is 12 MB (i.e., the size of each block set} would be 24 MB). 
In such a case, GenStore can specify the physical location of a 30-GB data structure by maintaining only the list of 1,250 (30 GB/24 MB) physical block addresses. 
This way, GenStore significantly reduces the size of the necessary mapping information from 300 MB (\omciv{with conventional} 4-KiB page mapping) to \omciv{only} 5 KB\omciv{ (1,250$\times$4 bytes), and} the saved internal DRAM space can be used for loading data structures in the accelerator mode.

\head{SSD Management Tasks} 
In accelerator mode, GenStore stops working as a regular SSD\omciv{.
It only reads} data structures \omciv{to perform filtering, and does not \jsiv{write} any} \jsiv{new data. 
Therefore, GenStore does not require any} write-related SSD-management tasks such as garbage collection~\cite{tavakkol2018flin,kim2020evanesco,cai2017error,park-dac-2016, park-dac-2019} and wear-leveling~\cite{chang2007efficient,cai2017error,cai-insidessd-2018}.

The other tasks necessary for ensuring data reliability, such as refreshing data to avoid \jsiv{uncorrectable errors due to }\omciv{data} retention and read disturb~\omciv{\cite{cai2017vulnerabilities, luo2018improving,luo2018heatwatch,cai2015read,cai2013error, cai2012flash, cai2017error, ha2015integrated, cai-insidessd-2018,luo2015warm}}, can be done before or after the filtering process\omciv{,} for two reasons. 
First, GenStore significantly limits the amount of data whose retention age would exceed the manufacturer-specified threshold for reliable operation (e.g., 1 year~\cite{micron3dnandflyer}) during the filtering process, since
GenStore’s filtering process takes a short time.\footnote{\omciv{E.g.,} less than 7 minutes for a 1TiB dataset even with a low-end SSD \omciv{(\ssdl)}\jsiv{~\cite{samsung860pro}}.}
GenStore simply refreshes such data~\cite{cai2013error,luo2018improving, cai2017error,cai2012flash} before starting the filtering process. 
Second, GenStore-FTL can easily avoid read disturbance errors for data with high read counts~\cite{cai2015read, ha2015integrated} since GenStore sequentially reads NAND flash blocks only \emph{once} during filtering.
After the filtering process, to prevent read disturb errors in \omciv{future} filtering processes \omciv{or regular accesses}, GenStore refreshes pages that store a data structure if the structure’s read count exceeds a certain threshold.

\section{Evaluation Methodology}
\label{sec-gs:methodology}

{
\head{Evaluated Systems}
We show the benefits of GenStore when it is integrated with the state-of-the-art software and hardware read mappers.
\vspace{-.2em}
 To this end, we evaluate the following systems:}
 \squishlist
     \item {\textbf{\base:} Minimap2~\cite{li2018minimap2} \omciv{is} a state-of-the-art software read mapper baseline for \jsr{both} short and long reads\omciv{.} GenCache~\cite{nag2019gencache} and Darwin~\cite{turakhia2018darwin} \omciv{are} state-of-the-art hardware read mappers for short and long reads, respectively.}
     \item {\textbf{\gsos:} \revmark{CQ4\\Part 1}
     \base integrated with an implementation of the GenStore filter without in-storage support \omciv{(Ext stands for \emph{external} to storage)}.
     \gsos concurrently filters reads while using \base to perform read mapping for unfiltered reads. 
     The goal of evaluating \gsos is to decouple the effects of \omcvii{GenStores's} two major benefits: 1) alleviating I/O bottlenecks via efficient in-storage processing and 2) reducing the workload of the read mapper by filtering reads with simple operations. \omciv{\gsos obtains the second benefit but not the first.} %
     For software read mappers, we \omciv{evaluate} a pure software implementation of GenStore ~\omciv{that} concurrently run\omcv{s} with \base.\footnote{\revp{We do not evaluate a separate \gsos configuration for long read software mapper since \base (Minimap2~\cite{li2018minimap2}) already incorporates the chaining filter used in GenStore-NM. GenStore-NM implements part of this chaining filter at low cost (enabled by the key observations in Section~4.3) to fit within the constraints of in-storage processing.}} For hardware read mappers, we \omciv{evaluate} a hardware implementation of GenStore \emph{outside} the SSD.}
     \item  {\textbf{\gs:} \base integrated with the {hardware} GenStore filter\omciv{ing accelerators, GenStore-EM and GenStore-NM, }as described in Sections~4.2 and 4.3\omciv{. GS} concurrently filters reads \emph{inside} the SSD while {using} \base to perform read mapping for unfiltered reads.}
 \squishend
 \vspace{-.2em}
\omciv{The source code of GenStore and scripts and datasets can be freely downloaded from \url{https://github.com/CMU-SAFARI/GenStore.}}

\head{Area and Power} 
We \hm{implement}  \rev{GenStore's logic components} in Verilog HDL. We synthesize our designs  using \hm{the} Synopsys Design Compiler~\cite{synopsysdc} \hm{with a} 65nm process {technology node} to estimate latency, area and power consumption.
We use \hm{the} SSD power values \hm{of the} Samsung 3D NAND flash-based SSD~\cite{samsung860pro}, and DRAM power values  based on DDR4 model~\cite{ddr4sheet, ghose2019demystifying}.

\head{Performance Model\omciv{ing}}  
{We evaluate \textbf{hardware} configurations using two state-of-the-art simulators} to analyze the performance of GenStore.
We model DRAM timing with \hm{the} DDR4 \rev{interface~\cite{ghose2019demystifying,ghose2018your}} in Ramulator~\cite{kim2016ramulator, ramulatorsource}, a widely-used, cycle-accurate DRAM simulator.
We model SSD performance using MQSim~\cite{tavakkol2018mqsim}\omcv{, a} widely-used \omcv{simulator for} modern SSDs. 
\revmark{A3}\revp{We {model} the end-to-end throughput of GenStore based on the throughput of each \hm{GenStore pipeline} stage: accessing NAND flash chips, accessing internal DRAM, accelerator \omciv{computation}, and transferring unfiltered data to the host.} \hm{We estimate the performance of GenCache and Darwin accelerators based on the data reported in the original works~\cite{nag2019gencache,turakhia2018darwin}.}

\omciv{\head{Real System Results}} {We use real systems to evaluate \omciv{all} \textbf{software} configurations.} For a given reference genome, separate \omciv{reference} indexes are generated for each sequencing technology using \omciv{the software read mapper's default settings for each technology~\cite{li2018minimap2}}.
All other \omciv{read mapping} parameters are kept at their default values. We perform all experiments on an AMD$^\text{\textregistered}$ EPYC$^\text{\textregistered}$ 7742 CPU with {1TB} DDR4 DRAM (available in user space). We measure power in these systems using AMD$^\text{\textregistered}$ \textmu{}Prof~\cite{microprof}.
We carefully evaluate GenStore’s benefits over the baseline in a conservative manner \hm{by {using optimized configurations} for all baselines}. Our baselines fetch data from the SSD by sequentially reading the data in batch\hm{es}~\cite{li2018minimap2}, \hm{while providing} sufficiently large DRAM \hm{capacity} to contain all data that gets reused \emph{during} {read mapping}. \rev{We \omcv{use} the number of threads that leads to each software configuration's best performance (i.e., execution time):  128 threads for \base and \revp{16 threads for \gsos}  in our experimental setup}.\footnote{\revp{Performance of \gsos saturates after 16 threads since it becomes bottlenecked by the external SSD bandwidth.}}

\rev{
\head{SSD Configurations} We analyze the benefits of GenStore \omciv{ on} three SSD configurations: 
1)~a low-end SSD (\ssdl)~\cite{inteldcs4500} with a SATA3 interface~\cite{SATA}, 
2)~a mid-end SSD (\ssdm)~\cite{samsung980pro} using a PCIe Gen3 M.2 interface~\cite{PCIE}, and
3)~a high-end SSD (\ssdh)~\cite{samsungPM1735} with a PCIe Gen4 interface~\cite{PCIE4}. 
}

\head{Datasets} 
\label{sec-gs:datasets}
For short read experiments, we use the \texttt{hg38} human reference genome~\cite{schneider2017evaluation}. {In Section~\ref{sec-gs:motivation},} \omciv{w}e use real short reads
(\texttt{SRR2052419}~\cite{zook2016extensive}, 19.6 GB). 
\rev{{In Section~\ref{sec-gs:results}}, to flexibly \omciv{and controllably} analyze the effect of read sets with different features,} we simulate \rev{read sets with various sizes (up to 440 GB) and different fractions of exactly-matching reads (75\% and 85\%)
} 
using \hm{the} Mason 2 \rev{genomic read simulator~\cite{holtgrewe2010mason}}. 
We generate reads with different exact-match rates by introducing sequence mutations \hm{(uniformly randomly drawn} from the gold\hm{-}standard mutation list for the \hm{human} sample \texttt{NA12878}~\cite{zook2016extensive}\hm{) to the reference genome}.
For the long read experiments, we use the reference genome and read set from  Table~\ref{table-gs:lr-profile} in Section~\ref{sec-gs:lr-overview}. \rev{To flexibly analyze the effect of data size, we generate larger read sets by concatenating the original read sets several times.}

\section{Evaluation}
\label{sec-gs:results}

\subsection{Area \atb{and Power} Analysis}
\label{sec-gs:area}

\omciv{We propose GenStore as an in-storage processing system that supports both accurate short reads and error-prone long reads with different \omcv{levels of} genetic variation \omcvi{between} \omcv{compared \omcvi{genomes}}}.
Table~\ref{tab-gs:area-power} shows the area and power \omcc{consumption} of each logic unit used in GenStore. In the first column, the table shows \omcvii{the} different logic units used in GenStore along with the width of each entry. \omcvii{The s}econd column shows the number of instances of each unit \omciv{in an 8-channel SSD}. We find the number of instances based on the frequency  of and the required throughput from each unit such that GenStore can process data concurrently read from all planes in all SSD channels. 
\omcv{GenStore's hardware units work at different clock frequencies, with the lowest frequency being 300 MHz (for each channel-level Chaining PE). 
While the hardware units can be designed to operate at \omcvi{higher} frequency, their throughput is sufficient when operating at lower frequencies since the end-to-end execution time of GenStore is bottlenecked by \omcvi{NAND} flash reads.} 
The third and fourth columns show the area and power of one instance of each unit.

\begin{table}[h]
        \centering
        \caption{%
        Area and power breakdown of GenStore's logic}
        \resizebox{0.9\columnwidth}{!}{%
        \begin{tabular}{c|c|c|c}
        \toprule
        \textbf{Logic unit} & \textbf{\# \omcvi{of \omcviii{instances}}} & \textbf{Area [mm$^2$]} & \textbf{Power [mW]}\\
        \midrule
        \midrule
        Comparator (64-bit) & \omciv{1 per SSD} &  0.0007 & 0.14\\
        $K$-mer Window ($10\times19\text{-bit}$) & 2 per  channel & 0.0018 & 0.27\\
        \omcviii{Hash} Accelerator (64-bit) & \omciv{2 per SSD} & 0.008 & 1.8\\
        Location Buffer \rev{($64\times64\text{-bit}$)} & 1 per  channel & \rev{0.00725} & \rev{0.37375}\\
        Chaining Buffer ($50\times(16\text{-bit}+64\text{-bit})$) & 1 per  channel & 0.008 & 0.95\\
        Chaining PE & 1 per  channel & 0.004& 0.98\\
        Control & \omciv{1 per SSD} & 0.0002 & 0.11\\
        \midrule
        \textit{Total for an 8-channel SSD} & - & 0.20 & 26.6\\
        \bottomrule
        \end{tabular}}
        \label{tab-gs:area-power}
\end{table}

\omcv{T}he \omciv{total hardware} area \omciv{needed for GenStore} is very small (0.20~mm$^2$ at 65~nm and 0.02~mm$^2$ at 14~nm\hm{,} only 0.006\% of a 14nm Intel Processor~\cite{wikichipcascade}).\footnote{Since lower technology nodes are not available publicly, we scale the area consumption down to {lower} process technolog\omciv{y nodes} using the methodology  in~\cite{stillmaker2017Scaling}.} 
The area overhead of GenStore hardware (0.06~mm$^2$ at 32~nm) is less than 9.5\% of the three 28nm ARM Cortex R4
processors~\omciv{\cite{cortexr4}  \omcv{in a SATA} SSD controller~\cite{samsung860pro}}.

\revp{The area and power\revmark{B2} of most logic units (except for the comparator, the hash64 accelerator, and the control unit) increase linearly with the channel count. Each instance of the comparator and hash64 accelerator supports multiple channels (up to 12 and 4 channels, respectively). Therefore, these units scale by $\lceil\frac{\#channels}{12}\rceil$ or $\lceil\frac{\#channels}{4}\rceil$, respectively. The area and power of GenStore's control unit remains the same across different channel counts.}

\subsection{GenStore-EM Analysis}

We analyze the benefits of GenStore-EM  \rev{for a 22-GB short read set} \revmark{E6}\revp{\omcvii{where} 80\% of  reads exactly match \omcvii{some subsequence in} the reference genome \omciv{(see Section~\ref{sec-gs:methodology} for input generation methodology)},} on a system with three different SSD configurations: \ssdl, \ssdm, \omcv{and} \ssdh.

\begin{figure}[!tbh]
\centering
 \includegraphics[width=0.9\linewidth]{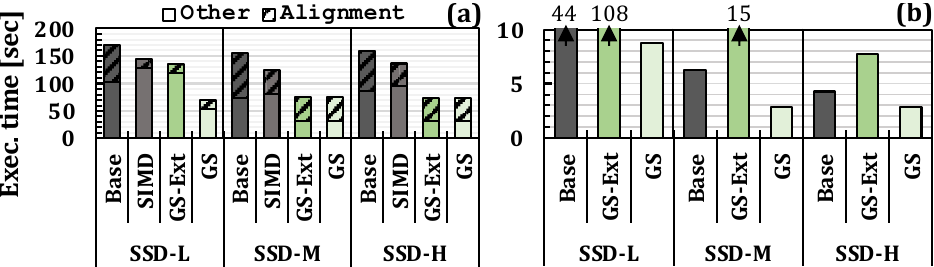}
\caption{\rev{\omciv{GenStore-EM} performance under different SSDs,} when integrated with a (a)~software read mapper (b)~hardware read mapper.}
\label{fig-gs:GS-SR-main}
\end{figure}

\head{\rev{Integration with a Software Read Mapper}}
Figure~\ref{fig-gs:GS-SR-main}a shows the execution time of \jsr{four read mapper configurations: 
1) } \base (Minimap2~\cite{li2018minimap2}),
2) \simd, an extension of \base with \hm{a} SIMD implementation of a baseline exactly-matching read filter \omciv{using 128-bit SIMD instructions}, 
\rev{3) \gsos}, and
4) \gs. 
\revp{We divide the execution time between \texttt{Alignment} (chaining and alignment's contribution to end-to-end execution time) and \texttt{Other} (file access, seeding, and exact match filtering's contribution to end-to-end execution time).\revmark{B3\\Part 1}}

We make four observations \jsr{from} Figure~\ref{fig-gs:GS-SR-main}a. 
\rev{First, for all SSD types\omciv{,} \gs significantly outperforms \base and \simd by $2.07$-$2.45\times$ and $1.66$-$2.09\times$, respectively\hm{, by alleviating} the cost of moving data from the SSD to the host CPU and remov\hm{ing} the burden of finding exactly-matching reads from the rest of the system (DRAM and the processor).
\revp{Second, \revmark{CQ4\\Part3}\gsos provides significant performance improvements over \omciv{both} \base and \simd in \ssdm and \ssdh. However, \gsos provides limited benefits over \simd in \ssdl since its performance is \jsr{bottlenecked} by the low external I/O bandwidth.}
Third, \hm{e}ven though \simd also filters out the same \hm{number} }\rev{of reads and reduces the alignment time similar\omciv{ly to} \gs and \gsos, it\hm{s} average performance benefit over \base ($1.19\times$) \omciv{is} quite limited compared to those of \gs ($2.23\times$) and \gsos ($1.83\times$). This is because %
1)~both \gs and \gsos reduce the number of memory accesses per read and convert the random memory accesses to more efficient streaming accesses, and 2)~\gs addresses the I/O bottlenecks due to }\rev{limited external SSD bandwidth.
Based on our observations, we draw two conclusions.
First, GenStore-EM leads to \omciv{large} performance improvements \omciv{in short read genomic analysis} by efficiently filtering large amounts of data in storage.
\revp{\omciv{Second, \revmark{CQ4\\Part4}even without in-storage processing support, the key idea of GenStore-EM can significantly improve the performance of the state-of-the-art software read mapper especially when using high-bandwidth SSDs (e.g., \ssdm and \ssdh).}}
}

\rev{
\head{Integration with a Hardware Read Mapper} 
Figure~\ref{fig-gs:GS-SR-main}b shows the execution time of \rev{three hardware read mapper configurations:
1)} \base (GenCache~\cite{nag2019gencache}),
\rev{2) \gsos, \omcvii{and}}
\jsr{3)} \gs.\footnote{\omciv{We do not show the execution breakdown of mapping in Figure~\ref{fig-gs:GS-SR-main}b} since we cannot obtain the execution time breakdown for the hardware read mapper from \cite{nag2019gencache}.}
\revp{Integrating GenStore with \revmark{D2}existing accelerators requires no architectural changes to GenStore and the accelerators because GenStore operates directly on the original data that is stored in the SSD, before any accelerator-specific operation starts.
The only factor that GenStore needs to consider is to generate \omciv{its} outputs (i.e., the information of unfiltered reads) in the format that the accelerator needs as its inputs. The host system is responsible for orchestrating the execution and data flows between GenStore and the accelerator.

}

We make \omciv{two} observations based on Figure ~\ref{fig-gs:GS-SR-main}b.
First, \gs significantly outperforms \base by $3.32\times$, $2.55\times$, and $1.52\times$ on systems with \ssdl, \ssdm, and \ssdh, respectively. 
Second, \gsos performs significantly slower than \texttt{Base} ($2.28$-$1.91\times$) on all systems since GenStore-EM requires accessing the large SSIndex data structure (Section~\ref{sec-gs:SRF}), while \gsos suffers from limited SSD external bandwidth.
We conclude that \omciv{\omcvii{the} in-storage processing approach in GenStore effectively addresses the I/O bottleneck, which becomes even more significant with hardware accelerators that address the computation bottleneck.}
}

\revp{\head{Effect of Read Set Features on Performance} \omciv{We study the benefits of GenStore-EM  depending on the characteristics of input read sets. 
To this end, we evaluate GenStore-EM while changing two key \omcv{characteristics}:
1) the input read set \revmark{G1\\Part1}size and 2) exactly-matching read rate. 
These two factors affect the data movement savings ($DM\_Saving$) of GenStore as governed by} Equation~\eqref{eq-gs:dm-saving}:
\begin{equation}
\label{eq-gs:dm-saving}
DM\_Saving = \frac{{Size_{Ref}+Size_{\omciv{ReadSet}}}}{Size_{Ref}+{Size_{\omciv{ReadSet}} \times (1 - Ratio_{Filter})}},
\end{equation}
where $Size_{Ref}$ is the size of the reference genome and its index (e.g., 7 GB for humans~\omciv{\cite{li2018minimap2}}),  $Size_{\omciv{ReadSet}}$ is the size of the read set, and $Ratio_{Filter}$ in GenStore-EM is the exact\omcvii{ly}-match\omcvii{ing read} rate of the read set.

Figure~\ref{fig-gs:GS-SR-sens}a shows the execution time of \base and \gs for the \omciv{baseline} software read mapper (Minimap2~\cite{li2018minimap2}) for input read sets with different sizes (1x, 10x, and 20x larger than \omciv{the size of our default} 22-GB short read set) \omciv{\jsr{and} different} exactly-matching read rates (75\% and 85\%) on a system with \ssdh. We make two observations.
First, \gs's performance benefit grows \omcv{as} input size \omcv{increases} (from $2.62\times$ to $4.75\times$) due to \omciv{larger} data movement savings.  Since $Size_{Ref}$ is constant, $DM\_Saving$ and the performance benefits of GenStore-EM increase with larger $Size_{\omciv{ReadSet}}$ (see Equation~\eqref{eq-gs:dm-saving}).
Second, \gs's performance benefit increases with higher exact\jsr{ly}-match\jsr{ing read} rates (from $3.46\times$ to $6.05\times$ for the largest read set) because with \omciv{a larger exactly-matching read rate (i.e., $Ratio_{Filter}$ in Equation~\eqref{eq-gs:dm-saving}), data movement saving ($DM\_Saving$)} increases, and the time spent on mapping the unfiltered reads decreases. Mapping the unfiltered reads is the key contributor to the end-to-end execution \omciv{time} of \gs since it \omciv{has larger} execution time compared to concurrently running GenStore-EM operations in the SSD. \omciv{We conclude that the benefits of GenStore-EM\omcv{, when integrated with the software read mapper,} increases with larger read sets and with \omcv{larger} exactly-matching read rates.}

\begin{figure}[h]
\centering
\includegraphics[width=0.9\linewidth]{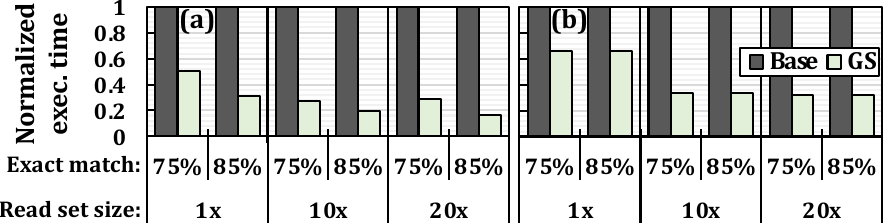}
\caption{\revp{GenStore-EM performance versus input size and exact match rate,} when integrated with (a)~software read mapper (b)~hardware read mapper.}
\label{fig-gs:GS-SR-sens}
\end{figure}

Figure~\ref{fig-gs:GS-SR-sens}b shows the execution time of \base and \gs for \omciv{a baseline} hardware read mapper  (GenCache~\cite{nag2019gencache}). We make two observations.
First, \gs's performance benefit increases with larger input sizes (from $1.52\times$ to $3.13\times$) due to its \omciv{larger data movement reduction}.  
Second,  \gs's performance benefit does \emph{not} increase with higher exact\jsr{ly}-match\jsr{ing read} rates. \jsr{This is because,} with the hardware read mapper, the end-to-end performance of \gs is dominated by the execution time of GenStore-EM's filter operations inside \omciv{the} SSD, which only depends on the input \omcviii{read set} size and not on \omciv{the} exact-match rate. \omciv{We conclude that the benefits of GenStore-EM, \omcv{when integrated} with the hardware read mapper, increases with larger \omcviii{input read set} sizes.}
}

\subsection{GenStore-NM Analysis}

\begin{figure}[h]
\centering
\includegraphics[width=0.9\linewidth]{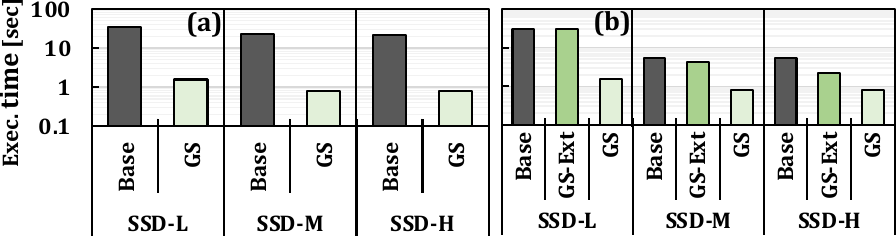}
\caption{\rev{\omciv{GenStore-NM} performance under different SSDs, when integrated with} (a)~software read mapper (b)~hardware read mapper.}
\label{fig-gs:GS-LR-main}
\end{figure}

\rev{We  analyze the benefits of GenStore-NM  \rev{for a 12.4-GB read set (\omcvi{the first \emph{No reference} use case} in Table~\ref{table-gs:lr-profile})} with 99.65\% of reads \emph{not} aligning on a system with three different SSD configurations: \ssdl, \ssdm, and \ssdh.

\head{Integration with a Software Read Mapper}
Figure~\ref{fig-gs:GS-LR-main}a shows the execution time of \jsr{two read mapper configurations: 
1) }\base (Minimap2~\cite{li2018minimap2}), which already incorporates the chaining filter, and 
\jsr{2)} \gs.\footnote{\revp{Unlike \fig{\ref{fig-gs:GS-SR-main}a}, we do not divide the execution time\omcv{s of} \texttt{Alignment} and \texttt{Other} in \fig{\ref{fig-gs:GS-LR-main}a} since the execution time of \gs is dominated by the filter operations of GenStore-NM. This is because, in this case, a large fraction \omcvi{(e.g., 99.65\% in our evaluated dataset)} of reads get filtered in the SSD and the execution time of mapping the unfiltered reads is very small.}}\revmark{B3\\Part2}
  We observe that \gs  outperforms \base by $22.4\times$, $29.0\times$, and $27.9\times$ on systems with \ssdl, \ssdm, and \ssdh, respectively.\footnote{In this case, \gs provides larger benefits for systems with \ssdm and \ssdh since these SSDs have larger internal bandwidth compared to \ssdl.} The reason is that \gs alleviates the cost of data movement from SSD to \omciv{the processor} and removes the burden of \omciv{mapping a large fraction of} reads from the rest of the system (DRAM and \omciv{the processor}).
}

\rev{
\head{Integration with a Hardware Read Mapper} 
Figure~\ref{fig-gs:GS-LR-main}b shows the execution time of \rev{three hardware read mapper configurations:
1)} \base (Darwin~\cite{turakhia2018darwin}), 
\rev{2) \gsos, and}
\jsr{3)} \gs. 
We make two observations based on Figure~\ref{fig-gs:GS-LR-main}b.
First, \gs significantly outperforms \base by $19.2\times$, $6.86\times$, and $6.85\times$ on systems with \ssdl, \ssdm, and \ssdh, respectively. The reason is that \gs alleviates the cost of data movement from SSD to the accelerator and removes the burden of \omciv{read mapping for the filtered reads} from the rest of the system (DRAM and the accelerator).
Second, \gsos does not provide significant performance benefits compared to \base in systems with \ssdl and \ssdm since the execution time of \gsos is dominated by the I/O overhead of bringing reads from SSD to the accelerator. \gsos performs $2.50\times$ faster than \base in systems with \ssdh since the I/O bottlenecks are partially alleviated with \ssdh. However, the benefits of \gsos are limited compared to \gs since \gs significantly alleviates the I/O overhead of bringing reads from SSD to the accelerator with all three SSD configurations.
}

\revp{\head{Effect of Read Set Features} \omciv{We study how the benefits of GenStore-NM vary depending on the characteristics of input read sets. 
To this end, we evaluate GenStore-NM while changing two key \omcv{characteristics}: 1) the input read set size\revmark{G1\\Part2} and 2) alignment rates (\omcv{\omcvii{the} fraction} of reads \omcv{in} the read set that align to the reference genome). 
We use input sets with different sizes (1$\times$, 10$\times$, and 20$\times$ larger than \omciv{the size of our default 12.5GB}  read set) \omciv{with} different alignment rates ($0.3\%$ and $37\%$, corresponding to \omcvi{the first and second \emph{No reference} use cases} in Table~4.1), on a system with \ssdh.} 
Figure~\ref{fig-gs:GS-LR-sens} shows the execution time of \base and \gs for the \omciv{baseline} software read mapper~\cite{li2018minimap2} (Figure~\ref{fig-gs:GS-LR-sens}a) and the \omciv{baseline} hardware read mapper~\cite{turakhia2018darwin} (Figure~\ref{fig-gs:GS-LR-sens}b)\omcvi{.}

\begin{figure}[h]
\centering
\includegraphics[width=0.9\linewidth]{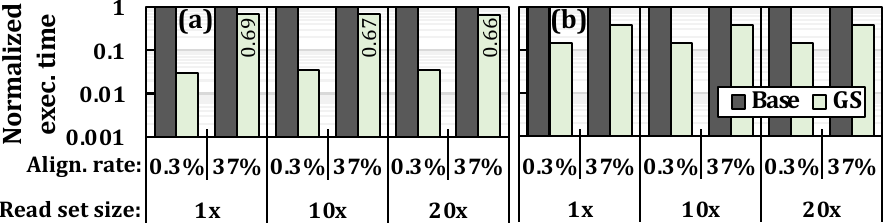}
\caption{\revp{GenStore-NM performance versus input size and \omciv{read alignment} rate, when integrated with} (a)~software read mapper (b)~hardware read mapper.}
\label{fig-gs:GS-LR-sens}
\end{figure}

\omciv{W}e make two \omcv{main} observations \omcvi{from \fig{\ref{fig-gs:GS-LR-sens}}}.
First, for both hardware and software read mappers, \gs's \omcv{performance benefits vary little as} the input size \omcv{changes.}
The reason is that $Size_{Ref}$ is very small (14.6~MB) for this experiment. Therefore, \omciv{data movement saving ($DM\_Saving$ in Equation~\eqref{eq-gs:dm-saving})} \omcv{mainly} correlates with \omciv{alignment rate (i.e., $1 - Ratio_{Filter}$)}. 
Second, for both software and hardware read mappers, \gs's performance benefit increase\omcv{s as} the ratio of non-aligning reads \omcv{increases (i.e., as alignment rate reduces)}, because with higher values of $Ratio_{Filter}$, both 1) $DM\_Saving$  increases and 2) the time spent on mapping unfiltered reads decreases. 
We conclude that GenStore-NM provides high performance benefits to both software and hardware read mappers with different input sizes and its benefits increase with lower \omciv{alignment} rates.
}

\subsection{Energy \omciv{Analysis}}

To \omcv{demonstrate the}  energy \omcv{benefits} of different GenStore modes, we \omciv{obtain} the energy of the host processor, the host-side DRAM, the DRAM inside the SSD, the communication between the SSD and the host, active and idle energy of the SSD, and the energy of the logic units used in GenStore (Section~\ref{sec-gs:area}). \omcv{We calculate the energy of each component based on its idle and dynamic power consumption and its execution time.}
We observe that by filtering out large amounts of data, GenStore reduces the end-to-end energy consumption of read mapping in all of our evaluations \omciv{compared to \omcv{\base (Minimap2~\cite{li2018minimap2})}}.
We measure the energy consumption for experiments on all SSD configurations. \omciv{By filtering exact matches \omcv{in a short read set with 80\% exactly-matching read rate (see Section~\ref{sec-gs:methodology})}}, GenStore-EM reduces the energy consumption by \omcv{on average (up to)} %
$3.92\times$ ($3.97\times$) \omcvi{across all storage configurations}. 
\omciv{By filtering non-matching reads in a long read set with \omcvi{a} read alignment rate of 0.35\% (\omcvi{the first \emph{No reference} use case} in Table~\ref{table-gs:lr-profile})}, GenStore-NM reduces the energy consumption by  \omcv{on average (up to)} $27.17\times$ ($29.25\times$)~\omcvi{across all storage configurations}.

\begin{figure}[h]
\centering
\includegraphics[width=0.75\linewidth]{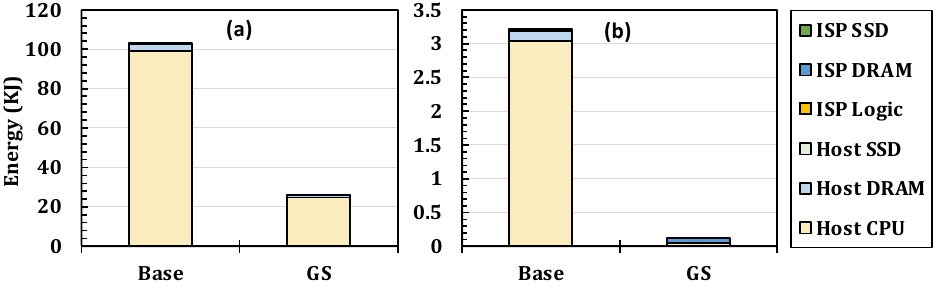}
\caption{\aooo{Energy breakdown of different systems for (a) short read mapping and (b) long read mapping.}}
\label{fig-gs:energy-brkdn-gs-1}
\end{figure}

\aooo{\fig{\ref{fig-gs:energy-brkdn-gs-1}} shows the energy breakdown of \base and \gs. We show energy breakdown between the host CPU, host DRAM, host SSD accesses, ISP logic units, in-storage DRAM, and SSD accesses during ISP. We observe that, by filtering a large number of low-reuse reads within the SSD, GenStore alleviates the energy burden on the rest of the system, particularly the host CPU which needs to perform the energy-intensive read mapping process.}

\begin{figure}[t]
\centering
\includegraphics[width=0.75\linewidth]{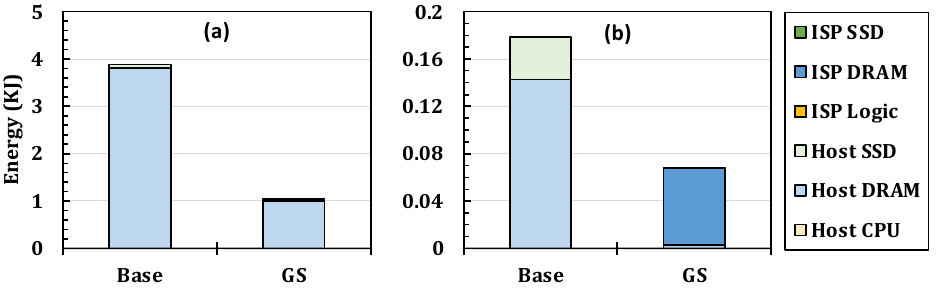}
\caption{\aooo{Energy breakdown of different systems, excluding the host CPU, for (a) short read mapping and (b) long read mapping.}}
\label{fig-gs:energy-brkdn-gs-2}
\end{figure}

\aooo{\fig{\ref{fig-gs:energy-brkdn-gs-2}} shows the energy breakdown of \base and \gs, excluding the host CPU, to isolate the energy distribution between main memory and storage accesses across configurations. This breakdown also serves as an idealized study of a scenario in which energy-efficient specialized hardware units replace general-purpose CPUs for computation. We make three observations. First, GenStore-EM (for short reads) significantly reduces DRAM energy consumption due to its read-size k-mer approach, which significantly reduces the number of memory accesses to the reference genome and its index structure. Second, GenStore-NM (for long reads) also significantly reduces overall DRAM energy consumption (host and internal DRAM combined)  due to its lightweight chaining filter that filters reads with a small number of seeds in a lightweight manner. Third, by filtering reads directly inside the SSD, GenStore also alleviates I/O energy overheads.}

\section{Discussion}
\label{sec-gs:discussion}

\omc{
\revmark{CQ3 and B5}As sequencing technologies develop in the future, we expect that GenStore will still play an important role in genomic sequence analysis.
We witness three major trends in sequencing technologies. First, the length and the accuracy of reads are expected to \omcv{increase~\cite{levy2016advancements,hu2021next,alser2020accelerating,morrison2020nanopore,nanopore2020}}. Second, short reads will continue to be widely used due to their \omcv{\emph{very}} high accuracy and low cost~\cite{quail2008large,pacbio2021,quail2012tale}\omcv{.} Third, DNA sequencing machines are increasingly adopting data processing capabilities to perform \omcv{\emph{both}} sequencing and genomic analysis within the same machine~\omcv{\cite{wang2021nanopore,dunn2021squigglefilter,loka2019reliable,zhang2021real,ardui2018single,kovaka2020targeted}}. 

Even with increases in long read accuracy (\emph{trend 1}),  GenStore-NM \omcvi{would} continue to filter large numbers of reads that do \omcvi{\emph{not}} align to \omcv{a} reference genome due to genetic differences between the \omcvi{compared genomes}~\cite{danko2021global,afshinnekoo2015geospatial,hsu2016urban,sherman2019assembly,li2021building,miga2021need}. 
\omcvi{Given that accurate short reads are essential for many processes in genome analysis (\emph{trend 2}), e.g.,}~for polishing~\cite{zhang2020comprehensive,firtina2020apollo,miga2020telomere} and validating~\cite{logsdon2021structure} long read sequences\omcvi{,} GenStore-EM can \omcv{continue to} accelerate short read mapping by filtering \omcv{out} exactly-matching reads.
With the push to provide DNA sequencing and preliminary genetic analysis on more portable integrated devices with limited compute resources and available DRAM capacity~\omcv{\cite{wang2021nanopore,alser2020accelerating,dunn2021squigglefilter,peresini2021nanopore}} (\emph{trend 3}), there will be a growing demand for \omcvi{data movement-minimizing} systems like GenStore to \omcvi{quickly} filter out a large fraction of reads at low cost and \omcv{high efficiency}~\cite{singh2021fpga,dunn2021squigglefilter,bhoyar2021high,ahmed2021pan,kovaka2020targeted,pomerantz2018real}.

With \omcvi{higher availability} of portable genome sequenc\omcvi{er\omcvii{s}}, genomic analyses will be performed more routinely~\cite{sunagawa2015structure,uk10k2015uk10k,danko2021global,alser2020accelerating,bloom2021massively,morrison2020nanopore,pomerantz2018real,clark2019diagnosis,farnaes2018rapid,sweeney2021rapid,alkan2009personalized,flores2013p4,ginsburg2009genomic,chin2011cancer,Ashley2016} \omcv{and will need to provide much faster results}~\cite{alser2020accelerating,singh2021fpga}. 
\omcv{I}n-storage processing is crucial for meeting the huge demands \omcv{for faster analyses} that scale well to large numbers of \omcvi{genomic samples}. 
Examples of these analyses include 
gene detection~\cite{lax2014longitudinal}, alignment of reads from rapidly mutating organisms such as SARS-CoV-2~\cite{bhoyar2021high,bloom2021massively}, and studies of as-of-yet undiscovered microbes~\cite{danko2021global,afshinnekoo2015geospatial,hsu2016urban, sunagawa2015structure}. 
In these analyses, GenStore-NM can effectively filter out $>$99.7\%~\cite{lax2014longitudinal,hsu2016urban}, $\sim$36.1\%~\cite{bhoyar2021high}, and $\sim$47.3\%~\cite{danko2021global,afshinnekoo2015geospatial,hsu2016urban} of the reads, respectively, \omcv{th\omcvi{ereby} greatly improving the end-to-end throughput \omcvi{and energy efficiency} of genome sequence analysis}.
We hope that our proposed techniques provide a foundation for future efforts \omcv{to} accelerat\omcv{e} genome analysis.
}

\section{Summary}
\label{sec-gs:conclusion}

We propose GenStore, a \omcv{new} in-storage \rev{processing system for genome sequence analysis}. \omcv{GenStore} can be integrated with \omcv{both} hardware and software read mappers to improve the end-to-end performance of \omcv{both} short \omcc{and} long read mapping.
We address the challenges of in-storage processing for genomic \omciv{read} filter\omciv{ing via}   
\omcc{new} hardware/software co-design\omciv{ed techniques and} develop \omciv{new} in-storage filtering accelerators for both \omciv{short and long} reads. 
\damlaa{\omciv{Our} evaluat\omciv{ions show that}}
GenStore provides \omcc{large performance and energy} 
improvements when integrated \omcc{in}to the state-of-the-art software and hardware \omciv{read mappers}. 

\subsection{\tomiii{Impact and Influence}}
\tomiii{
GenStore is published at the International Conference on Architectural Support for Programming Languages and Operating Systems (ASPLOS) in 2022~\cite{mansouri2022genstore}. It is fully open-sourced at~\cite{gssource}. An extended version of the ASPLOS 2022 paper is available on arXiv~\cite{arxivGS}. 

A significant amount of work (e.g.,~\cite{soysal2025mars,abakus23taco,zheng2025storage,kabra2025ciphermatch,chen2025reis,megis,grains,mansouri2026sage}) has already been influenced by GenStore. For example, a prior work~\cite{zheng2025storage} proposes additional in-storage filters for genomics; various other works extend the principles to other subdomains in bioinformatics~\cite{soysal2025mars,abakus23taco,megis,grains,mansouri2026sage}; and some work (e.g.,~\cite{chen2025reis}) extends the key ideas to other domains such as AI/ML.

We hope that the key ideas and approaches introduced in GenStore inspire further research in storage-centric designs in other domains related to health and life sciences. Such storage-centric designs can improve performance, energy efficiency, and cost-effectiveness of data-intensive applications needed to advance healthcare and life sciences, thereby facilitating their wider adoption and enabling solutions that were out of reach before.
}

\chapter{MegIS: Cooperative In-Storage Processing for Metagenomic Analysis}
\label{chap:megis}

\newcommand{\omi}[1]{#1}
\newcommand{\omii}[1]{#1}
\newcommand{\hhl}[1]{#1}
\newcommand{\bback}[1]{#1}
\newcommand{\gram}[1]{#1}
\renewcommand{\irev}[1]{#1}
\newcommand{\revh}[1]{#1}

\newcommand{\revid}[1]{}
\newcommand{\icut}[1]{}

\renewcommand{\proposal}{MegIS}
\renewcommand{\proposals}{\proposal{}'s}

\section{Motivational Analysis}
\label{sec-mg:motivation}

\subsection{Data Movement Overheads}
\label{sec-mg:motivation-ovhd}

\bback{We conduct experimental analysis to assess the storage system's impact on the performance of metagenomic analysis}.

 \head{Tools and Datasets} We analyze two state-of-the-art  tools 
 for presence/absence identification: 
1)~Kraken2~\cite{wood2019improved}, which queries \hhl{its} large database with random access patterns (\randomio),%
\footnote{We experiment with both techniques of accessing the database devised in the \randomio baseline~\cite{wood2014kraken} and report the best timing. The first technique uses mmap to access the database, while the second technique loads the entire database from the SSD to DRAM as the first step when the analysis starts. In this experiment, the second approach performs slightly better since, when analyzing our read set, the application accesses most parts of the database.}
and 2)~Metalign~\cite{lapierre2020metalign},
which exhibits \omii{mostly} sequential streaming accesses to \hhl{its} database (\streamio). We use the best-performing thread count for each \omi{tool}.
We use a query sample with 100 million reads
(CAMI-L, detailed in \sect{\ref{sec-mg:methodology}}) from the CAMI dataset~\cite{meyer2021critical}, commonly used for profiling metagenomic tools. 
We generate a database based on \hm{microbial genomes drawn from NCBI's databases}~\cite{ncbi2020,lapierre2020metalign} using default parameters for each tool. For Kraken2~\cite{wood2019improved}, this results in a 293 GB database. For Metalign~\cite{lapierre2020metalign}, this results in a 701 GB k-mer database
and \gram{a} 6.9~GB sketch tree. To show the impact of \omi{database size}, we also analyze larger k-mer databases (0.6~TB and 1.4~TB for Kraken2 and Metalign, respectively) that include more species.

\head{System Configurations} 
We use a high-end server with \hhl{an} AMD EPYC 7742 CPU~\cite{amdepyc} and 1.5-TB DDR4 DRAM\omii{~\cite{ddr4sheet}}. \omii{We n}ote that the DRAM size is larger than the size of all data accessed during the analysis by each tool. This way, we can analyze the fundamental I/O overhead of moving large amounts of low-reuse data from storage to the main memory without being limited by DRAM capacity.
We evaluate I/O \omii{overheads using}:
1)~a cost-optimized SSD (\ssdc)~\cite{samsung870evo} with a SATA3 interface~\cite{SATA}, 
2)~a performance-optimized SSD (\ssdp)~\cite{samsungPM1735} with a PCIe Gen4 interface~\cite{PCIE4}, and
3)~a hypothetical configuration with zero performance overhead due to storage I/O (\dram). 
\bback{\ssdp provides an order-of-magnitude higher sequential-read bandwidth than \ssdc \omii{~(detailed configurations in Table~\ref{table-mg:SSD_config})}. 
However, scaling up storage capacity only using performance-optimized SSDs is challenging due to their much higher prices \omiii{(e.g.,~\cite{PM1735price,PM9A3price,EVO870price})} and fewer PCIe slots compared to SATA \omii{slots available} on servers \omiii{(e.g.,~\cite{amdepyc})}}.

\begin{figure}[b]
    \centering
    \includegraphics[width=0.75\linewidth]{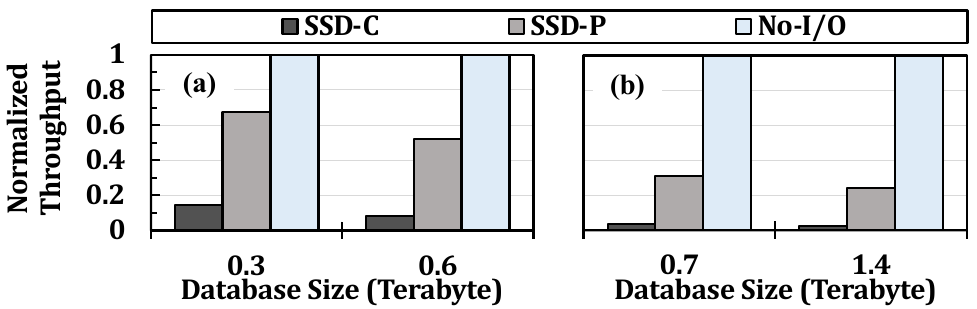}
    \caption{Performance of (a) \randomio and (b) \streamio under different storage configurations and database sizes.}
    \label{fig-mg:motivation}
\end{figure}

\head{Results and Analysis}
\fig{\ref{fig-mg:motivation}} shows the performance \omii{(throughput in terms of \#queries/sec)} of the tools normalized \omiii{to} \dram. We make three key observations. First, I/O overhead has a large impact on
performance \bback{for all cases}.
Compared to \ssdc (\ssdp), \dram leads to 9.4$\times$ (1.7$\times$) and 32.9$\times$ (3.6$\times$) better performance in \randomio and \streamio (average\omi{d} across both databases), respectively.   
While both baselines significantly suffer from large I/O overhead, we observe a relatively larger impact on \streamio due to its lower data reuse compared to \randomio. \omii{This is because the lower the data reuse, the less effectively the initial I/O cost can be amortized.}
Second, even using the \omii{costly} state-of-the-art SSD (\ssdp) does not alleviate this overhead, leaving large performance gaps between \ssdp and \dram in both tools.    
Third, I/O overhead increases as the databases grow. 
For example, in \randomio, the performance gap between \ssdc and \dram widens from 7.1$\times$ to 12.5$\times$ as the database expands from 0.3 TB to 0.6 TB. 
\bback{Based on these observations, we conclude that I/O accesses lead to large overheads in metagenomic \omii{analysis}, an issue expected to \omii{worsen} in the future}.

This I/O overhead, stemming from the need to move large amounts of low-reuse data\hhl{,} is a fundamental problem \omii{that is} hard to avoid. One might think it is possible to avoid this overhead by 1) \omii{using} sampling\revid{\label{rev:B6}B6} \irev{techniques} to shrink database \omii{sizes} \omi{(e.g.,}\omii{~\cite{kim2016centrifuge,wood2019improved,muller2017metacache,song2024centrifuger,Dilthey2019,Fan2021}}\omi{)} or 2) keeping all data required by metagenomic analysis \omii{completely and always} resident in main memory. Neither of these solutions is suitable. The first approach inevitably reduces accuracy\omii{~\cite{berger2023navigating,milanese2019microbial,meyer2021critical}} to levels unacceptable for many 
use cases \omi{(e.g.,}\omii{~\cite{milanese2019microbial,salzberg2016next,gihawi2023major,pockrandt2022metagenomic,berger2023navigating,Ackelsberg2015,Nasko2018,meyer2021critical}}\omi{)}.
The second approach is energy inefficient, costly, unscalable, and unsustainable due to two reasons. First, the sizes\revid{\label{rev:E1.3}E1.3} of metagenomic databases (which are already large, i.e., in some recent examples, exceeding a hundred terabytes~\cite{shiryev2023indexing,pebblescout}) have been increasing rapidly.
For example, recent trends show the \emph{doubling} of different important databases in only several months\omii{~\cite{enastats,ntdouble,Nasko2018}}. Second, 
regardless of \hhl{the} sizes\revid{\label{rev:E1.2}E1.2} \irev{of \omii{individual} databases}, different analyses need \emph{different databases}, \irev{with information from different sets of genomes or with varying parameters}. For example, a medical center may use various databases for its patients \irev{based on the patients' conditions\hhl{~\cite{schuele2021future}}} (e.g., for different viral infections~\cite{centrifuge_db,kim2016centrifuge}, sepsis~\cite{taxt2020rapid}, etc.).
Therefore, it is inefficient and unsustainable to maintain \emph{all} data required by \emph{all possible} analyses in DRAM at all times.\footnote{Ultimately, these are the same reasons that the metagenomic\omii{s} community has been investigating storage efficiency (e.g., the \omii{afore}mentioned sampling techniques\omii{~\cite{kim2016centrifuge,wood2019improved,muller2017metacache,song2024centrifuger,Dilthey2019,Fan2021}}) as opposed to merely relying on scaling the system's \omiii{main memory}~\cite{karasikov2020metagraph,pockrandt2022metagenomic,lemane2022kmtricks,alanko2023themisto,fan2023fulgor}.}

The I/O impact on end-to-end performance becomes even more prominent in emerging systems in which other bottlenecks%
\icut{(e.g., in computation or main memory)}
are alleviated. 
For example, while metagenomics can benefit from near-data processing at the main memory level, i.e., processing-in-memory (PIM)\omiii{~\cite{wu2021sieve,shahroodi2022krakenonmem,shahroodi2022demeter,hanhan2022edam,zou2022biohd,mutlu2022modern,ghose2019processing,mutlu2019processing,ghose2018enabling}}, these approaches still incur the overhead of moving the large, low reuse data from \omi{the} storage \omi{system}. In fact, by \omii{alleviating} other \omii{bottlenecks}, the impact of I/O on end-to-end performance increases. For example, for the 0.3-TB and 0.6-TB Kraken2 databases, 
\hhl{using}
a state-of-the-art PIM accelerator~\cite{wu2021sieve} of Kraken2,  \dram is on average 26.1$\times$ (3.0$\times$) faster than \ssdc (\ssdp). We conclude that while \omii{accelerating other bottlenecks in
metagenomic analysis (e.g., main memory bottlenecks)} can provide significant benefits, \omiii{doing so} does not alleviate the overhead\omii{s} of moving large, low-reuse data from the storage \omiii{system}.

\subsection{Our Goal}
\label{sec-mg:motivation-goal}

\irev{ISP 
can be a fundamental solution for reducing data movement. However,} designing an ISP system for metagenomics is challenging because none of the existing approaches can be directly implemented as an ISP system effectively due to \hhl{an} SSD's constrained hardware resources. 
\bback{Techniques such as \randomio hinder leveraging ISP’s large potential by preventing the full utilization of the SSD’s internal bandwidth due to costly conflicts in internal SSD resources~\cite{nadig2023venice,tavakkol2018flin,kim2022networked} \omii{caused by random accesses}. Techniques such as \streamio predominantly incur more suitable streaming accesses, but at the cost of more computation and main memory capacity requirements, posing challenges for ISP.
Therefore, directly adopting \omii{existing metagenomic analysis} approaches in storage incurs performance, energy, and lifetime overheads}.
\omi{\textbf{Our goal} in this work is to improve the performance \omiii{and efficiency} \omii{of metagenomic analysis} by reducing \omi{the large} data movement overhead \omi{from the storage system} in a cost-effective manner}.

\section{\proposal}
\label{sec-mg:mechanism}

\omii{W}e propose \proposal{}, the \emph{first} ISP system \omi{designed \omiii{for the end-to-end \mganalysis pipeline} to reduce \omiii{its} data movement overhead\omi{s \omii{from the storage system}}}.
\irev{\proposal{} is primarily designed as a system for accelerating\revid{\label{rev:D3.1}D3.1} metagenomic analysis}. \proposal{} extends the existing SSD controller and FTL \irev{\omii{\emph{without}} impacting the baseline SSD functionality. Therefore, when metagenomic acceleration is not in progress,} the SSD can be accessible for all other applications\irev{, similar to a general-purpose SSD}. 

We address the challenges \omi{of ISP for \mganalysis} via hardware/software co-design to enable \omii{what we call} \revid{\label{rev:D1.2}D1.2}\emph{cooperative ISP}. \omii{In other words, we do not solely focus on processing inside the storage system but, instead, we \omiii{exploit} the strengths of processing both inside and outside the storage system}. \irev{\proposal{} enables} an efficient pipeline between the host \omi{system} and \js{the storage system} to \js{maximally leverage and orchestrate the capabilities \omii{of both systems}}.

\irev{It\revid{\label{rev:D1.3}D1.3} is possible for MegIS’s ISP \omi{steps} to run on our lightweight specialized ISP accelerators or, alternatively, on the existing embedded cores in the SSD controller\footnote{\irev{These cores are available for \proposals{} ISP\revid{\label{rev:D3.3}D3.3}  since we envision that during metagenomic acceleration, \proposal{} is not used as a general-purpose SSD and does not run the baseline FTL. Instead, it runs \omii{\proposal{}~FTL}, which only performs lightweight and infrequent tasks during ISP \omii{(see \sect{\ref{sec-mg:mech-ftl}})}.}} or other general-purpose ISP systems (e.g., \cite{torabzadehkashi2019catalina,zou2022assasin,jun2015bluedbm,gu2016biscuit}). 
This is because, leveraging our optimizations, \proposals{} ISP \omi{steps} require only simple computation and small buffers.
Efficiently performing metagenomics on any of these underlying hardware units requires \proposals{} specialized task partitioning, data/computation flow coordination, storage technology-aware algorithmic optimizations, and data mapping. The ability to leverage existing hardware units (embedded SSD cores or general-purpose ISP \omi{systems}) \omii{helps with} \proposals{}\revid{\label{rev:E1.6}E1.6\\------\\\label{rev:B1.4}B1.4} ease of adoption. Ultimately, choosing between \omii{different} MegIS configurations (our specialized lightweight \omi{accelerators} or general-purpose hardware) is a design decision \omiii{that has} \omii{various} tradeoffs, with specialized accelerators achieving \omii{the highest} performance and power efficiency (\sect{\ref{sec-mg:eval-main}})}. 

\subsection{\irev{Overview}\revid{\label{rev:A2}A2}}
\label{sec-mg:mech-overview}

\begin{figure}[t]
    \centering
    \includegraphics[width=0.85\linewidth]{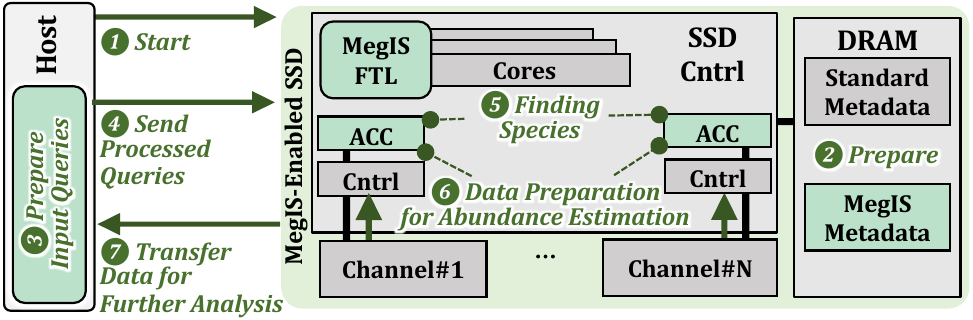}
    \caption{Overview of \proposal{}.}
    \label{fig-mg:metastore_overview}
\end{figure}

\fig{\ref{fig-mg:metastore_overview}} shows \irev{an overview of} \proposal{}'s \irev{steps.
We design \proposal{} as an efficient pipeline in the SSD and the host system.
We develop \omii{\proposal{}~FTL} \omi{(\sect{\ref{sec-mg:mech-ftl}})}, which is responsible for communication with the host system and data flow across the SSD hardware components (e.g., NAND flash chips, internal DRAM, and hardware accelerators) when running metagenomic analysis.}
Upon receiving \gram{a} notification from the host to initiate metagenomic analysis (\circled{1} \irev{in \fig{\ref{fig-mg:metastore_overview}}}), \proposal{} readies itself 
by loading the necessary \omii{\proposal{}~FTL} metadata
(\circled{2}).
\irev{After this preparation, \proposal{} starts its three-step execution.} 
In \rev{\textbf{Step 1} \omi{(\sect{\ref{sec-mg:mech-stage1}})}, the host processes the input read queries (\circled{3})} and transfers them in batches to the SSD (\circled{4}). 
In \textbf{Step 2} \omi{(\sect{\ref{sec-mg:mech-stage2}})}, the ISP units \omii{(ACC in \fig{\ref{fig-mg:metastore_overview}})} find the species present in the sample (\circled{5}). Steps 1 and 2 run in a pipelined manner.
In \textbf{Step 3} \omi{(\sect{\ref{sec-mg:mech-stage3}})}, \omii{\proposal{}} prepar\omii{es} (\circled{6}) and transfer\omii{s} (\circled{7}) the data needed for any further analysis. \omii{By doing so, \proposal{} facilitates integration with different abundance estimation approaches.}
\irev{\proposal{} leverages the SSD's \emph{full internal\revid{\label{rev:A4.2}A4.2} bandwidth} since it avoids channel conflicts (due to its specialized data/control flow) and frequent management tasks (by not requiring writes during its ISP steps)}.

\subsection{Step 1: Preparing the Input Queries}
\label{sec-mg:mech-stage1}

\omi{In this step, \proposal{} prepares the input read queries in a metagenomic sample for metagenomic analysis}. \proposal{} works with lexicographically-sorted data structures to avoid expensive random accesses to the SSD (similar to \streamio{}, described in \sect{\ref{sec:background-mg}}). Like many other metagenomic tools (\omi{e.g.,}\omii{~\cite{wood2014kraken,shahroodi2022krakenonmem,lapierre2020metalign,truong2015metaphlan2,kim2016centrifuge,song2024centrifuger,ounit2015clark,Dilthey2019,koslicki2016metapalette,shen2022kmcp,Marcelino2020,piro2016dudes,Fan2021,piro2020ganon,wood2019improved}}), we assume the sorted k-mer \emph{databases} are pre-built before the analysis. However, sorting k-mers extracted from the \emph{input query read set} 
is inefficient \omii{to perform \omii{offline}} due to the need to store a large data structure (sorted k-mer set) with each sample, potentially larger than the sample itself, causing significant storage capacity waste.
Therefore, to prepare the input queries, \proposal{} 1) extracts k-mers from the sample \omi{(\sect{\ref{sec-mg:mech-stage1-kmer-extraction}})}, 2) sorts the k-mers \omi{(\sect{\ref{sec-mg:mech-stage1-sorting}})}, and if needed, 3) prunes some k-mers \hhl{according to} user-defined criteria \omi{(\sect{\ref{sec-mg:kmer-exclusion}})}.

We execute this step in the host \omi{system} for three reasons. First, \omi{this step} benefits from the relatively larger DRAM and \omii{more powerful} comput\omii{ation} resources in the host. Second, due to the 
large \omiii{host-side} DRAM, performing this step in \gram{the} host leads to significantly fewer writes to the flash chips\omii{, positively impacting}
lifetime. For typical metagenomic read sets, storing k-mers extracted from reads within a sample takes tens of gigabytes (e.g., on average 60 GB with standard CAMI read sets \cite{meyer2021critical}). 
While generating and sorting k-mers inside the SSD 
is possible, it would necessitate frequent writes to flash chips \omii{or much larger DRAM}. Third, by leveraging the host \omi{system} for this step, we enable pipelining and overlapping Step 1 with Step 2 (which searches the large, low-reuse database).

\bback{To efficiently execute \omii{Step 1} on the host \omi{system}, we need to ensure two points. First, partitioning the application between the host system and the SSD should not incur significant overhead\omi{s} due to data transfer time. 
Second, while it is reasonable in most cases to expect the host DRAM to be large enough to contain all extracted k-mers from a sample, \proposal{} should accommodate scenarios where this is not the case and minimize the performance, lifetime, and endurance overheads of writes \omii{to flash chips} due to page swaps \omii{(i.e., moving data back-and-forth between the host DRAM and the SSD when the host DRAM is smaller than the application's working set size)}}.

\irev{\omii{The sequences in \proposal{}'s databases are encoded with two bits per character (i.e., \texttt{A}, \texttt{C}, \texttt{G}, \texttt{T} in DNA alphabet)} during their offline generation. For the read sets, \proposal{} is able to work with different formats. We \omii{perform} the first analysis step \omii{(Step~1)} in the host system so that any format conversion can be flexibly incorporated there (e.g., from ASCII \omii{or binary} to 2-bit encoding). 
\omii{The overhead of format conversion is negligible since it involves a straightforward transformation of the four nucleotide bases to the 2-bit encoded format}.
For the remainder of \proposal'{}s pipeline, we use \omiii{the} 2-bit encoding. 
}

\subsubsection{K-mer Extraction}
\label{sec-mg:mech-stage1-kmer-extraction}

\begin{figure}[t]
    \centering
    \includegraphics[width=.85\linewidth]{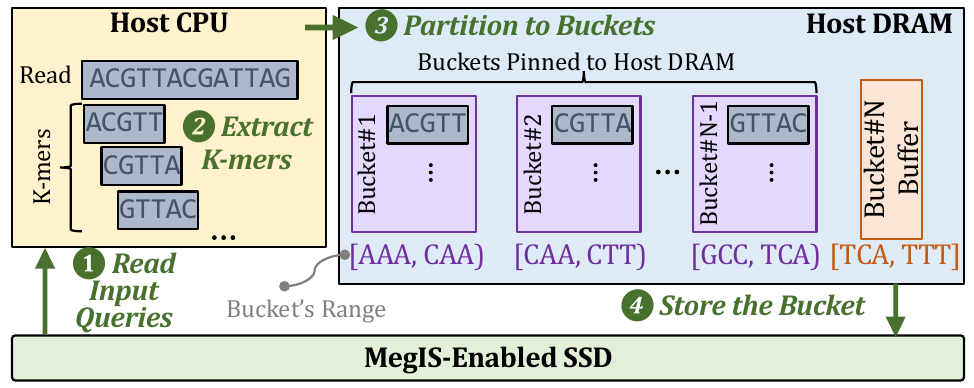}
    \caption{\omii{Overview of the} k-mer extraction \omii{process in \proposal{}}.}
    \label{fig-mg:kmer-gen}
\end{figure}

To reduce data transfer overhead between \omi{\omii{different parts of the} application that execute in the host system and in the storage system}, we propose a new input processing scheme by improving upon \hhl{the} input processing \omi{scheme} in KMC~\cite{kokot2017kmc3}. We partition the k-mers into buckets, each corresponding to a lexicographical range. 
This enables overlapping the k-mer sorting and transfer of a bucket \omi{to the SSD} with \omi{the} ISP \omi{operations of Step 2 (\sect{\ref{sec-mg:mech-stage2}})} on previous\omii{ly transferred} buckets. \omi{This is}
\bback{because the database k-mers are also sorted and can already be accessed within the 
corresponding range}.
\fig{\ref{fig-mg:kmer-gen}} shows an overview \omi{of \proposal{}'s k-mer extraction}. 
The host reads the input reads \omii{from the storage system} (\circled{1} in \fig{\ref{fig-mg:kmer-gen}}), extracts their k-mers (\circled{2}), and stores them in the buckets (\circled{3}).\footnote{\bback{To prevent bucket size imbalance, we initially create preliminary buckets for a small k-mer subset. In case of imbalance, we merge some buckets to \omiii{satisfy} a user-defined bucket count (default 512).}} 
In situations where a sample's extracted k-mers do not fit in the host DRAM, \proposal{} pins some buckets to \omii{the host} DRAM 
\bback{(e.g., Buckets $1$ to $N-1$ in \fig{\ref{fig-mg:kmer-gen}})} 
and uses the SSD to store the others. This way, k-mers belonging to buckets \omii{in the host DRAM} do \omii{\emph{not}} move back and forth between the host \omii{DRAM} and the SSD (\circled{4}). To reduce the overhead of accessing buckets \omii{in the SSD}, \proposal{} takes two measures. 
\omii{First, \proposal{} allocates buffers in the host DRAM specifically for buckets \omii{in the SSD}. Once these buffers are full, it efficiently transfers their contents to the SSD, maximizing the use of the sequential-write bandwidth}. Second, we map \omii{each} bucket's k-mers across SSD channels evenly for parallelism.
\bback{Since \proposal{} does not require writes to the flash chips after this step \omii{(\omiii{i.e.}, K-mer Extraction in Step 1)}, it can flush all of the FTL metadata for write-related management to free up internal DRAM for the next steps (details in \sect{\ref{sec-mg:mech-ftl}})}. 

\subsubsection{Sorting}
\label{sec-mg:mech-stage1-sorting}
\hm{After generating all k-mer buckets,}
\proposal{} proceeds to sort the k-mers within the individual buckets. As soon as a specific bucket $i$ is sorted, \proposal{} transfers this bucket to the DRAM inside the SSD in batches (to undergo \omi{Step 2}, as described in \sect{\ref{sec-mg:mech-stage2}}). 
Meanwhile, during the transfer of bucket $i$, \proposal{} advances to sort bucket $i+1$. 
 \proposal{} can orthogonally use a sorting accelerator \omii{(e.g.,~\cite{samardzic2020bonsai,qiao2022topsort,jayaraman2022hypersort})} to perform sorting.

\subsubsection{Excluding K-mers}
\label{sec-mg:kmer-exclusion}
\proposal{}, like various tools\omii{~\cite{benoit2016multiple,bovee2018finch,lapierre2020metalign,ounit2015clark,li2018minimap2,kokot2017kmc3}}, can exclude k-mers based on user-defined frequencies to improve accuracy. Users can exclude 1)~overly common \revh{(i.e., indiscriminative)} k-mers
and 2)~\omii{very infrequent} k-mers \omii{(e.g., those} that appear only once\omii{)}, which may represent 
\bback{sequencing} 
errors or low-abundance organisms \bback{that are hard to distinguish from random occurrences}.
\bback{Exclusion follows sorting, where k-mers are \omii{already} counted}. 
While the size of the extracted query k-mers (\sect{\ref{sec-mg:mech-stage1-kmer-extraction}}) can be large (on average 60~GB in our experiments),
the size of the k-mer set selected to go to \omi{Step 2} is much smaller (on average 6.5~GB) \bback{and is significantly smaller than the database that may reach several terabytes\omii{~\cite{ncbi2023,karasikov2020metagraph,shiryev2023indexing,pebblescout,lemane2023kmindex,marchet2023scalable}}}.

\subsection{Step 2: Finding Candidate \rev{Species}}
\label{sec-mg:mech-stage2}

In \omii{Step 2}, \proposal{} finds the \hm{species} present in the sample \bback{by 1) intersecting the query k-mers and the database \omi{k-mers},
and 2) finding the \omii{taxID}s of the intersecting k-mers}.
We perform this stage inside the SSD since it requires streaming the large database with low reuse and involves only lightweight computation. 
This enables \proposal{} to leverage the \omi{SSD's} large internal bandwidth 
 and alleviate the 
 \bback{overall}
 burden of moving\bback{/analyzing}
 large, low-reuse data from the rest of the system.

\bback{Considering the SSD's hardware limitations, \proposal{} should leverage the full internal bandwidth \omii{without} requir\omii{ing} expensive hardware \omi{resources} inside the SSD (e.g., large \omi{internal} DRAM \omii{size/}bandwidth and costly logic units). Performing this step effectively inside the SSD requires efficient coordination between the SSD and the host, mapping, hardware design, and storage technology-aware algorithmic optimizations}.

\subsubsection{Intersection Finding}
\label{sec-mg:mech-stage2-1}

In this step, \proposal{} finds the intersecting k-mers, i.e., k-mers present in both the query k-mer buckets arriving from the host \omi{system}
and the large k-mer database stored in the flash chips.

Relying \omii{solely} on the \omii{SSD's} internal DRAM \revh{to} 1) buffer the query k-mers arriving from the host \omi{system} and the database k-mers arriving from the SSD channels at full bandwidth and 2) stream through both to find their intersection \omii{can pressure the valuable internal DRAM} bandwidth.
For example, reading \omii{the database from the SSD channels} at full bandwidth in a high-end SSD can already exceed the LPDDR4 DRAM bandwidth used in current SSDs~\cite{zou2022assasin, samsung980pro,samsungPM1735} and even the \mbox{16-GB/s} DDR4 bandwidth~\cite{zou2022assasin,ddr4sheet}. To address this challenge, we adopt an approach similar to \cite{zou2022assasin} and 
\omii{operate on data fetched from flash chips without buffering them in the internal DRAM}.
Despite its benefits, this approach requires large buffers (64 KB for input and 64 KB for output) \emph{per channel}. 

To facilitate low-cost computation on flash data streams, we leverage two key features of \proposal{} to find the minimum required buffer size. First, the computation in this step is lightweight and does not require a large buffer for data awaiting computation. Second, data is uniformly spread across channels, with each compute unit handling data from one channel. 
Based on these, we directly read data from the flash \omii{chips} and include \emph{two} k-mer registers per channel. 
One register holds a k-mer as the computation input, 
\omii{while the other register \omiii{stores} the subsequent k-mer as it is read from the flash chips}.
This way, by only using two registers, \proposal{} directly compute\omii{s} on the flash data stream at low cost. 

\begin{figure}[h]
    \centering
    \includegraphics[width=0.85\linewidth]{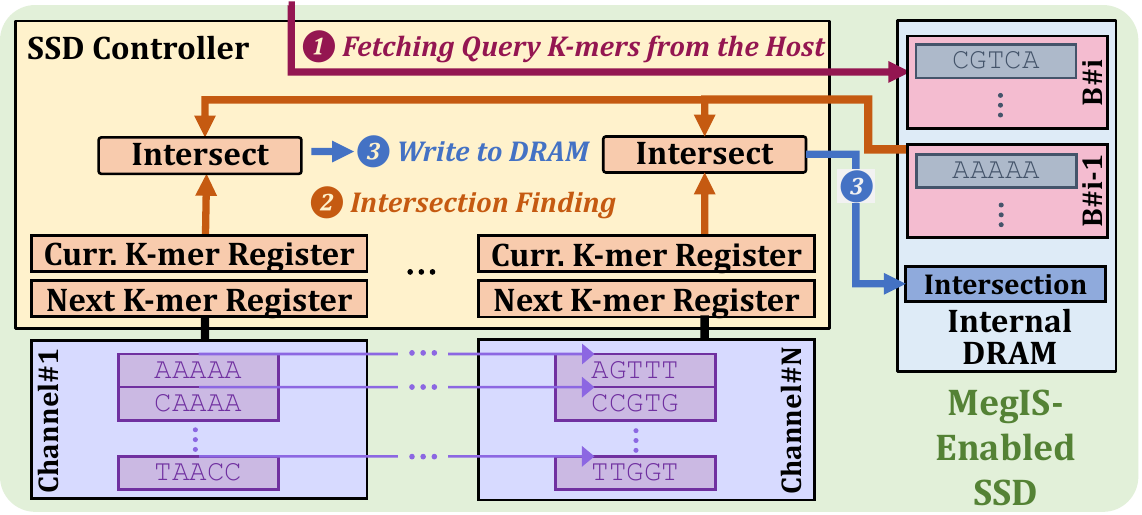}
    \caption{\omii{Overview of the} intersection finding \omii{process in \proposal{}}.}
    \label{fig-mg:intersection-finding}
\end{figure}

\fig{\ref{fig-mg:intersection-finding}} shows the \omi{overview of \proposals{} intersection finding \omiii{process}}.
First, \proposal{} reads the query k-mers to the internal DRAM in batches (\circled{1} in \fig{\ref{fig-mg:intersection-finding}}). Second, it concurrently reads both the sorted query k-mers \omii{(from the internal DRAM)}
and the sorted database k-mers \omii{(from the flash chips)},
performing a comparison to find their intersection (\circled{2}) using per-channel \emph{Intersect} units located \omii{in} the SSD controller.
Third, \proposal{} writes the intersecting k-mers to the internal DRAM for further analysis (\circled{3}).

\head{Fetching Query K-mers}
To efficiently use external bandwidth, \proposal{} moves buckets from the host \omi{system} to the internal DRAM in batches.
We manage two batches in the internal DRAM to overlap transfer and intersection finding.
For an SSD with 8 channels, 4 dies/channel, 2 planes/die, and 16-KiB pages, \proposal{} requires space for two 1-MiB batches\revid{\label{rev:A5.1}A5.1} \irev{(i.e., $B\#i-1$ and $B\#i$ in \fig{\ref{fig-mg:intersection-finding}})} in the internal DRAM.

\head{Intersection Finding}
\proposal{} reads the query k-mers from the internal DRAM and the database k-mers from the flash \omii{chips}. \omii{Intersection Finding} runs in a pipeline\omi{d manner} with \omii{Fetching Query K-mers}. 
\bback{We store the database evenly across different channels to leverage the full internal bandwidth when sequentially reading data using multi-plane operations.  \proposal{} finds the intersecting k-mers as follows:} 
If a database k-mer equals a query k-mer, \omi{\proposal{}} records \omii{the k-mer} as an \omii{intersecting k-mer}. If a query k-mer is larger (smaller), \proposal{} reads the next database (query) k-mer. 
\proposals{} Control Unit, located on the SSD controller, receives the comparison results and issues the control signals accordingly.\footnote{\omii{\figs{\ref{fig-mg:kmer-gen}, \ref{fig-mg:intersection-finding}, and \ref{fig-mg:taxid-finding}}} exclude Control Unit and its connections for readability.}

\head{Storing the Intersecting K-mers}
\proposal{} stores the intersecting k-mers in the \omii{SSD's} internal DRAM.\footnote{The \omii{intersecting k-mers do} not have a strict size requirement and can use the \irev{available} internal DRAM's space opportunistically. Usually, its small size allows it to fully fit in the internal DRAM. However, in a case where it does not, \proposal{} starts the \omii{taxID} retrieval (\sect{\ref{sec-mg:mech-stage2-2}}) for the already-found \omii{intersecting k-mers}; then resumes this step, overwriting the old \omii{intersecting k-mers}.} 
The internal DRAM needs to support 1) fetching the queries, 2) reading them out, 3) storing the intersection, and 4) \omii{reading} FTL metadata. Since the query k-mer set, the intersection, and the FTL metadata (FTL details in \sect{\ref{sec-mg:mech-ftl}}) are significantly smaller than the database, they can be accessed at a much smaller bandwidth than \omii{the bandwidth required for} reading the database. 
For example, for our datasets in \sect{\ref{sec-mg:methodology}}, when fully leveraging \ssdp's internal \omii{flash} bandwidth \omii{by reading the database from all flash channels}, \proposal{} requires only 2.4~GB/s of DRAM bandwidth \omi{to access all datasets stored in the internal DRAM}.

\subsubsection{Retrieving \omiii{TaxID}s}
\label{sec-mg:mech-stage2-2}

\omiii{\proposal{} finds the \omiii{taxID}s of the species corresponding to the intersecting k-mers by looking up the intersecting k-mers in a pre-built sketch database.
Each sketch is a small representative subset of k-mers associated with a given \omiii{taxID}. A sketch database stores the k-mer sketches and their associated \omiii{taxID}s for a given set of species}.
Similar to \cite{lapierre2020metalign}, we use CMash~\cite{liu2022cmash} to generate sketches. \proposal{} can \omiii{also} use other sketch \omiii{generation methods}. \proposal{} flexibly supports variable-sized k-mers \omii{in its sketch database}.  
As shown by prior works~\cite{liu2022cmash,kim2016centrifuge}, while longer k-mers \omiii{are more unique and} offer greater discrimination, they may result in missing matches between \omiii{the intersecting k-mers} and sketches. In such cases, users may also \omiii{search for} smaller k-mers \omiii{by looking up the prefixes of the intersecting k-mers in the sketch database}.
\omiii{This enables finding} additional matches \omiii{and increasing the true positive rate}.

\begin{figure}[t]
    \centering
    \includegraphics[width=0.9\linewidth]{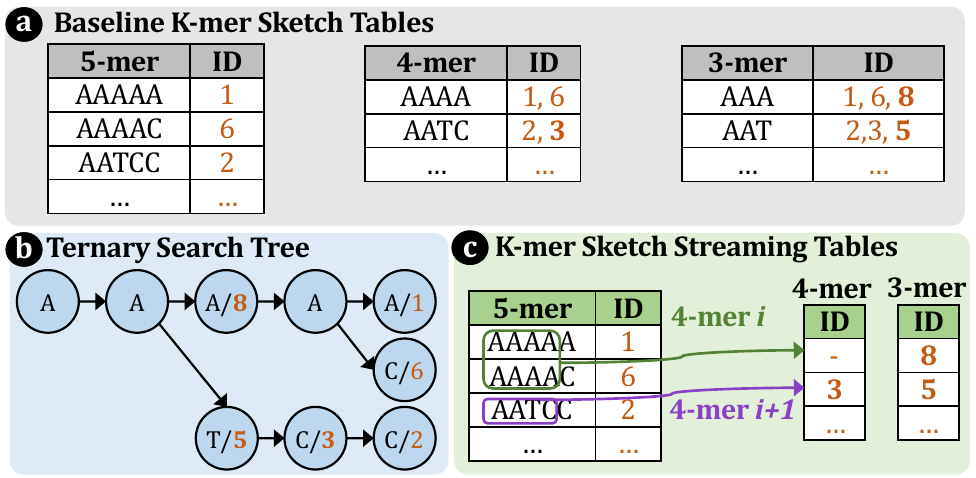}
    \caption{\omii{Overview of} sketch data structures.}
    \label{fig-mg:taxid-structures}
\end{figure}

Finding \omiii{taxID}s \omiii{for} variable-sized k-mers is challenging since it requires many pointer-chasing operations on a large data structure that may not fit in the \omiii{SSD's} internal DRAM.
To \omiii{support variable-sized k-mers}, some approaches \omiii{(e.g.,~\cite{marchet2021data,liu2022cmash,lapierre2020metalign,kim2016centrifuge,song2024centrifuger}}\omiii{)} \omiii{provide} data structures to encode the k-mer information in a space-efficient manner.
For example, CMash~\cite{liu2022cmash} encodes k-mers \gram{of} variable 
sizes in a ternary search tree. \fig{\ref{fig-mg:taxid-structures}} shows \omiii{sketch databases with variable-sized k-mers (k = 5, 4, and 3)} along\rev{side} 
their \omiii{taxID}s in \circled{a} separate tables, as used by some prior approaches~\cite{koslicki2016metapalette,weging2021taxonomic}, and \circled{b} in a ternary search tree. 
This tree structure is devised to 1) save space and 2) retrieve the \omiii{taxID}s for all k-mers with $k \leq k_{max}$ that are prefixes of \hm{a query} $k_{max}$-mer. 
\omiii{For example, as shown in \fig{\ref{fig-mg:taxid-structures}}, when traversing the tree to look up the 5-mer \texttt{AATCC}, we can look up the 4-mer \texttt{AATC} during the same traversal}. 
Despite its benefits, this approach requires up to $k_{max}$ pointer-chasing \gram{operations} for \emph{each lookup}. Performing these operations inside the SSD is challenging since the tree can be larger than the \omiii{SSD's} internal DRAM, and pointer chasing on flash arrays is expensive due to their significantly larger latency compared to DRAM.

While \proposal{} can \omii{perform \omiii{taxID} retrieval} in the host \omiii{system}, we identify a new optimization opportunity leveraging unique features of ISP (i.e., large internal bandwidth and storage capacity),
which avoids pointer-chasing at the cost of larger data structures. 
\fig{\ref{fig-mg:taxid-structures}} \circled{c} shows an overview of our approach, \emph{K-mer Sketch Streaming (KSS)}. 
For k-mers with \omii{k =} $k_{max}$, \proposal{} stores the k-mer sketches and their \omiii{taxID}s similar to \circled{a}. \proposal{}\omii{~keeps} this table \omiii{in a lexicographically-sorted order}.
For each smaller k-mer \omiii{(with k $< k_{max}$)}, \omiii{\proposal{}} only stores the \omiii{taxID}s that are \emph{not} attributed to their corresponding larger, more unique, k-mer. 
\omiii{For these smaller k-mers, \proposal{} does not store the k-mer itself and instead, uses the prefixes of the \mbox{$k_{max}$-mers} to retrieve the smaller k-mer\omiii{s}}.
\omiii{\proposal{}} allows for \omiii{taxID} retrieval by sequentially streaming through the intersecting k-mers  
\bback{(which are already sorted)}
and the KSS tables. 
While \gram{the} KSS data structure is larger than \omiii{the corresponding ternary search tree} \circled{b}, it is \omiii{much more} suitable for ISP due to \omiii{its streaming access feature}. KSS can also be \omiii{efficient} for \omiii{processing} outside \omiii{the} storage \omiii{system} with SSDs with \hm{high} external bandwidth (\sect{\ref{sec-mg:eval-main}}). KSS leads to 7.5$\times$ smaller data structures compared to the 107-GB \omi{data structure in} \circled{a}, and 2.1$\times$ larger compared to \circled{b} (dataset details in \sect{\ref{sec-mg:methodology}}).

\begin{figure}[t]
    \centering
    \includegraphics[width=0.9\linewidth]{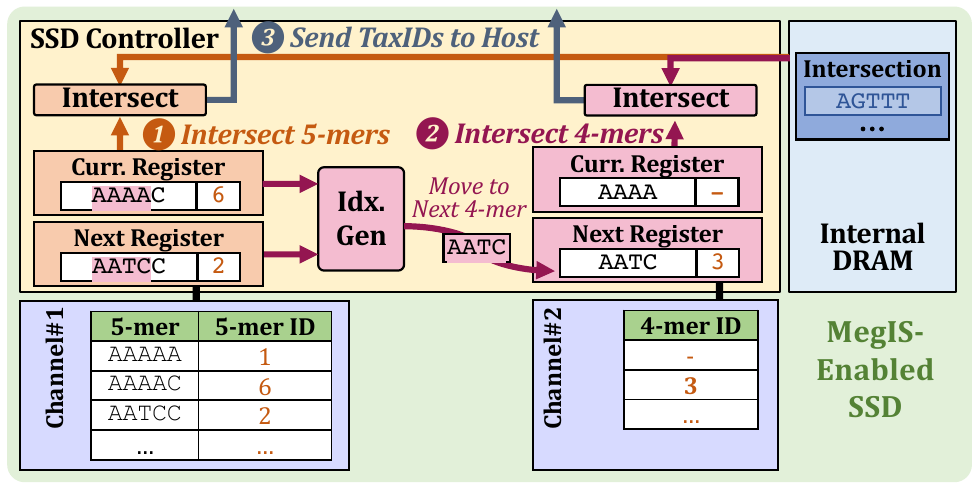}
    \caption{\omiii{Overview of the TaxID} retrieval \omiii{process in \proposal{}}.}
    \label{fig-mg:taxid-finding}
\end{figure}

\fig{\ref{fig-mg:taxid-finding}} shows \gram{an} overview of \proposals \omiii{taxID} retrieval \omiii{process}. As an example, we demonstrate retrieving 5- and 4-mers. First, \proposal{} reads the intersecting k-mers (i.e., 5-mers) from the internal DRAM and concurrently reads the 5-mer sketches and their IDs from an SSD channel to find their matches (using the same Intersect unit in \sect{\ref{sec-mg:mech-stage2-1}})  (\circled{1}). 
Second, to find 4-mer matches, \proposal{} compares the \emph{prefixes} of the intersecting 5-mers with the prefixes of the 5-mer sketches (\circled{2}). 
\proposal{}~\omiii{incorporates} a lightweight \emph{Index Generator}.
\omiii{It compares the 4-mer prefixes of each pair of consecutive 5-mers. When the prefixes differ (indicating the start of a new 4-mer), it identifies the new prefix as the new 4-mer and reads the next 4-mer \omiii{taxID} from a SSD channel.}
Third, \proposal{} sends the retrieved \omiii{taxID}s to the host (\circled{3}) as the IDs of the candidate species present in the sample.

\subsection{Step 3: Abundance Estimation}
\label{sec-mg:mech-stage3}

For applications \omiii{that require} abundance estimation, 
\bback{\proposal{} integrates further analysis on the candidate species identified as present \omiii{in the sample} at the end of Step 2}. 
\proposal{} can flexibly integrate with different approaches \omiii{to abundance estimation} used in various tools,
such as \rev{\inum{i}}~lightweight statistics \omiii{(e.g.,~\cite{lu2017bracken,dimopoulos2022haystac,koslicki2016metapalette})}
or \inum{ii}~more accurate and costly read mapping \omiii{(e.g.,}~\cite{lapierre2020metalign,kim2016centrifuge,milanese2019microbial}\omiii{)}, \bback{where the input read set is mapped to the reference genomes of candidate species \omiii{present in the sample}}. \omiii{Based on the relative number of reads that map to each species' reference genome, we can determine the occurrence frequencies of different species}.
\irev{\proposal{} can\revid{\label{rev:C3}C3} integrate with different existing statistical approaches or \omiii{read mapping, performed in the host or an accelerator,} specialized for short reads (e.g.,\hhl{~\cite{cali2020genasm,nag2019gencache,fujiki2018genax}}) or long reads (e.g.,\hhl{~\cite{turakhia2018darwin,cali2020genasm,nag2019gencache}}). \omiii{We note that Steps 1 and 2} of \proposal{} are based on k-mers extracted from the reads and do not depend on a specific read length}.

\begin{figure}[b]
    \centering
    \includegraphics[width=0.8\linewidth]{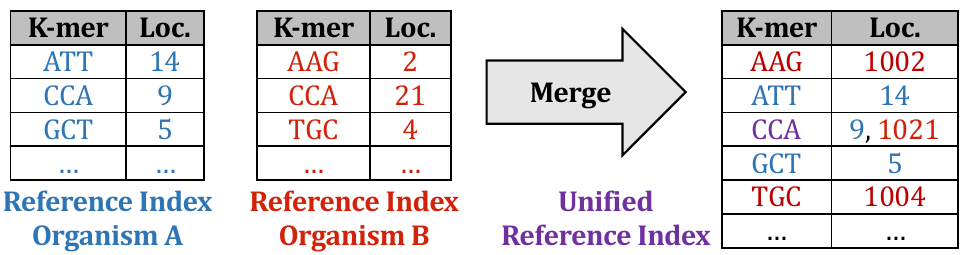}
    \caption{Merging the reference indexes \omiii{to facilitate read mapping during abundance estimation}.}
    \label{fig-mg:index-merger}
\end{figure}

\omiii{While the lightweight statistical approaches can work directly on the output of Step 2, \proposal{} requires additional data preparation to facilitate read mapping}.
The \omiii{read} mapper \omiii{requires} the query reads and a \omiii{unified} index of the reference genomes of the candidate species \omiii{present in the sample}~\cite{li2018minimap2}. \omiii{In comparison to using individual indexes for each species, the unified index eliminates the need to search through each index separately, thereby reducing the overheads of the \omiii{read} mapping process}.
\bback{Building \omiii{indexes for} individual species is a one-time task. Yet, creating a unified index for the initially unidentified species present in the sample cannot be done offline}.
\omiii{\proposal{}} facilitates \omiii{index generation for read mapping} by generating a unified 
index 
in the SSD.
\omiii{\fig{\ref{fig-mg:index-merger}} shows an example of the process of unified index generation in \proposal{}}.
\bback{Each index entry shows a k-mer and its location in that species' reference genome}.
\proposal{} reads each index \omiii{stored in the flash chips} sequentially and merges \omiii{their entries into a unified index}. 
\bback{When \omiii{\proposal{} finds} a common k-mer (e.g., \texttt{CCA} in \fig{\ref{fig-mg:index-merger}}), it stores \omiii{the corresponding location of the k-mer in both reference genomes,} adjusting the locations with appropriate offsets based on \omiii{the reference genome sizes}}. \omiii{After generating the unified index, \proposal{} transfers the index to the host system or an accelerator to perform read mapping for abundance estimation}.

\subsection{\omiii{\proposal{}~FTL}}
\label{sec-mg:mech-ftl}

\omiii{\proposal{}~FTL} needs simple changes to the baseline FTL to handle communication between the host and the SSD.

\omiii{\head{FTL Metadata}} At the beginning \omiii{of \proposals{} operation as a metagenomic acceleration framework}, \omiii{\proposal{}~FTL} maintains all metadata of the regular FTL in the internal DRAM.
For the only step \omiii{that} require\omiii{s} writes \omiii{to the NAND flash chips} (\sect{\ref{sec-mg:mech-stage1-kmer-extraction}}, \omiii{K-mer Extraction} in the host), \omiii{\proposal{}~FTL} uses the write-related metadata \omiii{(e.g., L2P, bad-block information)}. 
After \omiii{the K-mer Extraction step}, \proposal{} does not require writes \omiii{to the NAND flash chips}, so it flushes the regular L2P \omiii{metadata} and loads \omiii{\proposal{}~FTL}'s L2P \omiii{metadata} while still keeping the other metadata \omiii{of a regular FTL}.
 
\omiii{\proposal{} is designed to only access the underlying flash chips \emph{sequentially}, which inherently reduces the size of the required L2P mapping metadata.
In regular FTL, L2P mappings dominate the SSD's internal DRAM capacity due to the page-level granularity of mappings~\cite{mansouri2022genstore, kim-dac-2017, samsung870evo}. However, by accessing data sequentially, \omiii{\proposal{}~FTL} circumvents the need for such detailed page-level mappings. Instead, \omiii{\proposal{}~FTL} utilizes a more coarse-grained block-level mapping, which substantially reduces the size of L2P metadata.
Therefore, flushing regular L2P mapping metadata into flash chips and using \omiii{\proposal{}~FTL}'s metadata enables us to exploit most of the internal DRAM \omiii{bandwidth and capacity} during ISP}.

\begin{figure}[h]
    \centering
    \includegraphics[width=0.9\linewidth]{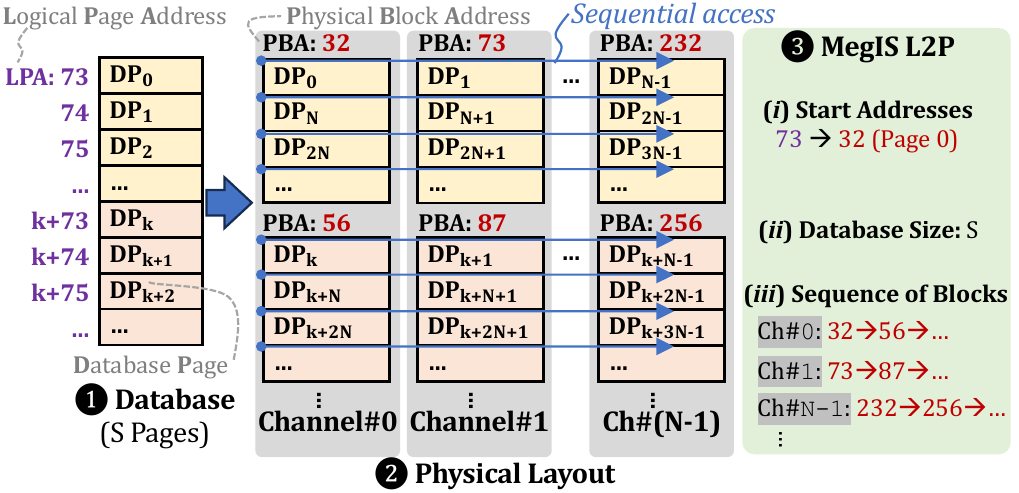}
    \caption{Data layout and mapping data structure in \proposal{}.}
    \label{fig-mg:ms-ftl}
\end{figure}

\omiii{\head{Data Placement}} \fig{\ref{fig-mg:ms-ftl}} shows how \omiii{\proposal{}~FTL} manages the target data \omiii{(i.e., databases\footnote{K-mer \omiii{databases (\sect{\ref{sec-mg:mech-stage2-1}})} and sketch databases \omiii{(\sect{\ref{sec-mg:mech-stage2-1}})} are the only data structures accessed from NAND flash memory during \proposals{} ISP operations.})} stored in NAND flash with reduced \omiii{L2P} metadata.
When storing a database \omiii{in the SSD} \circled{1}, \omiii{\proposal{}~FTL}  \emph{evenly and sequentially} distributes the data across all  channels \omiii{\circled{2}} while ensuring that every \emph{active} block~\cite{kawaguchi1995flash} 
\bback{(\omiii{i.e., blocks available for write operations in the SSD})}
in different channels has the same page offset.
\omiii{Since \proposal{} always accesses the database \emph{sequentially}, \omiii{\proposal{}~FTL}'s L2P mapping metadata} \circled{3} consists of \inum{i} the mapping between start logical page address \omiii{(LPA)} and physical page address \omiii{(PPA)}, \inum{ii} the database size, and \inum{iii} the sequence of physical block addresses storing the database.
As shown in \fig{\ref{fig-mg:ms-ftl}}, \omiii{\proposal{}~FTL} can sequentially read the stored database \omiii{from the starting LPA}
while performing
reads in a round-robin manner across channels. \omiii{To do so, \omiii{\proposal{}~FTL} just increments the PPA within a physical block and resets the PPA when reading the next block}.

Compared to the regular L2P, whose space overhead is 0.1\% of stored data (4 bytes per 4 KiB), \proposals{} L2P is very small. 
For example, \proposal{} only requires $\sim$1.3 MB to store a 4-TB database, assuming a physical block size of 12~MB: 4~bytes for each of the 349,525
used blocks (and a few bytes for the start L2P mapping and database size).
The only metadata other than L2P that must be kept during ISP is the per-block access count for read-disturbance management~\cite{cai2015read}, so the total \proposal{}\irev{-FTL}\revid{\label{rev:A5.3}A5.3} metadata size is up to 2.6~MB.

\omiii{\head{SSD Management Tasks}} \proposals{} ISP accelerators are located \omiii{in} the SSD controller and access data \emph{after} ECC. \irev{ECC\revid{\label{rev:C2}C2} does not restrict \proposals{} ISP performance. Modern SSDs are designed with ECC capabilities that match the full internal bandwidth \omiii{of the SSD} to support both I/O requests and internal data migrations due to management tasks like garbage collection\omiii{~\cite{kim2023decoupled,cai-insidessd-2018,cai2017error}}. 
\icut{Otherwise, ECC would limit the SSD’s performance, which would an unreasonable design. In many cases, modern SSDs incorporate per-channel ECC that is shared between multiple chips connected to a channel\hhl{\omiii{~\cite{kim2023decoupled,cai-insidessd-2018,cai2017error}}}. Even where multiple channels share an ECC, the ECC still needs to match the full internal bandwidth   to avoid bottlenecking the SSD’s performance.}}

\proposal{} performs other tasks for ensuring reliability (e.g., refresh to prevent uncorrectable errors\cite{cai2017vulnerabilities, luo2018improving,luo2018heatwatch,cai2015read,cai2013error, cai2012flash, cai2017error, ha2015integrated, cai-insidessd-2018,luo2015warm}) \emph{before or after} the ISP because 1) the duration of each \proposal{} process is significantly smaller than the manufacturer-specified threshold for reliable retention age (e.g., one year~\cite{micron3dnandflyer}), and 2) \proposal{} avoid\hhl{s} read disturbance errors~\cite{cai2015read} during ISP due to its sequential low-reuse accesses.

\subsection{\irev{Storage Interface Commands\revid{\label{rev:A7}A7}}}
\label{sec-mg:mech-interface}

\irev{\proposal{} requires three new NVMe commands. First, \textsf{MegIS\_Init} initiates the metagenomic analysis and communicates the size and starting address of the space in the host DRAM that is available for \proposals{} operations. \omiii{Upon receiving this command,} \proposal{} readies itself to work in the metagenomic acceleration mode \omiii{(\sect{\ref{sec-mg:mech-overview}})}. During metagenomic analysis steps, \omiii{\proposal{}~FTL} and \proposals{} FSM controller handle the data/control flow. 
Second, \textsf{MegIS\_Step} communicates the start and end of \omiii{each} step executed in the host to the SSD, enabling \proposal{} to manage control and data flow accordingly.  \omiii{\textsf{MegIS\_Step} specifies the step performed in the host system, such as k-mer extraction (\sect{\ref{sec-mg:mech-stage1-kmer-extraction}}) or sorting (\sect{\ref{sec-mg:mech-stage1-sorting}}), with an input argument. Each time this command with the same argument is sent, it alternates between marking the start and the end of a step}.
After completing the metagenomic analysis, \proposal{} switches back to operating as a baseline SSD. Third, \textsf{MegIS\_Write} is a specialized write operation that updates \omiii{\proposal{}~FTL}'s small mapping metadata whenever metagenomic data is written to the SSD}. \omiii{\textsf{MegIS\_Write} is similar to the regular NVMe write command, except that it updates mapping metadata in both the regular FTL and \omiii{\proposal{}~FTL}}.

\subsection{Multi-Sample Analysis}
\label{sec-mg:mech-multi-sample}

\newcommand\bases{\textsf{Base-S}\xspace}
\newcommand\basem{\textsf{Base-M}\xspace}
\newcommand\mss{\textsf{MS-S}\xspace}
\newcommand\msm{\textsf{MS-M}\xspace}
\newcommand\optm{\textsf{Opt-M}\xspace}

\begin{figure}[t]
    \centering
    \includegraphics[width=0.9\linewidth]{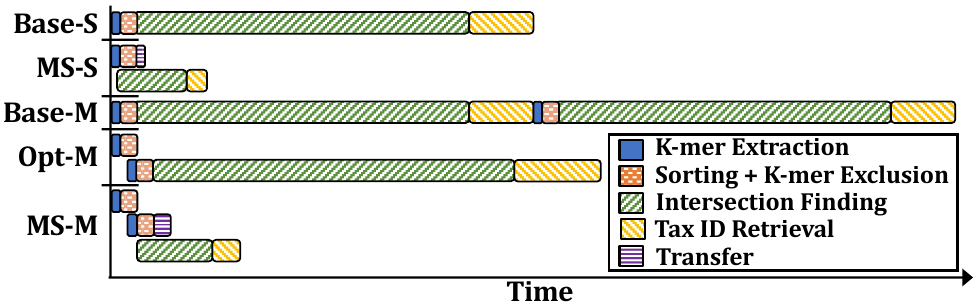}
    \caption{\new{Timeline of \omiii{analyzing single or multiple samples in the baseline and \proposal{}.}}}
    \label{fig-mg:multi-timeline}
\end{figure}

For some use cases \omiii{(e.g., globally tracing antimicrobial resistance\cite{danko2021global}, associating gut microbiomes to health status~\cite{Turnbaugh2007,hhrlich2011metahit})}, 
a metagenomic study can have multiple read sets (i.e., samples) that need to access the \omiii{same} database.
If the host's DRAM is larger than the k-mer sizes extracted from a sample, we use the available DRAM opportunistically to buffer k-mers extracted from \emph{several} samples. This way, \proposal{} streams through \omiii{one} database only once.
\fig{\ref{fig-mg:multi-timeline}} shows the timeline of analyzing \gram{a} single (\textsf{S}) or multiple (\textsf{M}) samples in the baseline (\textsf{Base}), in our proposed optimized approach in software (\textsf{Opt}), and in \proposal{} (\textsf{MS}).
\omiii{To accelerate input query processing (\sect{\ref{sec-mg:mech-stage1}}) when analyzing several input query samples,} \proposal{} can \omiii{be} flexibly integrate\omiii{d} with a sorting accelerator \omiii{(e.g.,~\cite{samardzic2020bonsai,qiao2022topsort,jayaraman2022hypersort})} \omiii{and} further improve end-to-end performance.

\section{Evaluation\revid{\label{rev:B3}B3} Methodology}
\label{sec-mg:methodology}

\head{Performance} 
We design a simulator that models all \hhl{of} \proposals{} components, including host operations, accessing flash chips, internal DRAM, in-storage accelerator, and host-SSD interfaces. We feed the latency and throughput of each component to this simulator.  
For the components in the \textbf{hardware-based} steps (e.g., ISP units in Steps 2 and 3):  We implement \proposals{} logic components in Verilog. We synthesize them using the Synopsys Design Compiler~\cite{synopsysdc} with a 65~nm library\cite{umcL65nm} and \irev{perform place-and-route using Cadence Innovus\hhl{~\cite{innovus}}}.
We use two state-of-the-art 
simulators, Ramulator~\cite{kim2016ramulator, ramulatorsource} to model SSD's internal DRAM, and MQSim\omii{~\cite{tavakkol2018mqsim,mqsimsource}} \omii{to model} SSD's internal operations.
For the components in the \textbf{software-based} step (e.g., host operations in Step 1)\omii{, we} measure performance on a real system, an AMD$^\text{\textregistered}$ EPYC$^\text{\textregistered}$ 7742 CPU\omii{~\cite{amdepyc}} with 128 physical cores \hhl{and} 1-TB DRAM (in all experiments unless stated otherwise). 
For the software baselines, we measure performance on this real system, with best-performing thread counts.
\omii{The source code of \proposal{}, scripts, and datasets can be freely downloaded from https://github.com/CMU-SAFARI/MegIS.}

\head{SSDs} We use \ssdc~\cite{samsung870evo} and \ssdp~\cite{samsungPM1735} as described in \sect{\ref{sec-mg:motivation-ovhd}} \irev{in our real system experiments. In our MQSim simulations for the ISP steps, we faithfully model the SSDs with the configurations summarized in \revid{\label{rev:A3}A3}Table~\ref{table-mg:SSD_config}}.

\begin{table}[h]
\vspace{5pt}
\centering
\footnotesize
\color{black}
\caption{\irev{SSD configurations.}}
\begin{tabular}{@{\hspace{-0.5pt}}c@{\hspace{-0.05pt}}|c|c@{\hspace{-0.05pt}}}
\toprule
\textbf{Specification} & \textbf{SSD-C} & \textbf{SSD-P} \\ 
\midrule
\midrule
\textbf{General}       & \multicolumn{2}{c}{\begin{tabular}[c]{@{}c@{}} 3D TLC NAND flash-based SSD\\ 4 TB capacity, 4 GB internal \omii{LPDDR4} DRAM\omii{~\cite{lpddr4}} \end{tabular}} \\ \midrule
\begin{tabular}[c]{@{}c@{}} \textbf{Bandwidth}\\ \textbf{(BW)} \end{tabular}  & \begin{tabular}[c]{@{}c@{}}600 MB/s interface  BW\\ (SATA3);\\ 560 MB/s sequential-read BW\\  1.2-GB/s channel I/O rate\end{tabular} & \begin{tabular}[c]{@{}c@{}}8 GB/s interface BW \\ (4-lane PCIe Gen4);\\ 7 GB/s sequential-read BW\\1.2-GB/s channel I/O rate\end{tabular} \\ \midrule
\begin{tabular}[c]{@{}c@{}} \textbf{NAND}\\ \textbf{Config} \end{tabular}  & 
\begin{tabular}[c]{@{}c@{}} 8 channels, 8 dies/channel,\\ 4 planes/die, 2,048 blocks/plane,\\ 196  WLs/block, 16 KiB/page \\ \textit{(4/8/16 channels in \fig{\ref{fig-mg:internal-bw-swp}})}\end{tabular}  &
\begin{tabular}[c]{@{}c@{}} 16 channels, 8 dies/channel,\\ 2 planes/die, 2,048 blocks/plane,\\ 196 WLs/block, 16 KiB/page \\ \textit{(8/16/32 channels in \fig{\ref{fig-mg:internal-bw-swp}})}\end{tabular}  \\ \midrule
\textbf{Latencies}     & \multicolumn{2}{c}{\begin{tabular}[c]{@{}c@{}} \omii{Read (tR): 52.5 $\mu$s,  Program (tPROG): 700 $\mu$s}\end{tabular}} \\ \midrule
\begin{tabular}[c]{@{}c@{}} \textbf{Embedded}\\ \textbf{Cores} \end{tabular}  & 3 ARM Cortex-R4 cores\omii{~\cite{cortexr4}}  & 4 ARM Cortex-R4 cores\omii{~\cite{cortexr4}}  \\ \midrule
\bottomrule
\end{tabular}
\label{table-mg:SSD_config}
\end{table}

\head{Area and Power} 
For logic components, we use the results from our Design Compiler synthesis. 
For SSD power, we use the values of a Samsung 3D NAND flash-based SSD~\cite{samsung860pro}. For DRAM power, we base the values on a DDR4 model~\cite{ddr4sheet, ghose2019demystifying}. For the CPU cores, we use AMD$^\text{\textregistered}$ \textmu{}Prof~\cite{microprof}.

\newcommand\popt{\textsf{P-Opt}\xspace}
\newcommand\aopt{\textsf{A-Opt}\xspace}
\head{Baseline Metagenomic Tools} We use a state-of-the-art performance-optimized (\textsf{P-Opt}) tool, Kraken2 + Bracken~\cite{wood2019improved}, and a state-of-the-art accuracy-optimized (\textsf{A-Opt}) tool, Metalign~\cite{lapierre2020metalign}. 
Particularly, for the presence/absence task, we use Kraken2 without Bracken, and Metalign without mapping \bback{(i.e., only KMC~\cite{kokot2017kmc3} + CMash~\cite{liu2022cmash})}. 
For abundance estimation, we use Kraken2 + Bracken, and full Metalign.
\omii{\aopt achieves significantly \omiii{higher} accuracy compared to \popt~\cite{meyer2021critical,lapierre2020metalign}. In particular, \aopt leads to \omiii{4.6--5.2}$\times$ higher F1 scores and \omiii{3--24}\% lower L1 norm error across \omiii{all tested} inputs. One major reason is that \aopt uses larger and richer databases compared to performance-optimized \popt.  
\proposals{} end-to-end accuracy matches the accuracy of \aopt because \proposals{} databases encode the same set of k-mers and sketches as \aopt.}

For both Metalign and \proposal{}, we use GenCache~\cite{nag2019gencache} for mapping. 
\bback{We use the mapping throughput as reported by the original paper~\cite{nag2019gencache}.
 \proposal{} can \omii{be} flexibly integrate\omii{d} with other mappers}. 
We also \irev{evaluate}
a state-of-the-art PIM k-mer matching accelerator~\cite{wu2021sieve} \hm{for} accelerat\hm{ing} Kraken2's pipeline. We use the k-mer matching performance as reported by the original paper~\cite{wu2021sieve}.

\head{Datasets} 
We use three query read sets from the commonly-used CAMI benchmark~\cite{sczyrba2017critical}, \bback{with low, medium, and high \hm{genetic} diversity} \omii{(i.e., CAMI-L, CAMI-M, and CAMI-H, respectively)}.
Each read set has 100 million reads.
We generate a database based on \hm{microbial genomes drawn from NCBI's databases}~\cite{ncbi2020,lapierre2020metalign}\irev{\revid{\label{rev:B5.2}B5.2} including 155,442 genomes for 52,961 microbial species.\icut{ (as listed in\hhl{~\cite{metalign_db}}).} For database generation, we use}  default parameters for each tool. For Kraken2~\cite{wood2019improved}, this results in a 293~GB database. For Metalign~\cite{lapierre2020metalign}, this results in a 701~GB k-mer database
and \gram{a} 6.9~GB sketch tree.
\proposal{} uses the same 701~GB k-mer database and a 14~GB sketch database for \omii{\proposals{}} \cmashopt~\omii{sketch database (\sect{\ref{sec-mg:mech-stage2-2}})}.

\section{Evaluation}
\label{sec-mg:eval}

\subsection{Presence/Absence Identification Analysis}
\label{sec-mg:eval-main}

\renewcommand\popt{\textsf{P-Opt}\xspace}
\renewcommand\aopt{\textsf{A-Opt}\xspace}
\newcommand\msnol{\textsf{MS-NOL}\xspace}
\newcommand\msext{\textsf{Ext-MS}\xspace}
\newcommand\mscc{\textsf{MS-CC}\xspace}
\newcommand\msfull{\textsf{MS}\xspace}

\bback{We use 1\gram{-}TB \omii{host} DRAM in this analysis (smaller than all datasets \omii{we evaluate})}. 
\omii{We examine seven metagenomic analysis configurations:}
1)~\popt,
2)~\aopt,
3)~{\aopt}+\cmashopt\omii{, where} \aopt leverages the software implementation of \omii{\proposals{}} \cmashopt approach (\sect{\ref{sec-mg:mech-stage2-2}}) instead of Metalign's CMash~\cite{lapierre2020metalign} \omii{for retrieving taxIDs}, 
\rev{\omii{4})~\msext: a \proposal{} implementation
without ISP, where the same accelerators used in \proposal{} are  outside the SSD},
\omii{5})~\msnol: a \proposal{} implementation \omii{\emph{without}} overlapping the host and SSD operations as enabled by \proposals{} bucketing (\sect{\ref{sec-mg:mech-stage1}}),  
\rev{6)~\mscc: a \proposal{} configuration
\omii{where} the SSD cores perform \proposals{} ISP tasks, and}
7)~\msfull: a \underline{M}egI\underline{S}~\rev{configuration \omii{where} the \omii{hardware} accelerators on the SSD controller perform the ISP tasks}.

\begin{figure}[h]
\centering
\vspace{0.2em}
 \includegraphics[width=0.85\linewidth]{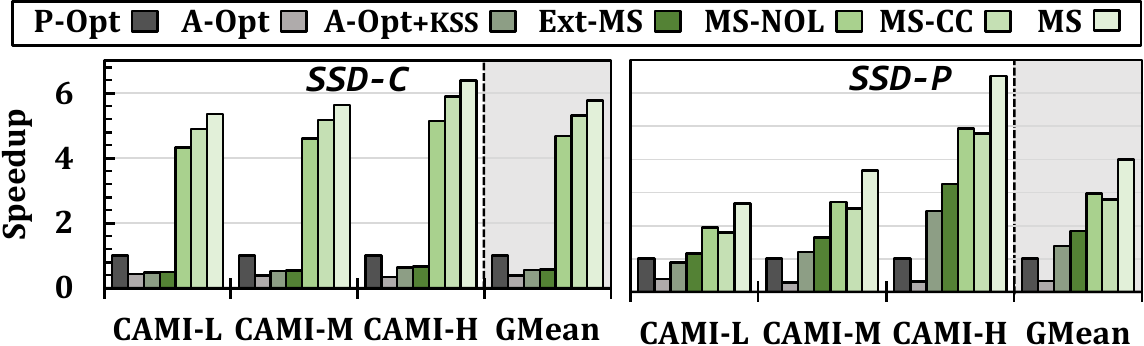}
\caption{\rev{Speedup for different SSDs and input sets.}}
\label{fig-mg:main-eval}
\end{figure}

\omii{\fig{\ref{fig-mg:main-eval}} shows the speedup of the seven configurations over \popt, on three read sets and with \ssdc and \ssdp}. 
\bback{We make six key observations}. 
First, \omii{\proposals{} full implementation (\msfull) achieves significant speedup compared to both performance-optimized (\popt) and accuracy-optimized (\aopt) baselines}. 
\omii{W}ith \ssdc (\ssdp), \msfull is 5.3--6.4$\times$ (2.7--6.5$\times$) faster compared to \popt, and 12.4--18.2$\times$ (6.9--20.4$\times$) faster compared to \aopt.
Second, \omii{{\aopt}+\cmashopt, which leverages \proposals{} taxID retrieval approach (\cmashopt) instead of {\aopt}'s baseline taxID retrieval approach,} improves \aopt's performance by 1.4$\times$ (4.2$\times$) on average on \ssdc (\ssdp). \omii{\proposals{} full implementation outperforms \aopt{}+\cmashopt by} 10.5$\times$ (2.9$\times$). This shows that while \omii{\proposals{}} \cmashopt approach\gram{,} even outside the SSD\gram{,} provides large benefits, \omii{\proposals{} full implementation} provides significant additional benefits by alleviating I/O overhead. 
Third, with \ssdc (\ssdp), \msfull leads to 23.5\% (34.9\%) \hm{greater} average speedup compared to \omiii{\proposals{} implementation without overlapping the steps (}\msnol\omiii{). This is} due to \proposals{} bucketing \omiii{scheme} that enables overlapping the steps.
\rev{Fourth, \msfull leads to 10.2$\times$ (2.2$\times$)
average speedup on \ssdc (\ssdp) compared to \omii{\proposals{} implementation outside the SSD} (\msext)  due to \omii{the benefits of \proposals{}} specialized ISP}.
\rev{Fifth,
while \mscc provides large speedup,
\msfull leads to 9\% (43\%) greater average speedup compared to \mscc on \ssdc (\ssdp). While both \proposal{} configurations provide large speedup, this shows \omii{that the hardware} accelerators \omii{are useful and their benefits improve} as the internal bandwidth grows.}
\omii{Sixth, \proposals{} speedup improves as the genetic diversity of the input read sets increases (from CAMI-L to CAMI-H). This is due to the presence of more species in more diverse read sets, which results in a greater number of sketch tree lookups \omiii{in the baseline} taxID \omiii{retrieval approach}. In contrast, \proposals{} \cmashopt efficiently retrieves all taxIDs in a single pass through the sketch tables}.

\begin{figure}[h]
\centering
 \includegraphics[width=0.9\linewidth]{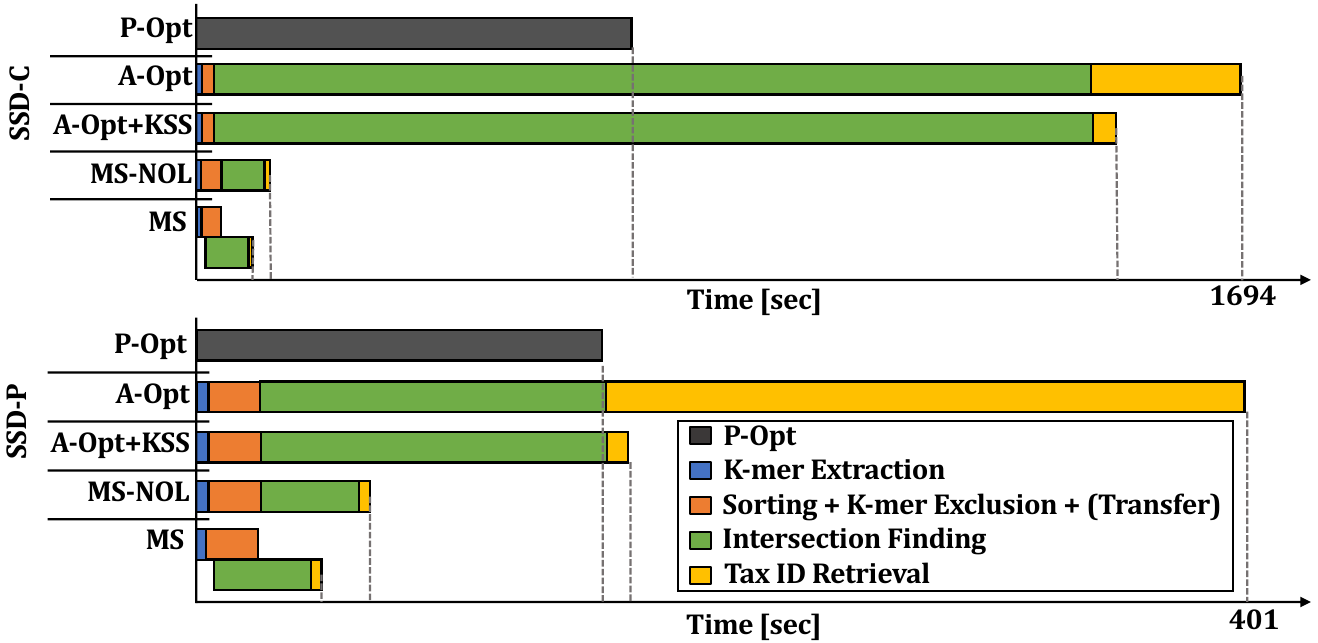}
\caption{\rev{Time breakdown with different SSDs for CAMI-L.}}
\label{fig-mg:exec-breakdown}
\end{figure}

\bback{To further demonstrate the benefits of \proposals{} optimizations}, 
\fig{\ref{fig-mg:exec-breakdown}}
shows the time breakdowns with \mbox{CAMI-L} as a representative \omii{input}. 
First, \omii{\cmashopt improves performance by reducing the execution time of taxID retrieval (as seen by \aopt{}+\cmashopt over \aopt).
Second, \proposal{} without overlapping improves performance over \aopt{}+\cmashopt by leveraging ISP to accelerate intersection finding and taxID retrieval  (as seen by \msnol over \aopt{}+\cmashopt).  
Third, adding overlapping in \proposals{} full implementation improves performance by overlapping the execution of sorting in the host system with intersection finding in the SSD (as seen by \msfull over \msnol).  
}

\pagebreak

\begin{figure}[h]
\centering
 \includegraphics[width=0.85\linewidth]{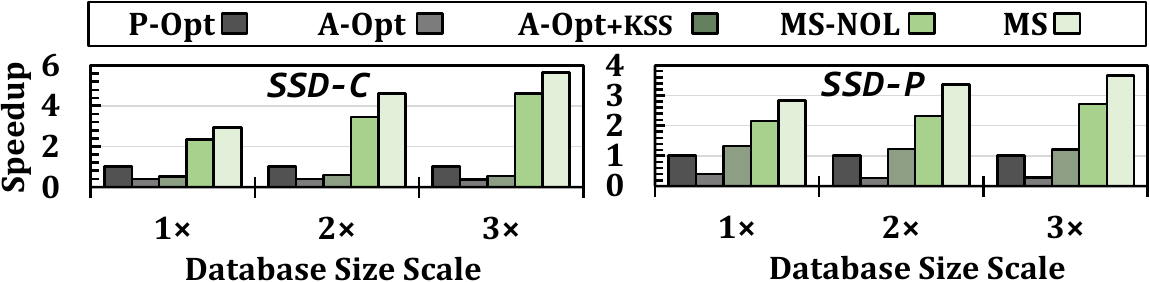}
\caption{\hm{Speedup with} different database sizes.}
\label{fig-mg:eval-db}
\end{figure}

\head{\omii{Effect} of Database Size} \fig{\ref{fig-mg:eval-db}} shows the \omii{effect} of database size, using CAMI-M as a representative input. The largest database size in each tool (marked by 3$\times$) equals the size mentioned in \sect{\ref{sec-mg:methodology}}.
We observe that \proposals{} speedups increase as the database size increases (up to 5.6$\times$/3.7$\times$ speedup compared to \popt 
on \ssdc/\ssdp~\omii{as database size grows to 3$\times$}).

\begin{figure}[b]
    \centering
    \includegraphics[width=0.85\linewidth]{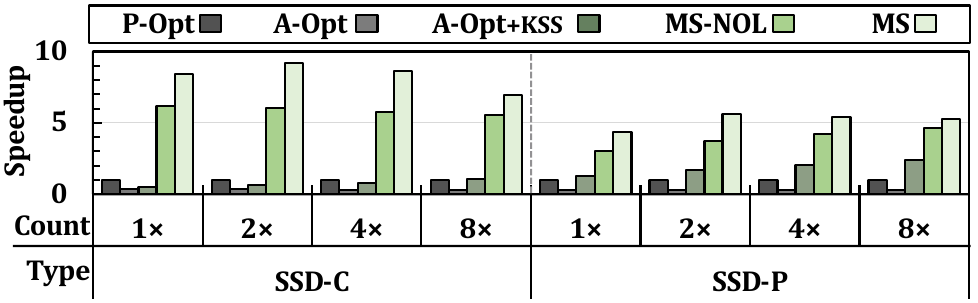}
    \caption{Speedup with different number of SSDs.}
    \label{fig-mg:eval-ssd}
\end{figure}

\head{\omii{Effect} of the Number of SSDs}
\proposal{} benefits from more SSDs in two ways.
First, mapping different databases to different SSDs allows \hhl{for} concurrent analyses, each benefiting from \proposal{}, as already shown \omii{(see \fig{\ref{fig-mg:main-eval}})}. \omii{Since} \proposals{} databases and queries \omii{are} sorted, the database can be disjointly split across SSDs. \fig{\ref{fig-mg:eval-ssd}} \omii{demonstrates} this case \omii{by showing the speedup of different configurations over \popt}.
We show that \proposal{} maintains its large speedup with many SSDs \omii{(i.e., up to eight)}. \omii{A}s the external bandwidth increases for the baselines \omii{(with the number of SSDs)}, internal bandwidth also increases for \proposal{}. Particularly, speedup over \popt increases until some point 
\bback{(two SSDs)}
because \proposal{} takes better advantage of the scaling due to its more efficient streaming accesses. 
Although there is a slight decrease in speedup when moving from two to eight SSDs, the speedup is still high (6.9$\times$/5.2$\times$ over eight SSD-Cs/SSD-Ps). This decrease is because in \proposal{}, due to the large internal bandwidth with 8 SSDs, the overall throughput becomes dependent on the host's sorting. Therefore, 
\bback{in systems with many SSDs},
\proposal{} can \omii{be} integrate\omii{d} with an accelerator for sorting \omii{(e.g.,~\cite{samardzic2020bonsai,qiao2022topsort,jayaraman2022hypersort})} for further speedup. We conclude that \proposal{} effectively leverages the increased internal bandwidth with more SSDs.
\irev{Due to this efficient use of multiple SSDs \revid{\label{rev:B5.1}B5.1}(owing to \proposals{} sorted database that can be disjointly partitioned), \proposal{} can efficiently scale up to very large databases that are distributed \omii{across} different SSDs.}

\begin{figure}[h]
\centering
 \includegraphics[width=0.85\linewidth]{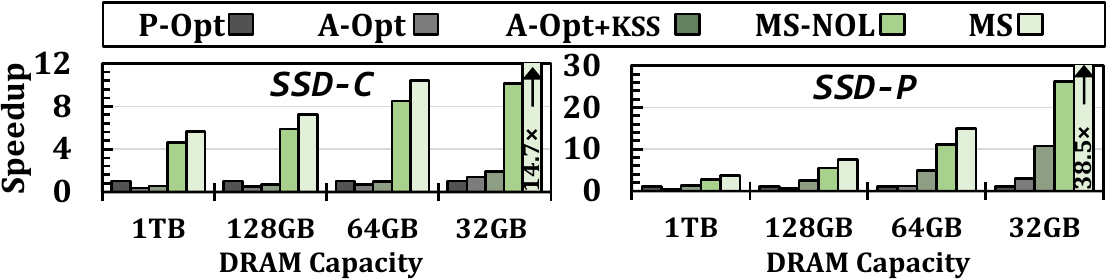}
\caption{\hm{Speedup} with different main memory capacities.}
\label{fig-mg:eval-mem}
\end{figure}

\head{\omii{Effect} of Main Memory Capacity} \fig{\ref{fig-mg:eval-mem}} \omii{demonstrates} the \omii{effect} of \omii{host} DRAM capacity \omii{by showing the speedup of all configurations over \popt}
with CAMI-M.\footnote{In all cases, \rev{except for the 32GB configuration,} all k-mer buckets extracted from the read set (\sect{\ref{sec-mg:mech-stage1-kmer-extraction}}) fit in \omii{the host} DRAM.}
\bback{To gain a fair understanding of I/O overheads when DRAM is smaller than the database, we reduce \omii{I/O} overheads as much as possible in software. We adopt an optimization~\cite{pockrandt2022metagenomic} to load and process \popt's database into chunks that fit in DRAM.\footnote{Note \aopt does not require this due to its streaming database accesses.} In this case, random accesses to the database in each chunk do not repeatedly access the SSD. However, two overheads still remain: 1) there is still the I/O cost of bringing all chunks from the SSD \omii{to the host DRAM}, and 2) for every \omii{database} chunk, all of the input sequences must be queried}.
We make three observations. 
First, \proposals{} speedup increases compared to \popt with smaller DRAM (e.g., up to \rev{38.5}$\times$ \omii{\omii{speedup with} 32GB of host DRAM}). This is because \popt's performance \revh{is} hindered by \omii{the host DRAM} capacity, while \proposal{} does not rely on large \omii{host} DRAM. Second, \aopt and {\aopt}+\cmashopt\ \revh{are not} affected by the small DRAM (except for the 32-GB configuration) due to their streaming database accesses.
But regardless of DRAM size, they suffer from I/O overhead.
Third, with the 32-GB DRAM\omii{, which is smaller than the extracted query k-mers in Step 1 (\sect{\ref{sec-mg:mech-stage1-kmer-extraction}})},
\msfull's speedup increases. \omii{This is because}~\proposals{} bucketing (\sect{\ref{sec-mg:mech-stage1-kmer-extraction}}) \omii{avoids unnecessary page swaps between the host DRAM and the SSD in this case}.
\bback{We conclude that \proposal{} enables fast and accurate analysis, \emph{without relying} on large DRAM or large SSD-external bandwidth}.

\begin{figure}[h]
\centering
 \includegraphics[width=0.85\linewidth]{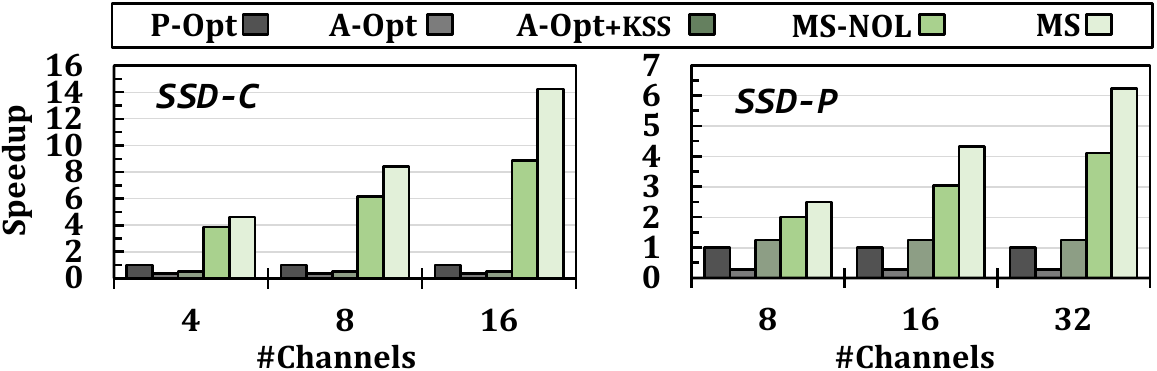}
\caption{\irev{Speedup with \omii{varying SSD internal bandwidth}}.}
\label{fig-mg:internal-bw-swp}
\end{figure}

\head{\omii{Effect} of Internal Bandwidth} \revid{\label{rev:A4.3}A4.3}\fig{\ref{fig-mg:internal-bw-swp}} shows the \omii{effect} of internal bandwidth (\omii{i.e.,} by varying the number of SSD channels) on \proposal{} with CAMI-M as a representative input. We observe that \proposals{} speedup increases as the internal bandwidth increases. 
On \ssdc (\ssdp), \proposal{} leads to 12.3--41.8x (8.6--21.6x) speedup over \aopt. \omii{The increased speedup is \omiii{due to the improved} performance of \proposals{} ISP steps as the internal bandwidth \omiii{increases}}.

\begin{figure}[b]
\centering
 \includegraphics[width=0.85\linewidth]{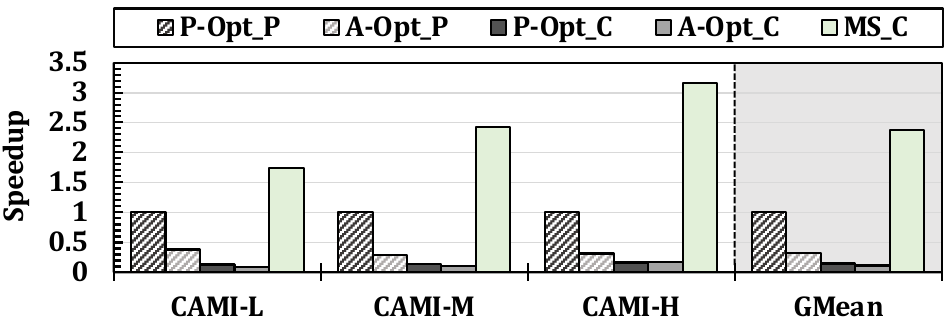}
\caption{Speedup of \proposal{} on a cost-optimized system over baselines on \omiii{cost-/}performance-optimized systems.}
\label{fig-mg:ms-ce}
\end{figure}

\newcommand\poptp{\textsf{P-Opt\_P}\xspace}
\newcommand\aoptp{\textsf{A-Opt\_P}\xspace}
\newcommand\poptc{\textsf{P-Opt\_C}\xspace}
\newcommand\aoptc{\textsf{A-Opt\_C}\xspace}
\newcommand\msc{\textsf{MS\_C}\xspace}

\head{Impact on System Cost Efficiency} 
\proposal{} increases system cost-efficiency because \inum{i}~it analyzes large amounts of data inside storage and removes a large part of the analysis burden from other parts of the system, and \inum{ii}~\hhl{it} does not rely on either high-bandwidth host-SSD interfaces or large DRAM.
\fig{\ref{fig-mg:ms-ce}}  compares \proposal{} on a cost-optimized system with \ssdc and 64-GB \omii{host} DRAM (\msc) to baselines \omiii{1) on the same system (\poptc and \aoptc) and 2) on} a performance-optimized system with  \ssdp and 1-TB \omii{host} DRAM (\poptp and \aoptp).\footnote{\omiii{For the performance-optimized system, we calculate the cost of 1TB DRAM to be roughly 7080 USD (8$\times$ 128GB modules~\cite{samsung128GBDDR4}) and the cost of \ssdp to be roughly 875 USD. For the performance-optimized system, we calculate the cost of 64GB DRAM to be roughly 312 USD (8$\times$ 8GB modules~\cite{samsung8GBDDR4}, assuming the same number of memory channels as the performance-optimized system) and the cost of \ssdc to be roughly 346 USD. Note that the cost of the \emph{total storage system} depends not only on the price of each SSD but also on the available interconnection slots in the systems, as systems typically have fewer PCIe slots (needed for \ssdp) than SATA slots (needed for \ssdc).}} \omiii{We make two key observations. First, \proposal{} on the cost-optimized system outperforms the baselines even when they run on the performance-optimized system.} \msc provides 2.4$\times$ and 7.2$\times$ average speedup compared to \poptp and \aoptp, respectively. Note that \msc provides the same accuracy as \aoptp and significantly higher accuracy than \poptp. 
\omiii{Second, baselines on the cost-optimized system experience significantly worse performance compared to when they run on the performance-optimized system. \poptc leads to 6.8$\times$ (7.7$\times$) average (maximum) slowdown over \poptp, and \aoptc leads to 2.8$\times$ (4.2$\times$) average (maximum) slowdown over \aoptp}. 
We conclude that \omiii{\proposal{} improves system cost-efficiency, while providing high performance and accuracy}.
This is critical \omii{to} both increasing the system cost-efficiency and enabling portable analysis, which is increasing\omii{ly} importan\omii{t due to} the advances of compact portable sequencers~\cite{minion21,jain2016oxford,cali2017nanopore} for on-site metagenomics\omii{~\cite{pomerantz2018real, chiang2019from,mutlu2023accelerating,alser2022molecules}}.

\begin{figure}[h]
\centering
 \includegraphics[width=0.85\linewidth]{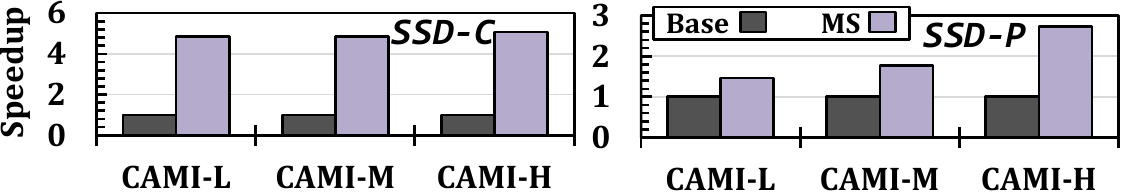}
\caption{\hm{Speedup over} a PIM-accelerated baseline~\cite{wu2021sieve}.}
\label{fig-mg:eval-pim}
\end{figure}

\head{Comparison to a PIM Accelerator} \fig{\ref{fig-mg:eval-pim}} compares \proposal{} to a PIM-accelerated baseline. We evaluate Kraken2's \emph{end-to-end} performance \bback{(i.e., including the I/O accesses to load data to the PIM accelerator, k-mer matching, sample classification, and other computation~\cite{wood2019improved})},
\hm{performing} k-mer matching \hm{on} a state-of-the-art
PIM system, Sieve~\cite{wu2021sieve}.\footnote{We do not use PIM for Metalign's k-mer matching as \omii{k-mer matching in Metalign} is bottlenecked only by I/O bandwidth, not main memory, due to its streaming accesses.} 
\proposal{} achieves 4.8-5.1$\times$ (1.5-2.7$\times$) speedup on \ssdc (\ssdp) \omii{over the PIM-accelerated baseline} while \omii{providing} significantly higher accuracy \omii{(4.8× higher F1 scores and 13\% lower L1 norm error)}.\footnote{A larger database for Kraken2 to encode richer information can increase accuracy \omii{for the PIM-accelerated baseline},
but with even large\hhl{r} I/O overhead.}

\subsection{Abundance Estimation Analysis}
\label{sec-mg:eval-abundance}

\begin{figure}[b]
\centering
 \includegraphics[width=0.85\linewidth]{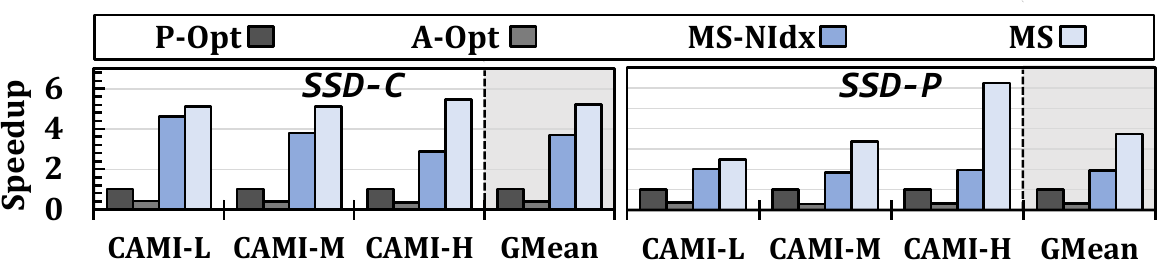}
\caption{\hm{Speedup} for abundance estimation.}
\vspace{0.3em}
\label{fig-mg:eval-abundance}
\end{figure}

\newcommand\msnidx{\textsf{MS-NIdx}\xspace}

\omii{We evaluate abundance estimation with four configurations:
1)~\popt,
2)~\aopt, 
3)~\msnidx: a \proposal{} implementation that does not leverage \proposals{} third step for generating a unified reference index (\sect{\ref{sec-mg:mech-stage3}}), and instead uses Minimap2~\cite{li2018minimap2} for index generation,
and 4)~\msfull: \proposals{} full implementation. 
\fig{\ref{fig-mg:eval-abundance}} shows speedups over \popt}.
We make two key observations. 
First, 
\omii{\proposals{} full implementation} leads to significant speedup \omii{compared to both performance- and accuracy-optimized baselines. \msfull provides}
5.1--5.5$\times$ (2.5--3.7$\times$) speedup on \ssdc (\ssdp) compared to \popt, and 12.0--15.3$\times$ (6.5--20.8$\times$) speedup compared to \popt. Second, \proposals{} full implementation achieves 65\% higher average speedup compared to \msnidx~\omii{due to \proposals{} efficient index generation}.

\subsection{Multi-Sample \omiii{Use Case}}
\label{sec-mg:eval-multi-sample}

\newcommand\mspipe{\textsf{MS-\omiii{Pipe}}\xspace}

\begin{figure}[h]
\centering
 \includegraphics[width=0.85\linewidth]{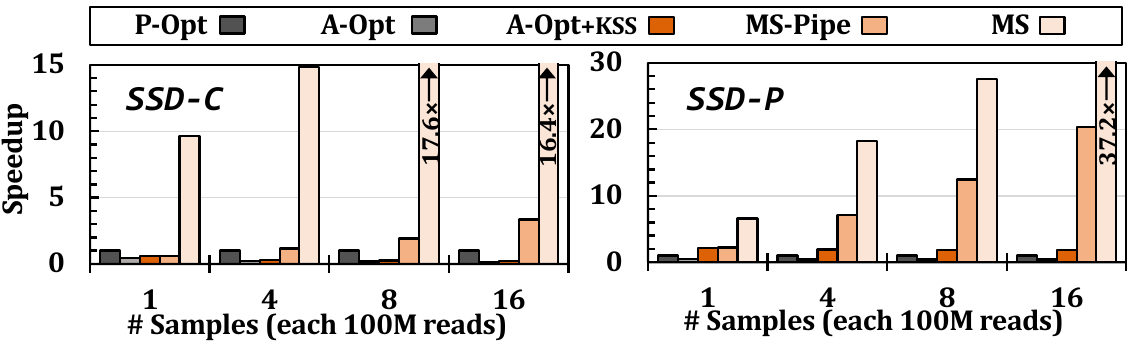}
\caption{\new{Speedup for multi-sample analysis.}}
\label{fig-mg:multi-sample}
\end{figure}

\fig{\ref{fig-mg:multi-sample}} shows speedup for \omiii{the} multi-sample use case \omiii{in which multiple samples need to access the same database} 
(\sect{\ref{sec-mg:mech-multi-sample}}). 
\omiii{We consider} 256-GB host DRAM in which we can buffer k-mers \hhl{from} 1--16 samples.
We show the performance of \omiii{\proposals{}} multi-sample pipelined optimization (as described in \sect{\ref{sec-mg:mech-multi-sample}}) in software (\mspipe) and in the full \proposal{} design (\msfull). In all configurations that require sorting \bback{(all except P-Opt)},
we use a state-of-the-art sorting accelerator~\cite{qiao2022topsort}.\footnote{\new{
We use the sorting throughput reported by the original paper~\cite{qiao2022topsort} and model the \omiii{data movement} time \omiii{between the sorting} accelerator \omiii{and other stages of \proposals{} pipeline}.}} 
First, \msfull achieves large speedup\omiii{s} of up to 37.2$\times$/100.2$\times$ \omiii{over} \popt/\aopt. Second, \mspipe leads to up to 20.5$\times$ (52.0$\times$) speedup \omiii{over} \aopt on \ssdc (\ssdp), and the speedup grows with the number of samples. 
\bback{We conclude that \omiii{\proposals{} pipeline optimization for the multi-sample use case in both software and hardware leads to large speedups over the baseline tools, and the hardware configuration leads to larger speedups compared to the software configuration by additionally leveraging ISP}}.

\subsection{Area and Power}
\label{sec-mg:eval-area}

\joel{Table~\ref{tab-mg:metastore_area_energy} shows the area and power consumption of \proposals{} \omii{hardware accelerator} units \new{at 300 MHz}.
While the\omii{se units} could be designed to operate at a higher frequency, their throughput is already sufficient since \proposal{} is bottlenecked by NAND flash read \omii{throughput}}.
\proposals{} hardware \omii{accelerator} area and power requirements are \omii{small}: only 0.04~mm$^2$ and 7.658~mW at 65~nm.
\irev{\omii{The accelerator} can be place-and-routed in a small space with 0.25mm $\times$ 0.25mm dimensions (0.0625 mm$^2$)}.
The area overhead of \omii{the accelerator} is  0.011 mm$^2$ at 32 nm,\footnote{We scale area to lower technology nodes using the methodology in~\cite{stillmaker2017Scaling}.}
which is 1.7\% of the three 28-nm ARM Cortex R4 cores~\cite{cortexr4} in a SATA SSD controller~\cite{samsung860pro}. 
\bback{\omii{While both the accelerator and the cores in the SSD controller can execute \proposals{} ISP tasks, the accelerator is} 26.85$\times$ more power-efficient than the cores}.

\begin{table}[t]
\centering
\vspace{0.5em}
\caption{Area and power consumption of \proposals{} logic.}
\label{tab-mg:metastore_area_energy}
\resizebox{0.85\columnwidth}{!}{%
\begin{tabular}{c|c|c|c}
\toprule
\textbf{Logic unit}                  & \textbf{\# of instances} & \textbf{Area [mm\textsuperscript{2}]} & \textbf{Power [mW]} \\ 
\midrule
\midrule
Intersect (120-bit)                 & 1 per channel              &    0.001361   &      0.284    \\
k-mer Registers (2$\times$ 120-bit)  & 1 per channel              &    0.002821   &      0.645    \\
Index Generator (64-bit)             & 1 per channel              &    0.000272   &      0.025    \\
Control Unit                         & 1 per SSD                  &    0.000188   &      0.026    \\\midrule
\textbf{Total for an 8-channel SSD}           & -                        & \textbf{0.04}    & \textbf{7.658} \\ \bottomrule
\end{tabular}
}
\end{table}

\icut{While\omn{We can remove this revision response since the topics are already covered in the previous sections.}\revid{\label{rev:B1.3}B1.3} \proposal{} introduces some changes to the system, its modifications are lightweight and we believe its large advantages justify the initial design effort. As discussed in \sect{\ref{sec-mg:mechanism}}, \proposals{} lightweight ISP functionality can even run on the existing embedded SSD cores (while still requiring an ISP-specialized firmware). At this low cost, \proposal{} not only improves performance, but it also improves cost-efficiency (\fig{\ref{fig-mg:ms-ce}}). This is critical because while sequencing cost has dropped significantly, computation cost has not been keeping up at the same rate\hhl{~\cite{berger2023navigating}}. Thus, \proposals{} contribution to high-performance, accurate, and cost-effective analysis paves the way for metagenomics’ broader adoption.}

\subsection{Energy}
\label{sec-mg:eval-energy}

\omii{We demonstrate the energy consumption of different \mganalysis tools by obtaining the energy of the host processor, the host DRAM, the accelerators, the SSD's internal DRAM, host/SSD communications, and the SSD accesses}. 
\bback{For each tool, we calculate the energy \omii{consumption of each part of the system} based on its active/idle power and execution time. 
We observe that} 
\proposal{} provides significant energy benefits \omii{over other software and hardware baselines} by alleviating I/O overhead and
reducing the burden \omii{of \mganalysis} in the system \omii{(the host processor and DRAM)}.
\bback{\rev{A}cross \omii{our evaluated} SSDs and datasets}, 
\proposal{} leads to 5.4$\times$ (9.8$\times$), 15.2$\times$ (25.7$\times$), \hm{and} 1.9$\times$ (3.5$\times$) average (maximum) energy reduction compared to \popt, \aopt, and the PIM-accelerated \popt \bback{~when finding species present in the sample}. \omiii{By eliminating the need to move the large databases outside the SSD, \proposal{} leads to I/O data movement reduction of 71.7$\times$ over \aopt and 30.1$\times$ over \popt and the PIM-accelerated \popt.}

\begin{figure}[h]
\centering
\includegraphics[width=\linewidth]{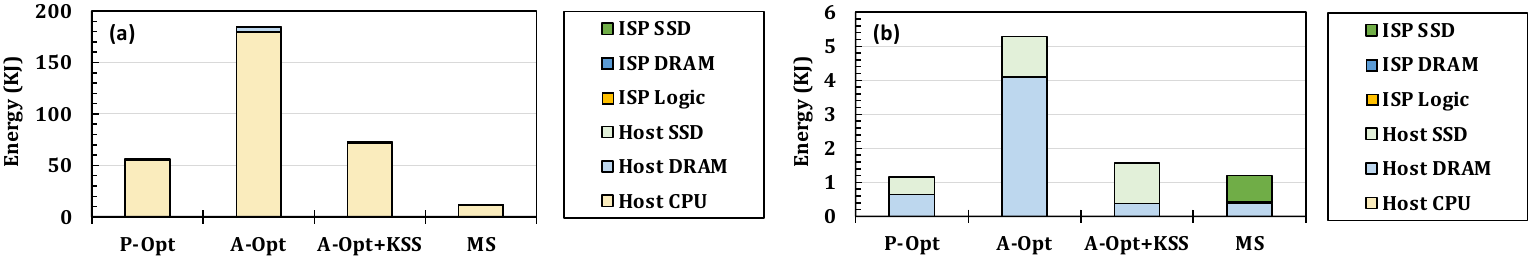}
\caption{\aooo{Energy breakdown of different systems (a) with and (b) without host CPU energy.}}
\label{fig-ms:energy-brkdn-ms-1}
\end{figure}

\aooo{\fig{\ref{fig-ms:energy-brkdn-ms-1}}(a) shows the energy breakdown of different systems for a representative input (CAMI-M). We show energy breakdown between the host CPU, host DRAM, host SSD accesses, ISP logic units, in-storage DRAM, and SSD accesses during ISP. We make two observations. First, by analyzing large amounts of data with lightweight logic units inside the storage system, \proposal{} leads to significantly lower energy consumption. Second, despite relying on a high-end host server node for computation, \aopt{}+KSS achieves large energy savings due to the shorter execution time of our proposed KSS approach and more efficient querying of the k-mer sketch database.}

\aooo{\fig{\ref{fig-ms:energy-brkdn-ms-1}}(b) shows the energy breakdown of different systems for a representative input (CAMI-M), excluding the host CPU, to isolate the energy distribution between main memory and storage accesses across configurations. This breakdown also serves as an idealized study of a scenario in which energy-efficient specialized hardware units replace general-purpose CPUs for computation. We make two observations. First, by analyzing data within the storage system, MegIS reduces the energy spent on accessing the storage system (please note that, as discussed in \sect{\ref{sec-mg:methodology}}, MegIS, \aopt, and \aopt{}+KSS use the same k-mer database size, which is much larger than the k-mer database used in \popt). Second, by more efficient access patterns to the k-mer sketch database in our proposed KSS approach, \aopt{}+KSS spends significantly less energy on host DRAM compared to \aopt.}

\section{Summary}
\label{sec-mg:conclusion}

We introduce \proposal{}, the \emph{first} \omii{in-storage processing system} designed to significantly reduce the data movement overhead of end-to-end metagenomic analysis.  To enable efficient \omii{in-storage processing} for metagenomics, we propose new 1) task partitioning,
\omiii{2})~data/computation flow coordination,
\omiii{3})~storage-aware algorithms,
\omiii{4})~data mapping, 
and \omiii{5)~lightweight in-storage accelerators}.  \omii{We demonstrate that \proposal{} greatly improves performance, energy consumption, \omiii{and system} cost efficiency at low area and power costs}.

\subsection{\tomiii{Impact and Influence}}

\tomiii{MegIS is published at the International Symposium on Computer Architecture (ISCA) in 2024~\cite{megis}. It is fully open-sourced at~\cite{megissource}. An extended version of the ISCA 2024 paper is available on arXiv~\cite{megisarxiv}.

As shown in \sect{\ref{sec-mg:eval-main}}, MegIS improves system cost-efficiency, while providing high performance and accuracy.
This is critical to enabling \inum{i}~wider adoption of metagenomic analysis and \inum{ii}~analysis on low-cost, portable devices, which is increasingly important due to the advances of compact portable sequencers~\cite{minion21,jain2016oxford,cali2017nanopore} for on-site metagenomics~\cite{pomerantz2018real, chiang2019from,mutlu2023accelerating,alser2022molecules}.

MegIS has already influenced several subsequent works (e.g.,~\cite{grains,mansouri2026sage,mansouri2026sagearxiv,chen2025reis,kabra2025ciphermatch}) in other areas of bioinformatics and other domains such as AI/ML. 

We hope that the key ideas and approaches introduced in MegIS inspire further research
in storage-centric designs in other data-intensive application domains in healthcare and life sciences, to facilitate their wider adoption and their execution on low-cost, portable devices.
}

\chapter{GRAINS: Storage-Aware Algorithm-Architecture Co-Design for Graph-Based Genomic and Metagenomic Analysis}
\label{chap:grains}

\renewcommand{\proposal}{GRAINS\xspace}

\newcommand\strings{\textsf{Strings}\xspace}
\newcommand\offsets{\textsf{Offsets}\xspace}
\newcommand\sizes{\textsf{Sizes}\xspace}
\newcommand\colors{\textsf{Colors}\xspace}

\renewcommand\todo[1]{}
\newcommand\cutcr[1]{}
\newcommand\omiiq[1]{}
\newcommand\omiiiq[1]{}
\renewcommand\omii[1]{{\color{black}{#1}}}
\renewcommand\omcr[1]{{\color{black}{#1}}}
\newcommand\ombarrier[0]{}

\section{Motivational Analysis}
\label{sec-gr:motivation}

We conduct experimental analyses to assess the impact of I/O overheads on graph-based genome analysis performance.

\subsection{Data Movement Overheads}
\label{sec-gr:motivation-dm}

\head{Tools and Datasets} We use two state-of-the-art tools for large genome graph analysis: \inum{i}~Fulgor~\cite{fan2023fulgor}, a node-centric tool, and \inum{ii}~the MetaGraph framework~\cite{karasikov2020metagraph,karasikov2022lossless,danciu2021topology,karasikov2019sparse}, an edge-centric tool. For each, we use the best-performing thread count and color encoding scheme. We evaluate k-mer set lookup, a fundamental task on genome graphs (\sect{\ref{sec:background-graph}}). 
Graphs are built from a pilot subsample of the global MetaSUB Consortium~\cite{danko2021global} dataset.
The resulting graph (with colors) is 659\,GB for Fulgor and 822\,GB for MetaGraph.\omiiiq{\tiny You asked whether we can evaluate even larger graphs. Constructing individual larger graphs is very challenging (needs much larger DRAM than our max 1.5TB DRAM nodes, and takes a very long time). But we can analyze databases of several graphs, which is also a common practice in reality. We already evaluate databases of multiple graphs in \sect{\ref{sec-gr:evaluation}}, but we can scale the overall size even further.\\}  
We evaluate queries with varying read counts, ranging from 100K reads\footnote{Note that smaller queries are also critical, as some applications (e.g., searching for antimicrobial resistance~\cite{danko2021global,karasikov2020metagraph,alipanahi2020metagenome,hunt2024allthebacteria} or for a single gene of interest~\cite{edgar2022petabase}) require searching a single or a few genes within a large graph.} to 10M reads (35 MB to 3.5 GB).
\sect{\ref{sec-gr:methodology}} provides further methodology details.

\head{System Configurations}
Our experiments run on a high-end server with an AMD EPYC 7742 CPU~\cite{amdepyc} and 1.5 TB of DDR4 DRAM~\cite{ddr4sheet}. DRAM capacity \emph{exceeds} the total size of all data accessed during the analysis. 
This ensures that \irevminor{we} capture the intrinsic I/O overhead of transferring large, low-reuse data from the storage system to main memory, without being constrained by memory capacity. We analyze
I/O overheads 
using two state-of-the-art SSDs: \inum{i}~\ssdm, with a PCIe Gen4 interface~\cite{samsungPM1735} and \inum{ii}~\ssdh, with a PCIe Gen5 interface~\cite{samsung9100PRO}.

\begin{figure}[h]
         \centering
         \includegraphics[width=0.8\columnwidth]{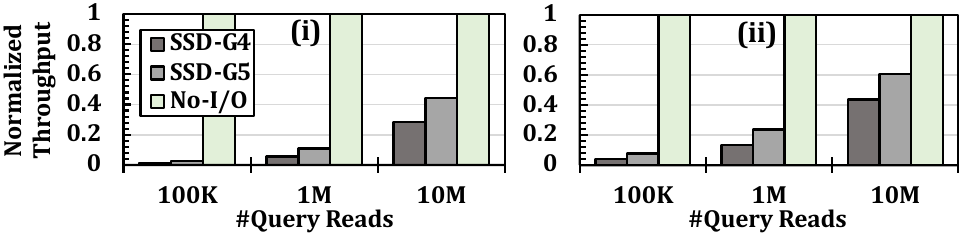}
         \caption{Normalized throughput of \inum{i}~Fulgor and \inum{ii}~MetaGraph under different storage configurations and query sizes.}
         \label{fig-gr:motivation1}
\end{figure}

\head{Observations}
\fig{\ref{fig-gr:motivation1}} shows throughput (\#queries/sec) of the tools, normalized to that of a hypothetical configuration with zero performance overhead due to storage I/O (\dram).  
We observe that in all cases, I/O leads to large overheads, even with the state-of-the-art SSDs. Compared to \ssdm (\ssdh), \dram leads to 16.7$\times$ (9.3$\times$) and 7.5$\times$ (4.5$\times$) better average performance in Fulgor and MetaGraph, respectively.

\begin{figure}[h]
         \centering
         \includegraphics[width=0.8\columnwidth]{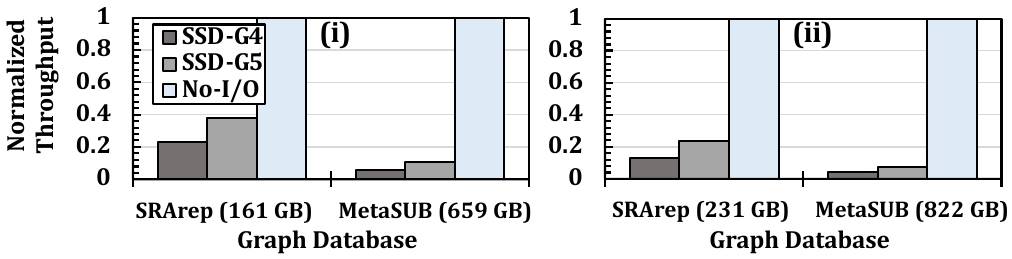}
         \caption{Normalized throughput of \inum{i}~Fulgor and \inum{ii}~MetaGraph under different storage configurations and graph sizes.}
         \label{fig-gr:motivation2}
\end{figure}

\fig{\ref{fig-gr:motivation2}} shows the throughput of the tools normalized to \dram, across different database sizes. The larger graph (G\_MetaSUB) is the one discussed earlier. The smaller graph (G\_SRArep) is generated from a representative subset of the Sequencing Read Archive's~\cite{katz2021sra} public portion. 
The resulting graph (with colors) is 161 GB for Fulgor and 231 GB for MetaGraph.  We observe that, in all cases, I/O overhead is substantial and increases with database size. For example, the speedup of \dram over \ssdh increases from 2.7$\times$ to 9.2$\times$ \omiii{(as the graph database size grows from 161 GB to 659 GB)} in Fulgor and from 4.3$\times$ to 13.4$\times$ \omiii{(as the graph database size grows from 231 GB to 822 GB)} in MetaGraph.

Based on these observations, we conclude that storage I/O leads to large overheads in graph-based genome analysis, an overhead that is expected to exacerbate in the future. This I/O overhead is due to moving large amounts of \emph{low-reuse} data from the storage system all the way to main memory, caches, and computational units. Due to its low reuse, caching this data in DRAM during analysis (even if DRAM is larger than the data size, as evaluated in our analyses) does not significantly amortize this overhead. While several works (e.g.,~\cite{\citegraphacc}) accelerate analysis with genome graphs, to our knowledge, they do not address storage I/O overheads. Although mitigating other bottlenecks (e.g., main memory or computation) leads to large benefits, \omiii{doing so does} not alleviate I/O overheads, whose impact on end-to-end performance becomes even larger as other bottlenecks are alleviated. 

This I/O overhead, due to moving large amounts of low-reuse data, is hard to avoid. One might think it is possible to avoid it by \inum{i}~producing smaller databases through sampling  (e.g.,~\cite{kim2016centrifuge,wood2019improved,muller2017metacache,song2024centrifuger,Dilthey2019,Fan2021}) or \inum{ii}~maintaining all required graph data completely and always resident in main memory. However, neither of these is suitable. The first approach necessarily reduces accuracy~\cite{berger2023navigating}, rendering it inapplicable for many use cases (e.g.,~\cite{\citegraphacc}). The second approach is energy-inefficient, unsustainable, costly, and unscalable for two reasons. First, the sizes of graph-based databases  (which are already large, i.e., tens to hundreds of TBs in some recent examples~\cite{karasikov2020metagraph,metagraphaws}) have been growing \omiii{rapidly}~\cite{katz2021sra,enastats,nhgri,stephens2015big}. Second, independent of the sizes of individual graphs, different analyses need different graphs, constructed from different sets of genomes and/or with varying parameters. For example, clinical studies may require graphs with patients' genomes kept separately for privacy~\cite{Berger2019}, while metagenomic studies and wastewater monitoring require highly diverse genome sets~\cite{gihawi2023major,song2024centrifuger,parkins2024wastewater,Blanco-Miguez2023}. Therefore, keeping all data for all possible analyses in DRAM at all times is prohibitively inefficient and ultimately unsustainable.

Several works (e.g.,~\cite{mansouri2022genstore,abakus23taco,megis,soysal2025mars}) have already demonstrated significant I/O overheads in non-graph-based genome analysis. While genome graphs enable a more compact representation of population-scale databases (thereby making it feasible to analyze databases that would be prohibitively large under traditional, linear representations), they still suffer from large I/O overheads. This is because the massive and continually growing scale of these databases still leads to very large graphs with low data reuse, and the graph topology introduces irregular, dependent access patterns, further reducing locality.

\vspace{-0.5em}
\subsection{\asp{Alleviating Data Movement Overheads}}
\label{sec-gr:motivation-ch-goals}

\subsubsection{Storage-Centric Computing (SCC)}
Processing data in the storage device can be a fundamental approach for reducing the I/O overheads for three reasons. First, SCC eliminates unnecessary data movement of low-reuse data across the system. Second, SCC reduces the computational burden imposed by low-reuse data on the rest of the system (e.g., main memory and computational units), \omii{such that these components can be used for other purposes or be turned off to save energy, \omiii{or can be made simpler or less costly~\cite{megis,boroumand2021google}}.} Third, as highlighted by many prior works (e.g.,\cite{li2023ecssd,mailthody2019deepstore,kang2021mithrilog,koo2017summarizer,mansouri2022genstore,wang2024beacongnn,jang2024smart,Kim2023optimstore,li2021glist}), SCC can leverage the high internal bandwidth of SSDs. In modern SSDs, the internal bandwidth often surpasses the external bandwidth. For example, a state-of-the-art SSD controller\cite{flashtecnvme5016} delivers 14~GB/s of external bandwidth and up to 57.6~GB/s of internal bandwidth (achieved via 16~channels, each with a peak of 3.6~GB/s). Over-provisioning internal bandwidth in modern SSDs is important to protect user-perceived external I/O performance, mitigating the effects of \inum{i} channel contention~\cite{nadig2023venice,kim2023decoupled,kim2022networked,tavakkol2014design} and \inum{ii} the SSD's internal data migration for management tasks~\cite{tavakkol2018flin,kim2020evanesco,cai2017error,park-dac-2019} and refresh~\cite{cai2013error,luo2018improving}. The effective internal bandwidth increases even further when processing directly in the NAND flash dies~\cite{chun2022pif,park2022flash,lee2025aif}. This is because in-flash processing exploits the aggregate bandwidth across all dies, a bandwidth that grows with the number of dies and significantly exceeds both \inum{i}~the bandwidth available to processing units on the SSD controller and \inum{ii}~the external SSD bandwidth.

\subsubsection{Challenges and Goal}
Designing a SCC system for graph-based genome analysis is challenging because none of the existing approaches can be directly implemented inside storage effectively due to modern SSDs' constrained hardware resources. Both node- and edge-centric representations of genome graphs require many random accesses during analysis. These random and irregular accesses cause costly contention in internal SSD components (e.g., channels and NAND flash chips~\cite{nadig2023venice,kim2022networked,tavakkol2018flin}), which hinder leveraging the SSD's internal bandwidth.
Therefore, directly adopting existing approaches inside storage leads to performance and energy overheads. 

\asp{Some SCC systems in other domains regularize accesses by performing sorting in either the SSD or the host. However, these approaches are not effective \omiii{for} graph-based genome analysis. Sorting the large number of accesses is impractical within the SSD due to resource constraints. Sorting on the host also fails to fully address access irregularity, as graph-based genome analysis involves dependent accesses. Coordinating sorting between the host and the SSD for each round of dependent accesses incurs significant data movement overhead.}

\textbf{Our goal} in this work is to improve the performance and efficiency of graph-based genome analysis by alleviating its data movement overheads from storage cost-effectively.

\section{\proposal}
\label{sec-gr:mech}

We propose \emph{\proposal}, a versatile storage-centric system for graph-based genome analysis. \proposal supports \omii{major} operations in analysis with genome graphs (e.g., k-mer set lookup and read mapping), 
and for alignment-based workflows, it flexibly integrates with existing alignment accelerators.
\proposal is primarily designed as a system to accelerate analysis with genome graphs. \proposal augments the existing SSD controller, flash dies, and FTL, and when not in the analysis acceleration mode, the SSD remains available for other applications, like a conventional general-purpose SSD.

We address the challenges of designing a SCC system for graph-based genome analysis via storage-aware algorithm architecture co-design based on our detailed examination of typical analysis pipelines that operate on genome graphs. To this end, we \inum{i}~make these pipelines more storage-friendly and \inum{ii}~further improve their performance, energy-efficiency, and cost-effectiveness via ISP and IFP. 

\proposal{}'s \omii{algorithm-architecture} co-design comprises three key aspects. \asp{First, we design \emph{\textbf{efficient batching and reordering techniques and pipelined execution flow}} specialized to the properties of genome graphs to reduce the number of random accesses to the graph}. 
\asp{Second, we devise \emph{\textbf{lightweight IFP units}} to process the needed parts of each page's graph data in the flash die. This IFP design avoids transferring low-reuse or unused parts of flash pages outside the die, thereby preventing bandwidth waste}.
\asp{Third, to fully exploit die-level parallelism during IFP, we design an \emph{\textbf{effective yet lightweight SCC scheduling technique}}, enabled by repurposing existing SSD structures. 
\proposal{}'s data placement (\sect{\ref{sec-gr:mech-ftl-ecc}}) drastically reduces the required in-DRAM mapping metadata, allowing us to use the freed-up internal DRAM space to maintain small per-die scheduling tables. 
We devise small ISP units on the SSD controller to interface with these tables and schedule IFP operations at low cost.}
\asp{To efficiently realize these aspects, while guaranteeing correctness and integrity, we apply simple changes to the FTL and adopt a lightweight ECC~\cite{lee2025aif}}.

\asp{Leveraging our optimizations, \proposal{}’s SCC steps require only simple operations and \omiii{small} buffer spaces, enabling execution on either \inum{i}~our lightweight, specialized IFP and ISP hardware units or \inum{ii}~general-purpose IFP (e.g.,\omiii{~\cite{chun2022pif,chen2024search,park2022flash,Chen2024aresflash}}) and ISP (e.g.,\omiii{~\cite{gu2016biscuit, kang2013enabling, wang2019project,acharya1998active,keeton1998case,riedel1998active,riedel2001active,merrikh2017high,tiwari2013active,tiwari2012reducing,boboila2012active,bae2013intelligent,torabzadehkashi2018compstor,kang2021iceclave,zou2022assasin,nadig2026conduit}}) units. Efficient genome graph analysis on both stems from \proposal{}’s specialized batching, scheduling, and data/computation flow coordination. Execution on special-purpose lightweight hardware leads to higher power efficiency (\sect{\ref{sec-gr:evals-area}}), while execution on existing \omiii{general-purpose} cores on the SSD controller or general-purpose \omiii{IFP and} ISP  units facilitates near-term adoption. Ultimately, choosing between these \proposal{} configurations is a design decision.}

\begin{figure}[h]
         \centering         \includegraphics[width=0.8\columnwidth]{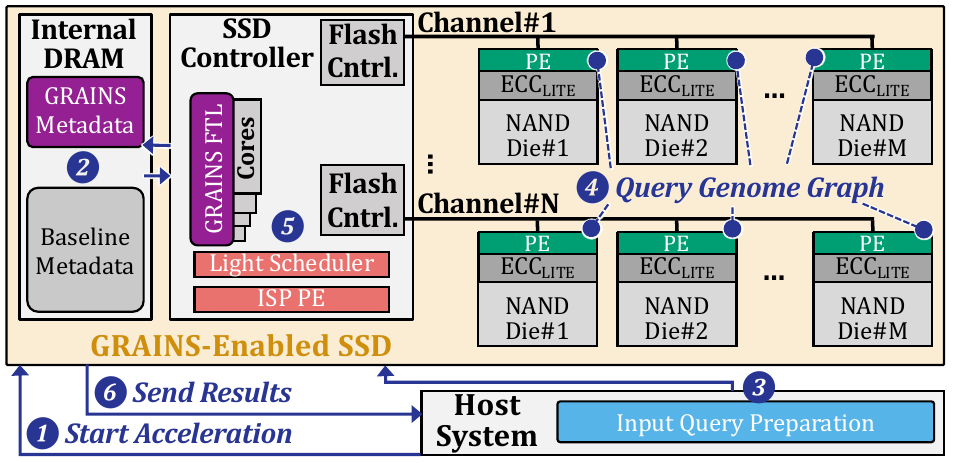}       
  \caption{Overview of \proposal.}
\label{fig-gr:grains-overview}
\end{figure}

\fig{\ref{fig-gr:grains-overview}} shows an overview of \proposal, with its lightweight IFP processing elements (PEs) on NAND flash dies and the lightweight ISP PE and scheduler on the SSD controller. \proposal FTL (\sect{\ref{sec-gr:mech-ftl-ecc}}) orchestrates host communications and data flow across the SSD components. 
After receiving a request through \proposal's interface commands (\sect{\ref{sec-gr:mech-interface}}) for accelerating graph-based genome analysis (\circled{1} in \fig{\ref{fig-gr:grains-overview}}), \proposal prepares (\circled{2}) for SCC execution by flushing the conventional FTL metadata and loading the \proposal FTL metadata (\sect{\ref{sec-gr:mech-ds-layout}}).
When execution starts, in the \emph{first step} (\sect{\ref{sec-gr:mech-step1}}), \proposal prepares input queries in the host via efficient batching and pipelined execution, and transfers the query batches to the SSD (\circled{3}). 
In the \emph{second step} (\sect{\ref{sec-gr:mech-step2}}), \proposal accesses graph data structures via its IFP (\circled{4}) and ISP (\circled{5}) units and sends the query results to the host (\circled{6}). These two steps are pipelined, with efficient coordination between resources to avoid writes during SCC. 
Through its efficient execution flow and by avoiding frequent management tasks (by not needing writes during SCC), \proposal effectively leverages the SSD's large internal bandwidth.

\section{Data Structures}
\label{sec-gr:mech-ds-layout}

\asp{\proposal uses a node-centric DBG representation based on an associative dictionary to compactly represent graph sequences and their associated colors. Similar to prior large-scale genome graph analysis tools (e.g.,~\cite{campanelli2025fast,fan2023fulgor,Pibiri2024macdbg,campanelli2024where}), we use SSHash~\cite{pibiri2022sparse} as our dictionary since, by exploiting the statistical features of k-mers,
it achieves a substantially better space-time trade-off over prior 
sequence dictionaries.\footnote{\asp{Since our techniques rely on general DBG properties, they are also applicable to other k-mer dictionaries that may be used as DBG backbones\omiii{~\cite{karasikov2020metagraph,fan2023fulgor,alanko2023themisto,rautiainen2020graphaligner,campanelli2025fast,Pibiri2024macdbg,campanelli2024where,Muggli2017SuccinctGraphs,turner2018integrating}}.}}}

\begin{figure}[h]
         \centering
         \includegraphics[width=0.83\columnwidth]{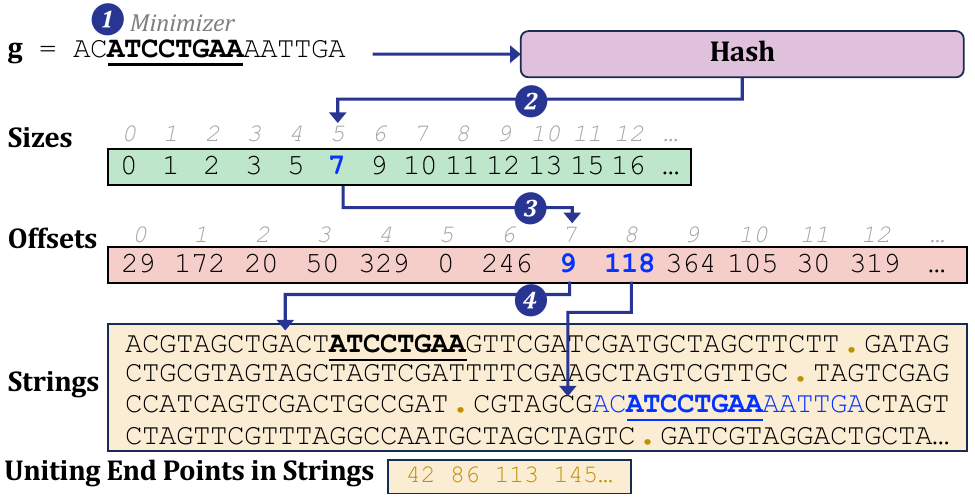}       
  \caption{Overview of the DBG structures and k-mer query.}
\label{fig-gr:example-sshash}
\end{figure}

\head{Graph Sequences} We store DBG's \emph{unitigs} (i.e., maximal non-branching paths) as contiguous
strings in a predetermined order, so that a k-mer occurring in any unitig can be quickly located using a minimal perfect hash function~\cite{pibiri2021pthash} built for the set of k-mers.\footnote{Note that in DBGs, edges between unitigs are represented implicitly via $(k-1)$-mer overlaps (\sect{\ref{sec:background-graph}}).}
\fig{\ref{fig-gr:example-sshash}} shows an overview of the DBG structures. The unitigs are stored in \emph{\strings}, and are indexed via \emph{\offsets} and \emph{\sizes}. Given a query read $R$, k-mers from the read are extracted and looked up.
Consider k-mer $g$. First, its \emph{minimizer} (i.e., the first $m$-mer with the smallest hash value) is extracted (\circled{1} in \fig{\ref{fig-gr:example-sshash}}). Second, with the minimizer's minimal perfect hash value ($h$), \sizes is indexed (\circled{2}). Third, with the \sizes values at locations $h$, \offsets is indexed (\circled{3}). $Sizes[h]$ from $Sizes[h+1]$ is then subtracted to find how many \offsets elements to read at 
$Sizes[h]$. Fourth, with the \offsets values, \strings is indexed (\circled{4}), \omiii{and} a window of length $k-m+1$~\cite{pibiri2022sparse} in \strings is checked to find the minimizer and check if the rest of the k-mer matches.

\aooo{We can easily convert the existing DBGs in the Graphical Fragment Assembly (GFA) format
to the SSHash-based format used in GRAINS. To this end, if only the canonical k-mers are
used during the construction of the GFA file, we can use the same procedure as done for
collections of sequences to generate an SSHash representation: we can extract the k-mers
encoded in the GFA, extract minimizers, calculate their hash values using SSHash, and encode
the offsets and strings accordingly. If instead the GFA file includes both orientations of some k-mers, an additional canonicalization and deduplication step is needed: we extract the k-mers,
canonicalize them, remove duplicates, and reassemble unitigs before proceeding with the same
SSHash construction.}

\begin{figure}[h]
         \centering
         \includegraphics[width=0.75\columnwidth]{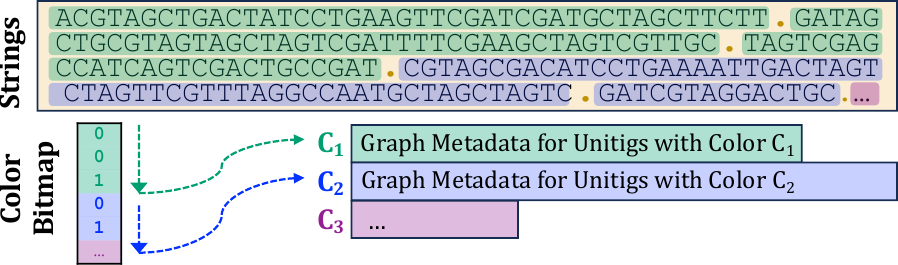}       
  \caption{Graph color \omiii{encoding used in \proposal}.}
  \vspace{-0.5em}
\label{fig-gr:colors}
\end{figure}

\head{Graph Colors} For storing graph colors, we adopt an approach similar to~\cite{fan2023fulgor}. As shown in \fig{\ref{fig-gr:colors}}, we sort the unitigs in \strings based on their colors (i.e., the ID pointing to the metadata for that graph unitig), and use a simple \emph{Color Bitmap} to mark the start of each new color. This encoding efficiently stores repeated colors at low \omiii{storage} overhead.    

\vspace{1em}
\section{Input Query Processing}
\label{sec-gr:mech-step1}

\asp{In this step, \proposal prepares reads for efficient graph queries in the next step (\sect{\ref{sec-gr:mech-step2}}). We perform two new optimizations inspired by the \omiii{characteristics} of genome graphs and queries.}

\asp{\head{Cross-Read K-Mer Batching} Instead of querying each read in a read set individually, we analyze k-mers across different reads together. This way, we can batch k-mers to reduce the number of random accesses to the graph. To enable this, \proposal populates a lightweight data structure associating each k-mer with the reads it originates from. When graph queries return the colors of matched k-mers, \proposal assigns colors to each read. We perform this step in the host since extracting k-mers from reads incurs frequent writes, which would reduce the SSD's lifetime if performed inside the SSD.}

\asp{\head{Genome-Graph-Aware Query Reordering} 
We leverage a common property of minimal-perfect-hash–based k-mer dictionaries in DBGs: 
The metadata required for identifying the small per-minimizer lookup ranges (i.e., the \sizes array, as shown in \fig{\ref{fig-gr:example-sshash}}) is much smaller than the graph’s \offsets and \strings.
Although the full graph is far too large and incurs large I/O overhead to access it in the host (as shown in \sect{\ref{sec-gr:motivation}}), \sizes occupies only a small fraction of the total space (e.g., $<$4\% in our large-scale graphs). This enables a powerful opportunity: We perform k-mer lookups on \sizes in the host and use the resulting \offsets indices to sort and batch k-mers before sending them to the SSD, thereby reducing the number of accesses and improving the access patterns to \offsets.\omiiiq{You asked "also Strings?"\\No, Strings needs other techniques, as discussed in the next sections.} To execute this step efficiently, as detailed below, we ensure that the time spent sorting the data and transmitting it to the SSD does not introduce significant overheads.}

\begin{figure}[h]
         \centering
         \includegraphics[width=0.93\columnwidth]{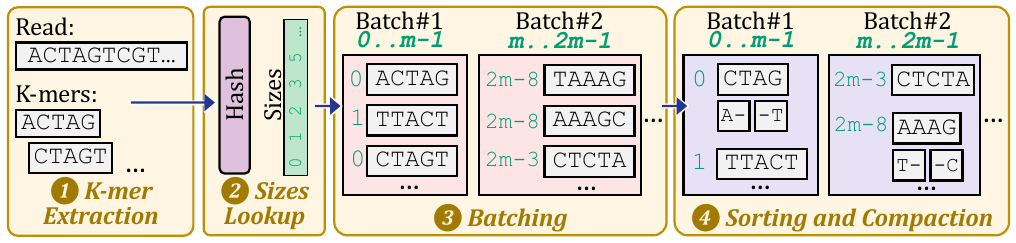}       
  \caption{Overview of \proposal's query processing.}
\label{fig-gr:grains-step1}
\end{figure}

\fig{\ref{fig-gr:grains-step1}} shows an overview of \proposal's query processing. \proposal starts by extracting k-mers from each read (\circled{1} in \fig{\ref{fig-gr:grains-step1}}) and looking them up in \sizes (\circled{2}). To reduce the overhead of sorting and transferring k-mers to the SSD, we take two measures. First, we propose a new k-mer processing scheme by improving upon k-mer processing in \cite{kokot2017kmc3,megis}. We \emph{partition the k-mers into batches corresponding to equal, disjoint ranges of \sizes values} (\circled{3}). Since these values serve as indices into \offsets, this partitioning enables an efficient pipeline: the sorting of one batch overlaps with the data transfer and \offsets lookups of the previous batch. Second, during sorting, we further compact k-mers by exploiting the fact that consecutive k-mers sorted by \sizes share the same minimizer (\circled{4}). Thus, we store the minimizer once and keep only the differences, thereby significantly reducing the transferred data size (e.g., by 2.3$\times$ on average across our evaluated datasets).

\section{Graph Query}
\label{sec-gr:mech-step2}

In this step, \proposal looks up the query k-mers in the graph and, for the k-mers that match a unitig, retrieves the associated colors. \proposal executes this step inside the SSD to benefit from its large internal bandwidth. In particular, \proposal performs IFP to avoid the need to transfer unnecessary parts of pages out of flash dies and to leverage multiple dies in parallel. 
To efficiently execute this step, \proposal needs to leverage this large internal bandwidth, without requiring expensive hardware resources in the SSD (e.g., costly logic units, or large internal DRAM capacity or bandwidth).
We demonstrate how \proposal achieves these when accessing genome graph \offsets (\sect{\ref{sec-gr:mech-offsets}}), \strings (\sect{\ref{sec-gr:mech-strings}}), and \colors (\sect{\ref{sec-gr:mech-colors}}). A lightweight FSM control unit on the SSD controller coordinates execution flow between these stages.

\subsubsection{Accessing Graph Offsets}
\label{sec-gr:mech-offsets}

\begin{figure}[h]
         \centering
         \includegraphics[width=0.9\columnwidth]{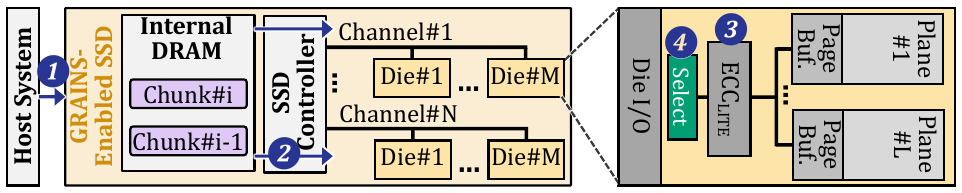}       
  \caption{Overview of \proposal{}’s \offsets lookups.}
\label{fig-gr:offsets-lookup}
\end{figure}

\fig{\ref{fig-gr:offsets-lookup}} provides an overview of \proposal{}’s \offsets lookups. \proposal receives \offsets indices and compacted k-mers from the host in chunks (\circled{1}). The internal DRAM holds two small chunks: one incoming from the host and one used for querying \offsets.
For an SSD
with 16 channels, 8 dies/channel, 4 planes/die, and 4-KiB pages,
\proposal needs two 2-MiB chunks.
\proposal reads the values from DRAM, maps them to physical addresses using its simple mapping (\sect{\ref{sec-gr:mech-ds-layout}}), and uses them to access \offsets (\circled{2}). Since the \sizes values are sorted by the host (\sect{\ref{sec-gr:mech-step1}}) and \offsets are uniformly placed across dies (as detailed in \sect{\ref{sec-gr:mech-ftl-ecc}}), the lookups lead to sequential accesses that fully exploit the internal bandwidth. 
After loading a page into a die’s page buffer, the lightweight ECC\textsubscript{LITE} scheme~\cite{lee2025aif} corrects errors (\circled{3}), after which the IFP unit selects and extracts only the targeted \offsets entry and sends it to the controller (\circled{4}), avoiding the transfer of full pages across the channel bus. The resulting values are cached in internal DRAM for subsequent \strings queries. Although full pages are read inside each die, only a small part of each page is written to internal DRAM, preventing the internal DRAM's bandwidth from becoming a bottleneck.

\subsubsection{Accessing Graph Strings}
\label{sec-gr:mech-strings}

Accesses to \strings must also fully exploit the SSD’s large internal bandwidth; however, unlike \offsets, the indices into \strings (i.e., the retrieved \offsets entries) arrive unsorted, preventing simple sequential streaming.

\head{Lightweight Scheduling} We introduce a lightweight scheduler that \inum{i}~avoids redundant page accesses, and \inum{ii}~maximizes die-level parallelism, without sorting accelerators in the SSD, which are inefficient, given the large number of elements and the limited bandwidth of internal DRAM\cite{zou2022assasin}.

\begin{figure}[h]
         \centering
         \includegraphics[width=0.98\columnwidth]{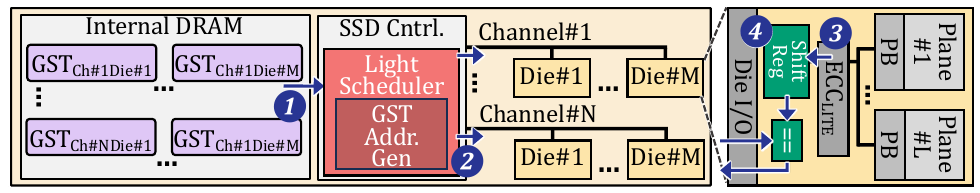}        
  \caption{Overview of \proposal's \strings lookups.}
\label{fig-gr:mech-strings}
\end{figure}

As shown in \fig{\ref{fig-gr:mech-strings}}, we design a small table, called \proposal Scheduler Table (\emph{GST}) for each die in the internal DRAM, with one row per \strings page. Each row stores the bit address within the page (as marked by \offsets entries), the k-mer minimizer and its suffixes/prefixes (\fig{\ref{fig-gr:grains-step1}}), a flag indicating whether the row is full (and if so, pointing to an extension table), and a small one-hot encoded bitmap marking the target plane in the die. This bitmap enables \proposal to issue \emph{multi-plane SSD operations}, allowing multiple planes of the same die to serve different accesses concurrently. 
When accessing \strings, \proposal schedules requests by simply reading one row per GST sequentially and issuing requests across all dies and planes in a round-robin manner. This naturally coalesces accesses to the same page, avoids redundant page accesses, and achieves high parallelism with minimal metadata and no complex logic.

\proposal can store GSTs in the internal DRAM because \inum{i}~its efficient address mapping (\sect{\ref{sec-gr:mech-ftl-ecc}}) frees most of the DRAM, and \inum{ii}~our compact k-mer representation (\sect{\ref{sec-gr:mech-step1}}) keeps GST entries small. Internal DRAM is sufficient even for large read sets and graphs (e.g., our large evaluated read set with $\sim$10 million query reads on a large-scale genome graph\footnote{Our evaluated graphs (\sect{\ref{sec-gr:methodology}}) are near the upper bounds of individual graph sizes. Graph-based databases that encode even larger volumes of data typically consist of several subgraphs that need to be queried. This is because graph construction time scales super-linearly, making construction of a single giant graph prohibitively long~\cite{Bowe2012SuccinctGraphs,pibiri2021pthash}.} requires only 2.9 GB, well within the 4-GB internal DRAM in a typical 4-TB SSD). Furthermore, to ensure efficient functionality even when usage exceeds the internal DRAM's capacity, \proposal keeps some \sizes batches (\sect{\ref{sec-gr:mech-step1}}) in host DRAM until the SSD finishes processing the current batch.

\fig{\ref{fig-gr:mech-strings}} shows an overview of \strings lookups. First, \proposal's scheduler reads \strings addresses and corresponding k-mers from each GST (\circled{1}) and sends the k-mers to the corresponding dies (\circled{2}). The scheduler performs simple increments to index the next row of GSTs. Second, \proposal reads the corresponding \strings page, and after ECC\textsubscript{LITE} (\circled{3}), the IFP unit performs a comparison (\circled{4}) between the \strings at the indexed window and the incoming k-mer from the controller. 
Finally, it sends the comparison result and the unitig IDs (for matching k-mers) back to the SSD controller.

\subsubsection{Accessing Graph Colors}
\label{sec-gr:mech-colors}

\begin{figure}[h]
         \centering
\includegraphics[width=0.98\columnwidth]{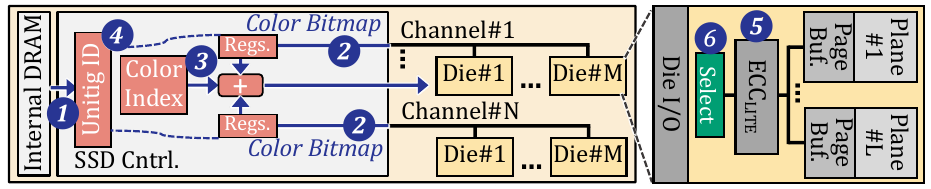}
  \caption{Overview of \proposal's Colors lookups.}
\label{fig-gr:mech-colors}
\end{figure}

\proposal retrieves the colors of the matched unitigs. Due to \proposal's efficient scheduling (\sect{\ref{sec-gr:mech-strings}}), unitig IDs are already retrieved in the internal DRAM in page order. Thus, given the underlying color encoding (\sect{\ref{sec-gr:mech-ds-layout}}), their colors can be obtained via efficient sequential scans of the Color Bitmap (\sect{\ref{sec-gr:mech-ds-layout}}).  \fig{\ref{fig-gr:mech-colors}} illustrates this. \proposal streams the next unitig ID into a small register (\circled{1}) while concurrently scanning the Color Bitmap from NAND flash dies (\circled{2}). Given that the bitmap is consumed immediately, \proposal{}'s ISP units directly operate on NAND flash streams (as in prior work~\cite{zou2022assasin}), without buffering them in the internal DRAM. This avoids bottlenecking the internal DRAM's bandwidth, which is especially important because this is the only stage that must read entire pages rather than small, selective portions. \proposal uses only two small, 32-bit registers per channel: one for the current and one for the incoming data. As the bitmap is processed, each time a ``\texttt{1}'' is encountered, \proposal increments \emph{Color Index} (\circled{3}). Once the bitmap position matches the unitig ID, \proposal uses the Color Index to index the \colors (\circled{5}). To access \colors, \proposal uses the same approach as \offsets (\sect{\ref{sec-gr:mech-offsets}}), where, after light ECC (\circled{6}), the IFP units select and transmit only the relevant portion of each page (\circled{7}).  This enables efficient, low-cost color accesses while fully leveraging the SSD’s internal bandwidth.

\section{FTL Integration and Reliability Support}
\label{sec-gr:mech-ftl-ecc}

\proposal FTL requires only minor changes to the baseline FTL. It designates each block as either genomic or non-genomic, and identifies genomic accesses through \proposal commands (\sect{\ref{sec-gr:mech-interface}}). For all non-genomic data, vendor-specific FTL features remain untouched, and the SSD behaves conventionally.  At the start of its SCC operations, since \proposal does not perform writes during them, it flushes the standard L2P metadata to the flash and loads the much smaller \proposal L2P metadata (as detailed below) into internal DRAM, while also keeping the other metadata of a conventional FTL. 

\head{Physical Allocation}
Since \proposal regularizes accesses, \proposal{}'s data placement is simple and enables \inum{i}~efficiently leveraging the SSD's internal bandwidth and \inum{ii}~minimizing mapping metadata. 
We uniformly distribute the blocks of the underlying graph data structures 
and the query reads across different channels, dies, and planes in a round-robin way. For efficient multi-plane operations, active blocks across different planes are aligned to the same page offset.

\head{Profiling-Guided Data Placement} \proposal{} adopts a uniform round-robin data placement across channels, dies, and planes. This data placement works synergistically with \proposal{}'s batching (\sect{\ref{sec-gr:mech-step1}}) and lightweight scheduling (\sect{\ref{sec-gr:mech-step2}}). Specifically, by uniformly distributing data across channels, dies, and planes, round-robin placement ensures that the sorted and batched \offsets queries produced by \proposal{}'s query processing translate directly into well-balanced, parallel accesses across all internal SSD resources. This uniform distribution is also what enables \proposal{}'s lightweight scheduler to \inum{i}~effectively merge accesses to repeated pages and \inum{ii}~efficiently leverage the SSD's internal bandwidth across all dies, without requiring workload-specific profiling. 

That said, for workloads with persistently skewed query distributions (e.g., when specific genomic regions are queried disproportionately often), profile-guided placement could offer additional load balancing benefits across SSD components. Such an optimization would be seamless to integrate with \proposal{}, as it would only affect the initial data layout without requiring changes to \proposal{}'s core batching, scheduling, or ISP/IFP logic. We consider this a promising future work.

\head{FTL Metadata Size} 
\proposal reduces the required FTL metadata size due to two reasons. 
First, instead of page-level L2P mappings (which, with 4 KiB pages, lead to $\sim$4~GB of metadata for a 4-TB SSD), \proposal stores L2P metadata at block granularity. This is possible because \proposal{} performs no writes during SCC, and thus, no fine-grained remapping is needed. For a 1-TB graph with 12-MB blocks, this requires only 0.35~MB. Second, \proposal{}'s uniform data placement makes physical locations computable from base addresses and stride, minimizing per-block metadata \omiii{needed for the SSD's reliable operations (e.g., read-disturbance counters)}. With per-block metadata, overall metadata size is 0.7~MB, freeing most of the internal DRAM for \proposal{}'s SCC and scheduling tables.

\head{Error Correction} 
For data accessed by \proposal{}’s ISP units on the SSD controller (e.g., Color Bitmap), accesses occur after the regular ECC. For structures accessed by \proposal's on-die hardware units, we adopt a lightweight on-die ECC technique, \omiii{\emph{ECC\textsubscript{LITE}}}, along with its associated \emph{Bias Error Encoding}~\cite{lee2025aif}.

\head{Other Management Tasks}
Since \omiii{in} SCC mode, \proposal does not incur writes,  it does not require write-related management such as GC and wear-leveling. 
\proposal performs other tasks for ensuring reliability before or after  SCC since 
\inum{i}~the duration of each \proposal process is significantly shorter than the manufacturer-specified threshold for reliable retention age (e.g., a year~\cite{micron3dnandflyer}), and
\inum{ii}~\proposal avoids read-disturbance errors\omiii{~\cite{cai2013error,cai_error_2012,cai2017error,cai2018errorsarxiv,cai-insidessd-2018,micheloni2010inside}} during SCC since, due to its efficient execution, each block is read at most once. After SCC steps, to prevent read disturb\omiii{ance} errors in the future, \proposal refreshes blocks whose read count exceeds a specific threshold.

\section{Interface Commands}
\label{sec-gr:mech-interface}

\proposal needs three new NVMe commands. 

\texttt{GRNS\_Start} initiates graph genome analysis acceleration. When \omiii{it receives} this command, \proposal prepares itself for \omiii{execution in} SCC mode (\sect{\ref{sec-gr:mech}}). During this phase, \proposal handles execution/data flow based on \proposal FTL and FSM controller.

\texttt{GRNS\_Steps} marks the start and end of each pipeline step and triggers \proposal{}'s lightweight FSM on the SSD controller, which drives the internal execution during SCC. 
The host uses this command to signal that a query batch is ready, then transfers the batch to the SSD's internal DRAM via the standard NVMe data path. Unlike conventional writes, these batches remain in the internal DRAM for processing and are not programmed to flash. The SSD then processes the batch through its internal pipeline with \proposal{}'s FSM on the SSD controller coordinating execution and data flow, and returns results to the host. This enables pipelining multiple batches: as the SSD processes the current batch, the host prepares the next one. No additional commands are needed because the FSM fully determines the internal execution flow once a batch arrives.  At the end of the last step, \proposal returns to operation as a conventional SSD.

\texttt{GRNS\_Write} is a specialized command for writing genome graphs to the SSD, which updates both the regular FTL’s L2P metadata and \proposal{}’s small L2P metadata.

\section{Deployment and Integration}

\proposal{} offers flexible deployment paths to minimize integration costs across both the hardware and software stacks.

\head{Flexible ISP/IFP Deployment} Due to our optimizations, \proposal{}'s ISP/IFP steps require only simple operations with small buffers (\sect{\ref{sec-gr:evals-area}}). Thus, they can execute on either our specialized, lightweight logic units or on general-purpose ISP/IFP (e.g.,~\cite{gu2016biscuit, kang2013enabling, wang2019project,acharya1998active,keeton1998case,riedel1998active,riedel2001active,merrikh2017high,tiwari2013active,tiwari2012reducing,boboila2012active,bae2013intelligent,torabzadehkashi2018compstor,kang2021iceclave,zou2022assasin,chun2022pif,chen2024search}). Specialized logic maximizes power efficiency (\sect{\ref{sec-gr:evals-area}}), while general-purpose ISP/IFP facilitates near-term adoption. Ultimately, choosing between these \proposal{} configurations is a design decision.

\head{Minimal and Non-Disruptive FTL Modifications} The required modifications to the FTL are small and non-disruptive, including simple changes to the baseline FTL, as described in \sect{\ref{sec-gr:mech-ftl-ecc}}. 
The FTL needs to support \proposal{}'s L2P mapping and garbage collection, which are lighter than the baseline (due to \proposal's data placement) and are already modeled in our evaluations by MQSim~\cite{tavakkol2018mqsim,mqsimsource} (\sect{\ref{sec-gr:methodology}}).
Most importantly, when \proposal{} is not operating in its SCC acceleration mode, the SSD operates identically to a standard, general-purpose SSD, which means standard storage deployments are intact.

\head{Seamless System-Level Integration} \proposal{} requires only three new NVMe commands for host-device communication (\sect{\ref{sec-gr:mech-interface}}). This lightweight protocol extension requires no physical interconnect modifications and, similar to prior works with added NVMe commands (e.g.,~\cite{lee2025aif,mansouri2022genstore,megis}), leverages standard vendor-specific NVMe extensions. Similarly, file system and driver modifications are strictly confined to identifying genome graph files and mapping these three commands. This ensures that software integration remains straightforward, low-cost, and non-disruptive to the existing storage stack.

\section{Methodology}
\label{sec-gr:methodology}

\newcommand\fg{\textsf{FG}\xspace}
\newcommand\mg{\textsf{MG}\xspace}
\newcommand\pim{\textsf{IdealAccMem}\xspace}
\newcommand\grn{\textsf{GRN}\xspace}
\newcommand\grnext{\textsf{GRN-Ext}\xspace}

\head{Performance} 
We design a simulator that models all of \proposal's components, including host operations, internal DRAM, accessing flash dies, in-storage hardware units (on flash dies and on the SSD controller), and SSD-host interfaces. We then feed the latency and throughput of each component to this simulator. Using this methodology, as also leveraged in prior SCC works (e.g.,~\cite{mansouri2022genstore,park2022flash,li2023ecssd,megis,Lee2024presto,mansouri2026sage})\todo{\tiny You recommended citing more. I need to look for more works with this methodology. OK to leave for arxiv?}, enables us to flexibly incorporate state-of-the-art system configurations in our evaluations. For \textbf{hardware-based} \omiii{components} (\omiii{e.g.,} when querying the graph):  We implement \proposal's ISP/IFP logic units in Verilog and synthesize them with a 22nm library~\cite{22gf} using the Synopsys Design Compiler~\cite{synopsysdc}. We use two state-of-the-art simulators, MQSim~\cite{tavakkol2018mqsim,mqsimsource} to model SSD's internal operations, and Ramulator~\cite{kim2016ramulator, ramulatorsource} \omiii{(new version introduced in \cite{luo2023ramulator,ramulator2source})} to model internal DRAM. For \textbf{software-based} \omiii{components} (e.g., when preparing queries), we measure their performance on a real system with 1.5-TB DRAM (in all experiments unless mentioned otherwise) and AMD$^\text{\textregistered}$ EPYC$^\text{\textregistered}$ 7742 CPU~\cite{amdepyc} with 128 physical cores. We use \ssdm~\cite{samsungPM1735} and \ssdh~\cite{samsung9100PRO} as described in \sect{\ref{sec-gr:motivation}} \omiii{and Table~\ref{table:SSD_config}} in our real system experiments. In our MQSim simulations for the ISP/IFP steps, we faithfully model these SSDs with configurations summarized in Table~\ref{table:SSD_config}. For the software baselines, we measure performance using the best-performing thread counts on \omiii{the same} real system \omiii{used for GRAINS's software-based components}.

\begin{table}[h]
\centering
\scriptsize
\caption{\omiii{Evaluated} SSD configurations.}
\begin{tabular}{@{\hspace{-0.5pt}}c@{\hspace{-0.05pt}}|c|c@{\hspace{-0.05pt}}}
\toprule
\textbf{Specification} & \textbf{SSD-G4} & \textbf{SSD-G5} \\ 
\midrule
\midrule
\textbf{General}       & \multicolumn{2}{c}{\begin{tabular}[c]{@{}c@{}}TLC NAND flash-based SSD\\ 4 TB capacity, 4 GB internal DRAM~\cite{lpddr4} \end{tabular}} \\ 
\midrule
\begin{tabular}[c]{@{}c@{}} \textbf{Bandwidth}\\ \textbf{(BW)} \end{tabular}  & 
\begin{tabular}[c]{@{}c@{}}PCIe Gen4 interface\omiii{~\cite{PCIE4}};\\ 7 GB/s sequential-read BW;\\ 1.2-GB/s channel I/O rate\end{tabular} & 
\begin{tabular}[c]{@{}c@{}}PCIe Gen5 interface\omiii{~\cite{PCIE5}};\\ 14.8 GB/s sequential-read BW;\\ 2.4-GB/s channel I/O rate\end{tabular} \\ 
\midrule
\textbf{NAND Config}  & 
\multicolumn{2}{c}{\begin{tabular}[c]{@{}c@{}}16 channels, 8 dies/channel, 4 planes/die
\end{tabular}} \\ 
\midrule
\bottomrule
\end{tabular}
\label{table:SSD_config}
\end{table}

To ensure correct composition between host-side and SSD-side timing (including pipelining), we explicitly model the interaction between stages. \proposal{}'s pipeline stages exchange data through well-defined chunk transfers across the host-SSD interface (as described in \sect{\ref{sec-gr:mech-step1}}), which we can faithfully model using the bandwidth and latency of the corresponding interfaces. We enforce stage dependencies via a producer-consumer abstraction; a stage can start only after the required data chunks are produced and transferred, while overlap between stages is naturally limited by the modeled communication bandwidth and device throughput. Doing so enables accurate timing composition of \proposal{}'s lightweight pipeline stages while faithfully accounting for the communication and synchronization costs between host and SSD.

\head{Area and Power}
For \proposal{}'s logic units, we use our Design Compiler synthesis results. 
SSD power is based on values of a Samsung 3D NAND flash-based SSD~\cite{samsung860pro}. DRAM power is based on values from a DDR4 model~\cite{ddr4sheet, ghose2019demystifying}. For the CPU cores, we measure power with AMD \textmu{}Prof~\cite{microprof}.

\head{Evaluated Systems} 
\asp{We evaluate three key tasks: \omiii{\inum{i}}~k-mer set lookup, \omiii{\inum{ii}} alignment-free read mapping, and \omiii{\inum{iii}} alignment-based read mapping. We evaluate the following tools for analysis with large-scale genome graphs:\footnote{\asp{As discussed in \sect{\ref{sec:background-graph}}, DBGs scale substantially better than other genome graph structures as the genetic diversity in the database increases. Other tools (e.g., VG~\cite{garrison2018vg} and minigraph~\cite{Li2020minigraph}) that use variation graphs as their graph structure do not scale as well and, hence, are not typically used in 
scenarios \omiii{with large-scale, genetically diverse databases}~\cite{karasikov2020metagraph,bvrinda2023efficient}.}}  
\inum{i}~Fulgor~\cite{fan2023fulgor} (\fg), as a state-of-the-art node-centric software tool (which uses SSHash~\cite{pibiri2022sparse}),
\inum{ii}~the MetaGraph framework~\cite{karasikov2020metagraph,karasikov2022lossless,danciu2021topology,karasikov2019sparse} (\mg) as a state-of-the-art edge-centric software tool; 
\inum{iii}~an \emph{idealized} hardware-accelerated baseline (\pim) where all genome graph query operations execute \emph{without} main memory overheads \omiii{(i.e., zero latency, infinite bandwidth, infinite capacity)}. This configuration can represent an \emph{ideal} PIM baseline as well. \pim still incurs the I/O overhead of bringing the large, low-reuse data from the storage system to main memory. Comparing against this baseline demonstrates the impact of storage I/O as a fundamental problem that persists even when main memory overhead is alleviated;
\inum{iv}~\proposal with all its optimizations, but with its lightweight hardware \omiii{as an accelerator} outside the SSD (\grnext), to show the benefits of \proposal{}'s optimizations without ISP/IFP, but with more storage-friendly execution/data flow. The lightweight hardware is connected to the host system via a PCIe interface with 16 GB/s bandwidth, and leverages the host DRAM to store its scheduling metadata (\sect{\ref{sec-gr:mech-strings}}). Similar to \proposal{}'s full implementation, \grnext performs input query processing on the host  (\sect{\ref{sec-gr:mech-step1}}), but instead of relying on its ISP/IFP logic units to query the graph (\sect{\ref{sec-gr:mech-step2}}), it uses the same lightweight \omiii{accelerator} logic units outside the SSD;
\inum{v}~\proposal{}'s full implementation, including ISP and IFP \omiii{(\grn)}.  
All these tools achieve the same accuracy since their underlying graphs are \emph{lossless} encodings of the database's k-mer sets~\cite{karasikov2020metagraph,fan2023fulgor,karasikov2022lossless,danciu2021topology,karasikov2019sparse}. 
For the alignment-based mapping, we integrate all tools with SeGraM, a state-of-the-art sequence-to-graph alignment accelerator~\cite{cali2022segram}.}

\head{Datasets}
We build graphs as \omiii{described} in \sect{\ref{sec-gr:motivation}}.
The graph (with colors) is 659 GB for \fg, \grn, and \pim, and 822 GB for \mg. We evaluate query read sets with 10M (QL), 1M (QM), and 100K (QS) reads. Note that smaller queries are also critical, as detailed in \sect{\ref{sec-gr:motivation}}.

\section{Evaluation}
\label{sec-gr:evaluation}

\subsection{Performance}

\begin{figure}[b]
    \centering
    \includegraphics[width=0.8\columnwidth]{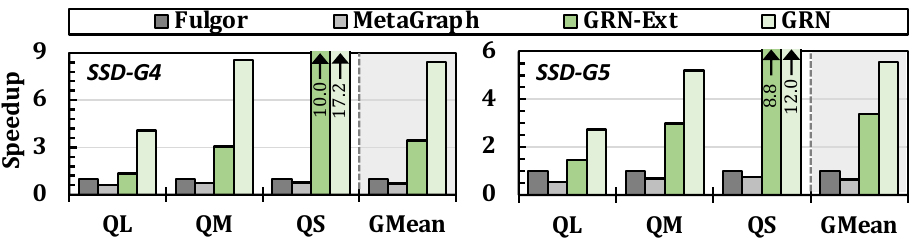}        
    \caption{Speedups with different input query sets and SSDs.}
    \label{fig-gr:grn-eval-main}
\end{figure}
\head{K-mer Set Lookup} \fig{\ref{fig-gr:grn-eval-main}} shows the speedups of different systems over \fg for k-mer set lookups. We make three key observations. First, \proposal provides significant speedups. In systems with \ssdm (\ssdh), \grn provides 8.4$\times$ (5.5$\times$) and 11.9$\times$ (8.5$\times$) average speedups over \fg and \mg, respectively. Second, by making the execution/data flow more storage-friendly, \proposal's implementation outside the SSD also provides large benefits. \grnext provides to 3.4$\times$ and 5.0$\times$ average speedup over \fg and \mg, respectively, across all inputs and SSDs. Third, \proposal's full implementation provides significantly larger benefits over \proposal's implementation outside SSD.
\grn~\omiii{provides}
2$\times$ average speedup over \grnext.
While \grnext exploits locality across queries by merging requests to the same regions of each graph structure and improving access patterns, the locality it captures cannot amortize the cost of transferring the large volume of low-reuse data across the storage-to-host interface. Thus, \proposal{}'s full implementation provides additional, significant benefits by alleviating \omiii{the large} I/O overheads through its SCC \omiii{design}.

\begin{figure}[h]
    \centering
    \includegraphics[width=0.85\columnwidth]{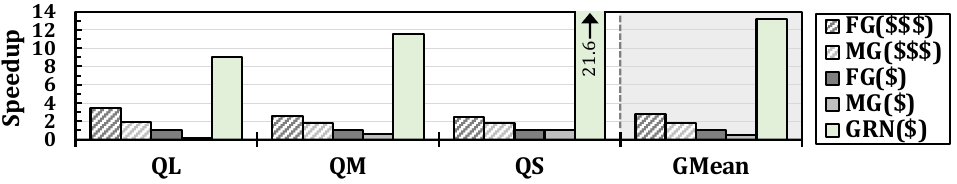}       
    \caption{Speedup of \proposal on a low-cost system, \omiii{i.e., \grn{}(\$)},
    over baselines on low-cost and performance-optimized systems.}
    \vspace{-0.5em}
    \label{fig-gr:grn-eval-cost}
\end{figure}

\head{Impact on Cost Efficiency}
\proposal improves system cost-efficiency since it analyzes large amounts of data in the SSD and does \emph{not} rely on either large host DRAM or high host-SSD interface bandwidth. We show this by evaluating two systems: a cost-optimized system (\$) with 64-GB \omiii{host} DRAM and \ssdm, and a costlier, performance-optimized system (\$\$\$) with 1.5-TB \omiii{host} DRAM and \ssdh. \fig{\ref{fig-gr:grn-eval-cost}} compares \proposal on the cost-optimized system with the baselines on both systems (with a representative query set, QM). To fairly evaluate the baselines when \omiii{host} DRAM is smaller than the graph, we reduce I/O overheads as much as possible in software for this scenario. To this end, 
we partition the graph such that each subgraph fits in the host DRAM, so random accesses to the graph do not repeatedly access the SSD. 
We partition the graph using 
database entry metadata
to guide partitioning choices, 
placing entries 
that likely share many k-mers
into the same partition.
Despite the benefits of this optimization, two sources of overhead remain: \inum{i}~the I/O overhead of transferring subgraphs,
and \inum{ii}~the need to query against each subgraph. 
We make two observations.
First, \proposal on the cost-optimized system significantly outperforms the baselines even on the performance-optimized system. \grn{}(\$) provides 4.7$\times$ and 5.2$\times$ average speedup over \fg{}(\$\$\$) and \mg{}(\$\$\$), respectively. 
Second, the performance of baselines on the cost-optimized system suffers significantly. 
When compared on the same cost-optimized system, \grn{}(\$) provides 13.2$\times$ and 26.9$\times$ average speedup over \fg{} and \mg{}, respectively.\omiiiq{GRN(\$\$\$) was evaluated in the previous figure (as GRN). I will add it to this fig and discussion as well in the arxiv version.} 
We conclude that \proposal improves \omiii{both} system cost-efficiency and \omiii{system} performance, which is critical for facilitating \inum{i}~the wide adoption of graph-based genome analysis, 
and \inum{ii}~analysis on low-cost, portable devices, a scenario rising in importance with the development of portable sequencing devices~\cite{MinIONMk1CO,palatnick2020igenomics,Ballard2018,Oehler2023} for on-site genome analysis~\cite{pomerantz2018real, chiang2019from,mutlu2023accelerating,alser2022molecules}.

\begin{figure}[t]
    \centering
    \includegraphics[width=0.88\columnwidth]{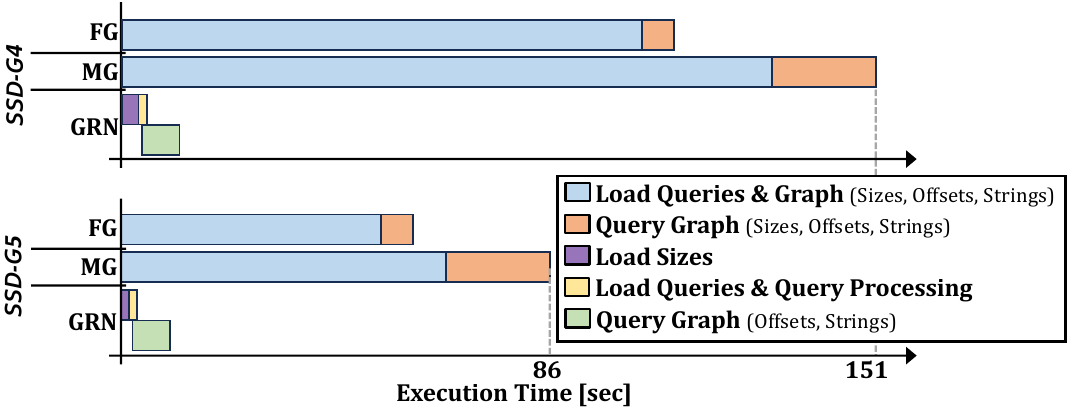}  
    \caption{Execution time breakdown.}
    \label{fig-gr:grn-eval-breakdown}
\end{figure}

\head{Execution Time Breakdown}
\fig{\ref{fig-gr:grn-eval-breakdown}} demonstrates the execution time breakdown of \fg, \mg, and \grn with different SSDs for a representative input query set (QM). In \fg and \mg, the graph and queries need to be first loaded from the storage system to main memory and processing units to be queried. In \grn, first, the \sizes data structure of the graph is loaded from the storage system \omiii{to the host DRAM}. Second, the input query processing step loads the input queries and batches them, as discussed in \sect{\ref{sec-gr:mech-step1}}. Third, as different batches of the input queries are processed, each small batch moves to the storage system in a pipelined manner to query the rest of the graph, without the need to move the data \omiii{(i.e., \offsets, \string, \colors)} outside the SSD, as discussed in \sect{\ref{sec-gr:mech-step2}}. Compared to \fg and \mg, \grn needs to move 31.4$\times$ and 39.1$\times$ less data outside the flash dies, respectively. This \omiii{reduction is due to} \proposal{}'s storage-friendly execution flow, analysis of large-scale graph structures where they originally reside, and efficient scheduling that enables leveraging die-level parallelism.

\newcommand\grnb{\textsf{GRN-B}\xspace}
\newcommand\grnbs{\textsf{GRN-B-S}\xspace}
\newcommand\grnbsscc{\textsf{GRN-B-S-SCC}\xspace}

 \begin{figure}[b]
    \centering
    \includegraphics[width=0.8\columnwidth]{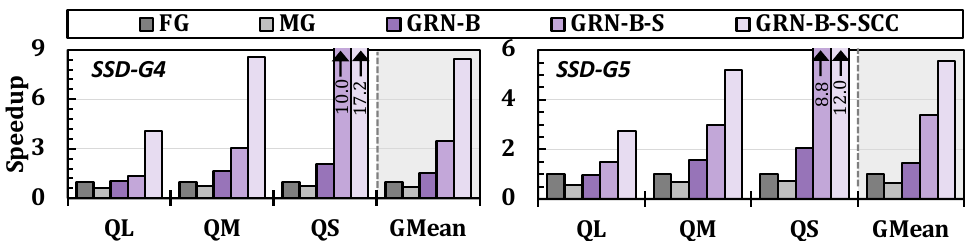}       
    \caption{Ablation tests of different \proposal optimizations.}
    \label{fig-gr:grn-eval-ablation}
\end{figure}

\head{Effect of Different \proposal Optimizations} \fig{\ref{fig-gr:grn-eval-ablation}} demonstrates the benefits of different \proposal optimizations. \grnb is a \proposal configuration including only the batching optimization (\sect{\ref{sec-gr:mech-step1}}). \grnbs is a \proposal configuration including both the batching and scheduling (\sect{\ref{sec-gr:mech-strings}}) optimizations (equivalent to \grnext, \sect{\ref{sec-gr:methodology}}). \grnbsscc is \proposal{}’s full configuration, including its SCC. Please note that we do not show a configuration with naive SCC and without batching and scheduling. This is because simply using SCC without these techniques leads to performance loss due to the irregular access patterns to genome graphs, 
\omiii{which are inefficient}
due to limitations of the SSD’s hardware \omiii{(e.g., costly conflicts in internal SSD resources such as channels and NAND flash chips~\cite{nadig2023venice,tavakkol2018flin,kim2022networked} during random and dependent accesses)}. We make three observations. First, \grnb provides 1.5$\times$ and 2.2$\times$ average speedup over \fg and \mg, respectively. This is due to more efficient accesses to the \offsets data structure. Second, \grnbs provides 2.3$\times$ average speedup over \grnb as it enables more storage-friendly accesses to \strings and \colors as well. Finally, through its efficient SCC, \grnbsscc provides 2.0$\times$ and 4.6$\times$ average speedup over \grnbs and \grnb, respectively.

\begin{figure}[h]
    \centering
    \includegraphics[width=0.8\columnwidth]{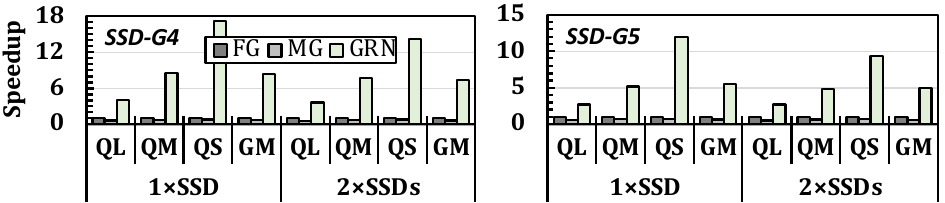}       
    \caption{Speedups with different numbers of SSDs.}
    \label{fig-gr:grn-eval-multi-ssd}
\end{figure}

\head{Effect of the Number of SSDs}
\proposal can benefit from multiple SSDs in two ways: First, different graphs can be stored on separate SSDs and analyzed concurrently, each benefiting from \proposal (as shown in \fig{\ref{fig-gr:grn-eval-main}}). 
Second, a single graph can be partitioned across SSDs to increase throughput. This is enabled by \proposal{}’s execution flow supporting efficient graph partitioning.
Since \proposal manages \offsets indices in the host, \offsets can be divided disjointly, each query being directed to the relevant SSD. \strings and \colors can also be partitioned: based on \offsets values, \proposal determines whether the referenced \strings entry resides locally or remotely.  
If remote, \proposal forwards values to the corresponding SSD. Given that \proposal{}'s IFP only selects and transfers relevant parts of \offsets, \proposal avoids read amplification. To further minimize communication overhead, \proposal batches \offsets values and maintains two 2-MB batches in the internal DRAM's freed-up space (\sect{\ref{sec-gr:mech-ds-layout}}) so that while one batch is transmitted, the other can be filled, thereby overlapping communication with SCC and efficiently utilizing internal and external bandwidths.
We evaluate this scenario in \fig{\ref{fig-gr:grn-eval-multi-ssd}}. We observe that 
\proposal maintains large speedups. This is because, as the external bandwidth increases for the baselines with more SSDs, the internal bandwidth also increases.
Although there is a slight decrease in speedup when moving to two SSDs in QS, the speedup is still significant (14.2$\times$ and 9.4$\times$ with \ssdm and \ssdh). This decrease is because, after increasing internal bandwidth, the impact of software query processing becomes relatively larger. Therefore, we can further accelerate these steps (\omiii{e.g.}, with a sorting accelerator) to increase benefits even further. 
We conclude that  \proposal can efficiently leverage multiple SSDs, making it also suitable for distributed systems.\omiiiq{We will add more SSDs for the extended version}

\begin{figure}[t]
    \centering
    \includegraphics[width=0.8\columnwidth]{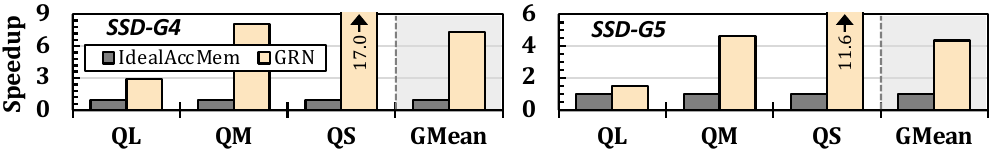}       
    \caption{Speedup over IdealAccMem.}
    \vspace{0.5em}
    \label{fig-gr:grn-eval-pim}
\end{figure}

\head{Comparison to IdealAccMem}  \fig{\ref{fig-gr:grn-eval-pim}} shows speedups over \pim. Despite \pim{}'s benefits, it still incurs I/O accesses to transfer data to main memory, even when DRAM is larger than data size (as in our evaluation). 
Through its storage-aware design, with efficient ISP/IFP operations, we observe that on the system with \ssdm (\ssdh), \proposal provides 7.4$\times$ (4.3$\times$) average speedup over \pim. 

\begin{figure}[h]
    \centering
    \includegraphics[width=0.8\columnwidth]{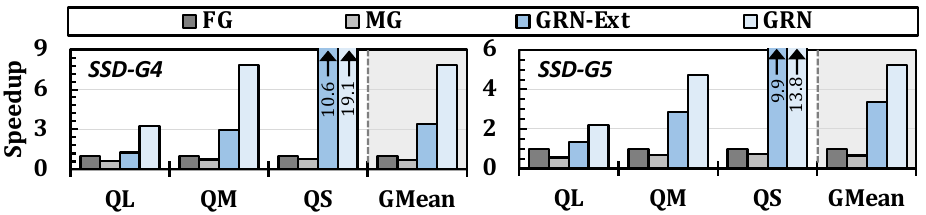}       
    \caption{Speedups for read mapping.}
    \label{fig-gr:grn-eval-mapping}
\end{figure}

\head{\asp{Alignment-Free} \omiii{Read} Mapping} \fig{\ref{fig-gr:grn-eval-mapping}}  shows speedups over \fg when performing \omiii{alignment-free} read mapping. \omiii{In this case, as discussed in \sect{\ref{sec-gr:mech-step1}}, \proposal sends the colors of all k-mers to the host system, and the host system determines the colors of each read based on the colors of its k-mers.} We observe that on systems with \ssdm (\ssdh), \grn provides 7.9$\times$ (5.2$\times$) and 11.2$\times$ (8.0$\times$) average speedups over \fg and \mg, respectively. 

\begin{figure}[ht]
    \centering
    \includegraphics[width=0.8\columnwidth]{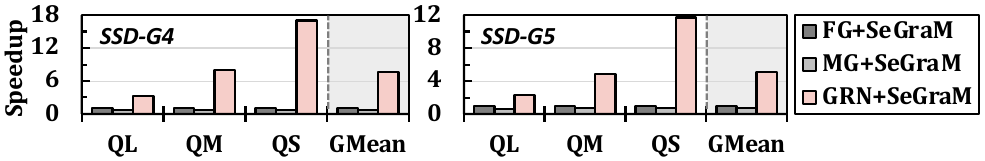}       
    \caption{Speedups for read mapping with alignment.}
    \label{fig-gr:grn-eval-alignment}
\end{figure}

\head{\asp{Alignment-Based Mapping}}
We integrate \grn, \fg, and \mg with SeGraM~\cite{cali2022segram} to perform alignment on the candidate graph regions identified by each tool (we use the throughput of alignment operations reported by \omiii{SeGraM}~\cite{cali2022segram}). \fig{\ref{fig-gr:grn-eval-alignment}} shows the speedups of different systems over \fg+SeGraM.
We observe that SeGraM+\grn provides 6.2$\times$ and 9.0$\times$ average speedup over SeGraM+\fg and SeGraM+\mg. 

\subsection{Area and Power \omiii{Overheads}}
\label{sec-gr:evals-area}
Table~\ref{table-gr:area-power} summarizes area and power for \proposal{}'s logic units at 333MHz. While these units can operate at higher frequencies, their throughput is already sufficient since \proposal's ISP/IFP pipelines are bottlenecked by flash reads. We use area/power values of ECC\textsubscript{LITE} from the original work~\cite{lee2025aif}.\footnote{Since we do not have access to the larger technology node used in \cite{lee2025aif}, we scale ECC\textsubscript{LITE}'s area to the lower node with the methodology in \cite{stillmaker2017Scaling}.}
\proposal{}'s hardware overhead is very small. On-controller units are 0.0025 mm\textsuperscript{2} (0.7\% of the four ARM cores~\cite{cortexr4} on an SSD controller). \proposal's on-die logic area is mainly (99\%) for ECC\textsubscript{LITE}, which, as reported by \cite{lee2025aif}, takes only 0.2\% of the flash chip area and fits well within its power budget.

\asp{As detailed in \sect{\ref{sec-gr:mech-step2}}, \proposal{}’s SCC can execute on either \inum{i}~our lightweight, specialized ISP/IFP units or \inum{ii}~general-purpose ISP (e.g.,~\cite{gu2016biscuit, kang2013enabling, wang2019project,acharya1998active,keeton1998case,riedel1998active,riedel2001active,merrikh2017high,tiwari2013active,tiwari2012reducing,boboila2012active,bae2013intelligent,torabzadehkashi2018compstor,kang2021iceclave,zou2022assasin}) and IFP (e.g.,~\cite{chun2022pif,chen2024search}). 
Choosing between these \proposal{} configurations is a design tradeoff: general\omiii{-purpose} units facilitate earlier adoption, while specialized units offer better area and power efficiency. For example, \proposal{}'s ISP units on the SSD controller consume two orders of magnitude lower area and 60.1$\times$ lower power than hardware units of a state-of-the-art general\omiii{-purpose} ISP.\footnote{\asp{Note that the power/area of \proposal{}'s IFP units are mainly dominated by ECC\textsubscript{LITE}, needed in both  \proposal and other IFP systems (e.g.,~\cite{chun2022pif, lee2025aif}).}}}

\begin{table}[t]
\centering
\caption{Area and power \omiii{consumption} of \proposal's logic units.}
\label{tab:metastore_area_energy}
\resizebox{0.45\columnwidth}{!}{%
\begin{tabular}{c|c|c}
\toprule
\textbf{Logic unit}                  & \textbf{Area [mm\textsuperscript{2}]} & \textbf{Power [mW]} \\ 
\midrule
\midrule
On SSD Controller                  &    0.0025    &      0.21    \\
On Die (ECC\textsubscript{LITE})   &    0.036    &      18.04    \\
On Die (Others)                    &    0.000093    &      0.01    \\\bottomrule
\end{tabular}
}
\vspace{-0.5em}
\label{table-gr:area-power}
\end{table}

\subsection{Energy} 
We find \omiii{the energy consumption of each evaluated graph-based genome analysis tool (when performing k-mer set lookups)} based on \omiii{the energy consumption of} different components (e.g., host processor, host DRAM, host-SSD communication, SSD components, and logic units). We calculate each component's energy based on its dynamic/idle power and execution time. 
We observe that across our evaluated input queries and SSDs, \proposal provides 4.4--21.0$\times$, 12.2--31.6$\times$, and 3.1--20.7$\times$ energy reduction over \fg, \mg, and \pim, respectively, when performing k-mer set lookups.\omiiiq{You mentioned the energy results are side-lined. I will add a figure to add more emphasis on them in the extended version. Please let me know if you suggest more actions.}

\begin{figure}[h]
    \centering
    \includegraphics[width=0.95\columnwidth]{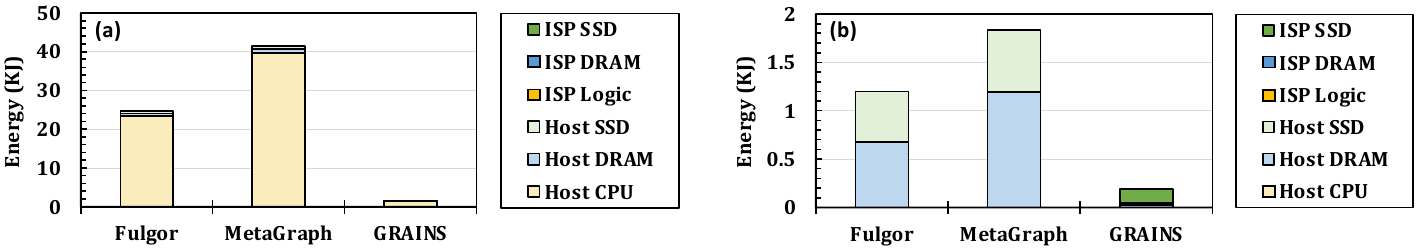}       
    \caption{Energy breakdown of different systems (a) with and (b) without host CPU energy.}
    \vspace{-0.5em}
    \label{fig-gr:grn-energy-breakdown1}
\end{figure}

\aooo{\fig{\ref{fig-gr:grn-energy-breakdown1}}(a) shows the energy breakdown of different systems for a representative query
set (QM). We show energy breakdown between the host CPU, host DRAM, host SSD accesses,
ISP logic units, in-storage DRAM, and SSD accesses during ISP. We observe that by analyzing
large amounts of low-reuse data using lightweight logic units inside the storage system and
reducing the overall execution time, GRAINS provides significant energy benefits compared
to FG and MG.}

\aooo{\fig{\ref{fig-gr:grn-energy-breakdown1}}(b) shows the energy breakdown of different systems for a representative query
set (QM), excluding the host CPU, to isolate the energy distribution between main memory
and storage accesses across configurations. This breakdown also serves as an idealized study
of a scenario in which energy-efficient specialized hardware units replace general-purpose
CPUs for computation. We make two observations. First, due to its improved access patterns,
GRAINS requires significantly fewer DRAM accesses, thereby reducing DRAM energy consumption. Second, through its efficient IFP, which eliminates the need to transfer unused or
low-reuse portions of each page outside the flash dies, GRAINS significantly reduces the storage system’s energy consumption.}

\vspace{-1em}
\section{Summary}
\label{sec-gr:conclusion}

We introduced GRAINS, the first storage-centric system designed to alleviate the I/O overhead of analysis with large genome graphs. Through detailed examination of these analysis pipelines, we \inum{i}~make them more
storage-friendly and \inum{ii}~enable efficient in-storage and in-flash processing. Our evaluations show that GRAINS improves performance, energy efficiency, and system cost-effectiveness of graph-based genome analysis at low area and power costs.

\vspace{-0.7em}
\subsection{\tomiii{Impact and Influence}}

\tomiii{GRAINS is recently published at the International Symposium on Computer Architecture (ISCA) in 2026~\cite{grains}. We plan to open-source GRAINS to facilitate future research. An extended version of the ISCA 2026 paper is available on arXiv~\cite{grainsextended}. 

As shown in \sect{\ref{sec-gr:evaluation}}, GRAINS improves system cost-efficiency and performance, which is critical for facilitating \inum{i}~the wide adoption of graph-based genome analysis, 
and \inum{ii}~analysis on low-cost, portable devices, a scenario rising in importance with the development of portable sequencing devices~\cite{MinIONMk1CO,palatnick2020igenomics,Ballard2018,Oehler2023} for on-site genome analysis~\cite{pomerantz2018real, chiang2019from,mutlu2023accelerating,alser2022molecules}. Wide adoption and portability are particularly critical factors in graph-based genome analysis due to key roles these analyses play in population-scale use cases. 

We hope GRAINS inspires further research into storage-centric designs across other data-intensive domains critical to population-scale settings (e.g., public health, precision medicine, and agriculture).
}

\chapter{SAGe: Algorithm-Architecture Co-Design for Highly-Compressed
Storage and High-Performance Access of Sequence Data}
\label{chap:sage}

\renewcommand\proposal{SAGe\xspace}
\renewcommand\proposals{SAGe's}

\section{Motivational Analysis}
\label{sec-sa:motivation}

\newcommand\pigz{\texttt{pigz}\xspace}
\newcommand\nspring{\texttt{(N)Spr}\xspace}
\newcommand\nsacc{\texttt{(N)SprAC}\xspace}
\newcommand\ideal{\texttt{Ideal}\xspace}

\nh{We show the impact of \nh{the} data preparation bottleneck in \sect{\ref{sec-sa:motivation-observations}} and discuss challenges of mitigating it in \sect{\ref{sec-sa:motivation-goal}}.}

\subsection{\nh{Data Preparation Bottleneck in Genome Sequence Analysis \oii{Accelerators}}}
\label{sec-sa:motivation-observations}

\nh{As shown in \fig{\ref{fig:intro-motivation}}, when genome sequence analysis is accelerated, data preparation emerges as a critical bottleneck. We perform experimental studies to understand the impact of this bottleneck when analyzing different real-world read sets.}

\head{Methodology} 
\omcr{We evaluate the end-to-end performance of \oii{a genome analysis} application, where execution includes both data preparation and genome analysis.}
For \emph{genome sequence analysis}, we use a state-of-the-art hardware accelerator~\cite{chen2023gem} for read mapping. 
For \emph{data preparation}, \nh{we consider the following configurations to decompress data into the desired uncompressed format:
\inum{i}~\pigz: A parallel version~\cite{adler2015pigz} of gzip, a commonly-used general compressor;
\inum{ii}~\nspring: Spring~\cite{chandak2018spring} and NanoSpring~\cite{Meng2023}, state-of-the-art \oii{software} compressors for short and long reads, respectively; and
\inum{iii}~\ideal: an idealized compressor with zero decompression time.} For our evaluated real-world read sets (\sect{\ref{sec-sa:methodology}}),
Spring and pigz achieve an average compression ratio of 16.9 and 5.4, respectively. 
\nh{We exclude other general-purpose compressors from our performance evaluations since, \nh{as detailed in \sect{\ref{sec:background-compression}}}, they achieve significantly worse compression ratios than genomic compressors, and thus do not align with the \oii{substantial need} to \oii{use} genomic compressors~\cite{hernaez2019genomic,dragenora}.
We do, however, include pigz \oii{(}parallel gzip\oii{)} since it is still widely used as a baseline comparison point in genomic compression research and literature.}

We evaluate \emph{end-to-end throughput} based on the throughput of data preparation and genome sequence analysis. We use the read mapping throughput reported by the original paper~\cite{chen2023gem}. 
For data preparation, we use a high-end server with 128 physical cores, as detailed in \sect{\ref{sec-sa:methodology}}. 
We use the best-performing thread count (i.e., after which adding more threads does not improve performance).
I/O operations (reading compressed data), decompression, and read mapping operate in a pipelined manner and in batches, which enables \emph{partial} overlapping \oii{of} these \oii{three} steps. Note that the decompressed data batch \oii{is} stream\oii{ed} directly to the \oii{read} mapper, without being written to the SSD. Further methodology details are in \sect{\ref{sec-sa:methodology}}.

\begin{figure}[h]
  \centering
  \includegraphics[width=0.85\columnwidth]{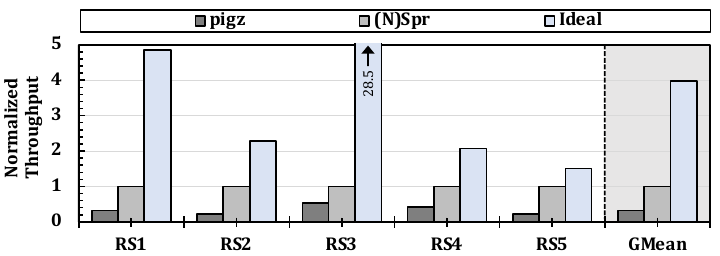}
  \caption{\nh{End-to-end throughput for different read sets.}}
  \label{fig:motivation-overhead} 
\end{figure}

\head{Observations} \fig{\ref{fig:motivation-overhead}} shows the end-to-end throughput normalized to the configuration with \nspring. 
\nh{We observe that data preparation greatly \oii{limits} end-to-end performance. \oii{If the data preparation bottleneck is eliminated, there would be 12.3$\times$ and 4.\oii{0}$\times$ average speedup for \pigz and \nspring.}}

\subsection{Challenges and Goal}
\label{sec-sa:motivation-goal}

\nh{\head{Requirements}} Our observations demonstrate that as accelerators dramatically speed up genome sequence analysis, data preparation emerges as a critical bottleneck in the execution pipeline. Note that storing read sets uncompressed is an inefficient and unsustainable way to mitigate this bottleneck, as they are typically 2-40$\times$ larger. Keeping ``hot'' data uncompressed is \emph{not} efficient either because genome analysis often involves many read sets that must be processed with the same priority (e.g., a cancer genomics study of whole-genome read sets can range from tens~\cite{drost2017use} to thousands of samples~\cite{weinstein2013cancer}, each around tens to hundreds of gigabytes compressed). As a result, even the ``hot'' data is too large to be stored efficiently uncompressed. 
Thus, it is common practice to store read sets compressed even when they are \emph{not} in an archival state, and to support integration with decompression (as seen in widely-used genome analysis software, e.g.,~\cite{li2018minimap2,berger2023navigating}).

\oii{Motivated by our observations, we conclude that} there is a critical need to effectively mitigate the data preparation bottleneck while meeting three key requirements: \inum{i}~\emph{high performance and energy efficiency}, \inum{ii}~\emph{high compression ratios}, comparable to state-of-the-art genomic compressors, and \inum{iii}~being \emph{lightweight} to enable seamless integration with a broad range of genome sequence analysis systems  (e.g., in FPGAs, GPUs, ASICs, portable devices, or NDP).

\head{Challenges} \nh{Meeting all three requirements is challenging.} This \oii{challenge is even larger in} genome analysis systems \oii{used} in hardware resource-constrained environments. Examples include portable genomics devices (e.g.,\oii{~\cite{MinIONMk1CO,palatnick2020igenomics,Ballard2018,Oehler2023,watsa2020portable,Quick2016,wang2021nanopore}}), which are critical to facilitating the wide adoption of genomics, or various NDP accelerators (e.g.,\oii{~\cite{wu2021sieve,shahroodi2022krakenonmem,shahroodi2022demeter,dashcam23micro,hanhan2022edam,zou2022biohd, cali2020genasm, huangfu2018radar, khatamifard2021genvom, gupta2019rapid, li2021pim, angizi2019aligns, zokaee2018aligner,Zhang_2023_alignerD,mansouri2022genstore,abakus23taco,megis,jun2016storage,soysal2025mars,kim2025nmp,zheng2025storage,cali2022segram,kim2018grim,kaplan2020bioseal,mao2022genpip,angizi2020pim}}), which lead to large benefits by mitigating data movement overheads of analyzing large-scale genomic data, but are typically implemented in constrained settings (e.g., within the memory system or storage device).

Some works accelerate computational kernels, such as BWT~\cite{qiao2019fpga,guo2013gpu,zhao2017streaming}, FM-index search~\cite{jiang2021exma,arram2015fpga}, or LZMA~\cite{chen2023efficient, leavline2013hardware}, which are widely used in  genomic compressors (e.g.,\ov{~\cite{lan2021genozip,Meng2023,kowalski2019pgrc,chandak2017compression,chen2023efficient}}). 
Despite their benefits, these works \oii{have two limitations. First, they have} high demands for resources like DRAM bandwidth~\cite{zhao2017streaming,wang2018accelerating,dragenora}, on‑chip buffers~\cite{qiao2019fpga}, DRAM capacity~\cite{jiang2021exma}, or compute resources~\cite{arram2015fpga,guo2013gpu,leavline2013hardware}. Efficiently meeting these \oii{hardware resource} demands is challenging, particularly in resource-constrained environments.\footnote{While some genomic (de)compression algorithms\ovi{~\cite{rajarajeswari2011dnabit,saada2016dna}} do not rely on expensive resources, they achieve poor compression ratios, averaging 5.3$\times$ lower than state-of-the-art genomic compressors~\cite{chandak2018spring,Meng2023} on our datasets.} \oii{Second, these works} accelerate only specific kernels and not the \oii{end-to-end} genomic decompression process (e.g.,~reconstructing the full reads from the consensus \oii{sequence} and mismatches). Thus, they do not fully mitigate the data preparation bottleneck (as shown in \sect{\ref{sec-sa:evals}}).

\nh{We use an example to elaborate on the limitations of \oii{resource-intensive data preparation techniques when} integrated with genome analysis systems in resource-constrained environments.}
Consider NDP genome analysis systems implemented inside the resource-constrained environment of the SSD (e.g.,\oii{~\cite{mansouri2022genstore,abakus23taco,megis,jun2016storage,zheng2025storage}}). To benefit from such systems, it is essential to efficiently perform data preparation inside the SSD, since moving the data outside the SSD for preparation \emph{completely undermines} the fundamental benefits of in-storage NDP. Unfortunately, 
performing large amounts of random accesses as needed for data preparation (\oii{matching patterns in large data
structures during decompression}) in the SSD  is inefficient. \oii{This is} due to costly \oii{resource contention within} the SSD (e.g.,~in channels and high-latency NAND flash chips\oii{~\cite{nadig2023venice,tavakkol2018flin,kim2022networked,cho2024aero,kim2025lazy}}). Although modern SSDs have an internal low-latency DRAM buffer, its capacity is small (e.g., 4 GB for a 4-TB SSD~\cite{samsung860pro}),
with over 95\% of it filled with mapping metadata, and its bandwidth is constrained by its \emph{single} channel~\cite{zou2022assasin}. 
\nh{Our motivational analysis (\sect{\ref{sec-sa:motivation-observations}}) emphasizes the \oii{mismatch} between the \oii{available resources} and data preparation's resource demands:} 
across our datasets, we observe that the state-of-the-art genomic decompressors~\cite{chandak2018spring,Meng2023} require random accesses to large amounts of data (up to 26 GB) and with high bandwidth. \oii{Consequently,} even on a high-end system \oii{used in our analysis}, with \emph{eight} DRAM channels, \oii{128 cores, \oiii{and 256 hardware threads,}} the performance of \oii{these genomic decompressors} saturates after \oii{32 threads} due to insufficient main memory bandwidth.

\nh{\textbf{Our goal} is to mitigate the data preparation bottleneck while achieving high performance and energy efficiency, high compression ratios, and a lightweight design. Doing so \oii{would} unlock the full potential of genome sequence analysis acceleration.}

\section{\proposal: Overview}
\label{sec-sa:mech-overview}

We propose \textbf{\proposal}, an algorithm-architecture co-design for highly-compressed \textbf{\underline{s}}torage and high-performance \textbf{\underline{a}}ccess of large-scale \textbf{\underline{ge}}nomic \nh{sequence} data.
\proposal{}
\inum{i}~mitigates the data preparation bottleneck,
\nh{\inum{ii}~achieves high compression ratios, and
\inum{iii}~is lightweight for seamless integration \nh{with a broad range of genome analysis systems}.}
\nh{\proposal is versatile, supporting data from different sequencing technologies and species.} 
\nh{\proposal's approach is based on the \textbf{key insight} that the information encoded in genomic sequence data follows specific trends, shaped by factors such as sequencing technology (e.g., error rates and read lengths) and common genetic phenomena (e.g., typical spatial distributions of genetic variations within genomes). By carefully taking these trends into account when synergistically co-designing algorithms and hardware, \proposal achieves high compression ratios, comparable to state-of-the-art genomic-specific compressors, while enabling \oiii{low} decompression \oiii{latency} using only lightweight hardware and efficient streaming accesses.} \proposal's synergistic co-design is the key enabler of SAGe’s lightweight, high-performance, and energy-efficient design.

\proposal's co-design comprises four aspects. 
First, during compression, \proposal encodes \nh{the} information of reads 
in hardware-friendly, lightweight \emph{data structures} 
and 
exploits genomic data properties in each read set to losslessly \emph{compress}  these structures \textbf{(\sect{\ref{sec-sa:mech-alg}})}.  
Second, we design \emph{lightweight hardware units} \textbf{(\sect{\ref{sec-sa:mech-hw}})} to efficiently interpret and decompress data. 
Third, we design an \emph{efficient data layout} \textbf{(\sect{\ref{sec-sa:mech-ftl}})} to leverage the storage system’s full bandwidth when accessing genomic data. Fourth, we design \emph{specialized interface commands} \textbf{(\sect{\ref{sec-sa:mech-interface}})} that are exposed to genome analysis applications \oiii{to access data and communicate with \proposal{}'s hardware to decompress the data to the
desired format.}

\begin{figure}[h]
  \centering
  \includegraphics[width=\columnwidth]{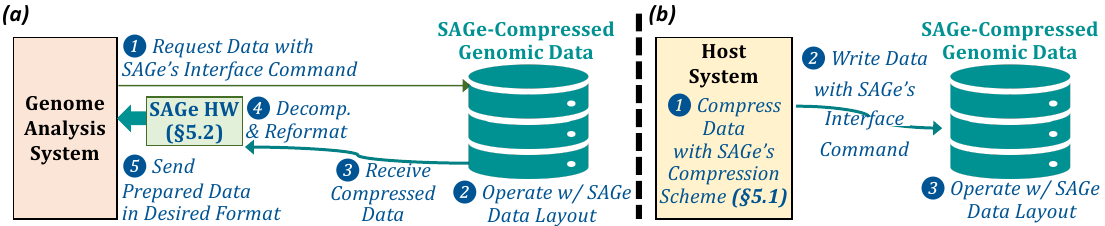}
  \caption{\oiii{High-level overview of \proposal{}'s (a) data preparation and (b) data compression and storage.}}
  \label{fig:sage-hl-overview} 
\end{figure}

\fig{\ref{fig:sage-hl-overview}(a)} shows \oiii{a} high-level overview of \proposal{}'s data preparation for a genome analysis system. When the genome analysis system requests data using \proposal's interface commands (\circled{1} in \fig{\ref{fig:sage-hl-overview}(a)}), \proposal starts operating based on its \oii{data layout} (\circled{2}).  
\proposal receives compressed data from the storage device (\circled{3}), decompresses it into the desired format using \proposal's hardware  (\circled{4}), and feeds it to the genome analysis system (\circled{5}). \proposal's lightweight design enables it to efficiently integrate with a broad range of genome analysis systems. In \sect{\ref{sec-sa:integration}}, we demonstrate case studies of \oiii{three ways \proposal can} integrate with different genome analysis systems. 

\fig{\ref{fig:sage-hl-overview}(b)} shows the high-level overview of how \proposal handles the compression and storage of genome sequence data. 
Since compression is not on the critical path of genome sequence analysis, we perform it on the host system.
The host compresses genome sequence data based on \proposal{}'s compression scheme (\circled{1} in \fig{\ref{fig:sage-hl-overview}(b)}). When writing the \proposal-compressed data, \proposal leverages its customized interface commands (\circled{2}) and data layout to facilitate efficient accesses later, during decompression (\circled{3}).

\section{\proposal: Detailed Design}
\label{sec-sa:mech-detailed-design}

\subsection{Compression Algorithm and Data Structures}
\label{sec-sa:mech-alg}

\newcommand\npaa{MMPA\xspace}
\newcommand\npb{MMPGA\xspace}

\begin{figure}[h]
  \centering
  \includegraphics[width=0.85\columnwidth]{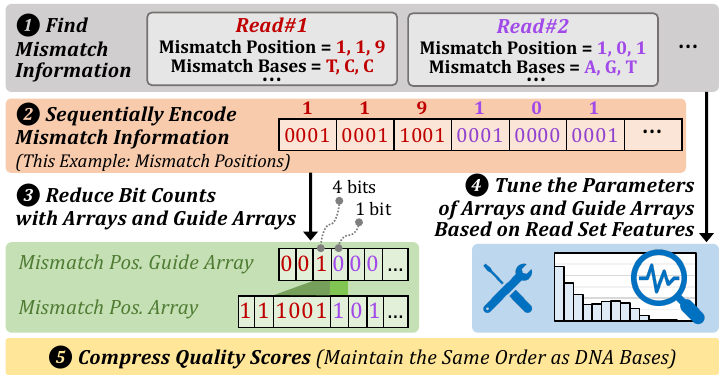}
  \caption{Overview of \proposal's compression mechanism.}
  \label{fig:overview} 
\end{figure}

\fig{\ref{fig:overview}} shows an overview of \proposal's lossless compression technique and its hardware-friendly encodings. 
\proposal{} \oiv{exploits} the consensus-based approach (\sect{\ref{sec:background-compression}}) widely used in genomic compression (e.g.,~\cite{chandak2017compression,dufort2021renano,kokot2022colord,kowalski2019pgrc}), storing each read's mismatch information relative to a consensus sequence (\circled{1}).
Similar to existing genomic compressors, \proposal identifies the mismatches during compression by mapping reads to the consensus sequence.\footnote{Note that this is independent from and does not eliminate the need for read mapping in the later genome analysis stages~\cite{Siren2024,Sedlazeck2018}.}
The key difference between \proposal's compression technique and existing genomics-specific compression technique\oiv{s} is the encoding of mismatch information: instead of using expensive backend general-purpose compressors used in genomic-specific compressors (e.g.,\oiv{~\cite{chandak2018spring,Meng2023,lan2021genozip,kowalski2019pgrc,chandak2017compression,chen2023efficient}}), which require costly
computational units, large buffers, or large DRAM bandwidth
or capacity (e.g., due to many random accesses to large data
structures for matching patterns),
\proposal \inum{i}~stores mismatch information in data structures that can be efficiently decoded by lightweight operations and streaming accesses, and \inum{ii}~minimizes their sizes by leveraging read set properties.

To this end, \proposal sequentially stores \oiv{different components of the} mismatch information of \oiv{all} reads in \oiv{the entire} read set contiguously in \oiv{three} arrays. \oiv{These arrays are dedicated to storing different parts of the reads' mismatch information, i.e., \inum{i}~mismatch positions, \inum{ii}~mismatch bases and types, and \inum{iii}~matching positions. \fig{\ref{fig:overview}} shows an example of the array used for storing mismatch positions (\circled{2}).} 
\proposal then adapts the bit width of each array entry and uses a \emph{guide array} to \oiv{indicate the number of bits used for each array entry} (\circled{3}). 
This is effective since the mismatch information in genomic datasets tends to follow specific trends, and by using a limited set of bit widths tailored to these trends, \proposal can significantly reduce data size. 
This approach is lossless since the arrays and guide arrays store a complete and exact representation of all mismatch information, which allows for the perfect reconstruction of each original read's DNA bases from the consensus sequence during decompression.
\nh{In the rest of this section, we discuss how sequencing technology and genetic phenomena influence the trends in each component of mismatch information, i.e., mismatch positions (\sects{\ref{sec-sa:mech-noise-pos-count}), mismatch bases and types (\ref{sec-sa:mech-bases-types}), and matching positions (\ref{sec-sa:mech-map-pos}}). Since these trends vary for each read set, during compression, \proposal tunes the parameters of arrays and guide arrays for \emph{each read set} (\circled{4}). 
The parameters are then encoded at the beginning of the compressed file.} 

The design of the arrays and guide arrays allows them to be interpreted via streaming accesses.
This enables \proposal to reconstruct reads using only lightweight operations during decompression.
As a result, \proposal eliminates the need for expensive hardware resources or frequent random accesses.

\nh{\proposal also performs lossless compression of quality scores (\circled{5}). This feature is optional and can be disabled by the user. This is because some accurate sequencers (e.g.,~\cite{fukasawa2020longqc}) do not report quality scores and print placeholders for format compatibility. 
Given the accuracy improvements in modern sequencing \oiv{technologies}, many workflows (e.g.,~\cite{li2018minimap2,wood2019improved,song2024centrifuger,kolmogorov2019assembly,colquhoun2021pandora}) no longer use quality scores. While some genomic compressors (e.g.,~\cite{Meng2023,vandamme2024tinted,karasikov2020metagraph}) do \emph{not} support quality score (de)compression, \proposal supports optional, lossless quality score (de)compression for broad applicability.}
\nh{\oiv{Since} quality scores do not have the same redundancy patterns
as DNA bases, \proposal compresses quality scores as a separate data stream from DNA bases, a common practice in genomic-specific compressors\oiv{~\cite{chandak2018spring,chandak2017compression,kokot2022colord,roguski2018fastore,kowalski2019pgrc,dufort2020enano,dragenora,chen2023efficient,Deorowicz2020,lan2021genozip,alyami2019lfastqc}}. \proposal maintains the same order for DNA bases and quality scores during compression.}
\nh{\proposal's quality score decompression runs on the host CPU because, as shown in \sect{\ref{sec-sa:mech-qual}}, the throughput of \oiv{\proposal's quality score decompression} is sufficient to avoid becoming a bottleneck in genome analysis pipelines. \sect{\ref{sec-sa:mech-qual}} provides further details about \proposal{}'s quality score compression and decompression.}

\subsubsection{\hm{Mismatch} Positions and Counts}
\label{sec-sa:mech-noise-pos-count}

\begin{figure}[h]
  \centering
  \includegraphics[width=\columnwidth]{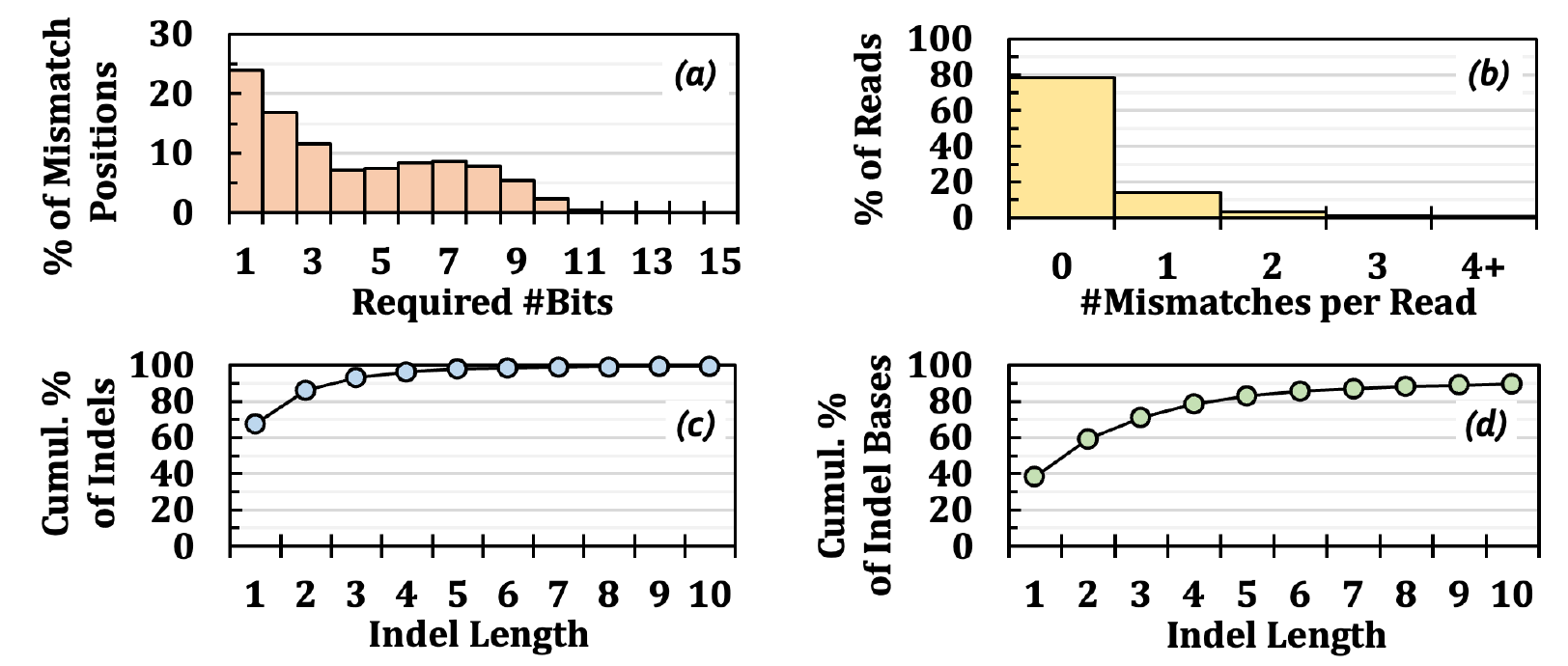}
  \caption{Distribution of  (a)~\#bits needed to store the delta-encoded \hm{mismatch} positions, and (b)~\oiv{mismatch counts per read}. Cumulative distribution
of (c)~indel block lengths, (d)~\#bases \new{in indel blocks}
of different lengths.}
  \label{fig:obsv-noise-pos} 
\end{figure}

\proposal stores \hm{mismatch} positions for reads in a read set sequentially in \oiv{the} Mismatching Position Array (\emph{\npaa}) and uses \oiv{the} Mismatching Position Guide Array (\emph{\npb}) to indicate the number of bits used for each \npaa entry and the number of \hm{mismatches} in each read. 
To reduce the sizes of \npaa and \npb, \proposal introduces optimizations based on the properties of read sets.

\fig{\ref{fig:obsv-noise-pos}(a)} shows the distribution of \#bits for the delta-encoded mismatch positions in a representative long read set (RS4 in Table~\ref{table-sa:eval-comp-ratio}). We observe that most delta‑encoded mismatch positions need only a few bits to store\sgprop{1}. This is due to two main factors that can lead to \oiv{spatially} nearby mismatches. First, genetic mutations tend to cluster in some regions of the genome\oiv{~\cite{Bourque2018,Tian2008,amos2013variation}}. Second, a \oiv{regional degradation} in sequencing quality can cause \nh{clusters of} incorrect bases in the read\oiv{~\cite{LaPierre2019,Delahaye2021sequencing,gleeson2021accurate}}. 
Based on this, \proposal performs two optimizations. 
First, during compression of each read set, \proposal tunes the number of distinct bit-counts used for \npaa and the specific values of each bit count. To this end, when finding the mismatch positions of reads, \proposal forms a histogram of bit-counts needed to represent the mismatch positions. \proposal then uses Algorithm~\ref{alg:bitcount} to systematically search for the \oiv{best} bit-count boundaries that minimize the total encoded size for that specific read set based on the information encoded in the histogram.
While the search for bit-counts is exhaustive, its space is limited, i.e.,  constrained by a small upper bound on the required bit-widths (typically $<$ 8 in our evaluation). As evaluated in \sect{\ref{sec-sa:comp-time}}, this makes the optimization cost \oiv{very small}. Second, \proposal exploits the higher frequency of smaller bit counts to further reduce the guide array size by assigning shorter representations to more common inputs\ov{~\cite{huffman2007method}}. For example, assuming four distinct bit counts, \proposal uses variable-length prefix codes 0, 10, 110, and 1110 to represent them (instead of 00, 01, 10, and 11). \oiv{\proposal stores the selected bit counts for the position array and their associated representation in the position guide array in a small \emph{Association Table}.}

\begin{algorithm}[ht]
\color{black}
\caption{Tuning Bit Counts}\label{alg:bitcount}
\begin{algorithmic}[1]
\footnotesize
\Require{Histogram of mismatch position bit counts $H$, where $|H|\leq32$, and a tuning convergence threshold $\varepsilon$.}
\Ensure{List of bit counts $W$ for encoding mismatch positions.}
\Statex
\State $\ell_\text{min}\gets\infty$ \Comment{Initialize min. encoding size}
\State $x_0\gets 0$
\ForAll{$d \in \{1,\ldots,8\}$} \Comment{Find optimal $|W|$.}
  \State $\ell_\text{last}\gets\ell_\text{min}$ \Comment{Best result from previous $d$ values}
  \ForAll{$(x_1,\ldots,x_d)\in|H|^d$ s.t.  $x_0<x_1<\cdots<x_d$}
    \State $\ell\gets$ total length of encoded mismatch positions and guide arrays s.t. positions with bit counts $\in(x_i,x_{i+1}]$ are encoded with $x_{i+1}$ bits.
    \If{$\ell < \ell_\text{min}$}
        \State $\ell_\text{min} = \ell$
        \State $W\gets (x_1,\ldots,x_d)$
    \EndIf
  \EndFor
  \If{$(\ell_\text{last}-\ell_\text{min})/\ell_\text{min} < \varepsilon$}
    \State\textbf{break} \Comment{Exit loop when $\ell_\text{min}$ converges, typically at $d<8$}
  \EndIf
\EndFor\\
\Return{$W$}
\end{algorithmic}
\end{algorithm}

\oiv{Since the number of mismatches can constitute a large fraction of mismatch information in short read sets, we analyze their properties.} \fig{\ref{fig:obsv-noise-pos}(b)} shows \#mismatches per read for a representative short read set (RS2 in Table~\ref{table-sa:eval-comp-ratio}). As also shown by prior works (e.g.,~\cite{nag2019gencache}), most short reads have \hm{no} or few \hm{mismatches} due to short read sequencing's low error rates\sgprop{2}.
We exploit this and use the variable-length encoding \oiv{described} above to assign shorter representations to more common inputs.

Given that \emph{indel blocks} (consecutive inserted or deleted bases) cause many mismatches in long reads, we analyze their properties in a representative read set (RS4 in Table~\ref{table-sa:eval-comp-ratio}).  \fig{\ref{fig:obsv-noise-pos}(c)} and \fig{\ref{fig:obsv-noise-pos}(d)} show the cumulative distributions of indel block lengths and the bases stored for different block lengths, respectively. We observe that \inum{i}~most indel blocks are of length one, a common trait in read sets~\cite{ono2020pbsim2,belyaeva2022best,magi2016characterization,wenger2019accurate,Garg2021}, and \inum{ii}~although single-base blocks are more frequent, larger blocks encompass most indel bases\sgprop{3}. 
\noindent Based on this, we implement two optimizations. First, we store the position of the first \hm{mismatch} and the length of the indel, rather than storing each mismatch position. Second, we use the indel lengths' bit-count distribution to minimize the \#bits needed for their encoding. To this end, we apply the same systematic bit-count tuning  of Algorithm~\ref{alg:bitcount}. Given the strong skew towards 1-bit indels~\cite{ono2020pbsim2,belyaeva2022best,magi2016characterization,wenger2019accurate,Garg2021}, we observe that the \oiv{best} configuration typically consists of two distinct bit counts: one for single-base indels and one for others. Accordingly, after detecting an indel (see \sect{\ref{sec-sa:mech-bases-types}}), we reserve one bit in \npb to indicate whether it is a single-base indel, and otherwise, we dedicate eight bits to encode the indel length, \oiv{as we observe these bit counts to be suitable fits across our evaluated datasets}. In the uncommon case where longer indels are more frequent, \proposal{} can \oiv{use Algorithm~\ref{alg:bitcount} to find other bit count configurations that minimize the encoded size.}

\ov{Based on all the optimizations discussed in this section, \proposal tunes the encoding of position arrays and position guide arrays to compress mismatch position information for all reads in a read set (as described in \fig{\ref{fig:overview}}). Given the lightweight structures of the position arrays and guide arrays, \proposal can decompress mismatch position information with simple operations and efficient access patterns}.

\begin{figure}[t]
  \centering
  \includegraphics[width=0.95\columnwidth]{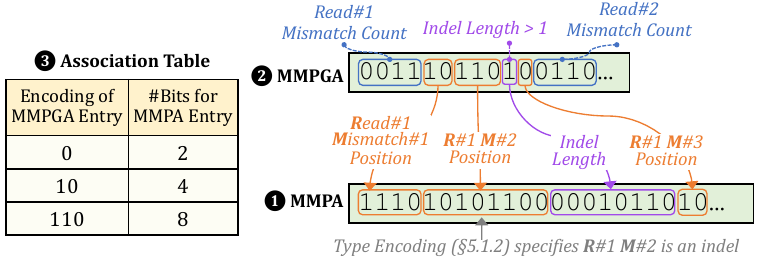}
  \caption{Example of decoding \hm{mismatch} positions for long reads in \proposal.}
  \label{fig:example-noise-pos} 
\end{figure}

\head{Example Walkthrough \oiv{of Mismatch Position Decompression}} \fig{\ref{fig:example-noise-pos}} shows an example of decompression of mismatch positions, 
demonstrating \oiv{the position array} \npaa (\circled{1}), \oiv{the position guide array} \npb (\circled{2}), and the small \oiv{Association Table}  (\circled{3}). \oiv{Based on the optimizations described in this section, during compression, the bit counts used for representing the \npaa elements are selected, and the associated representations used for these bit counts in the \npb are optimized and stored in the Association Table.} 
In the example of this figure, \oiv{three distinct bit counts are used for mismatch positions stored in the \npaa (i.e., 2, 4, and 8), and these bit count values are represented by 0, 10, and 110 in the \npb.}

To decompress mismatch positions,
\proposal first reads the \hm{mismatch} count of Read\#1 in \npb{} \oiv{(i.e., 0011)}. 
\new{It then scans \npb to obtain the bit counts for each of the \oiv{three} mismatch position\ov{s} \oiv{of Read\#1} and scans the specified number of bits from \npaa to decode these positions}. 
\oiv{For example, the first entry in \npb after the mismatch count is 10. Based on the Association Table, \proposal knows that it needs to read the next 4 bits in the \npaa{} (i.e., 1110) to decode the mismatch position.} 
If a \hm{mismatch} is an indel, \proposal checks \npb to determine if the indel is longer than one. If so, it reads the next eight bits from \npaa to determine its length.
After decoding all \hm{mismatch} positions of Read\#1, \proposal proceeds to the next read.

\subsubsection{\hm{Mismatch} Bases and Types}
\label{sec-sa:mech-bases-types}

\newcommand\nbta{\oiv{M}BTA\xspace}

\begin{figure}[t]
  \centering
  \includegraphics[width=0.8\columnwidth]{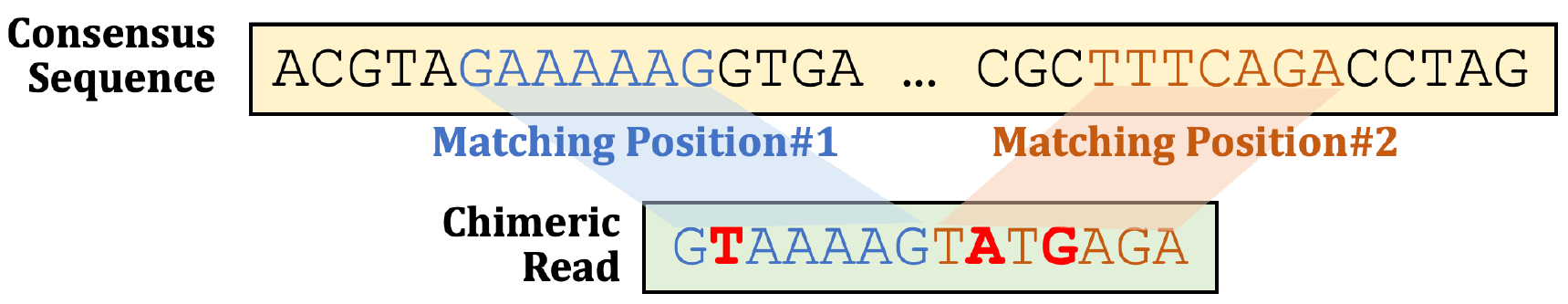}
  \caption{Example of a chimeric read.}
  \label{fig:example-chimeric} 
\end{figure}

\proposal stores \hm{mismatch} bases/types information of reads in a read set in a \hm{Mismatch} Base and Type Array (\emph{\nbta}) and reduces its size with two optimizations based on properties of genomic data. 
First, in long reads, we observe that a significant fraction of \hm{mismatch} bases (e.g., up to 80\% in our datasets) can originate from \emph{chimeric} reads, which are reads with sequences joined from different regions of the genome due to sequencing/library preparation errors, or structural variations~\cite{guan2016structural}\sgprop{4}. 
Parts of these reads map to different locations in the consensus. 
\fig{\ref{fig:example-chimeric}} 
shows a chimeric read with eight mismatches at matching position\#1 and nine at position\#2. 
Thus, considering only the \oiv{top} \hm{matching} position \oiv{(i.e., position\#1 in this example)}, as in prior works (e.g.,~\cite{dufort2021renano,kowalski2019pgrc}), results in many \hm{mismatches} \oiv{(i.e., eight)}. 
Instead of relying on expensive compressors to compress these mismatches, \proposal considers the top $N$ matching positions\footnote{We use N = 3 \oiv{as it led to the best results in our evaluated datasets}. \proposal can tune N to other values as well.} for chimeric reads. 
For example, considering both positions in \fig{\ref{fig:example-chimeric}} \oiv{(positions \#1 and \#2)} reduces mismatches to three, fewer than the \oiv{eight mismatches} at the \oiv{top matching} position.  
Since, in chimeric reads, the number of mismatches at individual positions can be substantial, reconstructing reads from multiple positions (and storing these positions) is more efficient than storing many mismatches from only the top position.

Second, since substitutions are the most common mismatch in short reads due to specifics of sequencing technologies~\cite{glenn2011field,goodwin2016coming,quail2012tale,kchouk2017generations,pfeiffer2018systematic}\sgprop{5}, we design a new encoding to avoid explicitly storing their type.
\proposal compares the mismatching base with the consensus sequence at the corresponding mismatch position. \oiv{If they differ, the mismatch is a substitution.} If they \oiv{are the same}, the mismatch cannot be a substitution and must be an indel. We then use a single bit to distinguish between an insertion and a deletion. This bypasses the need to explicitly store substitution types.

\subsubsection{\hm{Matching} Positions}
\label{sec-sa:mech-map-pos}

\newcommand\mpa{MPA\xspace}
\newcommand\mpb{MPGA\xspace}

\begin{figure}[t]
  \begin{center}
    \includegraphics[width=0.85\columnwidth]{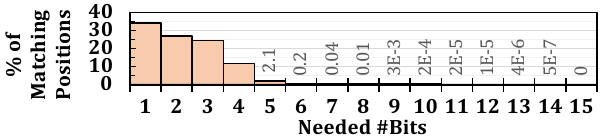}
  \end{center}
  \caption{Distribution of \#bits needed to store the delta-encoded matching position.}
  \label{fig:obsv-map-pos}
\end{figure}

Since sequencers read each part of the genome several times (to better allow later analysis to distinguish 
between biological signals and sequencing errors, \sect{\ref{sec:background-workflow}}), 
many reads in a read set match closely in the consensus sequence \sgprop{6}. 
Since each read in a read set can be \emph{mapped independently}~\cite{berger2023navigating}, it is possible to reorder them based on their matching position\oiv{s} in the consensus~\cite{chandak2017compression,chandak2018spring,roguski2018fastore,cogo2021genodedup}. Thus, consecutive reads require only a few bits to store their delta-encoded matching positions. 
\oiv{\fig{\ref{fig:obsv-map-pos}} shows this trend by demonstrating the distribution of the number of bits needed to store the delta-encoded matching positions in a representative real human read set (RS2 in Table~\ref{table-sa:eval-comp-ratio}).}
While a similar pattern of \oiv{skew towards specific bit counts} can be seen in other data types (e.g., images~\cite{crisan2021analyzing}), the possibility to \emph{reorder the reads} enables leveraging this pattern in genomic data more efficiently. 
Several genomic compression algorithms (e.g.,~\cite{chandak2017compression,chandak2018spring,roguski2018fastore,cogo2021genodedup}) also reorder reads for delta encoding and then apply entropy coding to further compress the matching positions.
However, their entropy decoding incurs expensive random lookups to match patterns during decompression. Instead, \proposal uses its hardware-friendly structures to significantly compress the matching positions. To this end,
\proposal sequentially encodes the delta-encoded matching positions of reads in a read set in the Matching Position Array (\mpa) and the Matching Position Guide Array (\mpb). \oiv{\proposal then uses the same approach used for reducing the sizes of position arrays and position guide arrays of mismatch positions in \sect{\ref{sec-sa:mech-noise-pos-count}} to reduce the sizes of \mpa and \mpb.}

\ov{\subsubsection{Corner Cases}
\label{sec-sa:mech-corner-cases}

Two corner cases require special handling. First, reads containing unidentified bases (\texttt{N}) expand the DNA alphabet to five characters, 
making 2-bit encoding impossible. Second, some reads include \emph{clips} (i.e., large insertion blocks at the beginning or end) that need reattachment during decompression. 
Adding indicator bits to \ovi{every compressed read to} specify these cases would significantly increase overhead (particularly for short reads). To avoid this, \ovi{\proposal marks a read as a \emph{corner case} by encoding a mismatch as} position~0. \ovi{\proposal} then use\ovi{s} a single bit in the MBTA to indicate whether a read has a mismatch at position~0, indeed, or is a corner case. \ovi{By doing so, \proposal}  efficiently identifies and handles \ovi{corner} cases without impacting standard reads.}

\subsubsection{Quality Scores}
\label{sec-sa:mech-qual}

\nh{\proposal \emph{losslessly} compresses quality scores. As discussed in \sect{\ref{sec-sa:mech-alg}},
this feature is optional and can be disabled.} 
\nh{\proposal compresses quality scores as a separate data stream from DNA bases, a common practice in genomic compressors (\sect{\ref{sec:background-compression}}). \proposal maintains the same order for DNA bases and quality scores when reordering the reads  (\sect{\ref{sec-sa:mech-map-pos}}).}

\proposal's quality score decompression runs on the host CPU because its throughput is sufficient to avoid becoming a bottleneck in genome analysis pipelines. This is because applications that require quality scores, such as variant calling (performed \oiv{on reads generated with sequencing techniques that report quality scores}),  typically only access scores for a small fraction of positions surrounding mismatches (as identified during the earlier read mapping step)\oiv{~\cite{Poplin2018,li2008mapping,Yu2015,Park2025}}. 
For example, across our evaluated read sets, on average only 0.03\% (maximum 10.7\%)\footnote{This ratio \oiv{approaches} upper limits of dissimilarity between an individual's genome and a reference genome (which determines the fraction of accessed quality scores) \oiv{typically tolerated in} genome analysis. At greater dissimilarity levels, \oiv{the efficacy of standard genome analysis is often compromised}~\cite{pearson2013introduction,joudaki2023aligning,prasad2022evaluating}.}
of quality score blocks (with a block size of 25 MB~\cite{grebnov2011libbsc}) are accessed.
This ratio is small because very few genomic locations are variable within a population~\cite{buffalo2021quantifying,10002015global}.
For these quality score blocks, we find that the time to decompress them on the host CPU is significantly shorter than mapping the entire read set using a read mapping accelerator~\cite{chen2023gem}. Therefore, given that reads can be analyzed in batches and in a pipeline\oiv{d manner}, the quality score decompression is \emph{not} on the critical path of the pipeline. \ov{For} our system configuration (\sect{\ref{sec-sa:methodology}}), quality score decompression on the host CPU would not become a bottleneck for cases where up to 17\% of quality scores are accessed. \oiv{This threshold provides a safe margin, ensuring that quality score decompression does not bottleneck execution in the vast majority of cases.}

\proposal's quality score (de)compression is based on the same software (de)compression used for quality scores in \cite{chandak2018spring} (its lossless mode). Since \proposal's quality score decompression runs on the host, it can also flexibly adopt other algorithms \oiv{(e.g.,~\cite{kokot2022colord,Meng2023,dufort2021renano,kowalski2019pgrc,dufort2020enano,dragenora,yang2025gpufastqlz,chen2023efficient,hach2012scalce,roguski2014dsrc2,Deorowicz2020,lan2021genozip,alyami2019lfastqc,cogo2021genodedup})}.

\subsection{Hardware for Data Preparation}
\label{sec-sa:mech-hw}

\newcommand\su{SU\xspace}
\newcommand\rcu{RCU\xspace}
\newcommand\cu{CU\xspace}

\subsubsection{Components}
\label{sec-sa:acc-components}
\begin{figure}[h]
  \centering
  \includegraphics[width=0.9\columnwidth]{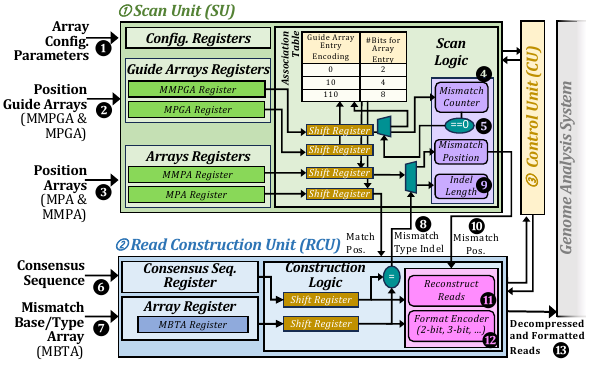}
  \caption{\oiv{Overview of \proposal's hardware.}}
  \label{fig:ssdg-acc-overview} 
\end{figure}

\fig{\ref{fig:ssdg-acc-overview}} shows the structure of \proposal's lightweight hardware units. 
\proposal consists of three components. \wcirc{1} Scan Unit (\emph{\su}) sequentially scans through \oiv{input position} arrays and \oiv{position} guide arrays to find mismatch information. 
\wcirc{2} Read Construction Unit (\emph{\rcu}) receives the mismatch information from the \su and reconstructs full reads by plugging the mismatches in\hm{to} the correct positions of the \oiv{input} consensus sequence. 
\wcirc{3}~Control Unit (\emph{\cu}) coordinates operations between the \su and the \rcu. 

\proposal enables decompression with streaming accesses, and thus, the \su and the \rcu do \emph{not} rely on large buffers, and instead only require small registers. 
Since \proposal stores mismatches in the order of appearance in each read, decoding is feasible via sequential accesses \oiv{to} the \oiv{position} arrays \oiv{and position guide arrays}. Similarly,  since \proposal stores the reads' positions based on their order in the consensus (similar to other compressors, e.g.~\cite{chandak2017compression,chandak2018spring,roguski2018fastore,cogo2021genodedup}, as explained in detail in \sect{\ref{sec-sa:mech-map-pos}}),
it only sequentially accesses the consensus \oiv{sequence}. 
The \oiv{registers used for buffering the position arrays and position guide arrays} each consist of eight bits since that is the largest element size used in the arrays in our compression scheme \oiv{(\sect{\ref{sec-sa:mech-alg}})}. \oiv{\proposal also uses an eight-bit register} since it loads the configuration parameters in eight-bit chunks. \oiv{\proposal uses a register of size 150~base pairs} since this is the largest read size in most short read sequencing datasets~\cite{glenn2011field,quail2012tale}. For short reads longer than this and for long reads, we reconstruct the reads in 150-base-pair chunks.\footnote{\oiv{As detailed in \sect{\ref{sec-sa:integration}}, in the third integration mode in \fig{\ref{fig:integration}}, \proposal requires two additional 64-bit registers for its operations.}} 
\proposal directly sends reads to the analysis system.
Since reads can be analyzed independently, analysis can start as soon as \oiv{\proposal{}'s output} arrives.

\subsubsection{Operations} 
\label{sec-sa:mech-acc-operations}
When \oiv{\proposal receives} a request to access a genomic read set, \oiv{it} starts preparing the reads in the read set. 
The \su reads the tuned configuration parameters of the read set (\circled{1} in \fig{\ref{fig:ssdg-acc-overview}}). 
This includes the \oiv{contents of the Association Table} (as shown in \fig{\ref{fig:example-noise-pos}}).
\oiv{The \su and \rcu then start operating concurrently to decode the mismatch information and reconstruct the full reads.
The \su decodes each read's matching position and mismatch positions by}
scanning through the position guide arrays (\circled{2}) and position arrays (\circled{3}), \oiv{as described in \sect{\ref{sec-sa:mech-alg}}}. Each time a new mismatch position is decoded, the mismatch count \oiv{is} decrement\oiv{ed} (\circled{4}), and when it reaches zero (\circled{5}), the \su reads a new mismatch count (for the next read).
\oiv{Meanwhile, the \rcu decodes the mismatch bases and types (as described in \sect{\ref{sec-sa:mech-bases-types}}) and reconstructs the full reads.}
\oiv{The \rcu scans through the consensus sequence (\circled{6}) and the \nbta (\circled{7})} and when it detects an indel \oiv{mismatch type}, it sends a signal (\circled{8}) to the \su to read the indel length (\circled{9}). 
\oiv{After decoding each read\ov{'}s matching position and its mismatch positions,} the \su sends them to the \rcu (\circled{\small10}). 
\oiv{Based on the information received from the \su, the \rcu reconstructs each read by applying the mismatches to the corresponding positions of the consensus sequence \circled{\small11}, and} 
flexibly formatting the reads (e.g.,~in 2-bit encoded, 3-bit encoded for reads with \texttt{N}, ASCII, etc.) as requested (\circled{\small12}).
\oiv{Finally, \proposal sends the decompressed and formatted reads to the genome analysis system (\circled{\small13})}.

\subsection{Data Layout}
\label{sec-sa:mech-ftl}

\nh{We \oiv{describe} \proposal's efficient data layout, which enables \oiv{it} to leverage the storage system's full bandwidth when accessing and preparing genomic data. We also discuss the feasibility and design efforts required to maintain this layout alongside the existing flash translation layer (FTL).} 
\proposal requires simple changes to the baseline FTL.
\proposal FTL designates each block as genomic or non-genomic.
The SSD recognizes genomic accesses through \proposal commands (\sect{\ref{sec-sa:mech-interface}}). For all other data, vendor-specific FTL features remain untouched, so the SSD behaves like a conventional SSD.

When writing a compressed genomic dataset, \proposal uniformly partitions data across the SSD channels. This is enabled by \proposal's sequential access patterns. 
Each partition of the consensus sequence, along with the compressed mismatch information of the reads matched to that partition, is placed in a separate channel.  
\proposal writes data in a round-robin fashion \oiv{across different channels} such that the active blocks in different channels have the same page offset. This enables \emph{multi-plane} read operations across all channels and leverages the SSD's \emph{full} bandwidth.

During garbage collection (GC)~\cite{tavakkol2018flin,kim2020evanesco,cai2017error,park-dac-2016, park-dac-2019}, we must select victim blocks such that we preserve \proposal{}'s capabilities for multi-plane operations. This requires maintaining the same page offset across all blocks in a parallel unit. Given that genomic read sets do not require partial updates\oiv{~\cite{leinonen2010sequence,cochrane2015international}} and are accessed sequentially, it is straightforward to realize this efficient GC. We perform GC in a grouped manner and select every block in the parallel unit as a group of victim blocks, which are then sequentially rewritten in the order they were originally written, as indicated by their logical address sequence.

\subsection{Interface Commands}
\label{sec-sa:mech-interface}

\proposal in\ov{t}roduces two new interface commands: \oiv{one to request genomic data in the desired format, and another to write compressed genomic data to the storage system}.

\noindent\textbf{\texttt{SAGe\_Read:}} A specialized read command to \inum{i}~access genomic data and \inum{ii}~specify the output format (e.g., 2-bit or 1-hot encoding) required by the genome analysis system. The other parameters (e.g., array/guide array parameters) are written at the beginning of each compressed file and are loaded into the \su (\sect{\ref{sec-sa:mech-hw}}) when accessing genomic data. Upon receiving this command, \proposal operates with its FTL (\sect{\ref{sec-sa:mech-ftl}}). 

\noindent\textbf{\texttt{SAGe\_Write:}} A specialized write command for writing genomic data to SSD and updating its FTL's mapping metadata.

\section{Case Studies of \proposal's Hardware Integration}
\label{sec-sa:integration}

\begin{figure}[h]
  \centering
  \includegraphics[width=0.85\columnwidth]{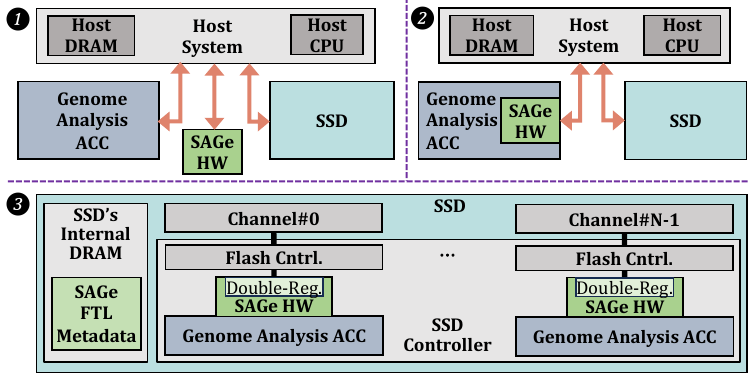}
  \caption{\micro{\proposal's integration with genome analysis systems.}}
  \label{fig:integration} 
\end{figure}

\fig{\ref{fig:integration}} shows three examples of \proposal's integration with genome analysis systems. 
In \circled{1}, 
\proposal{}'s hardware units are connected to the genome analysis system via a PCIe~\cite{PCIE} \oiv{or CXL~\cite{das2024introduction}} interface (i.e., similar to how individual accelerators or GPUs connect). 
In \circled{2}, \proposal's hardware is integrated on the same chip as the genome analysis system (which can be a stand-alone ASIC, or an FPGA, etc.), allowing for tighter integration and saving PCIe \oiv{or CXL \ov{ports},} which are scarce and can be contested by other components (e.g., SSDs, GPUs, or other accelerators). 
This is easily possible due to \proposal's low hardware cost. For example, \proposal's \oiv{logic units (\sect{\ref{sec-sa:eval-area-power}})} consume only 2.5\% of the lookup tables and 0.8\% of the flip-flops of a mid-range FPGA~\cite{fpgamidrange}.

\proposal's lightweight design enables its integration in resource-constrained environments. As an example, in \circled{3}, we demonstrate \proposal's integration with an in-storage NDP genome analysis system (e.g.,~\cite{mansouri2022genstore,abakus23taco,megis,jun2016storage}) on the SSD controller. Given that \proposal performs streaming accesses, \proposal's hardware operates on data fetched from NAND flash chips without needing to buffer them in the SSD's low-bandwidth, single-channel, internal DRAM~\cite{zou2022assasin}.
\proposal performs operations on flash data streams\ov{~\cite{zou2022assasin}} via lightweight double-buffering: one register to hold a chunk of data as computation input, and another to hold the subsequent data chunk arriving from the NAND flash chips. We use two 64-bit registers as they are sufficient to fully utilize the per-channel hardware components and SSD channel bandwidth.

\section{Methodology}
\label{sec-sa:methodology}

\renewcommand\pigz{\texttt{pigz}\xspace}
\renewcommand\nspring{\texttt{(N)Spr}\xspace}
\renewcommand\nsacc{\texttt{(N)SprAC}\xspace}
\newcommand\idec{\texttt{0TimeDec}\xspace}
\newcommand\psw{\texttt{SAGeSW}\xspace}
\newcommand\pmain{\texttt{SAGe}\xspace}
\newcommand\ssdgi{\texttt{SAGe\textsubscript{SSD}}\xspace}

\head{Evaluated Systems}
We evaluate the end-to-end performance of various genome analysis systems, where execution includes \emph{both} data preparation and genome analysis.
For \textbf{data preparation}, we use
\inum{i}~\pigz: \hm{A p}arallel version~\cite{adler2015pigz} of gzip, a commonly-used general compressor. \nh{We exclude other general-purpose compressors from our performance evaluations due to reasons detailed in \sect{\ref{sec-sa:motivation-observations}}.
We do, however, include pigz, i.e., parallel gzip, since it is still widely used as a baseline comparison point in genomic compression research and literature~\oiv{\cite{chandak2018spring,chandak2017compression,kokot2022colord,Meng2023,roguski2018fastore,dufort2020enano,dragenora,yang2025gpufastqlz,chen2023efficient,hach2012scalce,roguski2014dsrc2,Deorowicz2020,lan2021genozip,alyami2019lfastqc,cogo2021genodedup}}};
\inum{ii}~\nspring: Spring~\cite{chandak2018spring} and NanoSpring~\cite{Meng2023}, state-of-the-art compressors for short and long reads, respectively. \pigz and \nspring run on a high-end system, an AMD$^\text{\textregistered}$ EPYC$^\text{\textregistered}$ 7742 CPU~\cite{amdepyc} with 128 physical cores \oiv{(256 hardware threads)}. 
\inum{iii}~\nsacc: \nspring integrated with a BWT-accelerator. There are various accelerators for BWT (e.g.,~\cite{qiao2019fpga,guo2013gpu,zhao2017streaming}). We consider an idealized accelerator that can fully eliminate the BWT execution time from \nspring;
\inum{iv}~\idec: an idealized decompressor (with zero decompression time), but inefficient for integration \micro{in resource-constrained environments (e.g., in our evaluations, for integration with an in-storage NDP genome sequence analysis system)};\footnote{As a  conservative evaluation, \idec can also serve as an idealized representation of the closed-source industry software called ORA~\cite{dragenora}, recently developed for Illumina~\cite{illumina} short read (de)compression. ORA works similarly to other genomic compressors (\sect{\ref{sec:background-compression}}), but uses a different backend compressor for mismatch information, which still suffers from limitations for implementation in resource-constrained environments (detailed in \sect{\ref{sec-sa:motivation-goal}}). 
We cannot use ORA as our short-read decompression baseline due to its closed-source nature, \oiv{but its performance would be upper-bounded by \idec}.}
\inum{v}~\psw: \proposal with its decompression in software, \new{running on the host system (i.e., the same high-end, 128-core system used for \pigz and \nspring), to show the benefits of \proposal's algorithmic optimizations};
\inum{vi}~\pmain: \proposal's full implementation, with its decompression in hardware;
and \inum{vii}~\ssdgi: \pmain with its hardware implemented in the SSD, to integrate with an NDP genome analysis system on the same chip (mode \circled{3} in \fig{\ref{fig:integration}}). 
Decompressors other than \ssdgi connect to the analysis systems using a PCIe interface (\circled{1} in \fig{\ref{fig:integration}}). We do not explicitly \ov{evaluate} mode \circled{2}, \oiv{because given sufficient interface bandwidth, mode \circled{2}}  can perform the same as \circled{1}. As detailed in \sect{\ref{sec-sa:integration}}, \circled{1} enables standalone integration, while \circled{2} integrates on the same chip, thus saving PCIe slots.

We select Spring~\cite{chandak2018spring} and NanoSpring~\cite{Meng2023} as our genomics-specific software baselines because they provide a well-balanced and particularly suitable trade-off in the key aspects of compression ratio, decompression time, losslessness, and being open access. 
In contrast, other tools have less favorable characteristics in compression ratio (e.g.,~\cite{roguski2018fastore,chandak2017compression}), decompression time (e.g.,~\cite{kokot2022colord,dufort2020enano,dufort2021renano}), lossiness (e.g.,~\cite{chandak2017compression,roguski2018fastore}), or do not permit code access and/or modification (e.g.~\cite{dragenora,lan2021genozip}).\footnote{Being able to modify the code is critical for us since we need to adapt it to pass decompressed data batches to the analysis system (without merging data into a single file and writing it to disk).} As explained earlier, we use \idec as a strongly conservative and idealized representation of the closed-source tools.

For \textbf{genome sequence analysis}, we integrate all data preparation configurations with a state-of-the-art read mapping accelerator, GEM~\cite{chen2023gem}. To show \proposal's suitability for resource-constrained environments, we evaluate \proposal{}'s implementation inside the SSD to integrate with a state-of-the-art NDP genome analysis system, GenStore~\cite{mansouri2022genstore}. 
GenStore is an in-storage filter (ISF) that filters reads that do not require expensive read mapping directly inside the SSD, sending only the remaining reads to the mapper, thus alleviating the burden of moving and analyzing a large amount of low-reuse data from the rest of the system (i.e., external I/O, main memory, and compute units). 
\oiv{The resulting pipeline} performs data preparation $\rightarrow$ ISF $\rightarrow$ read mapping. \oiv{The benefits of this pipeline} arise from both \proposal's faster preparation and the ISF operations in \cite{mansouri2022genstore}. The key \oiv{to realizing this pipeline} is that \proposal is the only data preparation configuration that is lightweight enough for efficient implementation inside the SSD. Without \oiv{\proposal}, ISF would require genomic data to be stored uncompressed, which is inefficient, or to decompress data outside SSD, which undermines the fundamental benefits \oiv{of NDP}.

\head{Datasets} We evaluate real-world short- and long-read datasets. \new{Storage systems handle numerous read sets (e.g.,~in a medical center~\cite{Bick2024,Li2023whole} or in a cohort \cite{Bick2024,Li2023whole,10002015global,uk10k2015uk10k}), \hm{so} compressing each read set is essential to reduce the overall burden, regardless of their individual sizes. Thus, we analyze read sets of varying sizes, as \oiv{shown} in Table~\ref{table-sa:eval-comp-ratio}}.

\head{Performance}
We design a simulator that models all components involved during \proposal's execution, including accessing storage, hardware components (both for \proposal and genomic accelerators~\cite{chen2023gem,mansouri2022genstore}), \hm{the} SSD's internal DRAM (for the NDP accelerator~\cite{mansouri2022genstore}), host operations, and their interfaces. \new{Using this methodology, as also demonstrated in prior works (e.g.,~\cite{mansouri2022genstore,park2022flash,megis}), enables us to flexibly incorporate state-of-the-art system configurations in our analysis}. We feed the latency and throughput of each component to this simulator.  
For the components in \textbf{hardware-based} operations,  we implement \proposal's logic \oiv{units} in Verilog and synthesize using Design Compiler~\cite{synopsysdc} at 22~nm~\cite{22gf}.
We use 
Ramulator \oiv{1.0~\cite{kim2016ramulator, ramulatorsource,luo2023ramulator,ramulator2source}} to model \hm{the} SSD's internal DRAM, and MQSim~\cite{tavakkol2018mqsim,mqsimsource} to model the SSD's internal operations.
For the hardware mapper, we use the throughput reported by the original paper~\cite{chen2023gem}.
For the \textbf{software-based} operations (software decompressors), we measure performance with their best-performing thread counts, on a high-end real system, an AMD$^\text{\textregistered}$ EPYC$^\text{\textregistered}$ 7742 CPU~\cite{amdepyc} with 128 physical cores, 256 hardware threads, and 1.5-TB DRAM. 
We perform analysis with both a performance-optimized PCIe SSD~\cite{samsungPM1735} and a cost-optimized SATA SSD~\cite{samsung870evo}. 
Data communication between genome analysis accelerators, \proposal{}'s hardware units, and the storage system is modeled based on the bandwidth of the interfaces between them, depending on integration mode (\fig{\ref{sec-sa:integration}}).
I/O operations, decompression, and genome analysis execute on batches of genomic data in a pipelined manner, which enables partial overlapping \oiv{of} their execution. The synchronization between these stages is modeled via a producer-consumer abstraction.

\head{Prototype Feasibility and Real-System Overheads}
While any real-world prototype will introduce overheads not perfectly captured in simulation, we expect that system-level software changes in the I/O stack to have only a negligible impact. The required changes to the file system and the driver are confined to distinguishing genomics file types and supporting two new interface commands (\sect{\ref{sec-sa:mech-interface}}). Since this adds only simple new logic, we expect it to lead to negligible performance impact. The firmware needs to support \proposal{}’s L2P mapping and GC (\sect{\ref{sec-sa:mech-ftl}}), which are lighter than the baseline (due to \proposal{}'s uniform distribution of data across the SSD channels) and are already modeled by MQSim~\cite{tavakkol2018mqsim}.

\head{Area, Power, and Energy} 
For the accelerator logic units, we use the area and power values obtained from the Design Compiler synthesis~\cite{synopsysdc} of our Verilog HDL implementation of these units in a 22nm technology node~\cite{22gf}. For DRAM, we use the power values of a DDR4 model\oiv{~\cite{ddr4sheet,ghose2019demystifying,ghose2018your}}. For the CPU cores, we use power values obtained from AMD$^\text{\textregistered}$ \textmu{}Prof~\cite{microprof}.  
For \hm{the} SSD, we use the power values of a Samsung 3D NAND SSD~\cite{samsung860pro}. 
We calculate the energy of each component based on its idle and dynamic power and its execution time.
For end-to-end energy, we obtain the energy of the host processor, DRAM, \proposal{}'s logic units, and communication between them.

\section{Evaluation}
\label{sec-sa:evals}

\renewcommand\pigz{\texttt{pigz}}
\renewcommand\nspring{\texttt{(N)Spr}}
\renewcommand\nsacc{\texttt{(N)SprAC}}
\renewcommand\idec{\texttt{0TimeDec}}
\renewcommand\psw{\texttt{SAGeSW}}
\renewcommand\pmain{\texttt{SAGe}}
\renewcommand\ssdgi{\texttt{SAGe\textsubscript{SSD}}\xspace}

\subsection{Performance}
\label{sec-sa:evals-perf}

\begin{figure*}[h]
  \centering
    \includegraphics[width=\columnwidth]{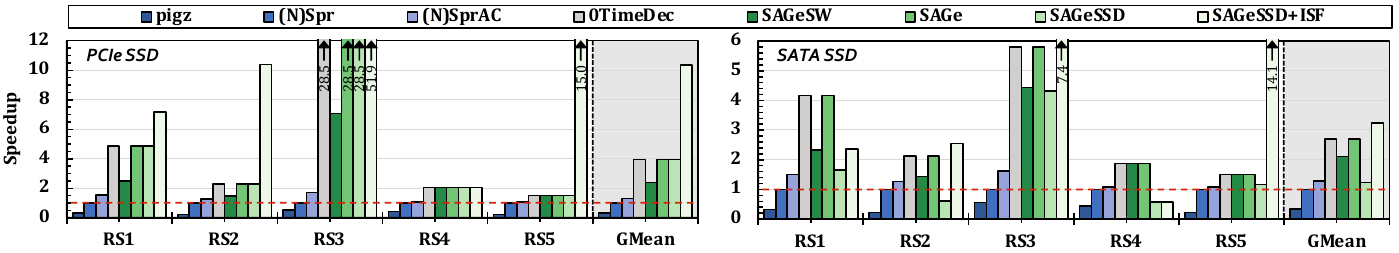}
        \vspace{-0.5em}
  \caption{End-to-end speedup for different read sets.}
  \label{fig:eval-new-full} 
\end{figure*}

\newcommand\map{\texttt{Map}\xspace}
\newcommand\isfmap{\texttt{ISF}\xspace}

\fig{\ref{fig:eval-new-full}} shows end-to-end \ov{performance}, where execution includes \emph{both} data preparation and genome analysis. 
Speedup is normalized to \nspring.
We make five key observations. 
First, \proposal provides significant speedups. \micro{On the system with the PCIe (SATA) SSD}, \pmain{} leads to 12.3$\times$ \micro{(8.1$\times$)},  3.9$\times$ \micro{(2.7$\times$)}, and 3.0$\times$ \micro{(2.1$\times$)} average speedup over \pigz{}, \nspring, and \nsacc, respectively. 
Second, \pmain{} matches \idec{} in performance because \pmain{} fully hides the decompression overhead in the execution pipeline. As explained in \sect{\ref{sec-sa:motivation}}, I/O, decompression, and read mapping are pipelined, so \ov{overall} throughput depends on the slowest stage. \idec{} and \pmain{} \ov{having} the same performance shows that decompression is no longer the slowest stage.
Third, due to its efficient access patterns, \proposal's implementation in software (\psw) also leads to speedups (2.3$\times$ on average) over \nspring{}. However, \psw's decompression still bottlenecks end-to-end performance. 
\psw{} leads to up to 4.0$\times$ slowdown over \pmain. 
The hardware implementation of \proposal decompression provides greater speedup than its software \ov{version} by handling bitwise operations (on arrays and guide arrays) more efficiently and managing data flow more effectively in a fine-grained manner. 
Ultimately, choosing between these two configurations (software or hardware) is a design decision. 
\proposal software facilitates its ease of adoption in \ov{the} near term, while \proposal hardware 1) provides \emph{better performance}, 2) has \emph{significantly better energy reduction} (across \emph{all} our datasets, as shown in \fig{\ref{fig:eval-energy}}), and 3) is very \emph{lightweight}, which makes it suitable for seamless integration with genome analysis hardware accelerators, even in resource-constrained environments. 
Fourth, by efficiently integrating \ov{\ssdgi{}} with \isfmap{} \ov{(i.e., the in-storage filter implemented inside the resource-constrained environment of the SSD),} \ssdgi{}+\isfmap leads to 7.8$\times$ \micro{(2.5$\times$)} average speedups over \nsacc{} on the system with the PCIe (SATA) SSD. 
\ssdgi{}+\isfmap outperforms \pmain{} in all cases, except when \inum{i}~the input and application do not \hm{largely take advantage} of \ov{\isfmap}{} \new{(i.e., in this case, \isfmap does not filter many reads in the read set)}, and \inum{ii}~\hm{the} SSD's limited external bandwidth bottlenecks performance (e.g., RS1 and RS4 with the SATA SSD). In these cases, \hm{the} \pmain~\new{configuration should be used} to decompress data outside the SSD to avoid moving larger decompressed data through the limited-bandwidth \oiv{storage} interface.
Fifth, regardless of how much decompression tools are optimized for performance, if \ov{their high resource requirements make them} unsuitable for adoption in resource-constrained environments, they miss out on the benefits of a wide range of genome analysis systems \ov{that are implemented in such constrained environments}. For example, \idec{} \ov{(i.e., an idealized decompressor with zero decompression time, but inefficient for integration in resource-constrained environments) cannot \ovi{efficiently and cost-effectively} integrate with \isfmap{} implemented inside the SSD and, as shown in our evaluations, it ends up being} on average 1.8$\times$ (up to  9.9$\times$) slower than \ssdgi{}+\isfmap.

\renewcommand\ssdgi{\texttt{SAGe\textsubscript{SSD}}\xspace}
\newcommand\ssdgo{\texttt{SG\textsubscript{out}}\xspace}

\begin{figure}[h]
  \centering
\includegraphics[width=0.85\columnwidth]{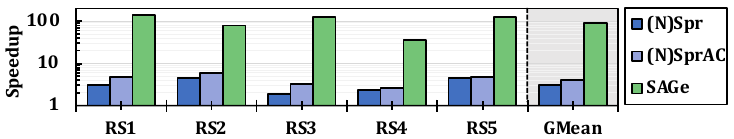}
  \caption{\new{Data preparation speedup.}}
  \label{fig:eval-prep-only} 
\end{figure}

\head{Data Preparation} 
\fig{\ref{fig:eval-prep-only}} shows \emph{only} data preparation throughput, normalized to \pigz{} (on \ov{a} system with the PCIe SSD). \pmain{} leads to 91.3$\times$, 29.5$\times$, and 22.3$\times$ average speedups over \pigz, \nspring, and \nsacc, respectively.

\begin{figure}[h]
  \centering

    \includegraphics[width=0.85\columnwidth]{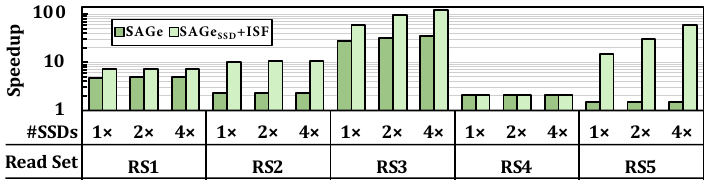}
  \caption{\new{End-to-end speedup with different SSD counts.}}
  \label{fig:multi-ssd} 
\end{figure}

\head{Multiple SSDs}
Since all data in \proposal are accessed in streams, they can be disjointly partitioned between multiple SSDs to enable concurrent access. 
\fig{\ref{fig:multi-ssd}} shows the speedups of \pmain{} and \ssdgi{}+\isfmap over \nspring~in a system with multiple PCIe SSDs. First, \pmain{} maintains its large speedup over \nspring. Second, \ov{with more SSDs,} \ssdgi{}+\isfmap's benefits \ov{increase} for some datasets (RS3 and RS5) since more SSDs improve \isfmap's performance, which was \ov{on} the critical path \ov{of the end-to-end execution}. 
\micro{We conclude that \proposal is able to effectively leverage multiple SSDs. Given this, and since each genomic read can be mapped independently, \proposal's design is also compatible for adoption in distributed storage systems.}

\subsection{Area and Power}
\label{sec-sa:eval-area-power}

Table~\ref{tab-sa:area-power} lists the area and power of \proposal's accelerator units at 1 GHz.
Although they can operate at a higher frequency, their throughput is already sufficient because \proposal's accelerator operations are bottlenecked by the NAND flash read throughput.
\proposal's accelerators consume \hm{a} small area and power of 0.002~mm$^2$ and \ov{0.49}~mW at 22~nm node.

\begin{table}[h]
\centering
\caption{Area and power consumption of \proposal's logic.}
\label{tab-sa:area-power}
\resizebox{0.9\columnwidth}{!}{%
\begin{tabular}{@{\hspace{-0.5pt}}c@{\hspace{-0.02pt}}|c|c|@{\hspace{-0.02pt}}c@{\hspace{-0.05pt}}}
\toprule
\textbf{Logic unit}                  & \textbf{\# of instances} & \textbf{Area [mm\textsuperscript{2}]} & \textbf{Power [mW]} \\ 
\midrule
\midrule
Scan Unit                 &   1 per channel     &   0.000045    &     0.014     \\
Read Construction Unit    &   1 per channel     &   0.000017    &      \ov{0.023}    \\
Double Registers & \multirow{2}*{1 per channel} & \multirow{2}*{0.00020} & \multirow{2}*{\ov{0.035}}\\
(for mode~\circled{3} in \fig{\ref{fig:integration}})    &        &       &          \\
Control Unit              &   1 per channel     &   0.000029    &      0.025    \\
\midrule
\textbf{Total for an 8-channel SSD}           & -                        & \textbf{0.002}    & \textbf{\ov{0.49} (+\ov{0.28} for  mode \circled{3})} \\ \bottomrule
\end{tabular}
\vspace{-0.5em}
}
\end{table}

\subsection{Energy}
\label{sec-sa:evals-energy}

\begin{figure}[h]
  \begin{center}
    \includegraphics[width=0.85\columnwidth]{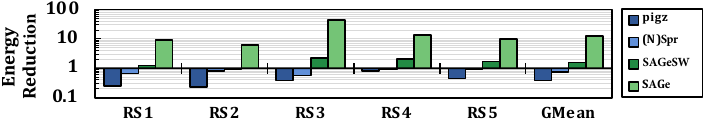}
  \end{center}
  \vspace{-0.5em}
  \caption{End-to-end energy reduction.}
  \vspace{-0.7em}
  \label{fig:eval-energy}
\end{figure}

\fig{\ref{fig:eval-energy}} shows the end-to-end energy reduction, \ov{where execution includes \emph{both} data preparation and genome analysis. Energy reduction is}  normalized to \nsacc{} (higher is better). 
We observ\hm{e that} \pmain{} leads to a significant average energy reduction of \nh{34.0$\times$, 16.9$\times$, and 13.0$\times$, over \pigz, \nspring, and \nsacc,} respectively.

\begin{figure}[h]
\centering
\includegraphics[width=0.7\linewidth]{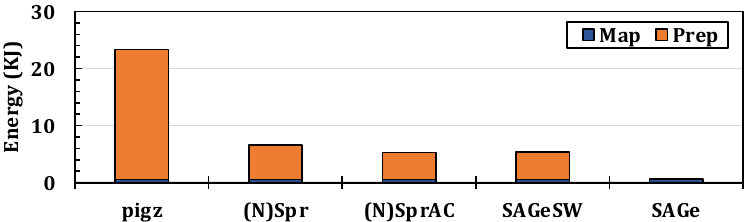}
\caption{\aooo{Energy breakdown of different systems.}}
\label{fig:energy-brkdn-sage-1}
\end{figure}

\aooo{\fig{\ref{fig:energy-brkdn-sage-1}} shows the energy breakdown of different systems. We show the energy breakdown between the data preparation and the analysis stages. We make two observations. First, in \pmain{}, the energy overhead of the preparation is significantly alleviated and no longer dominates overall energy consumption. Second, while \psw{} provides large performance benefits (as shown in \sect{\ref{sec-sa:evals-perf}}), it still spends a large amount of energy on data preparation as it relies on the energy-hungry host system (as detailed in our methodology in \sect{\ref{sec-sa:methodology}}).}

\newcommand\no{\texttt{NO}\xspace}
\newcommand\oone{\texttt{O1}\xspace}
\newcommand\otwo{\texttt{O2}\xspace}
\newcommand\othree{\texttt{O3}\xspace}
\newcommand\ofour{\texttt{O4}\xspace}

\subsection{Compression Ratio}
\label{sec-sa:evals-ratio}

Table~\ref{table-sa:eval-comp-ratio} shows the compression ratios of \pigz, \nspring, and \pmain. For DNA bases, \pmain~achieves 
\inum{i}~2.9$\times$ better average compression ratio than \pigz, 
and \inum{ii}~comparable compression ratios to \nspring, with a modest 4.6\% average reduction.

\begin{table}[h]
\centering
\caption{Compression ratios for different read sets.}
\label{table-sa:eval-comp-ratio}
\resizebox{0.8\columnwidth}{!}{%
\footnotesize
\begin{tabular}{@{\hspace{-0.5pt}}c|@{\hspace{-0.02pt}}c@{\hspace{-0.02pt}}|@{\hspace{-0.02pt}}c@{\hspace{-0.02pt}}|cc|cc|cc}
\hline
\toprule
\multirow{3}{*}{\textbf{Label}} & 
\multirow{3}{*}{\begin{tabular}[c]{@{}c@{}}\textbf{Read Set}\\ (\underline{S}hort, \underline{L}ong)\end{tabular}} & 
\multirow{3}{*}{\begin{tabular}[c]{@{}c@{}}\textbf{Uncomp.}\\ \textbf{Size (MB)}\\ DNA+Qual\end{tabular}} & 
\multicolumn{6}{c}{\textbf{Compression Ratio}} \\
\cline{4-9}
& & & \multicolumn{2}{c|}{\rule{0pt}{2.5ex}PigZ~\cite{adler2015pigz}} & 
\multicolumn{2}{c|}{\rule{0pt}{2.5ex}(Nano)Spring~\cite{Meng2023}} & 
\multicolumn{2}{c}{\rule{0pt}{2.5ex}\proposal} \\
& & & DNA & \multicolumn{1}{@{\hspace{-0.02pt}}c@{\hspace{-0.02pt}}|}{Qual.} & DNA & \multicolumn{1}{@{\hspace{-0.02pt}}c@{\hspace{-0.02pt}}|}{Qual.} & DNA & Qual. \\
\midrule \midrule
RS1 & SRR870667\_2~\cite{Motamayor2013} (S) & 10 000 & 3.39 & 2.23 & 24.8 & 2.80 & 22.8 & 2.80\\
RS2 & ERR194146\_1~\cite{eberle2017reference} (S) & 158 000 & 12.5 & 2.49 & 40.2 & 3.4 & 36.8 & 3.4\\
RS3 & SRR2052419\_1~\cite{zook2016extensive} (S) & 8 000 & 3.41 & 3.45 & 7.2 & 5.07 & 7.1 & 5.07\\
RS4 & PAO89685\_sampled~\cite{ontopendata} (L) & 24 000 & 3.93 & 1.79 & 4.8 & 2.19 & 4.5 & 2.19\\
RS5 & ERR5455028~\cite{Belser2021} (L) & 176 800 & 3.5 & 1.57 & 7.6 & 1.82 & 7.8 & 1.82\\
\bottomrule
\end{tabular}
}
\end{table}

\head{Effect of Input Data Characteristics}
\proposal can support datasets across different species because the consensus sequence is an approximation of the input sample's genomes (\sect{\ref{sec:background-compression}}). 
Given datasets \ov{from} the same species (e.g., RS2 and RS3 in our datasets, which are both \textit{Homo sapiens}), the larger one (RS2) achieves higher compression ratios. \ov{Since the storage overhead for the consensus sequence is constant, this one-time overhead is more effectively amortized as the number of compressed reads increases.}

\begin{figure}[b]
  \centering
    \includegraphics[width=0.8\columnwidth]{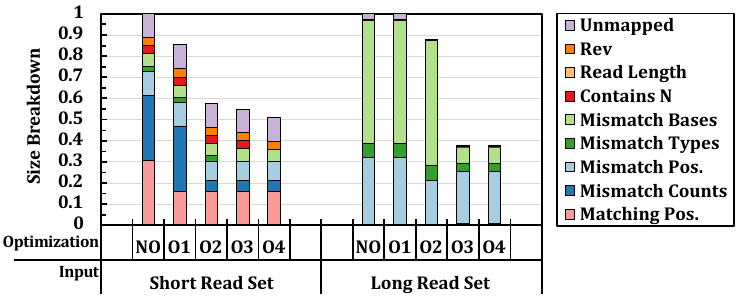}
  \caption{\ov{Effect} of different \proposal optimizations on \ov{the storage size \ovi{of mismatch information}}.}
  \label{fig:eval-step-by-step} 
\end{figure}

\head{\ovi{Effect of Different \proposal Optimizations}} To show the impact of different optimizations, \fig{\ref{fig:eval-step-by-step}}\footnote{\ov{\emph{Unmapped} refers to reads that do not match to the consensus sequence, and} \emph{Rev} refers to a bit marking if a read matches in reverse to the consensus.} presents the size breakdown of the reads' mismatch information for a short (RS2) and long read set (RS4) in five settings:
\inum{i}~\no, with no optimization on the raw mismatch information;
\inum{ii}~\oone, with matching position optimization (\sect{\ref{sec-sa:mech-map-pos}}) added to \no;
\inum{iii}~\otwo, with mismatch positions and count optimizations (\sect{\ref{sec-sa:mech-noise-pos-count}}) added to \oone;
\inum{iv}~\othree, with mismatching base and type optimizations (\sect{\ref{sec-sa:mech-bases-types}}) added to \otwo;
and \inum{v}~\ofour, with corner case optimizations \ov{(\sect{\ref{sec-sa:mech-corner-cases}})} added to \othree. 
We make six observations.
First, \oone significantly reduces the data size of the matching positions in short reads. Note that \oone is not critical for long reads, since matching positions do not constitute a large fraction of mismatch information in longer reads. 
Second, \otwo significantly reduces the data size of mismatch counts in short reads. 
This is because, as detailed in \sect{\ref{sec-sa:mech-noise-pos-count}}\ov{,} most reads in a short read dataset have \hm{no mismatches}, benefiting from \proposal's \hm{mismatch} count encoding.
Third, \otwo leads to a large reduction in the mismatch positions' data size in long reads. 
Fourth, \othree significantly reduces the bases for long reads by efficiently encoding chimeric reads \ov{(\sect{\ref{sec-sa:mech-bases-types}})}. However, the data size for mismatch positions increases in \othree due to the additional mismatches at the chimeric reads' new matching positions. This tradeoff is suitable for \proposal, as it can efficiently encode the greater number of mismatch positions.  
Fifth, \othree reduces the types' data size for short and long reads.
Sixth, \ofour reduces the data size required for labeling corner cases.

\nh{\head{Quality Scores} 
\pmain~achieves 
\inum{i}~33\% better average compression ratio than \pigz,
and \inum{ii}~the same ratios as \nspring. \ov{This is because, as mentioned in \sect{\ref{sec-sa:mech-qual}}, \proposal{}'s quality score (de)compression is based on the same
software (de)compression used for quality scores in \nspring.}}

\subsection{Compression Time}
\label{sec-sa:comp-time}

\begin{figure}[b]
  \centering
    \includegraphics[width=0.8\columnwidth]{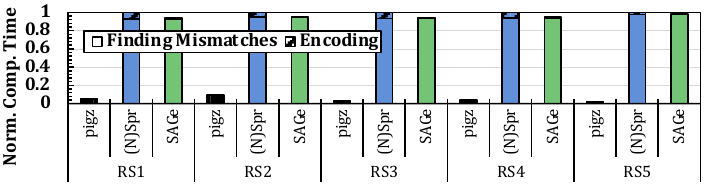}
    \vspace{-0.5em}
  \caption{Normalized compression time.}
  \label{fig:eval-comp-time} 
\end{figure}

Compression of genomic data is typically an offline and one-time task and is \emph{not} in the critical path of genome analysis. For completeness, \fig{\ref{fig:eval-comp-time}} reports the compression times \ov{of \pigz, \nspring, and \pmain}. In all cases, compression is performed on the host CPU. We make two observations. First, \proposal's compression is slightly faster than \nspring. Compression time\ov{s} of both \pmain{} and \nspring{} \ov{are} mainly dominated by finding mismatch information. After this step, \nspring{} uses back-end \ov{general-purpose} compression techniques to compress the mismatch information, while \pmain{} uses the procedure discussed in \sect{\ref{sec-sa:mech-alg}} to 
compress the mismatch information.
Second, genomics compressors (both \nspring{} and \pmain{}) have much longer compression times than general-purpose compressors due to the need to find mismatches. As mentioned in \sect{\ref{sec:background-compression}}, mismatch information enables genomics compressors to 
achieve substantially better compression ratios by
finding long-range similarity patterns that cannot be captured by general compressors.

\subsection{\ov{Resource Requirements}}
\label{sec-sa:eval-context}

Table~\ref{tab-sa:context} \ovi{quantifies} \proposal{}’s lightweightness and effectiveness \ov{by demonstrating the resource requirements of \proposal and other tools}. This table lists the state-of-the-art general-purpose and genomics-specific compression tools implemented on GPUs, FPGAs, ASICs, and CPUs.\footnote{We report compression ratio based on the ratio achieved by each tool’s compression algorithm on \emph{our} datasets. For decompression throughput, we report the best performance reported by the original works.}

\begin{table}[h]
\centering
\caption{Comparison \ovi{of} decompression tools.}
\label{tab-sa:context}
\resizebox{0.95\columnwidth}{!}{%
\color{black}
\begin{tabular}{m{8em}||m{5em}|m{3em}|m{2em}|m{12em}|m{5em}|m{6em}}
\toprule
\textbf{Tool}                  & \textbf{Genomics-specific} & \textbf{Avg. Comp. Ratio} & \textbf{End-to-end?} & \textbf{Hardware Requirements} & \textbf{Memory Footprint} & \textbf{Decomp. Throughput (GB/s)}\\
\midrule
\midrule
nvCOMP (DEFLATE)~\cite{nvcomp} & $\times$ & 5.3 & $\checkmark$ & \shortstack[l]{GPU\\ (NVIDIA A100~\cite{nvidiaa100})} & 1.5 GB & 50\\\midrule
Xilinx/AMD GZIP Engine~\cite{amdgzip} & $\times$ & 5.3 & $\checkmark$ & \shortstack[l]{FPGA\\ (AMD Alveo U50~\cite{amdalveou50})} & 80 KB & 0.7\\\midrule
xz~\cite{xz} & $\times$ & 6.7 & $\checkmark$ & \pbox{10em}{CPU\\ (AMD\textregistered{} EPYC\textregistered{}\\ 7742 CPU~\cite{amdepyc}\\ with 128 physical cores)} & 13 GB & 0.6\\\midrule
HW-accelerated ZStandard~\cite{choi2024hardware} & $\times$ & 6.7 & $\checkmark$ & \shortstack[l]{ASIC\\ (1.89 mm$^{2}$ \ov{in a}\\\ov{14nm \ovi{technology} node})} & 2--64 KB & 3.9\\\midrule
GPUFastqLZ~\cite{yang2025gpufastqlz} & $\checkmark$ & 5.8 & $\checkmark$ & \pbox{10em}{GPU\\ (A GPU server with four NVIDIA Tesla V100 GPUs ~\cite{nvidiateslav100})} & N/A\footnote{\ov{We cannot report this value since decompression performance is not reported by the paper and the FPGA implementation is not open-sourced.}} & 7.8\\\midrule
repaq~\cite{chen2023efficient} & $\checkmark$ & 17.1 & $\times$ & \shortstack[l]{FPGA\\ (AMD ALVEO U200~\cite{amdalveou200})} & 16 GB & N/A\\\midrule
Spring~\cite{chandak2018spring} \& NanoSpring~\cite{Meng2023} & $\checkmark$ & 16.9 & $\checkmark$ & \pbox{10em}{CPU\\ (AMD\textregistered{} EPYC\textregistered{}\\ 7742 CPU~\cite{amdepyc}\\ with 128 physical cores)} & 26 GB & 0.7\\\midrule
\proposal & $\checkmark$ & 15.8 & $\checkmark$ & \shortstack[l]{ASIC\\
(0.002 mm$^{2}$ \ov{in a}\\\ov{22nm \ovi{technology} node})} & \ov{128} B & 75.4\\
\bottomrule
\end{tabular}
}
\end{table}

We make three observations. 
First, leveraging its synergistic, data-aware co-design, \proposal yields the largest average decompression throughput. 
Second, \proposal achieves a significantly higher compression ratio than general-purpose decompressors and a comparable one to genomics-specific decompressors (and higher than the GPU-based genomics\ov{-specific decompressor~\cite{yang2025gpufastqlz}}). 
Third, \proposal eliminates the need for costly and power-hungry computational resources and large memory capacity or bandwidth since, unlike many \ov{de}compression techniques, it avoids frequent random accesses for matching patterns in large data structures.

\section{Summary}
\label{sec-sa:conclusion}

We propose \proposal, an algorithm-system co-design for \ov{highly-compressed} storage and \ov{high-performance} access of \ov{large-scale} genomic data, to \omcm{mitigate the data preparation
bottleneck \ov{in genome analysis}}. 
\ov{We leverage properties of genomic data to co-design \proposal{}'s
algorithm and architecture, such that highly-compressed data can be efficiently interpreted by lightweight hardware
and rapidly prepared for analysis.}
\ov{Due to its lightweight design, \proposal can be seamlessly integrated with a broad range of genome analysis systems to mitigate their data preparation bottlenecks and unlock the full potential of genome analysis acceleration.}
\ov{Our evaluations show that \proposal significantly improves the end-to-end performance and energy efficiency of state-of-the-art
genome sequence analysis accelerators, compared
to when the accelerators rely on state-of-the-art software
and hardware decompression tools.}

\subsection{\tomiii{Impact and Influence}}

\tomiii{SAGe is published at the IEEE International Symposium on High-Performance Computer Architecture (HPCA) in 2026~\cite{mansouri2026sage}. We plan to open-source SAGe to facilitate future research. An extended version of the HPCA 2026 paper is available on arXiv~\cite{mansouri2026sagearxiv}.

In this work, we show that addressing the data preparation bottleneck is crucial for unlocking the full potential of genome analysis acceleration. Our evaluations demonstrate that, due to its lightweight design, SAGe can be seamlessly integrated with a broad range of genome analysis systems to mitigate their data preparation bottlenecks and unlock the full potential of genome analysis acceleration. 

We hope that SAGe will inspire new studies and designs in this relatively unexplored research area.}

\chapter{Putting It All Together}
\label{chap:all-together}

In this dissertation, we introduce four novel storage-centric system designs for genomics and metagenomics: \tomii{GenStore, MegIS, GRAINS, and SAGe.} To alleviate the overheads of moving large amounts of low-reuse genomic and metagenomic data from the storage system to the rest of the system, we propose three storage-centric computing (SCC) designs that significantly reduce both storage data movement and the computational burden from the rest of the system: \inum{i}~GenStore: an in-storage processing system with low-cost and accurate in-storage filters, \inum{ii}~MegIS: an in-storage processing system designed to significantly reduce the data movement overhead of the end-to-end metagenomic analysis pipeline, and \inum{iii}~GRAINS: a system for analysis on large-scale genomic and metagenomic sequence graphs in storage. To mitigate the data preparation bottleneck in (meta)genomic analysis, we propose SAGe, an algorithm-architecture co-design for highly-compressed storage and high-performance access of large-scale genomic sequence data.

\sect{\ref{sec:together-motivation}} shows the motivation for a unified system that combines these designs. \sect{\ref{sec:together-design}} discusses the design of such a system, and \sect{\ref{sec:together-cost}} presents a cost-benefit analysis.

\section{Motivation for Combining the Proposed Designs into a Unified System}
\label{sec:together-motivation}

Combining the proposed designs into a unified system provides benefits in two key directions. We elaborate on these directions in \sect{\ref{sec:together-scc-compressed}} and \sect{\ref{sec:together-scc-adopt}}.

\subsection{Storage-Centric Computing on Compressed Data}
\label{sec:together-scc-compressed}
SAGe's lightweight design can be seamlessly implemented inside the SSD and integrated with SCC designs  to perform data preparation for them. To our knowledge, SAGe is the only data preparation approach that is high-performance, achieves high compression ratios (comparable to genomic-specific compression techniques), while being lightweight enough for seamless integration inside the SSD. Without SAGe, SCC systems would typically require data to be stored uncompressed, which is inefficient, or to decompress data outside the SSD, which undermines the fundamental benefits of SCC.

\subsection{Facilitating Wider Adoption}
\label{sec:together-scc-adopt}
Combining the proposed designs for genomics and metagenomics into one system can facilitate the wider adoption of these designs. This is due to two reasons. First, the combined design can have a broad applicability, as it can cover a range of tasks in both genomics and metagenomics, which are useful in many critical settings, such as clinical diagnostics, understanding disease outbreaks, agriculture, and more. Second, the combined design can reuse some common units (e.g., some registers used for k-mer double buffering next to each SSD channel), such that the area of the combined design is smaller than the sum of the area of all designs.

\section{Design of the Unified System}\todo{I think this section will benefit from a figure and more details}
\label{sec:together-design}

\subsection{Hardware Units} 
Due to their lightweight design, the proposed systems can \tomii{be} efficiently integrated within the same system. Since some hardware units (such as the double-buffering registers) are common across different designs, the unified system can share these components across the designs. For data preparation of read sets, SAGe hardware can be integrated in two ways. First, it can be integrated \tomii{into} the SSD controller (same as \circled{3} in \fig{\ref{fig:integration}}, as detailed in Chapter~\ref{chap:sage}), next to each channel, to prepare the data for cases where the read set is directly analyzed inside the SSD, such as GenStore. For cases where the query read sets need to be processed (i.e., via k-mer extraction and sorting) in the host system (e.g., MegIS and GRAINS), SAGe's hardware can be integrated \tomii{in} the host system such that the smaller compressed data moves from the SSD to the host system. In this case, SAGe can be integrated \tomii{in} the host system with configurations same as \circled{1} and \circled{2} in \fig{\ref{fig:integration}}, as detailed in Chapter~\ref{chap:sage}.

\subsection{FTL} 

For the proposed SCC systems (i.e., GenStore, MegIS, and GRAINS), an interface command\todo{be clear} (see \sect{\ref{sec:together-interface}}) can determine which FTL metadata should be loaded and used at the beginning of the execution of each acceleration \tomii{mode (i.e., GenStore, MegIS, GRAINS, and SAGe)}. Sections~\ref{sec-gs:ftl}, \ref{sec-mg:mech-ftl}, and \ref{sec-gr:mech-ftl-ecc} explain the FTL design of GenStore, MegIS, and GRAINS, respectively. Since these FTL metadata structures (which include the L2P address mapping of genomic and metagenomic databases and references) are small, they can co-exist with SAGe's lightweight FTL metadata (which includes the L2P address mapping of the compressed genomic and metagenomic read sets) in the SSD-internal DRAM.

\subsection{Interface Commands} 
\label{sec:together-interface}
The unified design can also simplify the interface commands. As opposed to having separate interface commands to mark the start of genomic or metagenomic acceleration modes (as discussed in Chapters~\ref{chap:genstore} to \ref{chap:grains}), the unified design has a unified command that signals to the SSD that it should start its operations in the storage-centric acceleration mode. A parameter in this command then specifies which exact acceleration mode \tomii{(i.e., GenStore, MegIS, GRAINS, and SAGe) should} be used. Regarding other commands, each design can orthogonally use its interface commands, as detailed in Sections~\ref{sec-gs:mechanism}, \ref{sec-mg:mech-interface}, \ref{sec-gr:mech-interface}, and \ref{sec-sa:mech-interface}.

\section{Cost-Benefit Analysis of the Unified System}
\label{sec:together-cost}

As mentioned in \sect{\ref{sec:together-scc-adopt}}, the area consumed by the unified design is smaller than the sum of the area consumed by each individual design. To find the area of the unified design, we convert the area of each design (as presented in Chapters~\ref{chap:genstore} to \ref{chap:sage}) to the same 32 nm technology node using the methodology presented in \cite{stillmaker2017Scaling}, and we consider that common units can be reused across the designs. The key common units are \inum{i}~the per-channel k-mer registers for double-buffering used in MegIS, GRAINS, and SAGe, and \inum{ii}~the per-channel comparators used in MegIS and GenStore. By accounting for these shared resources that can be reused across designs, we find that the area of the logic units on the SSD controller in the unified design is 0.08 mm$^2$, which is 6\% of the four ARM cores~\cite{arm_cortex_r8_ip} at 28nm technology node on an SSD controller~\cite{siliconmotion_sm2508_pb}. 

This lightweight unified design provides key benefits, including improved performance, energy efficiency, and overall system cost-effectiveness in genomic and metagenomic analysis (as shown for each design in this dissertation). 
As we evaluate in detail in \sect{\ref{sec-sa:evals-perf}}, SAGe's fast operations ensure that the pipeline of accessing NAND flash chips and performing data preparation is bottlenecked only by NAND flash throughput. Therefore, SAGe can be seamlessly integrated with GenStore, MegIS, and GRAINS to perform data preparation for them, enabling them to achieve their full benefits without requiring large-scale sequence data to be stored in uncompressed forms. 

\begin{figure}[h]
  \centering
    \includegraphics[width=\columnwidth]{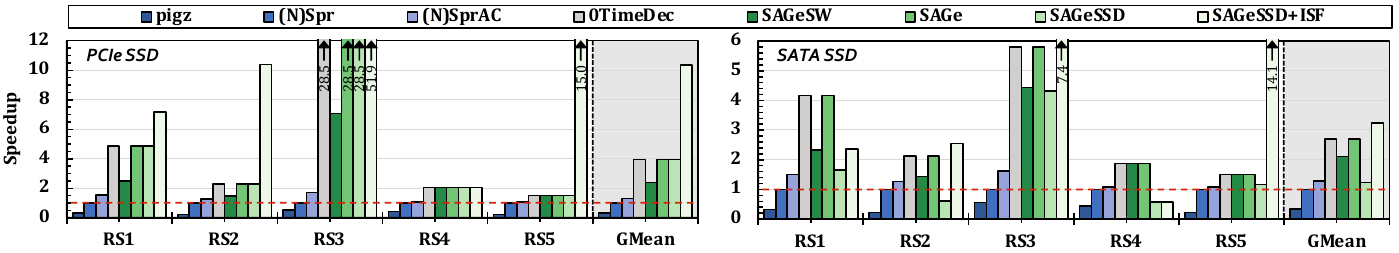}
  \caption{End-to-end speedup for different read sets.}
  \label{fig:eval-piat} 
\end{figure}

\aooo{As a case study, \fig{\ref{fig:eval-piat}} shows the performance evaluation of a configuration where SAGe is integrated with GenStore. 
In this figure, we demonstrate the end-to-end performance of various genome analysis systems, where execution includes \emph{both} data preparation and genome analysis.

For \textbf{data preparation}, we use
\inum{i}~\pigz: A parallel version~\cite{adler2015pigz} of gzip, a commonly-used general compressor;
\inum{ii}~\nspring: Spring~\cite{chandak2018spring} and NanoSpring~\cite{Meng2023}, state-of-the-art compressors for short and long reads, respectively. \pigz{} and \nspring{} run on a high-end system, an AMD$^\text{\textregistered}$ EPYC$^\text{\textregistered}$ 7742 CPU~\cite{amdepyc} with 128 physical cores (256 hardware threads);
\inum{iii}~\nsacc: \nspring integrated with a BWT-accelerator. There are various accelerators for BWT (e.g.,~\cite{qiao2019fpga,guo2013gpu,zhao2017streaming}). We consider an idealized accelerator that can fully eliminate the BWT execution time from \nspring;
\inum{iv}~\idec: an idealized decompressor (with zero decompression time), but inefficient for integration in resource-constrained environments (e.g., in our evaluations, for integration with an in-storage NDP genome sequence analysis system);
\inum{v}~\pmain: SAGe's full implementation, with its decompression in hardware;\footnote{For the analyses of the impact of different SAGe optimizations in isolation, please refer to Chapter~\ref{chap:sage}.}
and \inum{vi}~\ssdgi: \pmain with its hardware implemented in the SSD, to integrate with an NDP genome analysis system on the same chip (mode \circled{3} in \fig{\ref{fig:integration}}). 
Decompressors other than \ssdgi connect to the analysis systems using a PCIe interface (\circled{1} in \fig{\ref{fig:integration}}). We do not explicitly evaluate mode \circled{2}, because given sufficient interface bandwidth, mode \circled{2}  can perform the same as \circled{1}. As detailed in \sect{\ref{sec-sa:integration}}, \circled{1} enables standalone integration, while \circled{2} integrates on the same chip, thus saving PCIe slots. Chapter~\ref{chap:sage} provides more details behind the selection of data preparation baselines. 

For \textbf{genome sequence analysis}, we integrate all data preparation configurations with a state-of-the-art read mapping accelerator, GEM~\cite{chen2023gem}. To show SAGe's suitability for resource-constrained environments, we evaluate SAGe's implementation inside the SSD to integrate with GenStore's in-storage filters (ISFs). 
The resulting pipeline performs data preparation $\rightarrow$ ISF $\rightarrow$ read mapping. The benefits of this pipeline arise from both SAGe's faster preparation and the ISF operations in \cite{mansouri2022genstore}. The key to realizing this pipeline is that SAGe is the only data preparation configuration that is lightweight enough for efficient implementation inside the SSD. Without SAGe, ISF would require genomic data to be stored uncompressed, which is inefficient, or to decompress data outside SSD, which undermines the fundamental benefits of SCC. In this figure, speedup is normalized to \nspring. Chapter~\ref{chap:sage} provides further details about the system parameters and datasets.}

\aooo{We highlight four key observations. 
First, \proposal provides significant speedups. On the system with the PCIe (SATA) SSD, \pmain{} leads to 12.3$\times$ (8.1$\times$),  3.9$\times$ (2.7$\times$), and 3.0$\times$ (2.1$\times$) average speedup over \pigz{}, \nspring, and \nsacc, respectively. 
Second, \pmain{} matches \idec{} in performance because \pmain{} fully hides the decompression overhead in the execution pipeline. As explained in \sect{\ref{sec-sa:motivation}}, I/O, decompression, and read mapping are pipelined, so overall throughput depends on the slowest stage. \idec{} and \pmain{} having the same performance shows that decompression is no longer the slowest stage.
Third, by efficiently integrating \ssdgi{} with \isfmap{} (i.e., the in-storage filter implemented inside the resource-constrained environment of the SSD), \ssdgi{}+\isfmap leads to 7.8$\times$ (2.5$\times$) average speedups over \nsacc{} on the system with the PCIe (SATA) SSD. 
\ssdgi{}+\isfmap outperforms \pmain{} in all cases, except when \inum{i}~the input and application do not largely take advantage of \isfmap{} (i.e., in this case, \isfmap does not filter many reads in the read set), and \inum{ii}~the SSD's limited external bandwidth bottlenecks performance (e.g., RS1 and RS4 with the SATA SSD). In these cases, the \pmain~configuration should be used to decompress data outside the SSD to avoid moving larger decompressed data through the limited-bandwidth storage interface.
Fourth, regardless of how much decompression tools are optimized for performance, if their high resource requirements make them unsuitable for adoption in resource-constrained environments, they miss out on the benefits of a wide range of genome analysis systems that are implemented in such constrained environments. For example, \idec{} (i.e., an idealized decompressor with zero decompression time, but inefficient for integration in resource-constrained environments) cannot efficiently and cost-effectively integrate with \isfmap{} implemented inside the SSD and, as shown in our evaluations, it ends up being on average 1.8$\times$ (up to  9.9$\times$) slower than \ssdgi{}+\isfmap.}

Ultimately, we hope that the low-cost, unified storage-centric system consisting of different designs proposed in this dissertation would enable the wider adoption of accurate, data-intensive genomic and metagenomic analyses at low cost.\todo{add more comprehensive results}

\chapter{Conclusions and Future Directions}
\label{chap:conclusion}

In the pursuit of enabling high-performance, energy-efficient, and low-cost genomic and metagenomic analyses, this dissertation argues for storage-centric designs as a fundamental approach. 
In this dissertation, we identify that although there have been significant efforts to improve the analysis and storage of large-scale genomic and metagenomic data, there remain major outstanding problems in accessing stored sequence data and feeding it to the analysis units. The first problem is the overhead of \emph{moving large amounts of low-reuse data} from the storage system to main memory and computational units. We show that this overhead significantly hinders the end-to-end performance and energy-efficiency of (meta)genomic analysis due to storage I/O overheads and unnecessary computational burden on the rest of the system. The second problem we introduce and extensively analyze is the data preparation bottleneck, where compressed genomic sequence data needs to be first decompressed and formatted before it can be analyzed. We show that while hardware acceleration of genome analysis can potentially offer substantial performance benefits, a large part of these benefits gets amortized just by preparing the data and feeding it to the accelerators. 

To address these problems, we perform a set of research projects based on the thesis statement that \emph{“By designing customized storage-centric systems that efficiently 1) analyze genomic and metagenomic data inside the storage system, and 2) enable highly-compressed storage and high-performance access of large-scale sequence data, we can alleviate data movement overheads from the storage system, reduce the computational burden from the rest of the system, and mitigate the data preparation bottleneck, thereby significantly (e.g., by one to two orders of magnitude) improving system performance, energy efficiency, and cost of genomic and metagenomic analysis."} To this end, this dissertation introduces four novel storage-centric system designs for genomics and metagenomics.

First, we propose GenStore (Chapter~\ref{chap:genstore}), the first in-storage processing system designed for genomic analysis that significantly reduces both storage data movement and computational overheads of genomic analysis by exploiting low-cost and accurate in-storage filters. GenStore leverages hardware/software co-design to address the challenges of in-storage processing, supporting reads with 1) different properties such as read lengths and error rates, which highly depend on the sequencing technology, and 2) different degrees of genetic variation compared to the reference genome, which highly depend on the genomes that are being compared. Through rigorous analysis of read mapping processes of reads with different properties and degrees of genetic variation, we meticulously design low-cost hardware accelerators and data/computation flows inside the SSD. Our evaluation using a wide range of real genomic datasets shows that GenStore, when implemented in three modern NAND flash-based SSDs, significantly improves the read mapping
performance and energy efficiency of three state-of-the-art software (hardware) baselines, at low cost.  

Second, we propose MegIS (Chapter~\ref{chap:megis}), the first in-storage processing (ISP) system designed to significantly reduce the data movement overhead of the end-to-end metagenomic analysis pipeline. The key idea of MegIS is to enable cooperative ISP for metagenomics, where we do not solely focus on processing inside the storage system but, instead, we capitalize on the strengths of processing both inside and outside the storage system. We enable cooperative ISP via a synergistic hardware/software co-design between the storage system and the host system. We design MegIS as an efficient pipeline between the SSD and the host system to \inum{i}~leverage and \inum{ii}~orchestrate the capabilities of both. MegIS’s design is flexible, capable of supporting different types of metagenomic input datasets, and can be integrated into various metagenomic analysis pipelines. Our evaluation shows that MegIS significantly outperforms three state-of-the-art performance- and
accuracy-optimized software and hardware metagenomic tools, while matching the accuracy of the accuracy-optimized tool.

Third, we propose GRAINS (Chapter~\ref{chap:grains}), the first system for analysis on large-scale genomic and metagenomic sequence graphs in storage. Through our detailed examination of typical analysis pipelines on large-scale sequence graphs, we perform storage-aware algorithm-architecture co-design to \inum{i}~make the pipelines more storage-friendly and \inum{ii}~improve performance, energy-efficiency, and cost-effectiveness via in-storage and in-flash processing. GRAINS’s design is versatile and flexible as it supports major operations on sequence graphs and can be integrated into various analysis pipelines. Our evaluation shows that GRAINS provides significant speedup over the state-of-the-art software and hardware baselines, while providing significantly higher energy efficiency compared to these baselines.

Fourth, we propose SAGe (Chapter~\ref{chap:sage}) to mitigate the data preparation bottleneck, while maintaining high compression ratios (comparable to genomic-specific compression algorithms) and ensuring a lightweight design. To this end, we propose an algorithm-architecture co-design for highly-compressed storage and high-performance access of large-scale genomic sequence data. SAGe is designed based on the key insight that the information encoded in genomic data follows specific trends, shaped by factors such as sequencing technology (e.g., error rates and read lengths) and common genetic phenomena (e.g., typical spatial distributions of genetic variations within genomes). By carefully exploiting these characteristics to synergistically co-design algorithms and hardware, SAGe achieves high compression ratios, comparable to state-of-the-art genomic compressors, while enabling low decompression latencies, using only lightweight hardware and efficient streaming accesses. Due to its lightweight design, SAGe can be seamlessly integrated with a broad range of hardware accelerators for genome sequence analysis to mitigate their data preparation bottlenecks. Our results demonstrate that SAGe improves the average end-to-end performance and energy efficiency of two state-of-the-art genome sequence analysis accelerators compared to when the accelerators rely on state-of-the-art software and hardware decompression tools.

Across Chapters 4 to 7, we demonstrate that our storage-centric systems deliver these performance and energy benefits without relying on costly hardware resources throughout the system (e.g., external I/O bandwidth, large DRAM capacity, or extensive computational units). In Chapter~\ref{chap:all-together}, we demonstrate how these designs can be combined within a unified system to provide significant benefits in both genomic and metagenomic analyses, while maintaining low cost. We hope that the proposed designs make genomics and metagenomics more accessible for wider adoption.

\section{Future Research Directions}

In this dissertation, we make the case for storage-centric systems for genomics and metagenomics. We believe the ideas presented in this dissertation open up several promising avenues for future research. In this section, we provide an overview of potential directions for extending and enhancing the work presented in this dissertation.

\aooo{\subsection{\tomiv{Expanding} the Capabilities of In-Storage Filters for Genomic Data}}

\aooo{In Chapter~\ref{chap:genstore}, we present GenStore, which consists of new in-storage filters designed for filtering exactly-matching and non-matching reads. An important avenue for future research is to expand the capabilities of in-storage filters. For example, GenStore's exact-match filter (GenStore-EM) is currently designed for short reads, where a large fraction (e.g., on average 80\% of human short reads) exactly match the reference genome. GenStore-EM is not applicable to long reads because their greater length makes it extremely unlikely that an entire read will exactly match a subsequence of the reference genome, even with zero sequencing errors. However, as the accuracy of long read sequencing technologies continues to increase, large contiguous segments within a long read are likely to match the reference exactly. It would be beneficial to devise in-storage techniques that can identify and avoid moving these exactly-matching portions of long reads to the host system, transmitting only the segments that require approximate string matching. Realizing this filter poses new challenges: Since a full long read would not exactly match the reference, and the boundaries of exactly-matching regions are not known in advance, GenStore-EM cannot be adopted directly.  We believe that leveraging locality-sensitive hashing (LSH) can be a promising direction for efficiently realizing this filter.}

\aooo{\subsection{Expanding the Capabilities of Storage-Centric Systems for Metagenomic Analysis}}

\aooo{In Chapter~\ref{chap:megis}, we present MegIS, a host-SSD cooperative system for metagenomic analysis. In this chapter, we demonstrate and evaluate MegIS for two fundamental metagenomic tasks: presence/absence identification and abundance estimation, where the species present in a sample are not known in advance and must be identified by searching against a large reference database. Another important scenario arises when the goal is to detect the presence of specific (e.g., disease-causing) species, lineages, or alleles in a metagenomic sample. A prominent use case is wastewater-based epidemiology, where metagenomic sequencing of wastewater is used to monitor circulating pathogens and their variants within a community. For example, during the SARS-CoV-2 pandemic, wastewater-based epidemiology proved to be a cost-effective and rapid surveillance approach, capable of detecting emerging variants~\cite{polo2020making,bivins2020wastewater,gangwar2025wepp}. 
In these scenarios, the computational pipeline must efficiently search against a focused but frequently updated data structure to identify specific genomic signatures and their relative abundance. We believe it is a promising future direction to expand MegIS's capabilities to support such targeted metagenomic queries at a large scale. This would require designing new in-storage indexing structures optimized for these queries rather than exhaustive database scans, as well as mechanisms for accommodating frequent updates as new variants emerge. 
Tree-based approaches~\cite{turakhia2021ultrafast,gangwar2025wepp} have shown to provide significant benefits in lineage and variant identification in these scenarios, offering fine-grained resolution and compact data structures. We believe that techniques such as MegIS's KSS, as detailed in Chapter~\ref{chap:megis}, can further make these promising tree-based approaches more storage-friendly and better suited for storage-centric computing. KSS transforms dependent access patterns into sequential, streaming-friendly ones, which align well with the hardware constraints of storage-centric computing, particularly the long latency of NAND flash chips.}

\subsection{Storage-Centric Computing for Other Biological Data}

While our work introduces lightweight storage-centric computing (SCC) designs for analyzing genomic and metagenomic sequences, we believe SCC can also provide substantial benefits for analyzing other large-scale biological sequences (e.g., in transcriptomics~\cite{\citetranscriptomics} and single-cell analysis~\cite{\citesinglecell}). As demonstrated in this thesis, SCC can provide substantial benefits in improving the performance, energy efficiency, and system cost-effectiveness of genomics and metagenomics; factors that can ultimately facilitate the wider adoption of these data-intensive domains. As single-cell and spatial transcriptomics generate increasingly large, high-dimensional datasets, data movement overheads intensify, making SCC an important direction for investigation in these domains.

Effective SCC systems are enabled via algorithm-architecture co-design to optimize the algorithm and hardware synergistically based on unique properties of modern storage systems. Therefore, to design effective SCCs for other critical domains like transcriptomics and single-cell analysis, more research is needed to fundamentally examine the existing analysis pipelines and co-design the algorithm and hardware accordingly. We hope that the ideas presented in this work (e.g., storage-aware algorithmic optimizations, in-storage filters for low-reuse data, cooperative in-storage processing, and sequence data-aware encoding) can guide the design of future SCC systems for analyzing these other biological sequences.
For example, GenStore's in-storage filtering ideas can be expanded to filter RNA-seq reads against a reference genome, reducing unnecessary data movement and overall application burden.  

\subsection{Storage-Centric Designs for Constructing and Updating Large-Scale Sequence Databases and Graphs}

With the increasing rate of sequencing, the growth rate of (meta)genomic databases is increasing. This poses significant challenges in updating these databases, particularly in graph-based (meta)genomic analysis. Update to the underlying database or graph may currently require a full reconstruction of these in-storage data structures, which can incur significant preprocessing overhead in terms of time, energy, and system resource requirements (e.g., massive DRAM capacity). We suggest \tomii{that} future research explore storage-centric designs that support incremental and efficient updates to large-scale sequence databases and genome graphs without relying on expensive system resources. For example, co-designing storage data layouts and indexing schemes with the storage hardware could enable partial graph updates that leverage the internal bandwidth and parallelism of modern SSDs. The batching and scheduling techniques introduced by GRAINS, along with MegIS's cooperative in-storage processing approach, could serve as a foundation for designing such SCC systems for constructing and updating large-scale sequence databases and graphs.

\subsection{Highly-Compressed Storage and High-Performance Access of Other Biological Data Types}

We introduced an algorithm-architecture co-design for highly-compressed storage and high-performance access of large-scale genomic data. Given the large growth rate in other critical and emerging biological data types, we suggest future work in developing techniques for efficient storage and access of these other data types. We provide two key examples below.

\subsubsection{Efficiently Storing and Accessing Transcriptomic Data}

Transcriptomic data, generated by RNA sequencing (RNA-seq)~\cite{\citetranscriptomics}, shares several characteristics with genomic read data but also exhibits distinct features that pose new challenges and opportunities for compressed storage and high-performance access. For example, RNA-seq read sets have highly non-uniform coverage distributions, since transcript abundances vary widely across genes. Single-cell RNA-seq (scRNA-seq) introduces further complexities by collecting data from each cell in a tissue. Therefore, total data volumes from large-scale experiments (often comprising millions of cells) are growing rapidly. The key insight of our proposed design, SAGe, to exploit data-specific properties to co-design compression algorithms and lightweight hardware could be extended to transcriptomic data by leveraging the unique patterns of RNA-seq and scRNA-seq datasets to design new compression schemes that achieve high compression ratios while enabling fast, hardware-friendly decompression.

\subsubsection{Efficiently Storing and Accessing Raw Signal Genomic Data}

Sequencing technologies, such as Oxford Nanopore, generate raw electrical signals that are substantially larger than the base-called sequences derived from them. Raw signal genome analysis (RSGA), which directly analyzes these signals without basecalling, has emerged as a promising approach for genome analysis~\cite{\citesignalanalysis}.  Unlike basecalled reads encoded using a four-letter (A, C, G, T) nucleotide alphabet, raw signals consist of continuous electrical current measurements sampled at high frequency, producing floating-point time-series data with complex noise patterns. Extending SAGe's approach to raw signal data would require identifying and exploiting the unique statistical properties of these signals, such as characteristic current levels associated with different k-mers, temporal correlations between consecutive samples, and systematic noise patterns specific to the sequencing hardware. Such a co-designed compression and access scheme would be particularly impactful given the large and growing volumes of raw signal data and the increasing interest in RSGA.

\subsection{Genomic Privacy in the Era of Storage-Centric Computing}

Genomic data of an individual is among the most sensitive forms of information, as it can reveal not only the individual's identity but also their ancestry and family relationships. Strict privacy regulations are required to govern the storage, processing, and sharing of genomic data. These privacy requirements and regulations pose both challenges and opportunities for the adoption of storage-centric computing in genomics. 
On one hand, SCC offers a natural advantage for privacy-sensitive genomic analysis. In conventional systems, genomic data must be moved from storage to main memory and processors for analysis, exposing it across multiple system components and increasing the attack surface for potential data leakage. SCC fundamentally reduces this exposure by processing data where it resides, minimizing the movement of sensitive genomic information across the system. 
On the other hand, deploying SCC for genomic analysis in practice requires addressing several privacy-related challenges. First, regulatory frameworks may require auditability of data access within the storage device~\cite{\citeauditandprivacy}. Therefore, it is critical to explore how secure access logging can be integrated into SCC-enabled SSDs without compromising performance. Second, in many cases, it would be beneficial to store the data in an encrypted way. An efficient combination of SAGe's technique with encryption can further improve data preparation for encrypted genomic data. Third, in multi-tenant SSDs, it is critical to devise privacy-preserving mechanisms that prevent the SCC operations from corrupting other users' data. 

\tomiv{Recent works provide promising starting points toward private and secure storage-centric computing. For example, CIPHERMATCH~\cite{kabra2025ciphermatch} demonstrates that homomorphic-encryption-based exact string matching, including DNA string matching, can be efficiently accelerated using in-storage processing, thereby enabling computation directly on encrypted data. IceClave~\cite{kang2021iceclave} demonstrates a complementary direction by providing a lightweight trusted execution environment for in-storage computing. Building on these approaches, future work can explore co-designing privacy-preserving algorithms, trusted execution mechanisms, and storage-centric architectures to keep sensitive genomic data protected throughout storage and analysis.} Ultimately, we believe that fundamentally rethinking genomic data privacy in the era of SCC opens important avenues for future research to leverage SCC's unique opportunities while overcoming its challenges.

\tomii{\subsection{Facilitating Portable Genomic and Metagenomic Analysis} 

As sequencing technologies develop in the future, we expect that storage-centric design will play an important role in (meta)genomic  analysis.
We witness the trend that the DNA sequencing machines are increasingly adopting data processing capabilities to perform \omcv{\emph{both}} sequencing and analysis within the same machine~\omcv{\cite{wang2021nanopore,dunn2021squigglefilter}}. 
Portable devices enable (meta)genome analysis in the field, even in remote areas, and at the bedside for urgent personalized care. Examples of these analyses include agile viral detection (e.g., SARS-CoV-2) at remote locations or at the point-of-care~\cite{esbin2020overcoming}, nature conservation by analyzing the biodiversity of remote rain forests~\cite{pomerantz2018real}, acute medical or surgical management to prevent infant morbidity~\cite{farnaes2018rapid}.

With the push to provide DNA sequencing and preliminary genomic analysis on more portable integrated devices with limited compute resources and available DRAM capacity~\omcv{\cite{wang2021nanopore}}, there will be a growing demand for \omcvi{data movement-minimizing} systems like the storage-centric systems proposed in this thesis to \omcvi{quickly} filter out a large fraction of reads at low cost and \omcv{high efficiency}. 
Without requiring massive main memory capacity and a large number of processor cores, genome analysis can be done without relying on large-scale server nodes. For example, even mobile systems with small DRAM capacity can have large storage systems at low cost and can adopt SCC. Therefore, we believe leveraging SCC for portable (meta)genomic analysis system design is a promising future direction.}\todo{combining SCC and MCC}

\subsection{\tomiv{Combining Memory-Centric and Storage-Centric Computing for Efficient Biological Sequence Analysis}}

\tomiv{An important future direction is to synergistically combine memory-centric computing (MCC) and storage-centric computing (SCC) to reduce data movement throughout the system during biological sequence analysis. There are many possible ways to realize such systems using different technologies and integration approaches. We provide two key examples. First, it is possible to perform computation with different components present in modern SSDs, e.g., inside NAND flash chips, on the SSD controller, or using the internal DRAM. As shown by prior works (e.g.,~\cite{soysal2025mars,nadig2026conduit}), combining and efficiently coordinating these computation capabilities leads to significant benefits. Second, we can expand computation across the entire system, e.g., to the storage system, main memory, and cache hierarchy, so each part of the application runs where it benefits most. 
Future systems can explore heterogeneous pipelines that dynamically partition biological sequence analysis across the system based on factors such as data reuse, access patterns, locality, available bandwidth, and arithmetic intensity. We believe such hierarchical data-centric computing systems, which jointly exploit the capabilities of all parts of the system rather than optimizing either layer in isolation, provide a promising direction for fundamentally reducing data movement and enabling high-performance, energy-efficient, and cost-effective analysis of increasingly large and diverse biological datasets.}

\subsection{\tomiv{From New Technologies to New Capabilities in Health, Life Sciences, and Beyond}}

\tomiv{Emerging computing technologies can enable fundamentally new capabilities in healthcare and the life sciences that are impractical with conventional systems today. 
We suggest that future research explore how advances such as memory- and storage-centric computing, compute-capable memories, dense 3D integration, and tightly integrated sensing and computation can enable new capabilities (e.g., continuous and real-time biological analysis, richer personalized analyses, and sophisticated processing directly on portable or resource-constrained devices). We it is critical to co-design emerging technologies with applications, thereby identifying new analyses, workflows, and healthcare and life science capabilities that become possible when traditional bottlenecks on data movement, memory capacity, energy, and computation are fundamentally alleviated.}

\section{Concluding Remarks}

In this dissertation, we identify that although there have been significant efforts to improve the analysis and storage of large-scale genomic and metagenomic data, there are major outstanding problems in accessing stored sequence data and feeding it to the analysis units. These problems arise from \inum{i} the overhead of moving large amounts of low-reuse data from the storage system and the unnecessary burden on the rest of the system (e.g., main memory and computation units), and \inum{ii} the data preparation bottleneck, where compressed sequence data needs to be first decompressed and formatted before it can be analyzed. To alleviate data movement overheads and reduce the overall computational burden of low-reuse data,
we introduce three new storage-centric computing systems for (meta)genomics: GenStore (Chapter~\ref{chap:genstore}), MegIS (Chapter~\ref{chap:megis}), and GRAINS (Chapter~\ref{chap:grains}). To mitigate
the data preparation bottleneck, we propose SAGe (Chapter~\ref{chap:sage}), an algorithm-architecture co-design for highly-compressed storage and high-performance access of sequence data. We demonstrate that the proposed systems significantly improve system performance, energy efficiency, and cost-efficiency of (meta)genomic analysis. We hope that the storage-centric systems proposed in this dissertation facilitate the broader adoption of (meta)genomic analyses and inspire future research to fundamentally improve the performance, energy efficiency, and cost-effectiveness of other data-intensive application domains related to health and life sciences.\todo{add another point for portability}

\appendix
\cleardoublepage%
\chapter{Other Works of the Author}
\label{appendix:otherworks}

During the course of my graduate studies, I had the opportunity to work on various other projects across different fields. I acknowledge these works in four different categories.

\section{Architectures and Algorithms for Health and Life\linebreak Sciences}

\subsection{Other Work on Storage-Centric Computing for Genomics and Metagenomics} Beyond the work presented in this thesis, I had the pleasure of co-advising Melina Soysal in this direction. We developed \emph{MARS}~\cite{soysal2025mars}, a storage-centric computing system that leverages the heterogeneous resources available within modern storage systems (e.g., storage-internal DRAM, storage controller, flash chips) alongside their large storage capacity to tackle both data movement and computational overheads of raw signal genome analysis in an area-efficient and low-cost manner. 
I also led an overview paper\tomiv{~\cite{ghiasi2026enabling,ghiasi2026enablingarxiv}}, which summarizes our storage-centric designs for genomics and metagenomics and explains their key benefits. 

\subsection{Other Algorithm-Architecture Co-Designs} 
I had the pleasure of co-advising two works in this direction. I mentored Joel Lindegger in the development of \emph{Scrooge}, a fast and memory-frugal genomic sequence aligner~\cite{Lindegger2023scrooge}. I mentored Eirini
Tzermpou in architecting \tomiv{a new hardware/software} co-designed system for graph-based genome analysis, which is currently under submission. 
In collaboration with Damla Senol Cali, we proposed \emph{SeGraM}, a universal algorithm/hardware co-designed genomic mapping accelerator that can effectively and efficiently support both sequence-to-graph mapping and sequence-to-sequence mapping, for both short and long reads~\cite{cali2022segram}. 
In collaboration with Julien Eudine, we propose \emph{GenPairX}, a hardware-algorithm codesigned accelerator that efficiently minimizes the computational load of paired-end read mapping while enhancing the throughput of memory-intensive operations~\cite{eudine2026genpairx}.

\subsection{Algorithms and Software in Genomics} In collaboration with Can Firtina, we propose 1)~\emph{RawHash}, the first mechanism that can accurately and efficiently perform real-time analysis of nanopore raw signals for large genomes using a hash-based similarity search~\cite{firtina2023rawhash}, 
2)~\emph{Rawsamble}~\cite{firtina_rawsamble_2024}, the first mechanism that can identify regions of similarity between all raw signal pairs, known as all-vs-all overlapping, using a hash-based search mechanism,
and 3)~\emph{BLEND}, the first efficient and accurate mechanism that can identify both exact-matching and highly similar seeds with a single lookup of their hash values, called fuzzy seed matches~\cite{firtina_blend_2023}. 
\tomiv{With Joel Lindegger, we introduce \emph{RawAlign}, which is accurate, fast, and scalable raw nanopore signal mapping via combining seeding and alignment~\cite{lindegger2023rawalign}}
With Meryem Banu Cavlak, we propose \emph{TargetCall}, the first pre-basecalling filter to eliminate the wasted computation in basecalling~\cite{cavlak2022targetcall,cavlak_targetcall_2024}. 
In collaboration with Harun Mustafa, we propose \emph{MetaGraph-MLA}, label-guided alignment to variable-order de Bruijn graphs~\cite{Mustafa2024MLA}. In collaboration with Serghei Mangul, we 1) thoroughly examine the methodological advancements and current practices that have shaped metagenomic analysis~\cite{liu2025analysis} and 2) propose a framework to enable a decentralized future of open-science databases~\cite{sharma2026towards}.

\subsection{Architecture for Health (Arch4Health) Initiative}

I had the pleasure of leading the organization of  \textbf{Architecture for Health (Arch4Health)} initiative\tomiv{~\cite{ghiasi2026architecture,ghiasi2026architecturearxiv}}, which aims to (\emph{i})~identify and analyze key computational challenges in current and future health- and life science-related applications and (\emph{ii})~explore how computer architects and computing system designers can advance healthcare by addressing these challenges. We first present the motivations behind the Arch4Health initiative and, second, elaborate on its vision and goals, related topics, Arch4Health workshops, and future outlooks.

\subsubsection{Motivation}
\label{sec:motivation}

Recent advances in biotechnology and sensing technologies have enabled high-throughput, low-cost, and accurate biological data generation.
Modern sequencing platforms~\cite{turcatti_new_2008,fuller_method_2011,eid_real-time_2009,deamer_three_2016,zhang_single-molecule_2024} and other omics technologies~\cite{dai2022advances} can generate massive amounts of biological data (genomics~\cite{doblas2025smx,mutlu2023accelerating,alser2022molecules,shahroodi2023swordfish,subramaniyan2021accelerated,khatamifard2021genvom,li2021pim,turakhia2018darwin,chen2021high,haghi2021fpga,li2021pipebsw,ham2021accelerating,cali2020genasm,Zhang_2023_alignerD,soysal2025mars,kim2018grim,mao2022genpip,dphls2026,Walia2024talco,sadasivan2024genomic,Turakhia2025toward,Turakhia2019darwinwga,simon2026processing,eudine2026genpairx,Lindegger2023scrooge,pockrandt2022metagenomic,song2024centrifuger,Fan2021}, transcriptomics~\cite{chen2023hitchhikers,Sibbesen2023,haas_forensic_2021,clough2023ncbigeo}, proteomics~\cite{mallick2010proteomics,aslam2016proteomics,cho2007proteomics,patterson2003proteomics,graves2002molecular,pandey2000proteomics,Han2025lightnobel}, and metabolomics~\cite{alseekh2021mass,perez2019quantifying,liu2017metabolomics,zhang2012modern,johnson2012challenges}) at rapidly decreasing costs. Similarly, multimodal medical imaging technologies~\cite{marti2010multimodality,kasban2015comparative,shung2012principles,glover2011overview,kapoor2004introduction,townsend2004physical}
produce high-resolution data that capture complex physiological and pathological processes. Wearable and implantable sensing devices~\cite{pang2013recent,koydemir2018wearable,chan2012smart,lukowicz2004wearable,bonato2010wearable} continuously monitor physiological signals such as heart activity, glucose levels, and oxygen saturation in real time. Together, these advances have created an unprecedented volume of heterogeneous health- and life science-related data.

This wealth of data creates unique opportunities for advancing healthcare and biomedical discovery, such as precision medicine~\cite{clark2019diagnosis,farnaes2018rapid,sweeney2021rapid,alkan2009personalized,flores2013p4,ginsburg2009genomic,chin2011cancer,Ashley2016}, where treatments and therapeutic strategies can be tailored to the genetic and physiological characteristics of individual patients. Large-scale genomic and clinical datasets also advance 
personalized medicine~\cite{branco_bioinformatics_2021,quazi_artificial_2022,papadopoulou_application_2023,tafazoli_applying_2021,moon_precision_2022,hussen_emerging_2022,russell_pharmacogenomics_2021,verma_nanopore_2024,sweeney2021rapid}, tracking outbreaks of communicable diseases~\cite{li_application_2021,deng_integrated_2021,bloom2021massively,yelagandula2021multiplexed}, cancer research~\cite{jia_high-throughput_2022,liu_mrna-based_2023,van_de_sande_applications_2023,chakravarty_clinical_2021,cortes-ciriano_computational_2022,deveson_evaluating_2021,xiao_toward_2021,szustakowski_advancing_2021,lei_applications_2021,stadler_therapeutic_2021,tan_targeted_2022,degasperi_substitution_2022,xu_single-cell_2022,horak_comprehensive_2021,waarts_targeting_2022,xiao_tumor_2021,nandwani_lncrnas_2021,marchetti_error-corrected_2023,chen_next-generation_2021}, agriculture~\cite{campos_high_2021,gao_genome_2021,van_dijk_machine_2021,sun_twenty_2022,thudi_genomic_2021,shen_omics-based_2022,prasad2021soil,Mascher2024,Schreiber2024}, ensuring food safety~\cite{e002244,TONG2021130},  scientific discovery~\cite{urbanek2018degradation,edgar2022petabase,paoli2022biosynthetic}, and biodiversity conservation~\cite{Hogg2024,lewin2018earth}.
Continuous physiological monitoring via wearable sensors further enables proactive and preventive healthcare by enabling early detection of anomalies~\cite{sopic2018real} and timely clinical intervention~\cite{guk2019evolution,tyler2020real}.

Despite these opportunities, efficiently analyzing large-scale biological data poses significant challenges for conventional computing systems. First, these systems often cannot keep up with the high-throughput rate at which data is generated. For example, modern sequencing platforms can generate data at rates that create substantial computational bottlenecks in downstream analysis, including sequence alignment and variant calling~\cite{zhang2021real,firtina2023rawhash,kovaka2020targeted, mutlu2023accelerating,Payne2021,Bao2021Squigglenet,ulrich2022readbouncer}. Similarly, images are generated at a pace that exceeds the throughput of  image reconstruction analytics and machine learning-based inference. High-throughput processing is crucial in clinical settings, where real-time data processing can significantly impact patient outcomes by improving both response times in time-critical scenarios and the decision-making processes for therapeutic schemes~\cite{kakria2015real}.

Second, health- and life science-related applications suffer from data movement overheads~\cite{wu2021sieve,shahroodi2022krakenonmem,shahroodi2022demeter,dashcam23micro,hanhan2022edam,zou2022biohd,cali2020genasm, huangfu2018radar, khatamifard2021genvom, gupta2019rapid, li2021pim, angizi2019aligns,zokaee2018aligner,Zhang_2023_alignerD,cali2022segram,kim2018grim,kaplan2020bioseal,mao2022genpip,alser2022molecules,mutlu2023accelerating,mansouri2022genstore,abakus23taco,megis,jun2016storage,kim2025nmp,soysal2025mars,zheng2025storage,grains,grainsextended,mutlu2025memory,mutlu2022modern,mutlu2019processing,mutlu2019enabling,cali2017nanopore}. These workloads frequently involve irregular memory access patterns~\cite{yelick2020,langarita2022,kim2025nmp,robinson2021}, massive data transfer overheads~\cite{simon2026processing,ghose2019processing}, complex graph and statistical computations~\cite{li2022graph,yi2022graph}, and often rely on computationally intensive machine learning models~\cite{yuan2025ml}. As a result, data movement often dominates execution time and energy consumption. As datasets continue to grow in size and complexity, this bottleneck is further exacerbated~\cite{khan2020}. Growing adoption of deep learning models is also expected to amplify these challenges by requiring frequent movement of large model parameters and intermediate data across the computing stack~\cite{gholami2024}.

Third, systems must process sensitive patient data while satisfying strict privacy and regulatory requirements for health applications~\cite{rieke2020future,xu2021federated}. Hospitals, portable diagnostic devices and wearable platforms need to meet such requirements through secure data storage, trusted computation, federated learning, and privacy-preserving analytics~\cite{kaissis2020secure,sheller2020federated,zhang2025survey,kumar2025priv}. 

These challenges motivate the need for a closer collaboration between the health- and life science-related domain (e.g., clinical and precision medicine, omics research, genetics, computational biology, drug discovery, public health, and epidemiology), computer architecture, and computing systems design communities. Advancing healthcare applications requires rethinking computing system design across the entire stack, including hardware accelerators, memory hierarchies, storage systems, data movement mechanisms, distributed infrastructures, and privacy-aware architectures. 
Prior work has already demonstrated the promise of this direction. A growing body of computer architecture research has proposed algorithmic optimizations (e.g.,~\cite{li2018minimap2,marco2021fast,marcosola2023optimal,grootkoerkamp2024apa2,sadasivan2024genomic,tseng2025ultrafast,walia2025ultrafast,kim_fastremap_2022,Ashyralyyev2026gencore,ashyralyyev2026lcpan,alicioglu2024pairwise,manber1993suffix,siren2021pangenomics,Chandra2023,Ivanov2022,Ma2023,Hwang2025MEMO,Romain2023svjedi,simon2026processing,simon2025processing,altschul1990basic,breitwieser2022biodynamo,breitwieser2025design,breitwieser2023high,breitwieser2025teraagent}) or hardware accelerators (e.g.,~\cite{doblas2025smx,mutlu2023accelerating,alser2022molecules,shahroodi2023swordfish,subramaniyan2021accelerated,khatamifard2021genvom,li2021pim,turakhia2018darwin,chen2021high,haghi2021fpga,li2021pipebsw,ham2021accelerating,cali2020genasm,Zhang_2023_alignerD,soysal2025mars,kim2018grim,mao2022genpip,dphls2026,Walia2024talco,sadasivan2024genomic,Turakhia2025toward,Turakhia2019darwinwga,eudine2026genpairx,Lindegger2023scrooge,cali2022segram,Zhang2024Harp,Zeng2024asgdp,Shen2024128parallel,Li2024,Awan2021,Feng2021,Zhang2025,kim2025nmp,Huang2023meg2,Zhou2021,Sarkar2021,Galanos2021,Sinha2022,Chen2023,simon2026processing,simon2025processing,koliogeorgi2022gandafl,koliogeorgi2023profile,koliogeorgi2019dataflow,tsoutsouras2017exploration}), and/or reducing data movement overheads~\cite{mutlu2025memory,mutlu2022modern,mutlu2019processing,mutlu2019enabling,senol2021accelerating} via \emph{near-data processing} (e.g., in main memory~\cite{singh2021fpga,wu2021sieve,shahroodi2022krakenonmem,shahroodi2022demeter,dashcam23micro,hanhan2022edam,zou2022biohd,cali2020genasm, huangfu2018radar, khatamifard2021genvom, gupta2019rapid, li2021pim, angizi2019aligns, zokaee2018aligner,Zhang_2023_alignerD,soysal2025mars,cali2022segram,kim2018grim,kaplan2020bioseal,mao2022genpip,shahroodi2023swordfish,alonso2024bimsa,diab2022high} or storage~\cite{thesis,mansouri2022genstore,arxivGS,abakus23taco,megis,megisarxiv,jun2016storage,kim2025nmp,soysal2025mars,zheng2025storage,grains,grainsextended,mansouri2026sage,mansouri2026sagearxiv})
for workloads such as genome sequence analysis, metagenomic profiling, medical imaging, real-time physiological monitoring and biological simulation, providing substantial improvements in performance, energy efficiency, and scalability over conventional CPU- and GPU-based pipelines. 
Despite these promising advances, significant challenges remain unresolved and even bigger strides are necessary.
Computer architecture research, done collaboratively with experts in healthcare and life sciences, can play a key role in addressing these challenges and in enabling high-performance, energy-efficient, secure, and scalable computing systems for healthcare and life sciences.

\subsubsection{The Arch4Health Initiative}
\label{sec:arch4health}

\textbf{Vision and Goals.}
We introduce the \emph{Architecture for Health} \emph{(Arch4Health)} initiative, which aims to (i)~identify key computational challenges in current and future health- and life science-related applications and (ii)~explore how computer architects and computing system designers can advance healthcare by addressing these challenges. Since cross-disciplinary discussions are crucial for better identifying and solving challenges in real-world health- and life science-related applications, we aim to foster open discussions and cooperation between researchers with diverse backgrounds (i.e., from both computer architecture and health sciences communities, industry, and academia).

\noindent\textbf{Topics.}
Arch4Health invites contributions and collaborations on a broad range of topics at the intersection of computer architecture and health- and life science-related applications. These topics include, but are not limited to, computational biology (e.g., genomics, metagenomics, transcriptomics, single-cell analysis, spatial omics, proteomics,  drug design and discovery, gene editing, and other areas in precision medicine and public health), neuroscience (e.g., brain-machine interfaces and prosthetics), wearable systems for health, medical robotics (e.g., surgery and haptics), mental health, medical imaging (e.g., brain scans, radiology, and single-cell analysis), computational and digital pathology,
artificial intelligence and foundation models for biology and health (e.g., protein
language models, cell foundation models, and large language models
for clinical applications), agent-based simulations, medical privacy, and bio-sensors. We particularly welcome contributions that identify computational challenges in these domains and propose new ideas (architectures and systems together with algorithms) %
to improve performance, energy efficiency, cost-effectiveness, and privacy.

\noindent\textbf{Arch4Health Workshops.}
The primary instrument through which Arch4Health currently pursues its goals is a series of workshops co-located with various computer architecture and computing systems conferences. These workshops are designed to bring together researchers and practitioners from both the computer architecture and systems community and the broader biomedical and healthcare communities in a single venue, and to encourage substantive technical exchange and visibility. To this end, each workshop combines two complementary components. First, the workshops provide invited talks and keynotes that summarize (\emph{i})~a series of research in computing system designs for healthcare applications and (\emph{ii})~new ideas and directions in data-intensive healthcare applications. Second, the workshops invite researchers to submit their ongoing work on these topics, with a focus on new ideas. All workshops are livestreamed and available on YouTube~\cite{onurmutlulectures,firsta4h,seconda4h} to enable online, broad and unrestricted access to the entire world.

The first two editions of Arch4Health were held in conjunction with the IEEE/ACM International Symposium on Microarchitecture (MICRO) 2025 in Seoul and the IEEE International Symposium on High-Performance Computer Architecture (HPCA) 2026 in Sydney, and together they engaged a broad community spanning computer architecture, bioinformatics, and biomedical research. The MICRO 2025 edition~\cite{arch4health-micro2025,firsta4h} featured ten invited talks covering new algorithms, software, and hardware designs for brain-computer interfaces, medical wearables, genomics, metagenomics, proteomics, and agent-based simulation in life sciences. The HPCA 2026 edition~\cite{arch4health-hpca2026,seconda4h} expanded the program to a full-day format with ten talks covering new algorithms, software, and hardware designs for genomics and metagenomics, electrical signals in genomics, AI and algorithms in biology, and indexing and querying petabyte-scale biological sequences. Across both editions, speakers came from a diverse set of academic and research institutions around the world.

The next edition of Arch4Health~\cite{arch4health-ics2026} will be held in conjunction with the ACM International Conference on Supercomputing (ICS) 2026 in Belfast, Northern Ireland, United Kingdom. This edition further expands the workshop series by fostering interdisciplinary collaboration among computer architecture, high-performance computing, and health and life science communities. The workshop program features invited talks and discussions on emerging algorithms for genome analysis, scalable architectures, hardware-software co-design for proteomics and genomics and real-time monitoring in clinical environments. We hope this edition will further solidify the importance of the initiative and lead to exciting synergies and research directions.

Building on the success of the Arch4Health~\cite{arch4health-hpca2026,arch4health-micro2025} workshop series, we are expanding the scope toward the system software and system design domain by organizing the Sys4Health workshop~\cite{sys4health-sosp2026}, co-located with the 32nd Symposium on Operating Systems Principles conference (SOSP 2026). Sys4Health aims to bring together the systems and health and life sciences communities to explore how systems and software infrastructure can enable scalable, secure, and efficient health and life science applications.

\noindent\textbf{Future Outlook and a Call to Action}
The challenges and opportunities at the intersection of computer architecture and health- and life science-related domains are broad and pressing. Therefore, realizing the vision and goals of Arch4Health requires diverse and cross-disciplinary discussions. We invite computer architects and computing system designers to view health-related applications as a rich and important domain for architectural innovation, and we invite biomedical researchers and clinicians to engage with the architecture community in shaping the systems that will support the next generation of healthcare. 

Arch4Health is intended to grow as an open and inclusive initiative. We welcome contributions, ideas, and collaborations across multiple communities. Engagement and synergy between both academia and industry is essential to identify key challenges and pave the way toward impactful, practical solutions. Through continued workshops, joint research efforts, and community-building activities, we hope Arch4Health will help grow \emph{architecture for health} as a key area of computer architecture, whose advances translate into substantial impact in advancing healthcare and life sciences.

\section{Data-Centric Computing}

\subsection{Other Works on Storage-Centric Computing} 
I had the pleasure of co-advising Kangqi Chen on developing \emph{REIS}, a high-performance and energy-efficient retrieval system with in-storage processing~\cite{chen2025reis}. I also had the opportunity to collaborate with Rakesh Nadig on developing \emph{Conduit}, a general-purpose, programmer-transparent near-data processing (NDP) framework for SSDs that accelerates a broad range of workloads by leveraging available SSD computational resources~\cite{nadig2026conduit}. 

\subsection{Memory-Centric Computing and 3D Integration Technologies} I led the development of three works in this direction as the first author. 
In the first work, we introduce \emph{ALP}, a new programmer-transparent technique to leverage the performance benefits of NDP by alleviating the performance impact of inter-segment data movement between host and NDP units and enabling efficient partitioning of applications between host and NDP cores~\cite{mansourighiasi2023alp}.
In the second work, \emph{RevaMp3D}, we architect the processor core and cache hierarchy for systems with monolithically-integrated logic and memory~\cite{mansourighiasi2026revamp3d}. 
In the third work, we understand and model the performance landscape of ultra-Dense 3D memory systems across different dense 3D integration technologies. This work is currently ongoing. I presented an early version of this work in the ACM Student Research Competition (SRC) at PACT 2024.

I also had the opportunity to collaborate on various other works in this direction. In collaboration with Geraldo F. Oliveira and Nastaran Hajinazar, we propose \emph{SIMDRAM}, a flexible general-purpose processing-using DRAM framework that enables the efficient implementation of complex operations and provides a flexible mechanism to support the implementation of arbitrary user-defined operations~\cite{hajinazarsimdram}. 
\tomiv{With Ivan Fernandez, we propose \emph{MATSA}, an MRAM-based energy-efficient accelerator for time series analysis~\cite{fernandez2024matsa}}.
In collaboration with Nisa Bostanci and Konstantinos Kanellopoulos, we introduce \emph{IMPACT}, a set of high-throughput main memory-based timing attacks that leverage characteristics of processing-in-memory (PiM) architectures to establish covert and side channels~\cite{bostanci2025revisiting}.

\section{Optimizing Memory and Storage Systems} In collaboration with Yaohua Wang, we 1) design a new substrate, \emph{FIGARO}, which improves system performance via fine-grained in-DRAM data relocation and caching~\cite{wang2020figaro}, and 2) propose charge-level-aware look-ahead partial restoration (\emph{CAL}) to enable effective latency reduction for both activation and restoration operations in DRAM~\cite{wang2018reducing}. 
In collaboration with Arash Tavakkol, we propose the flash-level interference-aware scheduler (\emph{FLIN}) to alleviate unfairness in MQ-SSDs~\cite{tavakkol2018flin}.
In collaboration with Lois Orosa, we design CODIC, a low-cost substrate for enabling custom in-DRAM functionalities and optimizations~\cite{orosa_codic_2021}. In collaboration with Hasan Hassan, we introduce Copy-Row DRAM (\emph{CROW}), a flexible substrate that enables new mechanisms for improving DRAM performance, energy efficiency, and reliability~\cite{hassan2019crow}. In collaboration with Rakesh Nadig and Mohammad Sadrosadati, we propose \emph{Venice}, a new mechanism that introduces a low-cost interconnection network between the SSD controller and memory chips and utilizes the path diversity to intelligently resolve path conflicts in modern SSDs~\cite{nadig2023venice}.

\section{Other Algorithm-Architecture Co-Designs} 
In collaboration with Yu Liang, we introduce a hotness-aware and size-adaptive compressed swap scheme, \emph{Ariadne}, for mobile devices to mitigate relaunch latency and reduce CPU usage~\cite{liang2025ariadne}.
In collaboration with Konstantinos Kanellopoulos, we propose \emph{SMASH}~\cite{kanellopoulos2019smash}, a hardware-software cooperative mechanism that enables highly efficient indexing and storage of sparse matrices. 

\chapter{Complete List of the Author's Contributions}
\label{appendix:complete_list}
This section lists the author's contributions to the literature in reverse chronological order under four categories: 
(1)~major contributions that the author led, 
(2)~summary and initiative contributions led by the author,
(3)~co-supervised contributions by the author, 
and (4)~other contributions.

\section{Major Contributions Led by the Author}
\label{appendix:sec:first_author_contributions}

\begin{enumerate}

\item \textit{\textcolor{denim}{\textbf{GRAINS: Enabling High-Performance and Low-Cost Graph-Based Genome Analysis via Storage-Aware Algorithm-Architecture Co-Design}}}, \textbf{\footnotesize [ISCA, 2026]}
\\
\underline{N. Mansouri Ghiasi}, H. Mustafa, T. G\"uloglu, R. Nadig, K. Koliogeorgi, S. Ruiz, M. Rautmann, F. Eris, M. Sadrosadati, J. Park, O. Mutlu

\item \textit{\textcolor{denim}{\textbf{RevaMp3D: Architecting the Processor Core and Cache Hierarchy for Systems with Monolithically-Integrated Logic and Memory}}}, \textbf{\footnotesize [ACM TACO, 2026]}
\\
\underline{N. Mansouri Ghiasi}, M. Sadrosadati, G. F Oliveira, K. Kanellopoulos, R. Ausavarungnirun, J. Gómez Luna, A. Manglik, J. Ferreira, J. S Kim, C. Giannoula, N. Vijaykumar, J. Park, O. Mutlu

\item \textit{\textcolor{denim}{\textbf{SAGe: A Lightweight Algorithm-Architecture Co-Design for Mitigating the Data Preparation Bottleneck in Large-Scale Genome Sequence Analysis}}} \textbf{\footnotesize [HPCA, 2026]}
\\
\underline{N. Mansouri Ghiasi}, T. G\"uloglu, H. Mustafa, C. Firtina, K. Koliogeorgi, K. Kanellopoulos, H. Mao, R. Nadig, M. Sadrosadati, J. Park, O. Mutlu

\item \textit{\textcolor{denim}{\textbf{MegIS: High-Performance, Energy-Efficient, and Low-Cost Metagenomic Analysis with In-Storage Processing}}}, \textbf{\footnotesize [ISCA, 2024]}
\\
\underline{N. Mansouri Ghiasi}, M. Sadrosadati, H. Mustafa, A. Gollwitzer, C. Firtina, J. Eudine, H. Ma, J. Lindegger, M. Banu Cavlak, M. Alser, J. Park, O. Mutlu

\item \textit{\textcolor{denim}{\textbf{GenStore: A High-Performance In-Storage Processing System for Genome Sequence Analysis}}}, \textbf{\footnotesize [ASPLOS, 2022]}
\\
\underline{N. Mansouri Ghiasi}, J. Park, H. Mustafa, J. Kim, A. Olgun, A. Gollwitzer, D. Senol Cali, C. Firtina, H. Mao, N. Almadhoun Alserr, R. Ausavarungnirun, N. Vijaykumar, M. Alser, and O. Mutlu

\item \textit{\textcolor{denim}{\textbf{ALP: Alleviating CPU-Memory Data Movement Overheads in Memory-Centric Systems}}}, \textbf{\footnotesize [IEEE TETC, 2022]}
\\
\underline{N. Mansouri Ghiasi}, N. Vijaykumar, G. F Oliveira, L. Orosa, I. Fernandez, M. Sadrosadati, K. Kanellopoulos, N. Hajinazar, J. Gómez Luna, O. Mutlu

\end{enumerate}

\section{Summary and Initiative Papers Led by the Author}
\label{appendix:sec:summary-contributions}

\begin{enumerate}

\item \textit{\textcolor{denim}{\textbf{Architecture for Health Initiative (Arch4Health):
Computational Challenges in Health-Related Applications
and the Role of Computer Architecture in Addressing Them}}}, \textbf{\footnotesize [ICS Workshops, 2026]}
\\
\underline{N. Mansouri Ghiasi}*, K. Koliogeorgi*, O. Mutlu

\item \textit{\textcolor{denim}{\textbf{Enabling Fast, Efficient, and Low-Cost Genomic and Metagenomic Analyses via Storage-Centric System Designs}}}, \textbf{\footnotesize [ICS Workshops, 2026]}
\\
\underline{N. Mansouri Ghiasi}, O. Mutlu

\end{enumerate}

\section{Co-supervised Contributions}
\label{appendix:sec:co_supervised_contributions}

\begin{enumerate}

\item \textit{\textcolor{denim}{\textbf{\tomiv{A New} Software/Hardware Co-Design for
Graph-Based Genome Analysis}}} \textbf{\footnotesize [Under Submission, 2026, Blinded]}
\\
E. M. Tzermpou, K. Koliogeorgi, A. Ashyralyyev, S. Xydis, D. Soudris, \underline{N. Mansouri Ghiasi}, O. Mutlu

\item \textit{\textcolor{denim}{\textbf{REIS: A High-Performance and Energy-Efficient Retrieval System with In-Storage Processing}}} \textbf{\footnotesize [ISCA, 2025]}
\\
K. Chen, R. Nadig, A. Kosmas Kakolyris, M. Frouzakis, \underline{N. Mansouri Ghiasi}, Y. Liang, H. Mao, J. Park, M. Sadrosadati,  O. Mutlu

\item \textit{\textcolor{denim}{\textbf{MARS: Processing-In-Memory Acceleration of Raw Signal Genome Analysis Inside the Storage Subsystem}}} \textbf{\footnotesize [ICS, 2025]}
\\
M. Soysal, K. Koliogeorgi, C. Firtina, \underline{N. Mansouri Ghiasi}, R. Nadig, H. Mao, G. Francisco, Y. Liang, K. Zambaku, M. Sadrosadati,  O. Mutlu

\item \textit{\textcolor{denim}{\textbf{Scrooge: A Fast and Memory-Frugal Genomic Sequence Aligner for CPUs, GPUs, and ASICs}}}, \textbf{\footnotesize [Bioinformatics, 2023]}
\\
J. Lindegger, D. Senol Cali, M. Alser, J. Gómez-Luna, \underline{N. Mansouri Ghiasi}, O. Mutlu

\item \textit{\textcolor{denim}{\textbf{TargetCall: Eliminating the Wasted Computation in Basecalling via Pre-Basecalling Filtering}}}, \textbf{\footnotesize [APBC, 2023]}
\\
M. B. Cavlak, G. Singh, M. Alser, C. Firtina, J. Lindegger, M. Sadrosadati, \underline{N. Mansouri Ghiasi}, C. Alkan, O. Mutlu

\end{enumerate}

\section{Other Contributions}
\label{appendix:sec:other_contributions}

\begin{enumerate}

\item \textit{\textcolor{denim}{\textbf{Towards a decentralized future for  open-science databases}}}, \textbf{\footnotesize [Nature Genetics Comment, 2026]}
\\
G. Sharma, V. Munteanu, \underline{N. Mansouri Ghiasi}, U. Mahanta, J. Banerjee, S. Varma, L. Foschini, K. Ellrott, O. Mutlu, D. Ciorbă, R. A. Ophoff, V. Bostan, J. H. Moore, D. Sousoni, A. Krishnan, A. G. Lucaci, A. Tull, C. E. Mason, M. Dimian, G. Stolovitzky, F. G. Liberante, T. K. Oleksyk, S. Mangul

\item \textit{\textcolor{denim}{\textbf{Rawsamble: Overlapping and Assembling Raw Nanopore Signals using a Hash-based Seeding Mechanism}}}, \textbf{\footnotesize [Bioinformatics, 2026]}
\\
C. Firtina, M. Mordig, H. Mustafa, S. Goswami, \underline{N. Mansouri Ghiasi}, S. Mercogliano, F. Eris, J. Lindegger, A. Kahles, O. Mutlu

\item \textit{\textcolor{denim}{\textbf{Conduit: Programmer-Transparent Near-Data Processing Using Multiple Compute-Capable Resources in SSDs}}} \textbf{\footnotesize [HPCA, 2026]}
\\
R. Nadig, V. Arulchelvan, M. Kabra, H. Gupta, R. Bera, \underline{N. Mansouri Ghiasi}, N. Rao, Q. Jiang, A. Kakolyris, Y. Liang, M. Sadrosadati, O. Mutlu

\item \textit{\textcolor{denim}{\textbf{GenPairX: A Hardware-Algorithm Co-Designed Accelerator for Paired-End Read Mapping}}} \textbf{\footnotesize [HPCA, 2026]}
\\
Julien Eudine, Chu Li, Zhuo Cheng, Renzo Andri, Can Firtina, Mohammad Sadrosadati, \underline{N. Mansouri Ghiasi}, Konstantina Koliogeorgi, Anirban Nag, Arash Tavakkol, Haiyu Mao, Onur Mutlu, Shai Bergman, Ji Zhang

\item \textit{\textcolor{denim}{\textbf{Revisiting Main Memory-Based Covert and Side Channel Attacks in the Context of Processing-in-Memory}}} \textbf{\footnotesize [DSN, 2025]}
\\
N. Bostanci, K. Kanellopoulos, A. Olgun, A. G. Yaglikci, I. E. Yuksel, \underline{N. Mansouri Ghiasi}, Z. Bingol, M. Sadrosadati, O. Mutlu

\item \textit{\textcolor{denim}{\textbf{Analysis of metagenomic data}}}, \textbf{\footnotesize [Nature Reviews Methods Primers, 2025]}
\\
S. Liu, J. S. Rodriguez, V. Munteanu, C. Ronkowski, N. Kumar Sharma, M. Alser, F. Andreace, R. Blekhman, D. Błaszczyk, R. Chikhi, K. A. Crandall, K. Della Libera, D. Francis, A. Frolova, A. Shahar Gancz, N. E. Huntley, P. Jaiswal, T. Kosciolek, P. P. Łabaj, W. Łabaj, T. Luan, Ch. Mason, A. M. Moustafa, H. Subrahmaniam Muralidharan, O. Mutlu, \underline{N. Mansouri Ghiasi}, A. Rahnavard, F. Sun, S. Tian, B. T. Tierney, E. Van Syoc, R. Vicedomini, J. P. Zackular, A. Zelikovsky, K. Zielińska, E. Ganda, E. R. Davenport, M. Pop, D. Koslicki, S. Mangul

\item \textit{\textcolor{denim}{\textbf{Ariadne: Hotness-Aware and Size-Adaptive Compressed Swap Scheme for Mobile Devices}}}, \textbf{\footnotesize [HPCA, 2025]}
\\
Y. Liang, A. Shen, C. J. Xue, R. Pan, H. Mao, \underline{N. Mansouri Ghiasi}, Q. Jiang, R. Nadig, L. Li, R. Ausavarungnirun, M. Sadrosadati, O. Mutlu

\item \textit{\textcolor{denim}{\textbf{Label-Guided Seed-Chain-Extend Alignment on Annotated De Bruijn Graphs}}}, \textbf{\footnotesize [ISMB, 2024]}
\\
H. Mustafa, M. Karasikov, \underline{N. Mansouri Ghiasi}, G. Rätsch, A. Kahles

\item \textit{\textcolor{denim}{\textbf{RawAlign: Accurate, Fast, and Scalable Raw Nanopore Signal Mapping via Combining Seeding and Alignment}}}, \textbf{\footnotesize [IEEE Access, 2024]}
\\
J. Lindegger, C. Firtina, \underline{N. Mansouri Ghiasi}, M. Sadrosadati, M. Alser, O. Mutlu

\item \textit{\textcolor{denim}{\textbf{MATSA: An MRAM-Based Energy-Efficient Accelerator for Time Series Analysis}}}, \textbf{\footnotesize [IEEE Access, 2024]}
\\
I. Fernandez, C. Giannoula, A. Manglik, R. Quislant, N. Mansouri Ghiasi, J. Gomez-Luna, E. Gutierrez, O. Plata, O. Mutlu

\item \textit{\textcolor{denim}{\textbf{Venice: Improving Solid-State Drive Parallelism at Low Cost via Conflict-Free Accesses}}}, \textbf{\footnotesize [ISCA, 2023]}
\\
R. Nadig, M. Sadrosadati, H. Mao, \underline{N. Mansouri Ghiasi}, A. Tavakkol, J. Park, H. Sarbazi-Azad, J. Luna, O. Mutlu

\item \textit{\textcolor{denim}{\textbf{RawHash: Enabling Fast and Accurate Real-Time Analysis of Raw Nanopore Signals for Large Genomes}}}, \textbf{\footnotesize [ISMB/ECCB, 2023]}
\\
C. Firtina, \underline{N. Mansouri Ghiasi}, J. Lindegger, G. Singh, M. Banu Cavlak, H. Mao, and O. Mutlu

\item \textit{\textcolor{denim}{\textbf{BLEND: A Fast, Memory-Efficient, and Accurate Mechanism to Find Fuzzy Seed Matches}}}, \textbf{\footnotesize [NAR Genomics and Bioinformatics, 2023]}
\\
C. Firtina, J. Park, J. S. Kim, M. Alser, D. Senol Cali, T. Shahroodi, \underline{N. Mansouri Ghiasi}, G. Singh, K. Kanellopoulos, C. Alkan, O. Mutlu

\item \textit{\textcolor{denim}{\textbf{SeGraM: A Universal Hardware Accelerator for Genomic Sequence-to-Graph and Sequence-to-Sequence Mapping}}}, \textbf{\footnotesize [ISCA, 2022]}
\\
D. Senol Cali, K. Kanellopoulos, J. Lindegger, Z. Bingol, G. S. Kalsi, Z. Zuo, C. Firtina, M. B. Cavlak, J. Kim, \underline{N. Mansouri Ghiasi}, G. Singh, J. Gomez-Luna, N. Almadhoun Alserr, M. Alser, S. Subramoney, C. Alkan, S. Ghose, and O. Mutlu

\item \textit{\textcolor{denim}{\textbf{CODIC: A Low-Cost Substrate for Enabling Custom In-DRAM Functionalities and Optimizations}}}, \textbf{\footnotesize [ISCA, 2021]}
\\
L. Orosa, Y. Wang, M. Sadrosadati, J. S. Kim, M. Patel, I. Puddu, H. Luo, K. Razavi, J. G\'omez-Luna, H. Hassan, \underline{N. Mansouri Ghiasi}, S. Ghose, and O. Mutlu

\item \textit{\textcolor{denim}{\textbf{SIMDRAM: A Framework for Bit-Serial SIMD Processing Using DRAM}}}, \textbf{\footnotesize [ASPLOS, 2021]}
\\
N. Hajinazar, G. F. Oliveira, S. Gregorio, J. Dinis Ferreira, \underline{N. Mansouri Ghiasi}, M. Patel, M. Alser, S. Ghose, J. G\'omez-Luna, O. Mutlu

\item \textit{\textcolor{denim}{\textbf{FIGARO: Improving System Performance via Fine-Grained In-DRAM Data Relocation and Caching}}}, \textbf{\footnotesize [MICRO, 2020]}
\\
Y. Wang, L. Orosa, X. Peng, Y. Guo, S. Ghose, M. Patel, J. S. Kim, J. G\'omez-Luna, M. Sadrosadati, \underline{N. Mansouri Ghiasi}, O. Mutlu

\item \textit{\textcolor{denim}{\textbf{SMASH: Co-designing Software Compression and Hardware-Accelerated Indexing for Efficient Sparse Matrix Operations}}}, \textbf{\footnotesize [MICRO, 2019]}
\\
K. Kanellopoulos, N. Vijaykumar, C. Giannoula, R. AziziBarzoki, S. Koppula, \underline{N. Mansouri Ghiasi}, T. Shahroodi, J. G\'omez-Luna,  O. Mutlu

\item \textit{\textcolor{denim}{\textbf{CROW: A Low-Overhead Substrate for Improving DRAM Performance and Energy-Efficiency}}}, \textbf{\footnotesize [ISCA, 2019]}\\
H. Hassan, M. Patel, J. S. Kim, A. G. Yaglikci, N. Vijaykumar,\underline{N. Mansouri Ghiasi}, S. Ghose, O. Mutlu

\item \textit{\textcolor{denim}{\textbf{Reducing DRAM Latency via Charge-Level-Aware Look-Ahead Partial Restoration}}}, {\footnotesize\textbf{[MICRO, 2018]}}
\\
Y. Wang, A. Tavakkol, L. Orosa, S. Ghose, \underline{N. Mansouri Ghiasi}, M. Patel, J. S. Kim, H. Hassan, M. Sadrosadati, O. Mutlu

\item \textit{\textcolor{denim}{\textbf{FLIN: Enabling Fairness and Enhancing Performance in Modern NVMe Solid State Drives}}}, \textbf{\footnotesize [ISCA, 2018]}\\
A. Tavakkol, M. Sadrosadati, S. Ghose, J. S. Kim, Y. Luo, Y. Wang, \underline{N. Mansouri Ghiasi}, L. Orosa, J. G\'omez-Luna, O. Mutlu

\item \textit{\textcolor{denim}{\textbf{Efficient Critical Path Identification Based on Viability Analysis Method Considering Process Variations}}}, 
\textbf{\footnotesize [IEEE TVLSI, September 2017]}
\\
S. Abolmaali, \underline{N. Mansouri Ghiasi}, M. Kamal, A. Afzali-Kusha, M. Pedram

\end{enumerate}

\tomiv{\section{Open Source Tools and Repositories}

\begin{itemize}[align=left,leftmargin=*,widest={10}]
    \item \textbf{GenStore:} \href{https://github.com/CMU-SAFARI/GenStore}{https://github.com/CMU-SAFARI/GenStore}
    \item \textbf{MegIS:} \href{https://github.com/CMU-SAFARI/MegIS}{https://github.com/CMU-SAFARI/MegIS}
    \item \textbf{Ariadne:} \href{https://github.com/CMU-SAFARI/Ariadne}{https://github.com/CMU-SAFARI/Ariadne}
    \item \textbf{RawHash and Rawsamble:} \href{https://github.com/CMU-SAFARI/RawHash}{https://github.com/CMU-SAFARI/RawHash}
    \item \textbf{IMPACT:} \href{https://github.com/CMU-SAFARI/IMPACT}{https://github.com/CMU-SAFARI/IMPACT}
    \item \textbf{MLA:} \href{https://github.com/ratschlab/mla}{https://github.com/ratschlab/mla}
    \item \textbf{RawAlign:} \href{https://github.com/CMU-SAFARI/RawAlign}{https://github.com/CMU-SAFARI/RawAlign}
    \item \textbf{BLEND:} \href{https://github.com/CMU-SAFARI/BLEND}
    {https://github.com/CMU-SAFARI/BLEND}
    \item \textbf{Scrooge:} \href{https://github.com/CMU-SAFARI/Scrooge}{https://github.com/CMU-SAFARI/Scrooge}
    \item \textbf{TargetCall:} \href{https://github.com/CMU-SAFARI/TargetCall}{https://github.com/CMU-SAFARI/TargetCall}
    \item \textbf{SMASH:} \href{https://github.com/CMU-SAFARI/SMASH}{https://github.com/CMU-SAFARI/SMASH}
    \item \textbf{CROW:} \href{https://github.com/CMU-SAFARI/CROW}{https://github.com/CMU-SAFARI/CROW}
\end{itemize}}

\chapter{Curriculum Vitae of the Author}
\label{appendix:cv}

\newcommand{\paper}[4]{#1, \ifblank{#2}{}{\href{#2}}{\textit{``#3,''}} #4.}

\setlength{\parindent}{0pt}

\section*{Education}
\vspace{-0.5em}\hrule\vspace{0.8em}
\textbf{ETH Z\"urich}, Z\"urich, Switzerland \hfill{\textbf{\footnotesize{[2026]}}}\\
Ph.D., Department of Information Technology and Electrical Engineering  (D-ITET) \\
\underline{Advisor}: Prof. Onur Mutlu\\
\underline{Thesis}: Storage-Centric System Designs for Enabling Fast, Efficient, and Low-Cost Genomic and Metagenomic Analyses {[Defense Slides: \href{https://safari.ethz.ch/wp-content/uploads/nika-thesis-defense-talk-final.pdf}{pdf} \textbar \href{https://safari.ethz.ch/wp-content/uploads/nika-thesis-defense-talk-final.pptx}{~ppt}]}\\
\underline{Thesis Committee}: Prof. Onur Mutlu, Prof. Can Alkan, Prof. Reetuparna Das, Prof. Wen-mei Hwu, Prof. Jangwoo Kim, Prof. Yatish Turakhia\\

\noindent\textbf{ETH Z\"urich}, Z\"urich, Switzerland \hfill{\textbf{\footnotesize{[2019]}}}\\
{M.Sc., Department of Information Technology and Electrical Engineering (D-ITET)} \\
\underline{Advisor}: Prof. Onur Mutlu\\
\underline{Thesis}: Understanding the Opportunities and Bottlenecks in Near-Data Processing Engines\\

\noindent\textbf{University of Tehran}, Tehran, Iran \hfill{\textbf{\footnotesize{[2016]}}}\\
{B.Sc., School of Electrical and Computer Engineering}\\
\underline{Advisors}: Prof. Ali Afzali-Kusha, Prof. Mehdi Kamal\\
\underline{Thesis}: Finding Critical Paths in Digital Circuits in the Presence of Process Variations

\section*{Awards and Recognitions}
\vspace{-0.5em}\hrule\vspace{0.8em}
\begin{itemize}[align=left,leftmargin=*,widest={10}]
  \setlength\itemsep{0.3em}
    \item \tomiv{To be inducted into ISCA Hall of Fame}
    \hfill{\textbf{\footnotesize{[2026]}}}
    \item Selected to participate and present in the 13th Heidelberg Laureate Forum (HLF). Selected as one of the 20 young researchers to present at the forum
    \hfill{\textbf{\footnotesize{[2026]}}}
    \item Grand Finalist in ACM Student Research Competition held at the International Conference on Parallel Architectures and Compilation Techniques (PACT) 
    \hfill{\textbf{\footnotesize{[October 2024]}}}
    \item Recipient of the ETH Zurich Doc.Mobility Fellowship 
    \hfill{\textbf{\footnotesize{[September 2022]}}}
    \item Ranked among the top 10\% of the electrical engineering students of the University of Tehran 
    \hfill{\textbf{\footnotesize{[2016]}}}
    \item Awarded University of Tehran Fellowship for B.Sc.
    \hfill{\textbf{\footnotesize{[2011 - 2016]}}}
    \item Ranked top 0.1\% among more than 280,000 students in Iran's National University Entrance Exam
      \hfill{\textbf{\footnotesize{[2011]}}}
    \item Semifinalist at the Mathematics, Chemistry, and Literature National Olympiads 
    \hfill{\textbf{\footnotesize{[2010]}}}
    \item Member of the National Organization for Development of Exceptional Talents (NODET) during middle and high school

\end{itemize}

\section*{Research Experience}
\vspace{-0.5em}\hrule\vspace{0.8em}

\emph{Main areas of my research have been:}
\begin{itemize}[align=left,leftmargin=*,itemsep=0.1pt,topsep=0.5em]
  \item High-performance and energy-efficient computing systems for healthcare and life sciences\\
  \emph{[Under Submission, ISCA'26, HPCA'26, ICS'26, ICS'25, ISCA'24, Bioinformatics'23, ISCA'22,  ASPLOS'22]}
  \item Data-centric computing to overcome data movement bottlenecks by making storage and memory systems compute-capable: \\
  Storage-centric computing\\ \emph{[Under Submission, ISCA'26, HPCA'26, ICS'26, ISCA'25, ICS'25, ISCA'24, ASPLOS'22]},\\
  Memory-centric computing\\ \emph{[Under Submission, TACO'26, DSN'25, PACT ACM SRC'24, IEEE Access'24, TETC'22, ASPLOS'21]}
  \item Emerging technologies, in particular dense 3D integration\\
  \emph{[Under Submission, TACO'26, PACT ACM SRC'24]}
  \item Hardware-software co-design for AI/ML and mobile devices\\ 
  \emph{[Under Submission, ISCA'25, HPCA'25, MICRO'19]}
  \item Storage and memory system designs to improve performance, energy-efficiency, and reliability\\
  \emph{[Under Submission, ISCA'26, HPCA'26, ICS'26, TACO'26, ISCA'25, ICS'25, DSN'25, PACT ACM SRC'24, IEEE Access'24, ISCA'23, ASPLOS'22, TETC'22, ISCA'21, ASPLOS'21, MICRO'20, ISCA'19, MICRO'18, ISCA'18]}
  \item Bioinformatics algorithms and software\\ 
  \emph{[Nature Genetics'26, Bioinformatics'26, Nature Reviews Methods Primers'25, ISMB'24, IEEE Access'24, APBC'23, NAR Genomics \& Bioinformatics'23]}
\end{itemize}\vspace{1em}

\textbf{ETH Z\"urich}, ITET Department, SAFARI Research Group, advised by Prof. Onur Mutlu\\
\textbf{\textit{Senior Researcher}} \hfill{\textbf{\footnotesize{[2026 - present]}}}\\
\textbf{\textit{PhD Research Assistant}}
\hfill{\textbf{\footnotesize{[2020 - 2026]}}}
 
\begin{itemize}[align=left,leftmargin=*,noitemsep,topsep=0pt]
  \item High-performance and energy-efficient computing systems for healthcare and life sciences, specifically for genomics and metagenomics
  \item Storage- and memory-centric computing
  \item Emerging technologies, in particular dense 3D integration. Architecting the processor core and cache hierarchy for systems with dense 3D integration
  \item Hardware-software co-design for critical application domains, such as health and AI/ML
  \item Storage and memory system designs to improve performance, energy-efficiency, and reliability
  \item Bioinformatics algorithms and software, real-time genome analysis, population-scale genomics
\end{itemize}\vspace{1em}

\textbf{Stanford University}, EE and CS Departments, Robust Systems Group, advised by Prof. Subhasish Mitra\\ 
\textbf{\textit{Graduate Visiting Researcher}}
\hfill{\textbf{\footnotesize{[Sep 2023 - Feb 2024]}}}

\begin{itemize}[align=left,leftmargin=*,noitemsep,topsep=0pt]
    \item Understanding and modeling the performance landscape of ultra-dense 3D memory systems
    \item Exploring the potential of this emerging technology for improving the performance of data-intensive applications
\end{itemize}\vspace{1em}

\textbf{ETH Z\"urich}, ITET Department, SAFARI Research Group, advised by Prof. Onur Mutlu\\
\textbf{\textit{Master's Research Assistant}}
\hfill{\textbf{\footnotesize{[2017 - 2019]}}}
 
\begin{itemize}[align=left,leftmargin=*,noitemsep,topsep=0pt]
  \item Memory-centric computing
  \item Emerging technologies, in particular dense 3D integration. Architecting the processor core and cache hierarchy for systems with dense 3D integration
  \item Storage and memory system designs to improve performance, energy-efficiency, and reliability
\end{itemize}\vspace{1em}

\textbf{University of Tehran}, ECE Department, Low-Power High-Performance Nano-systems Laboratory, advised by Prof. Ali Afzali-Kusha, Prof. Mehdi Kamal

\textbf{\textit{Undergraduate Research Assistant}} 
\hfill{\textbf{\footnotesize{[Jul 2015 - Aug 2016]}}}
\begin{itemize}[align=left,leftmargin=*,noitemsep,topsep=0pt]
    \item Finding critical paths in digital circuits in the presence of process variation
\end{itemize}\vspace{1em}

\textbf{Texas Tech University}, CS Department,  
Data-Intensive Scalable Computing Laboratory, advised by Prof. Yong Chen\\
\textbf{\textit{Undergraduate Research Intern}}
\hfill{\textbf{\footnotesize{[Aug 2015 - Oct 2015]}}}
\begin{itemize}[align=left,leftmargin=*,noitemsep,topsep=0pt]
    \item RISC-V extension for data-intensive applications,
    Memory request coalescing for hybrid memory cube devices
\end{itemize}\vspace{1em}

\textbf{Polytechnic University of Turin}, Test Group, advised by Prof. Paolo Prinetto\\
\textbf{\textit{Undergraduate Research Intern}}
\hfill{\textbf{\footnotesize{[Jul 2014 - Sep 2014]}}}
\begin{itemize}[align=left,leftmargin=*,noitemsep,topsep=0pt]
    \item Design and synthesis of a hardware camera lens distortion core with High Level Synthesis
\end{itemize}\vspace{1em}

\textbf{University of Tehran}, ECE Department, CAD Research Group, advised by Prof. Zain Navabi\\
\textbf{\textit{Undergraduate Research Assistant}} 
\hfill{\textbf{\footnotesize{[Jul 2013 – Sep 2013]}}}
\begin{itemize}[align=left,leftmargin=*,noitemsep,topsep=0pt]
    \item Design and implementation of an educational extension board for FPGAs
\end{itemize}

\section*{First and Co-First Author Publications}
\vspace{-0.5em}\hrule\vspace{0.8em}

\textbf{Thesis Publications}

\begin{enumerate}[label={[P}\arabic*{]}]%

\item \paper{\underline{Nika Mansouri Ghiasi}, Harun Mustafa, Talu G\"uloglu, Rakesh Nadig, Konstantina Koliogeorgi, Susana Ruiz, Marc Rautmann, Furkan Eris, Mohammad Sadrosadati, Jisung Park, Onur Mutlu}{https://arxiv.org/pdf/2606.26468}{GRAINS: Enabling High-Performance and Low-Cost Graph-Based Genome Analysis via Storage-Aware Algorithm-Architecture Co-Design}{International Symposium on Computer Architecture (ISCA), 2026}

\item \paper{\underline{Nika Mansouri Ghiasi}, Talu G\"uloglu, Harun Mustafa, Can Firtina, Konstantina Koliogeorgi, Konstantinos Kanellopoulos, Haiyu Mao, Rakesh Nadig, Mohammad Sadrosadati, Jisung Park, Onur Mutlu}{https://arxiv.org/pdf/2504.03732}{SAGe: A Lightweight Algorithm-Architecture Co-Design for Mitigating the Data Preparation Bottleneck in Large-Scale Genome Sequence Analysis}{International Symposium on High-Performance Computer Architecture (HPCA), 2026}

\item \paper{\underline{Nika Mansouri Ghiasi}, Mohammad Sadrosadati, Harun Mustafa, Arvid Gollwitzer, Can Firtina, Julien Eudine, Haiyu Mao, Joel Lindegger, Meryem Banu Cavlak, Mohammed Alser, Jisung Park, Onur Mutlu}{https://arxiv.org/pdf/2406.19113}{MegIS: High-Performance, Energy-Efficient, and Low-Cost Metagenomic Analysis with In-Storage Processing}{International Symposium on Computer Architecture (ISCA), 2024}

\item \paper{\underline{Nika Mansouri Ghiasi}, Jisung Park, Harun Mustafa, Jeremie Kim, Ataberk Olgun, Arvid Gollwitzer, Damla Senol Cali, Can Firtina, Haiyu Mao, Nour Almadhoun Alserr, Rachata Ausavarungnirun, Nandita Vijaykumar, Mohammed Alser, and Onur Mutlu}{https://nikamgh.github.io/assets/papers/GenStore_asplos22-arxiv.pdf}{GenStore: A High-Performance In-Storage Processing System for Genome Sequence Analysis}{International Conference on Architectural Support for Programming Languages and Operating Systems (ASPLOS), 2022}

\end{enumerate}

\textbf{Other First and Co-First Author Publications}

\begin{enumerate}[resume, label={[P}\arabic*{]}]

\item \paper{\underline{Nika Mansouri Ghiasi}*, Konstantina Koliogeorgi*, Onur Mutlu}{https://arxiv.org/pdf/2606.22685}{Architecture for Health Initiative (Arch4Health):
Computational Challenges in Health-Related Applications
and the Role of Computer Architecture in Addressing Them}{ACM International Conference on Supercomputing (ICS)
 Workshops, 2026}

\item \paper{\underline{Nika Mansouri Ghiasi}, Onur Mutlu}{https://arxiv.org/pdf/2607.02552}{Enabling Fast, Efficient, and Low-Cost Genomic and Metagenomic Analyses via Storage-Centric System Designs}{ACM International Conference on Supercomputing (ICS)
 Workshops, 2026}

\item \paper{\underline{Nika Mansouri Ghiasi}, Mohammad Sadrosadati, Geraldo F. Oliveira, Konstantinos Kanellopoulos, Rachata Ausavarungnirun, Juan Gómez Luna, Joao Ferreira, Jeremie S Kim, Christina Giannoula, Nandita Vijaykumar, Jisung Park, Onur Mutlu}{https://dl.acm.org/doi/epdf/10.1145/3822173}{RevaMp3D: Architecting the Processor Core and Cache Hierarchy for Systems with Monolithically-Integrated Logic and Memory}{ACM Transactions on Architecture and Code Optimization
(TACO), 2026}

\item \paper{\underline{Nika Mansouri Ghiasi}, Nandita Vijaykumar, Geraldo F. Oliveira, Lois Orosa, Ivan Fernandez, Mohammad Sadrosadati, Konstantinos Kanellopoulos, Nastaran Hajinazar, Juan Gómez Luna, Onur Mutlu}{https://arxiv.org/pdf/2212.06292}{ALP: Alleviating CPU-Memory Data Movement Overheads in Memory-Centric Systems}{IEEE Transactions on Emerging Topics in Computing (IEEE TETC), 2022}
\end{enumerate}

\pagebreak

\section*{Co-Supervised Publications}
\vspace{-0.5em}\hrule\vspace{0.8em}

\begin{enumerate}[resume, label={[P}\arabic*{]}]

\item \paper{Eirni M. Tzermpou, Konstantina Koliogeorgi, Akmuhammet Ashyralyyev, Sotirios Xydis, Dimitrios Soudris, \underline{Nika Mansouri Ghiasi}, Onur Mutlu}{}{\tomiv{A New} Software/Hardware Co-Design for Graph-Based Genome Analysis}{Under Submission, 2026} [Blinded]

\item \paper{Mayank Kabra, Harshita Gupta, Eirini Maria Tzermpou, Phillip William Widdowson, \underline{Nika Mansouri Ghiasi}, Ataberk Olgun, Rakesh Nadig, Konstantinos Kanellopoulos, Rakesh Kumar, Abdullah Giray Yaglikci, Jisung Park, Onur Mutlu}{}{Efficient Processing on Homomorphically Encrypted Data via Storage-Centric \tomiv{Computing}}{Under Submission, 2026} [Blinded]

\item \paper{Kangqi Chen, Rakesh Nadig, Andreas Kosmas Kakolyris, Manos Frouzakis, \underline{Nika Mansouri Ghiasi}, Yu Liang, Haiyu Mao, Jisung Park, Mohammad Sadrosadati, Onur Mutlu}{https://arxiv.org/pdf/2506.16444}{REIS: A High-Performance and Energy-Efficient Retrieval System with In-Storage Processing}{International Symposium on Computer Architecture (ISCA), 2025}

\item \paper{Melina Soysal, Konstantina Koliogeorgi, Can Firtina, \underline{Nika Mansouri Ghiasi}, Rakesh Nadig, Haiyu Mao, Geraldo F. Oliveira, Yu Liang, Klea Zambaku, Mohammad Sadrosadati, Onur Mutlu}{https://arxiv.org/pdf/2506.10931}{MARS: Processing-In-Memory Acceleration of Raw Signal Genome Analysis Inside the Storage Subsystem}{ACM International Conference on Supercomputing (ICS), 2025}

\item \paper{Joel Lindegger, Can Firtina, \underline{Nika Mansouri Ghiasi}, Mohammad Sadrosadati, Mohammed Alser, Onur Mutlu}{https://arxiv.org/pdf/2310.05037}{RawAlign: Accurate, Fast, and Scalable Raw Nanopore Signal Mapping via Combining Seeding and Alignment}{IEEE Access, 2024}

\item \paper{Joël Lindegger, Damla Senol Cali, Mohammed Alser, Juan Gómez-Luna, \underline{Nika Mansouri Ghiasi}, Onur Mutlu}{https://arxiv.org/pdf/2208.09985.pdf}{Scrooge: A Fast and Memory-Frugal Genomic Sequence Aligner for CPUs, GPUs, and ASICs}{Bioinformatics, 2023}

\item \paper{Meryem Banu Cavlak, Gagandeep Singh, Mohammed Alser, Can Firtina, Joel Lindegger, Mohammad Sadrosadati, \underline{Nika Mansouri Ghiasi}, Can Alkan, Onur Mutlu}{https://arxiv.org/pdf/2212.04953.pdf}{TargetCall: Eliminating the Wasted Computation in Basecalling via Pre-Basecalling Filtering}{Asia Pacific Bioinformatics Conference (APBC), 2023}

\end{enumerate}
\pagebreak
\section*{Other Publications}
\vspace{-0.5em}\hrule\vspace{0.8em}

\begin{enumerate}[resume, label={[P}\arabic*{]}]

\item \paper{G. Sharma, V. Munteanu, \underline{N. Mansouri Ghiasi}, U. Mahanta, J. Banerjee, S. Varma, L. Foschini, K. Ellrott, O. Mutlu, D. Ciorbă, R. A. Ophoff, V. Bostan, J. H. Moore, D. Sousoni, A. Krishnan, A. G. Lucaci, A. Tull, C. E. Mason, M. Dimian, G. Stolovitzky, F. G. Liberante, T. K. Oleksyk, S. Mangul}{https://www.nature.com/articles/s41588-026-02606-x}{Towards a decentralized future for  open-science databases}{Nature Genetics, 2026}

\item \paper{Can Firtina, Maximilian Mordig, Harun Mustafa, Sayan Goswami, \underline{Nika Mansouri Ghiasi}, Stefano Mercogliano, Furkan Eris, Joël Lindegger, Andre Kahles, Onur Mutlu}{https://arxiv.org/pdf/2410.17801.pdf}{Rawsamble: Overlapping and Assembling Raw Nanopore Signals using a Hash-based Seeding Mechanism}{Bioinformatics, 2026}

\item \paper{Rakesh Nadig, Vamanan Arulchelvan, Mayank Kabra, Harshita Gupta, Rahul Bera, \underline{Nika Mansouri Ghiasi}, Nanditha Rao, Qingcai Jiang, Andreas Kosmas Kakolyris, Yu Liang, Mohammad Sadrosadati, Onur Mutlu}{https://www.arxiv.org/pdf/2601.17633}{Conduit: Programmer-Transparent Near-Data Processing Using Multiple Compute-Capable Resources in SSDs}{International Symposium on High-Performance Computer Architecture (HPCA), 2026}

\item \paper{Julien Eudine, Chu Li, Zhuo Cheng, Renzo Andri, Can Firtina, Mohammad Sadrosadati, \underline{Nika Mansouri Ghiasi}, Konstantina Koliogeorgi, Anirban Nag, Arash Tavakkol, Haiyu Mao, Onur Mutlu, Shai Bergman, Ji Zhang}{https://arxiv.org/pdf/2601.19384}{GenPairX: A Hardware-Algorithm Co-Designed Accelerator for Paired-End Read Mapping}{International Symposium on High-Performance Computer Architecture (HPCA), 2026}

\item \paper{Nisa Bostanci, Konstantinos Kanellopoulos, Ataberk Olgun, A. Giray Yaglikci, Ismail Emir Yuksel, \underline{Nika Mansouri Ghiasi}, Zulal Bingol, Mohammad Sadrosadati, Onur Mutlu}{https://arxiv.org/pdf/2404.11284}{Revisiting Main Memory-Based Covert and Side Channel Attacks in the Context of Processing-in-Memory}{Annual IEEE/IFIP International Conference on Dependable Systems and Networks (DSN), 2025}

\item \paper{Shaopeng Liu, Judith S. Rodriguez, Viorel Munteanu, Cynthia Ronkowski, Nitesh Kumar Sharma, Mohammed Alser, Francesco Andreace, Ran Blekhman, Dagmara Błaszczyk, Rayan Chikhi, Keith A. Crandall, Katja Della Libera, Dallace Francis, Alina Frolova, Abigail Shahar Gancz, Naomi E. Huntley, Pooja Jaiswal, Tomasz Kosciolek, Pawel P. Łabaj, Wojciech Łabaj, Tu Luan, Christopher Mason, Ahmed M. Moustafa, Harihara Subrahmaniam Muralidharan, Onur Mutlu, \underline{Nika Mansouri Ghiasi}, Ali Rahnavard, Fengzhu Sun, Shuchang Tian, Braden T. Tierney, Emily Van Syoc, Riccardo Vicedomini, Joseph P. Zackular, Alex Zelikovsky, Kinga Zielińska, Erika Ganda, Emily R. Davenport, Mihai Pop, David Koslicki, Serghei Mangul}{https://www.nature.com/articles/s43586-024-00376-6}{Analysis of metagenomic data}{Nature Reviews Methods Primers, 2025}

\item \paper{Yu Liang, Aofeng Shen, Chun Jason Xue, Riwei Pan, Haiyu Mao, \underline{Nika Mansouri Ghiasi}, Qingcai Jiang, Rakesh Nadig, Lei Li, Rachata Ausavarungnirun, Mohammad Sadrosadati, Onur Mutlu}{https://arxiv.org/pdf/2502.12826}{Ariadne: Hotness-Aware and Size-Adaptive Compressed Swap Scheme for Mobile Devices}{International Symposium on High-Performance Computer Architecture (HPCA), 2025}

\item \paper{Harun Mustafa, Mikhail Karasikov, \underline{Nika Mansouri Ghiasi}, Gunnar Rätsch, André Kahles}{https://academic.oup.com/bioinformatics/article/40/Supplement_1/i337/7700889}{Label-Guided Seed-Chain-Extend Alignment on Annotated De Bruijn Graphs}{ISMB, 2024}

\item \paper{Ivan Fernandez, Christina Giannoula, Aditya Manglik, Ricardo Quislant, \underline{Nika Mansouri Ghiasi}, Juan Gomez-Luna, Eladio Gutierrez, Oscar Plata, Onur Mutlu}{https://arxiv.org/pdf/2211.04369}{MATSA: An MRAM-Based Energy-Efficient Accelerator for Time Series Analysis}{IEEE Access, 2024}

\item \paper{Rakesh Nadig, Mohammad Sadrosadati, Haiyu Mao, \underline{Nika Mansouri Ghiasi}, Arash Tavakkol, Jisung Park, Hamid Sarbazi-Azad, Juan Gómez Luna, Onur Mutlu}{https://arxiv.org/pdf/2305.07768.pdf}{Venice: Improving Solid-State Drive Parallelism at Low Cost via Conflict-Free Accesses}{International Symposium on Computer Architecture (ISCA), 2023}

\item \paper{Can Firtina, \underline{Nika Mansouri Ghiasi}, Joel Lindegger, Gagandeep Singh, Meryem Banu Cavlak, Haiyu Mao, and Onur Mutlu}{https://arxiv.org/pdf/2301.09200.pdf}{RawHash: Enabling Fast and Accurate Real-Time Analysis of Raw Nanopore Signals for Large Genomes}{Annual Conference on Intelligent Systems for Molecular Biology and European Conference on Computational Biology (ISMB/ECCB), 2023}

\item \paper{Can Firtina, Jisung Park, Jeremie S. Kim, Mohammed Alser, Damla Senol Cali, Taha Shahroodi, \underline{Nika Mansouri Ghiasi}, Gagandeep Singh, Konstantinos Kanellopoulos, Can Alkan, Onur Mutlu}{https://arxiv.org/pdf/2112.08687.pdf}{BLEND: A Fast, Memory-Efficient, and Accurate Mechanism to Find Fuzzy Seed Matches}{NAR Genomics and Bioinformatics, 2023}

\item \paper{Damla Senol Cali, Konstantinos Kanellopoulos, Joel Lindegger, Zulal Bingol, Gurpreet S. Kalsi, Ziyi Zuo, Can Firtina, Meryem Banu Cavlak, Jeremie Kim, \underline{Nika Mansouri Ghiasi}, Gagandeep Singh, Juan Gomez-Luna, Nour Almadhoun Alserr, Mohammed Alser, Sreenivas Subramoney, Can Alkan, Saugata Ghose, and Onur Mutlu}{https://nikamgh.github.io/assets/papers/SeGraM_isca22.pdf}{SeGraM: A Universal Hardware Accelerator for Genomic Sequence-to-Graph and Sequence-to-Sequence Mapping}{International Symposium on Computer Architecture (ISCA), 2022}

\item \paper{Lois Orosa, Yaohua Wang, Mohammad Sadrosadati, Jeremie S. Kim, Minesh Patel, Ivan Puddu, Haocong Luo, Kaveh Razavi, Juan G\'omez-Luna, Hasan Hassan, \underline{Nika Mansouri Ghiasi}, Saugata Ghose, and Onur Mutlu}{https://arxiv.org/pdf/2106.05632}{CODIC: A Low-Cost Substrate for Enabling Custom In-DRAM Functionalities and Optimizations}{International Symposium on Computer Architecture (ISCA), 2021}

\item \paper{Nastaran Hajinazar, Geraldo F. Oliveira, Sven Gregorio, João Dinis Ferreira, \underline{Nika Mansouri Ghiasi}, Minesh Patel, Mohammed Alser, Saugata Ghose, Juan G\'omez-Luna, Onur Mutlu}{https://arxiv.org/pdf/2012.11890}{SIMDRAM: A Framework for Bit-Serial SIMD Processing Using DRAM}{International Conference on Architectural Support for Programming Languages and Operating Systems (ASPLOS), 2021}

\item \paper{Yaohua Wang, Lois Orosa, Xiangjun Peng, Yang Guo, Saugata Ghose, Minesh Patel, Jeremie S. Kim, Juan G\'omez-Luna, Mohammad Sadrosadati, \underline{Nika Mansouri Ghiasi}, Onur Mutlu}{https://arxiv.org/pdf/2009.08437}{FIGARO: Improving System Performance via Fine-Grained In-DRAM Data Relocation and Caching}{International Symposium on Microarchitecture (MICRO), 2020}

\item \paper{Konstantinos Kanellopoulos, Nandita Vijaykumar, Christina Giannoula, Roknoddin Azizi, Skanda Koppula, \underline{Nika Mansouri Ghiasi}, Taha Shahroodi, Juan G\'omez-Luna,  Onur Mutlu}{https://nikamgh.github.io/assets/papers/SMASH_micro19.pdf}{SMASH: Co-designing Software Compression and Hardware-Accelerated Indexing for Efficient Sparse Matrix Operations}{International Symposium on Microarchitecture (MICRO), 2019}

\item \paper{Hasan Hassan, Minesh Patel, Jeremie S. Kim, A. Giray Yaglikci, Nandita Vijaykumar, \underline{Nika Mansouri Ghiasi}, Saugata Ghose, Onur Mutlu}{https://nikamgh.github.io/assets/papers/CROW_isca19.pdf}{CROW: A Low-Overhead Substrate for Improving DRAM Performance and Energy-Efficiency}{International Symposium on Computer Architecture (ISCA), 2019}

\item \paper{Yaohua Wang, Arash Tavakkol, Lois Orosa, Saugata Ghose, \underline{Nika Mansouri Ghiasi}, Minesh Patel, Jeremie S. Kim, Hasan Hassan, Mohammad Sadrosadati, Onur Mutlu}{https://nikamgh.github.io/assets/papers/CAL_micro18.pdf}{Reducing DRAM Latency via Charge-Level-Aware Look-Ahead Partial Restoration}{International Symposium on Microarchitecture (MICRO), 2018}

\item \paper{Arash Tavakkol, Mohammad Sadrosadati, Saugata Ghose, Jeremie S. Kim, Yixin Luo, Yaohua Wang, \underline{Nika Mansouri Ghiasi}, Lois Orosa, Juan G\'omez-Luna, Onur Mutlu}{https://ieeexplore.ieee.org/document/8416843}{FLIN: Enabling Fairness and Enhancing Performance in Modern NVMe Solid State Drives}{International Symposium on Computer Architecture (ISCA), 2018}

\item \paper{Sheis Abolmaali, \underline{Nika Mansouri Ghiasi}, Mehdi Kamal, Ali Afzali-Kusha, Massoud Pedram}{https://ieeexplore.ieee.org/document/7934409}{Efficient Critical Path Identification Based on Viability Analysis Method Considering Process Variations}{IEEE Transactions on Very Large Scale Integration (VLSI) Systems, 2017}
\end{enumerate}

\section*{Manuscripts Under Submission}
\vspace{-0.5em}\hrule\vspace{0.8em}

\begin{enumerate}[label={[US}\arabic*{]}]

\item \paper{Eirni M. Tzermpou, Konstantina Koliogeorgi, Akmuhammet Ashyralyyev, Sotirios Xydis, Dimitrios Soudris, \underline{Nika Mansouri Ghiasi}, Onur Mutlu}{}{\tomiv{A New} Software/Hardware Co-Design for Graph-Based Genome Analysis}{2026} [Blinded]

\item \paper{Rakesh Nadig, Manos Frouzakis, \underline{Nika Mansouri Ghiasi}, Yonggon Park, Mayank Kabra, Sahand Divsalar, Andreas Kosmas Kakolyris, Harshita Gupta, Jisung Park, Mohammad Sadrosadati, Onur Mutlu}{}{Enabling Scalable In-Flash Processing in Solid-State Drives}{2026} [Blinded]

\item \paper{Mayank Kabra, Harshita Gupta, Eirini Maria Tzermpou, Phillip William Widdowson, \underline{Nika Mansouri Ghiasi}, Ataberk Olgun, Rakesh Nadig, Konstantinos Kanellopoulos, Rakesh Kumar, Abdullah Giray Yaglikci, Jisung Park, Onur Mutlu}{}{Efficient Processing on Homomorphically Encrypted Data via Storage-Centric \tomiv{Computing}}{2026} [Blinded]

\item \paper{Zhiheng Yue, Tianlang Zhao, Umut Baser, Yonggon Park, \underline{Nika Mansouri Ghiasi}, Harshita Gupta, Jikun Wang, Nisa Bostanci, Ismail Emir Yüksel, Andreas Kosmas Kakolyris, Mayank Kabra, Geraldo F. Oliveira, Mohammad Sadrosadati, Onur Mutlu}{}{Acceleration of AI Workloads with Hybrid Bonding-Based DRAM}{2026} [Blinded]

\item \paper{Konstantinos Sgouras, Ismail Emir Yuksel, Harsh Songara, F. Nisa Bostanci, Konstantina Koliogeorgi, \underline{Nika Mansouri Ghiasi}, Ataberk Olgun, Konstantinos Kanellopoulos, Onur Mutlu}{}{\tomiv{Flexible Data} Representations in Processing-Using-Memory}{2026} [Blinded]

\end{enumerate}

\section*{Conference and Workshop Talks}
\vspace{-0.5em}\hrule\vspace{0.8em}

\begin{itemize}[align=left,leftmargin=*,widest={10}]
    \item \textbf{GRAINS: Enabling High-Performance and Low-Cost Graph-Based Genome Analysis via Storage-Aware Algorithm-Architecture Co-Design}\\{[Slides: \href{https://nikamgh.github.io/assets/papers/GRAINS-isca26-talk.pdf}{pdf} \textbar \href{https://nikamgh.github.io/assets/papers/GRAINS-isca26-talk.pptx}{~ppt}]}
    \begin{itemize}[noitemsep,topsep=0pt]
        \item International Symposium on Computer Architecture (ISCA), 2026, Raleigh, USA 
    \end{itemize}
    \item \textbf{Storage-Centric System Designs for Enabling Fast, Efficient, and Low-Cost Genomic and Metagenomic Analyses}\\{[Slides: \href{https://nikamgh.github.io/assets/papers/MCCSys-ISCA26-Nika.pdf}{pdf} \textbar \href{https://nikamgh.github.io/assets/papers/MCCSys-ISCA26-Nika.pptx}{~ppt}]}
    \begin{itemize}[noitemsep,topsep=0pt] 
        \item MCCSys Workshop, in conjunction with the International Symposium on Computer Architecture (ISCA), 2026, Raleigh, USA
        \item Arch4Health Workshop, in conjunction with the ACM International Conference on Supercomputing (ICS), 2026, Belfast, Northern Ireland
        \item RECOMB-Arch Workshop, in conjunction with the  Research in Computational Molecular Biology (RECOMB), 2026, Thessaloniki, Greece 
        \item Arch4Health Workshop, in conjunction with the  International Symposium on High-Performance Computer Architecture (HPCA), 2026, Sydney, Australia
    \end{itemize}
    \item \textbf{Storage-Centric Computing for Genomics and Metagenomics}
    \begin{itemize}[noitemsep,topsep=0pt] 
        \item Arch4Health Workshop, in conjunction with the  International Symposium on Microarchitecture (MICRO), 2025, Seoul, South Korea
        \item Future of Memory and Storage (FMS), 2024, Santa Clara, USA
        \item CWIDCA Workshop, in conjunction with the  International Symposium on Microarchitecture (MICRO), 2024, Austin, USA
    \end{itemize}
    \item \textbf{SAGe: A Lightweight Algorithm-Architecture Co-Design for Mitigating the Data Preparation Bottleneck in Large-Scale Genome Sequence Analysis}\\{[Slides: \href{https://nikamgh.github.io/assets/papers/SAGe_hpca26-talk.pdf}{pdf} \textbar \href{https://nikamgh.github.io/assets/papers/SAGe_hpca26-talk.pptx}{~ppt}]}
    \begin{itemize}[noitemsep,topsep=0pt] 
        \item International Symposium on High-Performance Computer Architecture (HPCA), 2026, Sydney, Australia
        \item Future of Memory and Storage (FMS), 2026, Santa Clara, USA
    \end{itemize}
    \item \textbf{Understanding and Modeling the Performance Landscape of Ultra-Dense 3D Memory Systems}
    \begin{itemize}[noitemsep,topsep=0pt]
        \item ACM Student Research Competition held at the International Conference on Parallel Architectures and Compilation Techniques (PACT), 2024, Long Beach, USA\\
        \textit{Grand Finalist}
    \end{itemize}
    \item \textbf{MegIS: High-Performance, Energy-Efficient, and Low-Cost Metagenomic Analysis with In-Storage Processing}\\{[Slides: \href{https://nikamgh.github.io/assets/papers/MegIS-ISCA24-V6.pdf}{pdf} \textbar \href{https://nikamgh.github.io/assets/papers/MegIS-ISCA24-V6.pptx}{~ppt}]}
    \begin{itemize}[noitemsep,topsep=0pt]
        \item International Symposium on Computer Architecture (ISCA), 2024, Buenos Aires, Argentina
    \end{itemize}
    \item \textbf{GenStore: A High-Performance In-Storage Processing System for Genome Sequence Analysis}\\{[Slides: \href{https://nikamgh.github.io/assets/papers/GenStore_asplos22-talk.pdf}{pdf} \textbar \href{https://nikamgh.github.io/assets/papers/GenStore_asplos22-talk.pptx}{~ppt}]}
    \begin{itemize}[noitemsep,topsep=0pt]
        \item International Conference on Architectural Support for Programming Languages and Operating Systems (ASPLOS), 2022, Lausanne, Switzerland
        \item BIO-Arch Workshop, in conjunction with the  Research in Computational Molecular Biology (RECOMB), 2023, Istanbul, Turkey
        \item AACBB Workshop, in conjunction with International Symposium on Computer Architecture (ISCA), 2022, Virtual
    \end{itemize}
\end{itemize}

\section*{Invited Talks and Lectures}
\vspace{-0.5em}\hrule\vspace{0.8em}

\begin{itemize} [align=left,leftmargin=*,widest={10}]

\item \textbf{Storage-Centric Computing for Genomics and Metagenomics}
    \begin{itemize}[noitemsep,topsep=0pt] 
    \item University of Sydney, 2026, Sydney, Australia
    \item UNSW, 2026, Sydney, Australia
    \item KAIST, 2025, Daejeon, South Korea
    \item Global Data Storage Expert Forum, 2025, Suzhou, China
    \item Shanghai Jiao Tao University, 2025, Shanghai, China
    \item Tech Talk at AMD, 2024, Austin, Texas, USA
    \item UCLA, 2024, Los Angeles, California, USA
    \item UCI, 2024, Irvine, California, USA
    \item PIM tutorial, in conjunction with the International Symposium on Microarchitecture (MICRO), 2024, Austin, USA
    \item ACCESS Seminar Series, 2024, Online
\end{itemize}

\item \textbf{Enabling High-Performance, Energy-Efficient, and Cost-Effective Metagenomics via Storage-Centric Systems}
    \begin{itemize}[noitemsep,topsep=0pt] 
    \item International Conference in Clinical Metagenomics (ICCMg), 2025, Geneva, Switzerland
\end{itemize}

\item \textbf{Accelerating Biological Sequence Analysis via New Architectures and Algorithms}
\begin{itemize}[noitemsep,topsep=0pt]
    \item Lecture in the Computer Architecture course at ETH Zurich, 2025, Zurich, Switzerland
    \item Lecture in the Single Molecule Biosensors course at ETH Zurich, 2025, Zurich, Switzerland
\end{itemize}

\item \textbf{Architecting Systems with Ultra-Dense 3D Integrated Logic and Memory}
    \begin{itemize}[noitemsep,topsep=0pt]
    \item Talk at the RSG Seminar at Stanford University, 2023, California, USA
\end{itemize}

\item \textbf{GenStore: A High-Performance In-Storage Processing System for Genome Sequence Analysis}
    \begin{itemize}[noitemsep,topsep=0pt]
    \item In-Memory Processing Special Session at IEEE Computer Society Annual Symposium on VLSI (ISVSLI) 2022, Paphos, Cyprus
    \item Lectures in ETH Zurich P\&S Courses on Storage Systems, Mobile Genomics, and Processing-in-Memory, 2022 and 2023, Z\"urich, Switzerland
\end{itemize}
\end{itemize}

\section*{Teaching Experience} 
\vspace{-0.5em}\hrule\vspace{0.8em}

\textbf{Lecturer, ETH Z\"urich}
\hfill{\textbf{\footnotesize{[Spring 2026 - Present]}}}

\begin{itemize}
[noitemsep,topsep=0pt]
  \item \href{https://safari.ethz.ch/projects_and_seminars/spring2026/doku.php?id=arch_and_alg_for_health}{Architectures \& Algorithms for Health \& Life Sciences} [\href{https://www.youtube.com/playlist?list=PL5Q2soXY2Zi99HxZ9oq2DJ7-cx2wY1jkl}{Lecture videos}]
  \begin{itemize}[align=left,leftmargin=*,noitemsep,topsep=0pt]
    \item Spring 2026 
    \item Fall 2026 (\emph{Upcoming Semester})\vspace{0.3em}
  \end{itemize}
  \item \href{https://safari.ethz.ch/architecture/doku.php}{Computer Architecture}
    \begin{itemize}[align=left,leftmargin=*,noitemsep,topsep=0pt]
    \item Fall 2026 (\emph{Upcoming Semester})\vspace{0.5em}
  \end{itemize}
\end{itemize}

\textbf{Teaching Assistant, ETH Z\"urich}
\hfill{\textbf{\footnotesize{[Spring 2019 - Spring 2026]}}}
        \begin{itemize}[itemsep=0.2em,topsep=0.5em]
            \item \href{https://safari.ethz.ch/architecture/doku.php}{Computer Architecture} (Fall 2019 - Fall 2025)
            \item \href{https://safari.ethz.ch/digitaltechnik/doku.php}{Digital Design and Computer Architecture} (Spring 2019 - Spring 2026)
            \item \href{https://safari.ethz.ch/architecture_seminar/doku.php}{Seminar in Computer Architecture} (Spring 2019 - Spring 2025)
            \item \href{https://safari.ethz.ch/projects_and_seminars/doku.php?id=genome_seq_mobile}{Genome Sequencing on Mobile Devices} (Fall 2022 - Spring 2025)
            \item \href{https://safari.ethz.ch/projects_and_seminars/doku.php?id=bioinformatics}{Accelerating Genome Analysis with FPGAs, GPUs, and New Execution Paradigms} (Fall 2022 - Spring 2025)
            \item \href{https://safari.ethz.ch/projects_and_seminars/doku.php?id=processing_in_memory}{Data-Centric Architectures: Fundamentally Improving Performance and Energy} (Fall 2020 - Spring 2024)\vspace{0.5em}
        \end{itemize}

\textbf{University of Tehran, Teaching Assistant} 
\hfill{\textbf{\footnotesize{[Fall 2012 - Spring 2015]}}}
    \begin{itemize}[itemsep=0.2em,topsep=0.5em]
        \item Computer Architecture (Spring 2015)
        \item Digital Logic Design (Fall 2014)
        \item Introduction to Computing Systems and Programming (Fall 2014)
        \item Electronics I (Fall 2014)
        \item Probability and Statistics for Engineering (Spring 2014)
        \item FPGA Summer School with IEEE Student Branch (Summer 2013)
        \item Numerical Calculations (Fall 2012)
    \end{itemize}

\section*{Mentoring Experience}
\vspace{-0.5em}\hrule\vspace{0.8em}

\textbf{Graduate Students}
\begin{itemize}[align=left,leftmargin=*,itemsep=0.3pt, topsep=0.5em] 
    \item Eirini Tzermpou, Algorithm-architecture co-design for graph-based genome analysis (2026 - Ongoing)
    \begin{itemize}[itemsep=0pt, topsep=0pt]
        \item \paper{Eirni M. Tzermpou, Konstantina Koliogeorgi, Akmuhammet Ashyralyyev, Sotirios Xydis, Dimitrios Soudris, \underline{Nika Mansouri Ghiasi}, Onur Mutlu}{}{Topology-Aware Software/Hardware Co-Design for Graph-Based Genome Analysis}{Under Submission, 2026} [Blinded]\vspace{0.3em}
    \end{itemize}
        \item Mayank Kabra, Encryption and privacy in storage-centric computing (2026 - Ongoing)
    \begin{itemize}[itemsep=0pt, topsep=0pt]
        \item \paper{Mayank Kabra, Harshita Gupta, Eirini Maria Tzermpou, Phillip William Widdowson, \underline{Nika Mansouri Ghiasi}, Ataberk Olgun, Rakesh Nadig, Konstantinos Kanellopoulos, Rakesh Kumar, Abdullah Giray Yaglikci, Jisung Park, Onur Mutlu}{}{Efficient Query Processing on Homomorphically Encrypted Databases via Storage-Centric Algorithm-Hardware Co-Design}{Under Submission, 2026} [Blinded]\vspace{0.3em}
    \end{itemize}
    \item Timur Eke, Benchmarking and accelerating single-cell analysis (2026 - Ongoing)\vspace{0.3em}
    \item Harshita Gupta, Privacy guarantees in storage-centric computing (2025 - Ongoing)\vspace{0.3em}
    \item Kangqi Chen, Storage-centric retrieval systems (2025)
    \begin{itemize}[itemsep=0pt, topsep=0pt]
        \item \paper{Kangqi Chen, Rakesh Nadig, Andreas Kosmas Kakolyris, Manos Frouzakis, \underline{Nika Mansouri Ghiasi}, Yu Liang, Haiyu Mao, Jisung Park, Mohammad Sadrosadati, Onur Mutlu}{https://arxiv.org/pdf/2506.16444}{REIS: A High-Performance and Energy-Efficient Retrieval System with In-Storage Processing}{International Symposium on Computer Architecture (ISCA), 2025}\vspace{0.3em}
    \end{itemize}
    \item Melina Soysal, Storage-centric raw signal genome analysis (2024 - 2025)
    \begin{itemize}[itemsep=0pt, topsep=0pt]
        \item \paper{Melina Soysal, Konstantina Koliogeorgi, Can Firtina, \underline{Nika Mansouri Ghiasi}, Rakesh Nadig, Haiyu Mao, Geraldo Francisco, Yu Liang, Klea Zambaku, Mohammad Sadrosadati, Onur Mutlu}{https://arxiv.org/pdf/2506.10931}{MARS: Processing-In-Memory Acceleration of Raw Signal Genome Analysis Inside the Storage Subsystem}{ACM International Conference on Supercomputing (ICS), 2025}\vspace{0.3em}
    \end{itemize}
    \item Banu Cavlak, Improving the performance of basecalling genomic data (2023 - 2024)
    \begin{itemize}[itemsep=0pt, topsep=0pt]
        \item \paper{Meryem Banu Cavlak, Gagandeep Singh, Mohammed Alser, Can Firtina, Joel Lindegger, Mohammad Sadrosadati, \underline{Nika Mansouri Ghiasi}, Can Alkan, Onur Mutlu}{https://arxiv.org/pdf/2212.04953.pdf}{TargetCall: Eliminating the Wasted Computation in Basecalling via Pre-Basecalling Filtering}{Asia Pacific Bioinformatics Conference (APBC), 2023}\vspace{0.3em}
    \end{itemize}
    \item Joel Lindegger, Algorithm-architecture co-design for genome analysis (2023 - 2024)
    \begin{itemize}[itemsep=0pt, topsep=0pt]
        \item \paper{Joel Lindegger, Can Firtina, \underline{Nika Mansouri Ghiasi}, Mohammad Sadrosadati, Mohammed Alser, Onur Mutlu}{https://arxiv.org/pdf/2310.05037}{RawAlign: Accurate, Fast, and Scalable Raw Nanopore Signal Mapping via Combining Seeding and Alignment}{IEEE Access, 2024}
        \item \paper{Joël Lindegger, Damla Senol Cali, Mohammed Alser, Juan Gómez-Luna, \underline{Nika Mansouri Ghiasi}, Onur Mutlu}{https://arxiv.org/pdf/2208.09985.pdf}{Scrooge: A Fast and Memory-Frugal Genomic Sequence Aligner for CPUs, GPUs, and ASICs}{Bioinformatics, 2023}\vspace{1em}
    \end{itemize}
\end{itemize}

\textbf{Undergraduate Students}
\begin{itemize}[align=left,leftmargin=*,itemsep=0.3pt,topsep=0.5em] 
    \item Wonyoung Cheon, Benchmarking and accelerating single-cell analysis (2026 - Ongoing)
    \item Talu G\"uloglu, Storage-centric genomics, Alleviating data preparation bottlenecks in genome analysis (2023 - 2026)
    \begin{itemize}[itemsep=0pt, topsep=0pt]
        \item \paper{\underline{Nika Mansouri Ghiasi}, Harun Mustafa, Talu G\"uloglu, Rakesh Nadig, Konstantina Koliogeorgi, Susana Ruiz, Marc Rautmann, Furkan Eris, Mohammad Sadrosadati, Jisung Park, Onur Mutlu}{https://arxiv.org/pdf/2606.26468}{GRAINS: Enabling High-Performance and Low-Cost Graph-Based Genome Analysis via Storage-Aware Algorithm-Architecture Co-Design}{International Symposium on Computer Architecture (ISCA), 2026}
        \item \paper{\underline{Nika Mansouri Ghiasi}, Talu G\"uloglu, Harun Mustafa, Can Firtina, Konstantina Koliogeorgi, Konstantinos Kanellopoulos, Haiyu Mao, Rakesh Nadig, Mohammad Sadrosadati, Jisung Park, Onur Mutlu}{https://arxiv.org/pdf/2504.03732}{SAGe: A Lightweight Algorithm-Architecture Co-Design for Mitigating the Data Preparation Bottleneck in Large-Scale Genome Sequence Analysis}{International Symposium on High-Performance Computer Architecture (HPCA), 2026}\vspace{0.3em}
    \end{itemize}
    \item Marc Rautmann, Storage-centric genomics (2025 - 2026)
    \begin{itemize}[itemsep=0pt, topsep=0pt]
        \item \paper{\underline{Nika Mansouri Ghiasi}, Harun Mustafa, Talu G\"uloglu, Rakesh Nadig, Konstantina Koliogeorgi, Susana Ruiz, Marc Rautmann, Furkan Eris, Mohammad Sadrosadati, Jisung Park, Onur Mutlu}{https://arxiv.org/pdf/2606.26468}{GRAINS: Enabling High-Performance and Low-Cost Graph-Based Genome Analysis via Storage-Aware Algorithm-Architecture Co-Design}{International Symposium on Computer Architecture (ISCA), 2026}\vspace{0.3em}
    \end{itemize}
    \item Yeejoo Han, Understanding data movement overheads of genome assembly (2023)
\end{itemize}

\section*{\tomiv{Academic and Professional Service}}
\vspace{-0.5em}\hrule\vspace{0.8em}

\begin{itemize}[align=left,leftmargin=*,widest={10}]
    \item \textbf{Lead Organizer, Architecture for Health (Arch4Health) Initiative}
    \begin{itemize}[itemsep=0.3pt,topsep=0.2em]
        \item \href{https://events.safari.ethz.ch/micro25-arch4health/}{First Arch4Health Workshop}, held with the International Symposium on Microarchitecture (MICRO) 2025 [\href{https://www.youtube.com/live/lTc_gQzFNJI}{Videos}]
        \item \href{https://events.safari.ethz.ch/hpca26-arch4health/}{Second Arch4Health Workshop}, held with the International Symposium on High-Performance Computer Architecture (HPCA) 2026 [\href{https://www.youtube.com/watch?v=hSRSZCjnKVo}{Videos}]
        \item \href{https://events.safari.ethz.ch/ics26-arch4health/}{Third Arch4Health Workshop}, held with the International Conference on Supercomputing (ICS) 2026 [\href{https://www.youtube.com/watch?v=hSRSZCjnKVo}{Videos}]
        \item Fourth Arch4Health Workshop, to be held in conjunction with the International Symposium on Microarchitecture (MICRO) 2026
        \item \href{https://events.safari.ethz.ch/sosp26-sys4health/}{First Sys4Health Workshop}, to be held in conjunction with the Symposium on Operating Systems Principles (SOSP) 2026
    \end{itemize}

    \item \textbf{Program Committee Member}
    \begin{itemize}[itemsep=0.3pt,topsep=0.2em]
        \item International Symposium on High-Performance Computer Architecture (HPCA) 2027
        \item International Symposium on Microarchitecture (MICRO) 2026
    \end{itemize}
    \item \textbf{Reviewer}
    \begin{itemize}[noitemsep,topsep=0pt]
        \item Computer Architecture Letters (CAL) 2026
        \item Research in Computational Molecular Biology (RECOMB) 2026
        \item Pacific Symposium on Biocomputing (PSB) 2026
        \item Bioinformatics 2023 \& 2024
        \item IEEE Transactions on Computer-Aided Design of Integrated Circuits and Systems (TCAD) 2022
    \end{itemize}
    \item \textbf{Sub-Reviewer:} MICRO 2018-2025, ISCA 2018-2025, HPCA 2018-2020 and 2024-2025, ASPLOS 2023-2025, JSSC 2024, RECOMB 2022, DSN 2019, MSST 2019

    \item \textbf{Meet a Senior Student (MaSS)}, held with the International Symposium on Computer Architecture (ISCA), 2026, to offer research guidance and support to junior PhD students in navigating their PhD journey
\end{itemize}

\section*{Open Source Tools and Repositories}
\vspace{-0.5em}\hrule\vspace{0.8em}

\begin{itemize}[align=left,leftmargin=*,widest={10}]
    \item \textbf{GenStore:} \href{https://github.com/CMU-SAFARI/GenStore}{https://github.com/CMU-SAFARI/GenStore}
    \item \textbf{MegIS:} \href{https://github.com/CMU-SAFARI/MegIS}{https://github.com/CMU-SAFARI/MegIS}
    \item \textbf{Ariadne:} \href{https://github.com/CMU-SAFARI/Ariadne}{https://github.com/CMU-SAFARI/Ariadne}
    \item \textbf{RawHash and Rawsamble:} \href{https://github.com/CMU-SAFARI/RawHash}{https://github.com/CMU-SAFARI/RawHash}
    \item \textbf{IMPACT:} \href{https://github.com/CMU-SAFARI/IMPACT}{https://github.com/CMU-SAFARI/IMPACT}
    \item \textbf{MLA:} \href{https://github.com/ratschlab/mla}{https://github.com/ratschlab/mla}
    \item \textbf{RawAlign:} \href{https://github.com/CMU-SAFARI/RawAlign}{https://github.com/CMU-SAFARI/RawAlign}
    \item \textbf{BLEND:} \href{https://github.com/CMU-SAFARI/BLEND}
    {https://github.com/CMU-SAFARI/BLEND}
    \item \textbf{Scrooge:} \href{https://github.com/CMU-SAFARI/Scrooge}{https://github.com/CMU-SAFARI/Scrooge}
    \item \textbf{TargetCall:} \href{https://github.com/CMU-SAFARI/TargetCall}{https://github.com/CMU-SAFARI/TargetCall}
    \item \textbf{SMASH:} \href{https://github.com/CMU-SAFARI/SMASH}{https://github.com/CMU-SAFARI/SMASH}
    \item \textbf{CROW:} \href{https://github.com/CMU-SAFARI/CROW}{https://github.com/CMU-SAFARI/CROW}
\end{itemize}

\balance
\begin{singlespace}
\bibliographystyle{unsrt} %
\bibliography{
    backmatter/01_main
}

\begin{thebibliography}{1000}

\bibitem{srastats}
{{National Center for Biotechnology Information}}.
\newblock {SRA database growth}.
\newblock \url{https://www.ncbi.nlm.nih.gov/sra/docs/sragrowth/}, 2024.

\bibitem{stephens2015big}
Zachary~D. Stephens, Skylar~Y. Lee, Faraz Faghri, Roy~H. Campbell, Chengxiang Zhai, Miles~J. Efron, Ravishankar Iyer, Michael~C. Schatz, Saurabh Sinha, and Gene~E. Robinson.
\newblock {Big Data: Astronomical or Genomical?}
\newblock {\em PLOS Biology}, 2015.

\bibitem{moore1998cramming}
Gordon~E Moore.
\newblock {Cramming More Components onto Integrated Circuits}.
\newblock {\em Proceedings of the IEEE}, 1998.

\bibitem{wu2021sieve}
Lingxi Wu, Rasool Sharifi, Marzieh Lenjani, Kevin Skadron, and Ashish Venkat.
\newblock {Sieve: Scalable In-situ DRAM-based Accelerator Designs for Massively Parallel k-mer Matching}.
\newblock In {\em ISCA}, 2021.

\bibitem{alkan2009personalized}
Can Alkan, Jeffrey~M Kidd, Tomas Marques-Bonet, Gozde Aksay, Francesca Antonacci, Fereydoun Hormozdiari, Jacob~O Kitzman, Carl Baker, Maika Malig, Onur Mutlu, S~Cenk Sahinalp, Richard~A Gibbs, and Evan~E Eichler.
\newblock {Personalized Copy Number and Segmental Duplication Maps Using Next-Generation Sequencing}.
\newblock {\em Nature Genetics}, 2009.

\bibitem{lightbody_review_2019}
Gaye Lightbody, Valeriia Haberland, Fiona Browne, Laura Taggart, Huiru Zheng, Eileen Parkes, and Jaine~K Blayney.
\newblock Review of applications of high-throughput sequencing in personalized medicine: barriers and facilitators of future progress in research and clinical application.
\newblock {\em Briefings in Bioinformatics}, 2019.

\bibitem{morganti_next_2019}
Stefania Morganti, Paolo Tarantino, Emanuela Ferraro, Paolo D'Amico, Bruno~Achutti Duso, and Giuseppe Curigliano.
\newblock {Next Generation Sequencing (NGS): A Revolutionary Technology in Pharmacogenomics and Personalized Medicine in Cancer}.
\newblock In {\em Translational Research and Onco-Omics Applications in the Era of Cancer Personal Genomics}. 2019.

\bibitem{branco_bioinformatics_2021}
Iuliia Branco and Altino Choupina.
\newblock Bioinformatics: new tools and applications in life science and personalized medicine.
\newblock {\em Applied Microbiology and Biotechnology}, 2021.

\bibitem{quazi_artificial_2022}
Sameer Quazi.
\newblock Artificial intelligence and machine learning in precision and genomic medicine.
\newblock {\em Medical Oncology}, 2022.

\bibitem{aronson_building_2015}
Samuel~J. Aronson and Heidi~L. Rehm.
\newblock Building the foundation for genomics in precision medicine.
\newblock {\em Nature}, 2015.

\bibitem{f_lochel_comparative_2020}
Hannah F.~Löchel and Dominik Heider.
\newblock Comparative analyses of error handling strategies for next-generation sequencing in precision medicine.
\newblock {\em Scientific Reports}, 2020.

\bibitem{papadopoulou_application_2023}
Eirini Papadopoulou, Dimitra Bouzarelou, George Tsaousis, Athanasios Papathanasiou, Georgia Vogiatzi, Charalambos Vlachopoulos, Antigoni Miliou, Panagiota Papachristou, Efstathia Prappa, Georgios Servos, Konstantinos Ritsatos, Aristeidis Seretis, Alexandra Frogoudaki, and George Nasioulas.
\newblock Application of next generation sequencing in cardiology: current and future precision medicine implications.
\newblock {\em Frontiers in Cardiovascular Medicine}, 2023.

\bibitem{tafazoli_applying_2021}
Alireza Tafazoli, Henk-Jan Guchelaar, Wojciech Miltyk, Adam~J. Kretowski, and Jesse~J. Swen.
\newblock Applying {Next}-{Generation} {Sequencing} {Platforms} for {Pharmacogenomic} {Testing} in {Clinical} {Practice}.
\newblock {\em Frontiers in Pharmacology}, 2021.

\bibitem{gambardella_personalized_2020}
Valentina Gambardella, Noelia Tarazona, Juan~M. Cejalvo, Pasquale Lombardi, Marisol Huerta, Susana Roselló, Tania Fleitas, Desamparados Roda, and Andres Cervantes.
\newblock Personalized {Medicine}: {Recent} {Progress} in {Cancer} {Therapy}.
\newblock {\em Cancers}, 2020.

\bibitem{leary_development_2010}
Rebecca~J. Leary, Isaac Kinde, Frank Diehl, Kerstin Schmidt, Chris Clouser, Cisilya Duncan, Alena Antipova, Clarence Lee, Kevin McKernan, Francisco~M. De~La~Vega, Kenneth~W. Kinzler, Bert Vogelstein, Luis~A. Diaz, and Victor~E. Velculescu.
\newblock Development of {Personalized} {Tumor} {Biomarkers} {Using} {Massively} {Parallel} {Sequencing}.
\newblock {\em Science Translational Medicine}, 2010.

\bibitem{hamburg_margaret_a_path_2010}
{Hamburg Margaret A.} and {Collins Francis S.}
\newblock The {Path} to {Personalized} {Medicine}.
\newblock {\em New England Journal of Medicine}, 2010.

\bibitem{van_der_lee_technologies_2020}
Maaike van~der Lee, Marjolein Kriek, Henk-Jan Guchelaar, and Jesse~J. Swen.
\newblock Technologies for {Pharmacogenomics}: {A} {Review}.
\newblock {\em Genes}, 2020.

\bibitem{moon_precision_2022}
Dabin Moon, Hye~W. Park, Dongheon Surl, Dongju Won, Seung-Tae Lee, Saeam Shin, Jong~R. Choi, and Jinu Han.
\newblock Precision {Medicine} through {Next}-{Generation} {Sequencing} in {Inherited} {Eye} {Diseases} in a {Korean} {Cohort}.
\newblock {\em Genes}, 2022.

\bibitem{mohan_profiling_2020}
Sumitra Mohan, Victoria Foy, Mahmood Ayub, Hui~Sun Leong, Pieta Schofield, Sudhakar Sahoo, Tine Descamps, Bedirhan Kilerci, Nigel~K. Smith, Mathew Carter, Lynsey Priest, Cong Zhou, T.~Hedley Carr, Crispin Miller, Corinne Faivre-Finn, Fiona Blackhall, Dominic~G. Rothwell, Caroline Dive, and Gerard Brady.
\newblock Profiling of {Circulating} {Free} {DNA} {Using} {Targeted} and {Genome}-wide {Sequencing} in {Patients} with {SCLC}.
\newblock {\em Journal of Thoracic Oncology}, 2020.

\bibitem{chung_rapid_2020}
Claudia~C.Y. Chung, Gordon~K.C. Leung, Christopher~C.Y. Mak, Jasmine~L.F. Fung, Mianne Lee, Steven~L.C. Pei, Mullin~H.C. Yu, Vivian~C.C. Hui, Joshua~C.K. Chan, Jeffrey~F.T. Chau, Marcus~C.Y. Chan, Mandy~H.Y. Tsang, Wilfred~H.S. Wong, Joanna~Y.L. Tung, Kin~Shing Lun, Yiu~Ki Ng, Cheuk~Wing Fung, Mabel~S.C. Wong, Rosanna~M.S. Wong, Yu~Lung Lau, et~al.
\newblock Rapid whole-exome sequencing facilitates precision medicine in paediatric rare disease patients and reduces healthcare costs.
\newblock {\em The Lancet Regional Health – Western Pacific}, 2020.

\bibitem{bielinski_preemptive_2014}
Suzette~J. Bielinski, Janet~E. Olson, Jyotishman Pathak, Richard~M. Weinshilboum, Liewei Wang, Kelly~J. Lyke, Euijung Ryu, Paul~V. Targonski, Michael~D. Van~Norstrand, Matthew~A. Hathcock, Paul~Y. Takahashi, Jennifer~B. McCormick, Kiley~J. Johnson, Karen~J. Maschke, Carolyn~R. Rohrer~Vitek, Marissa~S. Ellingson, Eric~D. Wieben, Gianrico Farrugia, Jody~A. Morrisette, Keri~J. Kruckeberg, et~al.
\newblock Preemptive {Genotyping} for {Personalized} {Medicine}: {Design} of the {Right} {Drug}, {Right} {Dose}, {Right} {Time}—{Using} {Genomic} {Data} to {Individualize} {Treatment} {Protocol}.
\newblock {\em Mayo Clinic Proceedings}, 2014.

\bibitem{ho_enabling_2020}
Dean Ho, Stephen~R. Quake, Edward~R.B. McCabe, Wee~Joo Chng, Edward~K. Chow, Xianting Ding, Bruce~D. Gelb, Geoffrey~S. Ginsburg, Jason Hassenstab, Chih-Ming Ho, William~C. Mobley, Garry~P. Nolan, Steven~T. Rosen, Patrick Tan, Yun Yen, and Ali Zarrinpar.
\newblock Enabling {Technologies} for {Personalized} and {Precision} {Medicine}.
\newblock {\em Trends in Biotechnology}, 2020.

\bibitem{hussen_emerging_2022}
Bashdar~Mahmud Hussen, Sara~Tharwat Abdullah, Abbas Salihi, Dana~Khdr Sabir, Karzan~R. Sidiq, Mohammed~Fatih Rasul, Hazha~Jamal Hidayat, Soudeh Ghafouri-Fard, Mohammad Taheri, and Elena Jamali.
\newblock The emerging roles of {NGS} in clinical oncology and personalized medicine.
\newblock {\em Pathology - Research and Practice}, 2022.

\bibitem{russell_pharmacogenomics_2021}
Laura~E. Russell, Yitian Zhou, Ahmed~A. Almousa, Jasleen~K. Sodhi, Chukwunonso~K. Nwabufo, and Volker~M. Lauschke.
\newblock Pharmacogenomics in the era of next generation sequencing – from byte to bedside.
\newblock {\em Drug Metabolism Reviews}, 2021.

\bibitem{verma_nanopore_2024}
Renu Verma, Kesia~Esther da~Silva, Neesha Rockwood, Roeland~E. Wasmann, Nombuso Yende, Taeksun Song, Eugene Kim, Paolo Denti, Robert~J. Wilkinson, and Jason~R. Andrews.
\newblock A {Nanopore} {Sequencing}-based {Pharmacogenomic} {Panel} to {Personalize} {Tuberculosis} {Drug} {Dosing}.
\newblock {\em American Journal of Respiratory and Critical Care Medicine}, 2024.

\bibitem{clark2019diagnosis}
Michelle~M. Clark, Amber Hildreth, Sergey Batalov, Yan Ding, Shimul Chowdhury, Kelly Watkins, Katarzyna Ellsworth, Brandon Camp, Cyrielle~I. Kint, Calum Yacoubian, Lauge Farnaes, Matthew~N. Bainbridge, Curtis Beebe, Joshua J.~A. Braun, Margaret Bray, Jeanne Carroll, Julie~A. Cakici, Sara~A. Caylor, Christina Clarke, Mitchell~P. Creed, et~al.
\newblock {Diagnosis of Genetic Diseases in Seriously Ill Children by Rapid Whole-genome Sequencing and Automated Phenotyping and Interpretation}.
\newblock {\em Science Translational Medicine}, 2019.

\bibitem{farnaes2018rapid}
Lauge Farnaes, Amber Hildreth, Nathaly~M. Sweeney, Michelle~M. Clark, Shimul Chowdhury, Shareef Nahas, Julie~A. Cakici, Wendy Benson, Robert~H. Kaplan, Richard Kronick, Matthew~N. Bainbridge, Jennifer Friedman, Jeffrey~J. Gold, Yan Ding, Narayanan Veeraraghavan, David Dimmock, and Stephen~F. Kingsmore.
\newblock {Rapid Whole-genome Sequencing Decreases Infant Morbidity and Cost of Hospitalization}.
\newblock {\em NPJ Genomic Medicine}, 2018.

\bibitem{sweeney2021rapid}
Nathaly~M. Sweeney, Shareef~A. Nahas, Shimul Chowdhury, Sergey Batalov, Michelle Clark, Sara Caylor, Julie Cakici, John~J. Nigro, Yan Ding, Narayanan Veeraraghavan, Charlotte Hobbs, David Dimmock, and Stephen~F. Kingsmore.
\newblock {Rapid Whole Genome Sequencing Impacts Care and Resource Utilization in Infants with Congenital Heart Disease}.
\newblock {\em NPJ Genomic Medicine}, 2021.

\bibitem{flores2013p4}
Mauricio Flores, Gustavo Glusman, Kristin Brogaard, Nathan~D Price, and Leroy Hood.
\newblock {P4 Medicine: How Systems Medicine Will Transform the Healthcare Sector and Society}.
\newblock {\em Personalized Medicine}, 2013.

\bibitem{ginsburg2009genomic}
Geoffrey~S Ginsburg and Huntington~F Willard.
\newblock {Genomic and Personalized Medicine: Foundations and Applications}.
\newblock {\em Translational Research}, 2009.

\bibitem{chin2011cancer}
Lynda Chin, Jannik~N Andersen, and P~Andrew Futreal.
\newblock {Cancer Genomics: From Discovery Science to Personalized Medicine}.
\newblock {\em Nature Medicine}, 2011.

\bibitem{Ashley2016}
Euan~A Ashley.
\newblock {Towards Precision Medicine}.
\newblock {\em Nature Reviews Genetics}, 2016.

\bibitem{hansen2026complete}
Nancy~F. Hansen, Nathan Dwarshuis, Hyun~Joo Ji, Arang Rhie, Hailey Loucks, Glennis~A. Logsdon, Mitchell~R. Vollger, Jessica~M. Storer, Juhyun Kim, Eleni Adam, Nicolas Altemose, Dmitry Antipov, Mobin Asri, Sofia Barreira, Stephanie~C. Bohaczuk, Andrey~V. Bzikadze, Sara~A. Carioscia, Andrew Carroll, Kuan-Hao Chao, Yanan Chu, et~al.
\newblock A complete diploid human genome benchmark for personalized genomics.
\newblock {\em Cell}, 2026.

\bibitem{dunn2021squigglefilter}
Tim Dunn, Harisankar Sadasivan, Jack Wadden, Kush Goliya, Kuan-Yu Chen, David Blaauw, Reetuparna Das, and Satish Narayanasamy.
\newblock {SquiggleFilter: An Accelerator for Portable Virus Detection}.
\newblock In {\em MICRO}, 2021.

\bibitem{bertelli_rapid_2013}
C.~Bertelli and G.~Greub.
\newblock Rapid bacterial genome sequencing: methods and applications in clinical microbiology.
\newblock {\em Clinical Microbiology and Infection}, 2013.

\bibitem{arias_rapid_2016}
Armando Arias, Simon~J. Watson, Danny Asogun, Ekaete~Alice Tobin, Jia Lu, My~V.~T. Phan, Umaru Jah, Raoul Emeric~Guetiya Wadoum, Luke Meredith, Lucy Thorne, Sarah Caddy, Alimamy Tarawalie, Pinky Langat, Gytis Dudas, Nuno~R. Faria, Simon Dellicour, Abdul Kamara, Brima Kargbo, Brima~Osaio Kamara, Sahr Gevao, et~al.
\newblock Rapid outbreak sequencing of {Ebola} virus in {Sierra} {Leone} identifies transmission chains linked to sporadic cases.
\newblock {\em Virus Evolution}, 2016.

\bibitem{comin_investigation_2020}
Jessica Comin, Armando Chaure, Alberto Cebollada, Daniel Ibarz, Jesús Viñuelas, María~Asunción Vitoria, María~José Iglesias, and Sofía Samper.
\newblock Investigation of a rapidly spreading tuberculosis outbreak using whole-genome sequencing.
\newblock {\em Infection, Genetics and Evolution}, 2020.

\bibitem{Quick2016}
Joshua Quick, Nicholas~J. Loman, Sophie Duraffour, Jared~T. Simpson, Ettore Severi, Lauren Cowley, Joseph~Akoi Bore, Raymond Koundouno, Gytis Dudas, Amy Mikhail, Nobila Ou{\'e}draogo, Babak Afrough, Amadou Bah, Jonathan H.~J. Baum, Beate Becker-Ziaja, Jan~Peter Boettcher, Mar Cabeza-Cabrerizo, {\'A}lvaro Camino-S{\'a}nchez, Lisa~L. Carter, Juliane Doerrbecker, et~al.
\newblock {Real-time, portable genome sequencing for Ebola surveillance}.
\newblock {\em Nature}, 2016.

\bibitem{robinson_genomics_2013}
Esther~R. Robinson, Timothy~M. Walker, and Mark~J. Pallen.
\newblock Genomics and outbreak investigation: from sequence to consequence.
\newblock {\em Genome Medicine}, 2013.

\bibitem{fournier_clinical_2014}
Pierre-Edouard Fournier, Gregory Dubourg, and Didier Raoult.
\newblock Clinical detection and characterization of bacterial pathogens in the genomics era.
\newblock {\em Genome Medicine}, 2014.

\bibitem{koser_routine_2012}
Claudio~U. Köser, Matthew~J. Ellington, Edward J.~P. Cartwright, Stephen~H. Gillespie, Nicholas~M. Brown, Mark Farrington, Matthew T.~G. Holden, Gordon Dougan, Stephen~D. Bentley, Julian Parkhill, and Sharon~J. Peacock.
\newblock Routine {Use} of {Microbial} {Whole} {Genome} {Sequencing} in {Diagnostic} and {Public} {Health} {Microbiology}.
\newblock {\em PLOS Pathogens}, 2012.

\bibitem{eloit_diagnosis_2014}
Marc Eloit and marc lecuit.
\newblock The diagnosis of infectious diseases by whole genome next generation sequencing: a new era is opening.
\newblock {\em Frontiers in Cellular and Infection Microbiology}, 2014.

\bibitem{gardy_jennifer_l_whole-genome_2011}
{Gardy Jennifer L.}, {Johnston James C.}, {Sui Shannan J. Ho}, {Cook Victoria J.}, {Shah Lena}, {Brodkin Elizabeth}, {Rempel Shirley}, {Moore Richard}, {Zhao Yongjun}, {Holt Robert}, {Varhol Richard}, {Birol Inanc}, {Lem Marcus}, {Sharma Meenu K.}, {Elwood Kevin}, {Jones Steven J.M.}, {Brinkman Fiona S.L.}, {Brunham Robert C.}, and {Tang Patrick}.
\newblock Whole-{Genome} {Sequencing} and {Social}-{Network} {Analysis} of a {Tuberculosis} {Outbreak}.
\newblock {\em New England Journal of Medicine}, 2011.

\bibitem{taylor_angela_j_characterization_2015}
{Taylor Angela J.}, {Lappi Victoria}, {Wolfgang William J.}, {Lapierre Pascal}, {Palumbo Michael J.}, {Medus Carlota}, and {Boxrud David}.
\newblock Characterization of {Foodborne} {Outbreaks} of {Salmonella} enterica {Serovar} {Enteritidis} with {Whole}-{Genome} {Sequencing} {Single} {Nucleotide} {Polymorphism}-{Based} {Analysis} for {Surveillance} and {Outbreak} {Detection}.
\newblock {\em Journal of Clinical Microbiology}, 2015.

\bibitem{quainoo_scott_whole-genome_2017}
{Quainoo Scott}, {Coolen Jordy P. M.}, {van Hijum Sacha A. F. T.}, {Huynen Martijn A.}, {Melchers Willem J. G.}, {van Schaik Willem}, and {Wertheim Heiman F. L.}
\newblock Whole-{Genome} {Sequencing} of {Bacterial} {Pathogens}: the {Future} of {Nosocomial} {Outbreak} {Analysis}.
\newblock {\em Clinical Microbiology Reviews}, 2017.

\bibitem{goldberg_brittany_making_2015}
{Goldberg Brittany}, {Sichtig Heike}, {Geyer Chelsie}, {Ledeboer Nathan}, and {Weinstock George M.}
\newblock Making the {Leap} from {Research} {Laboratory} to {Clinic}: {Challenges} and {Opportunities} for {Next}-{Generation} {Sequencing} in {Infectious} {Disease} {Diagnostics}.
\newblock {\em mBio}, 2015.

\bibitem{besser_interpretation_2019}
John~M. Besser, Heather~A. Carleton, Eija Trees, Steven~G. Stroika, Kelley Hise, Matthew Wise, and Peter Gerner-Smidt.
\newblock Interpretation of {Whole}-{Genome} {Sequencing} for {Enteric} {Disease} {Surveillance} and {Outbreak} {Investigation}.
\newblock {\em Foodborne Pathogens and Disease}, 2019.

\bibitem{li_application_2021}
Weiwei Li, Qingpo Cui, Li~Bai, Ping Fu, Haihong Han, Jikai Liu, and Yunchang Guo.
\newblock Application of {Whole}-{Genome} {Sequencing} in the {National} {Molecular} {Tracing} {Network} for {Foodborne} {Disease} {Surveillance} in {China}.
\newblock {\em Foodborne Pathogens and Disease}, 2021.

\bibitem{deng_integrated_2021}
Yinhua Deng, Min Jiang, Patrick~S.L. Kwan, Chao Yang, Qiongcheng Chen, Yiman Lin, Yaqun Qiu, Yinghui Li, Xiaolu Shi, Liqiang Li, Yujun Cui, Qun Sun, and Qinghua Hu.
\newblock Integrated {Whole}-{Genome} {Sequencing} {Infrastructure} for {Outbreak} {Detection} and {Source} {Tracing} of {Salmonella} enterica {Serotype} {Enteritidis}.
\newblock {\em Foodborne Pathogens and Disease}, 2021.

\bibitem{kwong_whole_2015}
J.C. Kwong, N.~Mccallum, V.~Sintchenko, and B.P. Howden.
\newblock Whole genome sequencing in clinical and public health microbiology.
\newblock {\em Pathology}, 2015.

\bibitem{deurenberg_application_2017}
Ruud~H. Deurenberg, Erik Bathoorn, Monika~A. Chlebowicz, Natacha Couto, Mithila Ferdous, Silvia García-Cobos, Anna~M.D. Kooistra-Smid, Erwin~C. Raangs, Sigrid Rosema, Alida~C.M. Veloo, Kai Zhou, Alexander~W. Friedrich, and John~W.A. Rossen.
\newblock Application of next generation sequencing in clinical microbiology and infection prevention.
\newblock {\em Journal of Biotechnology}, 2017.

\bibitem{tang_infection_2017}
Patrick Tang, Matthew~A. Croxen, Mohammad~R. Hasan, William~W.L. Hsiao, and Linda~M. Hoang.
\newblock Infection control in the new age of genomic epidemiology.
\newblock {\em American Journal of Infection Control}, 2017.

\bibitem{croucher_application_2015}
Nicholas~J Croucher and Xavier Didelot.
\newblock The application of genomics to tracing bacterial pathogen transmission.
\newblock {\em Host–microbe interactions: bacteria • Genomics}, 2015.

\bibitem{bloom2021massively}
Joshua~S. Bloom, Laila Sathe, Chetan Munugala, Eric~M. Jones, Molly Gasperini, Nathan~B. Lubock, Fauna Yarza, Erin~M. Thompson, Kyle~M. Kovary, Jimin Park, Dawn Marquette, Stephania Kay, Mark Lucas, TreQuan Love, A.~Sina~Booeshaghi, Oliver~F. Brandenberg, Longhua Guo, James Boocock, Myles Hochman, Scott~W. Simpkins, et~al.
\newblock {Massively Scaled-up Testing for SARS-CoV-2 RNA via Next-generation Sequencing of Pooled and Barcoded Nasal and Saliva Samples}.
\newblock {\em Nature Biomedical Engineering}, 2021.

\bibitem{yelagandula2021multiplexed}
Ramesh Yelagandula, Aleksandr Bykov, Alexander Vogt, Robert Heinen, Ezgi {\"O}zkan, Marcus~Martin Strobl, Juliane~Christina Baar, Kristina Uzunova, Bence Hajdusits, Darja Kordic, Erna Suljic, Amina Kurtovic-Kozaric, Sebija Izetbegovic, Justine Schaeffer, Peter Hufnagl, Alexander Zoufaly, Tamara Seitz, Mariam Al-Rawi, Stefan Ameres, Juliane Baar, et~al.
\newblock {Multiplexed Detection of SARS-CoV-2 and Other Respiratory Infections in High Throughput by SARSeq}.
\newblock {\em Nature Communications}, 2021.

\bibitem{le2013selected}
Vien Thi~Minh Le and Binh~An Diep.
\newblock {Selected Insights from Application of Whole Genome Sequencing for Outbreak Investigations}.
\newblock {\em Current Opinion in Critical Care}, 2013.

\bibitem{nikolayevskyy2016whole}
Vlad Nikolayevskyy, Katharina Kranzer, Stefan Niemann, and Francis Drobniewski.
\newblock {Whole Genome Sequencing of Mycobacterium Tuberculosis for Detection of Recent Transmission and Tracing Outbreaks: A Systematic Review}.
\newblock {\em Tuberculosis}, 2016.

\bibitem{qiu2015whole}
Shaofu Qiu, Peng Li, Hongbo Liu, Yong Wang, Nan Liu, Chengyi Li, Shenlong Li, Ming Li, Zhengjie Jiang, Huandong Sun, Ying Li, Jing Xie, Chaojie Yang, Jian Wang, Hao Li, Shengjie Yi, Zhihao Wu, Leili Jia, Ligui Wang, Rongzhang Hao, et~al.
\newblock {Whole-genome Sequencing for Tracing the Transmission Link between Two ARD Outbreaks Caused by A Novel HAdV Serotype 7 Variant, China}.
\newblock {\em Scientific Reports}, 2015.

\bibitem{gilchrist2015whole}
Carol~A Gilchrist, Stephen~D Turner, Margaret~F Riley, William~A Petri, and Erik~L Hewlett.
\newblock {Whole-genome Sequencing in Outbreak Analysis}.
\newblock {\em Clinical Microbiology Reviews}, 2015.

\bibitem{lawrence_mutational_2013}
Michael~S. Lawrence, Petar Stojanov, Paz Polak, Gregory~V. Kryukov, Kristian Cibulskis, Andrey Sivachenko, Scott~L. Carter, Chip Stewart, Craig~H. Mermel, Steven~A. Roberts, Adam Kiezun, Peter~S. Hammerman, Aaron McKenna, Yotam Drier, Lihua Zou, Alex~H. Ramos, Trevor~J. Pugh, Nicolas Stransky, Elena Helman, Jaegil Kim, et~al.
\newblock Mutational heterogeneity in cancer and the search for new cancer-associated genes.
\newblock {\em Nature}, 2013.

\bibitem{vogelstein_cancer_2013}
Bert Vogelstein, Nickolas Papadopoulos, Victor~E. Velculescu, Shibin Zhou, Luis~A. Diaz, and Kenneth~W. Kinzler.
\newblock Cancer {Genome} {Landscapes}.
\newblock {\em Science}, 2013.

\bibitem{ramskold_full-length_2012}
Daniel Ramsköld, Shujun Luo, Yu-Chieh Wang, Robin Li, Qiaolin Deng, Omid~R Faridani, Gregory~A Daniels, Irina Khrebtukova, Jeanne~F Loring, Louise~C Laurent, Gary~P Schroth, and Rickard Sandberg.
\newblock Full-length {mRNA}-{Seq} from single-cell levels of {RNA} and individual circulating tumor cells.
\newblock {\em Nature Biotechnology}, 2012.

\bibitem{baslan_unravelling_2017}
Timour Baslan and James Hicks.
\newblock Unravelling biology and shifting paradigms in cancer with single-cell sequencing.
\newblock {\em Nature Reviews Cancer}, 2017.

\bibitem{shapiro_single-cell_2013}
Ehud Shapiro, Tamir Biezuner, and Sten Linnarsson.
\newblock Single-cell sequencing-based technologies will revolutionize whole-organism science.
\newblock {\em Nature Reviews Genetics}, 2013.

\bibitem{sakamoto_new_2020}
Yoshitaka Sakamoto, Sarun Sereewattanawoot, and Ayako Suzuki.
\newblock A new era of long-read sequencing for cancer genomics.
\newblock {\em Journal of Human Genetics}, 2020.

\bibitem{jia_high-throughput_2022}
Qingzhu Jia, Han Chu, Zheng Jin, Haixia Long, and Bo~Zhu.
\newblock High-throughput single-cell sequencing in cancer research.
\newblock {\em Signal Transduction and Targeted Therapy}, 2022.

\bibitem{lawson_tumour_2018}
Devon~A. Lawson, Kai Kessenbrock, Ryan~T. Davis, Nicholas Pervolarakis, and Zena Werb.
\newblock Tumour heterogeneity and metastasis at single-cell resolution.
\newblock {\em Nature Cell Biology}, 2018.

\bibitem{liu_mrna-based_2023}
Chuang Liu, Qiangqiang Shi, Xiangang Huang, Seyoung Koo, Na~Kong, and Wei Tao.
\newblock {mRNA}-based cancer therapeutics.
\newblock {\em Nature Reviews Cancer}, 2023.

\bibitem{van_de_sande_applications_2023}
Bram Van~de Sande, Joon~Sang Lee, Euphemia Mutasa-Gottgens, Bart Naughton, Wendi Bacon, Jonathan Manning, Yong Wang, Jack Pollard, Melissa Mendez, Jon Hill, Namit Kumar, Xiaohong Cao, Xiao Chen, Mugdha Khaladkar, Ji~Wen, Andrew Leach, and Edgardo Ferran.
\newblock Applications of single-cell {RNA} sequencing in drug discovery and development.
\newblock {\em Nature Reviews Drug Discovery}, 2023.

\bibitem{chakravarty_clinical_2021}
Debyani Chakravarty and David~B. Solit.
\newblock Clinical cancer genomic profiling.
\newblock {\em Nature Reviews Genetics}, 2021.

\bibitem{cortes-ciriano_computational_2022}
Isidro Cortés-Ciriano, Doga~C. Gulhan, Jake June-Koo Lee, Giorgio E.~M. Melloni, and Peter~J. Park.
\newblock Computational analysis of cancer genome sequencing data.
\newblock {\em Nature Reviews Genetics}, 2022.

\bibitem{deveson_evaluating_2021}
Ira~W. Deveson, Binsheng Gong, Kevin Lai, Jennifer~S. LoCoco, Todd~A. Richmond, Jeoffrey Schageman, Zhihong Zhang, Natalia Novoradovskaya, James~C. Willey, Wendell Jones, Rebecca Kusko, Guangchun Chen, Bindu~Swapna Madala, James Blackburn, Igor Stevanovski, Ambica Bhandari, Devin Close, Jeffrey Conroy, Michael Hubank, Narasimha Marella, et~al.
\newblock Evaluating the analytical validity of circulating tumor {DNA} sequencing assays for precision oncology.
\newblock {\em Nature Biotechnology}, 2021.

\bibitem{xiao_toward_2021}
Wenming Xiao, Luyao Ren, Zhong Chen, Li~Tai Fang, Yongmei Zhao, Justin Lack, Meijian Guan, Bin Zhu, Erich Jaeger, Liz Kerrigan, Thomas~M. Blomquist, Tiffany Hung, Marc Sultan, Kenneth Idler, Charles Lu, Andreas Scherer, Rebecca Kusko, Malcolm Moos, Chunlin Xiao, Stephen~T. Sherry, et~al.
\newblock Toward best practice in cancer mutation detection with whole-genome and whole-exome sequencing.
\newblock {\em Nature Biotechnology}, 2021.

\bibitem{bolton_cancer_2020}
Kelly~L. Bolton, Ryan~N. Ptashkin, Teng Gao, Lior Braunstein, Sean~M. Devlin, Daniel Kelly, Minal Patel, Antonin Berthon, Aijazuddin Syed, Mariko Yabe, Catherine~C. Coombs, Nicole~M. Caltabellotta, Mike Walsh, Kenneth Offit, Zsofia Stadler, Diana Mandelker, Jessica Schulman, Akshar Patel, John Philip, Elsa Bernard, et~al.
\newblock Cancer therapy shapes the fitness landscape of clonal hematopoiesis.
\newblock {\em Nature Genetics}, 2020.

\bibitem{szustakowski_advancing_2021}
Joseph~D. Szustakowski, Suganthi Balasubramanian, Erika Kvikstad, Shareef Khalid, Paola~G. Bronson, Ariella Sasson, Emily Wong, Daren Liu, J.~Wade~Davis, Carolina Haefliger, A.~Katrina~Loomis, Rajesh Mikkilineni, Hyun~Ji Noh, Samir Wadhawan, Xiaodong Bai, Alicia Hawes, Olga Krasheninina, Ricardo Ulloa, Alex~E. Lopez, Erin~N. Smith, et~al.
\newblock Advancing human genetics research and drug discovery through exome sequencing of the {UK} {Biobank}.
\newblock {\em Nature Genetics}, 2021.

\bibitem{navin_future_2011}
Nicholas Navin and James Hicks.
\newblock Future medical applications of single-cell sequencing in cancer.
\newblock {\em Genome Medicine}, 2011.

\bibitem{hong_rna_2020}
Mingye Hong, Shuang Tao, Ling Zhang, Li-Ting Diao, Xuanmei Huang, Shaohui Huang, Shu-Juan Xie, Zhen-Dong Xiao, and Hua Zhang.
\newblock {RNA} sequencing: new technologies and applications in cancer research.
\newblock {\em Journal of Hematology \& Oncology}, 2020.

\bibitem{lei_applications_2021}
Yalan Lei, Rong Tang, Jin Xu, Wei Wang, Bo~Zhang, Jiang Liu, Xianjun Yu, and Si~Shi.
\newblock Applications of single-cell sequencing in cancer research: progress and perspectives.
\newblock {\em Journal of Hematology \& Oncology}, 2021.

\bibitem{han_single-cell_2022}
Yingying Han, Dan Wang, Lushan Peng, Tao Huang, Xiaoyun He, Junpu Wang, and Chunlin Ou.
\newblock Single-cell sequencing: a promising approach for uncovering the mechanisms of tumor metastasis.
\newblock {\em Journal of Hematology \& Oncology}, 2022.

\bibitem{federici_variants_2020}
Giulia Federici and Silvia Soddu.
\newblock Variants of uncertain significance in the era of high-throughput genome sequencing: a lesson from breast and ovary cancers.
\newblock {\em Journal of Experimental \& Clinical Cancer Research}, 2020.

\bibitem{zhang_singlecell_2021}
Yijie Zhang, Dan Wang, Miao Peng, Le~Tang, Jiawei Ouyang, Fang Xiong, Can Guo, Yanyan Tang, Yujuan Zhou, Qianjin Liao, Xu~Wu, Hui Wang, Jianjun Yu, Yong Li, Xiaoling Li, Guiyuan Li, Zhaoyang Zeng, Yixin Tan, and Wei Xiong.
\newblock Single‐cell {RNA} sequencing in cancer research.
\newblock {\em Journal of Experimental \& Clinical Cancer Research}, 2021.

\bibitem{ren_understanding_2018}
Xianwen Ren, Boxi Kang, and Zemin Zhang.
\newblock Understanding tumor ecosystems by single-cell sequencing: promises and limitations.
\newblock {\em Genome Biology}, 2018.

\bibitem{tian_cicero_2020}
Liqing Tian, Yongjin Li, Michael~N. Edmonson, Xin Zhou, Scott Newman, Clay McLeod, Andrew Thrasher, Yu~Liu, Bo~Tang, Michael~C. Rusch, John Easton, Jing Ma, Eric Davis, Austyn Trull, J.~Robert Michael, Karol Szlachta, Charles Mullighan, Suzanne~J. Baker, James~R. Downing, David~W. Ellison, et~al.
\newblock {CICERO}: a versatile method for detecting complex and diverse driver fusions using cancer {RNA} sequencing data.
\newblock {\em Genome Biology}, 2020.

\bibitem{malone_molecular_2020}
Eoghan~R. Malone, Marc Oliva, Peter J.~B. Sabatini, Tracy~L. Stockley, and Lillian~L. Siu.
\newblock Molecular profiling for precision cancer therapies.
\newblock {\em Genome Medicine}, 2020.

\bibitem{tang_single-cell_2019}
Xiaoning Tang, Yongmei Huang, Jinli Lei, Hui Luo, and Xiao Zhu.
\newblock The single-cell sequencing: new developments and medical applications.
\newblock {\em Cell \& Bioscience}, 2019.

\bibitem{ellsworth_single-cell_2017}
Darrell~L. Ellsworth, Heather~L. Blackburn, Craig~D. Shriver, Shahrooz Rabizadeh, Patrick Soon-Shiong, and Rachel~E. Ellsworth.
\newblock Single-cell sequencing and tumorigenesis: improved understanding of tumor evolution and metastasis.
\newblock {\em Clinical and Translational Medicine}, 2017.

\bibitem{zhong_application_2021}
Yiming Zhong, Feng Xu, Jinhua Wu, Jeffrey Schubert, and Marilyn~M. Li.
\newblock Application of {Next} {Generation} {Sequencing} in {Laboratory} {Medicine}.
\newblock {\em Annals of Laboratory Medicine}, 2021.

\bibitem{stadler_therapeutic_2021}
Zsofia~K. Stadler, Anna Maio, Debyani Chakravarty, Yelena Kemel, Margaret Sheehan, Erin Salo-Mullen, Kaitlyn Tkachuk, Christopher~J. Fong, Bastien Nguyen, Amanda Erakky, Karen Cadoo, Ying Liu, Maria~I. Carlo, Alicia Latham, Hongxin Zhang, Ritika Kundra, Shaleigh Smith, Jesse Galle, Carol Aghajanian, Nadeem Abu-Rustum, et~al.
\newblock Therapeutic {Implications} of {Germline} {Testing} in {Patients} {With} {Advanced} {Cancers}.
\newblock {\em Journal of Clinical Oncology}, 2021.

\bibitem{tan_targeted_2022}
Aaron~C. Tan and Daniel~S.W. Tan.
\newblock Targeted {Therapies} for {Lung} {Cancer} {Patients} {With} {Oncogenic} {Driver} {Molecular} {Alterations}.
\newblock {\em Journal of Clinical Oncology}, 2022.

\bibitem{degasperi_substitution_2022}
Andrea Degasperi, Xueqing Zou, Tauanne Dias~Amarante, Andrea Martinez-Martinez, Gene Ching~Chiek Koh, João M.~L. Dias, Laura Heskin, Lucia Chmelova, Giuseppe Rinaldi, Valerie Ya~Wen Wang, Arjun~S. Nanda, Aaron Bernstein, Sophie~E. Momen, Jamie Young, Daniel Perez-Gil, Yasin Memari, Cherif Badja, Scott Shooter, Jan Czarnecki, Matthew~A. Brown, et~al.
\newblock Substitution mutational signatures in whole-genome–sequenced cancers in the {UK} population.
\newblock {\em Science}, 2022.

\bibitem{xu_single-cell_2022}
Junfen Xu, Yifeng Fang, Kelie Chen, Sen Li, Sangsang Tang, Yan Ren, Yixuan Cen, Weidong Fei, Bo~Zhang, Yuanming Shen, and Weiguo Lu.
\newblock Single-{Cell} {RNA} {Sequencing} {Reveals} the {Tissue} {Architecture} in {Human} {High}-{Grade} {Serous} {Ovarian} {Cancer}.
\newblock {\em Clinical Cancer Research}, 2022.

\bibitem{horak_comprehensive_2021}
Peter Horak, Christoph Heining, Simon Kreutzfeldt, Barbara Hutter, Andreas Mock, Jennifer Hüllein, Martina Fröhlich, Sebastian Uhrig, Arne Jahn, Andreas Rump, Laura Gieldon, Lino Möhrmann, Dorothea Hanf, Veronica Teleanu, Christoph~E. Heilig, Daniel~B. Lipka, Michael Allgäuer, Leo Ruhnke, Andreas Laßmann, Volker Endris, et~al.
\newblock Comprehensive {Genomic} and {Transcriptomic} {Analysis} for {Guiding} {Therapeutic} {Decisions} in {Patients} with {Rare} {Cancers}.
\newblock {\em Cancer Discovery}, 2021.

\bibitem{zhang_single-cell_2016}
Xiaoyan Zhang, Sadie~L. Marjani, Zhaoyang Hu, Sherman~M. Weissman, Xinghua Pan, and Shixiu Wu.
\newblock Single-{Cell} {Sequencing} for {Precise} {Cancer} {Research}: {Progress} and {Prospects}.
\newblock {\em Cancer Research}, 2016.

\bibitem{bruno_next_2020}
Rossella Bruno and Gabriella Fontanini.
\newblock Next {Generation} {Sequencing} for {Gene} {Fusion} {Analysis} in {Lung} {Cancer}: {A} {Literature} {Review}.
\newblock {\em Diagnostics}, 2020.

\bibitem{de_luca_fgfr_2020}
Antonella De~Luca, Riziero Esposito~Abate, Anna~M. Rachiglio, Monica~R. Maiello, Claudia Esposito, Clorinda Schettino, Francesco Izzo, Guglielmo Nasti, and Nicola Normanno.
\newblock {FGFR} {Fusions} in {Cancer}: {From} {Diagnostic} {Approaches} to {Therapeutic} {Intervention}.
\newblock {\em International Journal of Molecular Sciences}, 2020.

\bibitem{waarts_targeting_2022}
Michael~R. Waarts, Aaron~J. Stonestrom, Young~C. Park, and Ross~L. Levine.
\newblock Targeting mutations in cancer.
\newblock {\em The Journal of Clinical Investigation}, 2022.

\bibitem{lim_advancing_2020}
Bora Lim, Yiyun Lin, and Nicholas Navin.
\newblock Advancing {Cancer} {Research} and {Medicine} with {Single}-{Cell} {Genomics}.
\newblock {\em Cancer Cell}, 2020.

\bibitem{colomer_when_2020}
Ramon Colomer, Rebeca Mondejar, Nuria Romero-Laorden, Arantzazu Alfranca, Francisco Sanchez-Madrid, and Miguel Quintela-Fandino.
\newblock When should we order a next generation sequencing test in a patient with cancer?
\newblock {\em eClinicalMedicine}, 2020.

\bibitem{saadatpour_single-cell_2015}
Assieh Saadatpour, Shujing Lai, Guoji Guo, and Guo-Cheng Yuan.
\newblock Single-{Cell} {Analysis} in {Cancer} {Genomics}.
\newblock {\em Trends in Genetics}, 2015.

\bibitem{dizman_sequencing_2020}
Nazli Dizman, Zeynep~E. Arslan, Matthew Feng, and Sumanta~K. Pal.
\newblock Sequencing {Therapies} for {Metastatic} {Renal} {Cell} {Carcinoma}.
\newblock {\em Urologic Clinics}, 2020.

\bibitem{buzdin_rna_2020}
Anton Buzdin, Maxim Sorokin, Andrew Garazha, Alexander Glusker, Alex Aleshin, Elena Poddubskaya, Marina Sekacheva, Ella Kim, Nurshat Gaifullin, Alf Giese, Alexander Seryakov, Pavel Rumiantsev, Sergey Moshkovskii, and Alexey Moiseev.
\newblock {RNA sequencing for research and diagnostics in clinical oncology}.
\newblock {\em Seminars in Cancer Biology}, 2020.

\bibitem{xiao_tumor_2021}
Yi~Xiao and Dihua Yu.
\newblock Tumor microenvironment as a therapeutic target in cancer.
\newblock {\em Pharmacology \& Therapeutics}, 2021.

\bibitem{nandwani_lncrnas_2021}
Arun Nandwani, Shalu Rathore, and Malabika Datta.
\newblock {LncRNAs} in cancer: {Regulatory} and therapeutic implications.
\newblock {\em Cancer Letters}, 2021.

\bibitem{marchetti_error-corrected_2023}
Francesco Marchetti, Renato Cardoso, Connie~L. Chen, George~R. Douglas, Joanne Elloway, Patricia~A. Escobar, Tod Harper, Robert~H. Heflich, Darren Kidd, Anthony~M. Lynch, Meagan~B. Myers, Barbara~L. Parsons, Jesse~J. Salk, Raja~S. Settivari, Stephanie~L. Smith-Roe, Kristine~L. Witt, Carole~L. Yauk, Robert Young, Shaofei Zhang, and Sheroy Minocherhomji.
\newblock Error-corrected next generation sequencing – {Promises} and challenges for genotoxicity and cancer risk assessment.
\newblock {\em Mutation Research/Reviews in Mutation Research}, 2023.

\bibitem{chen_next-generation_2021}
Xiao Chen, Han Zhu, Chun Qiao, Sishu Zhao, Lu~Liu, Yan Wang, Huimin Jin, Sixuan Qian, and Yujie Wu.
\newblock Next-generation sequencing reveals gene mutations landscape and clonal evolution in patients with acute myeloid leukemia.
\newblock {\em Hematology}, 2021.

\bibitem{navin_first_2015}
Nicholas~E. Navin.
\newblock The first five years of single-cell cancer genomics and beyond.
\newblock {\em Genome Research}, 2015.

\bibitem{e002244}
{NIHR Global Health Research Unit on Genomic Surveillance of AMR}.
\newblock Whole-genome sequencing as part of national and international surveillance programmes for antimicrobial resistance: a roadmap.
\newblock {\em BMJ Global Health}, 2020.

\bibitem{TONG2021130}
Shanwei Tong, Luyao Ma, Jennifer Ronholm, William Hsiao, and Xiaonan Lu.
\newblock {Whole genome sequencing of Campylobacter in agri-food surveillance}.
\newblock {\em Current Opinion in Food Science}, 2021.

\bibitem{the_arabidopsis_genome_initiative_analysis_2000}
{The Arabidopsis Genome Initiative}.
\newblock Analysis of the genome sequence of the flowering plant {Arabidopsis} thaliana.
\newblock {\em Nature}, 2000.

\bibitem{zhu_applications_2020}
Haocheng Zhu, Chao Li, and Caixia Gao.
\newblock Applications of {CRISPR}–{Cas} in agriculture and plant biotechnology.
\newblock {\em Nature Reviews Molecular Cell Biology}, 2020.

\bibitem{choi_nanopore_2020}
Jae~Young Choi, Zoe~N. Lye, Simon~C. Groen, Xiaoguang Dai, Priyesh Rughani, Sophie Zaaijer, Eoghan~D. Harrington, Sissel Juul, and Michael~D. Purugganan.
\newblock Nanopore sequencing-based genome assembly and evolutionary genomics of circum-basmati rice.
\newblock {\em Genome Biology}, 2020.

\bibitem{stevens_sequence_2016}
Kristian~A Stevens, Jill~L Wegrzyn, Aleksey Zimin, Daniela Puiu, Marc Crepeau, Charis Cardeno, Robin Paul, Daniel Gonzalez-Ibeas, Maxim Koriabine, Ann~E Holtz-Morris, Pedro~J Martínez-García, Uzay~U Sezen, Guillaume Marçais, Kathy Jermstad, Patrick~E McGuire, Carol~A Loopstra, John~M Davis, Andrew Eckert, Pieter de~Jong, James~A Yorke, et~al.
\newblock Sequence of the {Sugar} {Pine} {Megagenome}.
\newblock {\em Genetics}, 2016.

\bibitem{campos_high_2021}
Maria~Doroteia Campos, Maria do~Rosário Félix, Mariana Patanita, Patrick Materatski, and Carla Varanda.
\newblock High throughput sequencing unravels tomato-pathogen interactions towards a sustainable plant breeding.
\newblock {\em Horticulture Research}, 2021.

\bibitem{gao_genome_2021}
Caixia Gao.
\newblock Genome engineering for crop improvement and future agriculture.
\newblock {\em Cell}, 2021.

\bibitem{van_dijk_machine_2021}
Aalt Dirk~Jan van Dijk, Gert Kootstra, Willem Kruijer, and Dick de~Ridder.
\newblock Machine learning in plant science and plant breeding.
\newblock {\em iScience}, 2021.

\bibitem{sun_twenty_2022}
Yanqing Sun, Lianguang Shang, Qian-Hao Zhu, Longjiang Fan, and Longbiao Guo.
\newblock Twenty years of plant genome sequencing: achievements and challenges.
\newblock {\em Trends in Plant Science}, 2022.

\bibitem{kim_application_2020}
Kyung~D. Kim, Yuna Kang, and Changsoo Kim.
\newblock Application of {Genomic} {Big} {Data} in {Plant} {Breeding}: {Past}, {Present}, and {Future}.
\newblock {\em Plants}, 2020.

\bibitem{thudi_genomic_2021}
Mahendar Thudi, Ramesh Palakurthi, James~C. Schnable, Annapurna Chitikineni, Susanne Dreisigacker, Emma Mace, Rakesh~K. Srivastava, C.~Tara Satyavathi, Damaris Odeny, Vijay~K. Tiwari, Hon-Ming Lam, Yan~Bin Hong, Vikas~K. Singh, Guowei Li, Yunbi Xu, Xiaoping Chen, Sanjay Kaila, Henry Nguyen, Sobhana Sivasankar, Scott~A. Jackson, et~al.
\newblock Genomic resources in plant breeding for sustainable agriculture.
\newblock {\em Journal of Plant Physiology}, 2021.

\bibitem{michael_building_2020}
Todd~P Michael and Robert VanBuren.
\newblock Building near-complete plant genomes.
\newblock {\em Current Opinion in Plant Biology}, 2020.

\bibitem{shen_omics-based_2022}
Yanting Shen, Guoan Zhou, Chengzhi Liang, and Zhixi Tian.
\newblock Omics-based interdisciplinarity is accelerating plant breeding.
\newblock {\em Current Opinion in Plant Biology}, 2022.

\bibitem{shahroodi2022demeter}
Taha Shahroodi, Mahdi Zahedi, Can Firtina, Mohammed Alser, Stephan Wong, Onur Mutlu, and Said Hamdioui.
\newblock {Demeter: A fast and energy-efficient food profiler using hyperdimensional computing in memory}.
\newblock {\em IEEE Access}, 2022.

\bibitem{prasad2021soil}
Shiv Prasad, Lal~Chand Malav, Jairam Choudhary, Sudha Kannojiya, Monika Kundu, Sandeep Kumar, and Ajar~Nath Yadav.
\newblock {Soil Microbiomes for Healthy Nutrient Recycling}.
\newblock {\em {Current Trends in Microbial Biotechnology for Sustainable Agriculture}}, 2021.

\bibitem{Mascher2024}
Martin Mascher, Murukarthick Jayakodi, Hyeonah Shim, and Nils Stein.
\newblock Promises and challenges of crop translational genomics.
\newblock {\em Nature}, 2024.

\bibitem{Schreiber2024}
Mona Schreiber, Murukarthick Jayakodi, Nils Stein, and Martin Mascher.
\newblock Plant pangenomes for crop improvement, biodiversity and evolution.
\newblock {\em Nature Reviews Genetics}, 2024.

\bibitem{urbanek2018degradation}
Aneta~K. Urbanek, Waldemar Rymowicz, and Aleksandra~M. Miro{\'{n}}czuk.
\newblock Degradation of plastics and plastic-degrading bacteria in cold marine habitats.
\newblock {\em {Applied Microbiology and Biotechnology}}, 2018.

\bibitem{edgar2022petabase}
Robert~C. Edgar, Jeff Taylor, Victor Lin, Tomer Altman, Pierre Barbera, Dmitry Meleshko, Dan Lohr, Gherman Novakovsky, Benjamin Buchfink, Basem Al-Shayeb, Jillian~F. Banfield, Marcos de~la Pe{\~{n}}a, Anton Korobeynikov, Rayan Chikhi, and Artem Babaian.
\newblock Petabase-scale sequence alignment catalyses viral discovery.
\newblock {\em Nature}, 2022.

\bibitem{paoli2022biosynthetic}
Lucas Paoli, Hans-Joachim Ruscheweyh, Clarissa~C. Forneris, Florian Hubrich, Satria Kautsar, Agneya Bhushan, Alessandro Lotti, Quentin Clayssen, Guillem Salazar, Alessio Milanese, Charlotte~I. Carlstr{\"o}m, Chrysa Papadopoulou, Daniel Gehrig, Mikhail Karasikov, Harun Mustafa, Martin Larralde, Laura~M. Carroll, Pablo S{\'a}nchez, Ahmed~A. Zayed, Dylan~R. Cronin, et~al.
\newblock Biosynthetic potential of the global ocean microbiome.
\newblock {\em Nature}, 2022.

\bibitem{Hogg2024}
Carolyn~J. Hogg.
\newblock Translating genomic advances into biodiversity conservation.
\newblock {\em {Nature Reviews Genetics}}, 2024.

\bibitem{lewin2018earth}
Harris~A. Lewin, Gene~E. Robinson, W.~John Kress, William~J. Baker, Jonathan Coddington, Keith~A. Crandall, Richard Durbin, Scott~V. Edwards, Félix Forest, M.~Thomas~P. Gilbert, Melissa~M. Goldstein, Igor~V. Grigoriev, Kevin~J. Hackett, David Haussler, Erich~D. Jarvis, Warren~E. Johnson, Aristides Patrinos, Stephen Richards, Juan~Carlos Castilla-Rubio, Marie-Anne van Sluys, et~al.
\newblock {Earth BioGenome Project: Sequencing life for the future of life}.
\newblock {\em {Proceedings of the National Academy of Sciences}}, 2018.

\bibitem{kanehisa_toward_2019}
Minoru Kanehisa.
\newblock Toward understanding the origin and evolution of cellular organisms.
\newblock {\em Protein Science}, 2019.

\bibitem{qing_whole_2022}
Jun Qing, Yi-De Meng, Feng He, Qing-Xin Du, Jian Zhong, Hong-Yan Du, Pan-Feng Liu, Lan-Ying Du, and Lu~Wang.
\newblock Whole genome re-sequencing reveals the genetic diversity and evolutionary patterns of {Eucommia} ulmoides.
\newblock {\em Molecular Genetics and Genomics}, 2022.

\bibitem{wittkopp_cis-regulatory_2012}
Patricia~J. Wittkopp and Gizem Kalay.
\newblock Cis-regulatory elements: molecular mechanisms and evolutionary processes underlying divergence.
\newblock {\em Nature Reviews Genetics}, 2012.

\bibitem{romero_comparative_2012}
Irene~Gallego Romero, Ilya Ruvinsky, and Yoav Gilad.
\newblock Comparative studies of gene expression and the evolution of gene regulation.
\newblock {\em Nature Reviews Genetics}, 2012.

\bibitem{wang_population_2020}
Xinchao Wang, Hu~Feng, Yuxiao Chang, Chunlei Ma, Liyuan Wang, Xinyuan Hao, A’lun Li, Hao Cheng, Lu~Wang, Peng Cui, Jiqiang Jin, Xiaobo Wang, Kang Wei, Cheng Ai, Sheng Zhao, Zhichao Wu, Youyong Li, Benying Liu, Guo-Dong Wang, Liang Chen, et~al.
\newblock Population sequencing enhances understanding of tea plant evolution.
\newblock {\em Nature Communications}, 2020.

\bibitem{hill_molecular_2021}
Mark~S. Hill, Pétra Vande~Zande, and Patricia~J. Wittkopp.
\newblock Molecular and evolutionary processes generating variation in gene expression.
\newblock {\em Nature Reviews Genetics}, 2021.

\bibitem{vaishnav_evolution_2022}
Eeshit~Dhaval Vaishnav, Carl~G. de~Boer, Jennifer Molinet, Moran Yassour, Lin Fan, Xian Adiconis, Dawn~A. Thompson, Joshua~Z. Levin, Francisco~A. Cubillos, and Aviv Regev.
\newblock The evolution, evolvability and engineering of gene regulatory {DNA}.
\newblock {\em Nature}, 2022.

\bibitem{zhang_haplotype-resolved_2021}
Xingtan Zhang, Shuai Chen, Longqing Shi, Daping Gong, Shengcheng Zhang, Qian Zhao, Dongliang Zhan, Liette Vasseur, Yibin Wang, Jiaxin Yu, Zhenyang Liao, Xindan Xu, Rui Qi, Wenling Wang, Yunran Ma, Pengjie Wang, Naixing Ye, Dongna Ma, Yan Shi, Haifeng Wang, et~al.
\newblock Haplotype-resolved genome assembly provides insights into evolutionary history of the tea plant {Camellia} sinensis.
\newblock {\em Nature Genetics}, 2021.

\bibitem{fay_evaluating_2008}
J~C Fay and P~J Wittkopp.
\newblock Evaluating the role of natural selection in the evolution of gene regulation.
\newblock {\em Heredity}, 2008.

\bibitem{kanzi_next_2020}
Aquillah~M. Kanzi, James~Emmanuel San, Benjamin Chimukangara, Eduan Wilkinson, Maryam Fish, Veron Ramsuran, and Tulio de~Oliveira.
\newblock Next {Generation} {Sequencing} and {Bioinformatics} {Analysis} of {Family} {Genetic} {Inheritance}.
\newblock {\em Frontiers in Genetics}, 2020.

\bibitem{wray_evolution_2003}
Gregory~A. Wray, Matthew~W. Hahn, Ehab Abouheif, James~P. Balhoff, Margaret Pizer, Matthew~V. Rockman, and Laura~A. Romano.
\newblock The {Evolution} of {Transcriptional} {Regulation} in {Eukaryotes}.
\newblock {\em Molecular Biology and Evolution}, 2003.

\bibitem{wu_one_2021}
Aiping Wu, Lulan Wang, Hang-Yu Zhou, Cheng-Yang Ji, Shang~Zhou Xia, Yang Cao, Jing Meng, Xiao Ding, Sarah Gold, Taijiao Jiang, and Genhong Cheng.
\newblock One year of {SARS}-{CoV}-2 evolution.
\newblock {\em Cell Host \& Microbe}, 2021.

\bibitem{signor_evolution_2018}
Sarah~A. Signor and Sergey~V. Nuzhdin.
\newblock The {Evolution} of {Gene} {Expression} in cis and trans.
\newblock {\em Trends in Genetics}, 2018.

\bibitem{whitehead_variation_2006}
Andrew Whitehead and Douglas~L. Crawford.
\newblock Variation within and among species in gene expression: raw material for evolution.
\newblock {\em Molecular Ecology}, 2006.

\bibitem{coolon_tempo_2014}
Joseph~D. Coolon, C.~Joel McManus, Kraig~R. Stevenson, Brenton~R. Graveley, and Patricia~J. Wittkopp.
\newblock Tempo and mode of regulatory evolution in {Drosophila}.
\newblock {\em Genome Research}, 2014.

\bibitem{ellegren2014genome}
Hans Ellegren.
\newblock {Genome Sequencing and Population Genomics in Non-Model Organisms}.
\newblock {\em Trends in Ecology \& Evolution}, 2014.

\bibitem{Prado-Martinez2013}
Javier Prado-Martinez, Peter~H. Sudmant, Jeffrey~M. Kidd, Heng Li, Joanna~L. Kelley, Belen Lorente-Galdos, Krishna~R. Veeramah, August~E. Woerner, Timothy~D. O'Connor, Gabriel Santpere, Alexander Cagan, Christoph Theunert, Ferran Casals, Hafid Laayouni, Kasper Munch, Asger Hobolth, Anders~E. Halager, Maika Malig, Jessica Hernandez-Rodriguez, Irene Hernando-Herraez, et~al.
\newblock {Great Ape Genetic Diversity and Population History}.
\newblock {\em Nature}, 2013.

\bibitem{Prohaska2019human}
Ana Prohaska, Fernando Racimo, Andrew~J Schork, Martin Sikora, Aaron~J Stern, Melissa Ilardo, Morten~Erik Allentoft, Lasse Folkersen, Alfonso Buil, J~Víctor Moreno-Mayar, Thorfinn Korneliussen, Daniel Geschwind, Andrés Ingason, Thomas Werge, Rasmus Nielsen, and Eske Willerslev.
\newblock {Human Disease Variation in the Light of Population Genomics}.
\newblock {\em Cell}, 2019.

\bibitem{danko2021global}
David Danko, Daniela Bezdan, Evan~E. Afshin, Sofia Ahsanuddin, Chandrima Bhattacharya, Daniel~J. Butler, Kern~Rei Chng, Daisy Donnellan, Jochen Hecht, Katelyn Jackson, Katerina Kuchin, Mikhail Karasikov, Abigail Lyons, Lauren Mak, Dmitry Meleshko, Harun Mustafa, Beth Mutai, Russell~Y. Neches, Amanda Ng, Olga Nikolayeva, et~al.
\newblock {A Global Metagenomic Map of Urban Microbiomes and Antimicrobial Resistance}.
\newblock {\em Cell}, 2021.

\bibitem{didelot2012transforming}
Xavier Didelot, Rory Bowden, Daniel~J Wilson, Tim~EA Peto, and Derrick~W Crook.
\newblock Transforming clinical microbiology with bacterial genome sequencing.
\newblock {\em Nature Reviews Genetics}, 2012.

\bibitem{marini2022towards}
Simone Marini, Rodrigo~A Mora, Christina Boucher, Noelle Robertson~Noyes, and Mattia Prosperi.
\newblock Towards routine employment of computational tools for antimicrobial resistance determination via high-throughput sequencing.
\newblock {\em Briefings in Bioinformatics}, 2022.

\bibitem{Houldcroft2017}
Charlotte~J. Houldcroft, Mathew~A. Beale, and Judith Breuer.
\newblock Clinical and biological insights from viral genome sequencing.
\newblock {\em Nature Reviews Microbiology}, 2017.

\bibitem{Jansz2024viral}
Natasha Jansz and Geoffrey~J. Faulkner.
\newblock Viral genome sequencing methods: benefits and pitfalls of current approaches.
\newblock {\em Biochemical Society Transactions}, 2024.

\bibitem{bentley_accurate_2008}
David~R. Bentley, Shankar Balasubramanian, Harold~P. Swerdlow, Geoffrey~P. Smith, John Milton, Clive~G. Brown, Kevin~P. Hall, Dirk~J. Evers, Colin~L. Barnes, Helen~R. Bignell, Jonathan~M. Boutell, Jason Bryant, Richard~J. Carter, R.~Keira~Cheetham, Anthony~J. Cox, Darren~J. Ellis, Michael~R. Flatbush, Niall~A. Gormley, Sean~J. Humphray, Leslie~J. Irving, et~al.
\newblock Accurate whole human genome sequencing using reversible terminator chemistry.
\newblock {\em Nature}, 2008.

\bibitem{margulies_genome_2005}
Marcel Margulies, Michael Egholm, William~E. Altman, Said Attiya, Joel~S. Bader, Lisa~A. Bemben, Jan Berka, Michael~S. Braverman, Yi-Ju Chen, Zhoutao Chen, Scott~B. Dewell, Lei Du, Joseph~M. Fierro, Xavier~V. Gomes, Brian~C. Godwin, Wen He, Scott Helgesen, Chun~He Ho, Gerard~P. Irzyk, Szilveszter~C. Jando, et~al.
\newblock Genome sequencing in microfabricated high-density picolitre reactors.
\newblock {\em Nature}, 2005.

\bibitem{shendure_accurate_2005}
Jay Shendure, Gregory~J. Porreca, Nikos~B. Reppas, Xiaoxia Lin, John~P. McCutcheon, Abraham~M. Rosenbaum, Michael~D. Wang, Kun Zhang, Robi~D. Mitra, and George~M. Church.
\newblock Accurate {Multiplex} {Polony} {Sequencing} of an {Evolved} {Bacterial} {Genome}.
\newblock {\em Science}, 2005.

\bibitem{harris_single-molecule_2008}
Timothy~D. Harris, Phillip~R. Buzby, Hazen Babcock, Eric Beer, Jayson Bowers, Ido Braslavsky, Marie Causey, Jennifer Colonell, James DiMeo, J.~William Efcavitch, Eldar Giladi, Jaime Gill, John Healy, Mirna Jarosz, Dan Lapen, Keith Moulton, Stephen~R. Quake, Kathleen Steinmann, Edward Thayer, Anastasia Tyurina, et~al.
\newblock Single-{Molecule} {DNA} {Sequencing} of a {Viral} {Genome}.
\newblock {\em Science}, 2008.

\bibitem{turcatti_new_2008}
Gerardo Turcatti, Anthony Romieu, Milan Fedurco, and Ana-Paula Tairi.
\newblock A new class of cleavable fluorescent nucleotides: synthesis and optimization as reversible terminators for {DNA} sequencing by synthesis †.
\newblock {\em Nucleic Acids Research}, 2008.

\bibitem{wu_termination_2007}
Weidong Wu, Brian~P. Stupi, Vladislav~A. Litosh, Dena Mansouri, Demetra Farley, Sidney Morris, Sherry Metzker, and Michael~L. Metzker.
\newblock Termination of {DNA} synthesis by {N6} -alkylated, not 3'- {O} -alkylated, photocleavable 2'-deoxyadenosine triphosphates.
\newblock {\em Nucleic Acids Research}, 2007.

\bibitem{fuller_rapid_2007}
Carl~W Fuller.
\newblock Rapid parallel nucleic acid analysis.
\newblock US Patent US7264934B2, 2007.

\bibitem{mckernan_reagents_2008}
Kevin McKernan, Alan Blanchard, Lev Kotler, and Gina Costa.
\newblock Reagents, methods, and libraries for bead-based sequencing.
\newblock US Patent US20090181860A1, 2008.

\bibitem{fuller_method_2011}
Carl~W Fuller and John~R Nelson.
\newblock Method for nucleic acid analysis.
\newblock US Patent US7871771B2, 2011.

\bibitem{eid_real-time_2009}
John Eid, Adrian Fehr, Jeremy Gray, Khai Luong, John Lyle, Geoff Otto, Paul Peluso, David Rank, Primo Baybayan, Brad Bettman, Arkadiusz Bibillo, Keith Bjornson, Bidhan Chaudhuri, Frederick Christians, Ronald Cicero, Sonya Clark, Ravindra Dalal, Alex deWinter, John Dixon, Mathieu Foquet, et~al.
\newblock Real-{Time} {DNA} {Sequencing} from {Single} {Polymerase} {Molecules}.
\newblock {\em Science}, 2009.

\bibitem{menestrina_ionic_1986}
Gianfranco Menestrina.
\newblock Ionic channels formed {byStaphylococcus} aureus alpha-toxin: {Voltage}-dependent inhibition by divalent and trivalent cations.
\newblock {\em The Journal of Membrane Biology}, 1986.

\bibitem{cherf_automated_2012}
Gerald~M Cherf, Kate~R Lieberman, Hytham Rashid, Christopher~E Lam, Kevin Karplus, and Mark Akeson.
\newblock Automated forward and reverse ratcheting of {DNA} in a nanopore at 5-Å precision.
\newblock {\em Nature Biotechnology}, 2012.

\bibitem{manrao_reading_2012}
Elizabeth~A Manrao, Ian~M Derrington, Andrew~H Laszlo, Kyle~W Langford, Matthew~K Hopper, Nathaniel Gillgren, Mikhail Pavlenok, Michael Niederweis, and Jens~H Gundlach.
\newblock Reading {DNA} at single-nucleotide resolution with a mutant {MspA} nanopore and phi29 {DNA} polymerase.
\newblock {\em Nature Biotechnology}, 2012.

\bibitem{laszlo_decoding_2014}
Andrew~H Laszlo, Ian~M Derrington, Brian~C Ross, Henry Brinkerhoff, Andrew Adey, Ian~C Nova, Jonathan~M Craig, Kyle~W Langford, Jenny~Mae Samson, Riza Daza, Kenji Doering, Jay Shendure, and Jens~H Gundlach.
\newblock Decoding long nanopore sequencing reads of natural {DNA}.
\newblock {\em Nature Biotechnology}, 2014.

\bibitem{deamer_three_2016}
David Deamer, Mark Akeson, and Daniel Branton.
\newblock Three decades of nanopore sequencing.
\newblock {\em Nature Biotechnology}, 2016.

\bibitem{kasianowicz_characterization_1996}
John~J. Kasianowicz, Eric Brandin, Daniel Branton, and David~W. Deamer.
\newblock Characterization of individual polynucleotide molecules using a membrane channel.
\newblock {\em Proceedings of the National Academy of Sciences}, 1996.

\bibitem{meller_rapid_2000}
Amit Meller, Lucas Nivon, Eric Brandin, Jene Golovchenko, and Daniel Branton.
\newblock Rapid nanopore discrimination between single polynucleotide molecules.
\newblock {\em Proceedings of the National Academy of Sciences}, 2000.

\bibitem{stoddart_single-nucleotide_2009}
David Stoddart, Andrew~J. Heron, Ellina Mikhailova, Giovanni Maglia, and Hagan Bayley.
\newblock Single-nucleotide discrimination in immobilized {DNA} oligonucleotides with a biological nanopore.
\newblock {\em Proceedings of the National Academy of Sciences}, 2009.

\bibitem{laszlo_detection_2013}
Andrew~H. Laszlo, Ian~M. Derrington, Henry Brinkerhoff, Kyle~W. Langford, Ian~C. Nova, Jenny~Mae Samson, Joshua~J. Bartlett, Mikhail Pavlenok, and Jens~H. Gundlach.
\newblock Detection and mapping of 5-methylcytosine and 5-hydroxymethylcytosine with nanopore {MspA}.
\newblock {\em Proceedings of the National Academy of Sciences}, 2013.

\bibitem{schreiber_error_2013}
Jacob Schreiber, Zachary~L. Wescoe, Robin Abu-Shumays, John~T. Vivian, Baldandorj Baatar, Kevin Karplus, and Mark Akeson.
\newblock Error rates for nanopore discrimination among cytosine, methylcytosine, and hydroxymethylcytosine along individual {DNA} strands.
\newblock {\em Proceedings of the National Academy of Sciences}, 2013.

\bibitem{butler_single-molecule_2008}
Tom~Z. Butler, Mikhail Pavlenok, Ian~M. Derrington, Michael Niederweis, and Jens~H. Gundlach.
\newblock Single-molecule {DNA} detection with an engineered {MspA} protein nanopore.
\newblock {\em Proceedings of the National Academy of Sciences}, 2008.

\bibitem{derrington_nanopore_2010}
Ian~M. Derrington, Tom~Z. Butler, Marcus~D. Collins, Elizabeth Manrao, Mikhail Pavlenok, Michael Niederweis, and Jens~H. Gundlach.
\newblock Nanopore {DNA} sequencing with {MspA}.
\newblock {\em Proceedings of the National Academy of Sciences}, 2010.

\bibitem{song_structure_1996}
Langzhou Song, Michael~R. Hobaugh, Christopher Shustak, Stephen Cheley, Hagan Bayley, and J.~Eric Gouaux.
\newblock Structure of {Staphylococcal} a-{Hemolysin}, a {Heptameric} {Transmembrane} {Pore}.
\newblock {\em Science}, 1996.

\bibitem{walker_pore-forming_1994}
Barbara Walker, John Kasianowicz, Musti Krishnasastry, and Hagan Bayley.
\newblock A pore-forming protein with a metal-actuated switch.
\newblock {\em Protein Engineering, Design and Selection}, 1994.

\bibitem{wescoe_nanopores_2014}
Zachary~L. Wescoe, Jacob Schreiber, and Mark Akeson.
\newblock Nanopores {Discriminate} among {Five} {C5}-{Cytosine} {Variants} in {DNA}.
\newblock {\em Journal of the American Chemical Society}, 2014.

\bibitem{lieberman_processive_2010}
Kate~R. Lieberman, Gerald~M. Cherf, Michael~J. Doody, Felix Olasagasti, Yvette Kolodji, and Mark Akeson.
\newblock Processive {Replication} of {Single} {DNA} {Molecules} in a {Nanopore} {Catalyzed} by phi29 {DNA} {Polymerase}.
\newblock {\em Journal of the American Chemical Society}, 2010.

\bibitem{bezrukov_dynamics_1996}
Sergey~M. Bezrukov, Igor Vodyanoy, Rafik~A. Brutyan, and John~J. Kasianowicz.
\newblock Dynamics and {Free} {Energy} of {Polymers} {Partitioning} into a {Nanoscale} {Pore}.
\newblock {\em Macromolecules}, 1996.

\bibitem{akeson_microsecond_1999}
Mark Akeson, Daniel Branton, John~J. Kasianowicz, Eric Brandin, and David~W. Deamer.
\newblock Microsecond {Time}-{Scale} {Discrimination} {Among} {Polycytidylic} {Acid}, {Polyadenylic} {Acid}, and {Polyuridylic} {Acid} as {Homopolymers} or as {Segments} {Within} {Single} {RNA} {Molecules}.
\newblock {\em Biophysical Journal}, 1999.

\bibitem{stoddart_nucleobase_2010}
David Stoddart, Andrew~J. Heron, Jochen Klingelhoefer, Ellina Mikhailova, Giovanni Maglia, and Hagan Bayley.
\newblock Nucleobase {Recognition} in {ssDNA} at the {Central} {Constriction} of the a-{Hemolysin} {Pore}.
\newblock {\em Nano Letters}, 2010.

\bibitem{ashkenasy_recognizing_2005}
Nurit Ashkenasy, Jorge Sánchez-Quesada, Hagan Bayley, and M.~Reza Ghadiri.
\newblock Recognizing a {Single} {Base} in an {Individual} {DNA} {Strand}: {A} {Step} {Toward} {DNA} {Sequencing} in {Nanopores}.
\newblock {\em Angewandte Chemie International Edition}, 2005.

\bibitem{stoddart_multiple_2010}
David Stoddart, Giovanni Maglia, Ellina Mikhailova, Andrew~J. Heron, and Hagan Bayley.
\newblock Multiple {Base}-{Recognition} {Sites} in a {Biological} {Nanopore}: {Two} {Heads} are {Better} than {One}.
\newblock {\em Angewandte Chemie International Edition}, 2010.

\bibitem{bezrukov_current_1993}
Sergey~M. Bezrukov and John~J. Kasianowicz.
\newblock Current noise reveals protonation kinetics and number of ionizable sites in an open protein ion channel.
\newblock {\em Physical Review Letters}, 1993.

\bibitem{zhang_single-molecule_2024}
Jia-Yuan Zhang, Yuning Zhang, Lele Wang, Fei Guo, Quanxin Yun, Tao Zeng, Xu~Yan, Lei Yu, Lei Cheng, Wei Wu, Xiao Shi, Junyi Chen, Yuhui Sun, Jingnan Yang, Rongrong Guo, Xianda Zhang, Liu’er Kong, Zong’an Wang, Junlei Yao, Yangsheng Tan, et~al.
\newblock A single-molecule nanopore sequencing platform.
\newblock {\em bioRxiv}, 2024.

\bibitem{cali2017nanopore}
Damla Senol~Cali, Jeremie~S Kim, Saugata Ghose, Can Alkan, and Onur Mutlu.
\newblock {Nanopore Sequencing Technology and Tools for Genome Assembly: Computational Analysis of the Current State, Bottlenecks and Future Directions}.
\newblock {\em Briefings in Bioinformatics}, 2018.

\bibitem{stoler2021sequencing}
Nicholas Stoler and Anton Nekrutenko.
\newblock {Sequencing error profiles of Illumina sequencing instruments}.
\newblock {\em NAR Genomics and Bioinformatics}, 2021.

\bibitem{goodwin2016coming}
Sara Goodwin, John~D McPherson, and W~Richard McCombie.
\newblock {Coming of Age: Ten Years of Next-generation Sequencing Technologies}.
\newblock {\em Nature Reviews Genetics}, 2016.

\bibitem{davis2021sequencerr}
Eric~M. Davis, Yu~Sun, Yanling Liu, Pandurang Kolekar, Ying Shao, Karol Szlachta, Heather~L. Mulder, Dongren Ren, Stephen~V. Rice, Zhaoming Wang, Joy Nakitandwe, Alexander~M. Gout, Bridget Shaner, Salina Hall, Leslie~L. Robison, Stanley Pounds, Jeffery~M. Klco, John Easton, and Xiaotu Ma.
\newblock {SequencErr: measuring and suppressing sequencer errors in next-generation sequencing data}.
\newblock {\em Genome Biology}, 2021.

\bibitem{sereika2022oxford}
Mantas Sereika, Rasmus~Hansen Kirkegaard, S{\o}ren~Michael Karst, Thomas~Yssing Michaelsen, Emil~Aarre S{\o}rensen, Rasmus~Dam Wollenberg, and Mads Albertsen.
\newblock {Oxford Nanopore R10.4 long-read sequencing enables the generation of near-finished bacterial genomes from pure cultures and metagenomes without short-read or reference polishing}.
\newblock {\em Nature Methods}, 2022.

\bibitem{jain2018nanopore}
Miten Jain, Sergey Koren, Karen~H. Miga, Josh Quick, Arthur~C. Rand, Thomas~A. Sasani, John~R. Tyson, Andrew~D. Beggs, Alexander~T. Dilthey, Ian~T. Fiddes, Sunir Malla, Hannah Marriott, Tom Nieto, Justin O'Grady, Hugh~E. Olsen, Brent~S. Pedersen, Arang Rhie, Hollian Richardson, Aaron~R. Quinlan, Terrance~P. Snutch, et~al.
\newblock {Nanopore Sequencing and Assembly of A Human Genome with Ultra-long Reads}.
\newblock {\em Nature Biotechnology}, 2018.

\bibitem{payne2018bulkvis}
Alexander Payne, Nadine Holmes, Vardhman Rakyan, and Matthew Loose.
\newblock {BulkVis: a graphical viewer for Oxford nanopore bulk FAST5 files}.
\newblock {\em Bioinformatics}, 2018.

\bibitem{amarasinghe2020opportunities}
Shanika~L Amarasinghe, Shian Su, Xueyi Dong, Luke Zappia, Matthew~E Ritchie, and Quentin Gouil.
\newblock {Opportunities and Challenges in Long-read Sequencing Data Analysis}.
\newblock {\em Genome Biology}, 2020.

\bibitem{hon2020highly}
Ting Hon, Kristin Mars, Greg Young, Yu-Chih Tsai, Joseph~W. Karalius, Jane~M. Landolin, Nicholas Maurer, David Kudrna, Michael~A. Hardigan, Cynthia~C. Steiner, Steven~J. Knapp, Doreen Ware, Beth Shapiro, Paul Peluso, and David~R. Rank.
\newblock {Highly Accurate Long-read HiFi Sequencing Data for Five Complex Genomes}.
\newblock {\em Scientific Data}, 2020.

\bibitem{ni2023benchmarking}
Ying Ni, Xudong Liu, Zemenu~Mengistie Simeneh, Mengsu Yang, and Runsheng Li.
\newblock {Benchmarking of Nanopore R10.4 and R9.4.1 flow cells in single-cell whole-genome amplification and whole-genome shotgun sequencing}.
\newblock {\em Computational and Structural Biotechnology Journal}, 2023.

\bibitem{wenger2019accurate}
Aaron~M. Wenger, Paul Peluso, William~J. Rowell, Pi-Chuan Chang, Richard~J. Hall, Gregory~T. Concepcion, Jana Ebler, Arkarachai Fungtammasan, Alexey Kolesnikov, Nathan~D. Olson, Armin T{\"o}pfer, Michael Alonge, Medhat Mahmoud, Yufeng Qian, Chen-Shan Chin, Adam~M. Phillippy, Michael~C. Schatz, Gene Myers, Mark~A. DePristo, Jue Ruan, et~al.
\newblock {Accurate Circular Consensus Long-read Sequencing Improves Variant Detection and Assembly of A Human Genome}.
\newblock {\em Nature Biotechnology}, 2019.

\bibitem{hhrlich2011metahit}
S.~Dusko Ehrlich and {The MetaHIT Consortium}.
\newblock {MetaHIT: The European Union Project on Metagenomics of the Human Intestinal Tract}.
\newblock In {\em Metagenomics of the Human Body}. 2011.

\bibitem{sunagawa2015structure}
Shinichi Sunagawa, Luis~Pedro Coelho, Samuel Chaffron, Jens~Roat Kultima, Karine Labadie, Guillem Salazar, Bardya Djahanschiri, Georg Zeller, Daniel~R. Mende, Adriana Alberti, Francisco~M. Cornejo-Castillo, Paul~I. Costea, Corinne Cruaud, Francesco d'Ovidio, Stefan Engelen, Isabel Ferrera, Josep~M. Gasol, Lionel Guidi, Falk Hildebrand, Florian Kokoszka, et~al.
\newblock {Structure and Function of the Global Ocean Microbiome}.
\newblock {\em Science}, 2015.

\bibitem{fierer2017embracing}
Noah Fierer.
\newblock Embracing the unknown: disentangling the complexities of the soil microbiome.
\newblock {\em Nature Reviews Microbiology}, 2017.

\bibitem{biodigs2026}
{The BioDIGS Consortium}, Tristen Alberts, Claude~F. Albritton, Rosa Alcazar, Zainab Aljabri, Maria Alvarez, Anish Aradhey, Mentewab Ayalew, Nareh Azizian, Yasmeen Balayah, Destiny~D. Ball, Efren Barragan, Corey Beshoar, Lyle Best, Emily Biggane, Joseph Biggane, Jesse Blick, Myron Blosser, Alex~Kenneth Brown, Michael~C. Campbell, et~al.
\newblock Unearthing soil biodiversity through collaborative genomic research and education.
\newblock {\em Nature Genetics}, 2026.

\bibitem{Ryon2022}
Krista~A. Ryon, Braden~T. Tierney, Alina Frolova, Andre Kahles, Christelle Desnues, Christos Ouzounis, Cynthia Gibas, Daniela Bezdan, Youping Deng, Ding He, Emmanuel Dias-Neto, Eran Elhaik, Evan Afshin, George Grills, Gregorio Iraola, Haruo Suzuki, Johannes Werner, Klas Udekwu, Lynn Schriml, Malay Bhattacharyya, et~al.
\newblock {A history of the MetaSUB consortium: Tracking urban microbes around the globe}.
\newblock {\em iScience}, 2022.

\bibitem{Zhu2025}
Congmin Zhu, Linwei Wu, Daliang Ning, Renmao Tian, Shuhong Gao, Bing Zhang, Jianshu Zhao, Ya~Zhang, Naijia Xiao, Yajiao Wang, Mathew~R. Brown, Qichao Tu, Dany Acevedo, Miriam Agullo-Barcelo, Juliana~Calabria de~Araujo, {\'E}rika~Ferreira de~Abreu Mac~Conell, Kevin Boehnke, Philip Bond, Charles~B. Bott, Patricia Bovio-Winkler, et~al.
\newblock Global diversity and distribution of antibiotic resistance genes in human wastewater treatment systems.
\newblock {\em Nature Communications}, 2025.

\bibitem{human2012structure}
Curtis Huttenhower, Dirk Gevers, Rob Knight, Sahar Abubucker, Jonathan~H. Badger, Asif~T. Chinwalla, Heather~H. Creasy, Ashlee~M. Earl, Michael~G. FitzGerald, Robert~S. Fulton, Michelle~G. Giglio, Kymberlie Hallsworth-Pepin, Elizabeth~A. Lobos, Ramana Madupu, Vincent Magrini, John~C. Martin, Makedonka Mitreva, Donna~M. Muzny, Erica~J. Sodergren, James Versalovic, et~al.
\newblock {Structure, Function and Diversity of the Healthy Human Microbiome}.
\newblock {\em Nature}, 2012.

\bibitem{kintz2017introducing}
Thomas~M. Kuntz and Jack~A. Gilbert.
\newblock {Introducing the Microbiome into Precision Medicine}.
\newblock {\em Trends in Pharmacological Sciences}, 2017.

\bibitem{dixon2020metagenomics}
Matthew Dixon, Maria Stefil, Michael McDonald, Truls~Erik Bjerklund-Johansen, Kurt Naber, Florian Wagenlehner, and Vladimir Mouraviev.
\newblock {Metagenomics in diagnosis and improved targeted treatment of UTI}.
\newblock {\em World Journal of Urology}, 2020.

\bibitem{mousa2024gut}
Walaa~K Mousa and Aya Al~Ali.
\newblock The gut microbiome advances precision medicine and diagnostics for inflammatory bowel diseases.
\newblock {\em International Journal of Molecular Sciences}, 2024.

\bibitem{tegegne2025g}
Henok~Ayalew Tegegne and Tor~C Savidge.
\newblock Gut microbiome metagenomics in clinical practice: bridging the gap between research and precision medicine.
\newblock {\em Gut Microbes}, 2025.

\bibitem{virgin2011metagenomics}
Herbert~W Virgin and John~A Todd.
\newblock Metagenomics and personalized medicine.
\newblock {\em Cell}, 2011.

\bibitem{zhao2024application}
Yu~Zhao, Wenhui Zhang, and Xin Zhang.
\newblock Application of metagenomic next-generation sequencing in the diagnosis of infectious diseases.
\newblock {\em Frontiers in Cellular and Infection Microbiology}, 2024.

\bibitem{taxt2020rapid}
Arne~M. Taxt, Ekaterina Avershina, Stephan~A. Frye, Umaer Naseer, and Rafi Ahmad.
\newblock Rapid identification of pathogens, antibiotic resistance genes and plasmids in blood cultures by nanopore sequencing.
\newblock {\em Scientific Reports}, 2020.

\bibitem{GRUMAZ2020405}
Christian Grumaz, Anne Hoffmann, Yevhen Vainshtein, Maria Kopp, Silke Grumaz, Philip Stevens, Sebastian~O. Decker, Markus~A. Weigand, Stefan Hofer, Thorsten Brenner, and Kai Sohn.
\newblock Rapid next-generation sequencing–based diagnostics of bacteremia in septic patients.
\newblock {\em The Journal of Molecular Diagnostics}, 2020.

\bibitem{gu2021rapid}
Wei Gu, Xianding Deng, Marco Lee, Yasemin~D. Sucu, Shaun Arevalo, Doug Stryke, Scot Federman, Allan Gopez, Kevin Reyes, Kelsey Zorn, Hannah Sample, Guixia Yu, Gurpreet Ishpuniani, Benjamin Briggs, Eric~D. Chow, Amy Berger, Michael~R. Wilson, Candace Wang, Elaine Hsu, Steve Miller, et~al.
\newblock Rapid pathogen detection by metagenomic next-generation sequencing of infected body fluids.
\newblock {\em Nature Medicine}, 2021.

\bibitem{charalampous2024routine}
Themoula Charalampous, Adela Alcolea-Medina, Luke~B. Snell, Christopher Alder, Mark Tan, Tom G.~S. Williams, Noor Al-Yaakoubi, Gul Humayun, Christopher I.~S. Meadows, Duncan L.~A. Wyncoll, Richard Paul, Carolyn~J. Hemsley, Dakshika Jeyaratnam, William Newsholme, Simon Goldenberg, Amita Patel, Fearghal Tucker, Gaia Nebbia, Mark Wilks, Meera Chand, et~al.
\newblock {Routine Metagenomics Service for ICU Patients with Respiratory Infection}.
\newblock {\em American Journal of Respiratory and Critical Care Medicine}, 2024.

\bibitem{Heitz2023}
Morgane Heitz, Albrice Levrat, Vladimir Lazarevic, Olivier Barraud, St{\'e}phane Bland, Emmanuelle Santiago-Allexant, Karen Louis, Jacques Schrenzel, and S{\'e}bastien Hauser.
\newblock {Metagenomics for the microbiological diagnosis of hospital-acquired pneumonia and ventilator-associated pneumonia (HAP/VAP) in intensive care unit (ICU): a proof-of-concept study}.
\newblock {\em Respiratory Research}, 2023.

\bibitem{Alcolea-Medina2025}
Adela Alcolea-Medina, Luke~B. Snell, Gul Humayun, Noor Al-Yaakoubi, Daniel Ward, Christopher Alder, Vishwa Patel, Fredrik Vivian, Christopher I.~S. Meadows, Duncan Wyncoll, Richard Paul, Nick Barratt, Rahul Batra, Jonathan Edgeworth, Gaia Nebbia, and James Whitehorn.
\newblock Rapid pan-microbial metagenomics for pathogen detection and personalised therapy in the intensive care unit: a single-centre prospective observational study.
\newblock {\em The Lancet Microbe}, 2025.

\bibitem{LIANG2023101898}
Yanxu Liang, Qingguo Feng, Kai Wei, Xiaoming Hou, Xiaotao Song, and Yuantao Li.
\newblock {Potential of metagenomic next-generation sequencing in detecting infections of ICU patients}.
\newblock {\em Molecular and Cellular Probes}, 2023.

\bibitem{Chien2022}
Jung-Yien Chien, Chong-Jen Yu, and Po-Ren Hsueh.
\newblock {Utility of Metagenomic Next-Generation Sequencing for Etiological Diagnosis of Patients with Sepsis in Intensive Care Units}.
\newblock {\em Microbiology Spectrum}, 2022.

\bibitem{Neyton2023}
Lucile P.~A. Neyton, Charles~R. Langelier, and Carolyn~S. Calfee.
\newblock {Metagenomic Sequencing in the ICU for Precision Diagnosis of Critical Infectious Illnesses}.
\newblock {\em Critical Care}, 2023.

\bibitem{GENG202181}
Shike Geng, Qing Mei, Chunyan Zhu, Xiaowei Fang, Tianjun Yang, Lei Zhang, Xiaoqin Fan, and Aijun Pan.
\newblock Metagenomic next-generation sequencing technology for detection of pathogens in blood of critically ill patients.
\newblock {\em International Journal of Infectious Diseases}, 2021.

\bibitem{Ren2021}
Di~Ren, Chao Ren, Renqi Yao, Lin Zhang, Xiaomin Liang, Guiyun Li, Jiaze Wang, Xinke Meng, Jia Liu, Yu~Ye, Haoli Li, Sha Wen, Yanhong Chen, Dan Zhou, Xisi He, Xiaohong Li, Kai Lai, Ying Li, and Shuiqing Gui.
\newblock The microbiological diagnostic performance of metagenomic next-generation sequencing in patients with sepsis.
\newblock {\em BMC Infectious Diseases}, 2021.

\bibitem{afshinnekoo2015geospatial}
Ebrahim Afshinnekoo, Cem Meydan, Shanin Chowdhury, Dyala Jaroudi, Collin Boyer, Nick Bernstein, Julia M. Maritz, Darryl Reeves, Jorge Gandara, Sagar Chhangawala, Sofia Ahsanuddin, Amber Simmons, Timothy Nessel, Bharathi Sundaresh, Elizabeth Pereira, Ellen Jorgensen, Sergios-Orestis Kolokotronis, Nell Kirchberger, Isaac Garcia, David Gandara, et~al.
\newblock {Geospatial Resolution of Human and Bacterial Diversity with City-scale Metagenomics}.
\newblock {\em Cell Systems}, 2015.

\bibitem{hsu2016urban}
Tiffany Hsu, Regina Joice, Jose Vallarino, Galeb Abu-Ali, Erica~M. Hartmann, Afrah Shafquat, Casey DuLong, Catherine Baranowski, Dirk Gevers, Jessica~L. Green, Xochitl~C. Morgan, John~D. Spengler, and Curtis Huttenhower.
\newblock {Urban Transit System Microbial Communities Differ by Surface Type and Interaction with Humans and the Environment}.
\newblock {\em Msystems}, 2016.

\bibitem{john2021next}
Goldin John, Nikhil~Shri Sahajpal, Ashis~K. Mondal, Sudha Ananth, Colin Williams, Alka Chaubey, Amyn~M. Rojiani, and Ravindra Kolhe.
\newblock {Next-Generation Sequencing (NGS) in COVID-19: A Tool for SARS-CoV-2 Diagnosis, Monitoring New Strains and Phylodynamic Modeling in Molecular Epidemiology}.
\newblock {\em Current Issues in Molecular Biology}, 2021.

\bibitem{nagy2021targeted}
Dorottya Nagy-Szakal, Mara Couto-Rodriguez, Heather~L. Wells, Joseph~E. Barrows, Marilyne Debieu, Kristin Butcher, Siyuan Chen, Agnes Berki, Courteny Hager, Robert~J. Boorstein, Mariah~K. Taylor, Colleen~B. Jonsson, Christopher~E. Mason, and Niamh~B. O’Hara.
\newblock {Targeted Hybridization Capture of SARS-CoV-2 and Metagenomics Enables Genetic Variant Discovery and Nasal Microbiome Insights}.
\newblock {\em Microbiology Spectrum}, 2021.

\bibitem{nieuwenhuijse2017metagenomic}
David~F. Nieuwenhuijse and Marion P.~G. Koopmans.
\newblock {Metagenomic Sequencing for Surveillance of Food- and Waterborne Viral Diseases}.
\newblock {\em Frontiers in Microbiology}, 2017.

\bibitem{downie2023surveillance}
Diane~L Downie, Preetika Rao, Corinne David-Ferdon, Sean Courtney, Justin Lee, Claire Quiner, Pia MacDonald, Keegan Barnes, Shelby~S Fisher, Joanne~D Andreadis, Jasmine Chaitram, Matthew~R Mauldin, Reyolds~M Salerno, Jarad Schiffer, and Adi Gundlapalli.
\newblock {1774. Surveillance for Emerging and Reemerging Pathogens Using Pathogen Agnostic Metagenomic Sequencing in the United States: A Critical Role for Federal Government Agencies}.
\newblock {\em Open Forum Infectious Diseases}, 2023.

\bibitem{hadfield2018nextstrain}
James Hadfield, Colin Megill, Sidney~M Bell, John Huddleston, Barney Potter, Charlton Callender, Pavel Sagulenko, Trevor Bedford, and Richard~A Neher.
\newblock {Nextstrain: Real-time Tracking of Pathogen Evolution}.
\newblock {\em Bioinformatics}, 2018.

\bibitem{parkins2024wastewater}
Michael~D Parkins, Bonita~E Lee, Nicole Acosta, Maria Bautista, Casey~RJ Hubert, Steve~E Hrudey, Kevin Frankowski, and Xiao-Li Pang.
\newblock {Wastewater-based surveillance as a tool for public health action: SARS-CoV-2 and beyond}.
\newblock {\em Clinical Microbiology Reviews}, 2024.

\bibitem{cdcpulsenet}
{Centers for Disease Control and Prevention}.
\newblock {Using the Latest Technology to Detect Outbreaks and Protect the Public’s Health}.
\newblock \url{https://www.cdc.gov/pulsenet/next-gen-wgs.html}, 01 2020.

\bibitem{Iqbal2012}
Zamin Iqbal, Mario Caccamo, Isaac Turner, Paul Flicek, and Gil McVean.
\newblock {De novo assembly and genotyping of variants using colored de Bruijn graphs}.
\newblock {\em Nature Genetics}, 2012.

\bibitem{marchet2021data}
Camille Marchet, Christina Boucher, Simon~J. Puglisi, Paul Medvedev, Mikaël Salson, and Rayan Chikhi.
\newblock Data structures based on k-mers for querying large collections of sequencing data sets.
\newblock {\em Genome Research}, 2021.

\bibitem{karasikov2020metagraph}
Mikhail Karasikov, Harun Mustafa, Daniel Danciu, Oleksandr Kulkov, Marc Zimmermann, Christopher Barber, Gunnar R{\"a}tsch, and Andr{\'e} Kahles.
\newblock Efficient and accurate search in petabase-scale sequence repositories.
\newblock {\em Nature}, 2025.

\bibitem{karasikov2022lossless}
Mikhail Karasikov, Harun Mustafa, Gunnar Rätsch, and André Kahles.
\newblock {Lossless indexing with counting de Bruijn graphs}.
\newblock {\em Genome Research}, 2022.

\bibitem{karasikov2019sparse}
Mikhail Karasikov, Harun Mustafa, Amir Joudaki, Sara Javadzadeh-no, Gunnar R\"{a}tsch, and Andr\'{e} Kahles.
\newblock {Sparse Binary Relation Representations for Genome Graph Annotation}.
\newblock {\em Journal of Computational Biology}, 2020.

\bibitem{danciu2021topology}
Daniel Danciu, Mikhail Karasikov, Harun Mustafa, André Kahles, and Gunnar Rätsch.
\newblock {{Topology-based sparsification of graph annotations}}.
\newblock {\em Bioinformatics}, 2021.

\bibitem{fan2023fulgor}
Jason Fan, Noor~Pratap Singh, Jamshed Khan, Giulio~Ermanno Pibiri, and Rob Patro.
\newblock {Fulgor: A Fast and Compact k-mer Index for Large-Scale Matching and Color Queries}.
\newblock In {\em WABI}, 2023.

\bibitem{bradley2019ultrafast}
Phelim Bradley, Henk~C Den~Bakker, Eduardo~PC Rocha, Gil McVean, and Zamin Iqbal.
\newblock {Ultrafast Search of All Deposited Bacterial and Viral Genomic Data}.
\newblock {\em Nature Biotechnology}, 2019.

\bibitem{siren2021pangenomics}
Jouni Sir{\'e}n, Jean Monlong, Xian Chang, Adam~M. Novak, Jordan~M. Eizenga, Charles Markello, Jonas~A. Sibbesen, Glenn Hickey, Pi-Chuan Chang, Andrew Carroll, Namrata Gupta, Stacey Gabriel, Thomas~W. Blackwell, Aakrosh Ratan, Kent~D. Taylor, Stephen~S. Rich, Jerome~I. Rotter, David Haussler, Erik Garrison, and Benedict Paten.
\newblock {Pangenomics enables genotyping of known structural variants in 5202 diverse genomes}.
\newblock {\em Science}, 2021.

\bibitem{Sherman2020}
Rachel~M. Sherman and Steven~L. Salzberg.
\newblock Pan-genomics in the human genome era.
\newblock {\em Nature Reviews Genetics}, 2020.

\bibitem{taylor2024beyond}
Dylan~J. Taylor, Jordan~M. Eizenga, Qiuhui Li, Arun Das, Katharine~M. Jenike, Eimear~E. Kenny, Karen~H. Miga, Jean Monlong, Rajiv~C. McCoy, Benedict Paten, and Michael~C. Schatz.
\newblock {Beyond the Human Genome Project: The Age of Complete Human Genome Sequences and Pangenome References}.
\newblock {\em Annual Review of Genomics and Human Genetics}, 2024.

\bibitem{eizenga2020pangenome}
Jordan~M. Eizenga, Adam~M. Novak, Jonas~A. Sibbesen, Simon Heumos, Ali Ghaffaari, Glenn Hickey, Xian Chang, Josiah~D. Seaman, Robin Rounthwaite, Jana Ebler, Mikko Rautiainen, Shilpa Garg, Benedict Paten, Tobias Marschall, Jouni Sirén, and Erik Garrison.
\newblock Pangenome graphs.
\newblock {\em Annual Review of Genomics and Human Genetics}, 2020.

\bibitem{Liao2023}
Wen-Wei Liao, Mobin Asri, Jana Ebler, Daniel Doerr, Marina Haukness, Glenn Hickey, Shuangjia Lu, Julian~K. Lucas, Jean Monlong, Haley~J. Abel, Silvia Buonaiuto, Xian~H. Chang, Haoyu Cheng, Justin Chu, Vincenza Colonna, Jordan~M. Eizenga, Xiaowen Feng, Christian Fischer, Robert~S. Fulton, Shilpa Garg, et~al.
\newblock A draft human pangenome reference.
\newblock {\em Nature}, 2023.

\bibitem{Armstrong2020cactus}
Joel Armstrong, Glenn Hickey, Mark Diekhans, Ian~T. Fiddes, Adam~M. Novak, Alden Deran, Qi~Fang, Duo Xie, Shaohong Feng, Josefin Stiller, Diane Genereux, Jeremy Johnson, Voichita~Dana Marinescu, Jessica Alf{\"o}ldi, Robert~S. Harris, Kerstin Lindblad-Toh, David Haussler, Elinor Karlsson, Erich~D. Jarvis, Guojie Zhang, et~al.
\newblock {Progressive Cactus is a multiple-genome aligner for the thousand-genome era}.
\newblock {\em Nature}, 2020.

\bibitem{Groza2024}
Cristian Groza, Carl Schwendinger-Schreck, Warren~A. Cheung, Emily~G. Farrow, Isabelle Thiffault, Juniper Lake, William~B. Rizzo, Gilad Evrony, Tom Curran, Guillaume Bourque, and Tomi Pastinen.
\newblock Pangenome graphs improve the analysis of structural variants in rare genetic diseases.
\newblock {\em Nature Communications}, 2024.

\bibitem{Holley2026}
Guillaume Holley, Hannes~P. Eggertsson, Snaedis Kristmundsdottir, Doruk Beyter, Astros~Th Skuladottir, Kristjan H.~S. Moore, Pall~I. Olason, Arnaldur Gylfason, Olafur~T. Magnusson, Asmundur Oddsson, Hreinn Stefansson, Agnar Helgason, Gisli Masson, Patrick Sulem, Daniel~F. Gudbjartsson, Kari Stefansson, and Bjarni~V. Halldorsson.
\newblock An icelandic pangenome reference.
\newblock {\em Nature}, 2026.

\bibitem{chiang2019from}
Augusto Dulanto~Chiang and John~P Dekker.
\newblock {From the Pipeline to the Bedside: Advances and Challenges in Clinical Metagenomics}.
\newblock {\em The Journal of Infectious Diseases}, 2019.

\bibitem{chiu2019clinical}
Charles~Y. Chiu and Steven~A. Miller.
\newblock Clinical metagenomics.
\newblock {\em Nature Reviews Genetics}, 2019.

\bibitem{kaplan2025pangenomicsbench}
Noah Kaplan, Jan-Niklas Schmelzle, Yufeng Gu, Erik Garrison, Christopher Batten, and Reetuparna Das.
\newblock {PangenomicsBench: A Benchmark Suite and Characterization of Pangenomics}.
\newblock In {\em IISWC}, 2025.

\bibitem{das2024systems}
Reetuparna Das and Satish Narayanasamy.
\newblock Systems challenges and opportunities for genomics.
\newblock {\em Computer}, 2024.

\bibitem{berger2023navigating}
Bonnie Berger and Yun~William Yu.
\newblock {Navigating bottlenecks and trade-offs in genomic data analysis}.
\newblock {\em Nature Reviews Genetics}, 2023.

\bibitem{wang2021nanopore}
Yunhao Wang, Yue Zhao, Audrey Bollas, Yuru Wang, and Kin~Fai Au.
\newblock {Nanopore Sequencing Technology, Bioinformatics and Applications}.
\newblock {\em Nature Biotechnology}, 2021.

\bibitem{hu2021next}
Taishan Hu, Nilesh Chitnis, Dimitri Monos, and Anh Dinh.
\newblock {Next-Generation Sequencing Technologies: An Overview}.
\newblock {\em Human Immunology}, 2021.

\bibitem{alser2022molecules}
Mohammed Alser, Joel Lindegger, Can Firtina, Nour Almadhoun, Haiyu Mao, Gagandeep Singh, Juan Gomez-Luna, and Onur Mutlu.
\newblock From molecules to genomic variations: Accelerating genome analysis via intelligent algorithms and architectures.
\newblock {\em Computational and Structural Biotechnology Journal}, 2022.

\bibitem{bokulich2020measuring}
Nicholas~A. Bokulich, Michal Ziemski, Michael~S. Robeson, and Benjamin~D. Kaehler.
\newblock Measuring the microbiome: Best practices for developing and benchmarking microbiomics methods.
\newblock {\em Computational and Structural Biotechnology Journal}, 2020.

\bibitem{schuele2021future}
Leonard Schuele, Hayley Cassidy, Nilay Peker, John~W.~A. Rossen, and Natacha Couto.
\newblock Future potential of metagenomics in microbiology laboratories.
\newblock {\em Expert Review of Molecular Diagnostics}, 2021.

\bibitem{chen2024improved}
Nae-Chyun Chen, Luis~F. Paulin, Fritz~J. Sedlazeck, Sergey Koren, Adam~M. Phillippy, and Ben Langmead.
\newblock Improved sequence mapping using a complete reference genome and lift-over.
\newblock {\em Nature Methods}, 2024.

\bibitem{Siren2024}
Jouni Sir{\'e}n, Parsa Eskandar, Matteo~Tommaso Ungaro, Glenn Hickey, Jordan~M. Eizenga, Adam~M. Novak, Xian Chang, Pi-Chuan Chang, Mikhail Kolmogorov, Andrew Carroll, Jean Monlong, and Benedict Paten.
\newblock Personalized pangenome references.
\newblock {\em Nature Methods}, 2024.

\bibitem{vaddadi2023minimizing}
Naga Sai~Kavya Vaddadi, Taher Mun, and Benjamin Langmead.
\newblock {Minimizing Reference Bias: The Impute-First Approach for Personalized Genome Analysis}.
\newblock In {\em BCB}, 2023.

\bibitem{aganezov2022complete}
Sergey Aganezov, Stephanie~M. Yan, Daniela~C. Soto, Melanie Kirsche, Samantha Zarate, Pavel Avdeyev, Dylan~J. Taylor, Kishwar Shafin, Alaina Shumate, Chunlin Xiao, Justin Wagner, Jennifer McDaniel, Nathan~D. Olson, Michael E.~G. Sauria, Mitchell~R. Vollger, Arang Rhie, Melissa Meredith, Skylar Martin, Joyce Lee, Sergey Koren, et~al.
\newblock A complete reference genome improves analysis of human genetic variation.
\newblock {\em Science}, 2022.

\bibitem{rhie2023complete}
Arang Rhie, Sergey Nurk, Monika Cechova, Savannah~J. Hoyt, Dylan~J. Taylor, Nicolas Altemose, Paul~W. Hook, Sergey Koren, Mikko Rautiainen, Ivan~A. Alexandrov, Jamie Allen, Mobin Asri, Andrey~V. Bzikadze, Nae-Chyun Chen, Chen-Shan Chin, Mark Diekhans, Paul Flicek, Giulio Formenti, Arkarachai Fungtammasan, Carlos Garcia~Giron, et~al.
\newblock {The complete sequence of a human Y chromosome}.
\newblock {\em Nature}, 2023.

\bibitem{nurk2022complete}
Sergey Nurk, Sergey Koren, Arang Rhie, Mikko Rautiainen, Andrey~V. Bzikadze, Alla Mikheenko, Mitchell~R. Vollger, Nicolas Altemose, Lev Uralsky, Ariel Gershman, Sergey Aganezov, Savannah~J. Hoyt, Mark Diekhans, Glennis~A. Logsdon, Michael Alonge, Stylianos~E. Antonarakis, Matthew Borchers, Gerard~G. Bouffard, Shelise~Y. Brooks, Gina~V. Caldas, et~al.
\newblock The complete sequence of a human genome.
\newblock {\em Science}, 2022.

\bibitem{kim2024airlifttcbb}
Jeremie~S. Kim, Can Firtina, Meryem~Banu Cavlak, Damla~Senol Cali, Nastaran Hajinazar, Mohammed Alser, Can Alkan, and Onur Mutlu.
\newblock {AirLift: A Fast and Comprehensive Technique for Remapping Alignments between Reference Genomes}.
\newblock {\em IEEE/ACM TCBB}, 2024.

\bibitem{illuminax}
Illumina.
\newblock {NovaSeq X Series Specifications}.
\newblock \url{https://emea.illumina.com/systems/sequencing-platforms/novaseq-x-plus/specifications.html}, 2023.

\bibitem{shokralla2015massively}
Shadi Shokralla, Teresita~M. Porter, Joel~F. Gibson, Rafal Dobosz, Daniel~H. Janzen, Winnie Hallwachs, G.~Brian Golding, and Mehrdad Hajibabaei.
\newblock {Massively parallel multiplex DNA sequencing for specimen identification using an Illumina MiSeq platform}.
\newblock {\em Scientific Reports}, 2015.

\bibitem{lapierre2020metalign}
Nathan LaPierre, Mohammed Alser, Eleazar Eskin, David Koslicki, and Serghei Mangul.
\newblock {Metalign: Efficient Alignment-based Metagenomic Profiling Via Containment Min Hash}.
\newblock {\em Genome Biology}, 2020.

\bibitem{katz2021sra}
Kenneth Katz, Oleg Shutov, Richard Lapoint, Michael Kimelman, J Rodney Brister, and Christopher O’Sullivan.
\newblock {The Sequence Read Archive: a decade more of explosive growth}.
\newblock {\em Nucleic Acids Research}, 2021.

\bibitem{enastats}
{{European Bioinformatics Institute}}.
\newblock {European Nucleotide Archive Statistics}.
\newblock \url{https://www.ebi.ac.uk/ena/browser/about/statistics}, 2023.

\bibitem{leinonen2010sequence}
Rasko Leinonen, Hideaki Sugawara, Martin Shumway, and International Nucleotide Sequence~Database Collaboration.
\newblock {The Sequence Read Archive}.
\newblock {\em Nucleic Acids Research}, 2010.

\bibitem{zhang2021real}
Haowen Zhang, Haoran Li, Chirag Jain, Haoyu Cheng, Kin~Fai Au, Heng Li, and Srinivas Aluru.
\newblock {Real-time Mapping of Nanopore Raw Signals}.
\newblock {\em Bioinformatics}, 2021.

\bibitem{firtina2023rawhash}
Can Firtina, Nika Mansouri~Ghiasi, Joel Lindegger, Gagandeep Singh, Meryem~Banu Cavlak, Haiyu Mao, and Onur Mutlu.
\newblock {RawHash: enabling fast and accurate real-time analysis of raw nanopore signals for large genomes}.
\newblock {\em Bioinformatics}, 2023.

\bibitem{kovaka2020targeted}
Sam Kovaka, Yunfan Fan, Bohan Ni, Winston Timp, and Michael~C Schatz.
\newblock {Targeted Nanopore Sequencing by Real-time Mapping of Raw Electrical Signal with UNCALLED}.
\newblock {\em Nature Biotechnology}, 2021.

\bibitem{mutlu2023accelerating}
Onur Mutlu and Can Firtina.
\newblock {Accelerating Genome Analysis via Algorithm-Architecture Co-Design}.
\newblock In {\em DAC}, 2023.

\bibitem{Payne2021}
Alexander Payne, Nadine Holmes, Thomas Clarke, Rory Munro, Bisrat~J. Debebe, and Matthew Loose.
\newblock Readfish enables targeted nanopore sequencing of gigabase-sized genomes.
\newblock {\em Nature Biotechnology}, 2021.

\bibitem{Bao2021Squigglenet}
Yuwei Bao, Jack Wadden, John~R. Erb-Downward, Piyush Ranjan, Weichen Zhou, Torrin~L. McDonald, Ryan~E. Mills, Alan~P. Boyle, Robert~P. Dickson, David Blaauw, and Joshua~D. Welch.
\newblock {SquiggleNet: real-time, direct classification of nanopore signals}.
\newblock {\em Genome Biology}, 2021.

\bibitem{ulrich2022readbouncer}
Jens-Uwe Ulrich, Ahmad Lutfi, Kilian Rutzen, and Bernhard~Y Renard.
\newblock {ReadBouncer: precise and scalable adaptive sampling for nanopore sequencing}.
\newblock {\em Bioinformatics}, 2022.

\bibitem{loose_real-time_2016}
Matthew Loose, Sunir Malla, and Michael Stout.
\newblock Real-time selective sequencing using nanopore technology.
\newblock {\em Nature Methods}, 2016.

\bibitem{senanayake_deepselectnet_2023}
Anjana Senanayake, Hasindu Gamaarachchi, Damayanthi Herath, and Roshan Ragel.
\newblock {DeepSelectNet}: deep neural network based selective sequencing for {Oxford} nanopore sequencing.
\newblock {\em BMC Bioinformatics}, 2023.

\bibitem{sam_kovaka_uncalled4_2024}
Sam Kovaka, Paul~W. Hook, Katharine~M. Jenike, Vikram Shivakumar, Luke~B. Morina, Roham Razaghi, Winston Timp, and Michael~C. Schatz.
\newblock {Uncalled4 improves nanopore DNA and RNA modification detection via fast and accurate signal alignment}.
\newblock {\em Nature Methods}, 2025.

\bibitem{lindegger_rawalign_2024}
Joël Lindegger, Can Firtina, Nika~Mansouri Ghiasi, Mohammad Sadrosadati, Mohammed Alser, and Onur Mutlu.
\newblock {RawAlign}: {Accurate}, {Fast}, and {Scalable} {Raw} {Nanopore} {Signal} {Mapping} via {Combining} {Seeding} and {Alignment}.
\newblock {\em IEEE Access}, 2024.

\bibitem{firtina2023rawhash2}
Can Firtina, Melina Soysal, Jo{\"e}l Lindegger, and Onur Mutlu.
\newblock {RawHash2}: Mapping raw nanopore signals using hash-based seeding and adaptive quantization.
\newblock {\em Bioinformatics}, 2024.

\bibitem{firtina_rawsamble_2024}
Can Firtina, Maximilian Mordig, Harun Mustafa, Sayan Goswami, Nika~Mansouri Ghiasi, Stefano Mercogliano, Furkan Eris, Joel Lindegger, André Kahles, and Onur Mutlu.
\newblock Rawsamble: Overlapping raw nanopore signals using a hash-based seeding mechanism.
\newblock {\em Bioinformatics}, 2026.

\bibitem{shih2023efficient}
Po~Jui Shih, Hassaan Saadat, Sri Parameswaran, and Hasindu Gamaarachchi.
\newblock Efficient real-time selective genome sequencing on resource-constrained devices.
\newblock {\em GigaScience}, 2023.

\bibitem{sadasivan2023rapid}
Harisankar Sadasivan, Jack Wadden, Kush Goliya, Piyush Ranjan, Robert~P Dickson, David Blaauw, Reetuparna Das, and Satish Narayanasamy.
\newblock {Rapid Real-time Squiggle Classification for Read until using RawMap}.
\newblock {\em Archives of Clinical and Biomedical Research}, 2023.

\bibitem{shivakumar_sigmoni_2024}
Vikram~S Shivakumar, Omar~Y Ahmed, Sam Kovaka, Mohsen Zakeri, and Ben Langmead.
\newblock Sigmoni: classification of nanopore signal with a compressed pangenome index.
\newblock {\em Bioinformatics}, 2024.

\bibitem{sadasivan_accelerated_2024}
Harisankar Sadasivan, Daniel Stiffler, Ajay Tirumala, Johnny Israeli, and {Satish Narayanasamy}.
\newblock Accelerated {Dynamic} {Time} {Warping} on {GPU} for {Selective} {Nanopore} {Sequencing}.
\newblock {\em Journal of Biotechnology and Biomedicine}, 2024.

\bibitem{gamaarachchi_gpu_2020}
Hasindu Gamaarachchi, Chun~Wai Lam, Gihan Jayatilaka, Hiruna Samarakoon, Jared~T. Simpson, Martin~A. Smith, and Sri Parameswaran.
\newblock {GPU} accelerated adaptive banded event alignment for rapid comparative nanopore signal analysis.
\newblock {\em BMC Bioinformatics}, 2020.

\bibitem{samarasinghe_energy_2021}
Suneth Samarasinghe, Pubudu Premathilaka, Wishma Herath, Hasindu Gamaarachchi, and Roshan Ragel.
\newblock Energy {Efficient} {Adaptive} {Banded} {Event} {Alignment} using {OpenCL} on {FPGAs}.
\newblock In {\em {ICIAfS}}, 2021.

\bibitem{rawbench}
Furkan Eris, Ulysse McConnell, Can Firtina, and Onur Mutlu.
\newblock {RawBench: A Comprehensive Benchmarking Framework for Raw Nanopore Signal Analysis Techniques}.
\newblock In {\em BCB}, 2025.

\bibitem{illumina}
Illumina.
\newblock {NovaSeq 6000 System Specifications}.
\newblock \url{https://emea.illumina.com/systems/sequencing-platforms/novaseq/specifications.html}, 2020.

\bibitem{jain2016oxford}
Miten Jain, Hugh~E. Olsen, Benedict Paten, and Mark Akeson.
\newblock {The Oxford Nanopore MinION: delivery of nanopore sequencing to the genomics community}.
\newblock {\em Genome Biology}, 2016.

\bibitem{pomerantz2018real}
Aaron Pomerantz, Nicol{\'a}s Pe{\~n}afiel, Alejandro Arteaga, Lucas Bustamante, Frank Pichardo, Luis~A Coloma, C{\'e}sar~L Barrio-Amor{\'o}s, David Salazar-Valenzuela, and Stefan Prost.
\newblock {Real-time DNA Barcoding in a Rainforest Using Nanopore Sequencing: Opportunities for Rapid Biodiversity Assessments and Local Capacity Building}.
\newblock {\em GigaScience}, 2018.

\bibitem{zhang2000greedy}
Zheng Zhang, Scott Schwartz, Lukas Wagner, and Webb Miller.
\newblock {A Greedy Algorithm for Aligning DNA Sequences}.
\newblock {\em Journal of Computational Biology}, 2000.

\bibitem{slater2005automated}
Guy St~C Slater and Ewan Birney.
\newblock {Automated Generation of Heuristics for Biological Sequence Comparison}.
\newblock {\em BMC Bioinformatics}, 2005.

\bibitem{li2018minimap2}
Heng Li.
\newblock {Minimap2: Pairwise Alignment for Nucleotide Sequences}.
\newblock {\em Bioinformatics}, 2018.

\bibitem{myers1999fast}
Gene Myers.
\newblock {A Fast Bit-vector Algorithm for Approximate String Matching Based on Dynamic Programming}.
\newblock {\em JACM}, 1999.

\bibitem{marco2021fast}
Santiago Marco-Sola, Juan~Carlos Moure, Miquel Moreto, and Antonio Espinosa.
\newblock {Fast Gap-affine Pairwise Alignment Using the Wavefront Algorithm}.
\newblock {\em Bioinformatics}, 2021.

\bibitem{marcosola2023optimal}
Santiago Marco-Sola, Jordan~M Eizenga, Andrea Guarracino, Benedict Paten, Erik Garrison, and Miquel Moreto.
\newblock {{Optimal gap-affine alignment in O(s) space}}.
\newblock {\em Bioinformatics}, 2023.

\bibitem{grootkoerkamp2024apa2}
Ragnar Groot~Koerkamp.
\newblock {A*PA2: Up to 19× Faster Exact Global Alignment}.
\newblock In {\em WABI}, 2024.

\bibitem{xin2013accelerating}
Hongyi Xin, Donghyuk Lee, Farhad Hormozdiari, Samihan Yedkar, Onur Mutlu, and Can Alkan.
\newblock {Accelerating Read Mapping with FastHASH}.
\newblock {\em BMC Genomics}, 2013.

\bibitem{xin2015shifted}
Hongyi Xin, John Greth, John Emmons, Gennady Pekhimenko, Carl Kingsford, Can Alkan, and Onur Mutlu.
\newblock {Shifted Hamming Distance: A Fast and Accurate SIMD-friendly Filter to Accelerate Alignment Verification in Read Mapping}.
\newblock {\em Bioinformatics}, 2015.

\bibitem{tseng2025ultrafast}
Yu-Hsiang Tseng, Sumit Walia, and Yatish Turakhia.
\newblock {Ultrafast and ultralarge multiple sequence alignments using TWILIGHT}.
\newblock {\em Bioinformatics}, 2025.

\bibitem{walia2025ultrafast}
Sumit Walia, Zexing Chen, Yu-Hsiang Tseng, and Yatish Turakhia.
\newblock {Ultrafast and Ultralarge Distance-Based Phylogenetics Using DIPPER}.
\newblock {\em bioRxiv}, 2025.

\bibitem{kim_fastremap_2022}
Jeremie~S Kim, Can Firtina, Meryem~Banu Cavlak, Damla Senol~Cali, Can Alkan, and Onur Mutlu.
\newblock {FastRemap}: a tool for quickly remapping reads between genome assemblies.
\newblock {\em Bioinformatics}, 2022.

\bibitem{Ashyralyyev2026gencore}
Akmuhammet Ashyralyyev, Ege Sirvan, Ecem {\.I}lg{\"u}n, Salem Malikic, Tu{\u g}kan Batu, S.~Cenk Sahinalp, and Can Alkan.
\newblock {GenCore: Genomic distance estimation using Locally Consistent Parsing}.
\newblock {\em bioRxiv}, 2026.

\bibitem{ashyralyyev2026lcpan}
Akmuhammet Ashyralyyev, Zülal Bingöl, Begüm~Filiz Öz, Kaiyuan Zhu, Salem Malikic, Uzi Vishkin, S.~Cenk Sahinalp, and Can Alkan.
\newblock {LCPan: efficient variation graph construction using Locally Consistent Parsing}.
\newblock {\em arXiv}, 2026.

\bibitem{alicioglu2024pairwise}
Ahmet~Cemal Alıcıoğlu and Can Alkan.
\newblock {Pairwise sequence alignment with block and character edit operations}.
\newblock {\em arXiv}, 2024.

\bibitem{kim2019airlift}
Jeremie~S Kim, Can Firtina, Meryem~Banu Cavlak, Damla~Senol Cali, Mohammed Alser, Nastaran Hajinazar, Can Alkan, and Onur Mutlu.
\newblock {AirLift: A Fast and Comprehensive Technique for Remapping Alignments between Reference Genomes}.
\newblock {\em arXiv}, 2019.

\bibitem{alser2025taming}
Mohammed Alser, Julien Eudine, and Onur Mutlu.
\newblock {Taming large-scale genomic analyses via sparsified genomics}.
\newblock {\em Nature Communications}, 2025.

\bibitem{rautiainen2020graphaligner}
Mikko Rautiainen and Tobias Marschall.
\newblock {GraphAligner: Rapid and Versatile Sequence-to-Graph Alignment}.
\newblock {\em Genome Biology}, 2020.

\bibitem{kim2019graph}
Daehwan Kim, Joseph~M Paggi, Chanhee Park, Christopher Bennett, and Steven~L Salzberg.
\newblock {Graph-based genome alignment and genotyping with HISAT2 and HISAT-genotype}.
\newblock {\em Nature Biotechnology}, 2019.

\bibitem{gao2020abpoa}
Yan Gao, Yongzhuang Liu, Yanmei Ma, Bo~Liu, Yadong Wang, and Yi~Xing.
\newblock {abPOA: an SIMD-based C library for fast partial order alignment using adaptive band}.
\newblock {\em Bioinformatics}, 2020.

\bibitem{jain2019pasgal}
Chirag Jain, Sanchit Misra, Haowen Zhang, Alexander Dilthey, and Srinivas Aluru.
\newblock {Accelerating Sequence Alignment to Graphs}.
\newblock In {\em IPDPS}, 2019.

\bibitem{Rautiainen2019}
Mikko Rautiainen, Veli M\"{a}kinen, and Tobias Marschall.
\newblock Bit-parallel sequence-to-graph alignment.
\newblock {\em Bioinformatics}, 2019.

\bibitem{Chandra2023}
Ghanshyam Chandra and Chirag Jain.
\newblock {Sequence to Graph Alignment Using Gap-Sensitive Co-linear Chaining}.
\newblock In {\em RECOMB}, 2023.

\bibitem{Ivanov2022}
Pesho Ivanov, Benjamin Bichsel, and Martin Vechev.
\newblock {Fast and Optimal Sequence-to-Graph Alignment Guided by Seeds}.
\newblock In {\em RECOMB}, 2022.

\bibitem{Ma2023}
Jun Ma, Manuel Cáceres, Leena Salmela, Veli M\"{a}kinen, and Alexandru~I Tomescu.
\newblock Chaining for accurate alignment of erroneous long reads to acyclic variation graphs.
\newblock {\em Bioinformatics}, 2023.

\bibitem{Darby2020vargas}
Charlotte~A Darby, Ravi Gaddipati, Michael~C Schatz, and Ben Langmead.
\newblock Vargas: heuristic-free alignment for assessing linear and graph read aligners.
\newblock {\em Bioinformatics}, 2020.

\bibitem{Hwang2025MEMO}
Stephen Hwang, Nathaniel~K. Brown, Omar~Y. Ahmed, Katharine~M. Jenike, Sam Kovaka, Michael~C. Schatz, and Ben Langmead.
\newblock Mem-based pangenome indexing for k-mer queries.
\newblock {\em Algorithms for Molecular Biology}, 2025.

\bibitem{Romain2023svjedi}
Sandra Romain and Claire Lemaitre.
\newblock {SVJedi-graph: improving the genotyping of close and overlapping structural variants with long reads using a variation graph}.
\newblock {\em Bioinformatics}, 2023.

\bibitem{Li2020minigraph}
Heng Li, Xiaowen Feng, and Chong Chu.
\newblock The design and construction of reference pangenome graphs with minigraph.
\newblock {\em Genome Biology}, 2020.

\bibitem{doblas2025smx}
Max Doblas, Po~Jui Shih, Oscar Lostes-Cazorla, Miquel Moreto, Christopher Batten, and Santiago Marco-Sola.
\newblock {Smx: Heterogeneous architecture for universal sequence alignment acceleration}.
\newblock In {\em MICRO}, 2025.

\bibitem{lou2020helix}
Qian Lou, Sarath~Chandra Janga, and Lei Jiang.
\newblock {Helix: Algorithm/architecture co-design for accelerating nanopore genome base-calling}.
\newblock In {\em PACT}, 2020.

\bibitem{lou2018brawl}
Qian Lou and Lei Jiang.
\newblock {Brawl: A spintronics-based portable basecalling-in-memory architecture for nanopore genome sequencing}.
\newblock {\em IEEE CAL}, 2018.

\bibitem{shahroodi2023swordfish}
Taha Shahroodi, Gagandeep Singh, Mahdi Zahedi, Haiyu Mao, Joel Lindegger, Can Firtina, Stephan Wong, Onur Mutlu, and Said Hamdioui.
\newblock {Swordfish: A Framework for Evaluating Deep Neural Network-based Basecalling using Computation-In-Memory with Non-Ideal Memristors}.
\newblock In {\em MICRO}, 2023.

\bibitem{markus2020benchmarking}
Ryan Marcus, Andreas Kipf, Alexander van Renen, Mihail Stoian, Sanchit Misra, Alfons Kemper, Thomas Neumann, and Tim Kraska.
\newblock Benchmarking learned indexes.
\newblock {\em Proceedings of the VLDB Endowment}, 2020.

\bibitem{subramaniyan2021accelerated}
Arun Subramaniyan, Jack Wadden, Kush Goliya, Nathan Ozog, Xiao Wu, Satish Narayanasamy, David Blaauw, and Reetuparna Das.
\newblock Accelerated seeding for genome sequence alignment with enumerated radix trees.
\newblock In {\em ISCA}, 2021.

\bibitem{huangfu2018radar}
Wenqin Huangfu, Shuangchen Li, Xing Hu, and Yuan Xie.
\newblock {RADAR: A 3D-ReRAM based DNA Alignment Accelerator Architecture}.
\newblock In {\em DAC}, 2018.

\bibitem{khatamifard2021genvom}
S~Karen Khatamifard, Zamshed Chowdhury, Nakul Pande, Meisam Razaviyayn, Chris~H Kim, and Ulya~R Karpuzcu.
\newblock {GeNVoM: Read Mapping Near Non-Volatile Memory}.
\newblock {\em IEEE/ACM TCBB}, 2021.

\bibitem{gupta2019rapid}
Saransh Gupta, Mohsen Imani, Behnam Khaleghi, Venkatesh Kumar, and Tajana Rosing.
\newblock {RAPID: A ReRAM Processing In-memory Architecture for DNA Sequence Alignment}.
\newblock In {\em ISLPED}, 2019.

\bibitem{li2021pim}
Xue-Qi Li, Guang-Ming Tan, and Ning-Hui Sun.
\newblock {PIM-Align: A Processing-in-Memory Architecture for FM-Index Search Algorithm}.
\newblock {\em Journal of Computer Science and Technology}, 2021.

\bibitem{angizi2019aligns}
Shaahin Angizi, Jiao Sun, Wei Zhang, and Deliang Fan.
\newblock {Aligns: A Processing-in-memory Accelerator for DNA Short Read Alignment Leveraging SOT-MRAM}.
\newblock In {\em DAC}, 2019.

\bibitem{zokaee2018aligner}
Farzaneh Zokaee, Hamid~R Zarandi, and Lei Jiang.
\newblock {Aligner: A Process-in-memory Architecture for Short Read Alignment in ReRAMs}.
\newblock {\em IEEE Computer Architecture Letters}, 2018.

\bibitem{turakhia2018darwin}
Yatish Turakhia, Gill Bejerano, and William~J Dally.
\newblock {Darwin: A Genomics Co-processor Provides up to 15,000 x Acceleration on Long Read Assembly}.
\newblock In {\em ASPLOS}, 2018.

\bibitem{fujiki2018genax}
Daichi Fujiki, Arun Subramaniyan, Tianjun Zhang, Yu~Zeng, Reetuparna Das, David Blaauw, and Satish Narayanasamy.
\newblock {Genax: A Genome Sequencing Accelerator}.
\newblock In {\em ISCA}, 2018.

\bibitem{madhavan2014race}
Advait Madhavan, Timothy Sherwood, and Dmitri Strukov.
\newblock {Race Logic: A Hardware Acceleration for Dynamic Programming Algorithms}.
\newblock {\em ACM SIGARCH Computer Architecture News}, 2014.

\bibitem{cheng2018bitmapper2}
Haoyu Cheng, Yong Zhang, and Yun Xu.
\newblock {Bitmapper2: A GPU-accelerated All-mapper Based on The Sparse Q-gram Index}.
\newblock {\em IEEE/ACM TCBB}, 2018.

\bibitem{houtgast2018hardware}
Ernst~Joachim Houtgast, Vlad-Mihai Sima, Koen Bertels, and Zaid Al-Ars.
\newblock {Hardware Acceleration of BWA-MEM Genomic Short Read Mapping for Longer Read Lengths}.
\newblock {\em Computational Biology and Chemistry}, 2018.

\bibitem{houtgast2017efficient}
Ernst~Joachim Houtgast, VladMihai Sima, Koen Bertels, and Zaid AlArs.
\newblock {An Efficient GPU-accelerated Implementation of Genomic Short Read Mapping with BWA-MEM}.
\newblock {\em ACM SIGARCH Computer Architecture News}, 2017.

\bibitem{zeni2020logan}
Alberto Zeni, Giulia Guidi, Marquita Ellis, Nan Ding, Marco~D Santambrogio, Steven Hofmeyr, Ayd{\i}n Bulu{\c{c}}, Leonid Oliker, and Katherine Yelick.
\newblock {Logan: High-performance GPU-based X-drop Long-read Alignment}.
\newblock In {\em IPDPS}, 2020.

\bibitem{ahmed2019gasal2}
Nauman Ahmed, Jonathan L{\'e}vy, Shanshan Ren, Hamid Mushtaq, Koen Bertels, and Zaid Al-Ars.
\newblock {GASAL2: A GPU Accelerated Sequence Alignment Library for High-Throughput NGS Data}.
\newblock {\em BMC Bioinformatics}, 2019.

\bibitem{nishimura2017accelerating}
Takahiro Nishimura, Jacir~L Bordim, Yasuaki Ito, and Koji Nakano.
\newblock {Accelerating the Smith-Waterman Algorithm Using Bitwise Parallel Bulk Computation Technique on GPU}.
\newblock In {\em IPDPSW}, 2017.

\bibitem{de2016cudalign}
Edans~Flavius de~Oliveira~Sandes, Guillermo Miranda, Xavier Martorell, Eduard Ayguade, George Teodoro, and Alba Cristina~Magalhaes Melo.
\newblock {CUDAlign 4.0: Incremental Speculative Traceback for Exact Chromosome-wide Alignment in GPU Clusters}.
\newblock {\em IEEE TPDS}, 2016.

\bibitem{liu2015gswabe}
Yongchao Liu and Bertil Schmidt.
\newblock {GSWABE: Faster GPU-accelerated Sequence Alignment with Optimal Alignment Retrieval for Short DNA Sequences}.
\newblock {\em Concurrency and Computation: Practice and Experience}, 2015.

\bibitem{liu2013cudasw++}
Yongchao Liu, Adrianto Wirawan, and Bertil Schmidt.
\newblock {CUDASW++ 3.0: Accelerating Smith-Waterman Protein Database Search by Coupling CPU and GPU SIMD Instructions}.
\newblock {\em BMC Bioinformatics}, 2013.

\bibitem{liu2009cudasw++}
Yongchao Liu, Douglas~L Maskell, and Bertil Schmidt.
\newblock {CUDASW++: Optimizing Smith-Waterman Sequence Database Searches for CUDA-enabled Graphics Processing Units}.
\newblock {\em BMC Research Notes}, 2009.

\bibitem{liu2010cudasw++}
Yongchao Liu, Bertil Schmidt, and Douglas~L Maskell.
\newblock {CUDASW++ 2.0: Enhanced Smith-Waterman Protein Database Search on CUDA-enabled GPUs Based on SIMT and Virtualized SIMD Abstractions}.
\newblock {\em BMC Research Notes}, 2010.

\bibitem{wilton2015arioc}
Richard Wilton, Tamas Budavari, Ben Langmead, Sarah~J Wheelan, Steven~L Salzberg, and Alexander~S Szalay.
\newblock {Arioc: High-throughput Read Alignment with GPU-accelerated Exploration of The Seed-and-extend Search Space}.
\newblock {\em PeerJ}, 2015.

\bibitem{goyal2017ultra}
Amit Goyal, Hyuk~Jung Kwon, Kichan Lee, Reena Garg, Seon~Young Yun, Yoon~Hee Kim, Sunghoon Lee, and Min~Seob Lee.
\newblock {Ultra-fast Next Generation Human Genome Sequencing Data Processing Using DRAGENTM Bio-IT Processor for Precision Medicine}.
\newblock {\em Open Journal of Genetics}, 2017.

\bibitem{chen2016spark}
Yu-Ting Chen, Jason Cong, Zhenman Fang, Jie Lei, and Peng Wei.
\newblock {When Spark Meets FPGAs: A Case Study for Next-Generation DNA Sequencing Acceleration}.
\newblock In {\em USENIX HotCloud}, 2016.

\bibitem{chen2014accelerating}
Peng Chen, Chao Wang, Xi~Li, and Xuehai Zhou.
\newblock {Accelerating the Next Generation Long Read Mapping with the FPGA-based System}.
\newblock {\em IEEE/ACM TCBB}, 2014.

\bibitem{chen2021high}
Yen-Lung Chen, Bo-Yi Chang, Chia-Hsiang Yang, and Tzi-Dar Chiueh.
\newblock {A High-Throughput FPGA Accelerator for Short-Read Mapping of the Whole Human Genome}.
\newblock {\em IEEE TPDS}, 2021.

\bibitem{fujiki2020seedex}
Daichi Fujiki, Shunhao Wu, Nathan Ozog, Kush Goliya, David Blaauw, Satish Narayanasamy, and Reetuparna Das.
\newblock {SeedEx: A Genome Sequencing Accelerator for Optimal Alignments in Subminimal Space}.
\newblock In {\em MICRO}, 2020.

\bibitem{banerjee2018asap}
Subho~Sankar Banerjee, Mohamed El-Hadedy, Jong~Bin Lim, Zbigniew~T Kalbarczyk, Deming Chen, Steven~S Lumetta, and Ravishankar~K Iyer.
\newblock {ASAP: Accelerated Short-read Alignment on Programmable Hardware}.
\newblock {\em IEEE TC}, 2019.

\bibitem{fei2018fpgasw}
Xia Fei, Zou Dan, Lu~Lina, Man Xin, and Zhang Chunlei.
\newblock {FPGASW: Accelerating Large-scale Smith--Waterman Sequence Alignment Application with Backtracking on FPGA Linear Systolic Array}.
\newblock {\em Interdisciplinary Sciences: Computational Life Sciences}, 2018.

\bibitem{waidyasooriya2015hardware}
Hasitha~Muthumala Waidyasooriya and Masanori Hariyama.
\newblock {Hardware-acceleration of Short-read Alignment Based on the Burrows-Wheeler Transform}.
\newblock {\em IEEE TPDS}, 2015.

\bibitem{chen2015novel}
Yu-Ting Chen, Jason Cong, Jie Lei, and Peng Wei.
\newblock {A Novel High-throughput Acceleration Engine for Read Alignment}.
\newblock In {\em FCCM}, 2015.

\bibitem{rucci2018swifold}
Enzo Rucci, Carlos Garcia, Guillermo Botella, Armando De~Giusti, Marcelo Naiouf, and Manuel Prieto-Matias.
\newblock {SWIFOLD: Smith-Waterman Implementation on FPGA with OpenCL for Long DNA Sequences}.
\newblock {\em BMC Systems Biology}, 2018.

\bibitem{haghi2021fpga}
Abbas Haghi, Santiago Marco-Sola, Lluc Alvarez, Dionysios Diamantopoulos, Christoph Hagleitner, and Miquel Moreto.
\newblock {An FPGA Accelerator of the Wavefront Algorithm for Genomics Pairwise Alignment}.
\newblock In {\em FPL}, 2021.

\bibitem{li2021pipebsw}
Luyi Li, Jun Lin, and Zhongfeng Wang.
\newblock {PipeBSW: A Two-Stage Pipeline Structure for Banded Smith-Waterman Algorithm on FPGA}.
\newblock In {\em ISVLSI}, 2021.

\bibitem{ham2020genesis}
Tae~Jun Ham, David Bruns-Smith, Brendan Sweeney, Yejin Lee, Seong~Hoon Seo, U~Gyeong Song, Young~H Oh, Krste Asanovic, Jae~W Lee, and Lisa~Wu Wills.
\newblock {Genesis: A Hardware Acceleration Framework for Genomic Data Analysis}.
\newblock In {\em ISCA}, 2020.

\bibitem{ham2021accelerating}
Tae~Jun Ham, Yejin Lee, Seong~Hoon Seo, U~Gyeong Song, Jae~W Lee, David Bruns-Smith, Brendan Sweeney, Krste Asanovic, Young~H Oh, and Lisa~Wu Wills.
\newblock {Accelerating Genomic Data Analytics With Composable Hardware Acceleration Framework}.
\newblock {\em IEEE Micro}, 2021.

\bibitem{wu2019fpga}
Lisa Wu, David Bruns-Smith, Frank~A. Nothaft, Qijing Huang, Sagar Karandikar, Johnny Le, Andrew Lin, Howard Mao, Brendan Sweeney, Krste Asanović, David~A. Patterson, and Anthony~D. Joseph.
\newblock {FPGA Accelerated Indel Realignment in the Cloud}.
\newblock In {\em HPCA}, 2019.

\bibitem{cali2020genasm}
Damla~Senol Cali, Gurpreet~S. Kalsi, Zülal Bingöl, Can Firtina, Lavanya Subramanian, Jeremie~S. Kim, Rachata Ausavarungnirun, Mohammed Alser, Juan Gomez-Luna, Amirali Boroumand, Anant Norion, Allison Scibisz, Sreenivas Subramoneyon, Can Alkan, Saugata Ghose, and Onur Mutlu.
\newblock {GenASM: A High-Performance, Low-Power Approximate String Matching Acceleration Framework for Genome Sequence Analysis}.
\newblock In {\em MICRO}, 2020.

\bibitem{Zhang_2023_alignerD}
Fan Zhang, Shaahin Angizi, Jiao Sun, Wei Zhang, and Deliang Fan.
\newblock {Aligner-D: Leveraging In-DRAM Computing to Accelerate DNA Short Read Alignment}.
\newblock {\em IEEE Journal on Emerging and Selected Topics in Circuits and Systems}, 2023.

\bibitem{soysal2025mars}
Melina Soysal, Konstantina Koliogeorgi, Can Firtina, Nika~Mansouri Ghiasi, Rakesh Nadig, Haiyu Mao, Geraldo~Francisco de~Oliveira~Junior, Yu~Liang, Klea Zambaku, Mohammad Sadrosadati, and Onur Mutlu.
\newblock {MARS: Processing-In-Memory Acceleration of Raw Signal Genome Analysis Inside the Storage Subsystem}.
\newblock In {\em ICS}, 2025.

\bibitem{kim2018grim}
Jeremie~S Kim, Damla~Senol Cali, Hongyi Xin, Donghyuk Lee, Saugata Ghose, Mohammed Alser, Hasan Hassan, Oguz Ergin, Can Alkan, and Onur Mutlu.
\newblock {GRIM-Filter: Fast Seed Location Filtering in DNA Read Mapping Using Processing-in-memory Technologies}.
\newblock {\em BMC Genomics}, 2018.

\bibitem{kaplan2020bioseal}
Roman Kaplan, Leonid Yavits, and Ran Ginosasr.
\newblock {BioSEAL: In-memory biological sequence alignment accelerator for large-scale genomic data}.
\newblock In {\em SYSTOR}, 2020.

\bibitem{mao2022genpip}
Haiyu Mao, Mohammed Alser, Mohammad Sadrosadati, Can Firtina, Akanksha Baranwal, Damla~Senol Cali, Aditya Manglik, Nour~Almadhoun Alserr, and Onur Mutlu.
\newblock {GenPIP: In-Memory Acceleration of Genome Analysis via Tight Integration of Basecalling and Read Mapping}.
\newblock In {\em MICRO}, 2022.

\bibitem{dphls2026}
Yingqi Cao, Anshu Guta, Jason Liang, and Yatish Turakhia.
\newblock {DP-HLS: A High-Level Synthesis Framework for Accelerating Dynamic Programming Algorithms in Bioinformatics}.
\newblock In {\em HPCA}, 2026.

\bibitem{wang20202}
Zhehong Wang, Tianjun Zhang, Daichi Fujiki, Arun Subramaniyan, Xiao Wu, Makoto Yasuda, Satoru Miyoshi, Masaru Kawaminami, Reetuparna Das, Satish Narayanasamy, and David Blaauw.
\newblock {A 2.46M Reads/s Seed-Extension Accelerator for Next-Generation Sequencing Using a String-Independent PE Array}.
\newblock {\em IEEE Journal of Solid-State Circuits}, 2021.

\bibitem{Walia2024talco}
Sumit Walia, Cheng Ye, Arkid Bera, Dhruvi Lodhavia, and Yatish Turakhia.
\newblock {TALCO: Tiling Genome Sequence Alignment Using Convergence of Traceback Pointers}.
\newblock In {\em HPCA}, 2024.

\bibitem{sadasivan2024genomic}
Harisankar Sadasivan, Artur Klauser, Juergen Hench, Yatish Turakhia, Gagandeep Singh, Alberto Zeni, Sarah Beecroft, Satish Narayanasamy, Jeff Nivala, Bob Robey, Onur Mutlu, Kristof Denolf, and Sriranjani Sitaraman.
\newblock {The Genomic Computing Revolution: Defining the Next Decades of Accelerating Genomics}.
\newblock In {\em HPEC}, 2024.

\bibitem{Turakhia2025toward}
Yatish Turakhia.
\newblock {Toward a Generalized Accelerator for Genome Sequence Analysis}.
\newblock {\em Communications of the ACM}, 2025.

\bibitem{Turakhia2019darwinwga}
Yatish Turakhia, Sneha~D. Goenka, Gill Bejerano, and WIlliam~J. Dally.
\newblock {Darwin-WGA: A Co-processor Provides Increased Sensitivity in Whole Genome Alignments with High Speedup}.
\newblock In {\em HPCA}, 2019.

\bibitem{simon2026processing}
William~Andrew Simon, Leonid Yavits, Konstantina Koliogeorgi, Yann Falevoz, Yoshihiro Shibuya, Dominique Lavenier, Irem Boybat, Klea Zambaku, Berkan Şahin, Mohammad Sadrosadati, Onur Mutlu, Abu Sebastian, Rayan Chikhi, {The BioPIM Consortium}, and Can Alkan.
\newblock {Processing-in-Memory for Genomics Workloads}.
\newblock {\em IEEE Micro}, 2026.

\bibitem{simon2025processing}
William~Andrew Simon, Leonid Yavits, Konstantina Koliogeorgi, Yann Falevoz, Yoshihiro Shibuya, Dominique Lavenier, Irem Boybat, Klea Zambaku, Berkan {\c{S}}ahin, Mohammad Sadrosadati, Onur Mutlu, Abu Sebastian, Rayan Chikhi, {The BioPIM Consortium}, and Can Alkan.
\newblock {Processing-in-memory for genomics workloads}.
\newblock arxiv, 2025.

\bibitem{diab_framework_2022}
Safaa Diab, Amir Nassereldine, Mohammed Alser, Juan Gómez-Luna, Onur Mutlu, and Izzat~El Hajj.
\newblock A framework for high-throughput sequence alignment using real processing-in-memory systems.
\newblock {\em arXiv}, 2022.

\bibitem{firtina2025enabling}
Can Firtina.
\newblock Enabling fast, accurate, and efficient real-time genome analysis via new algorithms and techniques.
\newblock {\em arXiv preprint arXiv:2503.02997}, 2025.

\bibitem{senol2021accelerating}
Damla Senol.
\newblock {\em {Accelerating Genome Sequence Analysis via Efficient Hardware/Algorithm Co-Design}}.
\newblock PhD thesis, Carnegie Mellon University, 2021.

\bibitem{koliogeorgi2023hardware}
Konstantina Koliogeorgi.
\newblock {\em Hardware acceleration techniques for computation and data intensive machine learning and bioinformatic applications}.
\newblock PhD thesis, National Technical University of Athens, 2023.

\bibitem{Lindegger2023scrooge}
Joël Lindegger, Damla Senol~Cali, Mohammed Alser, Juan Gómez-Luna, Nika~Mansouri Ghiasi, and Onur Mutlu.
\newblock {Scrooge: a fast and memory-frugal genomic sequence aligner for CPUs, GPUs, and ASICs}.
\newblock {\em Bioinformatics}, 2023.

\bibitem{eudine2026genpairx}
Julien Eudine, Chu Li, Zhuo Cheng, Renzo Andri, Can Firtina, Mohammad Sadrosadati, Nika~Mansouri Ghiasi, Konstantina Koliogeorgi, Anirban Nag, Arash Tavakkol, Haiyu Mao, Onur Mutlu, Shai Bergman, and Ji~Zhang.
\newblock {GenPairX: A Hardware-Algorithm Co-Designed Accelerator for Paired-End Read Mapping}.
\newblock In {\em HPCA}, 2026.

\bibitem{Pavon2024quetzal}
Julian Pavon, Ivan~Vargas Valdivieso, Carlos Rojas, Cesar Hernandez, Mehmet Aslan, Roger Figueras, Yichao Yuan, Joël Lindegger, Mohammed Alser, Francesc Moll, Santiago Marco-Sola, Oguz Ergin, Nishil Talati, Onur Mutlu, Osman Unsal, Mateo Valero, and Adrian Cristal.
\newblock {QUETZAL: Vector Acceleration Framework for Modern Genome Sequence Analysis Algorithms}.
\newblock In {\em ISCA}, 2024.

\bibitem{alonso2024bimsa}
Alejandro Alonso-Mar{\'\i}n, Ivan Fernandez, Quim Aguado-Puig, Juan G{\'o}mez-Luna, Santiago Marco-Sola, and Miquel Moreto.
\newblock {BIMSA: Accelerating Long Sequence Alignment Using Processing-In-Memory}.
\newblock {\em bioRxiv}, 2024.

\bibitem{firtina2024aphmm}
Can Firtina, Kamlesh Pillai, Gurpreet~S. Kalsi, Bharathwaj Suresh, Damla~Senol Cali, Jeremie~S. Kim, Taha Shahroodi, Meryem~Banu Cavlak, Jo\"{e}l Lindegger, Mohammed Alser, Juan~G\'{o}mez Luna, Sreenivas Subramoney, and Onur Mutlu.
\newblock {ApHMM: Accelerating Profile Hidden Markov Models for Fast and Energy-efficient Genome Analysis}.
\newblock {\em TACO}, 2024.

\bibitem{Doblas2023gmx}
Max Doblas, Oscar Lostes-Cazorla, Quim Aguado-Puig, Nick Cebry, Pau Fontova-Musté, Christopher~Frances Batten, Santiago Marco-Sola, and Miquel Moretó.
\newblock {GMX: Instruction Set Extensions for Fast, Scalable, and Efficient Genome Sequence Alignment}.
\newblock In {\em MICRO}, 2023.

\bibitem{kang2026lembas}
Seongyoung Kang, Se-Min Lim, and Sang-Woo Jun.
\newblock {Lembas: Cost-Efficient Genome Alignment with External Memory and FPGA Acceleration}.
\newblock In {\em ISCA}, 2026.

\bibitem{cali2022segram}
Damla~Senol Cali, Konstantinos Kanellopoulos, Jo{\"e}l Lindegger, Z{\"u}lal Bing{\"o}l, Gurpreet~S. Kalsi, Ziyi Zuo, Can Firtina, Meryem~Banu Cavlak, Jeremie Kim, Nika~Mansouri Ghiasi, Gagandeep Singh, Juan G{\'o}mez-Luna, Nour~Almadhoun Alserr, Mohammed Alser, Sreenivas Subramoney, Can Alkan, Saugata Ghose, and Onur Mutlu.
\newblock {SeGraM: A Universal Hardware Accelerator for Genomic Sequence-to-Graph and Sequence-to-Sequence Mapping}.
\newblock In {\em ISCA}, 2022.

\bibitem{Zhang2024Harp}
Yichi Zhang, Dibei Chen, Gang Zeng, Jianfeng Zhu, Zhaoshi Li, Longlong Chen, Shaojun Wei, and Leibo Liu.
\newblock {Harp: Leveraging Quasi-Sequential Characteristics to Accelerate Sequence-to-Graph Mapping of Long Reads}.
\newblock In {\em ASPLOS}, 2024.

\bibitem{Zeng2024asgdp}
Gang Zeng, Jianfeng Zhu, Yichi Zhang, Ganhui Chen, Zhenhai Yuan, Shaojun Wei, and Leibo Liu.
\newblock {A High-Performance Genomic Accelerator for Accurate Sequence-to-Graph Alignment Using Dynamic Programming Algorithm}.
\newblock {\em IEEE TPDS}, 2024.

\bibitem{Shen2024128parallel}
Zhe-Wei Shen, Jheng-Syun Huang, and Yi-Chang Lu.
\newblock {A Memory-Efficient Accelerator for 128-Parallel Sequence-to-Graph Alignment in Variant-Enriched Regions}.
\newblock In {\em BioCAS}, 2024.

\bibitem{Li2024}
Wen-Jun Li, Bhagwan~Narayan Rekadwad, Jian-Yu Jiao, and Nimaichand Salam.
\newblock {Exploring Microbial Dark Matter and the Status of Bacterial and Archaeal Taxonomy: Challenges and Opportunities in the Future}.
\newblock In {\em Modern Taxonomy of Bacteria and Archaea: New Methods, Technology and Advances}. 2024.

\bibitem{Mandal2020}
Kalikinkar Mandal, Bo~Yang, Guang Gong, and Mark Aagaard.
\newblock {Analysis and Efficient Implementations of a Class of Composited de Bruijn Sequences}.
\newblock {\em IEEE TC}, 2020.

\bibitem{Varma2013}
B.~Sharat~Chandra Varma, Kolin Paul, M.~Balakrishnan, and Dominique Lavenier.
\newblock {FAssem: FPGA Based Acceleration of De Novo Genome Assembly}.
\newblock In {\em FCCM}, 2013.

\bibitem{Awan2021}
Muaaz~Gul Awan, Steven Hofmeyr, Rob Egan, Nan Ding, Aydin Buluc, Jack Deslippe, Leonid Oliker, and Katherine Yelick.
\newblock {Accelerating large scale de novo metagenome assembly using GPUs}.
\newblock In {\em SC}, 2021.

\bibitem{Feng2021}
Zonghao Feng and Qiong Luo.
\newblock {Accelerating Sequence-to-Graph Alignment on Heterogeneous Processors}.
\newblock In {\em ICPP}, 2021.

\bibitem{Zhang2025}
Yichi Zhang, Jianfeng Zhu, Liangwei Li, Gang Zeng, Dibei Chen, Tairan Zhang, Yeyang Deng, Zhicheng Gong, Aoyang Zhang, Yang Liu, Shaojun Wei, and Leibo Liu.
\newblock {A 28-nm 239-bp/$\mu$J Agile Pangenome Analysis Accelerator for Multi-Scheme Read Mapping}.
\newblock {\em IEEE Journal of Solid-State Circuits}, 2025.

\bibitem{kim2025nmp}
Heewoo Kim, Sanjay Sri~Vallabh Singapuram, Haojie Ye, Joseph Izraelevitz, Trevor Mudge, Ronald Dreslinski, and Nishil Talati.
\newblock {NMP-PaK: Near-Memory Processing Acceleration of Scalable De Novo Genome Assembly}.
\newblock In {\em ISCA}, 2025.

\bibitem{Huang2023meg2}
Yu~Huang, Long Zheng, Haifeng Liu, Zhuoran Zhou, Dan Chen, Pengcheng Yao, Qinggang Wang, Xiaofei Liao, and Hai Jin.
\newblock {MeG2: In-Memory Acceleration for Genome Graphs Analysis}.
\newblock In {\em DAC}, 2023.

\bibitem{Angizi2020Panda}
Shaahin Angizi, Naima~Ahmed Fahmi, Deniz Najafi, Wei Zhang, and Deliang Fan.
\newblock {PANDA: Processing in Magnetic Random-Access Memory-Accelerated de Bruijn Graph-Based DNA Assembly}.
\newblock {\em Journal of Low Power Electronics and Applications}, 2024.

\bibitem{Qiu2017}
Shuang Qiu and Qiong Luo.
\newblock {Parallelizing Big De Bruijn Graph Construction on Heterogeneous Processors}.
\newblock In {\em ICDCS}, 2017.

\bibitem{Zhou2021}
Minxuan Zhou, Lingxi Wu, Muzhou Li, Niema Moshiri, Kevin Skadron, and Tajana Rosing.
\newblock {Ultra Efficient Acceleration for De Novo Genome Assembly via Near-Memory Computing}.
\newblock In {\em PACT}, 2021.

\bibitem{Sarkar2021}
Aritra Sarkar, Zaid Al-Ars, and Koen Bertels.
\newblock {QuASeR: Quantum Accelerated de novo DNA sequence reconstruction}.
\newblock {\em PLOS ONE}, 2021.

\bibitem{Varma2017}
B.~Sharat~Chandra Varma, Kolin Paul, M.~Balakrishnan, and Dominique Lavenier.
\newblock Hardware acceleration of de novo genome assembly.
\newblock {\em International Journal of Embedded Systems}, 2017.

\bibitem{Varma2016}
B.~Sharat~Chandra Varma, Kolin Paul, and M.~Balakrishnan.
\newblock {FPGA-Based Acceleration of De Novo Genome Assembly}.
\newblock In {\em Architecture Exploration of FPGA Based Accelerators for BioInformatics Applications}. 2016.

\bibitem{Goswami2018}
Sayan Goswami, Kisung Lee, Shayan Shams, and Seung-Jong Park.
\newblock {GPU-Accelerated Large-Scale Genome Assembly}.
\newblock In {\em IPDPS}, 2018.

\bibitem{Galanos2021}
Georgios Galanos, Pavlos Malakonakis, and Apostolos Dollas.
\newblock {An FPGA-Based Data Pre-Processing Architecture to Accelerate De-Novo Genome Assembly}.
\newblock In {\em BIBE}, 2021.

\bibitem{Angizi2020}
Shaahin Angizi, Naima~Ahmed Fahmi, Wei Zhang, and Deliang Fan.
\newblock {PIM-Assembler: A Processing-in-Memory Platform for Genome Assembly}.
\newblock In {\em DAC}, 2020.

\bibitem{Sinha2022}
Aman Sinha, Huei-Chun Yang, Pei-Yi Liu, Yen-Shi Kuo, Yuhao Fang, Tien-Shuo Chang, Ke-Han Li, and Bo-Cheng Lai.
\newblock {DSIM: Distributed Sequence Matching on Near-DRAM Accelerator for Genome Assembly}.
\newblock {\em IEEE Journal on Emerging and Selected Topics in Circuits and Systems}, 2022.

\bibitem{Meng2014}
Pingfan Meng, Matthew Jacobsen, Motoki Kimura, Vladimir Dergachev, Thomas Anantharaman, Michael Requa, and Ryan Kastner.
\newblock Hardware accelerated novel optical de novo assembly for large-scale genomes.
\newblock In {\em FPL}, 2014.

\bibitem{Hu2016}
Yuanqi Hu and Pantelis Georgiou.
\newblock {A Real-Time de novo DNA Sequencing Assembly Platform Based on an FPGA Implementation}.
\newblock {\em IEEE/ACM TCBB}, 2016.

\bibitem{Chen2023}
Yibo Chen, Jun-Han Huang, Yuhui Sun, Yong Zhang, Yuxiang Li, and Xun Xu.
\newblock {VRP Assembler: haplotype-resolvedde novoassembly of diploid and polyploid genomes using quantum computing}.
\newblock {\em Cell Reports Methods}, 2023.

\bibitem{Natarajan2018}
Santhi Natarajan, N.~KrishnaKumar, H.~V. Anuchan, Debnath Pal, and S.~K. Nandy.
\newblock {ReneGENE-Novo: Co-designed Algorithm-Architecture for Accelerated Preprocessing and Assembly of Genomic Short Reads}.
\newblock In {\em ARC}, 2018.

\bibitem{Ren2018}
Shanshan Ren, Nauman Ahmed, Koen Bertels, and Zaid Al-Ars.
\newblock {An Efficient GPU-Based de Bruijn Graph Construction Algorithm for Micro-Assembly}.
\newblock In {\em BIBE}, 2018.

\bibitem{li2024rapid}
Jiajie Li, Jan-Niklas Schmelzle, Yixiao Du, Simon Heumos, Andrea Guarracino, Giulia Guidi, Pjotr Prins, Erik Garrison, and Zhiru Zhang.
\newblock Rapid gpu-based pangenome graph layout.
\newblock In {\em SC24: International Conference for High Performance Computing, Networking, Storage and Analysis}, page 1–19. IEEE, November 2024.

\bibitem{alser2020technology}
Mohammed Alser, Jeremy Rotman, Dhrithi Deshpande, Kodi Taraszka, Huwenbo Shi, Pelin~Icer Baykal, Harry~Taegyun Yang, Victor Xue, Sergey Knyazev, Benjamin~D. Singer, Brunilda Balliu, David Koslicki, Pavel Skums, Alex Zelikovsky, Can Alkan, Onur Mutlu, and Serghei Mangul.
\newblock {Technology Dictates Algorithms: Recent Developments in Read Alignment}.
\newblock {\em Genome Biology}, 2021.

\bibitem{alser2020accelerating}
Mohammed Alser, Z{\"u}lal Bing{\"o}l, Damla~Senol Cali, Jeremie Kim, Saugata Ghose, Can Alkan, and Onur Mutlu.
\newblock {Accelerating Genome Analysis: A Primer on an Ongoing Journey}.
\newblock {\em IEEE Micro}, 2020.

\bibitem{singh2021fpga}
Gagandeep Singh, Mohammed Alser, Damla~Senol Cali, Dionysios Diamantopoulos, Juan G{\'o}mez-Luna, Henk Corporaal, and Onur Mutlu.
\newblock {FPGA-Based Near-Memory Acceleration of Modern Data-Intensive Applications}.
\newblock {\em IEEE Micro}, 2021.

\bibitem{nag2019gencache}
Anirban Nag, CN~Ramachandra, Rajeev Balasubramonian, Ryan Stutsman, Edouard Giacomin, Hari Kambalasubramanyam, and Pierre-Emmanuel Gaillardon.
\newblock {GenCache: Leveraging In-cache Operators for Efficient Sequence Alignment}.
\newblock In {\em MICRO}, 2019.

\bibitem{kim20111}
Jung-Sik Kim, Chi~Sung Oh, Hocheol Lee, Donghyuk Lee, Hyong-Ryol Hwang, Sooman Hwang, Byongwook Na, Joungwook Moon, Jin-Guk Kim, and Hanna Park.
\newblock {A 1.2 V 12.8 GB/s 2 Gb Mobile Wide-I/O DRAM With 4x 128 I/Os Using TSV Based Stacking}.
\newblock In {\em ISSCC}, 2011.

\bibitem{alser2017gatekeeper}
Mohammed Alser, Hasan Hassan, Hongyi Xin, O{\u{g}}uz Ergin, Onur Mutlu, and Can Alkan.
\newblock {GateKeeper: A New Hardware Architecture for Accelerating Pre-alignment in DNA Short Read Mapping}.
\newblock {\em Bioinformatics}, 2017.

\bibitem{alser2017magnet}
Mohammed Alser, Onur Mutlu, and Can Alkan.
\newblock {MAGNET: Understanding and Improving the Accuracy of Genome Pre-alignment Filtering}.
\newblock {\em IPSI TIR}, 2017.

\bibitem{alser2019shouji}
Mohammed Alser, Hasan Hassan, Akash Kumar, Onur Mutlu, and Can Alkan.
\newblock {Shouji: A Fast and Efficient Pre-alignment Filter for Sequence Alignment}.
\newblock {\em Bioinformatics}, 2019.

\bibitem{alser2020sneakysnake}
Mohammed Alser, Taha Shahroodi, Juan G{\'o}mez-Luna, Can Alkan, and Onur Mutlu.
\newblock {SneakySnake: A Fast and Accurate Universal Genome Pre-alignment Filter for CPUs, GPUs and FPGAs}.
\newblock {\em Bioinformatics}, 2020.

\bibitem{bingol2021gatekeeper}
Z{\"u}lal Bing{\"o}l, Mohammed Alser, Onur Mutlu, Ozcan Ozturk, and Can Alkan.
\newblock {GateKeeper-GPU: Fast and Accurate Pre-Alignment Filtering in Short Read Mapping}.
\newblock In {\em IPDPSW}, 2021.

\bibitem{hameed2021alpha}
Fazal Hameed, Asif~Ali Khan, and Jeronimo Castrillon.
\newblock {ALPHA: A Novel Algorithm-Hardware Co-design for Accelerating DNA Seed Location Filtering}.
\newblock {\em ITETC}, 2021.

\bibitem{guo2019hardware}
Licheng Guo, Jason Lau, Zhenyuan Ruan, Peng Wei, and Jason Cong.
\newblock {Hardware Acceleration of Long Read Pairwise Overlapping in Genome Sequencing: A Race between FPGA and GPU}.
\newblock In {\em FCCM}, 2019.

\bibitem{koslicki2016metapalette}
David Koslicki and Daniel Falush.
\newblock {MetaPalette: a k-mer Painting Approach for Metagenomic Taxonomic Profiling and Quantification of Novel Strain Variation}.
\newblock {\em mSystems}, 2016.

\bibitem{Marcelino2020}
Vanessa~R Marcelino, Philip~TLC Clausen, Jan~P Buchmann, Michelle Wille, Jonathan~R Iredell, Wieland Meyer, Ole Lund, Tania~C Sorrell, and Edward~C Holmes.
\newblock {CCMetagen: comprehensive and accurate identification of eukaryotes and prokaryotes in metagenomic data}.
\newblock {\em Genome Biology}, 2020.

\bibitem{piro2016dudes}
Vitor~C Piro, Martin~S Lindner, and Bernhard~Y Renard.
\newblock {DUDes: a top-down taxonomic profiler for metagenomics}.
\newblock {\em Bioinformatics}, 2016.

\bibitem{piro2020ganon}
Vitor~C Piro, Temesgen~H Dadi, Enrico Seiler, Knut Reinert, and Bernhard~Y Renard.
\newblock ganon: precise metagenomics classification against large and up-to-date sets of reference sequences.
\newblock {\em Bioinformatics}, 2020.

\bibitem{pockrandt2022metagenomic}
Christopher Pockrandt, Aleksey~V. Zimin, and Steven~L. Salzberg.
\newblock {Metagenomic classification with KrakenUniq on low-memory computers}.
\newblock {\em Journal of Open Source Software}, 2022.

\bibitem{wood2014kraken}
Derrick~E Wood and Steven~L Salzberg.
\newblock Kraken: ultrafast metagenomic sequence classification using exact alignments.
\newblock {\em Genome Biology}, 2014.

\bibitem{kim2016centrifuge}
Daehwan Kim, Li~Song, Florian~P Breitwieser, and Steven~L Salzberg.
\newblock {Centrifuge: Rapid and Sensitive Classification of Metagenomic Sequences}.
\newblock {\em Genome Research}, 2016.

\bibitem{wood2019improved}
Derrick~E Wood, Jennifer Lu, and Ben Langmead.
\newblock {Improved Metagenomic Analysis with Kraken 2}.
\newblock {\em Genome Biology}, 2019.

\bibitem{muller2017metacache}
Andr{\'e} M{\"u}ller, Christian Hundt, Andreas Hildebrandt, Thomas Hankeln, and Bertil Schmidt.
\newblock {MetaCache: context-aware classification of metagenomic reads using minhashing}.
\newblock {\em Bioinformatics}, 2017.

\bibitem{song2024centrifuger}
Li~Song and Ben Langmead.
\newblock {Centrifuger: lossless compression of microbial genomes for efficient and accurate metagenomic sequence classification}.
\newblock {\em Genome Biology}, 2024.

\bibitem{Dilthey2019}
Alexander~T Dilthey, Chirag Jain, Sergey Koren, and Adam~M Phillippy.
\newblock {Strain-level metagenomic assignment and compositional estimation for long reads with MetaMaps}.
\newblock {\em Nature Communications}, 2019.

\bibitem{Fan2021}
Jeremy Fan, Steven Huang, and Samuel~D Chorlton.
\newblock {BugSeq: a highly accurate cloud platform for long-read metagenomic analyses}.
\newblock {\em BMC Bioinformatics}, 2021.

\bibitem{jia2011metabing}
Peng Jia, Liming Xuan, Lei Liu, and Chaochun Wei.
\newblock {MetaBinG: Using GPUs to accelerate metagenomic sequence classification}.
\newblock {\em PLOS One}, 2011.

\bibitem{kobus2021metacache}
Robin Kobus, Andr{\'e} M{\"u}ller, Daniel J{\"u}nger, Christian Hundt, and Bertil Schmidt.
\newblock {MetaCache-GPU: ultra-fast metagenomic classification}.
\newblock In {\em ICPP}, 2021.

\bibitem{wang2023gpmeta}
Xuebin Wang, Taifu Wang, Zhihao Xie, Youjin Zhang, Shiqiang Xia, Ruixue Sun, Xinqiu He, Ruizhi Xiang, Qiwen Zheng, Zhencheng Liu, Jin’An Wang, Honglong Wu, Xiangqian Jin, Weijun Chen, Dongfang Li, and Zengquan He.
\newblock {GPMeta: a GPU-accelerated method for ultrarapid pathogen identification from metagenomic sequences}.
\newblock {\em Briefings in Bioinformatics}, 2023.

\bibitem{kobus2017accelerating}
Robin Kobus, Christian Hundt, Andr{\'e} M{\"u}ller, and Bertil Schmidt.
\newblock {Accelerating Metagenomic Read Classification on CUDA-enabled GPUs}.
\newblock {\em BMC Bioinformatics}, 2017.

\bibitem{Su2012}
Xiaoquan Su, Jian Xu, and Kang Ning.
\newblock {Parallel-META: efficient metagenomic data analysis based on high-performance computation}.
\newblock {\em BMC Systems Biology}, 2012.

\bibitem{su2013gpumetastorms}
Xiaoquan Su, Xuetao Wang, Gongchao Jing, and Kang Ning.
\newblock {GPU-Meta-Storms: computing the structure similarities among massive amount of microbial community samples using GPU}.
\newblock {\em Bioinformatics}, 2014.

\bibitem{Yano2014}
Masahiro Yano, Hiroshi Mori, Yutaka Akiyama, Takuji Yamada, and Ken Kurokawa.
\newblock {CLAST: CUDA implemented large-scale alignment search tool}.
\newblock {\em BMC Bioinformatics}, 2014.

\bibitem{saavedra2020mining}
Antonio Saavedra, Hans Lehnert, Cecilia Hern{\'a}ndez, Gonzalo Carvajal, and Miguel Figueroa.
\newblock {Mining discriminative k-mers in DNA sequences using sketches and hardware acceleration}.
\newblock {\em IEEE Access}, 2020.

\bibitem{zhang2023genomix}
Tianqi Zhang, Antonio Gonz{\'a}lez, Niema Moshiri, Rob Knight, and Tajana Rosing.
\newblock {GenoMiX: Accelerated Simultaneous Analysis of Human Genomics, Microbiome Metagenomics, and Viral Sequences}.
\newblock In {\em BioCAS}, 2023.

\bibitem{cervi2022metagenomic}
Gustavo~Henrique Cervi, Cec{\'\i}lia~Dias Flores, and Claudia~Elizabeth Thompson.
\newblock {Metagenomic Analysis: A Pathway Toward Efficiency Using High-Performance Computing}.
\newblock In {\em ICICT}, 2022.

\bibitem{shahroodi2022krakenonmem}
Taha Shahroodi, Mahdi Zahedi, Abhairaj Singh, Stephan Wong, and Said Hamdioui.
\newblock {KrakenOnMem: a memristor-augmented HW/SW framework for taxonomic profiling}.
\newblock In {\em ICS}, 2022.

\bibitem{dashcam23micro}
Zuher Jahshan, Itay Merlin, Esteban Garzón, and Leonid Yavits.
\newblock {DASH-CAM: Dynamic Approximate SearcH Content Addressable Memory for genome classification}.
\newblock In {\em MICRO}, 2023.

\bibitem{hanhan2022edam}
Robert Hanhan, Esteban Garz{\'o}n, Zuher Jahshan, Adam Teman, Marco Lanuzza, and Leonid Yavits.
\newblock Edam: edit distance tolerant approximate matching content addressable memory.
\newblock In {\em ISCA}, 2022.

\bibitem{zou2022biohd}
Zhuowen Zou, Hanning Chen, Prathyush Poduval, Yeseong Kim, Mahdi Imani, Elaheh Sadredini, Rosario Cammarota, and Mohsen Imani.
\newblock {BioHD: an efficient genome sequence search platform using HyperDimensional memorization}.
\newblock In {\em ISCA}, 2022.

\bibitem{zhu2013high}
Zexuan Zhu, Yongpeng Zhang, Zhen Ji, Shan He, and Xiao Yang.
\newblock {High-throughput DNA sequence data compression}.
\newblock {\em Briefings in Bioinformatics}, 2013.

\bibitem{Deorowicz2013}
Sebastian Deorowicz and Szymon Grabowski.
\newblock Data compression for sequencing data.
\newblock {\em Algorithms for Molecular Biology}, 2013.

\bibitem{giancarlo2013compressive}
Raffaele Giancarlo, Simona~E. Rombo, and Filippo Utro.
\newblock Compressive biological sequence analysis and archival in the era of high-throughput sequencing technologies.
\newblock {\em {Briefings in Bioinformatics}}, 2013.

\bibitem{Betschart2025}
Raphael~O. Betschart, Felix Thal{\'e}n, Stefan Blankenberg, Martin Zoche, Tanja Zeller, and Andreas Ziegler.
\newblock A benchmark study of compression software for human short-read sequence data.
\newblock {\em Scientific Reports}, 2025.

\bibitem{Walia2026}
Sumit Walia, Harsh Motwani, Yu-Hsiang Tseng, Kyle Smith, Russell Corbett-Detig, and Yatish Turakhia.
\newblock Compressive pangenomics using mutation-annotated networks.
\newblock {\em Nature Genetics}, 2026.

\bibitem{chandak2018spring}
Shubham Chandak, Kedar Tatwawadi, Idoia Ochoa, Mikel Hernaez, and Tsachy Weissman.
\newblock {SPRING: a next-generation compressor for FASTQ data}.
\newblock {\em Bioinformatics}, 2018.

\bibitem{Deorowicz2020}
Sebastian Deorowicz.
\newblock {FQSqueezer: k-mer-based compression of sequencing data}.
\newblock {\em Scientific Reports}, 2020.

\bibitem{lan2021genozip}
Divon Lan, Ray Tobler, Yassine Souilmi, and Bastien Llamas.
\newblock {Genozip: a universal extensible genomic data compressor}.
\newblock {\em Bioinformatics}, 2021.

\bibitem{alyami2019lfastqc}
Sultan Al~Yami and Chun-Hsi Huang.
\newblock {LFastqC: A lossless non-reference-based FASTQ compressor}.
\newblock {\em PLOS One}, 2019.

\bibitem{kowalski2019pgrc}
Tomasz~M Kowalski and Szymon Grabowski.
\newblock {PgRC: pseudogenome-based read compressor}.
\newblock {\em Bioinformatics}, 2019.

\bibitem{roguski2018fastore}
{\L{}}ukasz Roguski, Idoia Ochoa, Mikel Hernaez, and Sebastian Deorowicz.
\newblock {FaStore: a space-saving solution for raw sequencing data}.
\newblock {\em Bioinformatics}, 2018.

\bibitem{chandak2017compression}
Shubham Chandak, Kedar Tatwawadi, and Tsachy Weissman.
\newblock {Compression of genomic sequencing reads via hash-based reordering: algorithm and analysis}.
\newblock {\em Bioinformatics}, 2017.

\bibitem{cogo2021genodedup}
Vinicius Cogo, João Paulo, and Alysson Bessani.
\newblock {GenoDedup: Similarity-Based Deduplication and Delta-Encoding for Genome Sequencing Data}.
\newblock {\em IEEE TC}, 2021.

\bibitem{Meng2023}
Qingxi Meng, Shubham Chandak, Yifan Zhu, and Tsachy Weissman.
\newblock Reference-free lossless compression of nanopore sequencing reads using an approximate assembly approach.
\newblock {\em Scientific Reports}, 2023.

\bibitem{kokot2022colord}
Marek Kokot, Adam Gudy{\'{s}}, Heng Li, and Sebastian Deorowicz.
\newblock {CoLoRd: compressing long reads}.
\newblock {\em Nature Methods}, 2022.

\bibitem{dufort2020enano}
Guillermo Dufort~y {\'A}lvarez, Gadiel Seroussi, Pablo Smircich, José Sotelo, Idoia Ochoa, and {\'A}lvaro Mart{\'i}n.
\newblock {ENANO: Encoder for NANOpore FASTQ files}.
\newblock {\em Bioinformatics}, 2020.

\bibitem{dufort2021renano}
{Dufort y \'{A}lvarez, Guillermo and Seroussi, Gadiel and Smircich, Pablo and Sotelo-Silveira, Jos\'{e} and Ochoa, Idoia and Martín, \'{A}lvaro}.
\newblock {RENANO: a REference-based compressor for NANOpore FASTQ files}.
\newblock {\em Bioinformatics}, 2021.

\bibitem{vandamme2024tinted}
L{\'e}a Vandamme, Bastien Cazaux, and Antoine Limasset.
\newblock {K2R: Tinted de Bruijn Graphs implementation for efficient read extraction from sequencing datasets}.
\newblock {\em Bioinformatics Advances}, 2025.

\bibitem{dragenora}
{Illumina}.
\newblock {DRAGEN ORA Compression and Decompression}.
\newblock \url{https://support-docs.illumina.com/SW/DRAGEN_v38/Content/SW/DRAGEN/ORA_Compression_fDG_swHS.htm}, 2021.

\bibitem{yang2025gpufastqlz}
Taolue Yang, Youyuan Liu, Bo~Jiang, and Sian Jin.
\newblock {GpuFastqLZ: An Ultra Fast Compression Methodology for Fastq Sequence Data on GPUs}.
\newblock In {\em SC-W}, 2024.

\bibitem{chen2023efficient}
Shifu Chen, Yaru Chen, Zhouyang Wang, Wenjian Qin, Jing Zhang, Heera Nand, Jishuai Zhang, Jun Li, Xiaoni Zhang, Xiaoming Liang, and Mingyan Xu.
\newblock {Efficient sequencing data compression and FPGA acceleration based on a two-step framework}.
\newblock {\em Frontiers in Genetics}, 2023.

\bibitem{hach2012scalce}
Faraz Hach, Ibrahim Numanagić, Can Alkan, and S~Cenk Sahinalp.
\newblock {SCALCE: boosting sequence compression algorithms using locally consistent encoding}.
\newblock {\em Bioinformatics}, 2012.

\bibitem{roguski2014dsrc2}
Łukasz Roguski and Sebastian Deorowicz.
\newblock {DSRC 2—Industry-oriented compression of FASTQ files}.
\newblock {\em Bioinformatics}, 2014.

\bibitem{grabowski2022mbgc}
Szymon Grabowski and Tomasz~M Kowalski.
\newblock {MBGC: Multiple Bacteria Genome Compressor}.
\newblock {\em GigaScience}, 2022.

\bibitem{kowalski2026mbgc2}
Tomasz~M Kowalski.
\newblock {MBGC2: Boosting compression via efficient encoding of approximate matches in genome collections}.
\newblock {\em GigaScience}, 2026.

\bibitem{grabowski2026ffc}
Szymon Grabowski, Tomasz~M Kowalski, and Robert Susik.
\newblock {FFC: a scalable FASTA compressor}.
\newblock {\em Bioinformatics}, 2026.

\bibitem{deorowicz2023agc}
Sebastian Deorowicz, Agnieszka Danek, and Heng Li.
\newblock {AGC: compact representation of assembled genomes with fast queries and updates}.
\newblock {\em Bioinformatics}, 2023.

\bibitem{Kryukov2022}
Kirill Kryukov, Lihua Jin, and So~Nakagawa.
\newblock {Efficient compression of SARS-CoV-2 genome data using Nucleotide Archival Format}.
\newblock {\em Patterns}, 2022.

\bibitem{sousa2024jarvis3}
Maria J~P Sousa, Armando~J Pinho, and Diogo Pratas.
\newblock {JARVIS3: an efficient encoder for genomic data}.
\newblock {\em Bioinformatics}, 2024.

\bibitem{Sun2023}
Hui Sun, Yingfeng Zheng, Haonan Xie, Huidong Ma, Xiaoguang Liu, and Gang Wang.
\newblock {PMFFRC: a large-scale genomic short reads compression optimizer via memory modeling and redundant clustering}.
\newblock {\em BMC Bioinformatics}, 2023.

\bibitem{Nazari2025}
Foad Nazari, Sneh Patel, Melissa LaRocca, Alina Sansevich, Ryan Czarny, Giana Schena, and Emma~K. Murray.
\newblock {Lossless and reference-free compression of FASTQ/A files using GeneSqueeze}.
\newblock {\em Scientific Reports}, 2025.

\bibitem{collet2018zstandard}
Y.~Collet and M.~Kucherawy.
\newblock {RFC 8878: Zstandard Compression and the 'application/zstd' Media Type}, 2021.

\bibitem{pavlov20167}
Igor Pavlov.
\newblock {7-Zip}.
\newblock \url{https://7-zip.org/}, 2016.

\bibitem{Brotli}
Jyrki Alakuijala, Andrea Farruggia, Paolo Ferragina, Eugene Kliuchnikov, Robert Obryk, Zoltan Szabadka, and Lode Vandevenne.
\newblock {Brotli: A General-Purpose Data Compressor}.
\newblock {\em ACM TOIS}, 2018.

\bibitem{Katz1991US5051745A}
Phillip~W. Katz.
\newblock String searcher, and compressor using same.
\newblock US Patent US5051745A, 1991.

\bibitem{goyal2021dzip}
Mohit Goyal, Kedar Tatwawadi, Shubham Chandak, and Idoia Ochoa.
\newblock {DZip: Improved general-purpose loss less compression based on novel neural network modeling}.
\newblock In {\em DCC}, 2021.

\bibitem{goyal2018deepzip}
Mohit Goyal, Kedar Tatwawadi, Shubham Chandak, and Idoia Ochoa.
\newblock {DeepZip: Lossless Data Compression Using Recurrent Neural Networks}.
\newblock In {\em DCC}, 2019.

\bibitem{chen2024ha}
Xiang Chen, Tao Lu, Jiapin Wang, Yu~Zhong, Guangchun Xie, Xueming Cao, Yuanpeng Ma, Bing Si, Feng Ding, Ying Yang, Yunxin Huang, Yafei Yang, You Zhou, and Fei Wu.
\newblock {HA-CSD: Host and SSD Coordinated Compression for Capacity and Performance}.
\newblock In {\em IPDPS}, 2024.

\bibitem{bartik2015lz4}
Mat{\v{e}}j Bart{\'\i}k, Sven Ubik, and Pavel Kubalik.
\newblock {LZ4 compression algorithm on FPGA}.
\newblock In {\em ICECS}, 2015.

\bibitem{liu2018data}
Weiqiang Liu, Faqiang Mei, Chenghua Wang, Maire O’Neill, and Earl~E Swartzlander.
\newblock {Data compression device based on modified LZ4 algorithm}.
\newblock {\em IEEE TCE}, 2018.

\bibitem{fowers2015scalable}
Jeremy Fowers, Joo-Young Kim, Doug Burger, and Scott Hauck.
\newblock {A scalable high-bandwidth architecture for lossless compression on FPGAs}.
\newblock In {\em FCCM}, 2015.

\bibitem{chen2021fpga}
Jianyu Chen, Maurice Daverveldt, and Zaid Al-Ars.
\newblock {FPGA acceleration of zstd compression algorithm}.
\newblock In {\em IPDPSW}, 2021.

\bibitem{angerd2022gbdi}
Alexandra Angerd, Angelos Arelakis, Vasilis Spiliopoulos, Erik Sintorn, and Per Stenstr{\"o}m.
\newblock {GBDI: Going beyond base-delta-immediate compression with global bases}.
\newblock In {\em HPCA}, 2022.

\bibitem{gao2024beezip}
Ruihao Gao, Zhichun Li, Guangming Tan, and Xueqi Li.
\newblock {BeeZip: Towards An Organized and Scalable Architecture for Data Compression}.
\newblock In {\em ASPLOS}, 2024.

\bibitem{karandikar2023cdpu}
Sagar Karandikar, Aniruddha~N. Udipi, Junsun Choi, Joonho Whangbo, Jerry Zhao, Svilen Kanev, Edwin Lim, Jyrki Alakuijala, Vrishab Madduri, Yakun~Sophia Shao, Borivoje Nikolic, Krste Asanovic, and Parthasarathy Ranganathan.
\newblock {CDPU: Co-designing compression and decompression processing units for hyperscale systems}.
\newblock In {\em ISCA}, 2023.

\bibitem{9499902}
Yifan Yang, Joel~S. Emer, and Daniel Sanchez.
\newblock {SpZip: Architectural Support for Effective Data Compression In Irregular Applications}.
\newblock In {\em ISCA}, 2021.

\bibitem{abali2020data}
Bulent Abali, Bart Blaner, John Reilly, Matthias Klein, Ashutosh Mishra, Craig~B. Agricola, Bedri Sendir, Alper Buyuktosunoglu, Christian Jacobi, William~J. Starke, Haren Myneni, and Charlie Wang.
\newblock {Data compression accelerator on IBM POWER9 and z15 processors: Industrial product}.
\newblock In {\em ISCA}, 2020.

\bibitem{ncbi2023}
Eric~W Sayers, Evan~E Bolton, J Rodney Brister, Kathi Canese, Jessica Chan, Donald C Comeau, Catherine M Farrell, Michael Feldgarden, Anna~M Fine, Kathryn Funk, Eneida Hatcher, Sivakumar Kannan, Christopher Kelly, Sunghwan Kim, William Klimke, Melissa J Landrum, Stacy Lathrop, Zhiyong Lu, Thomas L Madden, Adriana Malheiro, et~al.
\newblock {Database resources of the National Center for Biotechnology Information in 2023}.
\newblock {\em Nucleic Acids Research}, 2022.

\bibitem{shiryev2023indexing}
Sergey~A. Shiryev and Richa Agarwala.
\newblock Indexing and searching petabase-scale nucleotide resources.
\newblock {\em Nature Methods}, 2024.

\bibitem{pebblescout}
{National Center for Biotechnology Information}.
\newblock {Introducing Pebblescout: Index and Search Petabyte-Scale Sequence Resources Faster than Ever}.
\newblock \url{https://ncbiinsights.ncbi.nlm.nih.gov/2023/09/14/introducing-pebblescout/}, 2023.

\bibitem{lemane2023kmindex}
T{\'e}o Lemane, Nolan Lezzoche, Julien Lecubin, Eric Pelletier, Magali Lescot, Rayan Chikhi, and Pierre Peterlongo.
\newblock {Indexing and real-time user-friendly queries in terabyte-sized complex genomic datasets with kmindex and ORA}.
\newblock {\em Nature Computational Science}, 2024.

\bibitem{marchet2023scalable}
Camille Marchet and Antoine Limasset.
\newblock {Scalable sequence database search using partitioned aggregated Bloom comb trees}.
\newblock {\em Bioinformatics}, 2023.

\bibitem{ntdouble}
National~Center for Biotechnology~Information.
\newblock {Re-evaluating the BLAST Nucleotide Database (nt)}.
\newblock \url{https://ncbiinsights.ncbi.nlm.nih.gov/2022/11/17/re-evaluating-blast-nucleotide-nt/}, 2022.

\bibitem{Lynch2010}
Michael Lynch.
\newblock {Evolution of the mutation rate}.
\newblock {\em Trends in Genetics}, 2010.

\bibitem{Nasko2018}
Daniel~J. Nasko, Sergey Koren, Adam~M. Phillippy, and Todd~J. Treangen.
\newblock {RefSeq database growth influences the accuracy of k-mer-based lowest common ancestor species identification}.
\newblock {\em Genome Biology}, 2018.

\bibitem{o2016reference}
Nuala~A O'Leary, Mathew~W Wright, J~Rodney Brister, Stacy Ciufo, Diana Haddad, Rich McVeigh, Bhanu Rajput, Barbara Robbertse, Brian Smith-White, Danso Ako-Adjei, et~al.
\newblock {Reference Sequence (RefSeq) Database at NCBI: Current Status, Taxonomic Expansion, and Functional Annotation}.
\newblock {\em Nucleic Acids Research}, 2016.

\bibitem{jiao2020microbial}
Jian-Yu Jiao, Lan Liu, Zheng-Shuang Hua, Bao-Zhu Fang, En-Min Zhou, Nimaichand Salam, Brian~P Hedlund, and Wen-Jun Li.
\newblock {Microbial dark matter coming to light: challenges and opportunities}.
\newblock {\em National Science Review}, 2020.

\bibitem{rautiainen2023telomere}
Mikko Rautiainen, Sergey Nurk, Brian~P. Walenz, Glennis~A. Logsdon, David Porubsky, Arang Rhie, Evan~E. Eichler, Adam~M. Phillippy, and Sergey Koren.
\newblock {Telomere-to-telomere assembly of diploid chromosomes with Verkko}.
\newblock {\em Nature Biotechnology}, 2023.

\bibitem{jarvis2022semi}
Erich~D. Jarvis, Giulio Formenti, Arang Rhie, Andrea Guarracino, Chentao Yang, Jonathan Wood, Alan Tracey, Francoise Thibaud-Nissen, Mitchell~R. Vollger, David Porubsky, Haoyu Cheng, Mobin Asri, Glennis~A. Logsdon, Paolo Carnevali, Mark J.~P. Chaisson, Chen-Shan Chin, Sarah Cody, Joanna Collins, Peter Ebert, Merly Escalona, et~al.
\newblock Semi-automated assembly of high-quality diploid human reference genomes.
\newblock {\em Nature}, 2022.

\bibitem{chen2023gem}
Longlong Chen, Jianfeng Zhu, Guiqiang Peng, Mingxu Liu, Shaojun Wei, and Leibo Liu.
\newblock {GEM: Ultra-Efficient Near-Memory Reconfigurable Acceleration for Read Mapping by Dividing and Predictive Scattering}.
\newblock {\em IEEE TPDS}, 2023.

\bibitem{xin2016optimal}
Hongyi Xin, Sunny Nahar, Richard Zhu, John Emmons, Gennady Pekhimenko, Carl Kingsford, Can Alkan, and Onur Mutlu.
\newblock {Optimal Seed Solver: Optimizing Seed Selection in Read Mapping}.
\newblock {\em Bioinformatics}, 2016.

\bibitem{mansouri2022genstore}
Nika Mansouri~Ghiasi, Jisung Park, Harun Mustafa, Jeremie Kim, Ataberk Olgun, Arvid Gollwitzer, Damla Senol~Cali, Can Firtina, Haiyu Mao, Nour Almadhoun~Alserr, Rachata Ausavarungnirun, Nandita Vijaykumar, Mohammed Alser, and Onur Mutlu.
\newblock {GenStore: A High-Performance In-Storage Processing System for Genome Sequence Analysis}.
\newblock In {\em ASPLOS}, 2022.

\bibitem{angizi2020pim}
Shaahin Angizi, Jiao Sun, Wei Zhang, and Deliang Fan.
\newblock {PIM-Aligner: A processing-in-MRAM platform for biological sequence alignment}.
\newblock In {\em DATE}, 2020.

\bibitem{vsovsic2017edlib}
Martin {\v{S}}o{\v{s}}i{\'c} and Mile {\v{S}}iki{\'c}.
\newblock {Edlib: A C/C++ Library for Fast, Exact Sequence Alignment Using Edit Distance}.
\newblock {\em Bioinformatics}, 2017.

\bibitem{needleman1970general}
Saul~B Needleman and Christian~D Wunsch.
\newblock {A General Method Applicable to the Search for Similarities in the Amino Acid Sequence of Two Proteins}.
\newblock {\em Journal of Molecular Biology}, 1970.

\bibitem{smith1981identification}
Temple~F Smith and Michael~S Waterman.
\newblock {Identification of Common Molecular Subsequences}.
\newblock {\em Journal of Molecular Biology}, 1981.

\bibitem{gotoh1982improved}
Osamu Gotoh.
\newblock {An Improved Algorithm for Matching Biological Sequences}.
\newblock {\em Journal of Molecular Biology}, 1982.

\bibitem{gssource}
Nika Mansouri~Ghiasi, Jisung Park, Harun Mustafa, Jeremie Kim, Ataberk Olgun, Arvid Gollwitzer, Damla Senol~Cali, Can Firtina, Haiyu Mao, Nour Almadhoun~Alserr, Rachata Ausavarungnirun, Nandita Vijaykumar, Mohammed Alser, and Onur Mutlu.
\newblock {GenStore Source Code}.
\newblock \url{https://github.com/CMU-SAFARI/GenStore}.

\bibitem{arxivGS}
Nika Mansouri~Ghiasi, Jisung Park, Harun Mustafa, Jeremie Kim, Ataberk Olgun, Arvid Gollwitzer, Damla Senol~Cali, Can Firtina, Haiyu Mao, Nour Almadhoun~Alserr, Rachata Ausavarungnirun, Nandita Vijaykumar, Mohammed Alser, and Onur Mutlu.
\newblock {GenStore: A High-Performance and Energy-Efficient In-Storage Computing System for Genome Sequence Analysis}.
\newblock In {\em arXiv}, 2022.

\bibitem{abakus23taco}
Lingxi Wu, Minxuan Zhou, Weihong Xu, Ashish Venkat, Tajana Rosing, and Kevin Skadron.
\newblock {Abakus: Accelerating k-mer Counting With Storage Technology}.
\newblock {\em TACO}, 2023.

\bibitem{zheng2025storage}
You-Kai Zheng, Ming-Liang Wei, Hsiang-Yun Cheng, Chia-Lin Yang, Ming-Hsiang Tsai, Chia-Chun Chien, Yuan-Hao Zhong, Po-Hao Tseng, and Hsiang-Pang Li.
\newblock {In-Storage Read-Centric Seed Location Filtering Using 3D-NAND Flash for Genome Sequence Analysis}.
\newblock In {\em ASPDAC}, 2025.

\bibitem{kabra2025ciphermatch}
Mayank Kabra, Rakesh Nadig, Harshita Gupta, Rahul Bera, Manos Frouzakis, Vamanan Arulchelvan, Yu~Liang, Haiyu Mao, Mohammad Sadrosadati, and Onur Mutlu.
\newblock {CIPHERMATCH: Accelerating Homomorphic Encryption-Based String Matching via Memory-Efficient Data Packing and In-Flash Processing}.
\newblock In {\em ASPLOS}, 2025.

\bibitem{chen2025reis}
Kangqi Chen, Rakesh Nadig, Manos Frouzakis, Nika~Mansouri Ghiasi, Yu~Liang, Haiyu Mao, Jisung Park, Mohammad Sadrosadati, and Onur Mutlu.
\newblock {REIS: A High-Performance and Energy-Efficient Retrieval System with In-Storage Processing}.
\newblock In {\em ISCA}, 2025.

\bibitem{megis}
Nika~Mansouri Ghiasi, Mohammad Sadrosadati, Harun Mustafa, Arvid Gollwitzer, Can Firtina, Julien Eudine, Haiyu Mao, Jo{\"e}l Lindegger, Meryem~Banu Cavlak, Mohammed Alser, Jisung Park, and Onur Mutlu.
\newblock {MegIS: High-Performance, Energy-Efficient, and Low-Cost Metagenomic Analysis with In-Storage Processing}.
\newblock In {\em ISCA}, 2024.

\bibitem{grains}
Nika Mansouri~Ghiasi, Harun Mustafa, Talu Güloglu, Rakesh Nadig, Konstantinos Kanellopoulos, Susana Rebolledo~Ruiz, Marc Rautmann, Furkan Eris, Mohammad Sadrosadati, Jisung Park, and Onur Mutlu.
\newblock {GRAINS: Enabling High-Performance and Low-Cost Graph-Based Genome Analysis via Storage-Aware Algorithm-Architecture Co-Design}.
\newblock In {\em ISCA}, 2026.

\bibitem{mansouri2026sage}
Nika~Mansouri Ghiasi, Talu Güloglu, Harun Mustafa, Can Firtina, Konstantina Koliogeorgi, Konstantinos Kanellopoulos, Haiyu Mao, Rakesh Nadig, Mohammad Sadrosadati, Jisung Park, and Onur Mutlu.
\newblock {SAGe: A Lightweight Algorithm-Architecture Co-Design for Mitigating the Data Preparation Bottleneck in Large-Scale Genome Sequence Analysis}.
\newblock In {\em HPCA}, 2026.

\bibitem{megissource}
Nika Mansouri~Ghiasi, Mohammad Sadrosadati, Harun Mustafa, Arvid Gollwitzer, Can Firtina, Julien Eudine, Haiyu Ma, Jo{\"e}l Lindegger, Meryem~Banu Cavlak, Mohammed Alser, Jisung Park, and Onur Mutlu.
\newblock {MegIS Source Code}.
\newblock \url{https://github.com/CMU-SAFARI/MegIS}.

\bibitem{megisarxiv}
Nika~Mansouri Ghiasi, Mohammad Sadrosadati, Harun Mustafa, Arvid Gollwitzer, Can Firtina, Julien Eudine, Haiyu Mao, Joël Lindegger, Meryem~Banu Cavlak, Mohammed Alser, Jisung Park, and Onur Mutlu.
\newblock {MegIS: High-Performance, Energy-Efficient, and Low-Cost Metagenomic Analysis with In-Storage Processing}.
\newblock {\em arXiv}, 2024.

\bibitem{mansouri2026sagearxiv}
Nika~Mansouri Ghiasi, Talu Güloglu, Harun Mustafa, Can Firtina, Konstantina Koliogeorgi, Konstantinos Kanellopoulos, Haiyu Mao, Rakesh Nadig, Mohammad Sadrosadati, Jisung Park, and Onur Mutlu.
\newblock {SAGe: A Lightweight Algorithm-Architecture Co-Design for Mitigating the Data Preparation Bottleneck in Large-Scale Genome Sequence Analysis}.
\newblock {\em arXiv}, 2026.

\bibitem{jun2016storage}
Sang-Woo Jun, Huy~T. Nguyen, Vijay Gadepally, and Arvind.
\newblock {In-storage Embedded Accelerator for Sparse Pattern Processing}.
\newblock In {\em HPEC}, 2016.

\bibitem{Wang2024ndsearch}
Yitu Wang, Shiyu Li, Qilin Zheng, Linghao Song, Zongwang Li, Andrew Chang, Hai~“Helen” Li, and Yiran Chen.
\newblock {NDSEARCH: Accelerating Graph-Traversal-Based Approximate Nearest Neighbor Search through Near Data Processing}.
\newblock In {\em ISCA}, 2024.

\bibitem{lee2022smartsage}
Yunjae Lee, Jinha Chung, and Minsoo Rhu.
\newblock {SmartSAGE: training large-scale graph neural networks using in-storage processing architectures}.
\newblock In {\em ISCA}, 2022.

\bibitem{Niu2024flashgnn}
Fuping Niu, Jianhui Yue, Jiangqiu Shen, Xiaofei Liao, and Hai Jin.
\newblock {FlashGNN: An In-SSD Accelerator for GNN Training}.
\newblock In {\em HPCA}, 2024.

\bibitem{Khadirsharbiyani2024smartgraph}
Soheil Khadirsharbiyani, Nima Elyasi, Armin~Haj Aboutalebi, Chun-Yi Liu, Changho Choi, and Mahmut~Taylan Kandemir.
\newblock {SmartGraph: A Framework for Graph Processing in Computational Storage}.
\newblock In {\em SoCC}, 2024.

\bibitem{Zhang2025taijigraph}
Xinmiao Zhang, Cheng Liu, Shengwen Liang, Hayden Kwok-Hay So, Ying Wang, Lei Zhang, Huawei Li, and Xiaowei Li.
\newblock {Taijigraph: an Out-Of-Core Graph Processing System Enhanced with Computational Storage}.
\newblock In {\em IPDPS}, 2025.

\bibitem{An2023baraddur}
Jiyoung An, Esmerald Aliaj, and Sang-Woo Jun.
\newblock {Barad-dur: Near-Storage Accelerator for Training Large Graph Neural Networks}.
\newblock In {\em PACT}, 2023.

\bibitem{Kang2024sting}
Seongyoung Kang and Sang-Woo Jun.
\newblock Sting: Near-storage accelerator framework for scalable triangle counting and beyond.
\newblock In {\em DAC}, 2024.

\bibitem{jun2018grafboost}
Sang-Woo Jun, Andy Wright, Sizhuo Zhang, Shuotao Xu, and Arvind.
\newblock {GraFBoost: Using Accelerated Flash Storage for External Graph Analytics}.
\newblock In {\em ISCA}, 2018.

\bibitem{matam2019graphssd}
Kiran~Kumar Matam, Gunjae Koo, Haipeng Zha, Hung-Wei Tseng, and Murali Annavaram.
\newblock {GraphSSD: graph semantics aware SSD}.
\newblock In {\em ISCA}, 2019.

\bibitem{grainsextended}
Nika Mansouri~Ghiasi, Harun Mustafa, Talu G{\"u}loglu, Rakesh Nadig, Konstantina Koliogeorgi, Susana Rebolledo~Ruiz, Marc Rautmann, Furkan Eris, Mohammad Sadrosadati, Jisung Park, and Onur Mutlu.
\newblock {GRAINS: Storage-Aware Algorithm-Architecture Co-Design Enabling High-Performance and Low-Cost Graph-Based Genome Analysis}.
\newblock In {\em arXiv}, 2026.

\bibitem{micheloni2010inside}
Rino Micheloni, Luca Crippa, and Alessia Marelli.
\newblock {\em {Inside NAND Flash Memories}}.
\newblock Springer Science \& Business Media, 2010.

\bibitem{cai-insidessd-2018}
Yu~Cai, Saugata Ghose, Erich~F. Haratsch, Yixin Luo, and Onur Mutlu.
\newblock {Reliability Issues in Flash-Memory-Based Solid-State Drives: Experimental Analysis, Mitigation, Recovery}.
\newblock In {\em Inside Solid State Drives (SSDs)}, 2018.

\bibitem{meza_revisiting_2015}
Justin Meza, Qiang Wu, Sanjeev Kumar, and Onur Mutlu.
\newblock Revisiting {Memory} {Errors} in {Large}-{Scale} {Production} {Data} {Centers}: {Analysis} and {Modeling} of {New} {Trends} from the {Field}.
\newblock In {\em {DSN}}, 2015.

\bibitem{meza_case_2013}
J.~Meza, Y.~Luo, S.~Khan, J.~Zhao, Y.~Xie, and O.~Mutlu.
\newblock A {Case} for {Efficient} {Hardware}-{Software} {Cooperative} {Management} of {Storage} and {Memory}.
\newblock In {\em {WEED}}, 2013.

\bibitem{meza2015large}
Justin Meza, Qiang Wu, Sanjev Kumar, and Onur Mutlu.
\newblock {A Large-Scale Study of Flash Memory Errors in the Field}.
\newblock {\em SIGMETRICS}, 2015.

\bibitem{cai_data_2015}
Yu~Cai, Yixin Luo, Erich~F Haratsch, Ken Mai, and Onur Mutlu.
\newblock Data {Retention} in {MLC} {NAND} {Flash} {Memory}: {Characterization}, {Optimization}, and {Recovery}.
\newblock In {\em {HPCA}}, 2015.

\bibitem{cai_error_2012}
Yu~Cai, Erich~F Haratsch, Onur Mutlu, and Ken Mai.
\newblock Error {Patterns} in {MLC} {NAND} {Flash} {Memory}: {Measurement}, {Characterization}, and {Analysis}.
\newblock In {\em {DATE}}, 2012.

\bibitem{cai2017error}
Yu~Cai, Saugata Ghose, Erich~F Haratsch, Yixin Luo, and Onur Mutlu.
\newblock {Error Characterization, Mitigation, and Recovery in Flash-Memory-Based Solid-State Drives}.
\newblock In {\em Proc. IEEE}, 2017.

\bibitem{cai2013error}
Yu~Cai, Gulay Yalcin, Onur Mutlu, Erich~F Haratsch, Adrian Cristal, Osman~S Unsal, and Ken Mai.
\newblock {Error Analysis and Retention-Aware Error Management for NAND Flash Memory}.
\newblock {\em Intel Technology Journal}, 2013.

\bibitem{cai_program_2013}
Yu~Cai, Onur Mutlu, Erich~F Haratsch, and Ken Mai.
\newblock Program {Interference} in {MLC} {NAND} {Flash} {Memory}: {Characterization}, {Modeling}, and {Mitigation}.
\newblock In {\em {ICCD}}, 2013.

\bibitem{cai2018errorsarxiv}
Yu~Cai, Saugata Ghose, Erich~F. Haratsch, Yixin Luo, and Onur Mutlu.
\newblock {Errors in Flash-Memory-Based Solid-State Drives: Analysis, Mitigation, and Recovery}.
\newblock {\em arXiv}, 2018.

\bibitem{cai_threshold_2013}
Yu~Cai, Erich~F Haratsch, Onur Mutlu, and Ken Mai.
\newblock Threshold {Voltage} {Distribution} in {MLC} {NAND} {Flash} {Memory}: {Characterization}, {Analysis}, and {Modeling}.
\newblock In {\em {DATE}}, 2013.

\bibitem{cai2017vulnerabilities}
Yu~Cai, Saugata Ghose, Yixin Luo, Ken Mai, Onur Mutlu, and Erich~F. Haratsch.
\newblock {Vulnerabilities in MLC NAND Flash Memory Programming: Experimental Analysis, Exploits, and Mitigation Techniques}.
\newblock In {\em HPCA}, 2017.

\bibitem{cai2012flash}
Yu~Cai, Gulay Yalcin, Onur Mutlu, Erich~F Haratsch, Adrian Cristal, Osman~S Unsal, and Ken Mai.
\newblock {Flash Correct-and-Refresh: Retention-Aware Error Management for Increased Flash Memory Lifetime}.
\newblock In {\em ICCD}, 2012.

\bibitem{luo2015warm}
Yixin Luo, Yu~Cai, Saugata Ghose, Jongmoo Choi, and Onur Mutlu.
\newblock {WARM: Improving NAND Flash Memory Lifetime with Write-Hotness Aware Retention Management}.
\newblock In {\em MSST}, 2015.

\bibitem{luo2018improving}
Yixin Luo, Saugata Ghose, Yu~Cai, Erich~F Haratsch, and Onur Mutlu.
\newblock {Improving 3D NAND Flash Memory Lifetime by Tolerating Early Retention Loss and Process Variation}.
\newblock {\em ACM POMACS}, 2018.

\bibitem{cho2024aero}
Sungjun Cho, Beomjun Kim, Hyunuk Cho, Gyeongseob Seo, Onur Mutlu, Myungsuk Kim, and Jisung Park.
\newblock {AERO: Adaptive Erase Operation for Improving Lifetime and Performance of Modern NAND Flash-Based SSDs}.
\newblock In {\em ASPLOS}, 2024.

\bibitem{nadig2026conduit}
Rakesh Nadig, Vamanan Arulchelvan, Mayank Kabra, Harshita Gupta, Rahul Bera, Nika~Mansouri Ghiasi, Nanditha Rao, Qingcai Jiang, Andreas~Kosmas Kakolyris, Yu~Liang, Mohammad Sadrosadati, and Onur Mutlu.
\newblock {Conduit: Programmer-Transparent Near-Data Processing Using Multiple Compute-Capable Resources in Solid State Drives}.
\newblock In {\em HPCA}, 2026.

\bibitem{nadig2023venice}
Rakesh Nadig, Mohammad Sadrosadati, Haiyu Mao, Nika~Mansouri Ghiasi, Arash Tavakkol, Jisung Park, Hamid Sarbazi-Azad, Juan~G{\'o}mez Luna, and Onur Mutlu.
\newblock {Venice: Improving Solid-State Drive Parallelism at Low Cost via Conflict-Free Accesses}.
\newblock In {\em ISCA}, 2023.

\bibitem{kim2020evanesco}
Myungsuk Kim, Jisung Park, Genhee Cho, Yoona Kim, Lois Orosa, Onur Mutlu, and Jihong Kim.
\newblock {Evanesco: Architectural Support for Efficient Data Sanitization in Modern Flash-Based Storage Systems}.
\newblock In {\em ASPLOS}, 2020.

\bibitem{park-dac-2016}
Jisung Park, Jaeyong Jeong, Sungjin Lee, Youngsun Song, and Jihong Kim.
\newblock {Improving Performance and Lifetime of {NAND} Storage Systems Using Relaxed Program Sequence}.
\newblock In {\em DAC}, 2016.

\bibitem{park-nvmsa-2018}
Jisung Park, Myungsuk Kim, Sungjin Lee, and Jihong Kim.
\newblock {Improving {I/O} Performance of Large-page Flash Storage Systems Using Subpage-parallel Reads}.
\newblock In {\em NVMSA}, 2018.

\bibitem{park2021reducing}
Jisung Park, Myungsuk Kim, Myoungjun Chun, Lois Orosa, Jihong Kim, and Onur Mutlu.
\newblock {Reducing Solid-State Drive Read Latency by Optimizing Read-Retry}.
\newblock In {\em ASPLOS}, 2021.

\bibitem{kim-dac-2017}
Myungsuk Kim, Jaehoon Lee, Sungjin Lee, Jisung Park, Youngsun Song, and Jihong Kim.
\newblock {Improving Performance and Lifetime of Large-page {NAND} Storages Using Erase-free Subpage Programming}.
\newblock In {\em DAC}, 2017.

\bibitem{tavakkol2018flin}
Arash Tavakkol, Mohammad Sadrosadati, Saugata Ghose, Jeremie Kim, Yixin Luo, Yaohua Wang, Nika~Mansouri Ghiasi, Lois Orosa, Juan G{\'o}mez-Luna, and Onur Mutlu.
\newblock {FLIN: Enabling Fairness and Enhancing Performance in Modern NVMe Solid State Drives}.
\newblock In {\em ISCA}, 2018.

\bibitem{cai2015read}
Yu~Cai, Yixin Luo, Saugata Ghose, and Onur Mutlu.
\newblock {Read Disturb Errors in MLC NAND Flash Memory: Characterization, Mitigation, and Recovery}.
\newblock In {\em IEEE/IFIP DSN}, 2015.

\bibitem{ha2015integrated}
Keonsoo Ha, Jaeyong Jeong, and Jihong Kim.
\newblock {An Integrated Approach for Managing Read Disturbs in High-density NAND Flash Memory}.
\newblock {\em IEEE TCAD}, 2015.

\bibitem{wang2014enhanced}
Jiadong Wang, Kasra Vakilinia, Tsung-Yi Chen, Thomas Courtade, Guiqiang Dong, Tong Zhang, Hari Shankar, and Richard Wesel.
\newblock {Enhanced Precision through Multiple Reads for LDPC Decoding in Flash Memories}.
\newblock {\em JSAC}, 2014.

\bibitem{zhao2013ldpc}
Kai Zhao, Wenzhe Zhao, Hongbin Sun, Xiaodong Zhang, Nanning Zheng, and Tong Zhang.
\newblock {{LDPC-in-SSD}: Making Advanced Error Correction Codes Work Effectively in Solid State Drives}.
\newblock In {\em FAST}, 2013.

\bibitem{dong2010use}
Guiqiang Dong, Ningde Xie, and Tong Zhang.
\newblock {On the Use of Soft-Decision Error-Correction Codes in NAND Flash Memory}.
\newblock {\em TCAS}, 2010.

\bibitem{kim2021performance}
Joonsung Kim, Kanghyun Choi, Wonsik Lee, and Jangwoo Kim.
\newblock {Performance Modeling and Practical Use Cases for Black-Box SSDs}.
\newblock {\em ACM TOS}, 2021.

\bibitem{kim2018ssdcheck}
Joonsung Kim, Pyeongsu Park, Jaehyung Ahn, Jihun Kim, Jong Kim, and Jangwoo Kim.
\newblock {SSDcheck: Timely and Accurate Prediction of Irregular Behaviors in Black-Box SSDs}.
\newblock In {\em MICRO}, 2018.

\bibitem{kim2017ssd}
Jihun Kim, Joonsung Kim, Pyeongsu Park, Jong Kim, and Jangwoo Kim.
\newblock {SSD Performance Modeling Using Bottleneck Analysis}.
\newblock {\em IEEE Computer Architecture Letters}, 2017.

\bibitem{cai2014neighbor}
Yu~Cai, Gulay Yalcin, Onur Mutlu, Erich~F Haratsch, Osman Unsal, Adrian Cristal, and Ken Mai.
\newblock {Neighbor-Cell Assisted Error Correction for MLC NAND Flash Memories}.
\newblock {\em SIGMETRICS}, 2014.

\bibitem{luo2016enabling}
Yixin Luo, Saugata Ghose, Yu~Cai, Erich~F Haratsch, and Onur Mutlu.
\newblock {Enabling Accurate and Practical Online Flash Channel Modeling for Modern MLC NAND Flash Memory}.
\newblock {\em JSAC}, 2016.

\bibitem{park2026experimental}
Yonggon Park, Hyunuk Cho, Onur Mutlu, Sungjin Lee, and Jisung Park.
\newblock {Experimental Study on System-Level Performance Impact of Read Disturbance in Modern SSDs}.
\newblock {\em Proceedings of the ACM on Measurement and Analysis of Computing Systems}, 2026.

\bibitem{shim2019exploiting}
Youngseop Shim, Myungsuk Kim, Myoungjun Chun, Jisung Park, Yoona Kim, and Jihong Kim.
\newblock {Exploiting Process Similarity of 3D Flash Memory for High Performance SSDs}.
\newblock In {\em MICRO}, 2019.

\bibitem{chun2026straw}
Myoungjun Chun, Jaeyong Lee, Inhyuk Choi, Jisung Park, Myungsuk Kim, and Jihong Kim.
\newblock {STRAW: Stress-Aware WL-Based Read Disturbance Management for High-Density NAND Flash Memory}.
\newblock In {\em ASPLOS}, 2026.

\bibitem{kim2024norns}
Earl Kim, Hyunuk Cho, Sungjun Cho, Myungsuk Kim, Jisung Park, Jaeyong Jeong, Eunkyoung Kim, and Sunghoi Hur.
\newblock {NORNS: Three Guides for Efficient Automatic Post-Fabrication Optimization of Modern NAND Flash Memory}.
\newblock In {\em ICCAD}, 2024.

\bibitem{wu2025ssdtrain}
Kun Wu, Jeongmin~Brian Park, Xiaofan Zhang, Mert Hidayeto{\u{g}}lu, Vikram~Sharma Mailthody, Sitao Huang, Steve Lumetta, and Wen-mei Hwu.
\newblock {SSDTrain: An Activation Offloading Framework to SSDs for Faster Large Language Model Training}.
\newblock In {\em DAC}, 2025.

\bibitem{qureshi2022bam}
Zaid Qureshi, Vikram~Sharma Mailthody, Isaac Gelado, Seungwon Min, Amna Masood, Jeongmin~Brian Park, Jinjun Xiong, Chris~J. Newburn, Dmitri Vainbrand, I-Hsin Chung, Michael Garland, William~J. Dally, and Wen-mei Hwu.
\newblock {BaM: A Case for Enabling Fine-grain High Throughput GPU-Orchestrated Access to Storage}.
\newblock {\em arXiv}, 2022.

\bibitem{qureshi2023gpu}
Zaid Qureshi, Vikram~Sharma Mailthody, Isaac Gelado, Seungwon Min, Amna Masood, Jeongmin Park, Jinjun Xiong, Chris~J. Newburn, Dmitri Vainbrand, I-Hsin Chung, Michael Garland, William Dally, and Wen-mei Hwu.
\newblock {GPU-Initiated On-Demand High-Throughput Storage Access in the BaM System Architecture}.
\newblock In {\em ASPLOS}, 2023.

\bibitem{son2026exploring}
Dowon Son, Yonggon Park, Hyunuk Cho, Hyungkyu Ham, Onur Mutlu, Sungjin Lee, Gwangsun Kim, and Jisung Park.
\newblock {Exploring High-Bandwidth Flash for Modern LLM Inference: Opportunities and Challenges}.
\newblock {\em CAL}, 2026.

\bibitem{chun2022pif}
Myungjun Chun, Jaeyong Lee, Sanggu Lee, Myungsuk Kim, and Jihong Kim.
\newblock {PiF: In-flash acceleration for data-intensive applications}.
\newblock In {\em HotStorage}, 2022.

\bibitem{chen2024search}
Yun-Chih Chen, Yuan-Hao Chang, and Tei-Wei Kuo.
\newblock {Search-In-Memory: Reliable, Versatile, and Efficient Data Matching in SSD’s NAND Flash Memory Chip for Data Indexing Acceleration}.
\newblock {\em IEEE TCAD}, 2024.

\bibitem{lee2025aif}
Jaeyong Lee, Hyeunjoo Kim, Sanghun Oh, Myoungjun Chun, Myungsuk Kim, and Jihong Kim.
\newblock {AiF: Accelerating On-Device LLM Inference Using In-Flash Processing}.
\newblock In {\em ISCA}, 2025.

\bibitem{Sun2025lincoln}
Weiyi Sun, Mingyu Gao, Zhaoshi Li, Aoyang Zhang, Iris~Ying Chou, Jianfeng Zhu, Shaojun Wei, and Leibo Liu.
\newblock {Lincoln: Real-Time 50~100B LLM Inference on Consumer Devices with LPDDR-Interfaced, Compute-Enabled Flash Memory}.
\newblock In {\em HPCA}, 2025.

\bibitem{Yu2024cambriconllm}
Zhongkai Yu, Shengwen Liang, Tianyun Ma, Yunke Cai, Ziyuan Nan, Di~Huang, Xinkai Song, Yifan Hao, Jie Zhang, Tian Zhi, Yongwei Zhao, Zidong Du, Xing Hu, Qi~Guo, and Tianshi Chen.
\newblock {Cambricon-LLM: A Chiplet-Based Hybrid Architecture for On-Device Inference of 70B LLM}.
\newblock In {\em MICRO}, 2024.

\bibitem{Kim2023optimstore}
Junkyum Kim, Myeonggu Kang, Yunki Han, Yang-Gon Kim, and Lee-Sup Kim.
\newblock {OptimStore: In-Storage Optimization of Large Scale DNNs with On-Die Processing}.
\newblock In {\em HPCA}, 2023.

\bibitem{choi2020flash}
Won~Ho Choi, Pi-Feng Chiu, Wen Ma, Gertjan Hemink, Tung~Thanh Hoang, Martin Lueker-Boden, and Zvonimir Bandic.
\newblock {An in-flash binary neural network accelerator with SLC NAND flash array}.
\newblock In {\em ISCAS}, 2020.

\bibitem{gao2021parabit}
Congming Gao, Xin Xin, Youyou Lu, Youtao Zhang, Jun Yang, and Jiwu Shu.
\newblock {ParaBit: Processing Parallel Bitwise Operations in NAND Flash Memory Based SSDs}.
\newblock In {\em MICRO}, 2021.

\bibitem{park2022flash}
Jisung Park, Roknoddin Azizi, Geraldo~F Oliveira, Mohammad Sadrosadati, Rakesh Nadig, David Novo, Juan G{\'o}mez-Luna, Myungsuk Kim, and Onur Mutlu.
\newblock {Flash-Cosmos: In-Flash Bulk Bitwise Operations Using Inherent Computation Capability of NAND Flash Memory}.
\newblock In {\em MICRO}, 2022.

\bibitem{chun2024rif}
Myoungjun Chun, Jaeyong Lee, Myungsuk Kim, Jisung Park, and Jihong Kim.
\newblock {RiF: Improving Read Performance of Modern SSDs Using an On-Die Early-Retry Engine}.
\newblock In {\em HPCA}, 2024.

\bibitem{Chen2024aresflash}
Jian Chen, Congming Gao, Youyou Lu, Yuhao Zhang, and Jiwu Shu.
\newblock {Ares-Flash: Efficient Parallel Integer Arithmetic Operations Using NAND Flash Memory}.
\newblock In {\em MICRO}, 2024.

\bibitem{Wong2025anvil}
Ryan Wong, Nikita Kim, Aniket Das, Kevin Higgs, Engin Ipek, Sapan Agarwal, Saugata Ghose, and Ben Feinberg.
\newblock {ANVIL: An In-Storage Accelerator for Name–Value Data Stores}.
\newblock In {\em ISCA}, 2025.

\bibitem{han2019novel}
Runze Han, Peng Huang, Yachen Xiang, Chen Liu, Zhen Dong, Zhiqiang Su, Yongbo Liu, Lu~Liu, Xiaoyan Liu, and Jinfeng Kang.
\newblock {A novel convolution computing paradigm based on NOR flash array with high computing speed and energy efficiency}.
\newblock {\em IEEE TCAS-I}, 2019.

\bibitem{wang2018three}
Panni Wang, Feng Xu, Bo~Wang, Bin Gao, Huaqiang Wu, He~Qian, and Shimeng Yu.
\newblock {Three-dimensional NAND flash for vector--matrix multiplication}.
\newblock {\em IEEE TVLSI}, 2018.

\bibitem{kang2021s}
Myeonggu Kang, Hyeonuk Kim, Hyein Shin, Jaehyeong Sim, Kyeonghan Kim, and Lee-Sup Kim.
\newblock {S-FLASH: A NAND flash-based deep neural network accelerator exploiting bit-level sparsity}.
\newblock {\em IEEE TC}, 2021.

\bibitem{lee2020neuromorphic}
Sung-Tae Lee and Jong-Ho Lee.
\newblock {Neuromorphic computing using NAND flash memory architecture with pulse width modulation scheme}.
\newblock {\em Frontiers in Neuroscience}, 2020.

\bibitem{kim2025crossbit}
Hyunjin Kim, Seunghwan Song, Sukhyun Choi, Jeongin Choe, Sanghyeok Han, Jisung Park, Jinho Lee, and Jae-Joon Kim.
\newblock {CrossBit: Bitwise Computing in NAND Flash Memory with Inter-Bitline Data Communication}.
\newblock In {\em MICRO}, 2025.

\bibitem{kultursay2013evaluating}
Emre K{\"u}lt{\"u}rsay, Mahmut Kandemir, Anand Sivasubramaniam, and Onur Mutlu.
\newblock {Evaluating STT-RAM as an Energy-Efficient Main Memory Alternative}.
\newblock In {\em ISPASS}, 2013.

\bibitem{meena2014overview}
Jagan~Singh Meena, Simon~Min Sze, Umesh Chand, and Tseung-Yuen Tseng.
\newblock {Overview Of Emerging Nonvolatile Memory Technologies}.
\newblock {\em Nanoscale Research Letters}, 2014.

\bibitem{lee2009architecting}
Benjamin~C Lee, Engin Ipek, Onur Mutlu, and Doug Burger.
\newblock {Architecting Phase Change Memory as a Scalable DRAM Alternative}.
\newblock In {\em ISCA}, 2009.

\bibitem{akinaga2010resistive}
Hiroyuki Akinaga and Hisashi Shima.
\newblock {Resistive Random Access Memory (ReRAM) Based on Metal Oxides}.
\newblock {\em Proc. IEEE}, 2010.

\bibitem{tehrani1999progress}
Said Tehrani, JM~Slaughter, E~Chen, M~Durlam, J~Shi, and M~DeHerren.
\newblock {Progress and Outlook for MRAM Technology}.
\newblock {\em IEEE MAG}, 1999.

\bibitem{qureshi2009scalable}
Moinuddin~K Qureshi, Vijayalakshmi Srinivasan, and Jude~A Rivers.
\newblock Scalable high performance main memory system using phase-change memory technology.
\newblock In {\em ISCA}, 2009.

\bibitem{yoon2014efficient}
Hanbin Yoon, Justin Meza, Naveen Muralimanohar, Norman~P Jouppi, and Onur Mutlu.
\newblock {Efficient Data Mapping and Buffering Techniques for Multilevel Cell Phase-Change Memories}.
\newblock {\em ACM TACO}, 2014.

\bibitem{meza2012enabling}
Justin Meza, Jichuan Chang, HanBin Yoon, Onur Mutlu, and Parthasarathy Ranganathan.
\newblock {Enabling Efficient and Scalable Hybrid Memories using Fine-granularity DRAM Cache Management}.
\newblock {\em CAL}, 2012.

\bibitem{hu2012exploring}
Yang Hu, Hong Jiang, Dan Feng, Lei Tian, Hao Luo, and Chao Ren.
\newblock {Exploring and Exploiting the Multilevel Parallelism Inside SSDs for Improved Performance and Endurance}.
\newblock {\em IEEE TC}, 2012.

\bibitem{gao2019parallel}
Congming Gao, Liang Shi, Chun~Jason Xue, Cheng Ji, Jun Yang, and Youtao Zhang.
\newblock {Parallel all the time: Plane Level Parallelism Exploration for High Performance SSDs}.
\newblock In {\em MSST}, 2019.

\bibitem{gao2020boosting}
Congming Gao, Liang Shi, Kai Liu, Chun~Jason Xue, Jun Yang, and Youtao Zhang.
\newblock {Boosting the Performance of SSDs via Fully Exploiting the Plane Level Parallelism}.
\newblock {\em IEEE TPDS}, 2020.

\bibitem{bjorling2017lightnvm}
Matias Bj{\o}rling, Javier Gonzalez, and Philippe Bonnet.
\newblock {LightNVM: the Linux open-channel SSD subsystem}.
\newblock In {\em FAST}, 2017.

\bibitem{park-dac-2019}
Jisung Park, Youngdon Jung, Jonghoon Won, Minji Kang, Sungjin Lee, and Jihong Kim.
\newblock {RansomeBlocker: a Low-Overhead Ransomware-Proof SSD}.
\newblock In {\em DAC}, 2019.

\bibitem{kim2023decoupled}
Jiho Kim, Myoungsoo Jung, and John Kim.
\newblock {Decoupled SSD: Rethinking SSD Architecture through Network-based Flash Controllers}.
\newblock In {\em ISCA}, 2023.

\bibitem{samsung860pro}
Samsung.
\newblock {Samsung SSD 860 PRO}.
\newblock \url{https://www.samsung.com/semiconductor/minisite/ssd/product/consumer/860pro/}, 2018.

\bibitem{inteldcs4500}
Intel.
\newblock {Intel SSD DC S4500 Series}.
\newblock \url{https://ark.intel.com/content/www/us/en/ark/products/120521/intel-ssd-dc-s4500-series-480gb-2-5in-sata-6gbs-3d1-tlc.html}, 2017.

\bibitem{samsung9100PRO}
Samsung.
\newblock {Samsung SSD 9100 PRO}.
\newblock \url{https://semiconductor.samsung.com/consumer-storage/internal-ssd/9100-pro/}, 2025.

\bibitem{anandcontroller}
AnandTech.
\newblock {New Enterprise SSD Controllers}.
\newblock \url{https://www.anandtech.com/show/16275/new-enterprise-ssd-controllers-from-silicon-motion-phison-fadu}.

\bibitem{flashtecnvme5016}
Microchip~Technology Incorporated.
\newblock {Flashtec$^\text{\textregistered}$ NVMe$^\text{\textregistered}$ 5106: Performance 16-Channel Gen 5 PCIe$^\text{\textregistered}$ Flash Controller}.
\newblock \url{https://ww1.microchip.com/downloads/aemDocuments/documents/DCS/ProductDocuments/Brochures/Flashtec-NVMe-5016-Sell-Sheet-00005529.pdf}, 2024.

\bibitem{chang2007efficient}
Li-Pin Chang.
\newblock {On Efficient Wear Leveling for Large-scale Flash-memory Storage Systems}.
\newblock In {\em ACM SAC}, 2007.

\bibitem{luo2018heatwatch}
Yixin Luo, Saugata Ghose, Yu~Cai, Erich~F. Haratsch, and Onur Mutlu.
\newblock {HeatWatch: Improving 3D NAND Flash Memory Device Reliability by Exploiting Self-Recovery and Temperature Awareness}.
\newblock In {\em HPCA}, 2018.

\bibitem{kao_naivebayescall_2010}
Wei-Chun Kao and Yun~S. Song.
\newblock {naiveBayesCall}: {An} {Efficient} {Model}-{Based} {Base}-{Calling} {Algorithm} for {High}-{Throughput} {Sequencing}.
\newblock In {\em RECOMB}, 2010.

\bibitem{cacho_base-calling_2018}
Ashley Cacho, Weixin Yao, and Xinping Cui.
\newblock Base-{Calling} {Using} a {Random} {Effects} {Mixture} {Model} on {Next}-{Generation} {Sequencing} {Data}.
\newblock {\em Statistics in Biosciences}, 2018.

\bibitem{erlich_alta-cyclic_2008}
Yaniv Erlich, Partha~P Mitra, Melissa delaBastide, W~Richard McCombie, and Gregory~J Hannon.
\newblock Alta-{Cyclic}: a self-optimizing base caller for next-generation sequencing.
\newblock {\em Nature Methods}, 2008.

\bibitem{wang_adaptive_2017}
Bo~Wang, Lin Wan, Anqi Wang, and Lei~M. Li.
\newblock An adaptive decorrelation method removes {Illumina} {DNA} base-calling errors caused by crosstalk between adjacent clusters.
\newblock {\em Scientific Reports}, 2017.

\bibitem{rougemont_probabilistic_2008}
Jacques Rougemont, Arnaud Amzallag, Christian Iseli, Laurent Farinelli, Ioannis Xenarios, and Felix Naef.
\newblock Probabilistic base calling of {Solexa} sequencing data.
\newblock {\em BMC Bioinformatics}, 2008.

\bibitem{shen_particlecall_2012}
Xiaohu Shen and Haris Vikalo.
\newblock {ParticleCall}: {A} particle filter for base calling in next-generation sequencing systems.
\newblock {\em BMC Bioinformatics}, 2012.

\bibitem{ji_bm-bc_2012}
Yuan Ji, Riten Mitra, Fernando Quintana, Alejandro Jara, Peter Mueller, Ping Liu, Yue Lu, and Shoudan Liang.
\newblock {BM}-{BC}: a {Bayesian} method of base calling for {Solexa} sequence data.
\newblock {\em BMC Bioinformatics}, 2012.

\bibitem{das_base_2013}
Shreepriya Das and Haris Vikalo.
\newblock Base calling for high-throughput short-read sequencing: dynamic programming solutions.
\newblock {\em BMC Bioinformatics}, 2013.

\bibitem{kircher_improved_2009}
Martin Kircher, Udo Stenzel, and Janet Kelso.
\newblock Improved base calling for the {Illumina} {Genome} {Analyzer} using machine learning strategies.
\newblock {\em Genome Biology}, 2009.

\bibitem{massingham_all_2012}
Tim Massingham and Nick Goldman.
\newblock All {Your} {Base}: a fast and accurate probabilistic approach to base calling.
\newblock {\em Genome Biology}, 2012.

\bibitem{ye_blindcall_2014}
Chengxi Ye, Chiaowen Hsiao, and Héctor Corrada~Bravo.
\newblock {BlindCall}: ultra-fast base-calling of high-throughput sequencing data by blind deconvolution.
\newblock {\em Bioinformatics}, 2014.

\bibitem{renaud_freeibis_2013}
Gabriel Renaud, Martin Kircher, Udo Stenzel, and Janet Kelso.
\newblock {freeIbis}: an efficient basecaller with calibrated quality scores for {Illumina} sequencers.
\newblock {\em Bioinformatics}, 2013.

\bibitem{das_onlinecall_2012}
Shreepriya Das and Haris Vikalo.
\newblock {OnlineCall}: fast online parameter estimation and base calling for illumina's next-generation sequencing.
\newblock {\em Bioinformatics}, 2012.

\bibitem{menges_totalrecaller_2011}
Fabian Menges, Giuseppe Narzisi, and Bud Mishra.
\newblock {TotalReCaller}: improved accuracy and performance via integrated alignment and base-calling.
\newblock {\em Bioinformatics}, 2011.

\bibitem{bravo_model-based_2010}
Héctor~Corrada Bravo and Rafael~A. Irizarry.
\newblock Model-{Based} {Quality} {Assessment} and {Base}-{Calling} for {Second}-{Generation} {Sequencing} {Data}.
\newblock {\em Biometrics}, 2010.

\bibitem{kao_bayescall_2009}
Wei-Chun Kao, Kristian Stevens, and Yun~S. Song.
\newblock {BayesCall}: {A} model-based base-calling algorithm for high-throughput short-read sequencing.
\newblock {\em Genome Research}, 2009.

\bibitem{alkan_limitations_2011}
Can Alkan, Saba Sajjadian, and Evan~E Eichler.
\newblock Limitations of next-generation genome sequence assembly.
\newblock {\em Nature Methods}, 2011.

\bibitem{firtina_genomic_2016}
Can Firtina and Can Alkan.
\newblock On genomic repeats and reproducibility.
\newblock {\em Bioinformatics}, 32(15):2243--2247, August 2016.

\bibitem{levene_zero-mode_2003}
M.~J. Levene, J.~Korlach, S.~W. Turner, M.~Foquet, H.~G. Craighead, and W.~W. Webb.
\newblock Zero-{Mode} {Waveguides} for {Single}-{Molecule} {Analysis} at {High} {Concentrations}.
\newblock {\em Science}, 299(5607):682--686, January 2003.

\bibitem{travers_flexible_2010}
Kevin~J. Travers, Chen-Shan Chin, David~R. Rank, John~S. Eid, and Stephen~W. Turner.
\newblock A flexible and efficient template format for circular consensus sequencing and {SNP} detection.
\newblock {\em Nucleic Acids Research}, 38(15):e159--e159, August 2010.

\bibitem{sharon_single-molecule_2013}
Donald Sharon, Hagen Tilgner, Fabian Grubert, and Michael Snyder.
\newblock A single-molecule long-read survey of the human transcriptome.
\newblock {\em Nature Biotechnology}, 31(11):1009--1014, November 2013.

\bibitem{firtina_hercules_2018}
Can Firtina, Ziv Bar-Joseph, Can Alkan, and A Ercument Cicek.
\newblock Hercules: a profile {HMM}-based hybrid error correction algorithm for long reads.
\newblock {\em Nucleic Acids Research}, 2018.

\bibitem{xiao_mecat_2017}
Chuan-Le Xiao, Ying Chen, Shang-Qian Xie, Kai-Ning Chen, Yan Wang, Yue Han, Feng Luo, and Zhi Xie.
\newblock {MECAT}: fast mapping, error correction, and de novo assembly for single-molecule sequencing reads.
\newblock {\em Nature Methods}, 2017.

\bibitem{stanojevic_telomere--telomere_2024}
Dominik Stanojević, Dehui Lin, Paola Florez~de Sessions, and Mile Šikić.
\newblock Telomere-to-telomere phased genome assembly using error-corrected {Simplex} nanopore reads.
\newblock {\em bioRxiv}, page 2024.05.18.594796, 2024.

\bibitem{salmela_lordec_2014}
Leena Salmela and Eric Rivals.
\newblock {LoRDEC}: accurate and efficient long read error correction.
\newblock {\em Bioinformatics}, 30(24):3506--3514, December 2014.

\bibitem{kang_hybrid-hybrid_2023}
Xiongbin Kang, Jialu Xu, Xiao Luo, and Alexander Schönhuth.
\newblock Hybrid-hybrid correction of errors in long reads with {HERO}.
\newblock {\em Genome Biology}, 24(1):275, December 2023.

\bibitem{holley_ratatosk_2021}
Guillaume Holley, Doruk Beyter, Helga Ingimundardottir, Peter~L. Møller, Snædis Kristmundsdottir, Hannes~P. Eggertsson, and Bjarni~V. Halldorsson.
\newblock Ratatosk: hybrid error correction of long reads enables accurate variant calling and assembly.
\newblock {\em Genome Biology}, 22(1):28, January 2021.

\bibitem{morisse_hybrid_2018}
Pierre Morisse, Thierry Lecroq, and Arnaud Lefebvre.
\newblock Hybrid correction of highly noisy long reads using a variable-order de {Bruijn} graph.
\newblock {\em Bioinformatics}, 34(24):4213--4222, December 2018.

\bibitem{wang_fmlrc_2018}
Jeremy~R. Wang, James Holt, Leonard McMillan, and Corbin~D. Jones.
\newblock {FMLRC}: {Hybrid} long read error correction using an {FM}-index.
\newblock {\em BMC Bioinformatics}, 19(1):50, February 2018.

\bibitem{zhu_lcat_2023}
Wufei Zhu and Xingyu Liao.
\newblock {LCAT}: an isoform-sensitive error correction for transcriptome sequencing long reads.
\newblock {\em Frontiers in Genetics}, 14, 2023.

\bibitem{salmela_accurate_2017}
Leena Salmela, Riku Walve, Eric Rivals, and Esko Ukkonen.
\newblock Accurate self-correction of errors in long reads using de {Bruijn} graphs.
\newblock {\em Bioinformatics}, 33(6):799--806, March 2017.

\bibitem{bao_halc_2017}
Ergude Bao and Lingxiao Lan.
\newblock {HALC}: {High} throughput algorithm for long read error correction.
\newblock {\em BMC Bioinformatics}, 18(1):204, April 2017.

\bibitem{haghshenas_colormap_2016}
Ehsan Haghshenas, Faraz Hach, S~Cenk Sahinalp, and Cedric Chauve.
\newblock {CoLoRMap}: {Correcting} {Long} {Reads} by {Mapping} short reads.
\newblock {\em Bioinformatics}, 32(17):i545--i551, September 2016.

\bibitem{goodwin_oxford_2015}
Sara Goodwin, James Gurtowski, Scott Ethe-Sayers, Panchajanya Deshpande, Michael~C. Schatz, and W.~Richard McCombie.
\newblock Oxford {Nanopore} sequencing, hybrid error correction, and de novo assembly of a eukaryotic genome.
\newblock {\em Genome Research}, 25(11):1750--1756, November 2015.

\bibitem{hackl_proovread_2014}
Thomas Hackl, Rainer Hedrich, Jörg Schultz, and Frank Förster.
\newblock proovread : large-scale high-accuracy {PacBio} correction through iterative short read consensus.
\newblock {\em Bioinformatics}, 30(21):3004--3011, November 2014.

\bibitem{hu_lscplus_2016}
Ruifeng Hu, Guibo Sun, and Xiaobo Sun.
\newblock {LSCplus}: a fast solution for improving long read accuracy by short read alignment.
\newblock {\em BMC Bioinformatics}, 17(1):451, November 2016.

\bibitem{salmela_correcting_2011}
Leena Salmela and Jan Schröder.
\newblock Correcting errors in short reads by multiple alignments.
\newblock {\em Bioinformatics}, 27(11):1455--1461, June 2011.

\bibitem{schroder_shrec_2009}
Jan Schröder, Heiko Schröder, Simon~J. Puglisi, Ranjan Sinha, and Bertil Schmidt.
\newblock {SHREC}: a short-read error correction method.
\newblock {\em Bioinformatics}, 25(17):2157--2163, September 2009.

\bibitem{salmela_correction_2010}
Leena Salmela.
\newblock Correction of sequencing errors in a mixed set of reads.
\newblock {\em Bioinformatics}, 26(10):1284--1290, May 2010.

\bibitem{koren_hybrid_2012}
Sergey Koren, Michael~C Schatz, Brian~P Walenz, Jeffrey Martin, Jason~T Howard, Ganeshkumar Ganapathy, Zhong Wang, David~A Rasko, W~Richard McCombie, Erich~D Jarvis, and Adam~M Phillippy.
\newblock Hybrid error correction and de novo assembly of single-molecule sequencing reads.
\newblock {\em Nature Biotechnology}, 30(7):693--700, July 2012.

\bibitem{au_improving_2012}
Kin~Fai Au, Jason~G. Underwood, Lawrence Lee, and Wing~Hung Wong.
\newblock Improving {PacBio} {Long} {Read} {Accuracy} by {Short} {Read} {Alignment}.
\newblock {\em PLOS ONE}, 7(10):e46679, October 2012.

\bibitem{samarakoon_leveraging_2024}
Hiruna Samarakoon, Yuk~Kei Wan, Sri Parameswaran, Jonathan Göke, Hasindu Gamaarachchi, and Ira~W. Deveson.
\newblock Leveraging {Basecaller}’s {Move} {Table} to {Generate} a {Lightweight} k-mer {Model}.
\newblock {\em bioRxiv}, 2024.

\bibitem{bhattacharya_molecular_2012}
Swati Bhattacharya, Ian~M. Derrington, Mikhail Pavlenok, Michael Niederweis, Jens~H. Gundlach, and Aleksei Aksimentiev.
\newblock Molecular {Dynamics} {Study} of {MspA} {Arginine} {Mutants} {Predicts} {Slow} {DNA} {Translocations} and {Ion} {Current} {Blockades} {Indicative} of {DNA} {Sequence}.
\newblock {\em ACS Nano}, 6(8):6960--6968, August 2012.

\bibitem{kawano_controlling_2009}
Ryuji Kawano, Anna E.~P. Schibel, Christopher Cauley, and Henry~S. White.
\newblock Controlling the {Translocation} of {Single}-{Stranded} {DNA} through a-{Hemolysin} {Ion} {Channels} {Using} {Viscosity}.
\newblock {\em Langmuir}, 25(2):1233--1237, January 2009.

\bibitem{cock2009sanger}
Peter J.~A. Cock, Christopher~J. Fields, Naohisa Goto, Michael~L. Heuer, and Peter~M. Rice.
\newblock {The Sanger FASTQ file format for sequences with quality scores, and the Solexa/Illumina FASTQ variants}.
\newblock {\em Nucleic Acids Research}, 2009.

\bibitem{Samarakoon2023accelerated}
Hiruna Samarakoon, James~M Ferguson, Hasindu Gamaarachchi, and Ira~W Deveson.
\newblock {Accelerated nanopore basecalling with SLOW5 data format}.
\newblock {\em Bioinformatics}, 2023.

\bibitem{cavlak2022targetcall}
Meryem~Banu Cavlak, Gagandeep Singh, Mohammed Alser, Can Firtina, Jo{\"e}l Lindegger, Mohammad Sadrosadati, Nika~Mansouri Ghiasi, Can Alkan, and Onur Mutlu.
\newblock {TargetCall: Eliminating the Wasted Computation in Basecalling via Pre-Basecalling Filtering}.
\newblock In {\em APBC}, 2023.

\bibitem{cavlak_targetcall_2024}
Meryem~Banu Cavlak, Gagandeep Singh, Mohammed Alser, Can Firtina, Joel Lindegger, Mohammad Sadrosadati, Nika~Mansouri Ghiasi, Can Alkan, and Onur Mutlu.
\newblock {TargetCall}: {Eliminating} the {Wasted} {Computation} in {Basecalling} via {Pre}-{Basecalling} {Filtering}.
\newblock {\em Frontiers in Genetics}, 2024.

\bibitem{xu2021fast}
Zhimeng Xu, Yuting Mai, Denghui Liu, Wenjun He, Xinyuan Lin, Chi Xu, Lei Zhang, Xin Meng, Joseph Mafofo, Walid~Abbas Zaher, and {others}.
\newblock Fast-bonito: {A} {Faster} {Deep} {Learning} {Based} {Basecaller} for {Nanopore} {Sequencing}.
\newblock {\em Artificial Intelligence in the Life Sciences}, 2021.

\bibitem{peresini2021nanopore}
Peter Pere\u{s}\'{i}ni, Vladim\'{i}r Bo\u{z}a, Bro\u{n}a Brejov\'{a}, and Tom\'{a}\u{s} Vina\u{r}.
\newblock {Nanopore Base Calling on the Edge}.
\newblock {\em Bioinformatics}, 2021.

\bibitem{boza_deepnano_2017}
Vladimír Boža, Broňa Brejová, and Tomáš Vinař.
\newblock {DeepNano}: {Deep} recurrent neural networks for base calling in {MinION} nanopore reads.
\newblock {\em PLOS One}, 2017.

\bibitem{boza_deepnano-blitz_2020}
Vladimír Boža, Peter Perešíni, Broňa Brejová, and Tomáš Vinař.
\newblock {DeepNano}-blitz: a fast base caller for {MinION} nanopore sequencers.
\newblock {\em Bioinformatics}, 2020.

\bibitem{oxford_nanopore_technologies_dorado_2024}
{Oxford Nanopore Technologies}.
\newblock Dorado, 2024.

\bibitem{oxford_nanopore_technologies_guppy_2017}
{Oxford Nanopore Technologies}.
\newblock Guppy, 2017.

\bibitem{lv_end--end_2020}
Xuan Lv, Zhiguang Chen, Yutong Lu, and Yuedong Yang.
\newblock An end-to-end {Oxford} nanopore basecaller using convolution-augmented transformer.
\newblock In {\em {BIBM}}, 2020.

\bibitem{singh2024rubicon}
Gagandeep Singh, Mohammed Alser, Kristof Denolf, Can Firtina, Alireza Khodamoradi, Meryem~Banu Cavlak, Henk Corporaal, and Onur Mutlu.
\newblock {RUBICON: a framework for designing efficient deep learning-based genomic basecallers}.
\newblock {\em Genome Biology}, 2024.

\bibitem{zhang_nanopore_2020}
Yao-zhong Zhang, Arda Akdemir, Georg Tremmel, Seiya Imoto, Satoru Miyano, Tetsuo Shibuya, and Rui Yamaguchi.
\newblock Nanopore basecalling from a perspective of instance segmentation.
\newblock {\em BMC Bioinformatics}, 2020.

\bibitem{xu_lokatt_2023}
Xuechun Xu, Nayanika Bhalla, Patrik Ståhl, and Joakim Jaldén.
\newblock Lokatt: a hybrid {DNA} nanopore basecaller with an explicit duration hidden {Markov} model and a residual {LSTM} network.
\newblock {\em BMC Bioinformatics}, 2023.

\bibitem{zeng_causalcall_2020}
Jingwen Zeng, Hongmin Cai, Hong Peng, Haiyan Wang, Yue Zhang, and Tatsuya Akutsu.
\newblock Causalcall: {Nanopore} {Basecalling} {Using} a {Temporal} {Convolutional} {Network}.
\newblock {\em Frontiers in Genetics}, 2020.

\bibitem{teng_chiron_2018}
Haotian Teng, Minh~Duc Cao, Michael~B Hall, Tania Duarte, Sheng Wang, and Lachlan J~M Coin.
\newblock Chiron: translating nanopore raw signal directly into nucleotide sequence using deep learning.
\newblock {\em GigaScience}, 2018.

\bibitem{konishi_halcyon_2021}
Hiroki Konishi, Rui Yamaguchi, Kiyoshi Yamaguchi, Yoichi Furukawa, and Seiya Imoto.
\newblock Halcyon: an accurate basecaller exploiting an encoder–decoder model with monotonic attention.
\newblock {\em Bioinformatics}, 2021.

\bibitem{yeh_msrcall_2022}
Yang-Ming Yeh and Yi-Chang Lu.
\newblock {MSRCall}: a multi-scale deep neural network to basecall {Oxford} {Nanopore} sequences.
\newblock {\em Bioinformatics}, 2022.

\bibitem{noordijk_baseless_2023}
Ben Noordijk, Reindert Nijland, Victor~J Carrion, Jos~M Raaijmakers, Dick de~Ridder, and Carlos de~Lannoy.
\newblock {baseLess}: lightweight detection of sequences in raw {MinION} data.
\newblock {\em Bioinformatics Advances}, 2023.

\bibitem{huang_sacall_2022}
Neng Huang, Fan Nie, Peng Ni, Feng Luo, and Jianxin Wang.
\newblock {SACall}: {A} {Neural} {Network} {Basecaller} for {Oxford} {Nanopore} {Sequencing} {Data} {Based} on {Self}-{Attention} {Mechanism}.
\newblock {\em IEEE/ACM TCBB}, 2022.

\bibitem{miculinic_mincall_2019}
Neven Miculinic, Marko Ratkovic, and Mile Sikic.
\newblock {MinCall} - {MinION} end2end convolutional deep learning basecaller.
\newblock {\em arXiv}, 2019.

\bibitem{loman_complete_2015}
Nicholas~J Loman, Joshua Quick, and Jared~T Simpson.
\newblock A complete bacterial genome assembled de novo using only nanopore sequencing data.
\newblock {\em Nature Methods}, 2015.

\bibitem{david_nanocall_2017}
Matei David, L~J Dursi, Delia Yao, Paul~C Boutros, and Jared~T Simpson.
\newblock Nanocall: an open source basecaller for {Oxford} {Nanopore} sequencing data.
\newblock {\em Bioinformatics}, 2017.

\bibitem{timp_dna_2012}
Winston Timp, Jeffrey Comer, and Aleksei Aksimentiev.
\newblock {DNA} {Base}-{Calling} from a {Nanopore} {Using} a {Viterbi} {Algorithm}.
\newblock {\em Biophysical Journal}, 2012.

\bibitem{schreiber_analysis_2015}
Jacob Schreiber and Kevin Karplus.
\newblock Analysis of nanopore data using hidden {Markov} models.
\newblock {\em Bioinformatics}, 2015.

\bibitem{ewing1998base}
Brent Ewing and Phil Green.
\newblock {Base-Calling of Automated Sequencer Traces Using Phred. II. Error Probabilities}.
\newblock {\em Genome Research}, 1998.

\bibitem{illuminafastq}
Illumina.
\newblock {Sequence file formats for a variety of data analysis options}.
\newblock \url{https://www.illumina.com/informatics/sequencing-data-analysis/sequence-file-formats.html}, 2024.

\bibitem{langmead2012fast}
Ben Langmead and Steven~L Salzberg.
\newblock {Fast gapped-read alignment with Bowtie 2}.
\newblock {\em Nature Methods}, 2012.

\bibitem{li2013aligningsequencereadsclone}
Heng Li.
\newblock {Aligning sequence reads, clone sequences and assembly contigs with BWA-MEM}.
\newblock {\em arXiv}, 2013.

\bibitem{altschul1990basic}
Stephen~F Altschul, Warren Gish, Webb Miller, Eugene~W Myers, and David~J Lipman.
\newblock {Basic Local Alignment Search Tool}.
\newblock {\em Journal of Molecular Biology}, 1990.

\bibitem{schleimer2003winnowing}
Saul Schleimer, Daniel~S Wilkerson, and Alex Aiken.
\newblock {Winnowing: Local Algorithms for Document Fingerprinting}.
\newblock In {\em ACM SIGMOD}, 2003.

\bibitem{roberts2004reducing}
Michael Roberts, Wayne Hayes, Brian~R Hunt, Stephen~M Mount, and James~A Yorke.
\newblock {Reducing Storage Requirements for Biological Sequence Comparison}.
\newblock {\em Bioinformatics}, 2004.

\bibitem{marccais2017improving}
Guillaume Mar{\c{c}}ais, David Pellow, Daniel Bork, Yaron Orenstein, Ron Shamir, and Carl Kingsford.
\newblock {Improving the Performance of Minimizers and Winnowing Schemes}.
\newblock {\em Bioinformatics}, 2017.

\bibitem{li2016minimap}
Heng Li.
\newblock {Minimap and Miniasm: Fast Mapping and De Novo Assembly for Noisy Long Sequences}.
\newblock {\em Bioinformatics}, 2016.

\bibitem{baeza-yates_new_1992}
Ricardo Baeza-Yates and Gaston~H. Gonnet.
\newblock A {New} {Approach} to {Text} {Searching}.
\newblock {\em Commun. ACM}, 1992.

\bibitem{papamichail_improved_2009}
Dimitris Papamichail and Georgios Papamichail.
\newblock Improved algorithms for approximate string matching (extended abstract).
\newblock {\em BMC Bioinformatics}, 2009.

\bibitem{suzuki_introducing_2018}
Hajime Suzuki and Masahiro Kasahara.
\newblock Introducing difference recurrence relations for faster semi-global alignment of long sequences.
\newblock {\em BMC Bioinformatics}, 2018.

\bibitem{waterman_biological_1976}
M.S Waterman, T.F Smith, and W.A Beyer.
\newblock Some biological sequence metrics.
\newblock {\em Advances in Mathematics}, 1976.

\bibitem{wu_onp_1990}
Sun Wu, Udi Manber, Gene Myers, and Webb Miller.
\newblock An {O}({NP}) sequence comparison algorithm.
\newblock {\em Information Processing Letters}, 1990.

\bibitem{wu_fast_1992}
Sun Wu and Udi Manber.
\newblock Fast text searching: allowing errors.
\newblock {\em Commun. ACM}, 1992.

\bibitem{wagner_string--string_1974}
Robert~A. Wagner and Michael~J. Fischer.
\newblock The {String}-to-{String} {Correction} {Problem}.
\newblock {\em J. ACM}, 1974.

\bibitem{sankoff_matching_1972}
David Sankoff.
\newblock Matching {Sequences} under {Deletion}/{Insertion} {Constraints}.
\newblock {\em Proceedings of the National Academy of Sciences}, 1972.

\bibitem{groot_koerkamp_exact_2024}
Ragnar Groot~Koerkamp and Pesho Ivanov.
\newblock Exact global alignment using {A}* with chaining seed heuristic and match pruning.
\newblock {\em Bioinformatics}, 2024.

\bibitem{sellers_theory_1974}
Peter~H. Sellers.
\newblock On the {Theory} and {Computation} of {Evolutionary} {Distances}.
\newblock {\em SIAM Journal on Applied Mathematics}, 1974.

\bibitem{ukkonen_algorithms_1985}
Esko Ukkonen.
\newblock Algorithms for approximate string matching.
\newblock {\em International Conference on Foundations of Computation Theory}, 1985.

\bibitem{10002015global}
{1000 Genomes Project Consortium}, Adam Auton, Lisa~D Brooks, Richard~M Durbin, Erik~P Garrison, Hyun~Min Kang, Jan~O Korbel, Jonathan~L Marchini, Shane McCarthy, Gil~A McVean, and Gon{\c c}alo~R Abecasis.
\newblock {A Global Reference for Human Genetic Variation}.
\newblock {\em Nature}, 2015.

\bibitem{uk10k2015uk10k}
{UK10K Consortium}, Klaudia Walter, Josine~L Min, Jie Huang, Lucy Crooks, Yasin Memari, Shane McCarthy, John R~B Perry, Changjiang Xu, Marta Futema, Daniel Lawson, Valentina Iotchkova, Stephan Schiffels, Audrey~E Hendricks, Petr Danecek, Rui Li, James Floyd, Louise~V Wain, In{\^e}s Barroso, Steve~E Humphries, et~al.
\newblock {The UK10K Project Identifies Rare Variants in Health and Disease}.
\newblock {\em Nature}, 2015.

\bibitem{alkan_genome_2011}
Can Alkan, Bradley~P. Coe, and Evan~E. Eichler.
\newblock Genome structural variation discovery and genotyping.
\newblock {\em Nature Reviews Genetics}, 2011.

\bibitem{Sedlazeck2018}
Fritz~J. Sedlazeck, Philipp Rescheneder, Moritz Smolka, Han Fang, Maria Nattestad, Arndt von Haeseler, and Michael~C. Schatz.
\newblock {Accurate detection of complex structural variations using single-molecule sequencing}.
\newblock {\em Nature Methods}, 2018.

\bibitem{poplin_scaling_2018}
Ryan Poplin, Valentin Ruano-Rubio, Mark~A. DePristo, Tim~J. Fennell, Mauricio~O. Carneiro, Geraldine~A. Van~der Auwera, David~E. Kling, Laura~D. Gauthier, Ami Levy-Moonshine, David Roazen, Khalid Shakir, Joel Thibault, Sheila Chandran, Chris Whelan, Monkol Lek, Stacey Gabriel, Mark~J Daly, Ben Neale, Daniel~G. MacArthur, and Eric Banks.
\newblock Scaling accurate genetic variant discovery to tens of thousands of samples.
\newblock {\em bioRxiv}, 2018.

\bibitem{weckx_novosnp_2005}
Stefan Weckx, Jurgen Del-Favero, Rosa Rademakers, Lieve Claes, Marc Cruts, Peter De~Jonghe, Christine Van~Broeckhoven, and Peter De~Rijk.
\newblock {novoSNP}, a novel computational tool for sequence variation discovery.
\newblock {\em Genome Research}, 2005.

\bibitem{kwok_comparative_1994}
Pui-Yan Kwok, Christopher Carlson, Thomas~D. Yager, Wendy Ankener, and Deborah~A. Nickerson.
\newblock Comparative {Analysis} of {Human} {DNA} {Variations} by {Fluorescence}-{Based} {Sequencing} of {PCR} {Products}.
\newblock {\em Genomics}, 1994.

\bibitem{nickerson_polyphred_1997}
Deborah~A. Nickerson, Vincent~O. Tobe, and Scott~L. Taylor.
\newblock {PolyPhred}: automating the detection and genotyping of single nucleotide substitutions using fluorescence-based resequencing.
\newblock {\em Nucleic Acids Research}, 1997.

\bibitem{marth_general_1999}
Gabor~T. Marth, Ian Korf, Mark~D. Yandell, Raymond~T. Yeh, Zhijie Gu, Hamideh Zakeri, Nathan~O. Stitziel, LaDeana Hillier, Pui-Yan Kwok, and Warren~R. Gish.
\newblock A general approach to single-nucleotide polymorphism discovery.
\newblock {\em Nature Genetics}, 1999.

\bibitem{Poplin2018}
Ryan Poplin, Pi-Chuan Chang, David Alexander, Scott Schwartz, Thomas Colthurst, Alexander Ku, Dan Newburger, Jojo Dijamco, Nam Nguyen, Pegah~T. Afshar, Sam~S. Gross, Lizzie Dorfman, Cory~Y. McLean, and Mark~A. DePristo.
\newblock {A universal SNP and small-indel variant caller using deep neural networks}.
\newblock {\em Nature Biotechnology}, 2018.

\bibitem{conrad_origins_2010}
Donald~F. Conrad, Dalila Pinto, Richard Redon, Lars Feuk, Omer Gokcumen, Yujun Zhang, Jan Aerts, T.~Daniel Andrews, Chris Barnes, Peter Campbell, Tomas Fitzgerald, Min Hu, Chun~Hwa Ihm, Kati Kristiansson, Daniel~G. MacArthur, Jeffrey~R. MacDonald, Ifejinelo Onyiah, Andy Wing~Chun Pang, Sam Robson, Kathy Stirrups, et~al.
\newblock Origins and functional impact of copy number variation in the human genome.
\newblock {\em Nature}, 2010.

\bibitem{mills_mapping_2011}
Ryan~E. Mills, Klaudia Walter, Chip Stewart, Robert~E. Handsaker, Ken Chen, Can Alkan, Alexej Abyzov, Seungtai~Chris Yoon, Kai Ye, R.~Keira Cheetham, Asif Chinwalla, Donald~F. Conrad, Yutao Fu, Fabian Grubert, Iman Hajirasouliha, Fereydoun Hormozdiari, Lilia~M. Iakoucheva, Zamin Iqbal, Shuli Kang, Jeffrey~M. Kidd, et~al.
\newblock Mapping copy number variation by population-scale genome sequencing.
\newblock {\em Nature}, 2011.

\bibitem{cooper_systematic_2008}
Gregory~M Cooper, Troy Zerr, Jeffrey~M Kidd, Evan~E Eichler, and Deborah~A Nickerson.
\newblock Systematic assessment of copy number variant detection via genome-wide {SNP} genotyping.
\newblock {\em Nature Genetics}, 2008.

\bibitem{eichler_widening_2006}
Evan~E Eichler.
\newblock Widening the spectrum of human genetic variation.
\newblock {\em Nature Genetics}, 2006.

\bibitem{cameron_comprehensive_2019}
Daniel~L. Cameron, Leon Di~Stefano, and Anthony~T. Papenfuss.
\newblock Comprehensive evaluation and characterisation of short read general-purpose structural variant calling software.
\newblock {\em Nature Communications}, 2019.

\bibitem{han_functional_2020}
Lide Han, Xuefang Zhao, Mary~Lauren Benton, Thaneer Perumal, Ryan~L. Collins, Gabriel~E. Hoffman, Jessica~S. Johnson, Laura Sloofman, Harold~Z. Wang, Matthew~R. Stone, Schahram Akbarian, Jaroslav Bendl, Michael Breen, Kristen~J. Brennand, Leanne Brown, Andrew Browne, Joseph~D. Buxbaum, Alexander Charney, Andrew Chess, Lizette Couto, et~al.
\newblock Functional annotation of rare structural variation in the human brain.
\newblock {\em Nature Communications}, 2020.

\bibitem{mandiracioglu_ecole_2024}
Berk Mandiracioglu, Furkan Ozden, Gun Kaynar, Mehmet~Alper Yilmaz, Can Alkan, and A.~Ercument Cicek.
\newblock {ECOLE}: {Learning} to call copy number variants on whole exome sequencing data.
\newblock {\em Nature Communications}, 2024.

\bibitem{dou_accurate_2020}
Yanmei Dou, Minseok Kwon, Rachel~E. Rodin, Isidro Cortés-Ciriano, Ryan Doan, Lovelace~J. Luquette, Alon Galor, Craig Bohrson, Christopher~A. Walsh, and Peter~J. Park.
\newblock Accurate detection of mosaic variants in sequencing data without matched controls.
\newblock {\em Nature Biotechnology}, 2020.

\bibitem{smolka_detection_2024}
Moritz Smolka, Luis~F. Paulin, Christopher~M. Grochowski, Dominic~W. Horner, Medhat Mahmoud, Sairam Behera, Ester Kalef-Ezra, Mira Gandhi, Karl Hong, Davut Pehlivan, Sonja~W. Scholz, Claudia M.~B. Carvalho, Christos Proukakis, and Fritz~J. Sedlazeck.
\newblock Detection of mosaic and population-level structural variants with {Sniffles2}.
\newblock {\em Nature Biotechnology}, 2024.

\bibitem{lin_svision_2022}
Jiadong Lin, Songbo Wang, Peter~A. Audano, Deyu Meng, Jacob~I. Flores, Walter Kosters, Xiaofei Yang, Peng Jia, Tobias Marschall, Christine~R. Beck, and Kai Ye.
\newblock {SVision}: a deep learning approach to resolve complex structural variants.
\newblock {\em Nature Methods}, 2022.

\bibitem{popic_cue_2023}
Victoria Popic, Chris Rohlicek, Fabio Cunial, Iman Hajirasouliha, Dmitry Meleshko, Kiran Garimella, and Anant Maheshwari.
\newblock Cue: a deep-learning framework for structural variant discovery and genotyping.
\newblock {\em Nature Methods}, 2023.

\bibitem{narzisi_genome-wide_2018}
Giuseppe Narzisi, André Corvelo, Kanika Arora, Ewa~A. Bergmann, Minita Shah, Rajeeva Musunuri, Anne-Katrin Emde, Nicolas Robine, Vladimir Vacic, and Michael~C. Zody.
\newblock Genome-wide somatic variant calling using localized colored de {Bruijn} graphs.
\newblock {\em Communications Biology}, 2018.

\bibitem{zheng_symphonizing_2022}
Zhenxian Zheng, Shumin Li, Junhao Su, Amy Wing-Sze Leung, Tak-Wah Lam, and Ruibang Luo.
\newblock Symphonizing pileup and full-alignment for deep learning-based long-read variant calling.
\newblock {\em Nature Computational Science}, 2022.

\bibitem{zhang_improved_2012}
Jin Zhang, Jiayin Wang, and Yufeng Wu.
\newblock An improved approach for accurate and efficient calling of structural variations with low-coverage sequence data.
\newblock {\em BMC Bioinformatics}, 2012.

\bibitem{layer_lumpy_2014}
Ryan~M. Layer, Colby Chiang, Aaron~R. Quinlan, and Ira~M. Hall.
\newblock {LUMPY}: a probabilistic framework for structural variant discovery.
\newblock {\em Genome Biology}, 2014.

\bibitem{karaoglanoglu_valor2_2020}
Fatih Karaoğlanoğlu, Camir Ricketts, Ezgi Ebren, Marzieh~Eslami Rasekh, Iman Hajirasouliha, and Can Alkan.
\newblock {VALOR2}: characterization of large-scale structural variants using linked-reads.
\newblock {\em Genome Biology}, 2020.

\bibitem{jiang_long-read-based_2020}
Tao Jiang, Yongzhuang Liu, Yue Jiang, Junyi Li, Yan Gao, Zhe Cui, Yadong Liu, Bo~Liu, and Yadong Wang.
\newblock Long-read-based human genomic structural variation detection with {cuteSV}.
\newblock {\em Genome Biology}, 2020.

\bibitem{ahsan_nanocaller_2021}
Mian~Umair Ahsan, Qian Liu, Li~Fang, and Kai Wang.
\newblock {NanoCaller} for accurate detection of {SNPs} and indels in difficult-to-map regions from long-read sequencing by haplotype-aware deep neural networks.
\newblock {\em Genome Biology}, 2021.

\bibitem{minoche_clinsv_2021}
Andre~E. Minoche, Ben Lundie, Greg~B. Peters, Thomas Ohnesorg, Mark Pinese, David~M. Thomas, Andreas Zankl, Tony Roscioli, Nicole Schonrock, Sarah Kummerfeld, Leslie Burnett, Marcel~E. Dinger, and Mark~J. Cowley.
\newblock {ClinSV}: clinical grade structural and copy number variant detection from whole genome sequencing data.
\newblock {\em Genome Medicine}, 2021.

\bibitem{baird_rapid_2008}
Nathan~A. Baird, Paul~D. Etter, Tressa~S. Atwood, Mark~C. Currey, Anthony~L. Shiver, Zachary~A. Lewis, Eric~U. Selker, William~A. Cresko, and Eric~A. Johnson.
\newblock Rapid {SNP} {Discovery} and {Genetic} {Mapping} {Using} {Sequenced} {RAD} {Markers}.
\newblock {\em PLOS ONE}, 2008.

\bibitem{zarate_parliament2_2020}
Samantha Zarate, Andrew Carroll, Medhat Mahmoud, Olga Krasheninina, Goo Jun, William~J Salerno, Michael~C Schatz, Eric Boerwinkle, Richard~A Gibbs, and Fritz~J Sedlazeck.
\newblock Parliament2: {Accurate} structural variant calling at scale.
\newblock {\em GigaScience}, 2020.

\bibitem{odonnell_mumco_2020}
Samuel O’Donnell and Gilles Fischer.
\newblock {MUM}\&{Co}: accurate detection of all {SV} types through whole-genome alignment.
\newblock {\em Bioinformatics}, 2020.

\bibitem{xu_smcounter2_2019}
Chang Xu, Xiujing Gu, Raghavendra Padmanabhan, Zhong Wu, Quan Peng, John DiCarlo, and Yexun Wang.
\newblock {smCounter2}: an accurate low-frequency variant caller for targeted sequencing data with unique molecular identifiers.
\newblock {\em Bioinformatics}, 2019.

\bibitem{heller_svim_2019}
David Heller and Martin Vingron.
\newblock {SVIM}: structural variant identification using mapped long reads.
\newblock {\em Bioinformatics}, 2019.

\bibitem{pedersen_cyvcf2_2017}
Brent~S Pedersen and Aaron~R Quinlan.
\newblock cyvcf2: fast, flexible variant analysis with {Python}.
\newblock {\em Bioinformatics}, 2017.

\bibitem{li_fermikit_2015}
Heng Li.
\newblock {FermiKit}: assembly-based variant calling for {Illumina} resequencing data.
\newblock {\em Bioinformatics}, 2015.

\bibitem{eisfeldt_tiddit_2017}
J~Eisfeldt, F~Vezzi, P~Olason, D~Nilsson, and A~Lindstrand.
\newblock {TIDDIT}, an efficient and comprehensive structural variant caller for massive parallel sequencing data [version 2; peer review: 2 approved].
\newblock {\em F1000Research}, 2017.

\bibitem{zheng_svsearcher_2023}
Yan Zheng, Xuequn Shang, and Wing-Kin Sung.
\newblock {SVsearcher}: {A} more accurate structural variation detection method in long read data.
\newblock {\em Computers in Biology and Medicine}, 2023.

\bibitem{medvedev_detecting_2010}
Paul Medvedev, Marc Fiume, Misko Dzamba, Tim Smith, and Michael Brudno.
\newblock Detecting copy number variation with mated short reads.
\newblock {\em Genome Research}, 2010.

\bibitem{garrison_haplotype-based_2012}
Erik Garrison and Gabor Marth.
\newblock Haplotype-based variant detection from short-read sequencing.
\newblock {\em arXiv}, 2012.

\bibitem{Olson2023}
Nathan~D. Olson, Justin Wagner, Nathan Dwarshuis, Karen~H. Miga, Fritz~J. Sedlazeck, Marc Salit, and Justin~M. Zook.
\newblock Variant calling and benchmarking in an era of complete human genome sequences.
\newblock {\em Nature Reviews Genetics}, 2023.

\bibitem{li2008mapping}
Heng Li, Jue Ruan, and Richard Durbin.
\newblock {Mapping short DNA sequencing reads and calling variants using mapping quality scores}.
\newblock {\em Genome Research}, 2008.

\bibitem{Yu2015}
Y.~William Yu, Deniz Yorukoglu, Jian Peng, and Bonnie Berger.
\newblock Quality score compression improves genotyping accuracy.
\newblock {\em Nature Biotechnology}, 2015.

\bibitem{Park2025}
Jimin Park, Daniel~E. Cook, Pi-Chuan Chang, Alexey Kolesnikov, Lucas Brambrink, Juan~Carlos Mier, Joshua Gardner, Brandy McNulty, Samuel Sacco, Ayse~G. Keskus, Asher Bryant, Tanveer Ahmad, Jyoti Shetty, Yongmei Zhao, Bao Tran, Giuseppe Narzisi, Adrienne Helland, Byunggil Yoo, Irina Pushel, Lisa~A. Lansdon, et~al.
\newblock {Accurate somatic small variant discovery for multiple sequencing technologies with DeepSomatic}.
\newblock {\em Nature Biotechnology}, 2025.

\bibitem{dunn2023n}
Tim Dunn, David Blaauw, Reetuparna Das, and Satish Narayanasamy.
\newblock n pore: n-polymer realigner for improved pileup-based variant calling.
\newblock {\em BMC bioinformatics}, 2023.

\bibitem{wu2020high}
Xiao Wu, Arun Subramaniyan, Zhehong Wang, Satish Narayanasamy, Reetuparna Das, and David Blaauw.
\newblock A high-throughput pruning-based pair-hidden-markov-model hardware accelerator for next-generation dna sequencing.
\newblock {\em IEEE Solid-State Circuits Letters}, 2020.

\bibitem{kececioglu_combinatorial_1995}
J.~D. Kececioglu and E.~W. Myers.
\newblock Combinatorial algorithms for {DNA} sequence assembly.
\newblock {\em Algorithmica}, 1995.

\bibitem{Cheng2021}
Haoyu Cheng, Gregory~T. Concepcion, Xiaowen Feng, Haowen Zhang, and Heng Li.
\newblock {Haplotype-resolved de novo assembly using phased assembly graphs with hifiasm}.
\newblock {\em Nature Methods}, 2021.

\bibitem{ekim_minimizer-space_2021}
Barış Ekim, Bonnie Berger, and Rayan Chikhi.
\newblock Minimizer-space de {Bruijn} graphs: {Whole}-genome assembly of long reads in minutes on a personal computer.
\newblock {\em Cell Systems}, 2021.

\bibitem{nurk_hicanu_2020}
Sergey Nurk, Brian~P. Walenz, Arang Rhie, Mitchell~R. Vollger, Glennis~A. Logsdon, Robert Grothe, Karen~H. Miga, Evan~E. Eichler, Adam~M. Phillippy, and Sergey Koren.
\newblock {HiCanu}: accurate assembly of segmental duplications, satellites, and allelic variants from high-fidelity long reads.
\newblock {\em Genome Research}, 2020.

\bibitem{fleischmann_whole-genome_1995}
Robert~D. Fleischmann, Mark~D. Adams, Owen White, Rebecca~A. Clayton, Ewen~F. Kirkness, Anthony~R. Kerlavage, Carol~J. Bult, Jean-Francois Tomb, Brian~A. Dougherty, Joseph~M. Merrick, Keith McKenney, Granger Sutton, Will FitzHugh, Chris Fields, Jeannine~D. Gocayne, John Scott, Robert Shirley, Li-lng Liu, Anna Glodek, Jenny~M. Kelley, et~al.
\newblock Whole-{Genome} {Random} {Sequencing} and {Assembly} of {Haemophilus} influenzae {Rd}.
\newblock {\em Science}, 1995.

\bibitem{myers_whole-genome_2000}
Eugene~W. Myers, Granger~G. Sutton, Art~L. Delcher, Ian~M. Dew, Dan~P. Fasulo, Michael~J. Flanigan, Saul~A. Kravitz, Clark~M. Mobarry, Knut H.~J. Reinert, Karin~A. Remington, Eric~L. Anson, Randall~A. Bolanos, Hui-Hsien Chou, Catherine~M. Jordan, Aaron~L. Halpern, Stefano Lonardi, Ellen~M. Beasley, Rhonda~C. Brandon, Lin Chen, Patrick~J. Dunn, et~al.
\newblock A {Whole}-{Genome} {Assembly} of {Drosophila}.
\newblock {\em Science}, 2000.

\bibitem{chin_phased_2016}
Chen-Shan Chin, Paul Peluso, Fritz~J Sedlazeck, Maria Nattestad, Gregory~T Concepcion, Alicia Clum, Christopher Dunn, Ronan O'Malley, Rosa Figueroa-Balderas, Abraham Morales-Cruz, Grant~R Cramer, Massimo Delledonne, Chongyuan Luo, Joseph~R Ecker, Dario Cantu, David~R Rank, and Michael~C Schatz.
\newblock Phased diploid genome assembly with single-molecule real-time sequencing.
\newblock {\em Nature Methods}, 2016.

\bibitem{chen_efficient_2021}
Ying Chen, Fan Nie, Shang-Qian Xie, Ying-Feng Zheng, Qi~Dai, Thomas Bray, Yao-Xin Wang, Jian-Feng Xing, Zhi-Jian Huang, De-Peng Wang, Li-Juan He, Feng Luo, Jian-Xin Wang, Yi-Zhi Liu, and Chuan-Le Xiao.
\newblock Efficient assembly of nanopore reads via highly accurate and intact error correction.
\newblock {\em Nature Communications}, 2021.

\bibitem{li_genome_2024}
Heng Li and Richard Durbin.
\newblock Genome assembly in the telomere-to-telomere era.
\newblock {\em Nature Reviews Genetics}, 2024.

\bibitem{kolmogorov2019assembly}
Mikhail Kolmogorov, Jeffrey Yuan, Yu~Lin, and Pavel~A. Pevzner.
\newblock {Assembly of long, error-prone reads using repeat graphs}.
\newblock {\em Nature Biotechnology}, 2019.

\bibitem{shafin_nanopore_2020}
Kishwar Shafin, Trevor Pesout, Ryan Lorig-Roach, Marina Haukness, Hugh~E. Olsen, Colleen Bosworth, Joel Armstrong, Kristof Tigyi, Nicholas Maurer, Sergey Koren, Fritz~J. Sedlazeck, Tobias Marschall, Simon Mayes, Vania Costa, Justin~M. Zook, Kelvin~J. Liu, Duncan Kilburn, Melanie Sorensen, Katy~M. Munson, Mitchell~R. Vollger, et~al.
\newblock Nanopore sequencing and the {Shasta} toolkit enable efficient de novo assembly of eleven human genomes.
\newblock {\em Nature Biotechnology}, 2020.

\bibitem{di_genova_efficient_2021}
Alex Di~Genova, Elena Buena-Atienza, Stephan Ossowski, and Marie-France Sagot.
\newblock Efficient hybrid de novo assembly of human genomes with {WENGAN}.
\newblock {\em Nature Biotechnology}, 2021.

\bibitem{bankevich_multiplex_2022}
Anton Bankevich, Andrey~V. Bzikadze, Mikhail Kolmogorov, Dmitry Antipov, and Pavel~A. Pevzner.
\newblock Multiplex de {Bruijn} graphs enable genome assembly from long, high-fidelity reads.
\newblock {\em Nature Biotechnology}, 2022.

\bibitem{Cheng2022}
Haoyu Cheng, Erich~D. Jarvis, Olivier Fedrigo, Klaus-Peter Koepfli, Lara Urban, Neil~J. Gemmell, and Heng Li.
\newblock {Haplotype-resolved assembly of diploid genomes without parental data}.
\newblock {\em Nature Biotechnology}, 2022.

\bibitem{ruan_fast_2020}
Jue Ruan and Heng Li.
\newblock Fast and accurate long-read assembly with wtdbg2.
\newblock {\em Nature Methods}, 2020.

\bibitem{cheng_scalable_2024}
Haoyu Cheng, Mobin Asri, Julian Lucas, Sergey Koren, and Heng Li.
\newblock Scalable telomere-to-telomere assembly for diploid and polyploid genomes with double graph.
\newblock {\em Nature Methods}, 2024.

\bibitem{eche_bos_2023}
Camille Eché, Carole Iampietro, Clément Birbes, Andreea Dréau, Claire Kuchly, Arnaud Di~Franco, Christophe Klopp, Thomas Faraut, Sarah Djebali, Adrien Castinel, Matthias Zytnicki, Erwan Denis, Mekki Boussaha, Cécile Grohs, Didier Boichard, Christine Gaspin, Denis Milan, and Cécile Donnadieu.
\newblock A {Bos} taurus sequencing methods benchmark for assembly, haplotyping, and variant calling.
\newblock {\em Scientific Data}, 2023.

\bibitem{vaser_time-_2021}
Robert Vaser and Mile Šikić.
\newblock Time- and memory-efficient genome assembly with {Raven}.
\newblock {\em Nature Computational Science}, 2021.

\bibitem{chen_accurate_2021}
Yu~Chen, Yixin Zhang, Amy~Y. Wang, Min Gao, and Zechen Chong.
\newblock Accurate long-read de novo assembly evaluation with {Inspector}.
\newblock {\em Genome Biology}, 2021.

\bibitem{pevzner_eulerian_2001}
Pavel~A. Pevzner, Haixu Tang, and Michael~S. Waterman.
\newblock An {Eulerian} path approach to {DNA} fragment assembly.
\newblock {\em Proceedings of the National Academy of Sciences}, 2001.

\bibitem{lin_assembly_2016}
Yu~Lin, Jeffrey Yuan, Mikhail Kolmogorov, Max~W. Shen, Mark Chaisson, and Pavel~A. Pevzner.
\newblock Assembly of long error-prone reads using de {Bruijn} graphs.
\newblock {\em Proceedings of the National Academy of Sciences}, 2016.

\bibitem{bonfield_new_1995}
James~K. Bonfield, Kathryn~F. Smith, and Rodger Staden.
\newblock A new {DNA} sequence assembly program.
\newblock {\em Nucleic Acids Research}, 1995.

\bibitem{peltola_seqaid_1984}
Hannu Peltola, Hans Söderlund, and Esko Ukkonen.
\newblock {SEQAID}: a {DNA} sequence assembling program based on a mathematical model.
\newblock {\em Nucleic Acids Research}, 1984.

\bibitem{kamath_hinge_2017}
Govinda~M. Kamath, Ilan Shomorony, Fei Xia, Thomas~A. Courtade, and David~N. Tse.
\newblock {HINGE}: long-read assembly achieves optimal repeat resolution.
\newblock {\em Genome Research}, 2017.

\bibitem{butler_allpaths_2008}
Jonathan Butler, Iain MacCallum, Michael Kleber, Ilya~A. Shlyakhter, Matthew~K. Belmonte, Eric~S. Lander, Chad Nusbaum, and David~B. Jaffe.
\newblock {ALLPATHS}: {De} novo assembly of whole-genome shotgun microreads.
\newblock {\em Genome Research}, 2008.

\bibitem{koren_canu_2017}
Sergey Koren, Brian~P. Walenz, Konstantin Berlin, Jason~R. Miller, Nicholas~H. Bergman, and Adam~M. Phillippy.
\newblock Canu: scalable and accurate long-read assembly via adaptive k-mer weighting and repeat separation.
\newblock {\em Genome Research}, 2017.

\bibitem{myers_fragment_2005}
Eugene~W. Myers.
\newblock The fragment assembly string graph.
\newblock {\em Bioinformatics}, 2005.

\bibitem{Haghshenas2020}
Ehsan Haghshenas, Hossein Asghari, Jens Stoye, Cedric Chauve, and Faraz Hach.
\newblock {HASLR: Fast Hybrid Assembly of Long Reads}.
\newblock {\em iScience}, 2020.

\bibitem{Zimin2017hybrid}
Aleksey~V. Zimin, Daniela Puiu, Ming-Cheng Luo, Tingting Zhu, Sergey Koren, Guillaume Marçais, James~A. Yorke, Jan Dvořák, and Steven~L. Salzberg.
\newblock {Hybrid assembly of the large and highly repetitive genome of Aegilops tauschii, a progenitor of bread wheat, with the MaSuRCA mega-reads algorithm}.
\newblock {\em Genome Research}, 2017.

\bibitem{gupta2025accurate}
Anshu Gupta, Siavash Mirarab, and Yatish Turakhia.
\newblock {Accurate, scalable, and fully automated inference of species trees from raw genome assemblies using ROADIES}.
\newblock {\em Proceedings of the National Academy of Sciences}, 2025.

\bibitem{firtina2020apollo}
Can Firtina, Jeremie~S Kim, Mohammed Alser, Damla Senol~Cali, A~Ercument Cicek, Can Alkan, and Onur Mutlu.
\newblock {Apollo: A Sequencing-technology-independent, Scalable and Accurate Assembly Polishing Algorithm}.
\newblock {\em Bioinformatics}, 2020.

\bibitem{truong2015metaphlan2}
Duy~Tin Truong, Eric~A Franzosa, Timothy~L Tickle, Matthias Scholz, George Weingart, Edoardo Pasolli, Adrian Tett, Curtis Huttenhower, and Nicola Segata.
\newblock {MetaPhlAn2 for Enhanced Metagenomic Taxonomic Profiling}.
\newblock {\em Nature Methods}, 2015.

\bibitem{ounit2015clark}
Rachid Ounit, Steve Wanamaker, Timothy~J Close, and Stefano Lonardi.
\newblock {CLARK: Fast and Accurate Classification of Metagenomic and Genomic Sequences Using Discriminative K-mers}.
\newblock {\em BMC Genomics}, 2015.

\bibitem{milanese2019microbial}
Alessio Milanese, Daniel~R. Mende, Lucas Paoli, Guillem Salazar, Hans-Joachim Ruscheweyh, Miguelangel Cuenca, Pascal Hingamp, Renato Alves, Paul~I. Costea, Luis~Pedro Coelho, Thomas S.~B. Schmidt, Alexandre Almeida, Alex~L. Mitchell, Robert~D. Finn, Jaime Huerta-Cepas, Peer Bork, Georg Zeller, and Shinichi Sunagawa.
\newblock {Microbial abundance, activity and population genomic profiling with mOTUs2}.
\newblock {\em Nature Communications}, 2019.

\bibitem{shen2022kmcp}
Wei Shen, Hongyan Xiang, Tianquan Huang, Hui Tang, Mingli Peng, Dachuan Cai, Peng Hu, and Hong Ren.
\newblock {KMCP: accurate metagenomic profiling of both prokaryotic and viral populations by pseudo-mapping}.
\newblock {\em Bioinformatics}, 2022.

\bibitem{sun2021challenges}
Zheng Sun, Shi Huang, Meng Zhang, Qiyun Zhu, Niina Haiminen, Anna~Paola Carrieri, Yoshiki V{\'a}zquez-Baeza, Laxmi Parida, Ho-Cheol Kim, Rob Knight, and Yang-Yu Liu.
\newblock Challenges in benchmarking metagenomic profilers.
\newblock {\em Nature Methods}, 2021.

\bibitem{lu2017bracken}
Jennifer Lu, Florian~P Breitwieser, Peter Thielen, and Steven~L Salzberg.
\newblock {Bracken: Estimating Species Abundance in Metagenomics Data}.
\newblock {\em PeerJ Computer Science}, 2017.

\bibitem{dimopoulos2022haystac}
Evangelos~A. Dimopoulos, Alberto Carmagnini, Irina~M. Velsko, Christina Warinner, Greger Larson, Laurent A.~F. Frantz, and Evan~K. Irving-Pease.
\newblock {HAYSTAC: A Bayesian framework for robust and rapid species identification in high-throughput sequencing data}.
\newblock {\em PLOS Computational Biology}, 2022.

\bibitem{meyer2021critical}
Fernando Meyer, Adrian Fritz, Zhi-Luo Deng, David Koslicki, Till~Robin Lesker, Alexey Gurevich, Gary Robertson, Mohammed Alser, Dmitry Antipov, Francesco Beghini, Denis Bertrand, Jaqueline~J. Brito, C.~Titus Brown, Jan Buchmann, Aydin Bulu{\c{c}}, Bo~Chen, Rayan Chikhi, Philip T. L.~C. Clausen, Alexandru Cristian, Piotr~Wojciech Dabrowski, et~al.
\newblock {Critical Assessment of Metagenome Interpretation: the second round of challenges}.
\newblock {\em Nature Methods}, 2022.

\bibitem{vasimuddin2019efficient}
Md~Vasimuddin, Sanchit Misra, Heng Li, and Srinivas Aluru.
\newblock {Efficient Architecture-aware Acceleration of BWA-MEM for Multicore Systems}.
\newblock In {\em IPDPS}, 2019.

\bibitem{khayat2021hidden}
Michael~M Khayat, Sayed Mohammad~Ebrahim Sahraeian, Samantha Zarate, Andrew Carroll, Huixiao Hong, Bohu Pan, Leming Shi, Richard~A Gibbs, Marghoob Mohiyuddin, Yuanting Zheng, and Fritz~J Sedlazeck.
\newblock {Hidden Biases in Germline Structural Variant Detection}.
\newblock {\em Genome Biology}, 2021.

\bibitem{kostlbacher2021pangenomics}
Stephan K{\"o}stlbacher, Astrid Collingro, Tamara Halter, Frederik Schulz, Sean~P Jungbluth, and Matthias Horn.
\newblock {Pangenomics Reveals Alternative Environmental Lifestyles among Chlamydiae}.
\newblock {\em Nature Communications}, 2021.

\bibitem{vandorp2020emergence}
Lucy {van Dorp}, Mislav Acman, Damien Richard, Liam~P. Shaw, Charlotte~E. Ford, Louise Ormond, Christopher~J. Owen, Juanita Pang, Cedric~C.S. Tan, Florencia~A.T. Boshier, Arturo~Torres Ortiz, and François Balloux.
\newblock {Emergence of genomic diversity and recurrent mutations in SARS-CoV-2}.
\newblock {\em Infection, Genetics and Evolution}, 2020.

\bibitem{Logsdon2025}
Glennis~A. Logsdon, Peter Ebert, Peter~A. Audano, Mark Loftus, David Porubsky, Jana Ebler, Feyza Yilmaz, Pille Hallast, Timofey Prodanov, DongAhn Yoo, Carolyn~A. Paisie, William~T. Harvey, Xuefang Zhao, Gianni~V. Martino, Mir Henglin, Katherine~M. Munson, Keon Rabbani, Chen-Shan Chin, Bida Gu, Hufsah Ashraf, et~al.
\newblock Complex genetic variation in nearly complete human genomes.
\newblock {\em Nature}, 2025.

\bibitem{Zheng2017alignment}
Xiangqun Zheng-Bradley, Ian Streeter, Susan Fairley, David Richardson, Laura Clarke, Paul Flicek, and {The 1000 Genomes Project Consortium}.
\newblock {Alignment of 1000 Genomes Project reads to reference assembly GRCh38}.
\newblock {\em GigaScience}, 2017.

\bibitem{wadden2022ultra}
Jack Wadden, Brandon~S Newell, Joshua Bugbee, Vishal John, Amy~K Bruzek, Robert~P Dickson, Carl Koschmann, David Blaauw, Satish Narayanasamy, and Reetuparna Das.
\newblock Ultra-rapid somatic variant detection via real-time targeted amplicon sequencing.
\newblock {\em Communications Biology}, 2022.

\bibitem{national2007new}
{{National Research Council}} et~al.
\newblock {Why Metagenomics?}
\newblock In {\em {The New Science of Metagenomics: Revealing the Secrets of Our Microbial Planet}}. 2007.

\bibitem{cdcamd}
{Centers for Disease Control and Prevention}.
\newblock {AMD: Developing Faster Tests}.
\newblock \url{https://www.cdc.gov/amd/what-we-do/faster-tests.html}, 10 2019.

\bibitem{CheckHayden2015}
Erika Check~Hayden.
\newblock Genome researchers raise alarm over big data.
\newblock {\em Nature}, 2015.

\bibitem{Muggli2017SuccinctGraphs}
Martin~D Muggli, Alexander Bowe, Noelle~R Noyes, Paul~S Morley, Keith~E Belk, Robert Raymond, Travis Gagie, Simon~J Puglisi, and Christina Boucher.
\newblock {Succinct colored de Bruijn graphs}.
\newblock {\em Bioinformatics}, 2017.

\bibitem{turner2018integrating}
Isaac Turner, Kiran~V Garimella, Zamin Iqbal, and Gil McVean.
\newblock {Integrating long-range connectivity information into de Bruijn graphs}.
\newblock {\em Bioinformatics}, 2018.

\bibitem{hunt2024allthebacteria}
Martin Hunt, Leandro Lima, Daniel Anderson, George Bouras, Michael~B Hall, Jane Hawkey, Oliver Schwengers, Wei Shen, John Lees, and Zamin Iqbal.
\newblock {AllTheBacteria - all bacterial genomes assembled, available and searchable}.
\newblock {\em bioRxiv}, 2025.

\bibitem{bingol2026debruijn}
Zülal Bingöl, Berkan Şahin, Klea Zambaku, Ricardo Roman-Brenes, Konstantina Koliogeorgi, Can Firtina, Onur Mutlu, and Can Alkan.
\newblock {De Bruijn graphs for pangenomics: in-depth performance benchmarking of de Bruijn graph-based tools for read mapping}.
\newblock {\em Briefings in Bioinformatics}, 2026.

\bibitem{Mustafa2024MLA}
Harun Mustafa, Mikhail Karasikov, Nika Mansouri~Ghiasi, Gunnar R\"{a}tsch, and André Kahles.
\newblock {Label-guided seed-chain-extend alignment on annotated De Bruijn graphs}.
\newblock {\em Bioinformatics}, 2024.

\bibitem{Negi2025}
Shloka Negi, Sarah~L. Stenton, Seth~I. Berger, Paolo Canigiula, Brandy McNulty, Ivo Violich, Joshua Gardner, Todd Hillaker, Sara~M. O'Rourke, Melanie~C. O'Leary, Elizabeth Carbonell, Christina Austin-Tse, Gabrielle Lemire, Jillian Serrano, Brian Mangilog, Grace VanNoy, Mikhail Kolmogorov, Eric Vilain, Anne O'Donnell-Luria, Emmanu{\`e}le D{\'e}lot, et~al.
\newblock Advancing long-read nanopore genome assembly and accurate variant calling for rare disease detection.
\newblock {\em The American Journal of Human Genetics}, 2025.

\bibitem{colquhoun2021pandora}
Rachel~M. Colquhoun, Michael~B. Hall, Leandro Lima, Leah~W. Roberts, Kerri~M. Malone, Martin Hunt, Brice Letcher, Jane Hawkey, Sophie George, Louise Pankhurst, and Zamin Iqbal.
\newblock Pandora: nucleotide-resolution bacterial pan-genomics with reference graphs.
\newblock {\em Genome Biology}, 2021.

\bibitem{alipanahi2021succinct}
Bahar Alipanahi, Alan Kuhnle, Simon~J Puglisi, Leena Salmela, and Christina Boucher.
\newblock {Succinct dynamic de Bruijn graphs}.
\newblock {\em Bioinformatics}, 2021.

\bibitem{weisberg2021genomic}
Alexandra~J. Weisberg, Niklaus~J. Grünwald, Elizabeth~A. Savory, Melodie~L. Putnam, and Jeff~H. Chang.
\newblock {Genomic Approaches to Plant-Pathogen Epidemiology and Diagnostics}.
\newblock {\em Annual Review of Phytopathology}, 2021.

\bibitem{alipanahi2020metagenome}
Bahar Alipanahi, Martin~D Muggli, Musa Jundi, Noelle~R Noyes, and Christina Boucher.
\newblock {{Metagenome SNP calling via read-colored de Bruijn graphs}}.
\newblock {\em Bioinformatics}, 2020.

\bibitem{Vieira2024}
Ana Vieira, Yu~Wan, Yan Ryan, Ho~Kwong Li, Rebecca~L. Guy, Maria Papangeli, Kristin~K. Huse, Lucy~C. Reeves, Valerie W.~C. Soo, Roger Daniel, Alessandra Harley, Karen Broughton, Chenchal Dhami, Mark Ganner, Marjorie~A. Ganner, Zaynab Mumin, Maryam Razaei, Emma Rundberg, Rufat Mammadov, Ewurabena~A. Mills, et~al.
\newblock {Rapid expansion and international spread of M1UK in the post-pandemic UK upsurge of Streptococcus pyogenes}.
\newblock {\em Nature Communications}, 2024.

\bibitem{gangwar2025wepp}
Pranav Gangwar, Pratik Katte, Manu Bhat, and Yatish Turakhia.
\newblock {WEPP: Phylogenetic Placement Achieves Near-Haplotype Resolution in Wastewater-Based Epidemiology}.
\newblock {\em medRxiv}, 2025.

\bibitem{polo2020making}
David Polo, Marcos Quintela-Baluja, Alexander Corbishley, Davey~L Jones, Andrew~C Singer, David~W Graham, and Jes{\'u}s~L Romalde.
\newblock {Making waves: wastewater-based epidemiology for COVID-19--approaches and challenges for surveillance and prediction}.
\newblock {\em Water Research}, 2020.

\bibitem{bloemen_development_2023}
Bram Bloemen, Mathieu Gand, Kevin Vanneste, Kathleen Marchal, Nancy~HC Roosens, and Sigrid~CJ De~Keersmaecker.
\newblock Development of a portable on-site applicable metagenomic data generation workflow for enhanced pathogen and antimicrobial resistance surveillance.
\newblock {\em Scientific Reports}, 2023.

\bibitem{mustafa2022algorithms}
Harun Mustafa.
\newblock {\em Algorithms for efficient sensitive search and sample comparison on petabase-scale genomics data}.
\newblock PhD thesis, ETH Zurich, 2022.

\bibitem{alanko2023themisto}
Jarno~N Alanko, Jaakko Vuohtoniemi, Tommi Mäklin, and Simon~J Puglisi.
\newblock {Themisto: a scalable colored k-mer index for sensitive pseudoalignment against hundreds of thousands of bacterial genomes}.
\newblock {\em Bioinformatics}, 2023.

\bibitem{cracco2023extremely}
Andrea Cracco and Alexandru~I Tomescu.
\newblock {Extremely fast construction and querying of compacted and colored de Bruijn graphs with GGCAT}.
\newblock {\em Genome Research}, 2023.

\bibitem{Baaijens2022}
Jasmijn~A. Baaijens, Paola Bonizzoni, Christina Boucher, Gianluca Della~Vedova, Yuri Pirola, Raffaella Rizzi, and Jouni Sir{\'e}n.
\newblock Computational graph pangenomics: a tutorial on data structures and their applications.
\newblock {\em Natural Computing}, 2022.

\bibitem{paten2017genome}
Benedict Paten, Adam~M Novak, Jordan~M Eizenga, and Erik Garrison.
\newblock Genome graphs and the evolution of genome inference.
\newblock {\em Genome Research}, 2017.

\bibitem{bvrinda2023efficient}
Karel B{\v{r}}inda, Leandro Lima, Simone Pignotti, Natalia Quinones-Olvera, Kamil Salikhov, Rayan Chikhi, Gregory Kucherov, Zamin Iqbal, and Michael Baym.
\newblock Efficient and robust search of microbial genomes via phylogenetic compression.
\newblock {\em Nature Methods}, 2025.

\bibitem{Chikhi2013}
Rayan Chikhi and Guillaume Rizk.
\newblock {Space-efficient and exact de Bruijn graph representation based on a Bloom filter}.
\newblock {\em Algorithms for Molecular Biology}, 2013.

\bibitem{marchet2020reindeer}
Camille Marchet, Zamin Iqbal, Daniel Gautheret, Mika{\"e}l Salson, and Rayan Chikhi.
\newblock {REINDEER: efficient indexing of k-mer presence and abundance in sequencing datasets}.
\newblock {\em Bioinformatics}, 2020.

\bibitem{almodaresi2018space}
Fatemeh Almodaresi, Hirak Sarkar, Avi Srivastava, and Rob Patro.
\newblock {A space and time-efficient index for the compacted colored de Bruijn graph}.
\newblock {\em Bioinformatics}, 2018.

\bibitem{alanko2023small}
Jarno~N. Alanko, Simon~J. Puglisi, and Jaakko Vuohtoniemi.
\newblock {Small Searchable $\kappa$-Spectra via Subset Rank Queries on the Spectral Burrows-Wheeler Transform}.
\newblock In {\em ACDA}, 2023.

\bibitem{pibiri2022sparse}
Giulio~Ermanno Pibiri.
\newblock Sparse and skew hashing of k-mers.
\newblock {\em Bioinformatics}, 2022.

\bibitem{Bowe2012SuccinctGraphs}
Alexander Bowe, Taku Onodera, Kunihiko Sadakane, and Tetsuo Shibuya.
\newblock {Succinct de Bruijn graphs}.
\newblock In {\em WABI}, 2012.

\bibitem{li2015megahit}
Dinghua Li, Chi-Man Liu, Ruibang Luo, Kunihiko Sadakane, and Tak-Wah Lam.
\newblock {MEGAHIT: an ultra-fast single-node solution for large and complex metagenomics assembly via succinct de Bruijn graph}.
\newblock {\em Bioinformatics}, 2015.

\bibitem{mangul2016reference}
Serghei Mangul and David Koslicki.
\newblock {Reference-free comparison of microbial communities via de Bruijn graphs}.
\newblock In {\em BCB}, 2016.

\bibitem{garrison2018vg}
Erik Garrison, Jouni Sir{\'{e}}n, Adam~M Novak, Glenn Hickey, Jordan~M Eizenga, Eric~T Dawson, William Jones, Shilpa Garg, Charles Markello, Michael~F Lin, Benedict Paten, and Richard Durbin.
\newblock {Variation Graph Toolkit Improves Read Mapping by Representing Genetic Variation in the Reference}.
\newblock {\em Nature Biotechnology}, 2018.

\bibitem{chang2025rapid}
Xian Chang, Adam~M. Novak, Jordan~M. Eizenga, Jouni Sir{\'e}n, Jean Monlong, Shloka Negi, Francesco Andreace, Sagorika Nag, Konstantinos Kyriakidis, Glenn Hickey, Stephen Hwang, Emmanu{\`e}le~C. D{\'e}lot, Andrew Carroll, Kishwar Shafin, Pi-Chuan Chang, Faith Okamoto, Benedict Paten, and {the Human Pangenome Reference Consortium}.
\newblock {Rapid, accurate long- and short-read mapping to large pangenome graphs with vg Giraffe}.
\newblock {\em bioRxiv}, 2025.

\bibitem{campanelli2024where}
Alessio Campanelli, Giulio~Ermanno Pibiri, Jason Fan, and Rob Patro.
\newblock {Where the Patterns Are: Repetition-Aware Compression for Colored de Bruijn Graphs}.
\newblock {\em Journal of Computational Biology}, 2024.

\bibitem{Holley2020bifrost}
Guillaume Holley and Páll Melsted.
\newblock {Bifrost: highly parallel construction and indexing of colored and compacted de Bruijn graphs}.
\newblock {\em Genome Biology}, 21(1), September 2020.

\bibitem{campanelli2025fast}
Alessio Campanelli, Giulio~Ermanno Pibiri, and Rob Patro.
\newblock {Fast Pseudoalignment Queries on Compressed Colored de Bruijn Graphs}.
\newblock In {\em WABI}, 2025.

\bibitem{cormen2009introduction}
Thomas~H. Cormen, Charles~E. Leiserson, Ronald~L. Rivest, and Clifford Stein.
\newblock {Introduction to Algorithms}, 2009.

\bibitem{hopcroft1973algorithm}
John Hopcroft and Robert Tarjan.
\newblock Algorithm 447: efficient algorithms for graph manipulation.
\newblock {\em Communications of the ACM}, 1973.

\bibitem{ligra}
Julian Shun and Guy~E Blelloch.
\newblock {Ligra: a Lightweight Graph Processing Framework for Shared Memory}.
\newblock In {\em PPoPP}, 2013.

\bibitem{merrill2012scalable}
Duane Merrill, Michael Garland, and Andrew Grimshaw.
\newblock {Scalable GPU graph traversal}.
\newblock In {\em PPoPP}, 2012.

\bibitem{chi2022accelerating}
Yuze Chi, Licheng Guo, and Jason Cong.
\newblock {Accelerating SSSP for power-law graphs}.
\newblock In {\em FPGA}, 2022.

\bibitem{Dadu2021polygraph}
Vidushi Dadu, Sihao Liu, and Tony Nowatzki.
\newblock {PolyGraph: Exposing the Value of Flexibility for Graph Processing Accelerators}.
\newblock In {\em ISCA}, 2021.

\bibitem{yao2022scalagraph}
Pengcheng Yao, Long Zheng, Yu~Huang, Qinggang Wang, Chuangyi Gui, Zhen Zeng, Xiaofei Liao, Hai Jin, and Jingling Xue.
\newblock {ScalaGraph: A Scalable Accelerator for Massively Parallel Graph Processing}.
\newblock In {\em HPCA}, 2022.

\bibitem{nhgri}
Kris~A. Wetterstrand.
\newblock {DNA Sequencing Costs: Data from the NHGRI Genome Sequencing Program (GSP)}.
\newblock \url{https://www.genome.gov/sequencingcostsdata}, 2024.

\bibitem{pedro2021integration}
João~Pedro {de Magalhães}, Cyril Lagger, and Robi Tacutu.
\newblock Integrative genomics of aging.
\newblock In {\em Handbook of the Biology of Aging}. Ninth edition, 2021.

\bibitem{Bick2024}
Alexander~G. Bick, Ginger~A. Metcalf, Kelsey~R. Mayo, Lee Lichtenstein, Shimon Rura, Robert~J. Carroll, Anjene Musick, Jodell~E. Linder, I.~King Jordan, Shashwat~Deepali Nagar, Shivam Sharma, Robert Meller, Melissa Basford, Eric Boerwinkle, Mine~S. Cicek, Kimberly~F. Doheny, Evan~E. Eichler, Stacey Gabriel, Richard~A. Gibbs, David Glazer, et~al.
\newblock {Genomic data in the All of Us Research Program}.
\newblock {\em Nature}, 2024.

\bibitem{Li2023whole}
Keren Carss, Bjarni~V. Halldorsson, Liping Hou, Jimmy Liu, Eleanor Wheeler, Yancy Lo, Kousik Kundu, Zhuoyi Huang, Ben Lacey, Ryan~S. Dhindsa, Diana Rajan, Jelena Randjelovic, Neil Marriott, Carol~E. Scott, Ahmet~Sinan Yavuz, Ian Johnston, Trevor Howe, Mary~Helen Black, Kari Stefansson, Robert Scott, et~al.
\newblock {Whole-genome sequencing of 490,640 UK Biobank participants}.
\newblock {\em Nature}, 2025.

\bibitem{Garrison2024}
Erik Garrison, Andrea Guarracino, Simon Heumos, Flavia Villani, Zhigui Bao, Lorenzo Tattini, J{\"o}rg Hagmann, Sebastian Vorbrugg, Santiago Marco-Sola, Christian Kubica, David~G. Ashbrook, Kaisa Thorell, Rachel~L. Rusholme-Pilcher, Gianni Liti, Emilio Rudbeck, Agnieszka~A. Golicz, Sven Nahnsen, Zuyu Yang, Moses~Njagi Mwaniki, Franklin~L. Nobrega, et~al.
\newblock Building pangenome graphs.
\newblock {\em Nature Methods}, 2024.

\bibitem{Minkin2020}
Ilia Minkin and Paul Medvedev.
\newblock {Scalable multiple whole-genome alignment and locally collinear block construction with SibeliaZ}.
\newblock {\em Nature Communications}, 2020.

\bibitem{Chen2025}
Xiufei Chen, Haiqi Xu, Xiao Shu, and Chun-Xiao Song.
\newblock Mapping epigenetic modifications by sequencing technologies.
\newblock {\em Cell Death {\&} Differentiation}, 2025.

\bibitem{Liu2025}
Tianyuan Liu and Ana Conesa.
\newblock Profiling the epigenome using long-read sequencing.
\newblock {\em Nature Genetics}, 2025.

\bibitem{Sigurpalsdottir2024}
Brynja~D. Sigurpalsdottir, Olafur~A. Stefansson, Guillaume Holley, Doruk Beyter, Florian Zink, Marteinn~Þ. Hardarson, Sverrir~Þ. Sverrisson, Nina Kristinsdottir, Droplaug~N. Magnusdottir, Olafur~Þ. Magnusson, Daniel~F. Gudbjartsson, Bjarni~V. Halldorsson, and Kari Stefansson.
\newblock {A comparison of methods for detecting DNA methylation from long-read sequencing of human genomes}.
\newblock {\em Genome Biology}, 2024.

\bibitem{Li2011}
Yuanyuan Li and Trygve~O. Tollefsbol.
\newblock {DNA Methylation Detection: Bisulfite Genomic Sequencing Analysis}.
\newblock {\em Epigenetics Protocols}, 2011.

\bibitem{Forcato2017}
Mattia Forcato, Chiara Nicoletti, Koustav Pal, Carmen~Maria Livi, Francesco Ferrari, and Silvio Bicciato.
\newblock {Comparison of computational methods for Hi-C data analysis}.
\newblock {\em Nature Methods}, 2017.

\bibitem{Pal2019}
Koustav Pal, Mattia Forcato, and Francesco Ferrari.
\newblock {Hi-C analysis: from data generation to integration}.
\newblock {\em Biophysical Reviews}, 2019.

\bibitem{lieberman2009comprehensive}
Erez Lieberman-Aiden, Nynke~L. van Berkum, Louise Williams, Maxim Imakaev, Tobias Ragoczy, Agnes Telling, Ido Amit, Bryan~R. Lajoie, Peter~J. Sabo, Michael~O. Dorschner, Richard Sandstrom, Bradley Bernstein, M.~A. Bender, Mark Groudine, Andreas Gnirke, John Stamatoyannopoulos, Leonid~A. Mirny, Eric~S. Lander, and Job Dekker.
\newblock {Comprehensive Mapping of Long-Range Interactions Reveals Folding Principles of the Human Genome}.
\newblock {\em Science}, 2009.

\bibitem{ncbi2025}
Eric W Sayers, Jeffrey Beck, Evan E Bolton, J Rodney Brister, Jessica Chan, Ryan Connor, Michael Feldgarden, Anna M Fine, Kathryn Funk, Jinna Hoffman, Sivakumar Kannan, Christopher Kelly, William Klimke, Sunghwan Kim, Stacy Lathrop, Aron Marchler-Bauer, Terence D Murphy, Chris O’Sullivan, Erin Schmieder, Yuriy Skripchenko, et~al.
\newblock {Database resources of the National Center for Biotechnology Information in 2025}.
\newblock {\em Nucleic Acids Research}, 2025.

\bibitem{thakur2023embl}
Matthew Thakur, Alex Bateman, Cath Brooksbank, Mallory Freeberg, Melissa Harrison, Matthew Hartley, Thomas Keane, Gerard Kleywegt, Andrew Leach, Mariia Levchenko, Sarah Morgan, Ellen M McDonagh, Sandra Orchard, Irene Papatheodorou, Sameer Velankar, Juan Antonio Vizcaino, Rick Witham, Barbara Zdrazil, and Johanna McEntyre.
\newblock {EMBL’s European Bioinformatics Institute (EMBL-EBI) in 2022}.
\newblock {\em Nucleic Acids Research}, 2023.

\bibitem{hernaez2019genomic}
Mikel Hernaez, Dmitri Pavlichin, Tsachy Weissman, and Idoia Ochoa.
\newblock {Genomic Data Compression}.
\newblock {\em Annual Review of Biomedical Data Science}, 2019.

\bibitem{intelqat}
Intel.
\newblock {Intel{\textregistered} QuickAssist Technology (Intel{\textregistered} QAT)}.
\newblock \url{https://www.intel.com/content/www/us/en/architecture-and-technology/intel-quick-assist-technology-overview.html}, 2024.

\bibitem{ibmzedc}
Mai Zeng, Marcelo~Lopes de~Moraes, Paul~W Novak, Pearlson Christopher, Ravinder Akula, and Vijayakumar Yeso.
\newblock {IBM zEnterprise Data Compression (zEDC): Implementation {\&} Exploitation Use Cases}.
\newblock Technical report, IBM, 2020.

\bibitem{xz}
The~Tukaani Project.
\newblock {XZ Utils}.
\newblock \url{https://tukaani.org/xz/}, 2024.

\bibitem{Ma2019}
Xiaotu Ma, Ying Shao, Liqing Tian, Diane~A. Flasch, Heather~L. Mulder, Michael~N. Edmonson, Yu~Liu, Xiang Chen, Scott Newman, Joy Nakitandwe, Yongjin Li, Benshang Li, Shuhong Shen, Zhaoming Wang, Sheila Shurtleff, Leslie~L. Robison, Shawn Levy, John Easton, and Jinghui Zhang.
\newblock {Analysis of error profiles in deep next-generation sequencing data}.
\newblock {\em Genome Biology}, 2019.

\bibitem{Abel2020}
Haley~J. Abel, David~E. Larson, Allison~A. Regier, Colby Chiang, Indraniel Das, Krishna~L. Kanchi, Ryan~M. Layer, Benjamin~M. Neale, William~J. Salerno, Catherine Reeves, Steven Buyske, Goncalo~R. Abecasis, Elizabeth Appelbaum, Julie Baker, Eric Banks, Raphael~A. Bernier, Toby Bloom, Michael Boehnke, Eric Boerwinkle, Erwin~P. Bottinger, et~al.
\newblock Mapping and characterization of structural variation in 17,795 human genomes.
\newblock {\em Nature}, 2020.

\bibitem{mmap}
Michael Kerrisk.
\newblock {\em {mmap(2) — Linux manual page}}, 2025.

\bibitem{papagiannis2020optimizing}
Anastasios Papagiannis, Giorgos Xanthakis, Giorgos Saloustros, Manolis Marazakis, and Angelos Bilas.
\newblock {Optimizing Memory-mapped I/O for Fast Storage Devices}.
\newblock In {\em USENIX ATC}, 2020.

\bibitem{papagiannis2018efficient}
Anastasios Papagiannis, Giorgos Saloustros, Pilar Gonz{\'a}lez-F{\'e}rez, and Angelos Bilas.
\newblock {An Efficient Memory-Mapped Key-Value Store for Flash Storage}.
\newblock In {\em Proceedings of the ACM Symposium on Cloud Computing}, 2018.

\bibitem{splice}
Michael Kerrisk.
\newblock {\em {splice(2) — Linux manual page}}, 2025.

\bibitem{sendfile}
Michael Kerrisk.
\newblock {\em {sendfile(2) — Linux manual page}}, 2025.

\bibitem{io_uring}
Michael Kerrisk.
\newblock {\em {io\_uring(7) — Linux manual page}}, 2020.

\bibitem{BlueField}
NVIDIA.
\newblock {NVIDIA BlueField-2 DPU}.
\newblock \url{https://www.nvidia.com/content/dam/en-zz/Solutions/Data-Center/documents/datasheet-nvidia-bluefield-2-dpu.pdf}, 2021.

\bibitem{NapatechF2070X}
Napatech.
\newblock {F2070X Data Processing Unit (DPU)}.
\newblock \url{https://www.napatech.com/support/resources/data-sheets/f2070x-data-processing-unit/}, 2023.

\bibitem{zhang2024dds}
Qizhen Zhang, Philip Bernstein, Badrish Chandramouli, Jiasheng Hu, and Yiming Zheng.
\newblock {DDS: DPU-Optimized Disaggregated Storage}.
\newblock {\em VLDB}, 2024.

\bibitem{zhong2024dpc}
Kan Zhong, Zhiwang Yu, Qiao Li, Xianqiang Luo, Linbo Long, Yujian Tan, Ao~Ren, and Duo Liu.
\newblock {DPC: DPU-accelerated High-Performance File System Client}.
\newblock In {\em Proceedings of the 53rd International Conference on Parallel Processing}, 2024.

\bibitem{zhu2025hidpu}
Wenbin Zhu, Zhaoyan Shen, Qian Wei, Renhai Chen, Xin Yao, Dongxiao Yu, and Zili Shao.
\newblock {HiDPU: A DPU-Oriented Hybrid Indexing Scheme for Disaggregated Storage Systems}.
\newblock In {\em FAST}, 2025.

\bibitem{gootzen2023dpfs}
Peter-Jan Gootzen, Jonas Pfefferle, Radu Stoica, and Animesh Trivedi.
\newblock {DPFS: DPU-Powered File System Virtualization}.
\newblock In {\em Proceedings of the 16th ACM International Conference on Systems and Storage}, 2023.

\bibitem{jeong2025mangoboost}
Heetaek Jeong, Wonsik Lee, Eunjin Baek, Changsu Kim, Changyeon Jo, Dongju Chae, Kanghyun Choi, Hamin Jang, Mohamed Elgammal, Sungmin Hong, Eriko Nurvitadhi, Dongup Kwon, and Jangwoo Kim.
\newblock {MangoBoost Alice: Extremely Fast, Seamless, and Versatile FPGA-Accelerated DPU Solutions}.
\newblock {\em IEEE Micro}, 2025.

\bibitem{Shainer2011}
Gilad Shainer, Ali Ayoub, Pak Lui, Tong Liu, Michael Kagan, Christian~R. Trott, Greg Scantlen, and Paul~S. Crozier.
\newblock {The development of Mellanox/NVIDIA GPUDirect over InfiniBand---a new model for GPU to GPU communications}.
\newblock {\em Computer Science - Research and Development}, 2011.

\bibitem{li2025managing}
Shaobo Li, Yirui~Eric Zhou, Yuqi Xue, Yuan Xu, and Jian Huang.
\newblock {Managing Scalable Direct Storage Accesses for GPUs with GoFS}.
\newblock In {\em SOSP}, 2025.

\bibitem{NVIDIA_GDS_Blog2022}
{NVIDIA}.
\newblock {GPUDirect Storage}: A direct path between storage and {GPU} memory.
\newblock NVIDIA Developer Blog, 2022.

\bibitem{Newburn2019GTC}
NVIDIA.
\newblock {How to Make Your Life Easier in the Age of Exascale Computing Using {NVIDIA GPUDirect} Technologies}.
\newblock NVIDIA GPU Technology Conference (GTC), 2019.

\bibitem{yang2017spdk}
Ziye Yang, James~R. Harris, Benjamin Walker, Daniel Verkamp, Changpeng Liu, Cunyin Chang, Gang Cao, Jonathan Stern, Vishal Verma, and Luse~E. Paul.
\newblock {SPDK: A Development Kit to Build High Performance Storage Applications}.
\newblock In {\em CloudCom}, 2017.

\bibitem{kim2016nvmedirect}
Hyeong-Jun Kim, Young-Sik Lee, and Jin-Soo Kim.
\newblock {NVMeDirect}: A user-space {I/O} framework for application-specific optimization on {NVMe} {SSDs}.
\newblock In {\em HotStorage}, 2016.

\bibitem{Sujay24BypassD}
Sujay Yadalam, Chloe Alverti, Vasileios Karakostas, Jayneel Gandhi, and Michael Swift.
\newblock {BypassD: Enabling fast userspace access to shared SSDs}.
\newblock In {\em ASPLOS}, 2024.

\bibitem{Kaesi2025LITESHIELD}
Kaesi Manakkal, Nathan Daughety, Marcus Pendleton, and Hui Lu.
\newblock {LITESHIELD: Secure Containers via Lightweight, Composable Userspace {$\mu$}Kernel Services}.
\newblock In {\em USENIX ATC}, 2025.

\bibitem{nvme2}
{NVM Express, Inc.}
\newblock {\em NVM Express Base Specification revision 2.0a}.
\newblock 2021.

\bibitem{min2023ezns}
Jaehong Min, Chenxingyu Zhao, Ming Liu, and Arvind Krishnamurthy.
\newblock {eZNS: An Elastic Zoned Namespace for Commodity ZNS SSDs}.
\newblock In {\em OSDI}, 2023.

\bibitem{SamsungFDP2023}
{Samsung Electronics}.
\newblock Getting started with {Flexible Data Placement (FDP)}.
\newblock Technical report, 2023.

\bibitem{wang2014efficient}
Peng Wang, Guangyu Sun, Song Jiang, Jian Ouyang, Shiding Lin, Chen Zhang, and Jason Cong.
\newblock {An efficient design and implementation of LSM-tree based key-value store on open-channel SSD}.
\newblock In {\em EuroSys}, 2014.

\bibitem{do2021better}
Jaeyoung Do, Ivan~Luiz Picoli, David Lomet, and Philippe Bonnet.
\newblock {Better database cost/performance via batched I/O on programmable SSD}.
\newblock {\em The VLDB Journal}, 2021.

\bibitem{song2025cam}
Ziyu Song, Jie Zhang, Jie Sun, Mo~Sun, Zihan Yang, Zheng Zhang, Xuzheng Chen, Fei Wu, Huajin Tang, and Zeke Wang.
\newblock {CAM: Asynchronous GPU-Initiated, CPU-Managed SSD Management for Batching Storage Access}.
\newblock In {\em ICDE}, 2025.

\bibitem{do2019improving}
Jaeyoung Do, David Lomet, and Ivan~Luiz Picoli.
\newblock {Improving CPU I/O performance via SSD controller FTL support for batched writes}.
\newblock In {\em Proceedings of the 15th International Workshop on Data Management on New Hardware}, 2019.

\bibitem{shen2013flashfq}
Kai Shen and Stan Park.
\newblock $\{$FlashFQ$\}$: A fair queueing $\{$I/O$\}$ scheduler for $\{$Flash-Based$\}$$\{$SSDs$\}$.
\newblock In {\em USENIX ATC}, 2013.

\bibitem{mao2017improving}
Bo~Mao, Suzhen Wu, and Lide Duan.
\newblock {Improving the SSD Performance by Exploiting Request Characteristics and Internal Parallelism}.
\newblock {\em IEEE Transactions on Computer-Aided Design of Integrated Circuits and Systems}, 2017.

\bibitem{yang2019cars}
Tianming Yang, Ping Huang, Weiying Zhang, Haitao Wu, and Longxin Lin.
\newblock {CARS: A Multi-layer Conflict-Aware Request Scheduler for NVMe SSDs}.
\newblock In {\em DATE}, 2019.

\bibitem{wang2013novel}
Hua Wang, Ping Huang, Shuang He, Ke~Zhou, Chunhua Li, and Xubin He.
\newblock {A novel I/O scheduler for SSD with improved performance and lifetime}.
\newblock In {\em MSST}, 2013.

\bibitem{nilakant2014prefedge}
Karthik Nilakant, Valentin Dalibard, Amitabha Roy, and Eiko Yoneki.
\newblock {PrefEdge: SSD Prefetcher for Large-Scale Graph Traversal}.
\newblock In {\em Proceedings of International Conference on Systems and Storage}, 2014.

\bibitem{chakraborttii2020learning}
Chandranil Chakraborttii and Heiner Litz.
\newblock {Learning I/O Access Patterns to Improve Prefetching in SSDs}.
\newblock In {\em Joint European Conference on Machine Learning and Knowledge Discovery in Databases}, 2020.

\bibitem{li2022pattern}
Jun Li, Xiaofei Xu, Zhigang Cai, Jianwei Liao, Kenli Li, Balazs Gerofi, and Yutaka Ishikawa.
\newblock {Pattern-Based Prefetching with Adaptive Cache Management Inside of Solid-State Drives}.
\newblock {\em ACM TOS}, 2022.

\bibitem{da2012parallel}
Randal~Burns Da~Zheng and Alexander~S Szalay.
\newblock {A Parallel Page Cache: IOPS and Caching for Multicore Systems}.
\newblock In {\em FAST}, 2012.

\bibitem{pham2024scalecache}
Kiet~Tuan Pham, Seokjoo Cho, Sangjin Lee, Lan~Anh Nguyen, Hyeongi Yeo, Ipoom Jeong, Sungjin Lee, Nam~Sung Kim, and Yongseok Son.
\newblock {ScaleCache: A Scalable Page Cache for Multiple Solid-State Drives}.
\newblock In {\em EuroSys}, 2024.

\bibitem{brokhman2019gaia}
Tanya Brokhman, Pavel Lifshits, and Mark Silberstein.
\newblock {GAIA: An OS Page Cache for Heterogeneous Systems}.
\newblock In {\em USENIX ATC}, 2019.

\bibitem{li2024streamcache}
Zhiyue Li and Guangyan Zhang.
\newblock {StreamCache: Revisiting Page Cache for File Scanning on Fast Storage Devices}.
\newblock In {\em USENIX ATC}, 2024.

\bibitem{canim2010ssd}
Mustafa Canim, George~A Mihaila, Bishwaranjan Bhattacharjee, Kenneth~A Ross, and Christian~A Lang.
\newblock {SSD Bufferpool Extensions for Database Systems}.
\newblock {\em Proceedings of the VLDB Endowment}, 2010.

\bibitem{wang2014cache}
Jianguo Wang, Eric Lo, Man~Lung Yiu, Jiancong Tong, Gang Wang, and Xiaoguang Liu.
\newblock {Cache Design of SSD-Based Search Engine Architectures: An Experimental Study}.
\newblock {\em TOIS}, 2014.

\bibitem{pekhimenko2016case}
Gennady Pekhimenko, Evgeny Bolotin, Nandita Vijaykumar, Onur Mutlu, Todd~C Mowry, and Stephen~W Keckler.
\newblock {A Case for Toggle-Aware Compression for GPU Systems}.
\newblock In {\em HPCA}, 2016.

\bibitem{pekhimenko2015energy}
Gennady Pekhimenko, Evgeny Bolotin, Mike O’Connor, Onur Mutlu, Todd~C Mowry, and Stephen~W Keckler.
\newblock {Energy-Efficient Data Compression for GPU Memory Systems}.
\newblock In {\em ASPLOS}, 2015.

\bibitem{pekhimenko2013linearly}
Gennady Pekhimenko, Vivek Seshadri, Yoongu Kim, Hongyi Xin, Onur Mutlu, Phillip~B Gibbons, Michael~A Kozuch, and Todd~C Mowry.
\newblock {Linearly Compressed Pages: A Low-Complexity, Low-Latency Main Memory Compression Framework}.
\newblock In {\em MICRO}, 2013.

\bibitem{pekhimenko2012base}
Gennady Pekhimenko, Vivek Seshadri, Onur Mutlu, Phillip~B Gibbons, Michael~A Kozuch, and Todd~C Mowry.
\newblock {Base-Delta-Immediate Compression: Practical Data Compression for On-Chip Caches}.
\newblock In {\em PACT}, 2012.

\bibitem{buyuktosunoglu2024enterprise}
Alper Buyuktosunoglu, David Trilla, Bulent Abali, Deanna Berger, Craig Walters, and Jang-Soo Lee.
\newblock {Enterprise-Class Cache Compression Design}.
\newblock In {\em HPCA}, 2024.

\bibitem{ekman2005robust}
Magnus Ekman and Per Stenstrom.
\newblock {A Robust Main-Memory Compression Scheme}.
\newblock In {\em ISCA}, 2005.

\bibitem{arelakis2014sc2}
Angelos Arelakis and Per Stenstrom.
\newblock {SC2: A Statistical Compression Cache Scheme}.
\newblock {\em ISCA}, 2014.

\bibitem{vijaykumar2015case}
Nandita Vijaykumar, Gennady Pekhimenko, Adwait Jog, Abhishek Bhowmick, Rachata Ausavarungnirun, Chita Das, Mahmut Kandemir, Todd~C Mowry, and Onur Mutlu.
\newblock {A Case for Core-Assisted Bottleneck Acceleration in GPUs: Enabling Flexible Data Compression with Assist Warps}.
\newblock {\em ISCA}, 2015.

\bibitem{park2022deepsketch}
Jisung Park, Jeonggyun Kim, Yeseong Kim, Sungjin Lee, and Onur Mutlu.
\newblock {$\{$DeepSketch$\}$: A New Machine $\{$Learning-Based$\}$ Reference Search Technique for $\{$Post-Deduplication$\}$ Delta Compression}.
\newblock In {\em FAST}, 2022.

\bibitem{ajdari2017scalable}
Mohammadamin Ajdari, Pyeongsu Park, Dongup Kwon, Joonsung Kim, and Jangwoo Kim.
\newblock {A Scalable HW-Based Inline Deduplication for SSD Arrays}.
\newblock {\em IEEE Computer Architecture Letters}, 2017.

\bibitem{ajdari2019cidr}
Mohammadamin Ajdari, Pyeongsu Park, Joonsung Kim, Dongup Kwon, and Jangwoo Kim.
\newblock {CIDR: A Cost-effective In-line Data Reduction System for Terabit-per-second Scale SSD Arrays}.
\newblock In {\em HPCA}, 2019.

\bibitem{ajdari2019fidr}
Mohammadamin Ajdari, Wonsik Lee, Pyeongsu Park, Joonsung Kim, and Jangwoo Kim.
\newblock {FIDR: A Scalable Storage System for Fine-Grain Inline Data Reduction with Efficient Memory Handling}.
\newblock In {\em MICRO}, 2019.

\bibitem{singh_sibyl_2022}
Gagandeep Singh, Rakesh Nadig, Jisung Park, Rahul Bera, Nastaran Hajinazar, David Novo, Juan Gómez-Luna, Sander Stuijk, Henk Corporaal, and Onur Mutlu.
\newblock Sibyl: {Adaptive} and {Extensible} {Data} {Placement} in {Hybrid} {Storage} {Systems} {Using} {Online} {Reinforcement} {Learning}.
\newblock In {\em {ISCA}}, 2022.

\bibitem{nadig2025harmonia}
Rakesh Nadig, Vamanan Arulchelvan, Rahul Bera, Taha Shahroodi, Gagandeep Singh, Andreas Kakolyris, Mohammad Sadrosadati, Jisung Park, and Onur Mutlu.
\newblock {Harmonia: A Multi-Agent Reinforcement Learning Approach to Data Placement and Migration in Hybrid Storage Systems}.
\newblock {\em arXiv}, 2025.

\bibitem{yang2024term}
Zhe Yang, Qing Wang, Xiaojian Liao, Youyou Lu, Keji Huang, and Jiwu Shu.
\newblock {TeRM: Extending RDMA-Attached Memory with SSD}.
\newblock In {\em FAST}, 2024.

\bibitem{niu2018hybrid}
Junpeng Niu, Jun Xu, and Lihua Xie.
\newblock {Hybrid Storage Systems: A Survey of Architectures and Algorithms}.
\newblock {\em IEEE Access}, 2018.

\bibitem{oliveira2023extending}
Geraldo~F Oliveira, Saugata Ghose, Juan G{\'o}mez-Luna, Amirali Boroumand, Alexis Savery, Sonny Rao, Salman Qazi, Gwendal Grignou, Rahul Thakur, Eric Shiu, et~al.
\newblock {Extending Memory Capacity in Modern Consumer Systems With Emerging Non-Volatile Memory: Experimental Analysis and Characterization Using the Intel Optane SSD}.
\newblock {\em IEEE Access}, 2023.

\bibitem{li2017utility}
Yang Li, Saugata Ghose, Jongmoo Choi, Jin Sun, Hui Wang, and Onur Mutlu.
\newblock {Utility-Based Hybrid Memory Management}.
\newblock In {\em CLUSTER}, 2017.

\bibitem{sun2013hybrid}
Guangyu Sun, Yongsoo Joo, Yibo Chen, Yiran Chen, and Yuan Xie.
\newblock {A Hybrid Solid-State Storage Architecture for the Performance, Energy Consumption, and Lifetime Improvement}.
\newblock In {\em Emerging Memory Technologies: Design, Architecture, and Applications}. Springer, 2013.

\bibitem{matsui2017design}
Chihiro Matsui, Chao Sun, and Ken Takeuchi.
\newblock {Design of Hybrid SSDs with Storage Class Memory and NAND Flash Memory}.
\newblock {\em Proceedings of the IEEE}, 2017.

\bibitem{ramos2011page}
Luiz~E Ramos, Eugene Gorbatov, and Ricardo Bianchini.
\newblock {Page Placement in Hybrid Memory Systems}.
\newblock In {\em ICS}, 2011.

\bibitem{das2024introduction}
Debendra Das~Sharma, Robert Blankenship, and Daniel Berger.
\newblock {An Introduction to the Compute Express Link (CXL) Interconnect}.
\newblock {\em ACM Computing Surveys}, 2024.

\bibitem{van2019hoti}
Stephen Van~Doren.
\newblock {Compute Express Link}.
\newblock In {\em HOTI}, 2019.

\bibitem{lim-isca09}
Kevin Lim, Jichuan Chang, Trevor Mudge, Parthasarathy Ranganathan, Steven~K. Reinhardt, and Thomas~F. Wenisch.
\newblock {Disaggregated Memory for Expansion and Sharing in Blade Servers}.
\newblock In {\em ISCA}, 2009.

\bibitem{nvmexpress_nvmeof}
{NVM Express, Inc.}
\newblock {\em {NVM Express over Fabrics Revision 1.1}}.
\newblock NVM Express, Inc., 2019.
\newblock Available at: \url{https://nvmexpress.org/specifications/}.

\bibitem{oh2024midas}
Seonggyun Oh, Jeeyun Kim, Soyoung Han, Jaeho Kim, Sungjin Lee, and Sam~H. Noh.
\newblock {MIDAS: Minimizing Write Amplification in {Log-Structured} Systems through Adaptive Group Number and Size Configuration}.
\newblock In {\em FAST}, 2024.

\bibitem{lanyue2016wisckey}
Lanyue Lu, Thanumalayan~Sankaranarayana Pillai, Andrea~C. Arpaci-Dusseau, and Remzi~H. Arpaci-Dusseau.
\newblock {WiscKey: Separating Keys from Values in SSD-conscious Storage}.
\newblock In {\em FAST}, 2016.

\bibitem{purandare2025valet}
Devashish~R Purandare, Peter Alvaro, Avani Wildani, Darrell~DE Long, and Ethan~L Miller.
\newblock {Valet: Efficient Data Placement on Modern SSDs}.
\newblock In {\em Proceedings of the 2025 ACM Symposium on Cloud Computing}, 2025.

\bibitem{zheng2026solidattention}
Xinrui Zheng, Dongliang Wei, Jianxiang Gao, Yixin Song, Zeyu Mi, and Haibo Chen.
\newblock {SolidAttention: Low-Latency SSD-based Serving on Memory-Constrained PCs}.
\newblock In {\em FAST}, 2026.

\bibitem{tavakkol2013nossd}
Arash Tavakkol, Mohammad Arjomand, and Hamid Sarbazi-Azad.
\newblock {Network-on-SSD: A Scalable and High-Performance Communication Design Paradigm for SSDs}.
\newblock {\em IEEE CAL}, 2013.

\bibitem{kim2022networked}
Jiho Kim, Seokwon Kang, Yongjun Park, and John Kim.
\newblock {Networked SSD: Flash Memory Interconnection Network for High-Bandwidth SSD}.
\newblock In {\em MICRO}, 2022.

\bibitem{yang2019reducing}
Pan Yang, Ni~Xue, Yuqi Zhang, Yangxu Zhou, Li~Sun, Wenwen Chen, Zhonggang Chen, Wei Xia, Junke Li, and Kihyoun Kwon.
\newblock {Reducing Garbage Collection Overhead in SSD Based on Workload Prediction}.
\newblock In {\em HotStorage}, 2019.

\bibitem{park2017method}
Jung~Kyu Park and Jaeho Kim.
\newblock {A method for reducing garbage collection overhead of SSD using machine learning algorithms}.
\newblock In {\em ICTC}, 2017.

\bibitem{jung2012taking}
Myoungsoo Jung, Ramya Prabhakar, and Mahmut~Taylan Kandemir.
\newblock {Taking Garbage Collection Overheads Off the Critical Path in SSDs}.
\newblock In {\em Middleware}, 2012.

\bibitem{ye2024achieving}
Min Ye, Qiao Li, Yina Lv, Jie Zhang, Tianyu Ren, Daniel Wen, Tei-Wei Kuo, and Chun~Jason Xue.
\newblock {Achieving Near-Zero Read Retry for 3D NAND Flash Memory}.
\newblock In {\em ASPLOS}, 2024.

\bibitem{kim2015subpage}
Jung-Hoon Kim, Sang-Hoon Kim, and Jin-Soo Kim.
\newblock {Subpage programming for extending the lifetime of NAND flash memory}.
\newblock In {\em DATE}, 2015.

\bibitem{kim2018utilizing}
Jung-Hoon Kim, Sang-Hoon Kim, and Jin-Soo Kim.
\newblock {Utilizing Subpage Programming to Prolong the Lifetime of Embedded NAND Flash-Based Storage}.
\newblock {\em IEEE Transactions on Consumer Electronics}, 2018.

\bibitem{compagnoni2017reviewing}
Christian~Monzio Compagnoni, Akira Goda, Alessandro~S Spinelli, Peter Feeley, Andrea~L Lacaita, and Angelo Visconti.
\newblock {Reviewing the Evolution of the NAND Flash Technology}.
\newblock {\em Proceedings of the IEEE}, 2017.

\bibitem{hedayati2019multi}
Mohammad Hedayati, Kai Shen, Michael~L Scott, and Mike Marty.
\newblock {Multi-Queue Fair Queuing}.
\newblock In {\em USENIX ATC}, 2019.

\bibitem{woo2021d2fq}
Jiwon Woo, Minwoo Ahn, Gyusun Lee, and Jinkyu Jeong.
\newblock {D2FQ: Device-Direct Fair Queueing for NVMe SSDs}.
\newblock In {\em FAST}, 2021.

\bibitem{li2023ecssd}
Siqi Li, Fengbin Tu, Liu Liu, Jilan Lin, Zheng Wang, Yangwook Kang, Yufei Ding, and Yuan Xie.
\newblock {ECSSD: Hardware/Data Layout Co-Designed In-Storage-Computing Architecture for Extreme Classification}.
\newblock In {\em ISCA}, 2023.

\bibitem{mailthody2019deepstore}
Vikram~Sharma Mailthody, Zaid Qureshi, Weixin Liang, Ziyan Feng, Simon~Garcia De~Gonzalo, Youjie Li, Hubertus Franke, Jinjun Xiong, Jian Huang, and Wen-mei Hwu.
\newblock {Deepstore: In-storage Acceleration for Intelligent Queries}.
\newblock In {\em MICRO}, 2019.

\bibitem{kang2021mithrilog}
Seongyoung Kang, Jiyoung An, Jinpyo Kim, and Sang-Woo Jun.
\newblock {Mithrilog: Near-storage accelerator for high-performance log analytics}.
\newblock In {\em MICRO}, 2021.

\bibitem{koo2017summarizer}
Gunjae Koo, Kiran~Kumar Matam, I~Te, HV~Krishna~Giri Narra, Jing Li, Hung-Wei Tseng, Steven Swanson, and Murali Annavaram.
\newblock {Summarizer: Trading Communication with Computing Near Storage}.
\newblock In {\em MICRO}, 2017.

\bibitem{wang2024beacongnn}
Yuyue Wang, Xiurui Pan, Yuda An, Jie Zhang, and Glenn Reinman.
\newblock {BeaconGNN: Large-Scale GNN Acceleration with Out-of-Order Streaming In-Storage Computing}.
\newblock In {\em HPCA}, 2024.

\bibitem{jang2024smart}
Hongsun Jang, Jaeyong Song, Jaewon Jung, Jaeyoung Park, Youngsok Kim, and Jinho Lee.
\newblock {Smart-Infinity: Fast Large Language Model Training using Near-Storage Processing on a Real System}.
\newblock In {\em HPCA}, 2024.

\bibitem{li2021glist}
Cangyuan Li, Ying Wang, Cheng Liu, Shengwen Liang, Huawei Li, and Xiaowei Li.
\newblock {GLIST: Towards In-Storage Graph Learning}.
\newblock In {\em USENIX ATC}, 2021.

\bibitem{lee2020smartssd}
Joo~Hwan Lee, Hui Zhang, Veronica Lagrange, Praveen Krishnamoorthy, Xiaodong Zhao, and Yang~Seok Ki.
\newblock {SmartSSD: FPGA Accelerated Near-Storage Data Analytics on SSD}.
\newblock {\em IEEE CAL}, 2020.

\bibitem{Jeong2025upp}
Ipoom Jeong, Jinghan Huang, Chuxuan Hu, Dohyun Park, Jaeyoung Kang, Nam~Sung Kim, and Yongjoo Park.
\newblock {UPP: Universal Predicate Pushdown to Smart Storage}.
\newblock In {\em ISCA}, 2025.

\bibitem{mahapatra2024instoragedomainspecificaccelerationserverless}
Rohan Mahapatra, Soroush Ghodrati, Byung~Hoon Ahn, Sean Kinzer, Shu-Ting Wang, Hanyang Xu, Lavanya Karthikeyan, Hardik Sharma, Amir Yazdanbakhsh, Mohammad Alian, and Hadi Esmaeilzadeh.
\newblock In-storage domain-specific acceleration for serverless computing.
\newblock In {\em ASPLOS}, 2024.

\bibitem{Lee2024presto}
Yunjae Lee, Hyeseong Kim, and Minsoo Rhu.
\newblock {PreSto: An In-Storage Data Preprocessing System for Training Recommendation Models}.
\newblock In {\em ISCA}, 2025.

\bibitem{gu2016biscuit}
Boncheol Gu, Andre~S. Yoon, Duck-Ho Bae, Insoon Jo, Jinyoung Lee, Jonghyun Yoon, Jeong-Uk Kang, Moonsang Kwon, Chanho Yoon, Sangyeun Cho, Jaeheon Jeong, and Duckhyun Chang.
\newblock {Biscuit: A Framework for Near-data Processing of Big Data Workloads}.
\newblock {\em ISCA}, 2016.

\bibitem{kang2013enabling}
Yangwook Kang, Yang-suk Kee, Ethan~L Miller, and Chanik Park.
\newblock {Enabling Cost-effective Data Processing with Smart SSD}.
\newblock In {\em MSST}, 2013.

\bibitem{acharya1998active}
Anurag Acharya, Mustafa Uysal, and Joel Saltz.
\newblock {Active Disks: Programming Model, Algorithms and Evaluation}.
\newblock {\em ASPLOS}, 1998.

\bibitem{keeton1998case}
Kimberly Keeton, David~A Patterson, and Joseph~M Hellerstein.
\newblock {A Case for Intelligent Disks (IDISKs)}.
\newblock {\em {SIGMOD Record}}, 1998.

\bibitem{riedel1998active}
Erik Riedel, Garth Gibson, and Christos Faloutsos.
\newblock {Active Storage for Large-Scale Data Mining and Multimedia Applications}.
\newblock {\em VLDB}, 1998.

\bibitem{riedel2001active}
Erik Riedel, Christos Faloutsos, Garth~A Gibson, and David Nagle.
\newblock {Active Disks for Large-Scale Data Processing}.
\newblock {\em Computer}, 2001.

\bibitem{tiwari2013active}
Devesh Tiwari, Simona Boboila, Sudharshan Vazhkudai, Youngjae Kim, Xiaosong Ma, Peter Desnoyers, and Yan Solihin.
\newblock {Active flash: Towards energy-efficient, in-situ data analytics on extreme-scale machines}.
\newblock In {\em FAST}, 2013.

\bibitem{tiwari2012reducing}
Devesh Tiwari, Sudharshan~S Vazhkudai, Youngjae Kim, Xiaosong Ma, Simona Boboila, and Peter~J Desnoyers.
\newblock {Reducing Data Movement Costs Using Energy-Efficient, Active Computation on SSD}.
\newblock In {\em HotPower}, 2012.

\bibitem{boboila2012active}
Simona Boboila, Youngjae Kim, Sudharshan~S Vazhkudai, Peter Desnoyers, and Galen~M Shipman.
\newblock {Active flash: Out-of-core data analytics on flash storage}.
\newblock In {\em MSST}, 2012.

\bibitem{bae2013intelligent}
Duck-Ho Bae, Jin-Hyung Kim, Sang-Wook Kim, Hyunok Oh, and Chanik Park.
\newblock {Intelligent SSD: a turbo for big data mining}.
\newblock In {\em CIKM}, 2013.

\bibitem{torabzadehkashi2018compstor}
Mahdi Torabzadehkashi, Siavash Rezaei, Vladimir Alves, and Nader Bagherzadeh.
\newblock {Compstor: An in-storage computation platform for scalable distributed processing}.
\newblock In {\em IPDPSW}, 2018.

\bibitem{kang2021iceclave}
Luyi Kang, Yuqi Xue, Weiwei Jia, Xiaohao Wang, Jongryool Kim, Changhwan Youn, Myeong~Joon Kang, Hyung~Jin Lim, Bruce Jacob, and Jian Huang.
\newblock {IceClave: A Trusted Execution Environment for In-Storage Computing}.
\newblock In {\em MICRO}, 2021.

\bibitem{zou2022assasin}
Chen Zou and Andrew~A Chien.
\newblock {ASSASIN: Architecture Support for Stream Computing to Accelerate Computational Storage}.
\newblock In {\em MICRO}, 2022.

\bibitem{jun2015bluedbm}
Sang-Woo Jun, Ming Liu, Sungjin Lee, Jamey Hicks, John Ankcorn, Myron King, Shuotao Xu, and Arvind.
\newblock {BlueDBM: An Appliance for Big Data Analytics}.
\newblock In {\em ISCA}, 2015.

\bibitem{jun2016bluedbm}
Sang-Woo Jun, Ming Liu, Sungjin Lee, Jamey Hicks, John Ankcorn, Myron King, and Shuotao Xu.
\newblock {BlueDBM: Distributed Flash Storage for Big Data Analytics}.
\newblock {\em ACM TOCS}, 2016.

\bibitem{torabzadehkashi2019catalina}
Mahdi Torabzadehkashi, Siavash Rezaei, Ali Heydarigorji, Hosein Bobarshad, Vladimir Alves, and Nader Bagherzadeh.
\newblock {Catalina: In-storage Processing Acceleration for Scalable Big Data Analytics}.
\newblock In {\em Euromicro PDP}, 2019.

\bibitem{cho2013xsd}
Benjamin~Y Cho, Won~Seob Jeong, Doohwan Oh, and Won~Woo Ro.
\newblock {XSD: Accelerating Mapreduce by Harnessing the GPU inside an SSD}.
\newblock In {\em Near-Data Processing}, 2013.

\bibitem{liang2019ins}
Shengwen Liang, Ying Wang, Cheng Liu, Huawei Li, and Xiaowei Li.
\newblock {InS-DLA: An In-SSD deep learning accelerator for near-data processing}.
\newblock In {\em FPL}, 2019.

\bibitem{Pan2025instattention}
Xiurui Pan, Endian Li, Qiao Li, Shengwen Liang, Yizhou Shan, Ke~Zhou, Yingwei Luo, Xiaolin Wang, and Jie Zhang.
\newblock {InstAttention: In-Storage Attention Offloading for Cost-Effective Long-Context LLM Inference}.
\newblock In {\em HPCA}, 2025.

\bibitem{Mahapatra2025isp_rag}
Rohan Mahapatra, Harsha Santhanam, Christopher Priebe, Hanyang Xu, and Hadi Esmaeilzadeh.
\newblock {In-Storage Acceleration of Retrieval Augmented Generation as a Service}.
\newblock In {\em ISCA}, 2025.

\bibitem{liang2019cognitive}
Shengwen Liang, Ying Wang, Youyou Lu, Zhe Yang, Huawei Li, and Xiaowei Li.
\newblock {Cognitive SSD: A Deep Learning Engine for In-Storage Data Retrieval}.
\newblock In {\em USENIX ATC}, 2019.

\bibitem{kim2020reducing}
Minsub Kim and Sungjin Lee.
\newblock {Reducing tail latency of DNN-based recommender systems using in-storage processing}.
\newblock In {\em APSys}, 2020.

\bibitem{lim2021lsm}
Minje Lim, Jeeyoon Jung, and Dongkun Shin.
\newblock {LSM-Tree Compaction Acceleration Using In-Storage Processing}.
\newblock In {\em ICCE-Asia}, 2021.

\bibitem{wang2016ssd}
Jianguo Wang, Dongchul Park, Yang-Suk Kee, Yannis Papakonstantinou, and Steven Swanson.
\newblock {SSD in-storage computing for list intersection}.
\newblock In {\em DaMoN}, 2016.

\bibitem{pei2019registor}
Shuyi Pei, Jing Yang, and Qing Yang.
\newblock {REGISTOR: A Platform for Unstructured Data Processing inside SSD Storage}.
\newblock {\em ACM TOS}, 2019.

\bibitem{do2013query}
Jaeyoung Do, Yang-Suk Kee, Jignesh~M Patel, Chanik Park, Kwanghyun Park, and David~J DeWitt.
\newblock {Query Processing on Smart SSDs: Opportunities and Challenges}.
\newblock In {\em ACM SIGMOD}, 2013.

\bibitem{seshadri2014willow}
Sudharsan Seshadri, Mark Gahagan, Sundaram Bhaskaran, Trevor Bunker, Arup De, Yanqin Jin, Yang Liu, and Steven Swanson.
\newblock {Willow: A User-Programmable SSD}.
\newblock In {\em USENIX OSDI}, 2014.

\bibitem{kim2016storage}
Sungchan Kim, Hyunok Oh, Chanik Park, Sangyeun Cho, Sang-Won Lee, and Bongki Moon.
\newblock {In-storage Processing of Database Scans and Joins}.
\newblock {\em Information Sciences}, 2016.

\bibitem{jeong2019react}
Won~Seob Jeong, Changmin Lee, Keunsoo Kim, Myung~Kuk Yoon, Won Jeon, Myoungsoo Jung, and Won~Woo Ro.
\newblock {REACT: Scalable and High-performance Regular Expression Pattern Matching Accelerator for In-storage Processing}.
\newblock {\em IEEE TPDS}, 2019.

\bibitem{liang2024hyqa}
Shengwen Liang, Ziming Yuan, Ying Wang, Dawen Xu, Huawei Li, and Xiaowei Li.
\newblock {HyQA: Hybrid Near-Data Processing Platform for Embedding Based Question Answering System}.
\newblock In {\em DATE}, 2024.

\bibitem{Hsu2024het3dnand}
Po-Kai Hsu, Vaidehi Garg, Anni Lu, and Shimeng Yu.
\newblock {A Heterogeneous Platform for 3D NAND-Based In-Memory Hyperdimensional Computing Engine for Genome Sequencing Applications}.
\newblock {\em IEEE TCAS-I}, 2024.

\bibitem{mutlu2025memory}
Onur Mutlu, Ataberk Olgun, and {\.I}smail~Emir Y{\"u}ksel.
\newblock {Memory-Centric Computing: Solving Computing's Memory Problem}.
\newblock In {\em IMW}, 2025.

\bibitem{mutlu2022modern}
Onur Mutlu, Saugata Ghose, Juan G{\'o}mez-Luna, and Rachata Ausavarungnirun.
\newblock {A Modern Primer on Processing in Memory}.
\newblock In {\em Emerging Computing: From Devices to Systems: Looking Beyond Moore and Von Neumann}. 2022.

\bibitem{mutlu2025modern}
Onur Mutlu, Saugata Ghose, Juan G{\'o}mez-Luna, Rachata Ausavarungnirun, Mohammad Sadrosadati, and Geraldo~F. Oliveira.
\newblock {A Modern Primer on Processing in Memory}.
\newblock In {\em arXiv}. 2025.

\bibitem{mutlu2019processing}
Onur Mutlu, Saugata Ghose, Juan G{\'o}mez-Luna, and Rachata Ausavarungnirun.
\newblock {Processing Data Where It Makes Sense: Enabling In-memory Computation}.
\newblock {\em Microprocessors and Microsystems}, 2019.

\bibitem{mutlu2019enabling}
Onur Mutlu, Saugata Ghose, Juan G{\'o}mez-Luna, and Rachata Ausavarungnirun.
\newblock {Enabling Practical Processing in and Near Memory for Data-Intensive Computing}.
\newblock In {\em DAC}, 2019.

\bibitem{ghose2019processing}
Saugata Ghose, Amirali Boroumand, Jeremie~S Kim, Juan G{\'o}mez-Luna, and Onur Mutlu.
\newblock {Processing-in-Memory: A Workload-Driven Perspective}.
\newblock {\em IBM Journal of Research and Development}, 2019.

\bibitem{mutlu2013memory}
Onur Mutlu.
\newblock {Memory Scaling: A Systems Architecture Perspective}.
\newblock In {\em IMW}, 2013.

\bibitem{mutlu2024memory}
Onur Mutlu, Ataberk Olgun, Geraldo~F Oliveira, and Ismail~E Yuksel.
\newblock {Memory-Centric Computing: Recent Advances in Processing-in-DRAM (Invited)}.
\newblock In {\em IEDM}, 2024.

\bibitem{yuksel2026memory}
Ismail~Emir Yuksel, Fatma~Nisa Bostanci, Ataberk Olgun, and Onur Mutlu.
\newblock {Memory-Centric Computing: Security Benefits and Challenges of Processing-in-DRAM}.
\newblock In {\em ICS Workshops}, 2026.

\bibitem{oliveira2022accelerating}
Geraldo~F Oliveira, Juan G{\'o}mez-Luna, Saugata Ghose, Amirali Boroumand, and Onur Mutlu.
\newblock {Accelerating Neural Network Inference with Processing-in-DRAM: From the Edge to the Cloud}.
\newblock {\em IEEE Micro}, 2022.

\bibitem{seshadri2015gather}
Vivek Seshadri, Thomas Mullins, Amirali Boroumand, Onur Mutlu, Phillip~B Gibbons, Michael~A Kozuch, and Todd~C Mowry.
\newblock {Gather-Scatter DRAM: In-DRAM Address Translation to Improve the Spatial Locality of Non-Unit Strided Accesses}.
\newblock In {\em MICRO}, 2015.

\bibitem{lee2015decoupled}
Donghyuk Lee, Lavanya Subramanian, Rachata Ausavarungnirun, Jongmoo Choi, and Onur Mutlu.
\newblock {Decoupled Direct Memory Access: Isolating CPU and IO Traffic by Leveraging a Dual-Data-Port DRAM}.
\newblock In {\em PACT}, 2015.

\bibitem{hashemi2016continuous}
Milad Hashemi, Onur Mutlu, and Yale~N Patt.
\newblock {Continuous Runahead: Transparent Hardware Acceleration for Memory Intensive Workloads}.
\newblock In {\em MICRO}, 2016.

\bibitem{hashemi2016accelerating}
Milad Hashemi, {Khubaib}, Eiman Ebrahimi, Onur Mutlu, and Yale~N. Patt.
\newblock {Accelerating Dependent Cache Misses with an Enhanced Memory Controller}.
\newblock In {\em ISCA}, 2016.

\bibitem{singh2020nero}
Gagandeep Singh, Dionysios Diamantopoulos, Christoph Hagleitner, Juan Gomez-Luna, Sander Stuijk, Onur Mutlu, and Henk Corporaal.
\newblock {NERO: A Near High-Bandwidth Memory Stencil Accelerator for Weather Prediction Modeling}.
\newblock In {\em FPL}, 2020.

\bibitem{asghari2016chameleon}
Hadi Asghari-Moghaddam, Young~Hoon Son, Jung~Ho Ahn, and Nam~Sung Kim.
\newblock {Chameleon: Versatile and Practical Near-DRAM Acceleration Architecture for Large Memory Systems}.
\newblock In {\em MICRO}, 2016.

\bibitem{sun2021abc}
Weiyi Sun, Zhaoshi Li, Shouyi Yin, Shaojun Wei, and Leibo Liu.
\newblock {ABC-DIMM: Alleviating the Bottleneck of Communication in DIMM-Based Near-Memory Processing with Inter-DIMM Broadcast}.
\newblock In {\em ISCA}, 2021.

\bibitem{ke2021near}
Liu Ke, Xuan Zhang, Jinin So, Jong-Geon Lee, Shin-Haeng Kang, Sukhan Lee, Songyi Han, YeonGon Cho, Jin~Hyun Kim, Yongsuk Kwon, KyungSoo Kim, Jin Jung, Ilkwon Yun, Sung~Joo Park, Hyunsun Park, Joonho Song, Jeonghyeon Cho, Kyomin Sohn, Nam~Sung Kim, and Hsien-Hsin~S. Lee.
\newblock {Near-Memory Processing in Action: Accelerating Personalized Recommendation With AxDIMM}.
\newblock {\em IEEE Micro}, 2022.

\bibitem{ke2019recnmp}
Liu Ke, Udit Gupta, Benjamin~Youngjae Cho, David Brooks, Vikas Chandra, Utku Diril, Amin Firoozshahian, Kim Hazelwood, Bill Jia, Hsien-Hsin~S. Lee, Meng Li, Bert Maher, Dheevatsa Mudigere, Maxim Naumov, Martin Schatz, Mikhail Smelyanskiy, Xiaodong Wang, Brandon Reagen, Carole-Jean Wu, Mark Hempstead, et~al.
\newblock {RecNMP: Accelerating Personalized Recommendation with Near-Memory Processing}.
\newblock In {\em ISCA}, 2020.

\bibitem{ahn2015scalable}
Junwhan Ahn, Sungpack Hong, Sungjoo Yoo, Onur Mutlu, and Kiyoung Choi.
\newblock {A Scalable Processing-in-memory Accelerator for Parallel Graph Processing}.
\newblock In {\em ISCA}, 2015.

\bibitem{drumond2017mondrian}
Mario Drumond, Alexandros Daglis, Nooshin Mirzadeh, Dmitrii Ustiugov, Javier Picorel, Babak Falsafi, Boris Grot, and Dionisios Pnevmatikatos.
\newblock {The Mondrian Data Engine}.
\newblock In {\em ISCA}, 2017.

\bibitem{boroum2019conda}
Amirali Boroumand, Saugata Ghose, Minesh Patel, Hasan Hassan, Brandon Lucia, Kevin Hsieh, Krishna~T. Malladi, Hongzhong Zheng, and Onur Mutlu.
\newblock {CoNDA: Enabling Efficient Near-Data Accelerator Communication by Optimizing Data Movement}.
\newblock {\em ISCA}, 2019.

\bibitem{boroumand2017lazypim}
Amirali Boroumand, Saugata Ghose, Minesh Patel, Hasan Hassan, Brandon Lucia, Kevin Hsieh, Krishna~T Malladi, Hongzhong Zheng, and Onur Mutlu.
\newblock {LazyPIM: An Efficient Cache Coherence Mechanism for Processing-in-memory}.
\newblock {\em CAL}, 2017.

\bibitem{NDC_ISPASS_2014}
Seth~H. Pugsley, Jeffrey Jestes, Huihui Zhang, Rajeev Balasubramonian, Vijayalakshmi Srinivasan, Alper Buyuktosunoglu, Al~Davis, and Feifei Li.
\newblock {NDC: Analyzing the impact of 3D-stacked memory+logic devices on MapReduce workloads}.
\newblock In {\em ISPASS}, 2014.

\bibitem{singh2019napel}
Gagandeep Singh, Giovanni ~, Geraldo~F Oliveira, Stefano Corda, Sander Stuijk, Onur Mutlu, and Henk Corporaal.
\newblock {NAPEL: Near-Memory Computing Application Performance Prediction via Ensemble Learning}.
\newblock In {\em DAC}, 2019.

\bibitem{azarkhish2016logic}
Erfan Azarkhish, Christoph Pfister, Davide Rossi, Igor Loi, and Luca Benini.
\newblock {Logic-Base Interconnect Design for Near Memory Computing in the Smart Memory Cube}.
\newblock {\em IEEE VLSI}, 2016.

\bibitem{azarkhish2018neurostream}
Erfan Azarkhish, Davide Rossi, Igor Loi, and Luca Benini.
\newblock {Neurostream: Scalable and Energy Efficient Deep Learning with Smart Memory Cubes}.
\newblock {\em TPDS}, 2018.

\bibitem{top-pim}
Dongping Zhang, Nuwan Jayasena, Alexander Lyashevsky, Joseph~L Greathouse, Lifan Xu, and Michael Ignatowski.
\newblock {TOP-PIM: Throughput-Oriented Programmable Processing in Memory}.
\newblock In {\em HPDC}, 2014.

\bibitem{RVU}
P.~C. Santos, G.~F. Oliveira, D.~G. Tomé, M.~A.~Z. Alves, E.~C. Almeida, and L.~Carro.
\newblock {Operand Size Reconfiguration for Big Data Processing in Memory}.
\newblock In {\em DATE}, 2017.

\bibitem{NIM}
Geraldo~F Oliveira, Paulo~C Santos, Marco~AZ Alves, and Luigi Carro.
\newblock {NIM: An HMC-Based Machine for Neuron Computation}.
\newblock In {\em ARC}, 2017.

\bibitem{gao2017tetris}
Mingyu Gao, Jing Pu, Xuan Yang, Mark Horowitz, and Christos Kozyrakis.
\newblock {TETRIS: Scalable and Efficient Neural Network Acceleration with 3D Memory}.
\newblock In {\em ASPLOS}, 2017.

\bibitem{hsieh2016transparent}
Kevin Hsieh, Eiman Ebrahimi, Gwangsun Kim, Niladrish Chatterjee, Mike O'Connor, Nandita Vijaykumar, Onur Mutlu, and Stephen~W Keckler.
\newblock {Transparent Offloading and Mapping (TOM): Enabling Programmer-Transparent Near-Data Processing in GPU Systems}.
\newblock In {\em ISCA}, 2016.

\bibitem{boroumand2021mitigating}
Amirali Boroumand, Saugata Ghose, Berkin Akin, Ravi Narayanaswami, Geraldo~F Oliveira, Xiaoyu Ma, Eric Shiu, and Onur Mutlu.
\newblock {Mitigating Edge Machine Learning Inference Bottlenecks: An Empirical Study on Accelerating Google Edge Models}.
\newblock arXiv:2103.00768 [cs.AR], 2021.

\bibitem{boroumand2021google}
Amirali Boroumand, Saugata Ghose, Berkin Akin, Ravi Narayanaswami, Geraldo~F Oliveira, Xiaoyu Ma, Eric Shiu, and Onur Mutlu.
\newblock {Google Neural Network Models for Edge Devices: Analyzing and Mitigating Machine Learning Inference Bottlenecks}.
\newblock In {\em PACT}, 2021.

\bibitem{boroumand2021polynesia}
Amirali Boroumand, Saugata Ghose, Geraldo~F Oliveira, and Onur Mutlu.
\newblock {Polynesia: Enabling Effective Hybrid Transactional/Analytical Databases with Specialized Hardware/Software Co-Design}.
\newblock arXiv:2103.00798 [cs.AR], 2021.

\bibitem{fernandez2020natsa}
Ivan Fernandez, Ricardo Quislant, Eladio Guti{\'e}rrez, Oscar Plata, Christina Giannoula, Mohammed Alser, Juan G{\'o}mez-Luna, and Onur Mutlu.
\newblock {NATSA: A Near-Data Processing Accelerator for Time Series Analysis}.
\newblock In {\em ICCD}, 2020.

\bibitem{LiM_3D_FFT_MM}
Qiuling Zhu, Berkin Akin, H.~Ekin Sumbul, Fazle Sadi, James~C. Hoe, Larry Pileggi, and Franz Franchetti.
\newblock {A 3D-stacked logic-in-memory accelerator for application-specific data intensive computing}.
\newblock In {\em 3DIC}, 2013.

\bibitem{akin2014hamlet}
Berkin Ak{\i}n, James~C Hoe, and Franz Franchetti.
\newblock {HAMLeT: Hardware Accelerated Memory Layout Transform within 3D-Stacked DRAM}.
\newblock In {\em HPEC}, 2014.

\bibitem{gao2016hrl}
Mingyu Gao and Christos Kozyrakis.
\newblock {HRL: Efficient and Flexible Reconfigurable Logic for Near-data Processing}.
\newblock In {\em HPCA}, 2016.

\bibitem{farmahini2015nda}
Amin Farmahini-Farahani, Jung~Ho Ahn, Katherine Morrow, and Nam~Sung Kim.
\newblock {NDA: Near-DRAM Acceleration Architecture Leveraging Commodity DRAM Devices and Standard Memory Modules}.
\newblock In {\em HPCA}, 2015.

\bibitem{boroumand2018google}
Amirali Boroumand, Saugata Ghose, Youngsok Kim, Rachata Ausavarungnirun, Eric Shiu, Rahul Thakur, Daehyun Kim, Aki Kuusela, Allan Knies, Parthasarathy Ranganathan, and Onur Mutlu.
\newblock {Google Workloads for Consumer Devices: Mitigating Data Movement Bottlenecks}.
\newblock In {\em ASPLOS}, 2018.

\bibitem{nai2017graphpim}
Lifeng Nai, Ramyad Hadidi, Jaewoong Sim, Hyojong Kim, Pranith Kumar, and Hyesoon Kim.
\newblock {GraphPIM: Enabling Instruction-Level PIM Offloading in Graph Computing Frameworks}.
\newblock In {\em HPCA}, 2017.

\bibitem{PEI}
Junwhan Ahn, Sungjoo Yoo, Onur Mutlu, and Kiyoung Choi.
\newblock {PIM-enabled Instructions: A Low-overhead, Locality-aware Processing-in-memory Architecture}.
\newblock In {\em ISCA}, 2015.

\bibitem{kwon202125}
Young-Cheon Kwon, Suk~Han Lee, Jaehoon Lee, Sang-Hyuk Kwon, Je~Min Ryu, Jong-Pil Son, O~Seongil, Hak-Soo Yu, Haesuk Lee, Soo~Young Kim, Youngmin Cho, Jin~Guk Kim, Jongyoon Choi, Hyun-Sung Shin, Jin Kim, BengSeng Phuah, HyoungMin Kim, Myeong~Jun Song, Ahn Choi, Daeho Kim, et~al.
\newblock {A 20nm 6GB Function-In-Memory DRAM, Based on HBM2 with a 1.2TFLOPS Programmable Computing Unit Using Bank-Level Parallelism, for Machine Learning Applications}.
\newblock In {\em ISSCC}, 2021.

\bibitem{lee2021hardware}
Sukhan Lee, Shin-haeng Kang, Jaehoon Lee, Hyeonsu Kim, Eojin Lee, Seungwoo Seo, Hosang Yoon, Seungwon Lee, Kyounghwan Lim, Hyunsung Shin, Jinhyun Kim, O~Seongil, Anand Iyer, David Wang, Kyomin Sohn, and Nam~Sung Kim.
\newblock {Hardware Architecture and Software Stack for PIM Based on Commercial DRAM Technology: Industrial Product}.
\newblock In {\em ISCA}, 2021.

\bibitem{lee2016simultaneous}
Donghyuk Lee, Saugata Ghose, Gennady Pekhimenko, Samira Khan, and Onur Mutlu.
\newblock {Simultaneous Multi-layer Access: Improving 3D-stacked Memory Bandwidth at Low Cost}.
\newblock {\em TACO}, 2016.

\bibitem{hsieh_accelerating_2016}
K.~Hsieh, S.~Khan, N.~Vijaykumar, K.~K. Chang, A.~Boroumand, S.~Ghose, and O.~Mutlu.
\newblock {Accelerating Pointer Chasing in 3D-Stacked Memory: Challenges, Mechanisms, Evaluation}.
\newblock In {\em {ICCD}}, 2016.

\bibitem{pattnaik2016scheduling}
Ashutosh Pattnaik, Xulong Tang, Adwait Jog, Onur Kayiran, Asit~K Mishra, Mahmut~T Kandemir, Onur Mutlu, and Chita~R Das.
\newblock {Scheduling Techniques for GPU Architectures With Processing-in-memory Capabilities}.
\newblock In {\em PACT}, 2016.

\bibitem{syncron}
Christina Giannoula, Nandita Vijaykumar, Nikela Papadopoulou, Vasileios Karakostas, Ivan Fernandez, Juan Gómez-Luna, Lois Orosa, Nectarios Koziris, Georgios Goumas, and Onur Mutlu.
\newblock {SynCron: Efficient Synchronization Support for Near-Data-Processing Architectures}.
\newblock In {\em HPCA}, 2021.

\bibitem{ghiasi2022alp}
Nika~Mansouri Ghiasi, Nandita Vijaykumar, Geraldo~F Oliveira, Lois Orosa, Ivan Fernandez, Mohammad Sadrosadati, Konstantinos Kanellopoulos, Nastaran Hajinazar, Juan~G{\'o}mez Luna, and Onur Mutlu.
\newblock {ALP: Alleviating CPU-Memory Data Movement Overheads in Memory-Centric Systems}.
\newblock {\em IEEE TETC}, 2022.

\bibitem{bostanci2025revisiting}
F.~Nisa Bostancı, Konstantinos Kanellopoulos, Ataberk Olgun, A.~Giray Yağlıkçı, İsmail Emir~Yüksel, Nika~Mansouri Ghiasi, Zülal Bingöl, Mohammad Sadrosadati, and Onur Mutlu.
\newblock {Revisiting Main Memory-Based Covert and Side Channel Attacks in the Context of Processing-in-Memory}.
\newblock In {\em DSN}, 2025.

\bibitem{HBM}
Dong~Uk Lee, Kyung~Whan Kim, Kwan~Weon Kim, Hongjung Kim, Ju~Young Kim, Young~Jun Park, Jae~Hwan Kim, Dae~Suk Kim, Heat~Bit Park, Jin~Wook Shin, et~al.
\newblock {A 1.2V 8Gb 8-Channel 128GB/s High-Bandwidth Memory (HBM) Stacked DRAM with Effective Microbump I/O Test Methods Using 29nm Process and TSV}.
\newblock In {\em ISSCC}, 2014.

\bibitem{hbm2}
{JEDEC Solid State Technology Assn.}
\newblock {JESD23-5D: High Bandwidth Memory (HBM) DRAM Standard}, March 2021.

\bibitem{hmc.spec.2.0}
{Hybrid Memory Cube Consortium}.
\newblock {HMC Specification 2.0}, 2014.

\bibitem{loh2008stacked}
Gabriel~H. Loh.
\newblock {3D-Stacked Memory Architectures for Multi-Core Processors}.
\newblock In {\em ISCA}, 2008.

\bibitem{gopireddy2019m3d}
B.~{Gopireddy} and J.~{Torrellas}.
\newblock {Designing Vertical Processors in Monolithic 3D}.
\newblock In {\em ISCA}, 2019.

\bibitem{mitra2018vlse}
S.~{Mitra}.
\newblock {Abundant-Data Computing: The N3XT 1,000X}.
\newblock In {\em VLSI-TSA}, 2018.

\bibitem{hwang2018cmos}
W.~{Hwang}, W.~{Wan}, S.~{Mitra}, and H.~.~P. {Wong}.
\newblock {Coming Up N3XT, After 2D Scaling of Si CMOS}.
\newblock In {\em ISCAS}, 2018.

\bibitem{mitra2015nano}
S.~{Mitra}.
\newblock {From Nanodevices to Nanosystems: The N3XT Information Technology}.
\newblock In {\em E3S}, 2015.

\bibitem{rich2020nano}
Dennis Rich, Andrew Bartolo, Carlo Gilardo, Binh Le, Haitong Li, Rebecca Park, Robert~M. Radway, Mohamed~M. Sabry~Aly, H.-S.~Philip Wong, and Subhasish Mitra.
\newblock {\em {Heterogeneous 3D Nano-Systems: The N3XT Approach?}}
\newblock 2020.

\bibitem{sabry2015abundant}
M.~M. {Sabry Aly}, M.~{Gao}, G.~{Hills}, C.~{Lee}, G.~{Pitner}, M.~M. {Shulaker}, T.~F. {Wu}, M.~{Asheghi}, J.~{Bokor}, F.~{Franchetti}, K.~E. {Goodson}, C.~{Kozyrakis}, I.~{Markov}, K.~{Olukotun}, L.~{Pileggi}, E.~{Pop}, J.~{Rabaey}, C.~{Ré}, H.~.~P. {Wong}, and S.~{Mitra}.
\newblock {Energy-Efficient Abundant-Data Computing: The N3XT 1,000x}.
\newblock {\em Computer}, 2015.

\bibitem{sabry2019n3xt}
M.~M. {Sabry Aly}, T.~F. {Wu}, A.~{Bartolo}, Y.~H. {Malviya}, W.~{Hwang}, G.~{Hills}, I.~{Markov}, M.~{Wootters}, M.~M. {Shulaker}, H.~. {Philip Wong}, and S.~{Mitra}.
\newblock {The N3XT Approach to Energy-Efficient Abundant-Data Computing}.
\newblock {\em Proc. IEEE}, 2019.

\bibitem{ghiasi2022revamp3d}
Nika~Mansouri Ghiasi, Mohammad Sadrosadati, Geraldo~F Oliveira, Konstantinos Kanellopoulos, Rachata Ausavarungnirun, Juan~G{\'o}mez Luna, Aditya Manglik, Jo{\\textasciitilde a}o Ferreira, Jeremie~S Kim, Christina Giannoula, et~al.
\newblock {RevaMp3D: Architecting the Processor Core and Cache Hierarchy for Systems with Monolithically-Integrated Logic and Memory}.
\newblock arXiv:2210.08508 [cs.AR], 2022.

\bibitem{skhynixpim}
S.~Lee, K.~Kim, S.~Oh, J.~Park, G.~Hong, D.~Ka, K.~Hwang, J.~Park, K.~Kang, J.~Kim, J.~Jeon, N.~Kim, Y.~Kwon, K.~Vladimir, W.~Shin, J.~Won, M.~Lee, H.~Joo, et~al.
\newblock {A 1ynm 1.25V 8Gb, 16Gb/s/pin GDDR6-based Accelerator-in-Memory Supporting 1TFLOPS MAC Operation and Various Activation Functions for Deep-Learning Applications}.
\newblock In {\em ISSCC}, 2022.

\bibitem{devaux2019true}
Fabrice Devaux.
\newblock {The True Processing in Memory Accelerator}.
\newblock In {\em Hot Chips}, 2019.

\bibitem{gomezluna2021benchmarking}
Juan G{\'o}mez-Luna, Izzat~El Hajj, Ivan Fernández, Christina Giannoula, Geraldo~F. Oliveira, and Onur Mutlu.
\newblock {Benchmarking a New Paradigm: An Experimental Analysis of a Real Processing-in-Memory Architecture}.
\newblock arXiv:2105.03814 [cs.AR], 2021.

\bibitem{gomez2021benchmarkingcut}
Juan G{\'o}mez-Luna, Izzat El~Hajj, Ivan Fernandez, Christina Giannoula, Geraldo~F Oliveira, and Onur Mutlu.
\newblock {Benchmarking Memory-Centric Computing Systems: Analysis of Real Processing-in-Memory Hardware}.
\newblock In {\em CUT}, 2021.

\bibitem{gomez2022benchmarking}
Juan G{\'o}mez-Luna, Izzat El~Hajj, Ivan Fernandez, Christina Giannoula, Geraldo~F Oliveira, and Onur Mutlu.
\newblock {Benchmarking a New Paradigm: Experimental Analysis and Characterization of a Real Processing-in-Memory System}.
\newblock {\em IEEE Access}, 2022.

\bibitem{gu2025pim}
Yufeng Gu, Alireza Khadem, Sumanth Umesh, Ning Liang, Xavier Servot, Onur Mutlu, Ravi Iyer, and Reetuparna Das.
\newblock {PIM is All You Need: A CXL-Enabled GPU-Free System for Large Language Model Inference}.
\newblock In {\em ASPLOS}, 2025.

\bibitem{chen2023simplepim}
Jinfan Chen, Juan G{\'o}mez-Luna, Izzat El~Hajj, Yuxin Guo, and Onur Mutlu.
\newblock {SimplePIM: A Software Framework for Productive and Efficient Processing-in-Memory}.
\newblock In {\em PACT}, 2023.

\bibitem{yang2026dcc}
Peiming Yang, Sankeerth Durvasula, Ivan Fernandez, Mohammad Sadrosadati, Onur Mutlu, Gennady Pekhimenko, and Christina Giannoula.
\newblock {DCC: Data-Centric Compilation of Machine Learning Kernels for Processing-In-Memory Architectures}.
\newblock In {\em 2026 ACM/IEEE 53rd Annual International Symposium on Computer Architecture (ISCA)}, pages 2582--2599. IEEE, 2026.

\bibitem{rhyner2024pimopt}
Steve Rhyner, Haocong Luo, Juan Gomez-Luna, Mohammad Sadrosadati, Jiawei Jiang, Ataberk Olgun, Harshita Gupta, Ce~Zhang, and Onur Mutlu.
\newblock {PIM-Opt: Demystifying Distributed Optimization Algorithms on a Real-World Processing-In-Memory System}.
\newblock In {\em PACT}, 2024.

\bibitem{giannoula2022sparsep}
Christina Giannoula, Ivan Fernandez, Juan~G{\'o}mez Luna, Nectarios Koziris, Georgios Goumas, and Onur Mutlu.
\newblock {SparseP: Towards Efficient Sparse Matrix Vector Multiplication on Real Processing-in-Memory Architectures}.
\newblock In {\em SIGMETRICS}, 2022.

\bibitem{barkhordar2025alpha}
Marzieh Barkhordar, Alireza Tabatabaeian, Mohammad Sadrosadati, Christina Giannoula, Juan~Gomez Luna, Izzat El~Hajj, Onur Mutlu, and Alaa~R Alameldeen.
\newblock {ALPHA-PIM: Analysis of Linear Algebraic Processing for High-Performance Graph Applications on a Real Processing-In-Memory System}.
\newblock In {\em IISWC}, 2025.

\bibitem{giannoula2024pygim}
Christina Giannoula, Peiming Yang, Ivan Fernandez, Jiacheng Yang, Sankeerth Durvasula, Yu~Xin Li, Mohammad Sadrosadati, Juan~Gomez Luna, Onur Mutlu, and Gennady Pekhimenko.
\newblock {PyGim: An Efficient Graph Neural Network Library for Real Processing-In-Memory Architectures}.
\newblock In {\em SIGMETRICS}, 2024.

\bibitem{gupta2023evaluating}
Harshita Gupta, Mayank Kabra, Juan G{\'o}mez-Luna, Konstantinos Kanellopoulos, and Onur Mutlu.
\newblock {Evaluating Homomorphic Operations on a Real-World Processing-In-Memory System}.
\newblock In {\em IISWC}, 2023.

\bibitem{zhao2026cosm}
Yilong Zhao, Fangxin Liu, Onur Mutlu, Mingyu Gao, Jian Liu, Haibing Guan, and Li~Jiang.
\newblock {COSM: A Cooperative Scheduling Framework for Concurrent PIM and CPU Execution on Mobile Devices}.
\newblock In {\em ISCA}, 2026.

\bibitem{chang2016low}
Kevin~K Chang, Prashant~J Nair, Donghyuk Lee, Saugata Ghose, Moinuddin~K Qureshi, and Onur Mutlu.
\newblock {Low-Cost Inter-Linked Subarrays (LISA): Enabling Fast Inter-Subarray Data Movement in DRAM}.
\newblock In {\em HPCA}, 2016.

\bibitem{seshadri2017ambit}
Vivek Seshadri, Donghyuk Lee, Thomas Mullins, Hasan Hassan, Amirali Boroumand, Jeremie Kim, Michael~A Kozuch, Onur Mutlu, Phillip~B Gibbons, and Todd~C Mowry.
\newblock {Ambit: In-Memory Accelerator for Bulk Bitwise Operations Using Commodity DRAM Technology}.
\newblock In {\em MICRO}, 2017.

\bibitem{hajinazarsimdram}
Nastaran Hajinazar, Geraldo~F. Oliveira, Sven Gregorio, Jo{\~a}o~Dinis Ferreira, Nika~Mansouri Ghiasi, Minesh Patel, Mohammed Alser, Saugata Ghose, Juan G{\'o}mez-Luna, and Onur Mutlu.
\newblock {SIMDRAM: A Framework for Bit-Serial SIMD Processing Using DRAM}.
\newblock In {\em ASPLOS}, 2021.

\bibitem{seshadri2013rowclone}
Vivek Seshadri, Yoongu Kim, Chris Fallin, Donghyuk Lee, Rachata Ausavarungnirun, Gennady Pekhimenko, Yixin Luo, Onur Mutlu, Phillip~B. Gibbons, Michael~A. Kozuch, and Todd~C. Mowry.
\newblock {RowClone: Fast and Energy-Efficient in-DRAM Bulk Data Copy and Initialization}.
\newblock In {\em MICRO}, 2013.

\bibitem{seshadri2019dram}
Vivek Seshadri and Onur Mutlu.
\newblock {In-DRAM Bulk Bitwise Execution Engine}.
\newblock arXiv:1905.09822 [cs.AR], 2019.

\bibitem{seshadri2016processing}
Vivek Seshadri and Onur Mutlu.
\newblock {The Processing Using Memory Paradigm: In-DRAM Bulk Copy, Initialization, Bitwise AND and OR}.
\newblock arXiv:1610.09603 [cs.AR], 2016.

\bibitem{seshadri.bookchapter17}
Vivek Seshadri and Onur Mutlu.
\newblock {Simple Operations in Memory to Reduce Data Movement}.
\newblock In {\em Advances in Computers}, volume 106. 2017.

\bibitem{seshadri2016buddy}
Vivek Seshadri, Donghyuk Lee, Thomas Mullins, Hasan Hassan, Amirali Boroumand, Jeremie Kim, Michael~A Kozuch, Onur Mutlu, Phillip~B Gibbons, and Todd~C Mowry.
\newblock {Buddy-RAM: Improving the Performance and Efficiency of Bulk Bitwise Operations Using DRAM}.
\newblock arXiv:1611.09988 [cs.AR], 2016.

\bibitem{seshadri2015fast}
Vivek Seshadri, Kevin Hsieh, Amirali Boroum, Donghyuk Lee, Michael~A Kozuch, Onur Mutlu, Phillip~B Gibbons, and Todd~C Mowry.
\newblock {Fast Bulk Bitwise AND and OR in DRAM}.
\newblock {\em CAL}, 2015.

\bibitem{angizi2019graphide}
Shaahin Angizi and Deliang Fan.
\newblock {GraphiDe: A Graph Processing Accelerator Leveraging In-DRAM-Computing}.
\newblock In {\em GLSVLSI}, 2019.

\bibitem{ferreira2021pluto}
Jo{\\textasciitilde a}o~Dinis Ferreira, Gabriel Falcao, Juan G{\'o}mez-Luna, Mohammed Alser, Lois Orosa, Mohammad Sadrosadati, Jeremie~S Kim, Geraldo~F Oliveira, Taha Shahroodi, Anant Nori, et~al.
\newblock {pLUTo: In-DRAM Lookup Tables to Enable Massively Parallel General-Purpose Computation}.
\newblock arXiv:2104.07699 [cs.AR], 2021.

\bibitem{mimdramextended}
Geraldo~F Oliveira, Ataberk Olgun, A.~Giray~Gregorio Yaglik{\c{c}}i, Nisa Bostanci, Juan G{\'o}mez-Luna, Saugata Ghose, and Onur Mutlu.
\newblock {MIMDRAM: An End-to-End Processing-Using-DRAM System for High-Throughput, Energy-Efficient and Programmer-Transparent Multiple-Instruction Multiple-Data Computing}.
\newblock In {\em HPCA}, 2024.

\bibitem{missingnot}
Ismail~Emir Yuksel, Yahya~Can Tu\u{g}rul, Ataberk Olgun, F.~Nisa Bostanci, A.~Giray Yaglik{\c{c}}i, Geraldo~F. Oliveira, Haocong Luo, Juan G{\'o}mez-Luna, Mohammad Sadrosadati, and Onur Mutlu.
\newblock {Functionally-Complete Boolean Logic in Real DRAM Chips: Experimental Characterization and Analysis}.
\newblock In {\em HPCA}, 2024.

\bibitem{yuksel2024simultaneous}
Ismail~Emir Yuksel, Yahya~Can Tugrul, F~Bostanci, Geraldo~F Oliveira, A~Giray Yaglikci, Ataberk Olgun, Melina Soysal, Haocong Luo, Juan G{\'o}mez-Luna, Mohammad Sadrosadati, et~al.
\newblock {Simultaneous Many-Row Activation in Off-the-Shelf DRAM Chips: Experimental Characterization and Analysis}.
\newblock In {\em DSN}, 2024.

\bibitem{olgun2022pidram}
Ataberk Olgun, Juan~Gomez Luna, Konstantinos Kanellopoulos, Behzad Salami, Hasan Hassan, Oguz Ergin, and Onur Mutlu.
\newblock {PiDRAM: A Holistic End-to-End FPGA-Based Framework for Processing-in-DRAM}.
\newblock {\em TACO}, 2022.

\bibitem{angizi2018pima}
S.~Angizi, Z.~He, and D.~Fan.
\newblock {PIMA-Logic: A Novel Processing-in-Memory Architecture for Highly Flexible and Energy-Efficient Logic Computation}.
\newblock In {\em DAC}, 2018.

\bibitem{angizi2018cmp}
S.~Angizi, A.~S. Rakin, and D.~Fan.
\newblock {CMP-PIM: An Energy-Efficient Comparator-Based Processing-in-Memory Neural Network Accelerator}.
\newblock In {\em DAC}, 2018.

\bibitem{levy.microelec14}
Yifat Levy, Jehoshua Bruck, Yuval Cassuto, Eby~G. Friedman, Avinoam Kolodny, Eitan Yaakobi, and Shahar Kvatinsky.
\newblock {Logic Operations in Memory Using a Memristive Akers Array}.
\newblock {\em Microelectronics Journal}, 2014.

\bibitem{kvatinsky.tcasii14}
S.~Kvatinsky, D.~Belousov, S.~Liman, G.~Satat, N.~Wald, E.~G. Friedman, A.~Kolodny, and U.~C. Weiser.
\newblock {MAGIC---Memristor-Aided Logic}.
\newblock {\em IEEE TCAS II: Express Briefs}, 2014.

\bibitem{Shafiee2016}
Ali Shafiee, Anirban Nag, Naveen Muralimanohar, Rajeev Balasubramonian, John~Paul Strachan, Miao Hu, R.~Stanley Williams, and Vivek Srikumar.
\newblock {ISAAC: A Convolutional Neural Network Accelerator with In-Situ Analog Arithmetic in Crossbars}.
\newblock In {\em ISCA}, 2016.

\bibitem{kvatinsky.iccd11}
S.~Kvatinsky, A.~Kolodny, U.~C. Weiser, and E.~G. Friedman.
\newblock {Memristor-Based IMPLY Logic Design Procedure}.
\newblock In {\em ICCD}, 2011.

\bibitem{kvatinsky.tvlsi14}
S.~Kvatinsky, G.~Satat, N.~Wald, E.~G. Friedman, A.~Kolodny, and U.~C. Weiser.
\newblock {Memristor-Based Material Implication (IMPLY) Logic: Design Principles and Methodologies}.
\newblock {\em TVLSI}, 2014.

\bibitem{gaillardon2016plim}
P.-E. Gaillardon, L.~Amaru, A.~Siemon, and et~al.
\newblock {The Programmable Logic-in-Memory (PLiM) Computer}.
\newblock In {\em DATE}, 2016.

\bibitem{bhattacharjee2017revamp}
D.~Bhattacharjee, R.~Devadoss, and A.~Chattopadhyay.
\newblock {ReVAMP: ReRAM Based VLIW Architecture for In-Memory Computing}.
\newblock In {\em DATE}, 2017.

\bibitem{hamdioui2015memristor}
S.~Hamdioui, L.~Xie, H.~A.~D. Nguyen, and et~al.
\newblock {Memristor Based Computation-in-Memory Architecture for Data-Intensive Applications}.
\newblock In {\em DATE}, 2015.

\bibitem{xie2015fast}
L.~Xie, H.~A.~D. Nguyen, M.~Taouil, and et~al.
\newblock {Fast Boolean Logic Papped on Memristor Crossbar}.
\newblock In {\em ICCD}, 2015.

\bibitem{hamdioui2017myth}
S.~Hamdioui, S.~Kvatinsky, and et~al. G.~Cauwenberghs.
\newblock {Memristor for Computing: Myth or Reality?}
\newblock In {\em DATE}, 2017.

\bibitem{yu2018memristive}
J.~Yu, H.~A.~D. Nguyen, L.~Xie, and et~al.
\newblock {Memristive Devices for Computation-in-Memory}.
\newblock In {\em DATE}, 2018.

\bibitem{yavits2021giraf}
Leonid Yavits, Roman Kaplan, and Ran Ginosar.
\newblock {GIRAF: General Purpose In-Storage Resistive Associative Framework}.
\newblock {\em TPDS}, 2021.

\bibitem{xi2020memory}
Yue Xi, Bin Gao, Jianshi Tang, An~Chen, Meng-Fan Chang, Xiaobo~Sharon Hu, Jan Van Der~Spiegel, He~Qian, and Huaqiang Wu.
\newblock {In-Memory Learning With Analog Resistive Switching Memory: A Review and Perspective}.
\newblock {\em Proc. IEEE}, 2020.

\bibitem{zheng2016tcam}
Le~Zheng, Sangho Shin, Scott Lloyd, Maya Gokhale, Kyungmin Kim, and Sung-Mo Kang.
\newblock {RRAM-Based TCAMs for Pattern Search}.
\newblock In {\em ISCAS}, 2016.

\bibitem{truong2021racer}
Minh~SQ Truong, Eric Chen, Deanyone Su, Liting Shen, Alexander Glass, L~Richard Carley, James~A Bain, and Saugata Ghose.
\newblock {RACER: Bit-Pipelined Processing Using Resistive Memory}.
\newblock In {\em MICRO}, 2021.

\bibitem{li2017drisa}
Shuangchen Li, Dimin Niu, Krishna~T Malladi, Hongzhong Zheng, Bob Brennan, and Yuan Xie.
\newblock {DRISA: A DRAM-Based Reconfigurable In-Situ Accelerator}.
\newblock In {\em MICRO}, 2017.

\bibitem{truong2022adapting}
Minh~SQ Truong, Liting Shen, Alexander Glass, Alison Hoffmann, L~Richard Carley, James~A Bain, and Saugata Ghose.
\newblock {Adapting the RACER Architecture to Integrate Improved In-ReRAM Logic Primitives}.
\newblock {\em JETCAS}, 2022.

\bibitem{ma20232}
Xiaoyang Ma, Shan Deng, Juejian Wu, Zijian Zhao, David Lehninger, Tarek Ali, Konrad Seidel, Sourav De, Xiyu He, Yiming Chen, Huazhong Yang, Vijaykrishnan Narayanan, Suman Datta, Thomas K{\"a}mpfe, Qing Luo, Kai Ni, and Xueqing Li.
\newblock {A 2-Transistor-2-Capacitor Ferroelectric Edge Compute-in-Memory Scheme With Disturb-Free Inference and High Endurance}.
\newblock {\em IEEE Electron Device Letters}, 2023.

\bibitem{slesazeck20192tnc}
Stefan Slesazeck, Taras Ravsher, Viktor Havel, Evelyn~T Breyer, Halid Mulaosmanovic, and Thomas Mikolajick.
\newblock {A 2TnC Ferroelectric Memory Gain Cell Suitable for Compute-in-Memory and Neuromorphic Application}.
\newblock In {\em IEDM}, 2019.

\bibitem{wang20211t2c}
Qiao Wang, Donglin Zhang, Yulin Zhao, Chao Liu, Qiao Hu, Xuanzhi Liu, Jianguo Yang, and Hangbing Lv.
\newblock {A 1T2C FeCAP-Based In-Situ Bitwise X(N)OR Logic Operation with Two-Step Write-Back Circuit for Accelerating Compute-in-Memory}.
\newblock {\em Micromachines}, 2021.

\bibitem{aga2017compute}
Shaizeen Aga, Supreet Jeloka, Arun Subramaniyan, Satish Narayanasamy, David Blaauw, and Reetuparna Das.
\newblock {Compute Caches}.
\newblock In {\em HPCA}, 2017.

\bibitem{eckert2018neural}
Charles Eckert, Xiaowei Wang, Jingcheng Wang, Arun Subramaniyan, Ravi Iyer, Dennis Sylvester, David Blaauw, and Reetuparna Das.
\newblock {Neural Cache: Bit-Serial In-Cache Acceleration of Deep Neural Networks}.
\newblock In {\em ISCA}, 2018.

\bibitem{dualitycache}
Daichi Fujiki, Scott Mahlke, and Reetuparna Das.
\newblock {Duality Cache for Data Parallel Acceleration}.
\newblock In {\em ISCA}, 2019.

\bibitem{kang2014energy}
Mingu Kang, Min-Sun Keel, Naresh~R Shanbhag, Sean Eilert, and Ken Curewitz.
\newblock {An Energy-Efficient VLSI Architecture for Pattern Recognition via Deep Embedding of Computation in SRAM}.
\newblock In {\em ICASSP}, 2014.

\bibitem{de2025proteus}
Geraldo~Francisco de~Oliveira~Junior, Mayank Kabra, Yuxin Guo, Kangqi Chen, Abdullah~Giray Yaglikci, Melina Soysal, Mohammad Sadrosadati, Joaquin Olivares~Bueno, Saugata Ghose, Juan G{\'o}mez-Luna, et~al.
\newblock {Proteus: Achieving High-Performance Processing-Using-DRAM with Dynamic Bit-Precision, Adaptive Data Representation, and Flexible Arithmetic}.
\newblock In {\em ICS}, 2025.

\bibitem{tokuda2026clutch}
Daichi Tokuda, Tatsuya Kubo, Ismail~Emir Yuksel, Ataberk Olgun, Haocong Luo, Tomoya Nagatani, Geraldo~F Oliveira, Abdullah~Giray Ya{\u{g}}l{\i}k{\c{c}}{\i}, Mohammad Sadrosadati, Onur Mutlu, et~al.
\newblock {Clutch: High Performance Vector-Scalar Comparison using DRAM via Chunked Temporal Coding}.
\newblock {\em ICS}, 2026.

\bibitem{yuksel2025pudhammer}
Ismail~Emir Yuksel, Akash Sood, Ataberk Olgun, O{\u{g}}uzhan Canpolat, Haocong Luo, Nisa Bostanci, Mohammad Sadrosadati, Giray Yaglikci, and Onur Mutlu.
\newblock {PuDHammer: Experimental Analysis of Read Disturbance Effects of Processing-using-DRAM in Real DRAM Chips}.
\newblock In {\em ISCA}, 2025.

\bibitem{tokuda2026pudghost}
Daichi Tokuda, {\.I}smail~Emir Y{\"u}ksel, Tatsuya Kubo, Ataberk Olgun, Haocong Luo, Nisa Bostanci, Jikun Wang, A~Giray Ya{\u{g}}l{\i}k{\c{c}}{\i}, Shinya Takamaeda-Yamazaki, and Onur Mutlu.
\newblock {PuDGhost: Experimental Analysis of Computation Result Corruption in Processing-using-DRAM Operations on Real DRAM Chips and Implications for Future Systems}.
\newblock {\em ISCA}, 2026.

\bibitem{deng2018dracc}
Quan Deng, Lei Jiang, Youtao Zhang, Minxuan Zhang, and Jun Yang.
\newblock {DrAcc: A DRAM Based Accelerator for Accurate CNN Inference}.
\newblock In {\em DAC}, 2018.

\bibitem{Chi2016}
Ping Chi, Shuangchen Li, Cong Xu, Tao Zhang, Jishen Zhao, Yongpan Liu, Yu~Wang, and Yuan Xie.
\newblock {PRIME: A Novel Processing-in-Memory Architecture for Neural Network Computation in ReRAM-Based Main Memory}.
\newblock In {\em ISCA}, 2016.

\bibitem{xin2020elp2im}
Xin Xin, Youtao Zhang, and Jun Yang.
\newblock {ELP2IM: Efficient and Low Power Bitwise Operation Processing in DRAM}.
\newblock In {\em HPCA}, 2020.

\bibitem{Song2018graphr}
Linghao Song, Youwei Zhuo, Xuehai Qian, Hai Li, and Yiran Chen.
\newblock {GraphR: Accelerating Graph Processing Using ReRAM}.
\newblock In {\em HPCA}, 2018.

\bibitem{song2017pipelayer}
Linghao Song, Xuehai Qian, Hai Li, and Yiran Chen.
\newblock {PipeLayer: A Pipelined ReRAM-Based Accelerator for Deep Learning}.
\newblock In {\em HPCA}, 2017.

\bibitem{gao2019computedram}
Fei Gao, Georgios Tziantzioulis, and David Wentzlaff.
\newblock {ComputeDRAM: In-Memory Compute Using Off-the-Shelf DRAMs}.
\newblock In {\em MICRO}, 2019.

\bibitem{Besta2021SISA}
Maciej Besta, Raghavendra Kanakagiri, Grzegorz Kwasniewski, Rachata Ausavarungnirun, Jakub Beránek, Konstantinos Kanellopoulos, Kacper Janda, Zur Vonarburg-Shmaria, Lukas Gianinazzi, Ioana Stefan, Juan~Gómez Luna, Jakub Golinowski, Marcin Copik, Lukas Kapp-Schwoerer, Salvatore Di~Girolamo, Nils Blach, Marek Konieczny, Onur Mutlu, and Torsten Hoefler.
\newblock {SISA: Set-Centric Instruction Set Architecture for Graph Mining on Processing-in-Memory Systems}.
\newblock In {\em MICRO}, 2021.

\bibitem{seshadri2018rowclone}
Vivek Seshadri, Yoongu Kim, Chris Fallin, Donghyuk Lee, Rachata Ausavarungnirun, Gennady Pekhimenko, Yixin Luo, Onur Mutlu, Phillip~B Gibbons, Michael~A Kozuch, et~al.
\newblock {RowClone: Accelerating Data Movement and Initialization Using DRAM}.
\newblock arXiv:1805.03502 [cs.AR], 2018.

\bibitem{li2016pinatubo}
Shuangchen Li, Cong Xu, Qiaosha Zou, Jishen Zhao, Yu~Lu, and Yuan Xie.
\newblock {Pinatubo: A Processing-in-Memory Architecture for Bulk Bitwise Operations in Emerging Non-Volatile Memories}.
\newblock In {\em DAC}, 2016.

\bibitem{imani2019floatpim}
Mohsen Imani, Saransh Gupta, Yeseong Kim, and Tajana Rosing.
\newblock {FloatPIM: In-Memory Acceleration of Deep Neural Network Training with High Precision}.
\newblock In {\em ISCA}, 2019.

\bibitem{he2020sparse}
Zhezhi He, Li~Yang, Shaahin Angizi, Adnan~Siraj Rakin, and Deliang Fan.
\newblock {Sparse BD-Net: A Multiplication-Less DNN with Sparse Binarized Depth-Wise Separable Convolution}.
\newblock {\em JETC}, 2020.

\bibitem{olgun2021quactrng}
Ataberk Olgun, Minesh Patel, Abdullah~Giray Ya\u{g}l{\i}k\c{c}{\i}, Haocong Luo, Jeremie~S. Kim, F.~Nisa Bostanc{\i}, Nandita Vijaykumar, O\u{g}uz Ergin, and Onur Mutlu.
\newblock {QUAC-TRNG: High-Throughput True Random Number Generation Using Quadruple Row Activation in Commodity DRAMs}.
\newblock In {\em ISCA}, 2021.

\bibitem{kim2019d}
Jeremie~S Kim, Minesh Patel, Hasan Hassan, Lois Orosa, and Onur Mutlu.
\newblock {D-RaNGe: Using Commodity DRAM Devices to Generate True Random Numbers With Low Latency and High Throughput}.
\newblock In {\em HPCA}, 2019.

\bibitem{kim2018dram}
Jeremie~S Kim, Minesh Patel, Hasan Hassan, and Onur Mutlu.
\newblock {The DRAM Latency PUF: Quickly Evaluating Physical Unclonable Functions by Exploiting the Latency-Reliability Tradeoff in Modern Commodity DRAM Devices}.
\newblock In {\em HPCA}, 2018.

\bibitem{bostanci2022dr}
F~Nisa Bostanc{\i}, Ataberk Olgun, Lois Orosa, A~Giray Ya{\u{g}}l{\i}k{\c{c}}{\i}, Jeremie~S Kim, Hasan Hassan, O{\u{g}}uz Ergin, and Onur Mutlu.
\newblock {DR-STRaNGe: End-to-End System Design for DRAM-Based True Random Number Generators}.
\newblock In {\em HPCA}, 2022.

\bibitem{ali2019memory}
Mustafa~F Ali, Akhilesh Jaiswal, and Kaushik Roy.
\newblock {In-Memory Low-Cost Bit-Serial Addition Using Commodity DRAM Technology}.
\newblock In {\em {TCAS-I}}, 2019.

\bibitem{li2018scope}
Shuangchen Li, Alvin~Oliver Glova, Xing Hu, Peng Gu, Dimin Niu, Krishna~T Malladi, Hongzhong Zheng, Bob Brennan, and Yuan Xie.
\newblock {SCOPE: A Stochastic Computing Engine for DRAM-Based In-Situ Accelerator}.
\newblock In {\em MICRO}, 2018.

\bibitem{subramaniyan2017parallel}
Arun Subramaniyan and Reetuparna Das.
\newblock {Parallel Automata Processor}.
\newblock In {\em ISCA}, 2017.

\bibitem{zha2020hyper}
Yue Zha and Jing Li.
\newblock {Hyper-AP: Enhancing Associative Processing Through A Full-Stack Optimization}.
\newblock In {\em ISCA}, 2020.

\bibitem{fujiki2018memory}
Daichi Fujiki, Scott Mahlke, and Reetuparna Das.
\newblock {In-Memory Data Parallel Processor}.
\newblock In {\em ASPLOS}, 2018.

\bibitem{orosa2021codic}
Lois Orosa, Yaohua Wang, Mohammad Sadrosadati, Jeremie Kim, Minesh Patel, Ivan Puddu, Haocong Luo, Kaveh Razavi, Juan G{\'o}mez-Luna, Hasan Hassan, Nika~Mansouri Ghiasi, Saugata Ghose, and Onur Mutlu.
\newblock {CODIC: A Low-Cost Substrate for Enabling Custom In-DRAM Functionalities and Optimizations}.
\newblock In {\em ISCA}, 2021.

\bibitem{sharad2013ultra}
Mrigank Sharad, Deliang Fan, and Kaushik Roy.
\newblock {Ultra Low Power Associative Computing with Spin Neurons and Resistive Crossbar Memory}.
\newblock In {\em DAC}, 2013.

\bibitem{rezaei2020nom}
Seyyed Hossein~SeyyedAghaei Rezaei, Mehdi Modarressi, Rachata Ausavarungnirun, Mohammad Sadrosadati, Onur Mutlu, and Masoud Daneshtalab.
\newblock {NoM: Network-on-Memory for Inter-Bank Data Transfer in Highly-Banked Memories}.
\newblock {\em CAL}, 2020.

\bibitem{simon2020blade}
William~Andrew Simon, Yasir~Mahmood Qureshi, Marco Rios, Alexandre Levisse, Marina Zapater, and David Atienza.
\newblock {BLADE: An In-Cache Computing Architecture for Edge Devices}.
\newblock {\em Trans. on Comp.}, 2020.

\bibitem{wang2019bit}
Xiaowei Wang, Jiecao Yu, Charles Augustine, Ravi Iyer, and Reetuparna Das.
\newblock {Bit Prudent In-Cache Acceleration of Deep Convolutional Neural Networks}.
\newblock In {\em HPCA}, 2019.

\bibitem{al2020towards}
Khalid Al-Hawaj, Olalekan Afuye, Shady Agwa, Alyssa Apsel, and Christopher Batten.
\newblock {Towards a Reconfigurable Bit-Serial/Bit-Parallel Vector Accelerator Using In-Situ Processing-in-SRAM}.
\newblock In {\em ISCAS}, 2020.

\bibitem{kim2021colonnade}
Hyunjoon Kim, Taegeun Yoo, Tony Tae-Hyoung Kim, and Bongjin Kim.
\newblock {Colonnade: A Reconfigurable SRAM-Based Digital Bit-Serial Compute-in-Memory Macro for Processing Neural Networks}.
\newblock {\em JSSC}, 2021.

\bibitem{jiang2020c3sram}
Zhewei Jiang, Shihui Yin, Jae-Sun Seo, and Mingoo Seok.
\newblock {C3SRAM: An In-Memory-Computing SRAM Macro Based on Robust Capacitive Coupling Computing Mechanism}.
\newblock {\em JSSC}, 2020.

\bibitem{jeloka201628}
Supreet Jeloka, Naveen~Bharathwaj Akesh, Dennis Sylvester, and David Blaauw.
\newblock {A 28 nm Configurable Memory (TCAM/BCAM/SRAM) Using Push-Rule 6T Bit Cell Enabling Logic-in-Memory}.
\newblock {\em JSSC}, 2016.

\bibitem{wang2023infinity}
Zhengrong Wang, Christopher Liu, Aman Arora, Lizy John, and Tony Nowatzki.
\newblock {Infinity Stream: Portable and Programmer-Friendly In-/Near-Memory Fusion}.
\newblock In {\em ASPLOS}, 2023.

\bibitem{kang2015energy}
Mingu Kang, Eric~P Kim, Min-sun Keel, and Naresh~R Shanbhag.
\newblock {Energy-Efficient and High Throughput Sparse Distributed Memory Architecture}.
\newblock In {\em ISCAS}, 2015.

\bibitem{imani2020dual}
Mohsen Imani, Saikishan Pampana, Saransh Gupta, Minxuan Zhou, Yeseong Kim, and Tajana Rosing.
\newblock {DUAL: Acceleration of Clustering Algorithms Using Digital-Based Processing in-Memory}.
\newblock In {\em MICRO}, 2020.

\bibitem{deng2019lacc}
Quan Deng, Youtao Zhang, Minxuan Zhang, and Jun Yang.
\newblock {LAcc: Exploiting Lookup Table-Based Fast and Accurate Vector Multiplication in DRAM-Based CNN Accelerator}.
\newblock In {\em DAC}, 2019.

\bibitem{sutradhar2021look}
Purab~Ranjan Sutradhar, Sathwika Bavikadi, Mark Connolly, Savankumar Prajapati, Mark~A Indovina, Sai Manoj~Pudukotai Dinakarrao, and Amlan Ganguly.
\newblock {Look-Up-Table Based Processing-in-Memory Architecture with Programmable Precision-Scaling for Deep Learning Applications}.
\newblock {\em TPDS}, 2021.

\bibitem{sutradhar2020ppim}
Purab~Ranjan Sutradhar, Mark Connolly, Sathwika Bavikadi, Sai Manoj~Pudukotai Dinakarrao, Mark~A Indovina, and Amlan Ganguly.
\newblock {pPIM: A Programmable Processor-in-Memory Architecture with Precision-Scaling for Deep Learning}.
\newblock {\em CAL}, 2020.

\bibitem{peng2023chopper}
Xiangjun Peng, Yaohua Wang, and Ming-Chang Yang.
\newblock {CHOPPER: A Compiler Infrastructure for Programmable Bit-Serial SIMD Processing Using Memory In DRAM}.
\newblock In {\em HPCA}, 2023.

\bibitem{sudarshan2022fefet}
Chirag Sudarshan, Taha Soliman, Thomas K{\"a}mpfe, Christian Weis, and Norbert Wehn.
\newblock {FeFET versus DRAM Based PIM Architectures: A Comparative Study}.
\newblock In {\em VLSI-SoC}, 2022.

\bibitem{sudarshan2022weighted}
Chirag Sudarshan, Taha Soliman, Jan Lappas, Christian Weis, Mohammad~Hassani Sadi, Matthias Jung, Andre Guntoro, and Norbert Wehn.
\newblock {A Weighted Current Summation Based Mixed Signal DRAM-PIM Architecture for Deep Neural Network Inference}.
\newblock {\em JETCAS}, 2022.

\bibitem{sudarshan2022optimization}
Chirag Sudarshan, Mohammad~Hassani Sadi, Christian Weis, and Norbert Wehn.
\newblock {Optimization of DRAM Based PIM Architecture for Energy-Efficient Deep Neural Network Training}.
\newblock In {\em ISCAS}, 2022.

\bibitem{sudarshan2021novel}
Chirag Sudarshan, Taha Soliman, Cecilia De~la Parra, Christian Weis, Leonardo Ecco, Matthias Jung, Norbert Wehn, and Andre Guntoro.
\newblock {A Novel DRAM-Based Process-in-Memory Architecture and Its Implementation for CNNs}.
\newblock In {\em ASP-DAC}, 2021.

\bibitem{ahmed2021pan}
Omar Ahmed, Massimiliano Rossi, Sam Kovaka, Michael~C Schatz, Travis Gagie, Christina Boucher, and Ben Langmead.
\newblock {Pan-genomic Matching Statistics for Targeted Nanopore Sequencing}.
\newblock {\em iScience}, 2021.

\bibitem{Ham2016Graphicionado}
Tae~Jun Ham, Lisa Wu, Narayanan Sundaram, Nadathur Satish, and Margaret Martonosi.
\newblock {Graphicionado: A high-performance and energy-efficient accelerator for graph analytics}.
\newblock In {\em MICRO}, 2016.

\bibitem{Rahman2020graphpulse}
Shafiur Rahman, Nael Abu-Ghazaleh, and Rajiv Gupta.
\newblock {GraphPulse: An Event-Driven Hardware Accelerator for Asynchronous Graph Processing}.
\newblock In {\em MICRO}, 2020.

\bibitem{Chen2022regraph}
Xinyu Chen, Yao Chen, Feng Cheng, Hongshi Tan, Bingsheng He, and Weng-Fai Wong.
\newblock {ReGraph: Scaling Graph Processing on HBM-enabled FPGAs with Heterogeneous Pipelines}.
\newblock In {\em MICRO}, 2022.

\bibitem{Yang2025IDGNN}
Jiaqi Yang, Hao Zheng, and Ahmed Louri.
\newblock {I-DGNN: A Graph Dissimilarity-based Framework for Designing Scalable and Efficient DGNN Accelerators}.
\newblock In {\em HPCA}, 2025.

\bibitem{Yan2025bingogcn}
Jiale Yan, Hiroaki Ito, Yuta Nagahara, Kazushi Kawamura, Masato Motomura, Thiem Van~Chu, and Daichi Fujiki.
\newblock {BingoGCN: Towards Scalable and Efficient GNN Acceleration with Fine-Grained Partitioning and SLT}.
\newblock In {\em ISCA}, 2025.

\bibitem{Peng2024maxkgnn}
Hongwu Peng, Xi~Xie, Kaustubh Shivdikar, Md~Amit Hasan, Jiahui Zhao, Shaoyi Huang, Omer Khan, David Kaeli, and Caiwen Ding.
\newblock {MaxK-GNN: Extremely Fast GPU Kernel Design for Accelerating Graph Neural Networks Training}.
\newblock In {\em ASPLOS}, 2024.

\bibitem{Yan2020hygcn}
Mingyu Yan, Lei Deng, Xing Hu, Ling Liang, Yujing Feng, Xiaochun Ye, Zhimin Zhang, Dongrui Fan, and Yuan Xie.
\newblock {HyGCN: A GCN Accelerator with Hybrid Architecture}.
\newblock In {\em HPCA}, 2020.

\bibitem{You2022gcod}
Haoran You, Tong Geng, Yongan Zhang, Ang Li, and Yingyan Lin.
\newblock {GCoD: Graph Convolutional Network Acceleration via Dedicated Algorithm and Accelerator Co-Design}.
\newblock In {\em HPCA}, 2022.

\bibitem{Hwang2023grow}
Ranggi Hwang, Minhoo Kang, Jiwon Lee, Dongyun Kam, Youngjoo Lee, and Minsoo Rhu.
\newblock {GROW: A Row-Stationary Sparse-Dense GEMM Accelerator for Memory-Efficient Graph Convolutional Neural Networks}.
\newblock In {\em HPCA}, 2023.

\bibitem{Sarkar2023flowgnn}
Rishov Sarkar, Stefan Abi-Karam, Yuqi He, Lakshmi Sathidevi, and Cong Hao.
\newblock {FlowGNN: A Dataflow Architecture for Real-Time Workload-Agnostic Graph Neural Network Inference}.
\newblock In {\em HPCA}, 2023.

\bibitem{Chen2022regnn}
Cen Chen, Kenli Li, Yangfan Li, and Xiaofeng Zou.
\newblock {ReGNN: A Redundancy-Eliminated Graph Neural Networks Accelerator}.
\newblock In {\em HPCA}, 2022.

\bibitem{Li2021gcnax}
Jiajun Li, Ahmed Louri, Avinash Karanth, and Razvan Bunescu.
\newblock {GCNAX: A Flexible and Energy-efficient Accelerator for Graph Convolutional Neural Networks}.
\newblock In {\em HPCA}, 2021.

\bibitem{Geng2020awbgcn}
Tong Geng, Ang Li, Runbin Shi, Chunshu Wu, Tianqi Wang, Yanfei Li, Pouya Haghi, Antonino Tumeo, Shuai Che, Steve Reinhardt, and Martin~C. Herbordt.
\newblock {AWB-GCN: A Graph Convolutional Network Accelerator with Runtime Workload Rebalancing}.
\newblock In {\em MICRO}, 2020.

\bibitem{Chen2023metanmp}
Dan Chen, Haiheng He, Hai Jin, Long Zheng, Yu~Huang, Xinyang Shen, and Xiaofei Liao.
\newblock {MetaNMP: Leveraging Cartesian-Like Product to Accelerate HGNNs with Near-Memory Processing}.
\newblock In {\em ISCA}, 2023.

\bibitem{Zhao2025mehyper}
Wenju Zhao, Pengcheng Yao, Dan Chen, Long Zheng, Xiaofei Liao, Qinggang Wang, Shaobo Ma, Yu~Li, Haifeng Liu, Wenjing Xiao, Yufei Sun, Bing Zhu, Hai Jin, and Jingling Xue.
\newblock {MeHyper: Accelerating Hypergraph Neural Networks by Exploring Implicit Dataflows}.
\newblock In {\em HPCA}, 2025.

\bibitem{Zhou2022GNNear}
Zhe Zhou, Cong Li, Xuechao Wei, Xiaoyang Wang, and Guangyu Sun.
\newblock {GNNear: Accelerating Full-Batch Training of Graph Neural Networks with near-Memory Processing}.
\newblock In {\em PACT}, 2022.

\bibitem{zhang2018graphp}
Mingxing Zhang, Youwei Zhuo, Chao Wang, Mingyu Gao, Yongwei Wu, Kang Chen, Christos Kozyrakis, and Xuehai Qian.
\newblock {GraphP: Reducing Communication for PIM-based Graph Processing with Efficient Data Partition}.
\newblock In {\em HPCA}, 2018.

\bibitem{Asiatici2021}
Mikhail Asiatici and Paolo Ienne.
\newblock {Large-Scale Graph Processing on FPGAs with Caches for Thousands of Simultaneous Misses}.
\newblock In {\em ISCA}, 2021.

\bibitem{Shin2025piccolo}
Changmin Shin, Jaeyong Song, Hongsun Jang, Dogeun Kim, Jun Sung, Taehee Kwon, Jae~Hyung Ju, Frank Liu, Yeonkyu Choi, and Jinho Lee.
\newblock {Piccolo: Large-Scale Graph Processing with Fine-Grained in-Memory Scatter-Gather}.
\newblock In {\em HPCA}, 2025.

\bibitem{Huang2022reflip}
Yu~Huang, Long Zheng, Pengcheng Yao, Qinggang Wang, Xiaofei Liao, Hai Jin, and Jingling Xue.
\newblock {Accelerating Graph Convolutional Networks Using Crossbar-based Processing-In-Memory Architectures}.
\newblock In {\em HPCA}, 2022.

\bibitem{Li2022hyperscale}
Shuangchen Li, Dimin Niu, Yuhao Wang, Wei Han, Zhe Zhang, Tianchan Guan, Yijin Guan, Heng Liu, Linyong Huang, Zhaoyang Du, Fei Xue, Yuanwei Fang, Hongzhong Zheng, and Yuan Xie.
\newblock {Hyperscale FPGA-as-a-service architecture for large-scale distributed graph neural network}.
\newblock In {\em ISCA}, 2022.

\bibitem{Wang2024motionaccel}
Shu-Ting Wang, Hanyang Xu, Amin Mamandipoor, Rohan Mahapatra, Byung~Hoon Ahn, Soroush Ghodrati, Krishnan Kailas, Mohammad Alian, and Hadi Esmaeilzadeh.
\newblock {Data Motion Acceleration: Chaining Cross-Domain Multi Accelerators}.
\newblock In {\em HPCA}, 2024.

\bibitem{Dai2023cegma}
Yue Dai, Youtao Zhang, and Xulong Tang.
\newblock {CEGMA: Coordinated Elastic Graph Matching Acceleration for Graph Matching Networks}.
\newblock In {\em HPCA}, 2023.

\bibitem{Kim2025eod}
Taehwan Kim, Yunki Han, Seohye Ha, Jiwan Kim, and Lee-Sup Kim.
\newblock {EOD: Enabling Low Latency GNN Inference via Near-Memory Concatenate Aggregation}.
\newblock In {\em ISCA}, 2025.

\bibitem{Li2024celeritas}
Yi~Li, Tsun-Yu Yang, Ming-Chang Yang, Zhaoyan Shen, and Bingzhe Li.
\newblock {Celeritas: Out-of-Core Based Unsupervised Graph Neural Network via Cross-Layer Computing}.
\newblock In {\em HPCA}, 2024.

\bibitem{esfahani2021locality}
Mohsen~Koohi Esfahani, Peter Kilpatrick, and Hans Vandierendonck.
\newblock Locality analysis of graph reordering algorithms.
\newblock In {\em IISWC}, 2021.

\bibitem{coleman2022graph}
Benjamin Coleman, Santiago Segarra, Alexander~J Smola, and Anshumali Shrivastava.
\newblock Graph reordering for cache-efficient near neighbor search.
\newblock {\em NeurIPS}, 2022.

\bibitem{kyrola2012graphchi}
Aapo Kyrola, Guy Blelloch, and Carlos Guestrin.
\newblock {GraphChi: Large-Scale graph computation on just a PC}.
\newblock In {\em OSDI}, 2012.

\bibitem{bulucc2016recent}
Ayd{\i}n Bulu{\c{c}}, Henning Meyerhenke, Ilya Safro, Peter Sanders, and Christian Schulz.
\newblock Recent advances in graph partitioning.
\newblock {\em Algorithm engineering: selected results and surveys}, 2016.

\bibitem{guo2013gpu}
GuiXin Guo, Shuang Qiu, ZhiQiang Ye, BingQiang Wang, Lin Fang, Mian Lu, Simon See, and Rui Mao.
\newblock {GPU-accelerated adaptive compression framework for genomics data}.
\newblock In {\em BigData}, 2013.

\bibitem{qiao2019fpga}
Weikang Qiao, Zhenman Fang, Mau-Chung~Frank Chang, and Jason Cong.
\newblock {An FPGA-based BWT accelerator for Bzip2 data compression}.
\newblock In {\em FCCM}, 2019.

\bibitem{zhao2017streaming}
Baofu Zhao, Yubin Li, Yu~Wang, and Huazhong Yang.
\newblock {Streaming sorting network based BWT acceleration on FPGA for lossless compression}.
\newblock In {\em ICFPT}, 2017.

\bibitem{jiang2021exma}
Lei Jiang and Farzaneh Zokaee.
\newblock Exma: A genomics accelerator for exact-matching.
\newblock In {\em HPCA}, 2021.

\bibitem{arram2015fpga}
James Arram, Moritz Pflanzer, Thomas Kaplan, and Wayne Luk.
\newblock {FPGA acceleration of reference-based compression for genomic data}.
\newblock In {\em FPT}, 2015.

\bibitem{wang2018accelerating}
Yuanrong Wang, Xueqi Li, Dawei Zang, Guangming Tan, and Ninghui Sun.
\newblock {Accelerating FM-index search for genomic data processing}.
\newblock In {\em ICPP}, 2018.

\bibitem{lim2025bancroft}
Se-Min Lim, Seongyoung Kang, and Sang-Woo Jun.
\newblock Bancroft: Genomics acceleration beyond on-device memory.
\newblock In {\em PACT}, 2025.

\bibitem{leavline2013hardware}
E~Jebamalar Leavline and DAAG Singh.
\newblock {Hardware implementation of LZMA data compression algorithm}.
\newblock {\em International Journal of Applied Information Systems}, 2013.

\bibitem{rajarajeswari2011dnabit}
Pothuraju Rajarajeswari and Allam Apparao.
\newblock {DNABIT compress--genome compression algorithm}.
\newblock {\em Bioinformation}, 2011.

\bibitem{saada2016dna}
Bacem Saada and Jing Zhang.
\newblock {DNA sequence compression technique based on modified DNABIT algorithm}.
\newblock In {\em Proceedings of the World Congress on Engineering, London}, 2016.

\bibitem{SATA}
{Serial ATA International Organization}.
\newblock {SATA revision 3.0 specifications}.
\newblock \url{https://www.sata-io.org}.

\bibitem{samsung980pro}
Samsung.
\newblock {Samsung SSD 980 PRO}.
\newblock \url{https://www.samsung.com/semiconductor/minisite/ssd/product/consumer/980pro/}, 2020.

\bibitem{PCIE}
PCI-SIG.
\newblock {PCI Express Base Specification Revision 3.0}.
\newblock \url{https://pcisig.com/specifications}.

\bibitem{samsungPM1735}
Samsung.
\newblock {Samsung SSD PM1735}.
\newblock \url{https://www.samsung.com/semiconductor/ssd/enterprise-ssd/MZPLJ3T2HBJR-00007/}, 2020.

\bibitem{PCIE4}
PCI-SIG.
\newblock {PCI Express Base Specification Revision 4.0, Version 1.0}.
\newblock \url{https://pcisig.com/specifications}.

\bibitem{amdepyc}
AMD.
\newblock {AMD$^\text{\textregistered}$ EPYC$^\text{\textregistered}$ 7742 CPU}.
\newblock \url{https://www.amd.com/en/products/cpu/amd-epyc-7742}.

\bibitem{micros9300pro}
Micron.
\newblock {Micron 9300 SSD}.
\newblock \url{https://www.micron.com/products/ssd/product-lines/9300}, 2019.

\bibitem{wdblue}
Western Digital.
\newblock {WD Blue SATA Internal SSD Hard Drive}.
\newblock \url{https://www.westerndigital.com/en-ca/products/internal-drives/wd-blue-sata-2-5-ssd#WDS400T2B0A}.

\bibitem{laguna2020seed}
Ann~Franchesca Laguna, Hasindu Gamaarachchi, Xunzhao Yin, Michael Niemier, Sri Parameswaran, and X~Sharon Hu.
\newblock {Seed-and-vote Based In-memory Accelerator for DNA Read Mapping}.
\newblock In {\em ICCAD}, 2020.

\bibitem{kaplan2018rassa}
Roman Kaplan, Leonid Yavits, and Ran Ginosar.
\newblock {RASSA: Resistive Prealignment Accelerator for Approximate DNA Long Read Mapping}.
\newblock {\em IEEE Micro}, 2018.

\bibitem{schneider2017evaluation}
Valerie~A Schneider, Tina Graves-Lindsay, Kerstin Howe, Nathan Bouk, Hsiu-Chuan Chen, Paul~A Kitts, Terence~D Murphy, Kim~D Pruitt, Fran{\c{c}}oise Thibaud-Nissen, Derek Albracht, et~al.
\newblock {Evaluation of GRCh38 and De Novo Haploid Genome Assemblies Demonstrates the Enduring Quality of the Reference Assembly}.
\newblock {\em Genome Research}, 2017.

\bibitem{dang2015secure}
Quynh Dang.
\newblock {Secure Hash Standard}.
\newblock \url{https://doi.org/10.6028/NIST.FIPS.180-4}, 2015.

\bibitem{rivest1992rfc1321}
Ronald Rivest.
\newblock {RFC1321: The MD5 Message-digest Algorithm}.
\newblock \url{https://datatracker.ietf.org/doc/rfc1321/}, 1992.

\bibitem{sims2014sequencing}
David Sims, Ian Sudbery, Nicholas~E Ilott, Andreas Heger, and Chris~P Ponting.
\newblock {Sequencing Depth and Coverage: Key Considerations in Genomic Analyses}.
\newblock {\em Nature Reviews Genetics}, 2014.

\bibitem{quail2012tale}
Michael~A Quail, Miriam Smith, Paul Coupland, Thomas~D Otto, Simon~R Harris, Thomas~R Connor, Anna Bertoni, Harold~P Swerdlow, and Yong Gu.
\newblock {A Tale of Three Next Generation Sequencing Platforms: Comparison of Ion Torrent, Pacific Biosciences and Illumina MiSeq Sequencers}.
\newblock {\em BMC Genomics}, 2012.

\bibitem{levy2016advancements}
Shawn~E Levy and Richard~M Myers.
\newblock {Advancements in Next-generation Sequencing}.
\newblock {\em Annual Review of Genomics and Human Genetics}, 2016.

\bibitem{lenovot740p}
Lenovo.
\newblock {ThinkPad T470p}.
\newblock \url{https://www.lenovo.com/ch/en/laptops/thinkpad/t-series/ThinkPad-T470p/p/22TP2TT470P}, 2016.

\bibitem{cheong-isscc-2018}
Wooseong Cheong, Chanho Yoon, Seonghoon Woo, Kyuwook Han, Daehyun Kim, Chulseung Lee, Youra Choi, Shine Kim, Dongku Kang, Geunyeong Yu, Jaehong Kim, Jaechun Park, Ki-Whan Song, Ki-Tae Park, Sangyeun Cho, Hwaseok Oh, Daniel~D.G. Lee, Jin-Hyeok Choi, and Jaeheon Jeong.
\newblock {A Flash Memory Controller for 15\textmu{}s Ultra-Low-Latency SSD Using High-Speed 3D NAND Flash with 3\textmu{}s Read Time}.
\newblock In {\em ISSCC}, 2018.

\bibitem{bhoyar2021high}
Rahul~C. Bhoyar, Abhinav Jain, Paras Sehgal, Mohit~Kumar Divakar, Disha Sharma, Mohamed Imran, Bani Jolly, Gyan Ranjan, Mercy Rophina, Sumit Sharma, Sanjay Siwach, Kavita Pandhare, Swayamprabha Sahoo, Maheswata Sahoo, Ananya Nayak, Jatindra~Nath Mohanty, Jayashankar Das, Sudhir Bhandari, Sandeep~K. Mathur, Anshul Kumar, et~al.
\newblock {High Throughput Detection and Genetic Epidemiology of SARS-CoV-2 Using COVIDSeq Next-generation Sequencing}.
\newblock {\em PLOS One}, 2021.

\bibitem{breitwieser2019human}
Florian~P Breitwieser, Mihaela Pertea, Aleksey~V Zimin, and Steven~L Salzberg.
\newblock {Human Contamination in Bacterial Genomes has Created Thousands of Spurious Proteins}.
\newblock {\em Genome Research}, 2019.

\bibitem{knight2018best}
Rob Knight, Alison Vrbanac, Bryn~C. Taylor, Alexander Aksenov, Chris Callewaert, Justine Debelius, Antonio Gonzalez, Tomasz Kosciolek, Laura-Isobel McCall, Daniel McDonald, Alexey~V. Melnik, James~T. Morton, Jose Navas, Robert~A. Quinn, Jon~G. Sanders, Austin~D. Swafford, Luke~R. Thompson, Anupriya Tripathi, Zhenjiang~Z. Xu, Jesse~R. Zaneveld, et~al.
\newblock {Best Practices for Analysing Microbiomes}.
\newblock {\em Nature Reviews Microbiology}, 2018.

\bibitem{sayers2021database}
Eric~W. Sayers, Jeffrey Beck, Evan~E. Bolton, Devon Bourexis, James~R. Brister, Kathi Canese, Donald~C. Comeau, Kathryn Funk, Sunghwan Kim, William Klimke, Aron Marchler-Bauer, Melissa Landrum, Stacy Lathrop, Zhiyong Lu, Thomas~L. Madden, Nuala O'Leary, Lon Phan, Sanjida~H. Rangwala, Valerie~A. Schneider, Yuri Skripchenko, et~al.
\newblock {Database resources of the National Center for Biotechnology Information}.
\newblock {\em Nucleic Acids Research}, 2021.

\bibitem{zook2016extensive}
Justin~M. Zook, David Catoe, Jennifer McDaniel, Lindsay Vang, Noah Spies, Arend Sidow, Ziming Weng, Yuling Liu, Christopher~E. Mason, Noah Alexander, Elizabeth Henaff, Alexa~B.R. McIntyre, Dhruva Chandramohan, Feng Chen, Erich Jaeger, Ali Moshrefi, Khoa Pham, William Stedman, Tiffany Liang, Michael Saghbini, et~al.
\newblock {Extensive Sequencing of Seven Human Genomes to Characterize Benchmark Reference Materials}.
\newblock {\em Scientific Data}, 2016.

\bibitem{clark2016genbank}
Karen Clark, Ilene Karsch-Mizrachi, David~J Lipman, James Ostell, and Eric~W Sayers.
\newblock {GenBank}.
\newblock {\em Nucleic Acids Research}, 2016.

\bibitem{wu2020new}
Fan Wu, Su~Zhao, Bin Yu, Yan-Mei Chen, Wen Wang, Zhi-Gang Song, Yi~Hu, Zhao-Wu Tao, Jun-Hua Tian, Yuan-Yuan Pei, Ming-Li Yuan, Yu-Ling Zhang, Fa-Hui Dai, Yi~Liu, Qi-Min Wang, Jiao-Jiao Zheng, Lin Xu, Edward~C. Holmes, and Yong-Zhen Zhang.
\newblock {A New Coronavirus Associated with Human Respiratory Disease in China}.
\newblock {\em Nature}, 2020.

\bibitem{sichtig2019fda}
Heike Sichtig, Timothy Minogue, Yi~Yan, Christopher Stefan, Adrienne Hall, Luke Tallon, Lisa Sadzewicz, Suvarna Nadendla, William Klimke, Eneida Hatcher, Martin Shumway, Dayanara~Lebron Aldea, Jonathan Allen, Jeffrey Koehler, Tom Slezak, Stephen Lovell, Randal Schoepp, and Uwe Scherf.
\newblock {FDA-ARGOS is a database with public quality-controlled reference genomes for diagnostic use and regulatory science}.
\newblock {\em Nature Communications}, 2019.

\bibitem{engel2014reference}
Stacia~R Engel, Fred~S Dietrich, Dianna~G Fisk, Gail Binkley, Rama Balakrishnan, Maria~C Costanzo, Selina~S Dwight, Benjamin~C Hitz, Kalpana Karra, Robert~S Nash, Shuai Weng, Edith~D Wong, Paul Lloyd, Marek~S Skrzypek, Stuart~R Miyasato, Matt Simison, and J~Michael Cherry.
\newblock {The Reference Genome Sequence of Saccharomyces Cerevisiae: Then and Now}.
\newblock {\em G3: Genes, Genomes, Genetics}, 2014.

\bibitem{berardini2015arabidopsis}
Tanya~Z Berardini, Leonore Reiser, Donghui Li, Yarik Mezheritsky, Robert Muller, Emily Strait, and Eva Huala.
\newblock {The Arabidopsis Information Resource: Making and Mining the “Gold Standard” Annotated Reference Plant Genome}.
\newblock {\em Genesis}, 2015.

\bibitem{larkin2020}
Aoife Larkin, Steven~J Marygold, Giulia Antonazzo, Helen Attrill, Gilberto dos Santos, Phani~V Garapati, Joshua L Goodman, L Sian Gramates, Gillian Millburn, Victor~B Strelets, Christopher~J Tabone, Jim Thurmond, and {FlyBase Consortium}.
\newblock {FlyBase: Updates to the Drosophila Melanogaster Knowledge Base}.
\newblock {\em Nucleic Acids Research}, 2020.

\bibitem{church2009lineage}
Deanna~M. Church, Leo Goodstadt, LaDeana~W. Hillier, Michael~C. Zody, Steve Goldstein, Xinwe She, Carol~J. Bult, Richa Agarwala, Joshua~L. Cherry, Michael DiCuccio, Wratko Hlavina, Yuri Kapustin, Peter Meric, Donna Maglott, Zoë Birtle, Ana~C. Marques, Tina Graves, Shiguo Zhou, Brian Teague, Konstantinos Potamousis, et~al.
\newblock {Lineage-specific Biology Revealed by a Finished Genome Assembly of the Mouse}.
\newblock {\em PLoS Biology}, 2009.

\bibitem{wang2007integer}
Thomas Wang.
\newblock {Integer Hash Function}.
\newblock \url{http://web.archive.org/web/20071223173210/http://www.concentric.net/~Ttwang/tech/inthash.htm}, 2007.

\bibitem{dobin2012}
Alexander Dobin, Carrie~A. Davis, Felix Schlesinger, Jorg Drenkow, Chris Zaleski, Sonali Jha, Philippe Batut, Mark Chaisson, and Thomas~R. Gingeras.
\newblock {{STAR: Ultrafast Universal RNA-seq Aligner}}.
\newblock {\em Bioinformatics}, 2012.

\bibitem{micron3dnandflyer}
Micron.
\newblock {Product Flyer: Micron 3D NAND Flash Memory}.
\newblock \url{https://www.micron.com/-/media/client/global/documents/products/product-flyer/3d_nand_flyer.pdf?la=en}, 2016.

\bibitem{synopsysdc}
{Synopsys, Inc.}
\newblock {Design Compiler}.
\newblock \url{https://www.synopsys.com/implementation-and-signoff/rtl-synthesis-test/design-compiler-graphical.html}.

\bibitem{ddr4sheet}
Micron~Technology Inc.
\newblock {4Gb: x4, x8, x16 DDR4 SDRAM Data Sheet}, 2016.

\bibitem{ghose2019demystifying}
Saugata Ghose, Tianshi Li, Nastaran Hajinazar, Damla~Senol Cali, and Onur Mutlu.
\newblock {Demystifying Complex Workload-DRAM Interactions: An Experimental Study}.
\newblock {\em ACM POMACS}, 2019.

\bibitem{ghose2018your}
Saugata Ghose, Abdullah~Giray Yaglik\c{c}i, Raghav Gupta, Donghyuk Lee, Kais Kudrolli, William~X. Liu, Hasan Hassan, Kevin~K. Chang, Niladrish Chatterjee, Aditya Agrawal, Mike O'Connor, and Onur Mutlu.
\newblock {What Your DRAM Power Models Are Not Telling You: Lessons from a Detailed Experimental Study}.
\newblock {\em POMACS}, 2018.

\bibitem{kim2016ramulator}
Yoongu Kim, Weikun Yang, and Onur Mutlu.
\newblock {Ramulator: A Fast and Extensible DRAM Simulator}.
\newblock {\em IEEE CAL}, 2015.

\bibitem{ramulatorsource}
Yoongu Kim, Weikun Yang, and Onur Mutlu.
\newblock {Ramulator 1.0 Source Code}.
\newblock \url{https://github.com/CMU-SAFARI/ramulator}.

\bibitem{tavakkol2018mqsim}
Arash Tavakkol, Juan G{\'o}mez-Luna, Mohammad Sadrosadati, Saugata Ghose, and Onur Mutlu.
\newblock {MQSim: A Framework for Enabling Realistic Studies of Modern Multi-queue SSD Devices}.
\newblock In {\em FAST}, 2018.

\bibitem{microprof}
{Advanced Micro Devices}.
\newblock {AMD{\textregistered} $\mu$Prof}.
\newblock \url{https://developer.amd.com/amd-uprof/}, 2021.

\bibitem{holtgrewe2010mason}
M.~Holtgrewe.
\newblock {Mason - A Read Simulator for Second Generation Sequencing Data}.
\newblock {\em Technical Report FU Berlin}, 2010.

\bibitem{wikichipcascade}
WikiChip.
\newblock {Cascade Lake SP - Intel}.
\newblock \url{https://en.wikichip.org/wiki/intel/cores/cascade\_lake\_sp}.

\bibitem{stillmaker2017Scaling}
Aaron Stillmaker and Bevan Baas.
\newblock {Scaling Equations for the Accurate Prediction of {{CMOS}} Device Performance from 180 Nm to 7 Nm}.
\newblock {\em Integration}, 2017.

\bibitem{cortexr4}
{{ARM Holdings}}.
\newblock {Cortex-R4}.
\newblock \url{https://developer.arm.com/ip-products/processors/cortex-r/cortex-r4}, 2011.

\bibitem{morrison2020nanopore}
Gretchen~A Morrison, Jianmin Fu, Grace~C Lee, Nathan~P Wiederhold, Connie~F Ca{\~n}ete-Gibas, Evelien~M Bunnik, and Brian~L Wickes.
\newblock {Nanopore Sequencing of the Fungal Intergenic Spacer Sequence as a Potential Rapid Diagnostic Assay}.
\newblock {\em Journal of Clinical Microbiology}, 2020.

\bibitem{nanopore2020}
Oxford~Nanopore Technologies.
\newblock {R10.3: the Newest Nanopore for High Accuracy Nanopore Sequencing – Now Available in Store}.
\newblock \url{https://nanoporetech.com/about-us/news/r103-newest-nanopore-high-accuracy-nanopore-sequencing-now-available-store/}, 2020.

\bibitem{quail2008large}
Michael~A Quail, Iwanka Kozarewa, Frances Smith, Aylwyn Scally, Philip~J Stephens, Richard Durbin, Harold Swerdlow, and Daniel~J Turner.
\newblock {A Large Genome Center's Improvements to the Illumina Sequencing System}.
\newblock {\em Nature Methods}, 2008.

\bibitem{pacbio2021}
PacBio Sequencing.
\newblock {Pacific Biosciences Closes Acquisition of Omniome and Establishes San Diego Presence}.
\newblock \url{https://www.pacb.com/press_releases/pacific-biosciences-closes-acquisition-of-omniome-and-establishes-san-diego-presence/}, 2021.

\bibitem{loka2019reliable}
Tobias~P Loka, Simon~H Tausch, and Bernhard~Y Renard.
\newblock {Reliable Variant Calling during Runtime of Illumina Sequencing}.
\newblock {\em Scientific Reports}, 2019.

\bibitem{ardui2018single}
Simon Ardui, Adam Ameur, Joris~R Vermeesch, and Matthew~S Hestand.
\newblock {Single Molecule Real-Time (SMRT) Sequencing Comes of Age: Applications and Utilities for Medical Diagnostics}.
\newblock {\em Nucleic Acids Research}, 2018.

\bibitem{sherman2019assembly}
Rachel~M. Sherman, Juliet Forman, Valentin Antonescu, Daniela Puiu, Michelle Daya, Nicholas Rafaels, Meher~Preethi Boorgula, Sameer Chavan, Candelaria Vergara, Victor~E. Ortega, Albert~M. Levin, Celeste Eng, Maria Yazdanbakhsh, James~G. Wilson, Javier Marrugo, Leslie~A. Lange, L.~Keoki Williams, Harold Watson, Lorraine~B. Ware, Christopher~O. Olopade, et~al.
\newblock {Assembly of A Pan-genome from Deep Sequencing of 910 Humans of African Descent}.
\newblock {\em Nature Genetics}, 2019.

\bibitem{li2021building}
Qiuhui Li, Shilin Tian, Bin Yan, Chi~Man Liu, Tak-Wah Lam, Ruiqiang Li, and Ruibang Luo.
\newblock {Building a Chinese Pan-genome of 486 Individuals}.
\newblock {\em Communications Biology}, 2021.

\bibitem{miga2021need}
Karen~H Miga and Ting Wang.
\newblock {The Need for A Human Pangenome Reference Sequence}.
\newblock {\em Annual Review of Genomics and Human Genetics}, 2021.

\bibitem{zhang2020comprehensive}
Haowen Zhang, Chirag Jain, and Srinivas Aluru.
\newblock {A Comprehensive Evaluation of Long Read Error Correction Methods}.
\newblock {\em BMC Genomics}, 2020.

\bibitem{miga2020telomere}
Karen~H. Miga, Sergey Koren, Arang Rhie, Mitchell~R. Vollger, Ariel Gershman, Andrey Bzikadze, Shelise Brooks, Edmund Howe, David Porubsky, Glennis~A. Logsdon, Valerie~A. Schneider, Tamara Potapova, Jonathan Wood, William Chow, Joel Armstrong, Jeanne Fredrickson, Evgenia Pak, Kristof Tigyi, Milinn Kremitzki, Christopher Markovic, et~al.
\newblock {Telomere-to-telomere Assembly of A Complete Human X Chromosome}.
\newblock {\em Nature}, 2020.

\bibitem{logsdon2021structure}
Glennis~A. Logsdon, Mitchell~R. Vollger, PingHsun Hsieh, Yafei Mao, Mikhail~A. Liskovykh, Sergey Koren, Sergey Nurk, Ludovica Mercuri, Philip~C. Dishuck, Arang Rhie, Leonardo~G. de~Lima, Tatiana Dvorkina, David Porubsky, William~T. Harvey, Alla Mikheenko, Andrey~V. Bzikadze, Milinn Kremitzki, Tina~A. Graves-Lindsay, Chirag Jain, Kendra Hoekzema, et~al.
\newblock {The structure, function and evolution of a complete human chromosome 8}.
\newblock {\em Nature}, 2021.

\bibitem{lax2014longitudinal}
Simon Lax, Daniel~P. Smith, Jarrad Hampton-Marcell, Sarah~M. Owens, Kim~M. Handley, Nicole~M. Scott, Sean~M. Gibbons, Peter Larsen, Benjamin~D. Shogan, Sophie Weiss, Jessica~L. Metcalf, Luke~K. Ursell, Yoshiki Vázquez-Baeza, Will~Van Treuren, Nur~A. Hasan, Molly~K. Gibson, Rita Colwell, Gautam Dantas, Rob Knight, and Jack~A. Gilbert.
\newblock {Longitudinal Analysis of Microbial Interaction between Humans and the Indoor Environment}.
\newblock {\em Science}, 2014.

\bibitem{ncbi2020}
Eric~W Sayers, Richa Agarwala, Evan~E Bolton, J Rodney Brister, Kathi Canese, Karen Clark, Ryan Connor, Nicolas Fiorini, Kathryn Funk, Timothy Hefferon, J Bradley Holmes, Sunghwan Kim, Avi Kimchi, Paul~A Kitts, Stacy Lathrop, Zhiyong Lu, Thomas~L Madden, Aron Marchler-Bauer, Lon Phan, Valerie~A Schneider, et~al.
\newblock {Database resources of the National Center for Biotechnology Information}.
\newblock {\em Nucleic Acids Research}, 2018.

\bibitem{samsung870evo}
Samsung.
\newblock {Samsung SSD 870 EVO}.
\newblock \url{https://www.samsung.com/semiconductor/minisite/ssd/product/consumer/870evo/}, 2021.

\bibitem{PM1735price}
{Samsung PM1735}.
\newblock \url{https://www.digitec.ch/en/s1/product/samsung-pm1735-3200-gb-pci-express-ssd-15678607}.

\bibitem{PM9A3price}
{Samsung PM9A3}.
\newblock \url{https://www.digitec.ch/en/s1/product/samsung-pm9a3-3840-gb-m2-22110-ssd-16404342}.

\bibitem{EVO870price}
{Samsung 870 EVO}.
\newblock \url{https://www.digitec.ch/en/s1/product/samsung-870-evo-4000-gb-25-ssd-14599189}.

\bibitem{salzberg2016next}
Steven~L. Salzberg, Florian~P. Breitwieser, Anupama Kumar, Haiping Hao, Peter Burger, Fausto~J. Rodriguez, Michael Lim, Alfredo Qui{\~n}ones-Hinojosa, Gary~L. Gallia, Jeffrey~A. Tornheim, Michael~T. Melia, Cynthia~L. Sears, and Carlos~A. Pardo.
\newblock Next-generation sequencing in neuropathologic diagnosis of infections of the nervous system.
\newblock {\em Neurology - Neuroimmunology Neuroinflammation}, 2016.

\bibitem{gihawi2023major}
Abraham Gihawi, Yuchen Ge, Jennifer Lu, Daniela Puiu, Amanda Xu, Colin~S. Cooper, Daniel~S. Brewer, Mihaela Pertea, and Steven~L. Salzberg.
\newblock Major data analysis errors invalidate cancer microbiome findings.
\newblock {\em mBio}, 2023.

\bibitem{Ackelsberg2015}
Joel Ackelsberg, Jennifer Rakeman, Scott Hughes, Jeannine Petersen, Paul Mead, Martin Schriefer, Luke Kingry, Alex Hoffmaster, and Jay E. Gee.
\newblock {Lack of Evidence for Plague or Anthrax on the New York City Subway}.
\newblock {\em Cell Systems}, 2015.

\bibitem{centrifuge_db}
Daehwan Kim, Li~Song, Florian~P Breitwieser, and Steven~L Salzberg.
\newblock {Centrifuge}.
\newblock \url{http://www.ccb.jhu.edu/software/centrifuge/}, 2020.

\bibitem{lemane2022kmtricks}
Téo Lemane, Paul Medvedev, Rayan Chikhi, and Pierre Peterlongo.
\newblock {kmtricks: efficient and flexible construction of Bloom filters for large sequencing data collections}.
\newblock {\em Bioinformatics Advances}, 2022.

\bibitem{ghose2018enabling}
Saugata Ghose, Kevin Hsieh, Amirali Boroumand, Rachata Ausavarungnirun, and Onur Mutlu.
\newblock {Enabling the Adoption of Processing-in-memory: Challenges, Mechanisms, Future Research Directions}.
\newblock {\em arXiv}, 2018.

\bibitem{kokot2017kmc3}
Marek Kokot, Maciej Długosz, and Sebastian Deorowicz.
\newblock {KMC 3: counting and manipulating k-mer statistics}.
\newblock {\em Bioinformatics}, 2017.

\bibitem{samardzic2020bonsai}
Nikola Samardzic, Weikang Qiao, Vaibhav Aggarwal, Mau-Chung~Frank Chang, and Jason Cong.
\newblock {Bonsai: High-performance adaptive merge tree sorting}.
\newblock In {\em ISCA}, 2020.

\bibitem{qiao2022topsort}
Weikang Qiao, Licheng Guo, Zhenman Fang, Mau-Chung~Frank Chang, and Jason Cong.
\newblock {TopSort: A High-Performance Two-Phase Sorting Accelerator Optimized on HBM-based FPGAs}.
\newblock In {\em FCCM}, 2022.

\bibitem{jayaraman2022hypersort}
Soundarya Jayaraman, Bingyi Zhang, and Viktor Prasanna.
\newblock {Hypersort: High-performance Parallel Sorting on HBM-enabled FPGA}.
\newblock In {\em ICFPT}, 2022.

\bibitem{benoit2016multiple}
Ga{\"e}tan Benoit, Pierre Peterlongo, Mahendra Mariadassou, Erwan Drezen, Sophie Schbath, Dominique Lavenier, and Claire Lemaitre.
\newblock Multiple comparative metagenomics using multiset k-mer counting.
\newblock {\em PeerJ Computer Science}, 2016.

\bibitem{bovee2018finch}
Roderick Bovee and Nick Greenfield.
\newblock {Finch: a tool adding dynamic abundance filtering to genomic MinHashing}.
\newblock {\em The Journal of Open Source Software}, 2018.

\bibitem{liu2022cmash}
Shaopeng Liu and David Koslicki.
\newblock {CMash: fast, multi-resolution estimation of k-mer-based Jaccard and containment indices}.
\newblock {\em Bioinformatics}, 2022.

\bibitem{weging2021taxonomic}
Silvio Weging, Andreas Gogol-Döring, and Ivo Grosse.
\newblock {Taxonomic analysis of metagenomic data with kASA}.
\newblock {\em Nucleic Acids Research}, 2021.

\bibitem{kawaguchi1995flash}
Atsuo Kawaguchi, Shingo Nishioka, and Hiroshi Motoda.
\newblock {A Flash-Memory Based File System}.
\newblock In {\em USENIX ATC}, 1995.

\bibitem{Turnbaugh2007}
Peter~J. Turnbaugh, Ruth~E. Ley, Micah Hamady, Claire~M. Fraser-Liggett, Rob Knight, and Jeffrey~I. Gordon.
\newblock The human microbiome project.
\newblock {\em Nature}, 2007.

\bibitem{umcL65nm}
{United Microelectronics Corporation}.
\newblock {UMK65LSCLLMVBBL\_A - UMC 65 nm Low-K 1.2V/1.0V Low Leakage LVT Tapless Standard Cell Library, version A02}, 2008.

\bibitem{innovus}
{Cadence Design Systems, Inc.}
\newblock {Innovus Implementation System}.
\newblock \url{https://www.cadence.com/en_US/home/tools/digital-design-and-signoff/soc-implementation-and-floorplanning/innovus-implementation-system.html}.

\bibitem{mqsimsource}
Arash Tavakkol, Juan G{\'o}mez-Luna, Mohammad Sadrosadati, Saugata Ghose, and Onur Mutlu.
\newblock {MQSim Source Code}.
\newblock \url{https://github.com/CMU-SAFARI/MQSim}.

\bibitem{lpddr4}
Samsung.
\newblock {LPDDR4}.
\newblock \url{https://semiconductor.samsung.com/dram/lpddr/lpddr4/}.

\bibitem{sczyrba2017critical}
Alexander Sczyrba, Peter Hofmann, Peter Belmann, David Koslicki, Stefan Janssen, Johannes Dr{\"o}ge, Ivan Gregor, Stephan Majda, Jessika Fiedler, Eik Dahms, Andreas Bremges, Adrian Fritz, Ruben Garrido-Oter, Tue~Sparholt J{\o}rgensen, Nicole Shapiro, Philip~D. Blood, Alexey Gurevich, Yang Bai, Dmitrij Turaev, Matthew~Z. DeMaere, et~al.
\newblock {Critical Assessment of Metagenome Interpretation—A Benchmark of Metagenomics Software}.
\newblock {\em Nature Methods}, 2017.

\bibitem{samsung128GBDDR4}
Samsung.
\newblock {Samsung 8 GB DRAM DDR4 8GB PC3200 UB 1Rx16 Samsung}.
\newblock \url{https://semiconductor.samsung.com/dram/module/udimm/m378a1g44ab0-cwe/}.

\bibitem{samsung8GBDDR4}
Samsung.
\newblock {Samsung 128 GB DDR4 3200 LRDIMM ECC Registred}.
\newblock \url{https://semiconductor.samsung.com/dram/module/lrdimm/m386aag40am3-cwe/}.

\bibitem{minion21}
{Oxford Nanopore Technologies}.
\newblock {MinION Mk1B IT Requirements}.
\newblock \url{https://community.nanoporetech.com/requirements_documents/minion-it-reqs.pdf}, 2021.

\bibitem{metagraphaws}
{Karasikov, Mikhail and Mustafa, Harun and Danciu, Daniel and Zimmermann, Marc and Barber, Christopher and R{\"a}tsch, Gunnar and Kahles, Andr{\'e}}.
\newblock {MetaGraph Amazon AWS S3 Bucket}.
\newblock \url{https://metagraph.s3.amazonaws.com/index.html}, 2025.

\bibitem{Berger2019}
Bonnie Berger and Hyunghoon Cho.
\newblock Emerging technologies towards enhancing privacy in genomic data sharing.
\newblock {\em Genome Biology}, 2019.

\bibitem{Blanco-Miguez2023}
Aitor Blanco-M{\'i}guez, Francesco Beghini, Fabio Cumbo, Lauren~J. McIver, Kelsey~N. Thompson, Moreno Zolfo, Paolo Manghi, Leonard Dubois, Kun~D. Huang, Andrew~Maltez Thomas, William~A. Nickols, Gianmarco Piccinno, Elisa Piperni, Michal Pun{\v{c}}och{\'a}{\v{r}}, Mireia Valles-Colomer, Adrian Tett, Francesca Giordano, Richard Davies, Jonathan Wolf, Sarah~E. Berry, et~al.
\newblock {Extending and improving metagenomic taxonomic profiling with uncharacterized species using MetaPhlAn 4}.
\newblock {\em Nature Biotechnology}, 2023.

\bibitem{tavakkol2014design}
Arash Tavakkol, Mohammad Arjomand, and Hamid Sarbazi-Azad.
\newblock {Design for scalability in enterprise SSDs}.
\newblock In {\em PACT}, 2014.

\bibitem{wang2019project}
Xiaohao Wang, Yifan Yuan, You Zhou, Chance~C Coats, and Jian Huang.
\newblock {Project Almanac: A Time-traveling Solid-state Drive}.
\newblock In {\em EuroSys}, 2019.

\bibitem{merrikh2017high}
Farnood Merrikh-Bayat, Xinjie Guo, Michael Klachko, Mirko Prezioso, Konstantin~K Likharev, and Dmitri~B Strukov.
\newblock High-performance mixed-signal neurocomputing with nanoscale floating-gate memory cell arrays.
\newblock {\em IEEE TNNLS}, 2017.

\bibitem{Pibiri2024macdbg}
Giulio~Ermanno Pibiri, Jason Fan, and Rob Patro.
\newblock {Meta-colored Compacted de Bruijn Graphs}.
\newblock In {\em RECOMB}, 2024.

\bibitem{pibiri2021pthash}
Giulio~Ermanno Pibiri and Roberto Trani.
\newblock {PTHash: Revisiting FCH Minimal Perfect Hashing}.
\newblock In {\em SIGIR}, 2021.

\bibitem{22gf}
Global Foundries.
\newblock {22nm CMOS FD-SOI technology}.
\newblock \url{https://gf.com/technology-platforms/fdx-fd-soi/}.

\bibitem{luo2023ramulator}
Haocong Luo, Yahya~Can Tu{\u{g}}rul, F~Nisa Bostanc{\i}, Ataberk Olgun, A~Giray Ya{\u{g}}l{\i}k{\c{c}}{\i}, and Onur Mutlu.
\newblock {Ramulator 2.0: A Modern, Modular, and Extensible DRAM Simulator}.
\newblock {\em IEEE CAL}, 2023.

\bibitem{ramulator2source}
Haocong Luo, Yahya~Can Tu{\u{g}}rul, F~Nisa Bostanc{\i}, Ataberk Olgun, A~Giray Ya{\u{g}}l{\i}k{\c{c}}{\i}, and Onur Mutlu.
\newblock {Ramulator 2.0 Source Code}.
\newblock \url{https://github.com/CMU-SAFARI/ramulator2}.

\bibitem{PCIE5}
PCI-SIG.
\newblock {PCI Express Base Specification Revision 5.0, Version 1.0}.
\newblock \url{https://pcisig.com/PCIExpress/Specs/Base/_5.0_1.0}.

\bibitem{MinIONMk1CO}
Oxford~Nanopore Technologies.
\newblock {MinION Mk1CO}.
\newblock \url{https://nanoporetech.com/document/requirements/minion-mk1c-spec}, 2024.

\bibitem{palatnick2020igenomics}
Aspyn Palatnick, Bin Zhou, Elodie Ghedin, and Michael~C Schatz.
\newblock {iGenomics: Comprehensive DNA sequence analysis on your Smartphone}.
\newblock {\em GigaScience}, 2020.

\bibitem{Ballard2018}
Zachary~S. Ballard, Calvin Brown, and Aydogan Ozcan.
\newblock {Mobile Technologies for the Discovery, Analysis, and Engineering of the Global Microbiome}.
\newblock {\em ACS Nano}, 2018.

\bibitem{Oehler2023}
Josephine~B. Oehler, Helen Wright, Zornitza Stark, Andrew~J. Mallett, and Ulf Schmitz.
\newblock The application of long-read sequencing in clinical settings.
\newblock {\em Human Genomics}, 2023.

\bibitem{adler2015pigz}
Mark Adler.
\newblock {pigz: A parallel implementation of gzip for modern multi-processor, multi-core machines}.
\newblock {\em Jet Propulsion Laboratory}, 2015.

\bibitem{drost2017use}
Jarno Drost, Ruben van Boxtel, Francis Blokzijl, Tomohiro Mizutani, Nobuo Sasaki, Valentina Sasselli, Joep de~Ligt, Sam Behjati, Judith~E. Grolleman, Tom van Wezel, Serena Nik-Zainal, Roland~P. Kuiper, Edwin Cuppen, and Hans Clevers.
\newblock {Use of CRISPR-modified human stem cell organoids to study the origin of mutational signatures in cancer}.
\newblock {\em Science}, 2017.

\bibitem{weinstein2013cancer}
John~N Weinstein, Eric~A Collisson, Gordon~B Mills, Kenna~R Shaw, Brad~A Ozenberger, Kyle Ellrott, Ilya Shmulevich, Chris Sander, and Joshua~M Stuart.
\newblock The cancer genome atlas pan-cancer analysis project.
\newblock {\em Nature Genetics}, 2013.

\bibitem{watsa2020portable}
Mrinalini Watsa, Gideon~A. Erkenswick, Aaron Pomerantz, and Stefan Prost.
\newblock Portable sequencing as a teaching tool in conservation and biodiversity research.
\newblock {\em PLOS Biology}, 2020.

\bibitem{kim2025lazy}
Hyeunjoo Kim, Sanghun Oh, Jaeyong Lee, and Jihong Kim.
\newblock {Lazy Discharge: A High-Speed Energy-Efficient Read Technique for NAND Flash Memory}.
\newblock {\em IEEE Design \& Test}, 2025.

\bibitem{fukasawa2020longqc}
Yoshinori Fukasawa, Luca Ermini, Hai Wang, Karen Carty, and Min-Sin Cheung.
\newblock {LongQC: A Quality Control Tool for Third Generation Sequencing Long Read Data}.
\newblock {\em G3: Genes, Genomes, Genetics}, 2020.

\bibitem{Bourque2018}
Guillaume Bourque, Kathleen~H. Burns, Mary Gehring, Vera Gorbunova, Andrei Seluanov, Molly Hammell, Micha{\"e}l Imbeault, Zsuzsanna Izsv{\'a}k, Henry~L. Levin, Todd~S. Macfarlan, Dixie~L. Mager, and C{\'e}dric Feschotte.
\newblock Ten things you should know about transposable elements.
\newblock {\em Genome Biology}, 2018.

\bibitem{Tian2008}
Dacheng Tian, Qiang Wang, Pengfei Zhang, Hitoshi Araki, Sihai Yang, Martin Kreitman, Thomas Nagylaki, Richard Hudson, Joy Bergelson, and Jian-Qun Chen.
\newblock Single-nucleotide mutation rate increases close to insertions/deletions in eukaryotes.
\newblock {\em Nature}, 2008.

\bibitem{amos2013variation}
William Amos.
\newblock {Variation in Heterozygosity Predicts Variation in Human Substitution Rates between Populations, Individuals and Genomic Regions}.
\newblock {\em PLOS One}, 2013.

\bibitem{LaPierre2019}
Nathan LaPierre, Rob Egan, Wei Wang, and Zhong Wang.
\newblock De novo nanopore read quality improvement using deep learning.
\newblock {\em BMC Bioinformatics}, 2019.

\bibitem{Delahaye2021sequencing}
Clara Delahaye and Jacques Nicolas.
\newblock {Sequencing DNA with nanopores: Troubles and biases}.
\newblock {\em PLOS One}, 2021.

\bibitem{gleeson2021accurate}
Josie Gleeson, Adrien Leger, Yair D~J Prawer, Tracy~A Lane, Paul~J Harrison, Wilfried Haerty, and Michael~B Clark.
\newblock {Accurate expression quantification from nanopore direct RNA sequencing with NanoCount}.
\newblock {\em Nucleic Acids Research}, 2021.

\bibitem{huffman2007method}
David~A Huffman.
\newblock {A Method for the Construction of Minimum-Redundancy Codes}.
\newblock {\em Proceedings of the IRE}, 2007.

\bibitem{ono2020pbsim2}
Yukiteru Ono, Kiyoshi Asai, and Michiaki Hamada.
\newblock {PBSIM2: a simulator for long-read sequencers with a novel generative model of quality scores}.
\newblock {\em Bioinformatics}, 2020.

\bibitem{belyaeva2022best}
Anastasiya Belyaeva, Andrew Carroll, Daniel Cook, Daniel Liu, Kishwar Shafin, and Pi-Chuan Chang.
\newblock {Best: A Tool for Characterizing Sequencing Errors}.
\newblock {\em bioRxiv}, 2022.

\bibitem{magi2016characterization}
Alberto Magi, Betti Giusti, and Lorenzo Tattini.
\newblock {Characterization of MinION nanopore data for resequencing analyses}.
\newblock {\em Briefings in Bioinformatics}, 2016.

\bibitem{Garg2021}
Shilpa Garg, Arkarachai Fungtammasan, Andrew Carroll, Mike Chou, Anthony Schmitt, Xiang Zhou, Stephen Mac, Paul Peluso, Emily Hatas, Jay Ghurye, Jared Maguire, Medhat Mahmoud, Haoyu Cheng, David Heller, Justin~M. Zook, Tobias Moemke, Tobias Marschall, Fritz~J. Sedlazeck, John Aach, Chen-Shan Chin, et~al.
\newblock {Chromosome-scale, haplotype-resolved assembly of human genomes}.
\newblock {\em Nature Biotechnology}, 2021.

\bibitem{guan2016structural}
Peiyong Guan and Wing-Kin Sung.
\newblock {Structural variation detection using next-generation sequencing data: A comparative technical review}.
\newblock {\em Methods}, 2016.

\bibitem{glenn2011field}
Travis~C Glenn.
\newblock {Field Guide to Next-Generation DNA Sequencers}.
\newblock {\em Molecular Ecology Resources}, 2011.

\bibitem{kchouk2017generations}
Mehdi Kchouk, Jean-Francois Gibrat, and Mourad Elloumi.
\newblock {Generations of Sequencing Technologies: from First to Next Generation}.
\newblock {\em Biology and Medicine}, 2017.

\bibitem{pfeiffer2018systematic}
Franziska Pfeiffer, Carsten Gr{\"o}ber, Michael Blank, Kristian H{\"a}ndler, Marc Beyer, Joachim~L Schultze, and G{\"u}nter Mayer.
\newblock {Systematic Evaluation of Error Rates and Causes in Short Samples in Next-generation Sequencing}.
\newblock {\em Scientific Reports}, 2018.

\bibitem{crisan2021analyzing}
Diana Crișan, Alexandru Irimia, Dan Gota, Liviu Miclea, Adela Puscasiu, Ovidiu Stan, and Honoriu Valean.
\newblock {Analyzing Benford’s law’s powerful applications in image forensics}.
\newblock {\em Applied Sciences}, 2021.

\bibitem{pearson2013introduction}
William~R. Pearson.
\newblock {An Introduction to Sequence Similarity (“Homology”) Searching}.
\newblock {\em Current Protocols in Bioinformatics}, 2013.

\bibitem{joudaki2023aligning}
Amir Joudaki, Alexandru Meterez, Harun Mustafa, Ragnar Groot~Koerkamp, André Kahles, and Gunnar Rätsch.
\newblock Aligning distant sequences to graphs using long seed sketches.
\newblock {\em Genome Research}, 2023.

\bibitem{prasad2022evaluating}
Aparna Prasad, Eline~D Lorenzen, and Michael~V Westbury.
\newblock Evaluating the role of reference-genome phylogenetic distance on evolutionary inference.
\newblock {\em Molecular Ecology Resources}, 2022.

\bibitem{grebnov2011libbsc}
Ilya Grebnov.
\newblock libbsc: A high performance data compression library, 2011.

\bibitem{buffalo2021quantifying}
Vince Buffalo.
\newblock {Quantifying the relationship between genetic diversity and population size suggests natural selection cannot explain Lewontin’s Paradox}.
\newblock {\em eLife}, 2021.

\bibitem{cochrane2015international}
Guy Cochrane, Ilene Karsch-Mizrachi, Toshihisa Takagi, and International~Nucleotide Sequence Database Collaboration.
\newblock {The International Nucleotide Sequence Database Collaboration}.
\newblock {\em Nucleic Acids Research}, 2015.

\bibitem{fpgamidrange}
AMD/Xilinx.
\newblock {Kintex Ultrascale+ FPGA (KU15P)}.
\newblock \url{https://www.amd.com/en/products/adaptive-socs-and-fpgas/fpga/kintex-ultrascale-plus.html%7D%7D }.

\bibitem{Motamayor2013}
Juan~C. Motamayor, Keithanne Mockaitis, Jeremy Schmutz, Niina Haiminen, Donald~Livingstone III, Omar Cornejo, Seth~D. Findley, Ping Zheng, Filippo Utro, Stefan Royaert, Christopher Saski, Jerry Jenkins, Ram Podicheti, Meixia Zhao, Brian~E. Scheffler, Joseph~C. Stack, Frank~A. Feltus, Guiliana~M. Mustiga, Freddy Amores, Wilbert Phillips, et~al.
\newblock The genome sequence of the most widely cultivated cacao type and its use to identify candidate genes regulating pod color.
\newblock {\em Genome Biology}, 2013.

\bibitem{eberle2017reference}
Michael~A. Eberle, Epameinondas Fritzilas, Peter Krusche, Morten Källberg, Benjamin~L. Moore, Mitchell~A. Bekritsky, Zamin Iqbal, Han-Yu Chuang, Sean~J. Humphray, Aaron~L. Halpern, Semyon Kruglyak, Elliott~H. Margulies, Gil McVean, and David~R. Bentley.
\newblock A reference data set of 5.4 million phased human variants validated by genetic inheritance from sequencing a three-generation 17-member pedigree.
\newblock {\em Genome Research}, 2017.

\bibitem{ontopendata}
Andrea Talenti.
\newblock {Sequencing Genome in a Bottle samples}.
\newblock \url{https://labs.epi2me.io/giab-2023.05/}, 2023.

\bibitem{Belser2021}
Caroline Belser, Franc-Christophe Baurens, Benjamin Noel, Guillaume Martin, Corinne Cruaud, Benjamin Istace, Nabila Yahiaoui, Karine Labadie, Eva H{\v{r}}ibov{\'a}, Jaroslav Dole{\v{z}}el, Arnaud Lemainque, Patrick Wincker, Ang{\'e}lique D'Hont, and Jean-Marc Aury.
\newblock Telomere-to-telomere gapless chromosomes of banana using nanopore sequencing.
\newblock {\em Communications Biology}, 2021.

\bibitem{nvcomp}
NVIDIA.
\newblock {NVIDIA nvCOMP}.
\newblock \url{https://docs.nvidia.com/cuda/nvcomp/}, 2024.

\bibitem{nvidiaa100}
NVIDIA.
\newblock {NVIDIA A100 Tensor Core GPU}.
\newblock \url{https://www.nvidia.com/en-us/data-center/a100/}, 2021.

\bibitem{amdgzip}
AMD.
\newblock {AMD GZIP Compression \& Decompression}.
\newblock \url{https://www.amd.com/en/developer/resources/alveo-apps/amd-gzip-compression-decompression.html}.

\bibitem{amdalveou50}
AMD.
\newblock {AMD Alveo\texttrademark{} U50 Data Center Accelerator Card}.
\newblock \url{https://www.amd.com/en/products/accelerators/alveo/u50/a-u50-p00g-pq-g.html}, 2019.

\bibitem{choi2024hardware}
Junsun Choi.
\newblock A hardware accelerator generator for zstandard decompression.
\newblock Master's thesis, University of California, Berkeley, 2024.

\bibitem{nvidiateslav100}
NVIDIA.
\newblock {NVIDIA Tesla V100}.
\newblock \url{https://www.nvidia.com/en-gb/data-center/tesla-v100/}, 2017.

\bibitem{amdalveou200}
AMD.
\newblock {AMD Alveo\texttrademark{} U200 Data Center Accelerator Card (Active)}.
\newblock \url{https://www.amd.com/en/products/accelerators/alveo/u200/a-u200-a64g-pq-g.html}, 2018.

\bibitem{arm_cortex_r8_ip}
{ARM}.
\newblock {Cortex-R8 Product Support: Technical Specifications}.
\newblock \url{https://developer.arm.com/Processors/Cortex-R8}.

\bibitem{siliconmotion_sm2508_pb}
{Silicon Motion}.
\newblock {M2508 Superior Performance with Low Power PCIe Gen5 x4 NVMe 2.0 SSD Controller}.
\newblock Technical report, 2024.

\bibitem{bivins2020wastewater}
Aaron Bivins, Devin North, Arslan Ahmad, Warish Ahmed, Eric Alm, Frederic Been, Prosun Bhattacharya, Lubertus Bijlsma, Alexandria~B. Boehm, Joe Brown, Gianluigi Buttiglieri, Vincenza Calabro, Annalaura Carducci, Sara Castiglioni, Zeynep Cetecioglu~Gurol, Sudip Chakraborty, Federico Costa, Stefano Curcio, Francis~L. de~los Reyes, Jeseth Delgado~Vela, et~al.
\newblock {Wastewater-Based Epidemiology: Global Collaborative to Maximize Contributions in the Fight Against COVID-19}.
\newblock {\em Environmental Science \& Technology}, 2020.

\bibitem{turakhia2021ultrafast}
Yatish Turakhia, Bryan Thornlow, Angie~S Hinrichs, Nicola De~Maio, Landen Gozashti, Robert Lanfear, David Haussler, and Russell Corbett-Detig.
\newblock {Ultrafast Sample placement on Existing tRees (UShER) enables real-time phylogenetics for the SARS-CoV-2 pandemic}.
\newblock {\em Nature genetics}, 2021.

\bibitem{wang2009rna}
Zhong Wang, Mark Gerstein, and Michael Snyder.
\newblock {RNA-Seq: a revolutionary tool for transcriptomics}.
\newblock {\em Nature Reviews Genetics}, 2009.

\bibitem{lowe2017transcriptomics}
Rohan Lowe, Neil Shirley, Mark Bleackley, Stephen Dolan, and Thomas Shafee.
\newblock Transcriptomics technologies.
\newblock {\em PLOS Computational Biology}, 2017.

\bibitem{angerer2017single}
Philipp Angerer, Lukas Simon, Sophie Tritschler, F.~Alexander Wolf, David Fischer, and Fabian~J. Theis.
\newblock {Single cells make big data: New challenges and opportunities in transcriptomics}.
\newblock {\em Current Opinion in Systems Biology}, 2017.

\bibitem{Stark2019}
Rory Stark, Marta Grzelak, and James Hadfield.
\newblock {RNA sequencing: the teenage years}.
\newblock {\em Nature Reviews Genetics}, 2019.

\bibitem{chen2023hitchhikers}
Jiung-Wen Chen, Lisa Shrestha, George Green, André Leier, and Tatiana~T Marquez-Lago.
\newblock {The hitchhikers’ guide to RNA sequencing and functional analysis}.
\newblock {\em Briefings in Bioinformatics}, 2023.

\bibitem{weirather2017comprehensive}
Jason~L Weirather, Mariateresa de~Cesare, Yunhao Wang, Paolo Piazza, Vittorio Sebastiano, Xiu-Jie Wang, David Buck, and Kin~Fai Au.
\newblock {Comprehensive Comparison of Pacific Biosciences and Oxford Nanopore Technologies and Their Applications to Transcriptome Analysis}.
\newblock {\em F1000Research}, 2017.

\bibitem{Sibbesen2023}
Jonas~A. Sibbesen, Jordan~M. Eizenga, Adam~M. Novak, Jouni Sirén, Xian Chang, Erik Garrison, and Benedict Paten.
\newblock Haplotype-aware pantranscriptome analyses using spliced pangenome graphs.
\newblock {\em Nature Methods}, 2023.

\bibitem{haas_forensic_2021}
Cordula Haas, Jacqueline Neubauer, Andrea~Patrizia Salzmann, Erin Hanson, and Jack Ballantyne.
\newblock Forensic transcriptome analysis using massively parallel sequencing.
\newblock {\em Forensic Science International: Genetics}, 2021.

\bibitem{liu_desalt_2019}
Bo~Liu, Yadong Liu, Junyi Li, Hongzhe Guo, Tianyi Zang, and Yadong Wang.
\newblock {deSALT}: fast and accurate long transcriptomic read alignment with de {Bruijn} graph-based index.
\newblock {\em Genome Biology}, 2019.

\bibitem{lachmann2018massive}
Alexander Lachmann, Denis Torre, Alexandra~B. Keenan, Kathleen~M. Jagodnik, Hoyjin~J. Lee, Lily Wang, Moshe~C. Silverstein, and Avi Ma'ayan.
\newblock {Massive mining of publicly available RNA-seq data from human and mouse}.
\newblock {\em Nature Communications}, 2018.

\bibitem{clough2023ncbigeo}
Emily Clough, Tanya Barrett, Stephen~E Wilhite, Pierre Ledoux, Carlos Evangelista, Irene F Kim, Maxim Tomashevsky, Kimberly A Marshall, Katherine H Phillippy, Patti M Sherman, Hyeseung Lee, Naigong Zhang, Nadezhda Serova, Lukas Wagner, Vadim Zalunin, Andrey Kochergin, and Alexandra Soboleva.
\newblock {NCBI GEO: archive for gene expression and epigenomics data sets: 23-year update}.
\newblock {\em Nucleic Acids Research}, 2023.

\bibitem{bray2016near}
Nicolas~L Bray, Harold Pimentel, P{\'a}ll Melsted, and Lior Pachter.
\newblock {Near-optimal probabilistic RNA-seq quantification}.
\newblock {\em Nature biotechnology}, 2016.

\bibitem{tang2009mrnaseq}
Fuchou Tang, Catalin Barbacioru, Yangzhou Wang, Ellen Nordman, Clarence Lee, Nanlan Xu, Xiaohui Wang, John Bodeau, Brian~B. Tuch, Asim Siddiqui, Kaiqin Lao, and M.~Azim Surani.
\newblock {mRNA-Seq whole-transcriptome analysis of a single cell}.
\newblock {\em Nature Methods}, 2009.

\bibitem{lahnemann2020eleven}
David L{\"a}hnemann, Johannes K{\"o}ster, Ewa Szczurek, Davis~J. McCarthy, Stephanie~C. Hicks, Mark~D. Robinson, Catalina~A. Vallejos, Kieran~R. Campbell, Niko Beerenwinkel, Ahmed Mahfouz, Luca Pinello, Pavel Skums, Alexandros Stamatakis, Camille Stephan-Otto Attolini, Samuel Aparicio, Jasmijn Baaijens, Marleen Balvert, Buys~de Barbanson, Antonio Cappuccio, Giacomo Corleone, et~al.
\newblock Eleven grand challenges in single-cell data science.
\newblock {\em Genome Biology}, 2020.

\bibitem{wen2022singlecell}
Lu~Wen, Guoqiang Li, Tao Huang, Wei Geng, Hao Pei, Jialiang Yang, Miao Zhu, Pengfei Zhang, Rui Hou, Geng Tian, Wentao Su, Jian Chen, Dake Zhang, Pingan Zhu, Wei Zhang, Xiuxin Zhang, Ning Zhang, Yunlong Zhao, Xin Cao, Guangdun Peng, et~al.
\newblock {Single-cell technologies: From research to application}.
\newblock {\em The Innovation}, 2022.

\bibitem{trapnell2014dynamics}
Cole Trapnell, Davide Cacchiarelli, Jonna Grimsby, Prapti Pokharel, Shuqiang Li, Michael Morse, Niall~J. Lennon, Kenneth~J. Livak, Tarjei~S. Mikkelsen, and John~L. Rinn.
\newblock The dynamics and regulators of cell fate decisions are revealed by pseudotemporal ordering of single cells.
\newblock {\em Nature Biotechnology}, 2014.

\bibitem{wolf2018scanpy}
F.~Alexander Wolf, Philipp Angerer, and Fabian~J. Theis.
\newblock {SCANPY: large-scale single-cell gene expression data analysis}.
\newblock {\em Genome Biology}, 2018.

\bibitem{stuart2019comprehensive}
Tim Stuart, Andrew Butler, Paul Hoffman, Christoph Hafemeister, Efthymia Papalexi, William~M. Mauck~III, Yuhan Hao, Marlon Stoeckius, Peter Smibert, and Rahul Satija.
\newblock {Comprehensive Integration of Single-Cell Data}.
\newblock {\em Cell}, 2019.

\bibitem{hafemeister2019normalization}
Christoph Hafemeister and Rahul Satija.
\newblock {Normalization and variance stabilization of single-cell RNA-seq data using regularized negative binomial regression}.
\newblock {\em Genome Biology}, 2019.

\bibitem{Pattengale2020Decentralized}
Nicholas~D Pattengale and Corey~M Hudson.
\newblock Decentralized genomics audit logging via permissioned blockchain ledgering.
\newblock {\em BMC Medical Genomics}, 2020.

\bibitem{ma2020efficient}
Shuaicheng Ma, Yang Cao, and Li~Xiong.
\newblock Efficient logging and querying for blockchain-based cross-site genomic dataset access audit.
\newblock {\em BMC Medical Genomics}, 2020.

\bibitem{bonomi2020privacy}
Luca Bonomi, Yingxiang Huang, and Lucila Ohno-Machado.
\newblock Privacy challenges and research opportunities for genomic data sharing.
\newblock {\em Nature Genetics}, 2020.

\bibitem{akgun2015privacy}
Mete Akg{\"u}n, A~Osman Bayrak, Bugra Ozer, and M~{\c{S}}amil Sa{\u{g}}{\i}ro{\u{g}}lu.
\newblock Privacy preserving processing of genomic data: A survey.
\newblock {\em Journal of Biomedical Informatics}, 2015.

\bibitem{esbin2020overcoming}
Meagan~N. Esbin, Oscar~N. Whitney, Shasha Chong, Anna Maurer, Xavier Darzacq, and Robert Tjian.
\newblock {Overcoming the bottleneck to widespread testing: a rapid review of nucleic acid testing approaches for COVID-19 detection}.
\newblock {\em RNA}, 2020.

\bibitem{ghiasi2026enabling}
Nika~Mansouri Ghiasi and Onur Mutlu.
\newblock {Enabling Fast, Efficient, and Low-Cost Genomic and Metagenomic Analyses via Storage-Centric System Designs}.
\newblock In {\em ICS Workshops}, 2026.

\bibitem{ghiasi2026enablingarxiv}
Nika~Mansouri Ghiasi and Onur Mutlu.
\newblock {Enabling Fast, Efficient, and Low-Cost Genomic and Metagenomic Analyses via Storage-Centric System Designs}.
\newblock In {\em arXiv}, 2026.

\bibitem{firtina_blend_2023}
Can Firtina, Jisung Park, Mohammed Alser, Jeremie~S Kim, Damla~Senol Cali, Taha Shahroodi, Nika~Mansouri Ghiasi, Gagandeep Singh, Konstantinos Kanellopoulos, Can Alkan, and Onur Mutlu.
\newblock {BLEND: A Fast, Memory-Efficient, and Accurate Mechanism to Find Fuzzy Seed Matches in Genome Analysis}.
\newblock {\em NAR Genomics and Bioinformatics}, 2023.

\bibitem{lindegger2023rawalign}
Jo{\"e}l Lindegger, Can Firtina, Nika~Mansouri Ghiasi, Mohammad Sadrosadati, Mohammed Alser, and Onur Mutlu.
\newblock Rawalign: Accurate, fast, and scalable raw nanopore signal mapping via combining seeding and alignment.
\newblock {\em arXiv preprint arXiv:2310.05037}, 2023.

\bibitem{liu2025analysis}
Shaopeng Liu, Judith~S. Rodriguez, Viorel Munteanu, Cynthia Ronkowski, Nitesh~Kumar Sharma, Mohammed Alser, Francesco Andreace, Ran Blekhman, Dagmara B{\l}aszczyk, Rayan Chikhi, Keith~A. Crandall, Katja Della~Libera, Dallace Francis, Alina Frolova, Abigail~Shahar Gancz, Naomi~E. Huntley, Pooja Jaiswal, Tomasz Kosciolek, Pawel~P. {\L}abaj, Wojciech {\L}abaj, et~al.
\newblock {Analysis of metagenomic data}.
\newblock {\em Nature Reviews Methods Primers}, 2025.

\bibitem{sharma2026towards}
Gaurav Sharma, Viorel Munteanu, Nika~Mansouri Ghiasi, Utkarsha Mahanta, Jineta Banerjee, Susheel Varma, Luca Foschini, Kyle Ellrott, Onur Mutlu, Dumitru Ciorb{\u{a}}, Roel~A. Ophoff, Viorel Bostan, Jason~H. Moore, Despoina Sousoni, Arunkumar Krishnan, Alexander~G. Lucaci, Alba Tull, Christopher~E. Mason, Mihai Dimian, Gustavo Stolovitzky, et~al.
\newblock {Towards a decentralized future for open-science databases}.
\newblock {\em Nature Genetics}, 2026.

\bibitem{ghiasi2026architecture}
Nika~Mansouri Ghiasi, Konstantina Koliogeorgi, and Onur Mutlu.
\newblock {Architecture for Health Initiative (Arch4Health): Computational Challenges in Health-Related Applications and the Role of Computer Architecture in Addressing Them}.
\newblock In {\em ICS Workshops}, 2026.

\bibitem{ghiasi2026architecturearxiv}
Nika~Mansouri Ghiasi, Konstantina Koliogeorgi, and Onur Mutlu.
\newblock {Architecture for Health Initiative (Arch4Health): Computational Challenges in Health-Related Applications and the Role of Computer Architecture in Addressing Them}.
\newblock In {\em arXiv}, 2026.

\bibitem{dai2022advances}
Xiaofeng Dai and Li~Shen.
\newblock {Advances and Trends in Omics Technology Development}.
\newblock {\em Frontiers in Medicine}, 2022.

\bibitem{mallick2010proteomics}
Parag Mallick and Bernhard Kuster.
\newblock Proteomics: a pragmatic perspective.
\newblock {\em Nature Biotechnology}, 2010.

\bibitem{aslam2016proteomics}
Bilal Aslam, Madiha Basit, Muhammad~Atif Nisar, Mohsin Khurshid, and Muhammad~Hidayat Rasool.
\newblock {Proteomics: Technologies and Their Applications}.
\newblock {\em Journal of Chromatographic Science}, 2017.

\bibitem{cho2007proteomics}
William~CS Cho.
\newblock Proteomics technologies and challenges.
\newblock {\em Genomics, Proteomics \& Bioinformatics}, 2007.

\bibitem{patterson2003proteomics}
Scott~D. Patterson and Ruedi~H. Aebersold.
\newblock {Proteomics: the first decade and beyond}.
\newblock {\em Nature Genetics}, 2003.

\bibitem{graves2002molecular}
Paul~R. Graves and Timothy A.~J. Haystead.
\newblock {Molecular Biologist's Guide to Proteomics}.
\newblock {\em Microbiology and Molecular Biology Reviews}, 2002.

\bibitem{pandey2000proteomics}
Akhilesh Pandey and Matthias Mann.
\newblock {Proteomics to study genes and genomes}.
\newblock {\em Nature}, 2000.

\bibitem{Han2025lightnobel}
Seunghee Han, Soongyu Choi, and Joo-Young Kim.
\newblock {LightNobel: Improving Sequence Length Limitation in Protein Structure Prediction Model via Adaptive Activation Quantization}.
\newblock In {\em ISCA}, 2025.

\bibitem{alseekh2021mass}
Saleh Alseekh, Asaph Aharoni, Yariv Brotman, K{\'e}vin Contrepois, John D'Auria, Jan Ewald, Jennifer C.~Ewald, Paul~D. Fraser, Patrick Giavalisco, Robert~D. Hall, Matthias Heinemann, Hannes Link, Jie Luo, Steffen Neumann, Jens Nielsen, Leonardo Perez~de Souza, Kazuki Saito, Uwe Sauer, Frank~C. Schroeder, Stefan Schuster, et~al.
\newblock {Mass spectrometry-based metabolomics: a guide for annotation, quantification and best reporting practices}.
\newblock {\em Nature Methods}, 2021.

\bibitem{perez2019quantifying}
Yasset Perez-Riverol, Andrey Zorin, Gaurhari Dass, Manh-Tu Vu, Pan Xu, Mihai Glont, Juan~Antonio Vizca{\'i}no, Andrew~F. Jarnuczak, Robert Petryszak, Peipei Ping, and Henning Hermjakob.
\newblock {Quantifying the impact of public omics data}.
\newblock {\em Nature Communications}, 2019.

\bibitem{liu2017metabolomics}
Xiaojing Liu and Jason~W. Locasale.
\newblock {Metabolomics: A Primer}.
\newblock {\em Trends in Biochemical Sciences}, 2017.

\bibitem{zhang2012modern}
Aihua Zhang, Hui Sun, Ping Wang, Ying Han, and Xijun Wang.
\newblock {Modern analytical techniques in metabolomics analysis}.
\newblock {\em The Analyst}, 2012.

\bibitem{johnson2012challenges}
Caroline~H. Johnson and Frank~J. Gonzalez.
\newblock {Challenges and opportunities of metabolomics}.
\newblock {\em Journal of Cellular Physiology}, 2012.

\bibitem{marti2010multimodality}
Luis Mart{\'\i}-Bonmat{\'\i}, Ram{\'o}n Sopena, Paula Bartumeus, and Pablo Sopena.
\newblock Multimodality imaging techniques.
\newblock {\em Contrast Media \& Molecular Imaging}, 2010.

\bibitem{kasban2015comparative}
Hany Kasban, Mohsen A.~M. El-Bendary, and Dina~H. Salama.
\newblock {A Comparative Study of Medical Imaging Techniques}.
\newblock {\em International Journal of Information Science and Intelligent System}, 2015.

\bibitem{shung2012principles}
K.~Kirk Shung, Michael~B. Smith, and Benjamin M.~W. Tsui.
\newblock {\em {Principles of Medical Imaging}}.
\newblock Academic Press, 2012.

\bibitem{glover2011overview}
Gary~H Glover.
\newblock Overview of functional magnetic resonance imaging.
\newblock {\em Neurosurgery Clinics of North America}, 2011.

\bibitem{kapoor2004introduction}
Vibhu Kapoor, Barry~M McCook, and Frank~S Torok.
\newblock {An introduction to PET-CT imaging}.
\newblock {\em Radiographics}, 2004.

\bibitem{townsend2004physical}
David~W. Townsend.
\newblock {Physical Principles and Technology of Clinical PET Imaging}.
\newblock {\em Annals of the Academy of Medicine, Singapore}, 2004.

\bibitem{pang2013recent}
Changhyun Pang, Chanseok Lee, and Kahp-Yang Suh.
\newblock Recent advances in flexible sensors for wearable and implantable devices.
\newblock {\em Journal of Applied Polymer Science}, 2013.

\bibitem{koydemir2018wearable}
Hatice~Ceylan Koydemir and Aydogan Ozcan.
\newblock Wearable and implantable sensors for biomedical applications.
\newblock {\em Annual Review of Analytical Chemistry}, 2018.

\bibitem{chan2012smart}
Marie Chan, Daniel Est{\`e}ve, Jean-Yves Fourniols, Christophe Escriba, and Eric Campo.
\newblock {Smart wearable systems: Current status and future challenges}.
\newblock {\em Artificial Intelligence in Medicine}, 2012.

\bibitem{lukowicz2004wearable}
Paul Lukowicz, Tunde Kirstein, and Gerhard Tr{\"o}ster.
\newblock {Wearable Systems for Health Care Applications}.
\newblock {\em Methods of Information in Medicine}, 2004.

\bibitem{bonato2010wearable}
Paolo Bonato.
\newblock {Wearable Sensors and Systems}.
\newblock {\em IEMBDE}, 2010.

\bibitem{sopic2018real}
Dionisije Sopic, Amin Aminifar, Amir Aminifar, and David Atienza.
\newblock Real-time event-driven classification technique for early detection and prevention of myocardial infarction on wearable systems.
\newblock {\em IEEE TBioCAS}, 2018.

\bibitem{guk2019evolution}
Kyeonghye Guk, Gaon Han, Jaewoo Lim, Keunwon Jeong, Taejoon Kang, Eun-Kyung Lim, and Juyeon Jung.
\newblock Evolution of wearable devices with real-time disease monitoring for personalized healthcare.
\newblock {\em Nanomaterials}, 2019.

\bibitem{tyler2020real}
Jonathan Tyler, Sung~Won Choi, and Muneesh Tewari.
\newblock Real-time, personalized medicine through wearable sensors and dynamic predictive modeling: a new paradigm for clinical medicine.
\newblock {\em Current Opinion in Systems Biology}, 2020.

\bibitem{kakria2015real}
Priyanka Kakria, NK~Tripathi, and Peerapong Kitipawang.
\newblock A real-time health monitoring system for remote cardiac patients using smartphone and wearable sensors.
\newblock {\em International Journal of Telemedicine and Applications}, 2015.

\bibitem{yelick2020}
Katherine Yelick, Aydın Buluç, Muaaz Awan, Ariful Azad, Benjamin Brock, Rob Egan, Saliya Ekanayake, Marquita Ellis, Evangelos Georganas, Giulia Guidi, Steven Hofmeyr, Oguz Selvitopi, Cristina Teodoropol, and Leonid Oliker.
\newblock {The Parallelism Motifs of Genomic Data Analysis}.
\newblock {\em Philosophical Transactions of the Royal Society A: Mathematical, Physical and Engineering Sciences}, 2020.

\bibitem{langarita2022}
Ruben Langarita, Adria Armejach, Javier Setoain, Pablo Ibanez-Marin, Jesus Alastruey-Benede, and Miquel Moreto.
\newblock { Compressed Sparse FM-Index: Fast Sequence Alignment Using Large K-Steps }.
\newblock {\em IEEE/ACM TCBB}, 2022.

\bibitem{robinson2021}
Tony Robinson, Jim Harkin, and Priyank Shukla.
\newblock {Hardware Acceleration of Genomics Data Analysis: Challenges and Opportunities}.
\newblock {\em Bioinformatics}, 2021.

\bibitem{li2022graph}
Michelle~M. Li, Kexin Huang, and Marinka Zitnik.
\newblock {Graph Representation Learning in Biomedicine and Healthcare}.
\newblock {\em Nature Biomedical Engineering}, 2022.

\bibitem{yi2022graph}
Hai-Cheng Yi, Zhu-Hong You, De-Shuang Huang, and Chee~Keong Kwoh.
\newblock {Graph Representation Learning in Bioinformatics: Trends, Methods and Applications}.
\newblock {\em Briefings in Bioinformatics}, 2022.

\bibitem{yuan2025ml}
Han Yuan.
\newblock {Overcoming Computational Resource Limitations in Deep Learning for Healthcare: Strategies Targeting Data, Model, and Computing}.
\newblock {\em Medicine Advances}, 2025.

\bibitem{khan2020}
Kamil Khan, Sudeep Pasricha, and Ryan~Gary Kim.
\newblock A survey of resource management for processing-in-memory and near-memory processing architectures.
\newblock {\em Journal of Low Power Electronics and Applications}, 2020.

\bibitem{gholami2024}
Amir Gholami, Zhewei Yao, Sehoon Kim, Coleman Hooper, Michael~W. Mahoney, and Kurt Keutzer.
\newblock {AI and Memory Wall}.
\newblock {\em IEEE Micro}, 2024.

\bibitem{rieke2020future}
Nicola Rieke, Jonny Hancox, Wenqi Li, Fausto Milletar{\`i}, Holger~R. Roth, Shadi Albarqouni, Spyridon Bakas, Mathieu~N. Galtier, Bennett~A. Landman, Klaus Maier-Hein, S{\'e}bastien Ourselin, Micah Sheller, Ronald~M. Summers, Andrew Trask, Daguang Xu, Maximilian Baust, and M.~Jorge Cardoso.
\newblock {The Future of Digital Health with Federated Learning}.
\newblock {\em npj Digital Medicine}, 2020.

\bibitem{xu2021federated}
Jie Xu, Benjamin~S. Glicksberg, Chang Su, Peter Walker, Jiang Bian, and Fei Wang.
\newblock {Federated Learning for Healthcare Informatics}.
\newblock {\em Journal of Healthcare Informatics Research}, 2021.

\bibitem{kaissis2020secure}
Georgios~A. Kaissis, Marcus~R. Makowski, Daniel R{\"u}ckert, and Rickmer~F. Braren.
\newblock {Secure, Privacy-Preserving and Federated Machine Learning in Medical Imaging}.
\newblock {\em Nature Machine Intelligence}, 2020.

\bibitem{sheller2020federated}
Micah~J. Sheller, Brandon Edwards, G.~Anthony Reina, Jason Martin, Sarthak Pati, Aikaterini Kotrotsou, Mikhail Milchenko, Walter Xu, Daniel Marcus, Rivka~R. Colen, and Spyridon Bakas.
\newblock {Federated Learning in Medicine: Facilitating Multi-Institutional Collaborations Without Sharing Patient Data}.
\newblock {\em Scientific Reports}, 2020.

\bibitem{zhang2025survey}
Bonan Zhang, Chao Chen, Ickjai Lee, Kyungmi Lee, and Kok-Leong Ong.
\newblock {A Survey on Security and Privacy Issues in Wearable Health Monitoring Devices}.
\newblock {\em Computers \& Security}, 2025.

\bibitem{kumar2025priv}
K.A.~Sathish Kumar, Leema Nelson, and Betshrine~Rachel Jibinsingh.
\newblock {Systematic Review of Privacy-Preserving Federated Learning in Decentralized Healthcare Systems}.
\newblock {\em Franklin Open}, 2025.

\bibitem{manber1993suffix}
Udi Manber and Gene Myers.
\newblock {Suffix Arrays: A New Method for On-Line String Searches}.
\newblock {\em SIAM Journal on Computing}, 1993.

\bibitem{breitwieser2022biodynamo}
Lukas Breitwieser, Ahmad Hesam, Jean de~Montigny, Vasileios Vavourakis, Alexandros Iosif, Jack Jennings, Marcus Kaiser, Marco Manca, Alberto Di~Meglio, Zaid Al-Ars, Fons Rademakers, Onur Mutlu, and Roman Bauer.
\newblock {BioDynaMo: a modular platform for high-performance agent-based simulation}.
\newblock {\em Bioinformatics}, 2022.

\bibitem{breitwieser2025design}
Lukas~Johannes Breitwieser.
\newblock {\em {Design and Analysis of an Extreme-Scale, High-Performance, and Modular Agent-Based Simulation Platform}}.
\newblock PhD thesis, ETH Zurich, 2025.

\bibitem{breitwieser2023high}
Lukas Breitwieser, Ahmad Hesam, Fons Rademakers, Juan G{\'o}mez-Luna, and Onur Mutlu.
\newblock {High-Performance and Scalable Agent-Based Simulation with BioDynaMo}.
\newblock In {\em PPoPP}, 2023.

\bibitem{breitwieser2025teraagent}
Lukas Breitwieser, Ahmad Hesam, A.~Giray Ya{\u{g}}l{\i}k{\c{c}}{\i}, Mohammad Sadrosadati, Fons Rademakers, and Onur Mutlu.
\newblock {TeraAgent: A Distributed Agent-Based Simulation Engine for Simulating Half a Trillion Agents}.
\newblock {\em arXiv preprint arXiv:2509.24063}, 2025.

\bibitem{koliogeorgi2022gandafl}
Konstantina Koliogeorgi, Sotirios Xydis, Georgi Gaydadjiev, and Dimitrios Soudris.
\newblock {GANDAFL: Dataflow Acceleration for Short Read Alignment on NGS Data}.
\newblock {\em IEEE TC}, 2022.

\bibitem{koliogeorgi2023profile}
Konstantina Koliogeorgi, Dimitrios Soudris, and Sotirios Xydis.
\newblock {Profile-Driven Banded Smith-Waterman acceleration for Short Read Alignment}.
\newblock In {\em DAC}, 2023.

\bibitem{koliogeorgi2019dataflow}
Konstantina Koliogeorgi, Nils Voss, Sotiria Fytraki, Sotirios Xydis, Georgi Gaydadjiev, and Dimitrios Soudris.
\newblock {Dataflow Acceleration of Smith-Waterman with Traceback for High Throughput Next Generation Sequencing}.
\newblock In {\em FPL}, 2019.

\bibitem{tsoutsouras2017exploration}
Vasileios Tsoutsouras, Konstantina Koliogeorgi, Sotirios Xydis, and Dimitrios Soudris.
\newblock {An Exploration Framework for Efficient High-Level Synthesis of Support Vector Machines: Case Study on ECG Arrhythmia Detection for Xilinx Zynq SoC}.
\newblock {\em Journal of Signal Processing Systems}, 2017.

\bibitem{diab2022high}
Safaa Diab, Amir Nassereldine, Mohammed Alser, Juan G{\'o}mez-Luna, Onur Mutlu, and Izzat~El Hajj.
\newblock {High-Throughput Pairwise Alignment with the Wavefront Algorithm Using Processing-in-Memory}.
\newblock arXiv:2204.02085 [cs.AR], 2022.

\bibitem{thesis}
Nika Mansouri~Ghiasi.
\newblock {\em {Storage-Centric System Designs for Enabling Fast, Efficient, and Low-Cost Genomic and Metagenomic Analyses}}.
\newblock PhD thesis, ETH Zurich, 2026.

\bibitem{onurmutlulectures}
{Onur Mutlu Lectures}.
\newblock {Onur Mutlu Lectures}.
\newblock \url{https://www.youtube.com/@OnurMutluLectures}.
\newblock YouTube Channel.

\bibitem{firsta4h}
{The First Workshop on Architecture for Health (Arch4Health)}.
\newblock YouTube Livestream, 2025.
\newblock \url{https://www.youtube.com/watch?v=lTc_gQzFNJI}.

\bibitem{seconda4h}
{The Second Workshop on Architecture for Health (Arch4Health)}.
\newblock YouTube Livestream, 2026.
\newblock \url{https://www.youtube.com/watch?v=hSRSZCjnKVo&t}.

\bibitem{arch4health-micro2025}
{The First Workshop on Architecture for Health (Arch4Health)}.
\newblock In conjunction with the 58th IEEE/ACM International Symposium on Microarchitecture (MICRO), 2025.
\newblock \url{https://events.safari.ethz.ch/micro25-arch4health/}.

\bibitem{arch4health-hpca2026}
{The Second Workshop on Architecture for Health (Arch4Health)}.
\newblock In conjunction with the IEEE International Symposium on High-Performance Computer Architecture (HPCA), 2026.
\newblock \url{https://events.safari.ethz.ch/hpca26-arch4health/}.

\bibitem{arch4health-ics2026}
{The Third Workshop on Architecture for Health (Arch4Health)}.
\newblock In conjunction with the ACM International Conference on Supercomputing 2026 (ICS), 2026.
\newblock \url{https://events.safari.ethz.ch/ics26-arch4health/}.

\bibitem{sys4health-sosp2026}
{The First Workshop on Systems for Health (Sys4Health)}.
\newblock In conjunction with the 32nd Symposium on Operating Systems Principles (SOSP), 2026.
\newblock \url{https://events.safari.ethz.ch/sosp26-sys4health/}.

\bibitem{mansourighiasi2023alp}
Nika~Mansouri Ghiasi, Nandita Vijaykumar, Geraldo~F. Oliveira, Lois Orosa, Ivan Fernandez, Mohammad Sadrosadati, Konstantinos Kanellopoulos, Nastaran Hajinazar, Juan~Gómez Luna, and Onur Mutlu.
\newblock {ALP: Alleviating CPU-Memory Data Movement Overheads in Memory-Centric Systems}.
\newblock {\em IEEE TETC}, 2023.

\bibitem{mansourighiasi2026revamp3d}
Nika~Mansouri Ghiasi, Mohammad Sadrosadati, Geraldo~F. Oliveira, Konstantinos Kanellopoulos, Rachata Ausavarungnirun, Juan~Gómez Luna, João Ferreira, Jeremie~S. Kim, Christina Giannoula, Nandita Vijaykumar, Jisung Park, and Onur Mutlu.
\newblock {RevaMp3D: Architecting the Processor Core and Cache Hierarchy for Systems with Monolithically-Integrated Logic and Memory}.
\newblock {\em TACO}, 2026.

\bibitem{fernandez2024matsa}
Ivan Fernandez, Christina Giannoula, Aditya Manglik, Ricardo Quislant, Nika~Mansouri Ghiasi, Juan G{\'o}mez-Luna, Eladio Gutierrez, Oscar Plata, and Onur Mutlu.
\newblock {MATSA: An MRAM-Based Energy-Efficient Accelerator for Time Series Analysis}.
\newblock {\em IEEE Access}, 2024.

\bibitem{wang2020figaro}
Yaohua Wang, Lois Orosa, Xiangjun Peng, Yang Guo, Saugata Ghose, Minesh Patel, Jeremie~S. Kim, Juan~Gómez Luna, Mohammad Sadrosadati, Nika~Mansouri Ghiasi, and Onur Mutlu.
\newblock {FIGARO: Improving System Performance via Fine-Grained In-DRAM Data Relocation and Caching}.
\newblock In {\em MICRO}, 2020.

\bibitem{wang2018reducing}
Yaohua Wang, Arash Tavakkol, Lois Orosa, Saugata Ghose, Nika Mansouri~Ghiasi, Minesh Patel, Jeremie~S. Kim, Hasan Hassan, Mohammad Sadrosadati, and Onur Mutlu.
\newblock {Reducing DRAM Latency via Charge-Level-Aware Look-Ahead Partial Restoration}.
\newblock In {\em MICRO}, 2018.

\bibitem{orosa_codic_2021}
L.~Orosa, Y.~Wang, M.~Sadrosadati, J.~S. Kim, M.~Patel, I.~Puddu, H.~Luo, K.~Razavi, J.~Gomez-Luna, H.~Hassan, N.~Mansouri-Ghiasi, S.~Ghose, and O.~Mutlu.
\newblock {CODIC}: {A} {Low}-{Cost} {Substrate} for {Enabling} {Custom} {In}-{DRAM} {Functionalities} and {Optimizations}.
\newblock In {\em {ISCA}}, 2021.

\bibitem{hassan2019crow}
Hasan Hassan, Minesh Patel, Jeremie~S. Kim, A.~Giray Yaglikci, Nandita Vijaykumar, Nika~Mansouri Ghiasi, Saugata Ghose, and Onur Mutlu.
\newblock {CROW: A Low-Cost Substrate for ImprovingDRAM Performance, Energy Efficiency, and Reliability}.
\newblock In {\em ISCA}, 2019.

\bibitem{liang2025ariadne}
Yu~Liang, Aofeng Shen, Chun~Jason Xue, Riwei Pan, Haiyu Mao, Nika~Mansouri Ghiasi, Qingcai Jiang, Rakesh Nadig, Lei Li, Rachata Ausavarungnirun, Mohammad Sadrosadati, and Onur Mutlu.
\newblock {Ariadne: A Hotness-Aware and Size-Adaptive Compressed Swap Technique for Fast Application Relaunch and Reduced CPU Usage on Mobile Devices}.
\newblock In {\em HPCA}, 2025.

\bibitem{kanellopoulos2019smash}
Konstantinos Kanellopoulos, Nandita Vijaykumar, Christina Giannoula, Roknoddin Azizi, Skanda Koppula, Nika Mansouri~Ghiasi, Taha Shahroodi, Juan Gomez-Luna, and Onur Mutlu.
\newblock {SMASH: Co-Designing Software Compression and Hardware-Accelerated Indexing for Efficient Sparse Matrix Operations}.
\newblock In {\em MICRO}, 2019.

\end{thebibliography}
\end{singlespace}

\bookmarksetup{startatroot}
\end{document}